How stars drive seismicity on planets and moons

# Global coupling mechanism of Sun resonant forcing of Mars, Moon, and Earth seismicity

M. Omerbashich[1*]

[1)] Geophysics Online, 3501 Jack Northrop Ave, Ste. 6172, Los Angeles CA 90250

Global seismicity on all three solar system bodies with in situ measurements (Earth, Moon, and Mars) is mainly due to the mechanical Rieger resonance (RR) of macroscopic flapping of the solar wind, driven by the well-known $P_{Rg}$=~154-day Rieger period and commonly detected in most heliophysical data types and the interplanetary magnetic field (IMF). Thus, InSight mission marsquakes rates are periodic with $P_{Rg}$ as characterized by a very high (≫12) fidelity $\Phi=2.8\cdot10^6$ and by being the only ≥99%-significant spectral peak in the 385.8–64.3-nHz (1–180-day) band of highest planetary energies; the longest-span (v.9) release of raw data revealed the entire RR, excluding a tectonically active Mars. To check this, I analyzed the rates of the October 2015–February 2019, $M_w$5.6+ earthquakes, and all (1969–1977) Apollo program moonquakes. To decouple the magnetospheric and IMF effects, I analyzed the Earth and Moon seismicity during the traversals of the Earth's magnetotail vs. IMF. The analysis showed with ≥99–67% confidence and Φ≫12 fidelity that (an unspecified majority of) moonquakes and $M_w$5.6+ earthquakes also recur at RR periods. Approximately half of the spectral peaks split but also into clusters that average into the usual Rieger periodicities, where magnetotail reconnecting clears the signal. Moonquakes are mostly forced at times of solar-wind resonance and not just during tides, as previously and simplistically believed. There is no significant dependence of sun-driven seismicity recurrence on solar cycles. Earlier claims that solar plasma dynamics could be seismogenic due to electrical surging or magnetohydrodynamic interactions between magnetically trapped plasma and water molecules embedded within solid matter or for reasons unknown are corroborated. This first conclusive recovery of the global coupling mechanism of solar-planetary seismogenesis calls for a reinterpretation of the seismicity phenomenon and reliance on global seismic magnitude scales. The predictability of solar-wind macroscopic dynamics is now within reach, which paves the way for long-term, physics-based seismic and space weather prediction and the safety of space missions. Gauss–Vaníček Spectral Analysis revolutionizes geophysics by computing nonlinear global dynamics directly (renders approximating of dynamics obsolete).

HIGHLIGHTS

- Mars, Moon, & Earth seismicity is due primarily to macroscopic dynamics of the solar wind at Rieger resonance; affected little by solar cycles
- Seismicity phenomenon is also astrophysical in origin instead of being exclusively geophysical as believed by some
- Marsquakes are practically exclusively forced by the solar wind, confirming a generally held view that Mars is tectonically inactive
- Moonquakes are mostly forced at the solar wind's resonance times and not just during tides, which was a simplistic view
- The connection between geomagnetism (aurorae activity) and strong seismicity, known for over a century, is explained for the first time
- First application of rigorous Gauss–Vaníček Spectral Analysis (GVSA) by least squares in planetary seismology & space physics
- GVSA revolutionizes seismology & geophysics by computing nonlinear global dynamics directly (makes approximating of dynamics obsolete).

## 1. Introduction

While the solar-wind speed does not appear to be a (direct) triggering mechanism of earthquakes as based on a statistical analysis by Love and Thomas (2013), the solar-wind dynamic at the interface with the Earth's magnetosphere, where it causes aurorae, is associated with terrestrial seismicity recordings (Ringler et al., 2020) (Tape et al., 2020). At the same time, plate tectonics — invoked commonly to explain the global seismicity phenomenon — is not the ultimate Earth theory (Jacoby, 2001), and classical approaches based on the heat transfer geophysical hypothesis and from it deriving mantle convection models

[*)] Correspondence to: omerbashich@geophysics.online, hm@royalfamily.ba.

failed to help us understand why the Earth has plate tectonics in the first place (Stevenson, 2008). From the viewpoint of fluid dynamics, for example, mantle convection is a simple phenomenon: a typical velocity of this convective flow is only a few cm/yr, while kinetic energy is comparable to the kinetic energy of a car running on a freeway, and therefore negligibly small compared with the rate of total potential energy release (Ogawa, 2007). Besides defying basic logic thus, mantle convection models contradict crucial data known for a long time, like the uppermost mantle's very-high $M_w$6+ velocities of over 8.5 km/s (Yegorkin and Chernyshov, 1983). In simplest terms, mantle convection models by design break down if one can demonstrate a coupling between continents and flows in the mantle — for example, a widely held model by Richter and McKenzie (1977), whose authors admit this explicitly even themselves.

To demonstrate one such continent-mantle global coupling mechanism, Omerbashich (2020a) approached the problem from the mechanical resonance perspective instead, attempting to extract Earth's body (mechanical) superharmonic resonance from global tectonic earthquake occurrences. That approach resulted in the first successful extraction of the globally controlling mechanism for initiating and triggering seismic events and sequences. Ferrazzini and Aki (1987) previously demonstrated such a resonant mechanism of seismogenesis but only for tremors (Gupta, 2011). This successful extension of a local seismicity mechanism to global scales indicated global resonance plate tectonics, which Omerbashich (2022) then demonstrated from continuous GPS measurements as actual waves in the solid Earth and on continental scales. Besides, a separate demonstration of Moon-body resonance from global moonquakes by Omerbashich (2020b) provided data-based spatiotemporally independent proof of global resonance tectonics and its universal (astrophysical) character. This demonstration of the astrophysical nature of the seismicity phenomenon confirmed an earlier theoretical demonstration by Omerbashich (2006a) that extended this *georesonation* concept to unspecified energies and times by successfully re-proportionating the Newtonian gravitational proportion, and thus expressing gravity, via the speed of light (also taken as a proportion), on macroscopic and quantum scales. That the mechanical resonance tectonics is astrophysical in character is supported observationally by globally present km-scale Faraday latticing photographed in recent years on numerous bodies throughout the solar system — notably abundantly on Ceres and in the form of polygonal morphology, mostly hexagonal (Omerbashich, 2020b).

The mechanical resonance of a body is a physical phenomenon that occurs when the body's eigenperiod of vibration coincides with ("resonates to") another physical system's eigenperiod or its fractional multiple, resulting in an additional vibration ("resonance"). The nature of a matched period, i.e., whether subharmonic or superharmonic, governs the resonance. Solids exposed to a resonance tend to crumble in a structural failure. Body resonances also arise via magnified (planetary) vibration due to *frequency demultiplication* as one of the rarest macroscopic physical phenomena and one that can magnify by 100s of times the energy injected at the fundamental disturbing frequency (Den Hartog, 1985; Omerbashich, 2007). As a deterministic process, a mechanical resonance makes resonance seismotectonics a deterministic process also. This relation forbids stochastic processes such as mantle convection from playing a dynamically significant role on the global level. Namely, plate tectonics is not an autonomous activity — and has continued for billions of years on Earth with the help of some agents that break the lithosphere from the outside, resulting in continental collisions and hot mantle plumes — where continental collision is a stochastic process, while plate behavior is expected to be rather stochastic as well (Ogawa, 2007). Stochastic dynamic processes cannot drive a deterministic dynamic process; only the reverse is physically plausible. Thus the present study attempts to detect such external agents of resonance tectonics in the solar system using all *in situ* seismicity measurements available from solid bodies: Mars, Moon, and Earth.

Most promising candidates for the external geodynamics agents that affect seismicity are presumably in the body's immediate vicinity, mainly the interplanetary magnetic field (IMF) carried around the heliosphere by the solar wind, whose own dynamic is locally transient. Since the rupturing process is not of interest and focusing on it can obscure any results of such studies of planetary interface dynamics, seismic energies (or seismic magnitudes) are better suited for this purpose than seismic moments are, where any probability of reaching physically or statistically false conclusions becomes constrained by methodology, i.e., analyzing spatiotemporally independent *in situ* data from three different astronomical bodies. Here it is implied as physically and geologically valid that the nature of the time series analyzed is such that they represent discrete samples of seismic energies (here in seismic magnitudes) as a continuous function of time. Namely, Omerbashich (2004) previously demonstrated from earthquakes and using (preliminary) geophysical Earth models that there exists a stunning similarity between spectra of time series of seismic events (expressed in seismic energies vs. seismic magnitudes) and that they possess precursory qualities. Finally, that there is a global budget of seismic energy between consecutive natural seismic events and not just immediately following them was shown by Omerbashich (2007), revealing a tidal modulation of the global gravitational field. A continuous nature of seismic energy is also discernable from the data plots along the temporal axis of the process depicted in Figs. 1–3.

On the other hand, the solar system often exhibits different periodicities in electromagnetic radiation and energetic particle events, ranging from the ~11-year sunspot cycle to the 27-day rotational period (Chowdhury et al., 2008; Chowdhury et al., 2016), including various mid-range periods (Bai, 2003). During solar cycle 21, Rieger et al. (1984) found the dominant (longest) such quasi-period, $P_{Rg}$=~154 days, in 139 γ-ray and >500 soft X-ray flares recorded by the Gamma-Ray Spectrometer aboard the Solar Maximum Mission. Since then, $P_{Rg}$ has been detected in the IMF in the Earth's vicinity (Cane et al., 1998) and virtually all heliophysical data like various flare types, photospheric magnetic flux, and group sunspot numbers. Those studies also reported $P_{Rg}$-related mid-term periodicities of ~128, ~102, ~78, and ~51 days, referred to as Ri*eger-type periodicities* (Dimitropoulou et al., 2008), which arise as modulations of $P_{Rg}$: 5/6 $P_{Rg}$, 2/3 $P_{Rg}$, 1/2 $P_{Rg}$, and 1/3 $P_{Rg}$, respectively. Also, various longer periodicities and their modulations were reported in the past from almost all heliophysics data types. These include sunspot numbers, solar flare index, solar radio flux, proton speed, and others, except for the coronal index and 10.7 cm solar flux (Forgács-Dajka and Borkovits, 2007). The well-known *Rieger period* (Chowdhury et al., 2009), $P_{Rg}$, thus can be regarded as the most obvious candidate for the external agent for shaping surfaces of terrestrial bodies so, for instance, during penumbra transitions, at ~153-day it most strongly affects the dynamics of satellites observing the Earth's global gravity field (Tzamali and Pagiatakis, 2021). Furthermore, Simpson (1967) previously postulated that IMF variations could cause seismic fracturing through surging crustal currents or magnetohydrodynamic (Alfvén, 1942) coupling.

The Rieger period originates in (is clocked by) the Sun, which sets it as the guiding periodicity of the polar (resonantly emitted) solar wind in the Sun's lowest global energies that, at the same time, are the highest planetary energies. This ubiquitous period then arises as a folded offshoot from the global anti-resonance simultaneously from the northerly and southerly polar (fast) solar winds, resulting in the period's well-known power and prominence on macroscopic scales (Omerbashich, 2023a). Then the process responsible for the IMF disturbance at the Rieger-type periodicities primarily reflects the flapping of solar wind, plasma, and other types of mechanical waves (mostly made up of electrons) as they are pushed outwards globally resonantly as well as by coronal mass ejections. Therefore, and as observed commonly in mechanical resonances occurring when energetic waves hit insurmountable obstacles, the Rieger-type periodicities likely are naturally occurring resonance harmonics of the electromagnetic blanket's macroscopic dynamics as influenced by solar-system bodies and their physical fields, gravitational and magnetic primarily (ibid.). Besides reflecting the Sun resonances that are mainly rotationally induced (Singh and Badruddin, 2019), mechanical resonances in the solar wind also arise because the IMF experiences various influences in addition to gravitation and magnetism of solar-system bodies, including excessive emissions of free electrons when affected by the solar mass ejections of protons.

In addition to its detection in the IMF, the *Rieger resonance* (RR) — comprising the Rieger period $P_{Rg}$ and Rieger-type periodicities — was found in most types of heliophysics data during solar cycles 19–24, like the already mentioned flares, photospheric magnetic flux, group sunspot numbers, and proton speed. In the past, the Rieger period and Rieger-type periodicities have been reported in different ranges depending on the data, location, epoch, and methodology, as 155–160 days, 160–165 days, 175–188 days, and 180–190 days; see, e.g., Gurgenashvili et al. (2017). Most of those studies indicate a leading (longest) periodicity ranging from 152- to 158-day, which seems to be particularly dominant in the time phase from ~1979 to 1983, corresponding to the solar activity maximum (Chowdhury et al., 2008). The same agents that force solar activity could at least be responsible for some geomagnetic and seismic activity (Odintsov et al., 2006).

The present study is integrative and unique in its methodological approach of comparative analysis of seismicity data from different astronomical bodies at different epochs, so this text is neither intended for nor prepared with any particular geoscientific community in mind.

## 2. Methodology & Data

### 2.1 *Methodology*

To analyze the seismicity occurrences, I apply the rigorous Gauss–Vaníček method of spectral analysis (GVSA) by Vaníček (1969, 1971) and compute spectra by a least-squares fit of data periodicity and trigonometric functions. A GVSA spectrum, $s_j$, is obtained at a spectral resolution $k$ (here 1000 spectral values throughout), for $k$ corresponding periods $T_j$ or frequencies $\omega_j$ and output with spectral magnitudes $M_j$, as:

$$s_j(T_j, M_j);\ j = 1 \dots k \wedge j \in \mathbb{Z} \wedge k \in \aleph. \quad (1)$$

In its simplest form, i.e., when there is no *a priori* knowledge on data constituents such as datum offsets, linear trends, and instrumental drifts, a GVSA spectrum $\boldsymbol{s}$ is computed for $k$ corresponding frequencies $\omega_j$ and output with spectral magnitudes $M_j$, as (Omerbashich, 2004):

$$\boldsymbol{s}(\omega_j, M_j) = \frac{\boldsymbol{l}^{\mathrm{T}} \cdot \boldsymbol{p}(\omega_j)}{\boldsymbol{l}^{\mathrm{T}} \cdot \boldsymbol{l}}, \quad (2)$$

obtained after two orthogonal projections. First, of the vector of $m$ observations, $\boldsymbol{l}$, onto the manifold $Z(\boldsymbol{\Psi})$ spanned by different base functions (columns of $\mathbf{A}$ matrix) at a time instant $t$, $\boldsymbol{\Psi} = [\cos\omega t, \sin\omega t]$, to obtain the best fitting approximant $\boldsymbol{p} = \sum_{i=1}^{m} \hat{c}_i\, \boldsymbol{\Psi}_i$ to $\boldsymbol{l}$ such that the residuals $\hat{\boldsymbol{v}} = \boldsymbol{l} - \boldsymbol{p}$ are minimized in the least-squares sense for $\hat{\boldsymbol{c}} = (\boldsymbol{\Psi}^{\mathrm{T}}\mathbf{C}_l^{-1}\boldsymbol{\Psi})^{-1} \cdot \boldsymbol{\Psi}^{\mathrm{T}}\mathbf{C}_l^{-1}\boldsymbol{l}$. The second projection, of $\boldsymbol{p}$ onto $\boldsymbol{l}$, enables us to obtain the spectral value, Eq. (2). Vectors $\boldsymbol{u}_j = \boldsymbol{\Psi}^{\mathrm{T}}\, \boldsymbol{\Psi}_{\mathrm{NK}+1}$ and $\boldsymbol{v}_j = \boldsymbol{\Psi}^{\mathrm{T}}\, \boldsymbol{\Psi}_{\mathrm{NK}+2}$, $j$=1, 2…NK∈ℵ, compose columns of the matrix $\mathbf{A}_{\mathrm{NK,NK}} = \boldsymbol{\Psi}^{\mathrm{T}}\, \boldsymbol{\Psi}$. Note here that the vectors of known constituents compose matrix $\hat{\mathbf{A}}_{\mathrm{m,m}} = \hat{\boldsymbol{\Psi}}^{\mathrm{T}}\, \hat{\boldsymbol{\Psi}}$, in which case the base functions that span the manifold $Z(\boldsymbol{\Psi})$ get expanded by known-constituent base functions, $\hat{\boldsymbol{\Psi}}$, to $\boldsymbol{\Psi} = [\hat{\boldsymbol{\Psi}}, \cos\omega t, \sin\omega t]$. For a detailed treatment of GVSA with known data constituents, see, e.g., Wells et al. (1985). Subsequently, the method got simplified into non-rigorous (strictly non-least-squares) formats like the Lomb-Scargle technique created to lower computational burden, but which is no longer an issue. At the same time, the conventional Fourier transform and spectrum are just special cases of a more general least-squares formulation (Craymer, 1998).

GVSA provides absolute ("to data resolution") accuracy in extracting periodicities from any natural datasets (Omerbashich, 2021, 2020a, 2020b, 2007, 2006b). Fed raw data, i.e., analyzing data without preprocessing — including filtering, windowing, tapering, and padding gaps with zeros — GVSA outputs spectral peaks against linear background noise levels. This kind of processing enables relative spectral computations whose results are directly energy-stratified for physical systems, allowing for the direct separation of resonance forcers from corresponding harmonics and other systematic signal contents. Here input data are utilized in their raw form, i.e., without performing any post-processing that some use to enhance spectra. GVSA revolutionizes physical sciences by enabling direct computations of global nonlinear dynamics (Omerbashich, 2023b), rendering approximative approaches like spherical decomposition obsolete (Omerbashich, 2023a).

GVSA is strict in that, besides estimating a uniform spectrum-wide statistical significance in var% for the desired level, say 95%, in a spectrum from a time series with *q* data values and *w* known constituents as $1–0.95^{2/(q–w–2)}$ (Steeves, 1981) (Wells et al., 1985), it also imposes an additional constraint for determining the validity of each significant peak individually — the fidelity or realism, Φ. In advanced statistics, fidelity is a general information measure based on the coordinate-independent cumulative distribution and critical yet previously neglected symmetry considerations (Kinkhabwala, 2013). In communications theory, fidelity measures how undesirable it is (according to some fidelity criterion we devise) to receive one piece of information when another is transmitted (Shannon, 1948). In GVSA, fidelity thus is defined in terms of the theory of spectral analysis as a measure of how undesirable it is for two frequencies to coincide (occupy the same frequency space of a sample). Then a value of GVSA fidelity is obtained as that time interval (in units of the timescale of the time series analyzed) by which the period of a significant spectral peak must be elongated or shortened to be π-phase-shiftable within the length of that time series. As such, Φ measures the unresolvedness between two consecutive significant spectral peaks (those that cannot be π-phase-shifted). When periods of such spectral peaks differ by more than the fidelity value of the former, those peaks are resolvable. As the *degree of a spectral peak's dependence or tendency to cluster*, this criterion reveals whether a spectral peak can share a systematic nature with another spectral peak, e.g., be part of a batch or an underlying dynamical process like resonance or reflection. The spectral peaks that meet this criterion are in the LSSA software output listed amongst insignificant and the rest amongst significant (hereafter: *physically-statistically significant peaks* or just (fully) *significant peaks* for short).

Namely, GVSA comes with a complete statistical package, LSSA, in which statistical testing, significance, and fidelity estimates are all performed in parallel with the computation of spectra and refer to variance as the most natural descriptor of noise in a physical system. Thus the spectral magnitude of a peak in a variance spectrum represents the contribution of that specific frequency to the data variance, expressed in var% (percentage variance). In addition to var%, GVSA computes spectra in classically used dB (Pagiatakis, 1999).

The spectral contents beyond the low end of the 30–180-day spectral band of interest are in the intermediate or mid-term band (Forgács–Dajka and Borkovits, 2007) and, as such, were of no interest. Namely, numerous studies over the past three decades have identified and confirmed the Rieger period as a genuine driver of (Rieger-type) periodicities or harmonics in the same band, so spectral periods longer than 180 days are possibly reflections or harmonics of lower drivers still. On the other hand, spectral contents beyond the high end of the band of interest likely are related to the Sun's rotational period of ~27 days as the driver.

2.1 *Mars data*

Seismic data collected *in situ* reveal fundamental physical properties of a heavenly body. After Viking 2 in 1976 did not succeed in placing its seismometer on the ground (Lorenz et al., 2017), the second attempt to learn significantly about the main physical characteristics of Mars came about with the InSight mission (*Interior Exploration using Seismic Investigations, Geodesy and Heat Transport*) that landed in late 2018. Preliminary analyses of 2019-2020 InSight seismic data elucidated that bulk of events recorded are of high-frequency energy, are distributed spatially unevenly, and occur unusually far from the lander. At the same time, unlike earthquakes, the event rate fluctuates annually. In addition, marsquakes are a highly localized phenomenon, while only a few show discernable seismic phase arrivals. All such marsquakes occur 1,800 km from the lander, on the sunken Cerberus Fossae Plateau, one of the youngest geological structures on Mars, possibly caused by subsidence or extensional faulting. Marsquakes are fundamentally different from earthquakes in several ways: for example, they are considerably smaller by seismic energy released, the strongest being the event recorded at teleseismic distances with a magnitude of ~3.6. Both local and seismic background noise on Mars can be significantly lower than on Earth without the constant tremors induced by mechanical resonances. (Ceylan et al., 2021a; Ceylan et al., 2021b; van Driel et al., 2021)

Many candidate-causes of Mars seismicity were looked into in the past, such as tidal force, fault activity, atmospheric coupling, landslide, and meteoroid impact; see, e.g., Knapmeyer et al. (2021), who favor the Sun declinations as the regime under which most marsquakes occur, i.e., between dusk and midnight. And since interplanetary magnetism drastically increases at dusk (Walsh et al., 2014), in what follows, I use the InSight seismic data to spectrally probe the 30–180-day band of highest areophysical energies to determine if external forcing contributes to the peculiarities of Martian seismicity as indicated by the Knapmeyer et al. (2021) proposal. Here the working assumption is that a periodic signature of such forcing could expose the process. Because such a process excludes plate tectonics, here the physical hypothesis is that Mars has no active tectonics and thus no continuous seismicity, which is tested by spectrally analyzing occurrences of continuous (high-frequency) marsquakes detected with InSight as they make up more than 80% of marsquakes and are therefore the best temporal representative of Martian seismicity. The 30–180-day band is also the range of RR — the dominant periodic forcing in the solar wind that carries the IMF; see, e.g., Russell (2001).

To investigate possible external drivers of Martian planetary dynamics, I first spectrally analyze the 26 March 2019–28 March 2020 marsquake occurrences from the Marsquakes Catalog by Clinton et al. (2021). The time interval spanned one solar minimum. As cataloged for 2019–2020, the InSight data reveal the planetary dynamics of Mars, primarily characterized by temporal clustering in the rupturing process and an orbital (here external forcing is assumed tacitly as Mars samples such forcing orbitally) variation in the event frequency, Fig. 1 and Table 1. The highly localized spatial clustering of the events, and their tendency to cluster in time, indicate an external planetary forcing with a lock to a dominant forcer active during the sampling. Also used in the present study is the InSight raw seismic data release v.9 of 1 April 2022 (for more on data see statements), containing $q_{mars}$=1755 mixed-quality events spanning 13 January 2019–30 September 2021.

Importantly, InSight magnetometer data are useless for investigating possibilities of a correlation between the marsquakes occurrences and Mars magnetic field variations primarily because the magnetometer was static while the events were occurring across various remote locations. Also, any variability in the wind noise should not be confused with variability in the marsquake source processes. Given that any new catalog version of marsquake data intends to replace any previous version and not only extend it, an affirmative result from the most recent catalog would also have an added methodological robustness. This expanded reliability is not standalone but adds to the correctness of the main result — the physical hypothesis that macroscopic dynamics of astrophysical magnetic fields cause seismicity — if following from three spatiotemporally independent data sets (different astronomical bodies), which is the most rigorous methodological criterion possible in the present study.

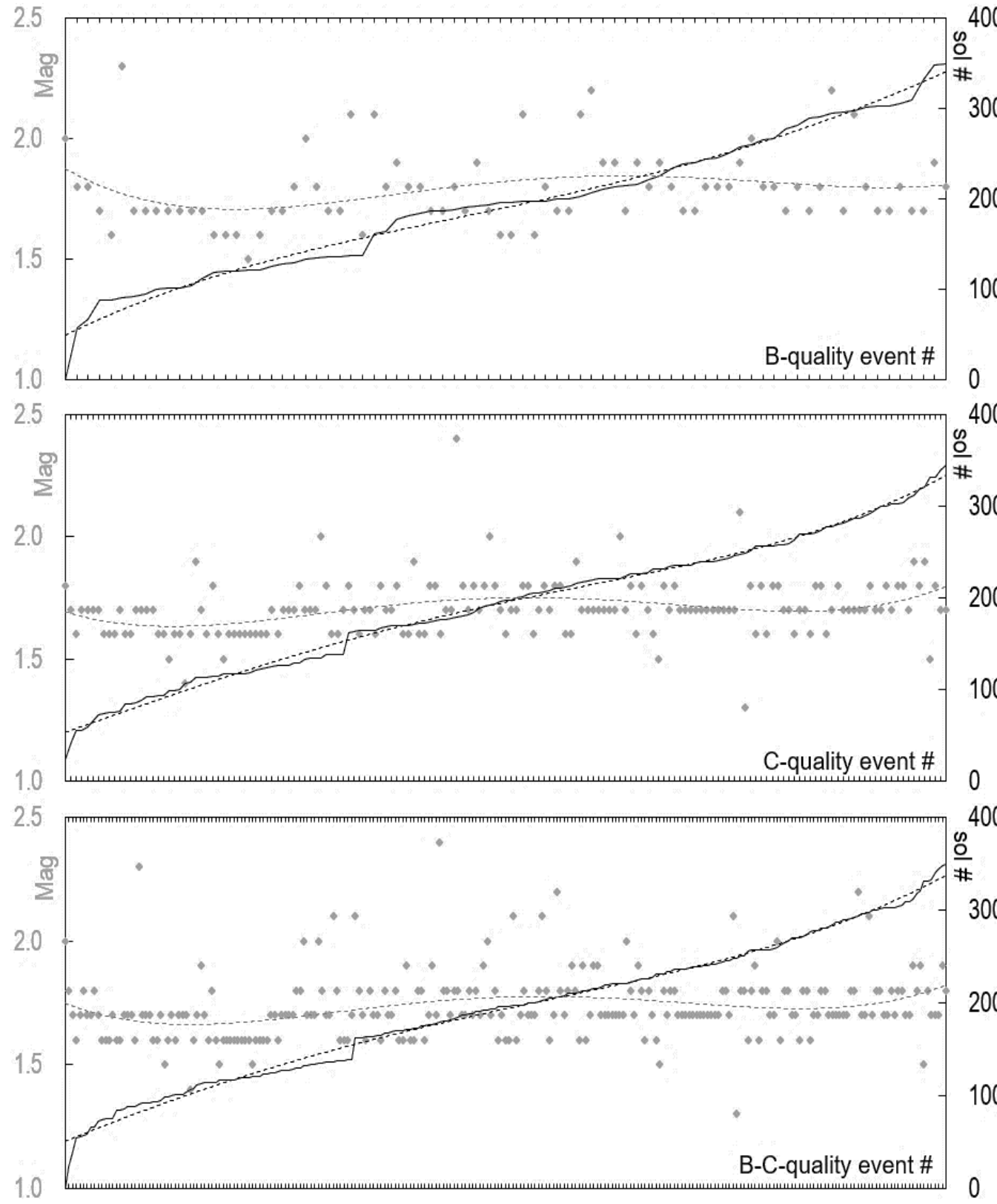

Figure 1. Marsquakes in seismic magnitudes (gray rhombs) vs. times of occurrence (solid black lines). A comparison of 78 B-quality (top panel), 163 C-quality (mid), and 241 B–C-combined-quality events (bottom) suggests that the InSight record consistently maintains the rate of events and associated trends regardless of data quality. This conclusion is supported by 4th-order polynomial trends (dashed lines), indicating an orbitally induced variation in the event rate. The data were high-frequency events from Supplementary Table 2 of Clinton et al. (2021). D-quality events omitted as only weakly observed or strongly contaminated by noise (Clinton et al., 2021). Data plotted indiscriminately for a more natural depiction of driving processes. Usually, time series get depicted along artificially equipaced (presumably uniformly paced) time axes, but such representation masks nonlinear background dynamics that can only reveal themselves on their natural timescales as the only ones useful for visual inspection (remaining very involved for mathematical modeling, including linearization). The opposite applies too, and thus nonlinear timescales like the one depicted cannot reveal equipaced variations like seasonality that cannot then be readily seen in Fig. 1 even if present in the record as claimed by Knapmeyer et al. (2021). The spectral analyses in the present study did not confirm seasonality either. However, it is possible to misinterpret portions of RR taken out of the context as a seasonality. Data given in Table 1 in an analysis-ready format, times reduced to 0-origin. Spectral computations refer to the Coordinated Universal Time (UTC), but the same data in Local Mean Solar Time (LMST) produce identical results.

| T(s) | Mag. | T(s) | Mag. | T(s) | Mag. | T(s) | Mag. | T(s) | Mag. | T(s) | Mag. |
|---|---|---|---|---|---|---|---|---|---|---|---|
| 0 | 2.0 | 10132068 | 1.6 | 14491858 | 1.6 | 17509753 | 1.6 | 20698692 | 1.7 | 24610254 | 1.7 |
| 2050374 | 1.8 | 10308278 | 1.5 | 14494136 | 1.8 | 17511843 | 1.7 | 20704022 | 1.8 | 24801334 | 1.8 |
| 3572810 | 1.7 | 10310875 | 1.6 | 14501012 | 1.7 | 17518967 | 1.8 | 20870526 | 1.8 | 24870502 | 1.8 |
| 4812338 | 1.6 | 10312163 | 1.6 | 14572869 | 1.7 | 17675603 | 1.7 | 20874880 | 1.7 | 24972553 | 1.7 |
| 4904288 | 1.7 | 10316335 | 1.6 | 14580402 | 1.6 | 17683420 | 1.7 | 20877211 | 1.7 | 25155245 | 1.7 |
| 5075450 | 1.8 | 10321957 | 1.6 | 14763037 | 1.8 | 17692020 | 1.7 | 20880027 | 1.7 | 25410899 | 1.7 |
| 5161947 | 1.7 | 10485517 | 1.6 | 14848567 | 1.7 | 17866968 | 1.8 | 21066749 | 1.7 | 25426471 | 1.7 |
| 5779227 | 1.7 | 10494966 | 1.6 | 14966838 | 1.7 | 17967797 | 2.1 | 21142775 | 1.7 | 25603268 | 1.7 |
| 5878271 | 1.8 | 10584241 | 1.6 | 15012806 | 1.8 | 18047870 | 1.8 | 21146374 | 1.7 | 25665939 | 1.7 |
| 6413098 | 1.7 | 10585270 | 1.6 | 15018582 | 1.6 | 18117072 | 1.6 | 21153727 | 1.7 | 25755404 | 1.8 |
| 6500182 | 1.6 | 10669504 | 1.5 | 15023559 | 1.6 | 18216427 | 1.7 | 21233348 | 1.7 | 25854575 | 1.8 |
| 6582386 | 1.6 | 10671291 | 1.6 | 15132978 | 1.9 | 18299921 | 2.2 | 21236859 | 1.7 | 26035050 | 2.2 |
| 6588730 | 1.6 | 10683600 | 1.6 | 15195263 | 1.6 | 18301772 | 1.8 | 21331924 | 1.7 | 26214695 | 1.7 |
| 6680268 | 1.7 | 10842664 | 1.6 | 15284658 | 1.6 | 18401523 | 1.7 | 21406808 | 1.7 | 26219052 | 1.7 |
| 7409355 | 1.6 | 10927313 | 1.6 | 15312201 | 1.8 | 18583132 | 1.8 | 21491158 | 1.7 | 26293840 | 2.1 |
| 7422934 | 1.6 | 11014791 | 1.7 | 15492982 | 1.8 | 18585420 | 1.8 | 21596016 | 1.8 | 26574096 | 1.7 |
| 7506335 | 1.7 | 11024119 | 1.7 | 15545143 | 1.6 | 18654356 | 1.9 | 21679222 | 1.8 | 26579469 | 1.8 |
| 7738409 | 1.7 | 11112124 | 1.6 | 15635454 | 1.7 | 18737628 | 1.6 | 21775456 | 1.7 | 26737437 | 1.8 |
| 7777430 | 1.7 | 11197591 | 1.7 | 15649801 | 1.9 | 18836384 | 1.9 | 21947735 | 2.1 | 26748274 | 1.7 |
| 7827265 | 1.6 | 11206961 | 1.7 | 15757755 | 1.7 | 18840891 | 1.6 | 21970752 | 1.3 | 26751900 | 1.7 |
| 8018620 | 2.3 | 11298235 | 1.7 | 15830656 | 2.4 | 19020543 | 1.7 | 22198657 | 1.8 | 26758695 | 1.8 |
| 8088183 | 1.7 | 11387441 | 1.7 | 15985391 | 1.8 | 19101867 | 1.9 | 22217701 | 1.8 | 26824093 | 1.7 |
| 8095439 | 1.7 | 11457837 | 1.8 | 16006071 | 1.8 | 19102914 | 1.9 | 22654834 | 1.6 | 26914981 | 1.8 |
| 8098918 | 1.7 | 11460462 | 1.8 | 16091954 | 1.7 | 19181513 | 1.7 | 22731395 | 1.8 | 27015340 | 1.8 |
| 8271066 | 1.6 | 11734406 | 2.0 | 16255998 | 1.8 | 19374619 | 1.7 | 22735939 | 1.9 | 27355817 | 1.7 |
| 8273040 | 1.6 | 11741737 | 1.7 | 16267207 | 1.8 | 19458791 | 1.7 | 22741959 | 1.6 | 27361675 | 1.7 |
| 8358961 | 1.7 | 11821575 | 1.7 | 16519762 | 1.7 | 19542347 | 1.7 | 22748821 | 1.8 | 27558139 | 1.9 |
| 8708349 | 1.5 | 11827854 | 1.7 | 16524261 | 1.7 | 19545243 | 1.7 | 22820619 | 1.8 | 28174209 | 1.8 |
| 8715566 | 1.6 | 11919729 | 2.0 | 16608920 | 1.8 | 19548644 | 1.7 | 22825208 | 1.7 | 28435955 | 1.9 |
| 8809619 | 1.7 | 11928997 | 1.8 | 16616394 | 1.7 | 19555671 | 1.7 | 22926489 | 1.7 | 29319442 | 1.5 |
| 8890883 | 1.6 | 12000441 | 1.7 | 16883240 | 1.8 | 19574081 | 2.0 | 23000475 | 2 | 29321206 | 1.8 |
| 8899290 | 1.7 | 12084161 | 1.7 | 16912114 | 1.7 | 19640094 | 1.8 | 23279193 | 1.6 | 29416415 | 1.7 |
| 8982738 | 1.7 | 12099084 | 2.1 | 16963561 | 1.9 | 19814530 | 1.7 | 23442420 | 1.8 | 30098143 | 1.7 |
| 9153164 | 1.7 | 12172435 | 1.8 | 17003677 | 2.0 | 19903187 | 1.9 | 23624199 | 1.8 | 30481708 | 1.7 |
| 9428420 | 1.4 | 12177059 | 1.6 | 17143474 | 1.7 | 19991954 | 1.8 | 23887952 | 1.7 | 30812773 | 1.9 |
| 9523945 | 1.6 | 12185404 | 1.6 | 17166670 | 1.8 | 19995896 | 1.6 | 23893351 | 1.7 | 30985847 | 1.8 |
| 9866331 | 1.7 | 12269018 | 1.6 | 17321799 | 1.6 | 20002911 | 1.8 | 23904357 | 1.6 | | |
| 9885475 | 1.9 | 12274371 | 1.7 | 17332121 | 1.7 | 20157020 | 1.7 | 23995679 | 1.8 | | |
| 10044592 | 1.7 | 14307834 | 2.1 | 17335276 | 1.6 | 20438513 | 1.6 | 24247736 | 1.8 | | |
| 10055291 | 1.6 | 14310432 | 1.8 | 17408558 | 1.6 | 20445137 | 1.5 | 24543478 | 1.6 | | |
| 10128868 | 1.8 | 14401746 | 1.7 | 17435002 | 2.1 | 20518412 | 1.8 | 24601478 | 1.7 | | |

Table 1. Mars data, Fig. 1, reduced to 0-origin. From the Marsquakes Catalog by Clinton et al. (2021), their corrected Supplementary Table 2.

### 2.3 *Moon data*

The moonquake data consisted of 13058 seismic events from the *Levent.1008* Moonquakes Catalog (Nakamura et al., 1981), last updated in 2008. The samplings were timestamped at a once-per-minute rate and spanned the time interval 1969–1977, as collected within the Apollo Program PSE/ALSEP (Passive Seismic Experiment/Apollo Lunar Surface Experiments Package), missions 11–16. Previously, Omerbashich (2020b) had used those data in their raw form to demonstrate a super-harmonic resonance mechanism for generating lunar seismicity. Here I use the same data, but after removing meteoroid, rocket, and Lunar Module impacts and events with duplicate timestamps to enable the time-series monotony. The removal resulted in a dataset of $q_{\mathrm{moon}}$=10815 genuine natural and unclassified moonquakes, Table A-1 in Supplement A, plotted in Fig. 2 indiscriminately for the most realistic depiction of any driving processes. The data interval spanned the entire Apollo mission's duration, including solar minima and maxima.

Since there is still no uniform definition of the moonquake magnitude, the lunar events were assigned generic random magnitudes as values of seismic magnitudes in undeclared units and generated randomly from within a range delimited by two fixed select values [5.5, 7.5] (Omerbashich, 2020b). Namely, from the statistics point of view, seismic magnitudes are irrelevant for the present study because the study is concerned not with a seismic investigation into rupturing processes (sources, sizes, locations) and the associated seismic moments but with frequency analyses of timings alone. So here, GVSA establishes periodicity of neither the variation in seismic magnitudes nor moments; instead, it models periodicity in the timings of the events, and most directly so.

When modeling relies on real-world data, as does here, that modeling is physical by definition, and any numerical analyses of those data are physically meaningful too. Then the main task in the present study is to find a 1–on–1 systematic correspondence between internal and external dynamics (not to be confused with a statistical correlation since a spectral analysis is not from the realm of statistical analysis but that of numerical analysis). Thus, usual approaches to statistical analyses of quakes (most commonly earthquakes), which begin with the determination of the magnitude of completeness, and the restriction of the data set to events stronger than that threshold, are irrelevant since, as mentioned, GVSA is a numerical rather than statistical method. Besides, spectral analyses reported in the present study were already verified statistically, with periods tested for statistical and physical significance within the LSSA scientific software package. However, seismic magnitudes are not critical information when a widely reported ensemble of physical periods gets detected (alone, at that) — as was the case with resonance trains reported in the present study. The significance in such cases becomes immediately physical (an equivalent of 100% statistical significance) since the extracted information instantly identifies as a physical process. Systematic dynamical processes are always deterministic and only presumed stochastic until and if such positive detection materializes. Thus, even random magnitudes from some physically realistic range will suffice for such extractions, as only timings are critical in such cases (Omerbashich, 2020a, 2020b).

From the physics point of view, as already mentioned, the randomness of seismic magnitudes does not affect the result in relative spectral analyses of time series when only previously widely reported physical periodicities are found (Omerbashich, 2020b). It is the train as a whole that is at least 67%-significant by the sheer fact that it got retrieved practically entirely. Such an approach is the best we can do when we lack seismic magnitudes on a uniform seismic scale, as in the Moon case. In other words, unless the body of interest already acts as a closed and integrated physical system (as a minimum condition for any such detection to be statistically 67%-significant at least), the dynamics of such a body cannot sustain any resonance train on global scales either. This virtually ideal situation for nonlinear oscillations to arise then results in the computed statistical fidelity values exceeding the >12 thresholds (for them to be considered reflective of a physical process; Omerbashich, 2006b, 2007b) and by many magnitudes of order at that, Fig. 3. Additional criteria for validating results via fidelity exist (Omerbashich, 2021, 2020b).

## 2.4 *Earth data*

The data represented the strong terrestrial seismicity, sampled at a 1Hz rate, in moment magnitudes $M_w$ as the most realistic (physics-based) depiction of seismicity (Kanamori, 1977) (Dziewonski et al., 1981). The data set consisted of occurrence times of all 866 earthquakes of $M_w$5.6+, spanning $\Delta t_{earth}$=1211 days from 01 October 2015–02 February 2019. This time-interval is chosen for its recency and completeness and because outside solar maxima or minima so that the earthquake data also serve to investigate possible effects of solar activity independently of the moonquake data that spanned both maxima and minima and of marsquake data that spanned one minimum, thus making the analysis methodologically overall most rigorous (temporally diverse) possible. Here the time interval should suffice for the 1–6-month band of interest; see Supplement B. The robust (outliers M** discarded) mean values from the USGS (United States Geological Survey), EMSC (European-Mediterranean Seismological Centre), and GFZ (German Research Centre for Geosciences) moment magnitudes are analyzed spectrally as:

$$\overline{M_i}\big|_{i=1}^{q} = \frac{1}{3}\left[M_i^{USGS} + M_i^{EMSC} + M_i^{GFZ}\right]; \qquad i \in \aleph \wedge q \in \Re \wedge \overline{M_i}\big|_{i=1}^{q} \Leftrightarrow M_i \neq M^{**}. \quad (3)$$

Here, similarly to considerations from Omerbashich (2020a), where the data compilation is from, it is tacitly assumed that any excess events — which were temporally and spatially relatively close to the most energetic event for a corresponding time cluster — would overrepresent this specific seismic response to the external forcing of nonlinearity (as by the RR process in this case) instead of enriching its representation. Besides, declustering (classically: removing data for a better understanding of an underlying physical process) here is negligible and thus spectrally insignificant when using a spectral analysis method blind to gaps in data, as was done here. Therefore, I decluster the record to eliminate redundancies and consequently exclude the 21 events that had occurred within minutes of both the time and the geographic location of another event, thereby keeping the most energetic events per time cluster, if any. This declustering resulted in recompilation of $q_{earth}$=845 (occurrences of) $M_w$5.6+ events, Table B-1 in Supplement B. Previously, Omerbashich (2020a) had used the same data to demonstrate the superharmonic resonance mechanism for generating Earth seismicity. Note that the goal of declustering here is not to exclude aftershocks as understood classically but events based on their timing, as the present study (a time-series analysis) is concerned with event timing alone and neither rupturing processes nor source sizes. The $M_w$5.6 magnitude cutoff sits at twice the lower limit of the $M_w$6.2 used for the ±5% cutoff magnitude by Omerbashich (2020a), here then expanded for higher data density while still reflecting the seismicity generation mechanism as that revealed from the same data by *ibid*.

The present study examines the effects of Moon-Earth magnetotail reconnecting on both moonquakes and earthquakes. Any possible effects of geomagnetism on earthquakes here get ignored or cannot be verified because geomagnetic field variations in the band of interest are unknown or unreliable at best. Namely, the quality and resolution of the existing geomagnetic field observations are available as assessed in the literature for high-frequency bands only, i.e., the spectral bands of periods shorter than the semidiurnal tide or frequencies higher than 0.825 μHz. Besides, unlike in heliophysics, the range of frequencies within which RR resides largely remains unexplored in geophysics.

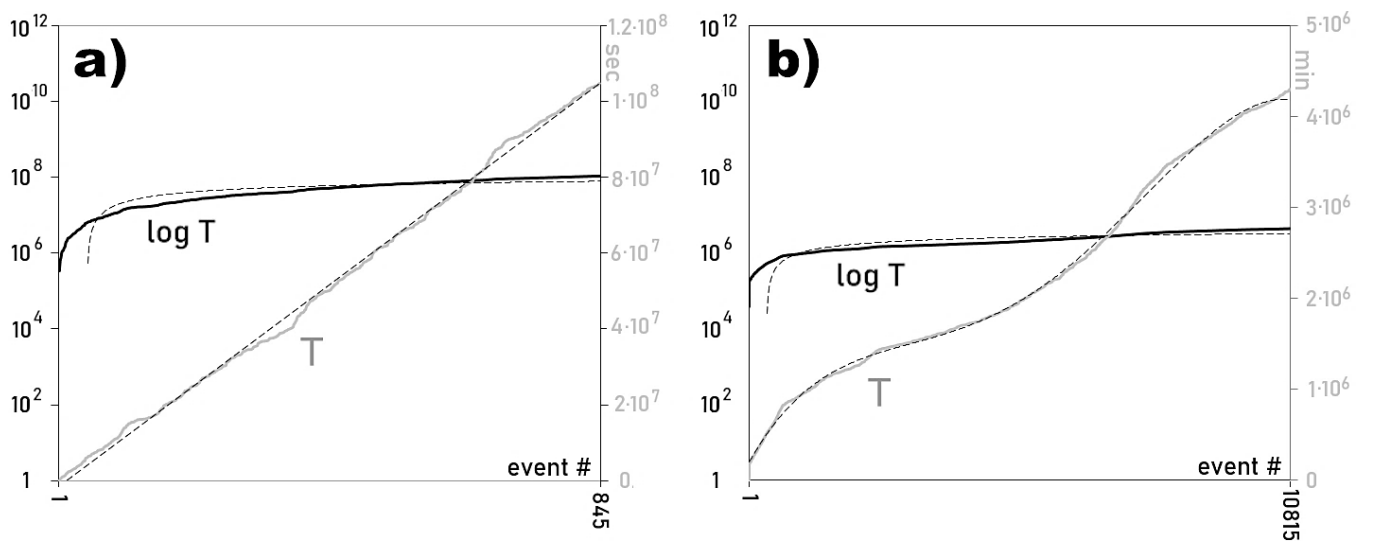

Figure 2. Occurrences, T, of 845 consecutive $M_w$5.6+ strong earthquakes from 2015–2019 (panel a) and 10815 consecutive natural moonquakes from 1969-1977, i.e., excluding meteoroid, S-IVB (Saturn V and Saturn IB rockets) impacts, and Lunar Module impacts, as well as events with duplicate sampling timestamp (panel b), plotted indiscriminately for a most natural representation of driving processes. Time-series data commonly are plotted along artificial equipaced (presumably uniformly paced) time axes, which, as such, mask any nonlinear background dynamics. Both panels: solid gray line represents T in seconds of Earth time (panel a) and minutes (panel b); the solid black line is the plot of log T. Also shown are corresponding trends (gray dashed): linear — revealing a nonlinear background process as it attempts to drive terrestrial seismicity but gets suppressed (panel a), and the $4^{th}$ order polynomial trend — revealing a free nonlinear background process as it intensely forces lunar seismicity periodically in that process's own (unspecified) timeframe (panel b). Superimposed ideal logarithmic trends (black dashed) highlight temporal inflection points of the process. Note shown physical processes and trends depend on neither the event source nor rupturing size, indicating an externally induced seismicity, more so in the Moon case. Assuming IMF variations commonly generate seismicity on solid-surface bodies, the revealed trends mean the RR process affects the Moon immensely and Earth to an extent. Event times reduced to 0-origin. See Supplements for data sets. Note linear trend in panel (a) is for reference only, and is used to show that a nonlinear trend exists about it, as seen from the visual inspection since such relationships are too involved to be described mathematically and are beyond the scope of the present study.

## 2.5 *Earth data and Moon data separation*

Due to the inherent instability of magnetism and assuming that the RR-forced seismicity is not Mars-specific, one cannot expect the RR process to reflect in the same manner at a given instant on both magnetospheres and the respective local IMF simultaneously. So if a planetary magnetosphere oscillates with a certain Rieger periodicity, the IMF at the body does not necessarily coincide exactly with this period (or it oscillates with a different Rieger period), and vice versa.

Harada et al. (2010) used 2007–2009 data from the SELENE (SELenological and ENgineering Explorer KAGUYA) lunar mission to demonstrate the magnetic influence of the Earth's magnetic tail on the Moon. A strong electric field was detected near the lunar surface when the Moon crosses the magnetotail, in which the plasma conditions differ significantly from those within the IMF. This relatively intensive and mutable electrical system occurs when the Moon, otherwise freed from magnetism, passes the plasma sheet, which sits centrally in the magnetic tail.

To study the effects of the geomagnetic tail on terrestrial and lunar seismicity, I make use of the blindness of GVSA to data gaps and divide natural moonquakes into segments when the Moon was in the magnetotail (2865 events) or within three days of the Full Moon (Phillips, 2008), vs. IMF (7950 events), i.e., the remainder of the data set. This separation of the earthquake data has resulted in time series with 210 and 635 events, respectively. This procedure revealed the zonal periodicity mismatch described above, Fig. 3. As seen from verifying the comparisons of Fig. 3 a & c vs. Fig. 2 a & b, respectively, against the comparisons amongst Fig. 3 b & d vs. Fig. 2 a & b, respectively, the Earth magnetotail is indeed responsible for most of the mismatch of the lunar seismic response and RR-forcing (as represented by a polynomial trend), while the process got exposed as inherent to the IMF, as expected. Note that the dataset size from before vs. after the separation matters little since the revealed separation effect only arises when we plot the event times indiscriminately. The only relevant factor is the process behind the nonlinearity, as exposed by this separation and in the process's own timeframe of which we know nothing, thus making any statistical analyses of the above process impossible. Besides, change in dataset size due to data separation does not affect the result (or on Figs. 2 & 3) since GVSA is impervious to data gaps always in the same way. So the relative accuracy of estimated spectra is maintained since (relative) spectral analyses are the core of the present study.

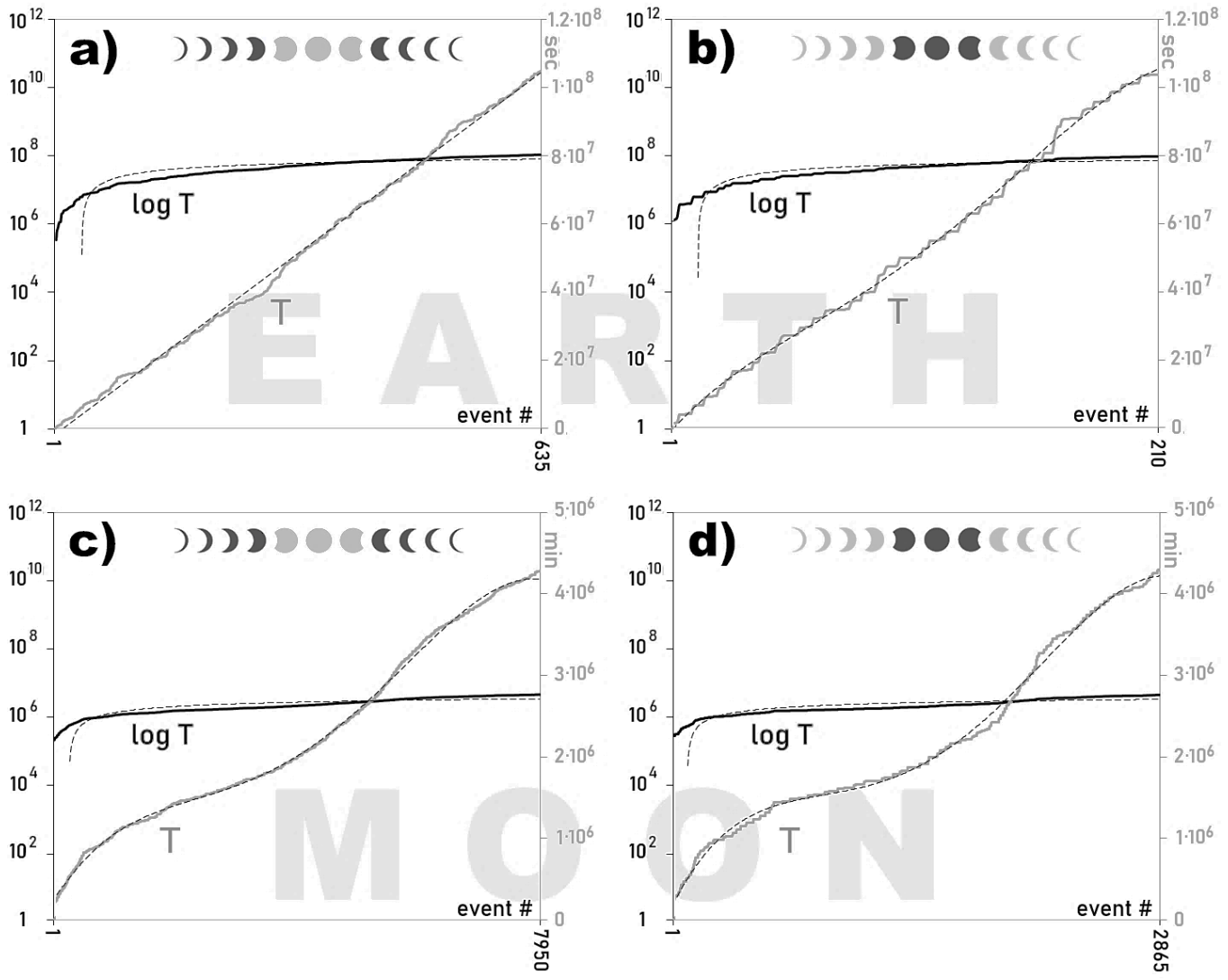

Figure 3. Indiscriminately plotted $M_w$5.6+ strong terrestrial (top row) and lunar (bottom) events occurrences, T, from Fig. 2, separated into events during the Moon traversals of the Earth magnetotail (panels b & d), vs. while crossing the IMF (panels a & c). As in Fig. 2, a nonlinear background driver of strong terrestrial seismicity is seen as suppressed (panel a) but then resumes nonlinearity (as fitted weakly by the $4^{th}$ order polynomial) as the Moon crosses the tail. This switch from a suppressed to free nonlinearity of terrestrial seismicity occurrences when the Moon moves from the IMF to magnetotail shows that the lunar opposition disturbs a dynamic of the Earth-Moon system (body magnetism has reconnected). This dynamic is then seismogenic itself. In addition, the nonlinearity of the dominant forcing process of terrestrial seismicity is weaker than the nonlinearity of the lunar seismicity's forcing process, so different nonlinear dynamics dominate the two bodies. A significantly better correspondence of moonquake occurrences with the $4^{th}$ order polynomial in panel c vs. panel d revealed that the lunar seismicity occurrences are most faithfully nonlinear as the Moon traverses the IMF, and the magnetotail affected the nonlinearity. This causality, in turn, shows that the main driver of lunar seismicity lies in the IMF, as this seismicity perhaps arises via a magnetohydrodynamical mechanism (Simpson, 1967; Alfvén, 1942) whereby the response of the Moon to IMF variability intensifies on each reconnection with the Earth magnetotail. Note that a magnetotail crossing here begins three days before a lunar opposition (Full Moon) and lasts for six days (Phillips, 2008), so the Moon spends about ¼ of its orbit in the Earth magnetotail. The line representation as in Fig. 2. See Supplements for data sets.

# 3. RESULTS

## 3.1 *Mars results*

Spectral analysis of the Marsquakes Catalog has revealed externally forced dynamics of Mars seismicity, Table 2 and Fig. 4. The longest periodicity detected is the widely reported Rieger period $P_{Rg}$=154.48 days at 4.8 var%, found to be the Mars system driver as it absorbed all power at the 99%-significance and has the highest statistical fidelity found in the present study, of staggering $\phi=2.9\cdot10^6$. This detection confirms conclusively that the external forcing is dominant in Martian dynamics, as a fidelity of >12 is considered reflective of a physical process (Omerbashich, 2006b, 2007b). Other significant periods are 42.07-day at 2.5 var% with ≥95% significance and again very high $\phi=2.1\cdot10^5$, then 54.51-day at 2.1 var% with ≥89% significance and very high $\phi=3.6\cdot10^5$, and 73.87-day at 1.5 var% with ≥67% significance and very high $\phi=6.6\cdot10^5$.

When also using D-quality events, periodicity estimates got largely preserved to within a few days, where the driver estimate stayed the same to ±1 day, Fig. 5. Significance levels went down somewhat but not below the resonance; this drop was due to poor quality of D events and the use of random magnitudes since D-quality marsquakes came without seismic magnitudes. This overall drop in quality also manifested as the lowering of statistical fidelity by five orders of magnitude from the spectrum of B–C-quality events combined, Fig. 4. This is an example of GVSA not requiring real-world variations to estimate or verify a known set of frequencies, as long as random magnitudes are constrained by some realistic range, here 1.0–2.5. In this case, a train of well-known natural resonance got extracted from a record of real-world seismic occurrences from another planet with 43% of magnitudes selected randomly, where the driver estimate preserved the $\phi>12$ fidelity threshold indicative of a physical process. The recovered ensemble's power stayed within the 20–40 dB range from B–C events combined (Fig. 5 callout vs. Fig. 4 callout). The thus demonstrated ability to extract systematic information from sparse and incomplete records of real-world data (here missing nearly half of total variations), taken *in situ* in another world, recommends GVSA highly not just to planetary seismologists but everyone else interested in learning a whole lot about systematic dynamics when very little system information is available. Note that GVSA requires just three values to estimate a spectrum.

The above recovery of RR from the (Clinton et al., 2021) Marsquake Catalog was partial. Since this could not be due to data span, to examine if it was due to heavy editing of raw data by *ibid*., I next spectrally analyze the most recent/complete (at the time of writing the present study) release of InSight raw seismic data, v.9, of 1 April 2022 (see statements on data). Since the events were of varying and unknown quality while lacking seismic magnitudes on a unique scale, and given the promising analysis of events with random magnitudes, Fig. 5, I assigned seismic magnitudes randomly from within a range delimited by two fixed select values [1.0, 3.0]. This extended range appeared realistic for this (v.9) release of raw data because events increased somewhat in strength since Clinton et al. (2021) published their results. If — unlike the shorter data edited by *ibid.* and revealing partial RR in Figs. 4 & 5 — raw data with random seismic magnitudes now expose the complete RR, then Martian seismicity likely is due to RR entirely, which would corroborate a generally held view of Mars as a tectonically inactive planet. This result would also refute contrarian claims, such as recently by Banerdt et al. (2020), which Stähler et al. (2022) too subsequently refuted. If so, then marsquakes classification by quality or another RR-unrelated criterion loses any physical meaning also.

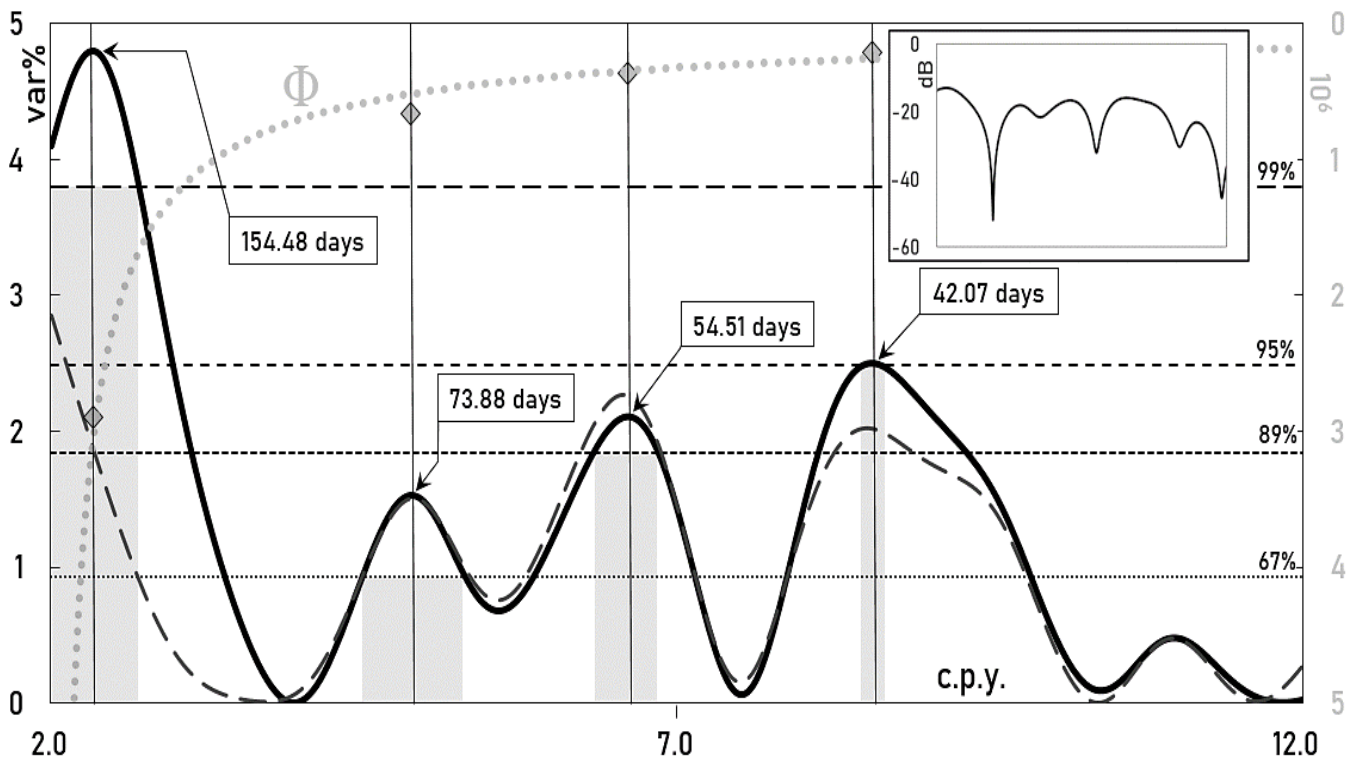

Figure 4. GVSA spectrum (solid curve) of data from the Marsquakes Catalog (Clinton et al., 2021) in the 1–6 months band of the highest planetary energies that exceed the environmental noise levels by magnitudes of order, so noise is methodologically filtered out and cannot be implicated in the spectrum to a significant degree. Since a data span of twice the spectral band's lower end suffices for credible spectral analyses in that band, and the time series here spanned the selected bandwidth twice, the B–C time-series with combined quality proved sufficient for extracting the dominant forcer. Thus the Rieger period $P_{Rg}$=154.48 days was returned as the system driver as it absorbed all power at the 99% level of significance, and this period has the highest statistical fidelity found in the present study, of $\phi=2.9\cdot10^6$. At the same time, the Rieger-type periodicities (harmonics) dominate the record across lower significance levels, but also with $\phi\gg12$, Table 2, where the fidelity of 12 and higher reflects a physical process (Omerbashich, 2006b); here $\phi$-curve represents a power-trend fit of the fidelity values on significant spectral peaks. Note the clustering on the $1/3\cdot P_{Rg}$ spectral peak. The spectrum in the Rieger band does not indicate any other systematic contents (including environmental seasonality, and therefore no need to model or "enforce" it further; for details see Wells et al. 1985), other than the Rieger and Rieger-type periods, i.e., the Rieger resonance train. Of all high-frequency events from Clinton et al. (2021), the B- and C-quality marsquake series separately were insufficient to provide spectra with significant periodicity (not shown), so the combined B/C time series was used, as justified by the consistent maintenance of event rates, Fig. 1. The underlying spectrum (dashed curve) of the same data but with $P_{Rg}$ mathematically suppressed ("enforced"; effectively removed from spectra during computation, Wells et al., 1985), indicates neither additional forcing dynamics nor areophysical systematic processes. Statistical fidelity values per spectral peak plotted with the power trend on the secondary axis (gray axis). The area of the gray boxes represents the power absorption capacity of a spectral peak at the corresponding significance level; the local spectral span above the respective significance level determines the box width and, together with a variance percentage achieved by the corresponding peak, describes the absorption capacity. Callout: the power spectrum of the same data but in dB (Pagiatakis, 1999). Pronounced and broad lobes indicate significant spectral peaks comprise practically all the power also. Note non-sharp resonance peaks reflect low data resolution as limited by scant Mars seismicity. Frequencies are in cycles per year (c.p.y.), with periods indicated on significant spectral peaks as well. Note also that immediate outer bands above the high band-end belong to Sun resonance due to the rotational rate of ~27 days (Singh and Badruddin, 2019), and those below the low band-end to poorly understood mid-term periodicities (Forgács–Dajka and Borkovits, 2007). Therefore, looking into the outer bands would be counterintuitive for the present study.

| Period | Significance level | T [days] | $T_{solar}$ [days] | ΔT [%] | Φ | Mag. [var%] | Power [dB] |
|---|---|---|---|---|---|---|---|
| $P_{Rg}$ | ≥99% (3.80 var%) | 154.485 | 154 | −0.3 | $2.9\cdot10^6$ | 4.8 | -12.97 |
| $1/2\cdot P_{Rg}$ | ≥67% (0.93 var%) | 73.878 | 78 | +5.3 | $6.6\cdot10^5$ | 1.5 | -18.09 |
| $1/3\cdot P_{Rg}$ | ≥89% (1.84 var%) | 54.507 | 51 | −6.9 | $3.6\cdot10^5$ | 2.1 | -16.67 |
| $3/11\cdot P_{Rg}$ | ≥95% (2.49 var%) | 42.073 | | −0.1 | $2.1\cdot10^5$ | 2.5 | -15.91 |

Table 2. Results of GVSA spectral analysis, Fig. 4, of the Marsquakes Catalog, Fig. 1. The previously reported $P_{Rg}$, $1/2\cdot P_{Rg}$, and $1/3\cdot P_{Rg}$ got matched, while the previously reported $5/6\cdot P_{Rg}$ and $2/3\cdot P_{Rg}$ periods got absorbed by $P_{Rg}$. Note the degeneration of the $1/3\cdot P_{Rg}$ within the Mars vicinity into a $3/11\cdot P_{Rg}$ split. Also, note that the above-given agreement (ΔT) with previously reported such periods $T_{solar}$ is the upper limit (the worse) of such matchings, while $1/2\cdot P_{Rg}$ and $1/3\cdot P_{Rg}$ were also previously reported as in the present study, e.g., by Özgüç and Ataç (1994). The $T_{solar}$ value for $P_{Rg}$ is per Chowdhury et al. (2009) and for the $1/2\cdot P_{Rg}$ and $1/3\cdot P_{Rg}$ harmonics per Dimitropoulou et al. (2008).

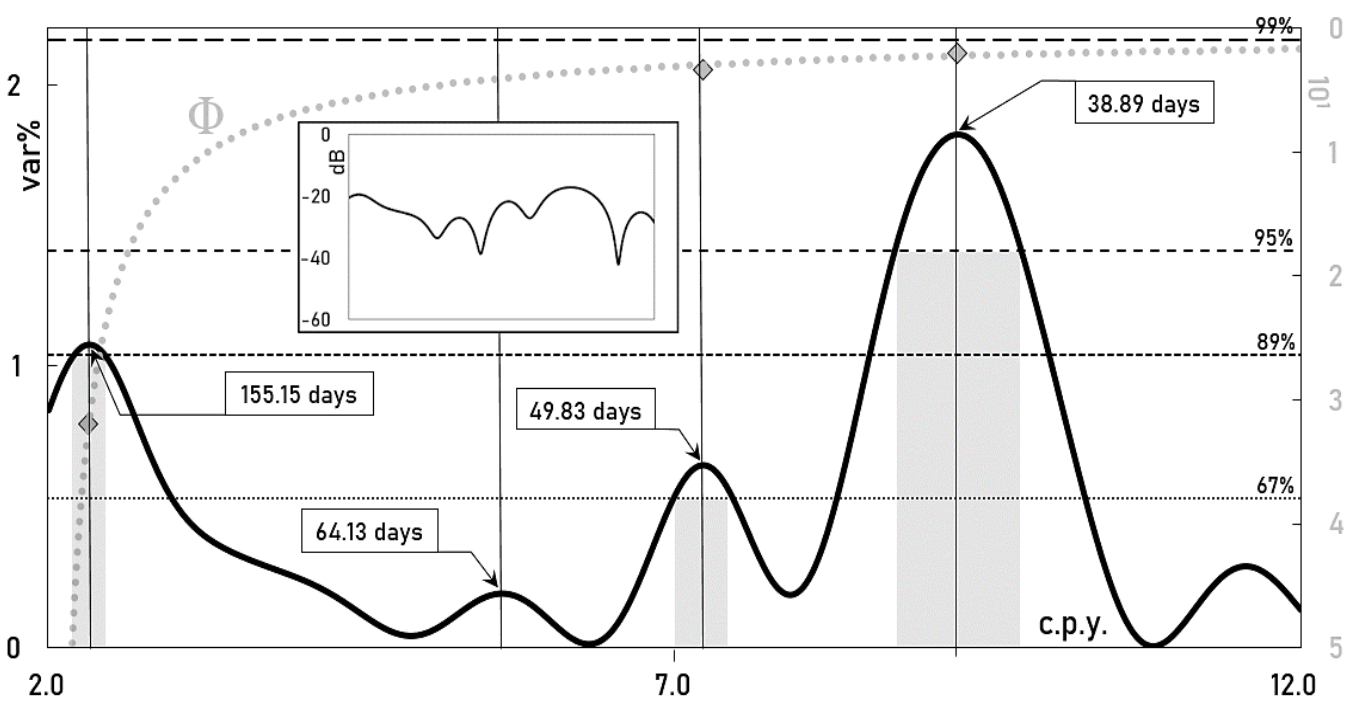

Figure 5. Same as Fig. 4, but of all 424 B–C–D-quality high-frequency marsquakes combined, thus including the 183 D-quality events with random seismic magnitudes. Periodicity estimates were largely preserved (to within a few days). Also plotted is the corresponding power spectrum (callout), revealing that the $P_{Rg}$ driver still absorbs the most power. Note that the detected resonance peaks did somewhat sharpen compared to Fig. 4, and more so on harmonics, as expected when increasing data resolution (herein by roughly doubling it).

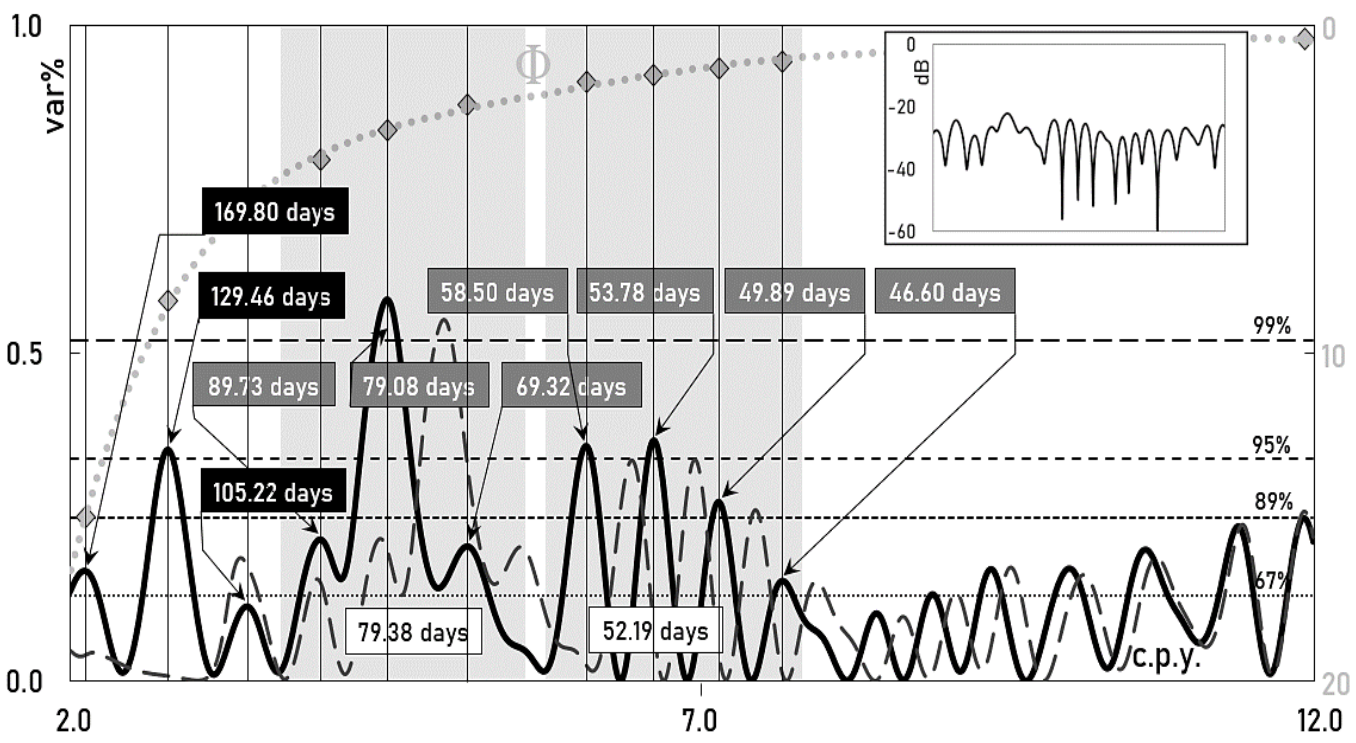

Figure 6. GVSA spectrum (solid curve) of all 1755 events from the 1 April 2022 release v.9 of the InSight raw seismic data in the 1–6 months band of the highest planetary energies. RR periods highlighted in black. The dashed solid curve represents the spectrum of the same data but after enforcing (mathematically removing the effects of; see, e.g., Wells et al., 1985) $P_{Rg}$. The enforcing reshuffled power in spectral peaks somewhat and introduced a slight phase shift in the high-frequency end of the spectra, but the train configuration is largely preserved — as another indication that the extracted periods represent a single physical process. Gray boxes mark subbands of spectral peak clustering, with the values of clustering periods highlighted in gray and cluster-averaged values in white. Callout: power spectrum of the same data. The analyzed time series of v.9 measurement is in Supplement Table C.

As seen from Fig. 6, GVSA of raw data from the v.9-release in the 30–180 days band has indeed recovered the complete RR: $P_{Rg}$ as 169.8-day with most of the power (callout) and Φ>12, and all of the commonly reported harmonics (Dimitropoulou et al., 2008): ~128-day as 129.5-day, ~102-day as 105.2-day, ~78-day as 79.4-day, and ~51-day as 52.2-day, Table 3. The harmonics were reported from data collected throughout the solar system and are thus not exclusive to Mars, so they are external and all that more plausible as the causal mechanism of marsquakes. On the other hand, the GVSA-extracted values of RR harmonics are all phase-shifted in the same sense — by ~1 day from the respective commonly reported values, Table 3 — thus confirming the high quality of GVSA as a method and indicating Mars-only features that affect RR harmonics in such a uniform way. The previous point is emphasized by serial anisotropic peak splitting, but to clusters that average to RR periodicities, Fig. 6. The above values are at least 67%-significant, except for the 102.5-day period that is significant at or above 63%, but this is of no concern given that the period is part of a well-known ensemble of ≥67%-significant harmonics.

$P_{Rg}$ shifted from the 154.5-day, Fig. 4, to 169.8-day, Fig. 6, or the previously unreported 165–175-day range (Gurgenashvili et al., 2017). This arbitrary shifting was likely due to a combination of the transient nature of $P_{Rg}$ and, to a minor extent, mixed quality of v.9-release raw data since the shift occurred in the same sense as when B/C events got expanded with D-quality events, Fig. 5 vs. Fig. 4. For a test (not shown), extending the spectral band to 1–9 months (30–270 days) resulted in an insignificantly varied $P_{Rg}$, of 170.39-day at ≥67% significance. Such robustness confirmed GVSA qualities rooted in this method's features, like the unique handling of spectral leakages and the non-dependence on the Nyquist frequency (Craymer, 1998).

| Period | Significance level | T [days] | $T_{solar}$ [days] | ΔT [%] | Φ | Mag. [var%] | Power [dB] |
|---|---|---|---|---|---|---|---|
| $P_{Rg}$ | ≥67% (0.13 var%) | 169.80 | 154 | -10.3 | 15.0 | 0.2 | -27.74 |
| $5/6 \cdot P_{Rg}$ | ≥95% (0.34 var%) | 129.46 | 128 | -1.1 | 8.4 | 0.4 | -24.51 |
| $2/3 \cdot P_{Rg}$ | ≥63% (0.11 var%) | 105.22 | 102 | -3.2 | 5.6 | 0.1 | 29.44 |
| $1/2 \cdot P_{Rg}$ | ≥99% (0.52 var%) | (79.38) | 78 | -1.8 | 3.2 | 0.6 | -22.32 |
| $1/3 \cdot P_{Rg}$ | ≥95% (0.34 var%) | (52.19) | 51 | -2.3 | 1.7 | 0.4 | -24.44 |

Table 3. Results of GVSA spectral analysis, Fig. 6, of the release v.9 of InSight raw seismic data of 1 April 2022. The $T_{solar}$ common values for $P_{Rg}$ and harmonics are as in Table 2. Averaged values from clusters of split spectral peaks are in parentheses, while other respective values refer to the majority of spectral peaks in a spectral cluster, Fig. 6.

## 3.2 *Moon results*

As found above for Mars, celestial bodies sample the IMF-resident RR process rotationally. Thus, as it orbits Earth, the magnetism-stripped Moon scans the Earth's magnetotail for ~1/4 and the IMF for ~3/4 of a revolution or synchronous rotation. As a result of the constant mutual interference by the smaller and intrinsic-magnetism-freed Moon against the more massive and magnetically shielded Earth, half of the RR periodicities in earthquakes occurrences and five of six periodicities in moonquakes occurrences split as the Moon traverses the IMF — but into clusters that in all cases average to RR periodicities as well, Fig. 7, same as for Mars, Fig. 6. The identical result from two astronomical bodies without inherent magnetism is stunning and it validates the starting physical hypothesis on the IMF causing seismicity in solid bodies.

The Moon-Earth magnetic reconnection significantly reduces the interference (i.e., it clears the signal), thereby reducing the total number of split spectral peaks in the terrestrial seismic response to the RR process from 3 to 2, cf. Fig. 7-c vs. Fig. 7-b, and in the lunar response, from 5 to 3, cf. Fig. 7-f vs. Fig. 7-e. Note that spectral magnitudes in lunar seismicity spectra are here also, as in (Omerbashich 2020a, 2020b), an order of magnitude smaller than in terrestrial seismicity spectra, as was expected since the energy emissions from moonquakes are considerably lower than from earthquakes. This energy-proportional response, sensed naturally by variance spectra, is reflected in the power spectra as narrower and less pronounced gray lobes, see left vs. right column of Fig. 7. Also, note again that a significance level of or above 67% suffices for validating physical period ensembles that get frequently reported, as is the case here.

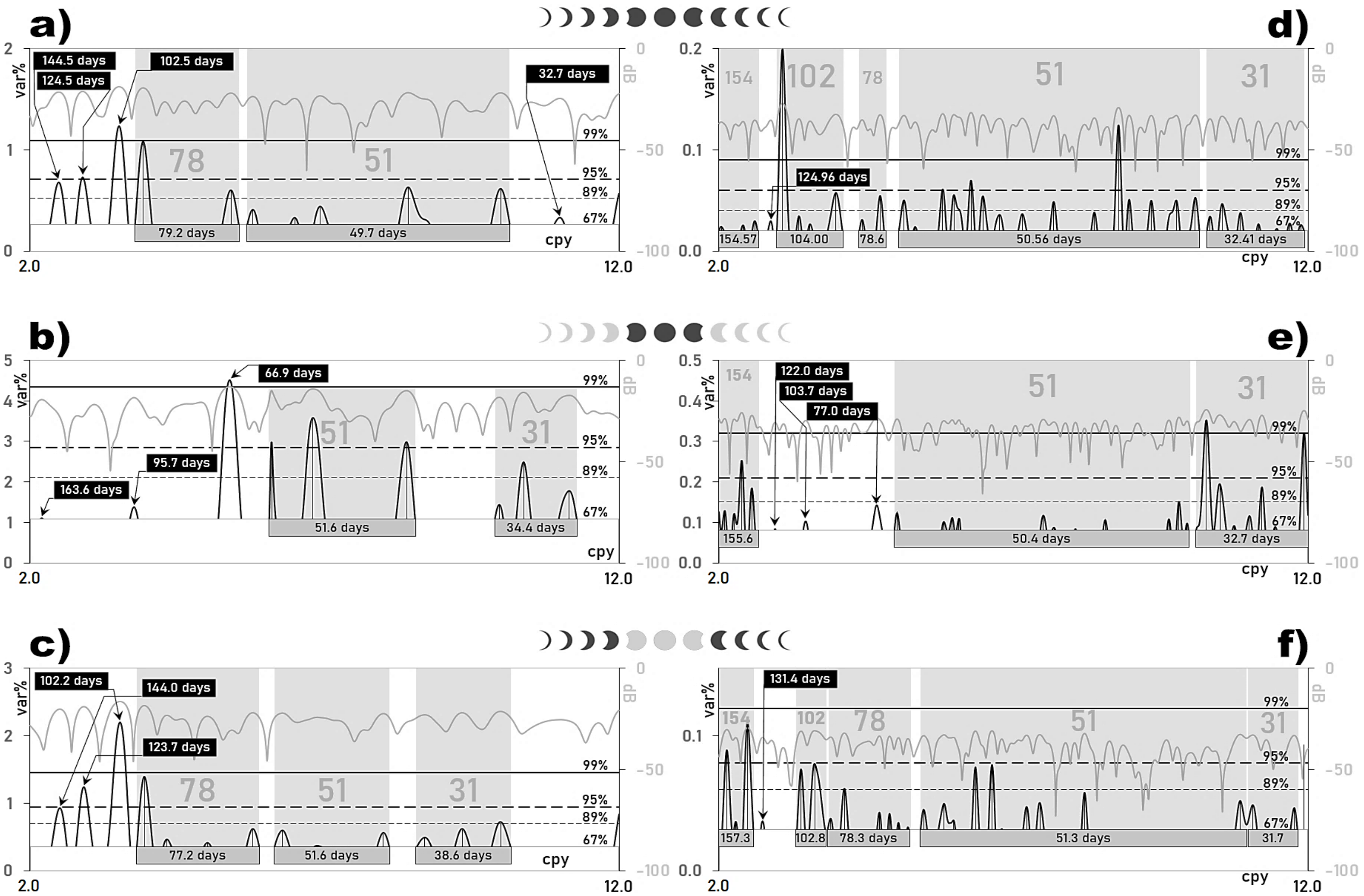

Figure 7. Significant peaks of GVSA variance spectra (dark line), in var%, and power spectra (gray), in dB — of terrestrial (left-column panels) and lunar (right) seismicity occurrences, Figs 2 & 3 and Supplements A & B. Shown are spectra of all events in Supplement Tables A-1 & B-1 (top row, left-to-right), events in Tables A-2 & B-2 while the Moon was traversing Earth's magnetotail (middle), and events in Tables A-3 & B-3 while the Moon was crossing the IMF (bottom). In all cases, events responded with all RR periodicities, except when the Moon traversed the magnetotail, which led to a suppression of the 128-day primary harmonics in terrestrial seismicity (panel b). Despite the splitting into clusters (gray boxes) within relatively powerless spectral subbands (gray lobes become narrower or less pronounced), all clusters too average to RR periodicities, Table 4. Statistical fidelity for the above spectra is astounding ($\Phi \gg 12$), ranging from low to high frequencies as $7.5\cdot10^6$–$3.8\cdot10^4$ and $7.5\cdot10^3$–$2.2\cdot10^2$ for terrestrial and lunar seismicity, respectively, where already $\Phi>12$ is considered reflective of a physical process (Omerbashich, 2006b). RR periodicities reported most often are 154-, 128-, 102-, 78-, 51-, and 31-day, but, depending on data type/epoch and processing methodology, the periods were reported in other ranges as well; see, e.g., Gurgenashvili et al. (2017). Note that non-sharp resonance peaks reflect low data resolution, as limited by the scantiness of Moon seismicity and $M_w$5.6+ Earth seismicity. Frequencies are in cycles per year (c.p.y.), band 30–180 days with periods indicated on significant peaks also.

The non-stratified lunar seismicity shows no clear orbital preference; thus, moonquakes occurred more frequently neither when the Moon was in the Earth's magnetotail nor the IMF. The interference of RR waves by the magnetotail has increased the power both band-wide (seen as the widened or otherwise more prominent lobes in power spectra) and in GVSA-typical representations of field energy levels against the linear variance background (Omerbashich, 2007), seen here as spectral peaks doubling in magnitudes, middle row of Fig. 7.

The difference in the spectral frequencies of global seismic responses of the Moon and Earth to the IMF-resident RR process, Fig. 8 and Table 5, measures the mismatch of individual RR periodicities due to different dynamics that influence the ecliptic-covering blanket of solar ejecta and solar wind. This mismatch can serve as a future gauge for dynamically estimating the wavelengths of the RR and mapping its change with time.

The main reason prior spectral investigations into earthquake and moonquake occurrences returned no RR periodicities lies in their analyzing events during traversals of IMF and magnetotail combined. For instance, Bulow et al. (2007) examined time series of deep moonquakes only, which, since the present study relied on the seisms mostly, should have given similar results had those authors also separated the events in the above way. Instead, *ibid*. erroneously concluded that deep moonquakes occur at periods corresponding to the draconic and anomalistic but not the synodic month. Namely, when all the events are analyzed combined, the resulting spectra encompass the RR process, albeit masked behind combined magnetotail and magnetosphere interferences, so that moonquakes spectra return clustered periods only, Fig. 7-d. In earthquakes, such spectra cluster for most of the spectral band, and the lowest RR frequencies get significantly mismatched due to other geodynamical drivers sharing the respective high-energies bands.

| T [days] | $T_{Rg}$ [days] | $T'_{Rg}$ [days] | ΔT [days] | ΔT [%] |
|---|---|---|---|---|
| From 845 (total) earthquakes: | | | | |
| 144.550 | 144.5 | 154 | 9.45 | 6.1% |
| 124.529 | 124.5 | 128 | 3.47 | 2.7% |
| 102.520 | 102.5 | 102 | -0.52 | -0.5% |
| 91.792 | | | | |
| 66.625 | 79.2 | 78 | -1.21 | -1.5% |
| 62.243 | | | | |
| 55.432 | | | | |
| 51.986 | | | | |
| 42.774 | | | | |
| 36.043 | 49.7 | 51 | 1.30 | 2.6% |
| 32.730 | 32.7 | 31 | -1.73 | -5.6% |
| From 210 magnetotail earthquakes: | | | | |
| 163.621 | 163.6 | 154 | -9.62 | -6.2% |
| 95.700 | 95.7 | 102 | 6.30 | 6.2% |
| 66.872 | 66.9 | 78 | 11.13 | 14.3% |
| 58.977 | | | | |
| 52.904 | | | | |
| 42.927 | 51.6 | 51 | -0.60 | -1.2% |
| 36.116 | | | | |
| 34.688 | | | | |
| 32.289 | 34.4 | 31 | -3.36 | -10.9% |
| From 635 IMF earthquakes: | | | | |
| 143.971 | 144.0 | 154 | 10.03 | 6.5% |
| 123.673 | 123.7 | 128 | 4.33 | 3.4% |
| 102.229 | 102.2 | 102 | -0.23 | -0.2% |
| 91.326 | | | | |
| 83.289 | | | | |
| 71.813 | | | | |
| 62.243 | 77.2 | 78 | 0.83 | 1.1% |
| 57.286 | | | | |
| 52.365 | | | | |
| 45.023 | 51.6 | 51 | -0.56 | -1.1% |
| 41.347 | | | | |
| 38.555 | | | | |
| 36.043 | 38.6 | 31 | -7.65 | -24.7% |

| T [days] | $T_{Rg}$ [days] | $T'_{Rg}$ [days] | ΔT [days] | ΔT [%] |
|---|---|---|---|---|
| From 10815 (total) natural moonquakes: | | | | |
| 176.467 | | | | |
| 149.352 | | | | |
| 137.899 | 154.6 | 154 | -0.57 | -0.4% |
| 124.962 | 125.0 | 128 | 3.04 | 2.4% |
| 116.842 | | | | |
| 107.099 | | | | |
| 101.651 | | | | |
| 90.407 | 104.0 | 102 | -2.00 | -2.0% |
| 81.220 | | | | |
| 76.066 | 78.6 | 78 | -0.64 | -0.8% |
| 69.996 | | | | |
| 67.500 | | | | |
| 62.135 | | | | |
| 60.161 | | | | |
| 57.377 | | | | |
| 55.517 | | | | |
| 53.217 | | | | |
| 50.384 | | | | |
| 46.840 | | | | |
| 42.978 | | | | |
| 40.971 | | | | |
| 40.147 | | | | |
| 38.555 | | | | |
| 36.894 | | | | |
| 35.686 | 50.6 | 51 | 0.44 | 0.9% |
| 34.822 | | | | |
| 34.128 | | | | |
| 33.183 | | | | |
| 32.260 | | | | |
| 31.388 | | | | |
| 30.796 | | | | |
| 30.303 | 32.4 | 31 | -1.41 | -4.6% |

| T [days] | $T_{Rg}$ [days] | $T'_{Rg}$ [days] | ΔT [days] | ΔT [%] |
|---|---|---|---|---|
| From 2865 magnetotail natural moonquakes: | | | | |
| 171.420 | | | | |
| 159.274 | | | | |
| 151.236 | | | | |
| 140.594 | 155.6 | 154 | -1.63 | -1.1% |
| 121.995 | 122.0 | 128 | 6.01 | 4.7% |
| 103.702 | 103.7 | 102 | -1.70 | -1.7% |
| 77.044 | 77.0 | 78 | 0.96 | 1.2% |
| 71.670 | | | | |
| 64.940 | | | | |
| 61.921 | | | | |
| 60.464 | | | | |
| 59.268 | | | | |
| 47.965 | | | | |
| 46.840 | | | | |
| 45.709 | | | | |
| 44.687 | | | | |
| 42.073 | | | | |
| 37.354 | | | | |
| 36.705 | | | | |
| 36.188 | 50.4 | 51 | 0.55 | 1.1% |
| 35.025 | | | | |
| 34.291 | | | | |
| 33.492 | | | | |
| 32.700 | | | | |
| 32.088 | | | | |
| 31.470 | | | | |
| 30.176 | 32.7 | 31 | -1.75 | -5.6% |

| T [days] | $T_{Rg}$ [days] | $T'_{Rg}$ [days] | ΔT [days] | ΔT [%] |
|---|---|---|---|---|
| From 7950 IMF natural moonquakes: | | | | |
| 169.004 | | | | |
| 157.875 | | | | |
| 145.133 | 157.3 | 154 | -3.34 | -2.2% |
| 131.351 | 131.4 | 128 | -3.35 | -2.6% |
| 106.151 | | | | |
| 99.403 | 102.8 | 102 | -0.78 | -0.8% |
| 92.500 | | | | |
| 87.122 | | | | |
| 76.227 | | | | |
| 73.576 | | | | |
| 71.385 | | | | |
| 69.188 | 78.3 | 78 | -0.33 | -0.4% |
| 65.772 | | | | |
| 61.815 | | | | |
| 60.161 | | | | |
| 56.654 | | | | |
| 54.261 | | | | |
| 52.982 | | | | |
| 49.825 | | | | |
| 48.352 | | | | |
| 43.869 | | | | |
| 37.509 | | | | |
| 33.183 | 51.3 | 51 | -0.31 | -0.6% |
| 32.552 | | | | |
| 31.889 | | | | |
| 30.587 | 31.7 | 31 | -0.68 | -2.2% |

Table 4. Values for the GVSA spectra, Fig. 7. Listed are all significant periods, T, in terrestrial (left column) and lunar seismicity (right). Spectral peaks at RR periodicities and cluster mean values, $T_{Rg}$, are compared against respective most often reported values of RR periods, $T_{Rg}'$. The comparison consisted in calculating differences, ΔT, between $T_{Rg}'$ and $T_{Rg}$.

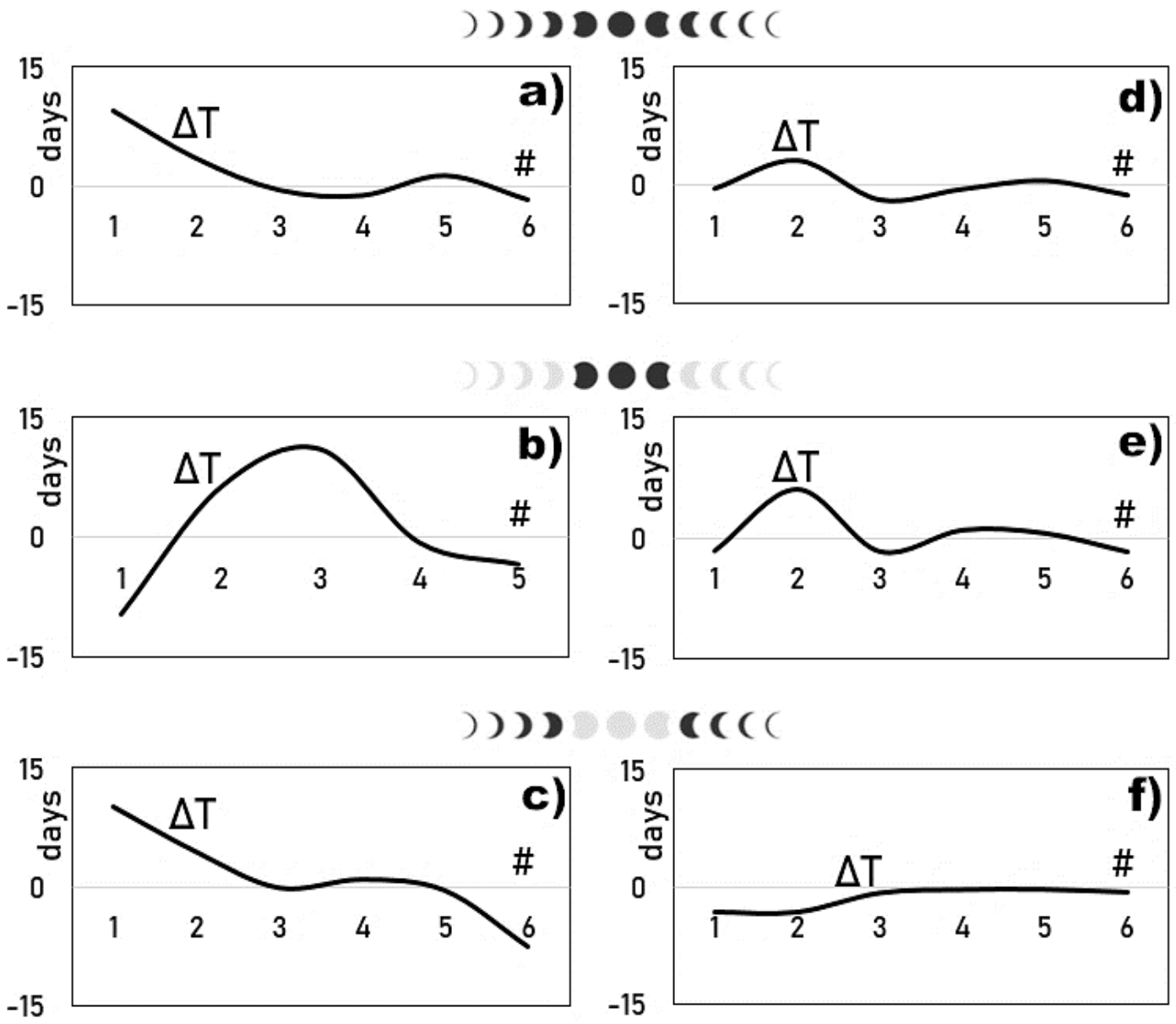

Figure 8. Matches, ΔT, of the most commonly reported RR periodicities to significant spectral peaks of earthquakes (left column) and moonquakes (right) occurrences, from Fig. 7, as differences between all GVSA-estimated significant periods from Table 1, including averages of split peaks, and the most commonly reported values of the nearest (longest six) RR periodicities, respectively: difference from ~154 days (value #1), from ~128 days (#2), from ~102 days (#3), from ~78 days (#4), from ~51 days (#5), and ~31 days (#6). Moon's passages through the geomagnetic tail cause the change in sign of the low-end mismatch as obtained from the terrestrial seismic response (to the RR process), panel b vs. panel c. Similarly, the Earth magnetotail also interferes with the lunar seismic response in lower frequencies, panel e vs. panel f. Finally, the matches for the de-magnetized Moon as it freely traverses the IMF are practically absolute across the spectral band, panel f, revealing that the IMF affects lunar seismicity more than the Earth magnetosphere could.

| | $P_i$ within tail [days] | $P_i$ within IMF [days] | $P_i^{average}$ [days] | $P_i^{common}$ [days] | Δ [days] | Δ [%] |
|---|---|---|---|---|---|---|
| Earth | 163.6 | 144.0 | 153.8 | 154 | 0.2 | 0.1% |
| Moon | 155.6 | 157.3 | 156.4 | 154 | -2.4 | -1.6% |
| Earth | x | 123.7 | 123.7 | 128 | 4.3 | 3.4% |
| Moon | 122.0 | 131.4 | 126.7 | 128 | 1.3 | 1.0% |
| Earth | 95.7 | 102.2 | 99.0 | 102 | 3 | 2.9% |
| Moon | 103.7 | 102.8 | 103.2 | 102 | -1.2 | -1.2% |
| Earth | 66.9 | 77.2 | 72.1 | 78 | 5.9 | 7.6% |
| Moon | 77.0 | 78.3 | 77.6 | 78 | 0.4 | 0.5% |
| Earth | 51.6 | 51.6 | 51.6 | 51 | -0.6 | -1.2% |
| Moon | 50.4 | 51.3 | 50.9 | 51 | 0.1 | 0.2% |
| Earth | 34.4 | 38.6 | 36.5 | 31 | -5.5 | -17.7% |
| Moon | 32.7 | 31.7 | 32.2 | 31 | -1.2 | -3.9% |

Table 5. The match of RR periodicities as determined from the spectra of earthquakes and moonquakes occurrences, Fig. 7 and Table 4, and the most commonly reported respective values, Fig. 8. The averaged Rieger period, $P_{Rg}$, is closer to the most often-reported value for Earth than for Moon. All averaged Rieger-type periodicities (resonance harmonics) agree better with respective commonly reported values for the Moon than for Earth. This comprehensive inconsistency indicates that different processes dominate seismicity on the two bodies, Figs. 2 & 3. When the Moon was in the Earth magnetotail, interference suppressed the 128-day (primary harmonic), allowing the power to shift to other periods whose peak magnitudes then doubled, revealing that lunar influence on earthquakes is primarily magnetic and alternating. This suppression of a long RR period (the strongest harmonic) in the Earth's seismic response to the RR process during magnetotail-Moon reconnection shows that the Earth is only slightly better protected from the magnetically induced seismicity than the Moon is. Then the Earth's magnetosphere offers a weak shield so that even the trace magnetism of the Moon can switch off the most potent ones of terrestrial seismicity harmonics. On the other hand, the global magnetism-free Moon responds to the entire RR process yieldingly.

### 3.3 *Earth results*

The $M_w$5.6+ earthquake occurrences show the Rieger period and weak and significantly offset periods from $P_{Rg}$, primarily and consistently in the low-frequency (most energetic) subband of the 1–6-month band, Fig. 7-a, b, c. Such a better sensitivity of earthquakes to the RR process in higher frequencies is not only expected because the Earth is magnetically shielded but also because the lunar orbit ($T_{Moon}$ and $1/2 \cdot T_{Moon}$) coincides with the $1/5 \cdot P_{Rg}$ and $1/10 \cdot P_{Rg}$ resonance harmonics, respectively — so that the tidal resonance (global seismicity generation mechanism) of the Earth demonstrated by Omerbashich (2022, 2020a) comes combined with magnetic influences of the IMF and the Moon.

Hence, any geodynamical impacts of the RR process also can be understood as coincidental. Stronger (primarily gravitational-tidal) processes act on Earth resonantly (*ibid.*) and with frequencies that are modifications of the Rieger period. On the other hand, the Earth's magnetosphere generally appears inadequate to (entirely) filter out cosmic radiation. Thus the RR process constantly injects energy that gets resonantly magnified, contributing a significant force of geodynamics through lower parts of the atmosphere. Then this excess in irregular variability of electrons in the Earth's atmosphere is due to the Rieger forcing of the Earth and contributes significantly to intermittently observed variations in the total electron content (TEC). Under this scenario, the inherent instability of the magnetic shielding and the transient nature of RR help explain why the precursory qualities of TEC are only rarely and irregularly, i.e., randomly associated with earthquakes: they are locally transient, mimicking the dynamics of their RR forcer.

## 4. Discussion

Depending on data, location, epoch, and methodology, RR periodicities were reported previously in varying ranges: 155–160 days, 160–165 days, 175–188 days, and 180–190 days; see, e.g., Gurgenashvili et al. (2017). Most of those studies indicated a leading (longest) periodicity in the 152–158-day range, which appears to be dominant, especially in the time phase from ~1979–1983, corresponding to the solar activity maximum (Chowdhury et al., 2008). The commonly reported value of this (longest) periodicity — the well-known Rieger period — is ~154 days (Chowdhury et al., 2009).

There are various proposals on the origin of the resonance process underlying the $P_{Rg}$ period in the dynamics of Sun-ejected particles and its modulations and harmonics, including possible influences of planetary constellations on the Sun, e.g., by Kurochkin (1998) and Abreu et al. (2012). One of the few undisputed claims of this kind comes from Bai and Cliver (1990), who suggested that the $P_{Rg}$ response could be simulated with a damped, periodically forced nonlinear oscillator exhibitting periodic and chaotic behavior. A more comprehensive explanation of $P_{Rg}$ and its modulations is also possible so that many of the arguments put forward in the past are simultaneously justified and that the solar system harbors an interplay of resonant feedback loops amongst individual planets and moons with the Sun. Again, the same agents that force solar activity could be responsible for at least some geomagnetic and seismic activity (Odintsov et al., 2006).

Even though seismic magnitudes for Mars are not available on a uniform seismic scale, which is a situation similar to that for the Moon, they are not of critical importance when a widely reported ensemble of physical periods is detected (and more so if detected alone) such as the resonance train reported in the present study not once but three times. The significance in such a case becomes immediately physical (corresponding to the statistical significance level of 100%) since the extracted information identifies itself as a congruent natural process. Such processes are always deterministic, as opposed to having been presumably stochastic until such a positive detection. Thus, even random magnitudes from some physically realistic range will suffice for such extractions, as only timings are critical in such cases (Omerbashich, 2020a, 2020b). If the process were not deterministic, seismic magnitudes would matter when estimating the statistical significance of periods detected. This causality is seen best in Figs. 5 & 6 for marsquakes and Fig. 7-right panels for moonquakes.

The longer-spanning v.9. release of InSight raw seismic data has revealed the entire RR, Fig. 6, thus excluding a tectonically active Mars. A general view is that Mars does not have active plate tectonics, so it should have no continuous seismicity either. The generation of InSight-sampled incessant seismicity externally (tidally-resonantly) by an electromagnetic phenome-non is supported additionally by the fact that mechanical resonance induces both small earthquakes (Ferrazzini and Aki, 1987) (Gupta, 2011) and small marsquakes (Suemoto et al., 2020; van Driel et al., 2021), while Flambaum and Pavlov (2016) speculated it could induce tectonic seismicity as well. Also, tides are not a negligible geodynamical force when the Earth, the oceans particularly, get excited not too far from resonance (Kundt and Jessner, 1986). Besides, both strong ($M_w$5.6+) earthquakes (Omerbashich, 2020a) and (mainly deep) moonquakes (Omerbashich, 2020b) can occur at tidal resonance periods. In addition, externally forced variations in the IMF can lead to surges of currents in the crust, which, in turn, can trigger seismicity (Simpson, 1967). The recent MAVEN (Mars Atmosphere and Volatile Evolution) mission results by Ramstad et al. (2020) support that scenario by detecting an externally forced electromagnetism on Mars.

At the same time, Love and Thomas (2013) reached their conclusion (on the insignificance of the solar-terrestrial triggering of seismicity) based on a statistical analysis of possible relationships between solar and regional-seismic activities. However, since a specific earthquake preparation area can respond to external action in various ways (from earthquake triggering to insensitivity), the simple correlation analysis of earthquake catalogs covering all (both sensitive and insensitive) regional seismic events cannot provide reliable results and conclusions on global significance. Besides, solar plasma macroscopic dynamics drive the Earth's magnetosphere so that they can be associated with terrestrial seismicity recordings, as found by Ringler et al. (2020) and Tape et al. (2020). However, as seen in the present study from the spatiotemporally independent astrophysical triple-confirmation of the physical hypothesis on solar wind-seismicity connection, what those authors detected was, to an extent, a seismic signal rather than noise — thus contrary to those authors classically (conservatively) assuming that any global seismic mechanisms previously unrecognized as seismogenic must outright be noise. Similarly erroneous was the conclusion by Tape et al. (2020) that seismometers record magnetic field variations only due to ferromagnetic materials used in their construction and not because those geomagnetic field variations are, to some extent, associated with ground motions genuinely, as herein pinpointed to an exact macroscopic dynamic process in the IMF — that of Rieger resonance. Notably, the GRACE-FO satellites that continuously measure the Earth's global gravity field get dynamically affected most strongly at the Rieger period, of ~153 days, as deduced recently by Tzamali and Pagiatakis (2021), who also used GVSA.

An alternative mechanism consists of the planetary angular velocity changes induced by a magnetohydrodynamic (MHD) coupling between the solar and planetary magnetic fields (Simpson, 1967; Alfvén, 1942). However, the lack of an intrinsic global magnetic field on Mars, with just a sporadic presence of local magnetic fields, makes this scenario less likely to occur naturally on Mars. Therefore, external magnetic fields like the IMF are required to create sustained couplings in the Mars case. The Martian environment can sustain such facilitation: while Mars already possesses a weak transient magnetic field confined to the very old and heavily cratered south highlands (Connerney et al., 2001), the solar wind–Mars interaction already strongly depends on the solar activity, solar zenith angle, and the altitude of interaction (Trotignon et al., 2000). When taken together, these facts greatly resemble the seismogenic environment implied by InSight and therefore agree with the outcome of this study on the solar wind being seismogenic.

Since Mars also reacts significantly at minima, Fig. 2 and cf. reanalysis of Viking 2 data by Lorenz et al. (2017), while the Sun only responds at maxima, Mars is closer to the source of the primary process, which then could lie in the upstream solar wind. An extensive data analysis by Simpson (1967) supports this conclusion by showing that the frequency of non-declustered 22561 global M5.5+ earthquakes spanning 13.5 years and their weaker foreshocks and aftershocks also peaked at a solar minimum. As indicated above, the heliophysics community itself largely remains in the dark about the time and location of the RR periodicities, just as the geophysical MHD community examines mainly subdiurnal spectral bands when investigating magnetospheric coupling, e.g., Denskat et al. (1983), rather than focusing dynamics studies on the band of highest interplanetary energy as in the present study.

Finally, the solar wind-induced seismicity on Mars, via solar-wind interactions with the atmosphere, translates into a premise that day-night variations in the geophysical properties of a planet are not necessarily associated with thermospheric heating (including associated wind speeds) alone. A recent report by Knapmeyer et al. (2021) supports this premise by finding that Mars seismicity is driven under the Sun declination, as the absolute majority of (high-frequency) InSight events occurred between local ~17h and midnight. Mean current densities in the Earth magnetopause are 18 and 27 nA $m^{-2}$ for dawn and duskside (the maximum current density in magnetotail is 10 nA $m^{-2}$ for dawn vs. 25 nA $m^{-2}$ for duskside), respectively. These densities are possibly due to transport via kinetic Alfvén waves during exposure to northward IMF, where the higher current density reveals the higher duskside magnetic field that also energizes faster (Walsh et al., 2014). Thus, while MHD simulations suggest that dawn-dusk asymmetry in solar wind entry by reconnection is related to ionospheric conductance (Li et al., 2008), the similarity between magnetized Earth's ionosphere and non-magnetized Mars's ionosphere (Burrell et al., 2020) makes it plausible that similar proportions of the dawn-dusk asymmetry should hold for Mars as well.

Unlike statistical correlations, which contain no physical meaning and can arise purely by chance, significant peaks in the spectrum of a time series of data describing a physical system reveal the (systematic) dynamics (physics) of that system. As with any mechanical resonance, which usually occurs and can even amplify system energy due to obstacles in propagation paths and other reasons, the results of the present study show that the RR process is a state of continuous mechanical resonance of the solar ejecta within the solar wind that blankets the ecliptic. The RR process gets modified as the wind emitted in the solar system — with the Sun as the largest emitter — encounters obstacles mainly in the form of rocky planets and moons (and probably also gaseous giants), as well as the occasional co-inclined visitor bodies such as comets that spend months or longer near the ecliptic.

While thermal moonquakes are events caused by the day-night thermal expansion of rock at sunset, sunlight also tenuously electrifies and ionizes the Moon (Halekas et al., 2018), so a continuous magnetohydrodynamic and electrical interaction (of a weak electromagnetic field with trapped water molecules) could also be rupturing lunar rock — instead of or in combination with the thermal seismogenesis. The continuity of the interplay is supported further by a recent finding by Shang et al. (2020) that the Earth's magnetotail does not constantly shield the Moon from external particles.

Solar-wind particles could also be heating the Moon's surface, as around half of the energy emitted in a solar flare is in the form of solar wind protons. The magnetic trapping of such a particle blast in the Earth magnetotail could temporarily lead to very high particle densities so that, when crossing this energized region, the surface of the Moon could become severely scorched. In addition, the IMF bound up with the solar particle blast wave could act as a magnetic bottle, retarding the dispersal of this region of high particle density on its way from the Sun. (LaViolette, 1983)

Furthermore, modeling suggests that lunar hydrogen gets cumulatively depleted as the Moon traverses the Earth magnetotail (Tucker et al., 2021). However, the hydration part of the interplay is supported further by the recent Chandrayaan-1 mission results showing that the Earth replenishes the lunar water supply via Earth magnetotail particles during each traversal — when a particle bridge gets established between the Earth and its natural satellite (Wang et al., 2021). The Earth's magnetosphere thus has the ability needed to reconnect with lunar magnetism regardless of the weakness and instability of the latter. Finally, the Moon's ionosphere extends when shielded by the Earth, which causes the lunar plasma to measurably perturb the plasma coming from the Sun and the Earth, leading to observable variations in electrical currents and the spatial distribution of electrons even (Halekas et al., 2018).

Then, as demonstrated in the present study, the electromagnetic induction of lunar seismicity is comparable to the mechanism for inducing seismicity on Mars. However, the surging crustal currents and magnetohydrodynamic coupling mechanisms for an external generation of seismicity could influence the bodies concurrently, but this could be to a considerably different degree individually. Furthermore, since Mars and Moon feature no active plate tectonics and probably no significant tectonics of any kind, i.e., any significant strain is absent, the mechanism of seismogenesis elucidated herein is not simply a trigger mechanism. Likewise, given the absence of precipitation and large bodies of water on Mars and the Moon, lubrication is excluded or likely unrelated also, and this mechanism probably rests on interactions among rock-embedded water molecules and magnetically trapped plasma.

Note that the here-presented statistically significant physical correspondence between internal and external lunar dynamics is more general (and therefore probably more plausible too) than to vaguely claim that lunar seismicity gets induced tidally — as often found in the literature. As mentioned above, while the RR train includes periods very close to the tidal periods (Omerbashich, 2020a) and gravitational co-phenomena cannot be excluded when considering solar-wind-induced seismicity, those two processes can couple and thus together affect selenodynamics at shared frequencies. This scenario is especially plausible given that the tidal processes concentrate stress repeatedly, and the tides pump fluids cyclically between suitably oriented shear zones (Frohlich and Nakamura, 2009), which could aid such coupling in the driving of selenodynamics both tidally and magnetohydrodynamically.

Obtaining a realistic description of IMF-induced seismicity to enable global estimates of RR wavelengths and the associated seismicity prediction in near real-time necessitates switching to advanced global seismic magnitude scales. A suitable choice is the DAS scale (Das et al., 2019; Das and Meneses, 2021), $M_{wg}$, which is based on global seismicity and can thus average rotational sampling of IMF-residing Rieger resonant process more faithfully than the locally tied $M_w$ could, all while offering a better representation of the Earth seismic energy budget and event emissions. Since derived from already first few cycles of P-waves, the DAS global scale is fully connected to source processes and therefore is a natural descriptor of global seismogenic processes like the Rieger's. As a function of both body magnitudes, $m_b$, and moment magnitudes, $M_o$, DAS is physically refined and closely related to high- and low-frequency ground motions. Being both a subtle and global indiscriminate descriptor of seismicity, DAS is pretied into tectonics and recurrence processes. Its magnitudes, $M_{wg}$, are globally optimized, non-saturating, and provide a range-wide measure of quake size. These advantages make DAS a preferred scale for planetary and lunar seismology.

The astonishing success of the here applied methodological approach, in which seismicity from three (all available) bodies produced the same result, renders it unnecessary to discuss in detail the limitations of the present study and the employed theory, methods, and argument.

## 5. Conclusions

Spectra of Mars, Moon, and Earth seismicity events occurrences in the highest planetary energies — which, unlike statistical correlations, measure relative system dynamics directly by establishing a physical correlation between two different dynamics (an equivalent of their statistical correlation at 100%) — are significantly periodic and only so with the well-known Rieger resonance (RR) of macroscopic dynamics of the solar wind. This result follows from the spatiotemporally and solar-cycle independent measurements from those only three bodies in our solar system for which we have *in situ* data. The result has thus confirmed (without speculating on causal mechanisms) previous claims that the solar wind/plasma dynamics are seismogenic and could be due to a magnetohydrodynamic interplay between magnetically trapped plasma and water molecules embedded within solid matter, or other forms of interaction. Thus RR, as a primary underlying resonant process in the heliosphere and the coupling mechanism of Sun-caused planetary seismicity, drives areophysics, selenophysics, and geophysics, while causing the tidal-resonant fracture response that the InSight and Apollo missions and terrestrial observatories had sampled. The seismicity phenomenon is not only due to local and regional but also global and astrophysical reasons.

Since the three data sets — marsquakes during a solar minimum, moonquakes during minima and maxima, and earthquakes outside either minima or maxima — refer to different and mutually non-overlapping intervals, astro-physically induced seismicity appears to occur independently of solar activity. Therefore, locally transient undulations in the solar wind suffice for the mechanism that the present study has demonstrated beyond doubt based on all available data. For example, a noticeable degeneration of 1/3 $P_{Rg}$ into a 3/11 $P_{Rg}$ split peak from marsquakes, with the second-highest significance level despite subband clustering, indicates a varying anisotropy and thus a surging interplanetary magnetic field (IMF) as felt in the vicinity of Mars. This indication and the Moon and Earth results call for landing seismic probes on other planets and moons. Such seismic missions to Europa and Titan were proposed recently (Lorenz and Panning, 2018) and could be used to map the extent of RR and estimate its wavelengths in near real-time.

Because only external electromagnetic forcing controls the band of highest energies in the Mars dynamic system and causes the only measurable continuous seismicity (as the indicator of tectonic activity), the present analysis supports the generally held view that Mars is tectonically inactive in terms of plate tectonics. The discovery of an external and principal causal mechanism of natural seismicity requires a reinterpretation of the phenomenon of seismicity to encompass the physics of a body-wide fracturing response of solid matter to variations in the interplanetary and other magnetic fields. For this, advanced global seismic scales are needed, such as the DAS magnitude scale, which is globally optimized and physically balanced and can thus represent reality better while implying easy decoupling from internal (tectonic) and external (within the IMF and rotational-inertial) physical processes in the Earth vicinity. Finally, based on the Mars and Moon results, the Earth result could be extendable to at least some $M_w$5.7- terrestrial seismicity, while the discovered mechanism, due to its astrophysical nature, likely affects gaseous giants as well. In addition, the main result indirectly provides a means to creating crude maps of the (planetary critical) solar wind's macroscopic dynamics shown here as entirely predictable. Then pending quality estimates of each RR component's wavelength, this result means that the solar wind as a threat is predictable. The predictability of solar-wind macroscopic dynamics enables a significant improvement in physics-based (space) weather and seismicity predictions for Earth and beyond, as well as higher safety of space missions and settlements and other installations on planets and moons.

The present study, as a reproducible strict computation from all (millions of) raw rates of seismicity occurrences from the Earth and both space missions that collected *in situ* seismometer measurements (on the Moon and Mars) for years, has revealed a previously reported mechanical resonance of the interplanetary magnetic field. Any discovery of a known dynamic in any global astrophysical dynamics (here of highest planetary energy levels), which also was successfully twice reproduced, confirms the associated computations as correct beyond doubt ("by definition") and overrides or completes, and thereby redefines, all considerations at lower energy levels at once, including those in disagreement with the result.

## Acknowledgments

This paper benefited from discussions with and suggestions by Ralph Lorenz (Johns Hopkins University) and Yosio Nakamura (the University of Texas at Austin). Barbara Pope (NASA) helped with the data. Comments and suggestions from the peer reviewers Victor Novikov (Russian Academy of Sciences) and Dimitar Ouzounov (Chapman University) are appreciated. The here used scientific software LSSA v.5.0, based on the rigorous method of spectral analysis by Vaníček (1969, 1971), was from: http://www2.unb.ca/gge/Research/GRL/LSSA/sourceCode.html.

## Supplementary materials

See supplementary materials A&B&C for complete diverse (spatiotemporally & solar-cycle independent) seismicity time series from the Moon, Earth, and Mars, whose Gauss–Vaníček variance spectra reveal are all significantly periodic only with the Rieger period and its resonance harmonics (Rieger-type periodicities) of the solar wind's macroscopic dynamics.

## Availability of Data

Moonquakes time series are in the Supplement A Tables (originally at: https://hdl.handle.net/2152/65671). Earthquakes time series are in the Supplement B Tables (originally at: https://n2t.net/ark:/88439/x020219). Marsquakes time series are in Table 1 (originally at: https://doi.org/10.1016/j.pepi.2020.106595, the corrected Supplementary Table 2 of their Marsquakes Catalog). The v.9 InSight data are in the Supplement C Table (originally at: https://pds-geosciences.wustl.edu/insight/urn-nasa-pds-insight_seis/data_derived/). All data are in the public domain.

Supplement A – moonquakes Tables

| T(min) | Mag' |
|---|---|
| 0.0 | 5.9 |
| 841.0 | 5.8 |
| 37461.0 | 6.8 |
| 39252.0 | 5.6 |
| 40362.0 | 5.9 |
| 171792.0 | 6.9 |
| 175872.0 | 6.8 |
| 177312.0 | 6.0 |
| 179112.0 | 5.6 |
| 179287.0 | 6.6 |
| 180192.0 | 7.1 |
| 180372.0 | 7.2 |
| 180720.0 | 5.7 |
| 181167.0 | 7.0 |
| 181512.0 | 7.4 |
| 181812.0 | 5.9 |
| 182104.0 | 5.6 |
| 182896.0 | 7.0 |
| 183012.0 | 6.8 |
| 188261.0 | 5.7 |
| 189885.0 | 6.4 |
| 192012.0 | 6.7 |
| 194115.0 | 6.2 |
| 195295.0 | 6.0 |
| 195704.0 | 6.3 |
| 195972.0 | 7.3 |
| 196092.0 | 6.9 |
| 196507.0 | 5.6 |
| 196997.0 | 6.9 |
| 197352.0 | 7.2 |
| 197772.0 | 7.2 |
| 198418.0 | 6.2 |
| 198912.0 | 7.3 |
| 199152.0 | 5.6 |
| 199629.0 | 5.6 |
| 199866.0 | 6.2 |
| 200081.0 | 7.3 |
| 200292.0 | 6.9 |
| 200412.0 | 7.2 |
| 201510.0 | 6.4 |
| 201596.0 | 6.7 |
| 202414.0 | 7.5 |
| 206112.0 | 6.0 |
| 210283.0 | 7.2 |
| 213372.0 | 6.1 |
| 218606.0 | 6.9 |
| 219432.0 | 6.0 |
| 219712.0 | 7.4 |
| 220452.0 | 7.4 |
| 220752.0 | 5.8 |
| 220812.0 | 5.9 |
| 221019.0 | 6.5 |
| 221892.0 | 6.2 |
| 223127.0 | 7.1 |
| 226152.0 | 6.0 |
| 228810.0 | 6.7 |
| 229512.0 | 6.2 |
| 229657.0 | 6.7 |
| 229734.0 | 6.4 |
| 231092.0 | 6.7 |
| 231126.0 | 6.5 |
| 231375.0 | 6.1 |
| 231677.0 | 5.7 |
| 233472.0 | 7.3 |
| 234100.0 | 5.6 |
| 234610.0 | 6.8 |
| 234852.0 | 6.8 |
| 235272.0 | 6.4 |
| 235400.0 | 6.5 |
| 236221.0 | 5.7 |
| 236412.0 | 6.4 |
| 236941.0 | 7.3 |
| 237172.0 | 6.5 |
| 237672.0 | 6.9 |
| 237732.0 | 6.1 |
| 237972.0 | 6.7 |
| 238032.0 | 6.4 |
| 239292.0 | 6.1 |
| 239412.0 | 5.9 |
| 240438.0 | 5.9 |
| 240612.0 | 7.2 |
| 241967.0 | 5.6 |
| 242693.0 | 7.0 |
| 246492.0 | 6.7 |
| 254367.0 | 6.7 |
| 255132.0 | 7.0 |
| 255662.0 | 6.0 |
| 256056.0 | 6.3 |
| 256812.0 | 6.1 |
| 257238.0 | 6.9 |
| 259009.0 | 5.7 |
| 259392.0 | 6.0 |
| 260772.0 | 6.8 |
| 262794.0 | 7.4 |
| 262979.0 | 6.9 |
| 266429.0 | 6.8 |
| 267430.0 | 5.6 |
| 268212.0 | 5.8 |
| 270702.0 | 6.9 |
| 270912.0 | 6.4 |
| 270972.0 | 7.1 |
| 272292.0 | 6.5 |
| 272831.0 | 6.5 |
| 273042.0 | 6.1 |
| 273092.0 | 6.1 |
| 273429.0 | 7.2 |
| 274727.0 | 7.4 |
| 274841.0 | 7.2 |
| 276181.0 | 6.1 |
| 276292.0 | 6.1 |
| 276732.0 | 6.3 |
| 276972.0 | 6.4 |
| 277364.0 | 6.3 |
| 277479.0 | 6.5 |
| 278412.0 | 6.5 |
| 279432.0 | 5.7 |
| 279552.0 | 5.9 |
| 280191.0 | 7.4 |
| 282732.0 | 6.8 |
| 284112.0 | 7.4 |
| 287866.0 | 6.5 |
| 292973.0 | 5.7 |
| 295241.0 | 5.7 |
| 295392.0 | 6.7 |
| 295632.0 | 5.9 |
| 296090.0 | 6.8 |
| 296142.0 | 6.8 |
| 296680.0 | 6.3 |
| 296832.0 | 5.7 |
| 297132.0 | 6.2 |
| 297192.0 | 6.1 |
| 297647.0 | 7.2 |
| 298392.0 | 6.9 |
| 298717.0 | 5.9 |
| 298837.0 | 7.4 |
| 298907.0 | 6.0 |
| 299064.0 | 6.7 |
| 299189.0 | 6.2 |
| 299243.0 | 7.2 |
| 299476.0 | 5.9 |
| 300240.0 | 5.6 |
| 300663.0 | 7.3 |
| 302712.0 | 5.8 |
| 305652.0 | 5.9 |
| 307723.0 | 6.5 |
| 312160.0 | 7.0 |
| 312185.0 | 7.4 |
| 312552.0 | 6.9 |
| 312732.0 | 5.8 |
| 312972.0 | 6.4 |
| 313151.0 | 6.9 |
| 313413.0 | 6.7 |
| 313816.0 | 7.2 |
| 314412.0 | 5.6 |
| 314472.0 | 7.3 |
| 314740.0 | 5.8 |
| 314864.0 | 6.9 |
| 315612.0 | 5.5 |
| 317834.0 | 6.4 |
| 317962.0 | 6.1 |
| 318237.0 | 6.1 |
| 318312.0 | 6.1 |
| 318732.0 | 7.3 |
| 319546.0 | 5.7 |
| 320438.0 | 6.3 |
| 321067.0 | 6.4 |
| 326649.0 | 6.8 |
| 331752.0 | 6.5 |
| 332331.0 | 6.5 |
| 333648.0 | 7.5 |
| 334152.0 | 6.1 |
| 336012.0 | 7.1 |
| 338233.0 | 5.6 |
| 338712.0 | 6.5 |
| 339362.0 | 5.7 |
| 340152.0 | 7.0 |
| 341592.0 | 6.9 |
| 342462.0 | 5.9 |
| 342972.0 | 7.0 |
| 345076.0 | 6.7 |
| 345612.0 | 7.1 |
| 347671.0 | 6.1 |
| 352447.0 | 6.4 |
| 352660.0 | 7.0 |
| 352732.0 | 6.6 |
| 353447.0 | 6.0 |
| 353529.0 | 6.3 |
| 353583.0 | 7.2 |
| 353872.0 | 7.2 |
| 354120.0 | 7.1 |
| 354219.0 | 6.2 |
| 354930.0 | 7.1 |
| 355251.0 | 6.0 |
| 355720.0 | 5.7 |
| 355752.0 | 6.4 |
| 355872.0 | 7.2 |
| 356052.0 | 7.5 |
| 356172.0 | 6.8 |
| 356232.0 | 7.4 |
| 356415.0 | 6.7 |
| 358155.0 | 5.7 |
| 359221.0 | 6.8 |
| 361265.0 | 6.6 |
| 362879.0 | 5.9 |
| 363562.0 | 5.7 |
| 369317.0 | 5.7 |
| 371530.0 | 7.1 |
| 371652.0 | 7.4 |
| 371892.0 | 6.2 |
| 371952.0 | 7.2 |
| 373854.0 | 6.3 |
| 373937.0 | 5.7 |
| 374242.0 | 5.7 |
| 374472.0 | 7.1 |
| 375054.0 | 6.1 |
| 376587.0 | 6.7 |
| 377532.0 | 6.1 |
| 377987.0 | 5.7 |
| 378792.0 | 5.7 |
| 379092.0 | 7.4 |
| 379212.0 | 6.6 |
| 379613.0 | 6.7 |
| 379840.0 | 7.4 |
| 380207.0 | 5.5 |
| 381972.0 | 6.8 |
| 384267.0 | 7.0 |
| 386734.0 | 6.7 |
| 389800.0 | 7.4 |
| 390123.0 | 6.3 |
| 391872.0 | 6.0 |
| 392112.0 | 6.7 |
| 392172.0 | 5.8 |
| 392301.0 | 7.4 |
| 392561.0 | 7.4 |
| 392604.0 | 5.7 |
| 392933.0 | 7.4 |
| 393372.0 | 5.8 |
| 393734.0 | 7.2 |
| 394427.0 | 6.9 |
| 394692.0 | 6.7 |
| 395220.0 | 7.5 |
| 395425.0 | 6.6 |
| 396252.0 | 7.3 |
| 397128.0 | 7.1 |
| 397572.0 | 5.5 |
| 397812.0 | 6.5 |
| 399072.0 | 7.1 |
| 399252.0 | 6.8 |
| 399372.0 | 7.3 |
| 402372.0 | 6.7 |
| 402462.0 | 7.2 |
| 405400.0 | 5.9 |
| 408248.0 | 7.0 |
| 411377.0 | 5.9 |
| 412876.0 | 5.7 |
| 413352.0 | 6.5 |
| 413712.0 | 6.2 |
| 413772.0 | 7.2 |
| 414912.0 | 7.0 |
| 415534.0 | 6.8 |
| 415700.0 | 5.8 |
| 416878.0 | 6.1 |
| 417012.0 | 7.1 |
| 417792.0 | 5.5 |
| 417912.0 | 6.3 |
| 418152.0 | 7.3 |
| 418902.0 | 6.6 |
| 419292.0 | 5.6 |
| 419412.0 | 7.4 |
| 419759.0 | 6.9 |
| 420852.0 | 7.1 |
| 421833.0 | 7.5 |
| 422552.0 | 7.2 |
| 427244.0 | 5.6 |
| 430573.0 | 7.1 |
| 430872.0 | 5.9 |
| 430992.0 | 7.2 |
| 431360.0 | 5.7 |
| 431519.0 | 6.7 |
| 431844.0 | 6.9 |
| 432020.0 | 5.6 |
| 432252.0 | 5.9 |
| 432759.0 | 7.4 |
| 432794.0 | 7.0 |
| 433404.0 | 6.8 |
| 434184.0 | 6.0 |
| 436512.0 | 5.8 |
| 436632.0 | 6.8 |
| 436812.0 | 6.4 |
| 452412.0 | 6.9 |
| 452833.0 | 6.5 |
| 452915.0 | 7.4 |
| 453259.0 | 7.4 |
| 454272.0 | 7.5 |
| 455675.0 | 6.9 |
| 456372.0 | 6.1 |
| 456912.0 | 7.2 |
| 457846.0 | 7.0 |
| 458412.0 | 6.1 |
| 459083.0 | 6.1 |
| 459552.0 | 6.1 |
| 459912.0 | 6.8 |
| 460027.0 | 5.5 |
| 461112.0 | 6.4 |
| 463727.0 | 6.1 |
| 465252.0 | 7.2 |
| 467342.0 | 5.8 |
| 467998.0 | 6.8 |
| 469812.0 | 5.9 |
| 470377.0 | 5.7 |
| 470739.0 | 7.1 |
| 470922.0 | 7.2 |
| 471918.0 | 7.0 |
| 472003.0 | 5.6 |
| 472115.0 | 6.4 |
| 472314.0 | 7.2 |
| 472512.0 | 6.0 |
| 473230.0 | 6.5 |
| 473308.0 | 6.7 |
| 473734.0 | 5.7 |
| 473952.0 | 6.8 |
| 474490.0 | 5.5 |
| 475872.0 | 6.3 |
| 475933.0 | 6.9 |
| 476692.0 | 7.2 |
| 477977.0 | 5.9 |
| 485352.0 | 5.9 |
| 488791.0 | 7.5 |
| 489317.0 | 5.5 |
| 490032.0 | 5.6 |
| 491412.0 | 5.8 |
| 492024.0 | 5.7 |
| 493690.0 | 6.4 |
| 494627.0 | 5.9 |
| 494790.0 | 6.8 |
| 494850.0 | 7.3 |
| 497901.0 | 5.5 |
| 498492.0 | 5.9 |
| 498672.0 | 6.4 |
| 499284.0 | 6.8 |
| 499932.0 | 5.7 |
| 500869.0 | 6.7 |
| 502872.0 | 6.0 |
| 503412.0 | 5.8 |
| 504962.0 | 5.8 |
| 507072.0 | 5.5 |
| 507312.0 | 5.8 |
| 507738.0 | 6.2 |
| 508187.0 | 5.6 |
| 508512.0 | 6.5 |
| 508752.0 | 7.4 |
| 508812.0 | 6.1 |
| 509557.0 | 6.6 |
| 509706.0 | 7.0 |
| 509832.0 | 7.4 |
| 510072.0 | 6.1 |
| 511245.0 | 7.3 |
| 511692.0 | 5.8 |
| 513594.0 | 6.1 |
| 514452.0 | 7.5 |
| 514769.0 | 7.2 |
| 514796.0 | 6.0 |
| 516842.0 | 6.0 |
| 518712.0 | 5.6 |
| 520924.0 | 6.5 |
| 527112.0 | 6.1 |
| 527352.0 | 7.4 |
| 528462.0 | 6.4 |
| 528552.0 | 6.3 |
| 528732.0 | 7.0 |
| 529800.0 | 7.2 |
| 530472.0 | 5.8 |
| 530674.0 | 6.5 |
| 531381.0 | 5.6 |
| 532992.0 | 6.5 |
| 533172.0 | 6.9 |
| 534044.0 | 6.1 |
| 534053.0 | 6.6 |
| 534372.0 | 6.6 |
| 534552.0 | 7.0 |
| 534672.0 | 5.6 |
| 534792.0 | 5.6 |
| 536521.0 | 5.7 |
| 537372.0 | 5.6 |
| 538217.0 | 6.5 |
| 538752.0 | 6.7 |
| 538932.0 | 7.3 |
| 544160.0 | 7.2 |
| 544392.0 | 6.8 |
| 544632.0 | 6.3 |
| 544872.0 | 6.0 |
| 544950.0 | 7.0 |
| 545029.0 | 5.8 |
| 546012.0 | 7.2 |
| 546635.0 | 6.9 |
| 546870.0 | 7.2 |
| 547916.0 | 6.1 |
| 548306.0 | 6.4 |
| 548458.0 | 7.5 |
| 548482.0 | 7.0 |
| 548597.0 | 6.2 |
| 549290.0 | 6.9 |
| 549364.0 | 6.0 |
| 549700.0 | 7.3 |
| 549920.0 | 5.8 |
| 549950.0 | 5.9 |
| 550272.0 | 7.1 |
| 551198.0 | 7.3 |
| 551552.0 | 7.1 |
| 552597.0 | 6.2 |
| 552930.0 | 6.0 |
| 553917.0 | 7.5 |
| 554772.0 | 6.2 |
| 555175.0 | 6.7 |
| 555469.0 | 6.8 |
| 558730.0 | 7.4 |
| 559272.0 | 6.9 |
| 559530.0 | 5.9 |
| 560194.0 | 5.9 |
| 561912.0 | 5.7 |
| 563292.0 | 6.1 |
| 566639.0 | 7.3 |
| 567672.0 | 6.6 |
| 567912.0 | 6.4 |
| 568222.0 | 7.0 |
| 569772.0 | 6.8 |
| 570040.0 | 6.8 |
| 570552.0 | 6.1 |
| 570672.0 | 7.1 |
| 570732.0 | 6.9 |
| 571140.0 | 5.9 |
| 571932.0 | 7.0 |
| 572482.0 | 6.6 |
| 572592.0 | 6.3 |
| 572810.0 | 7.3 |
| 572987.0 | 5.8 |
| 573432.0 | 7.2 |
| 574144.0 | 6.1 |
| 574590.0 | 6.4 |
| 575052.0 | 5.9 |
| 575495.0 | 6.3 |
| 577333.0 | 6.0 |
| 579896.0 | 6.7 |
| 579965.0 | 7.2 |
| 582952.0 | 6.3 |
| 585322.0 | 6.0 |
| 586740.0 | 6.3 |
| 587204.0 | 7.5 |
| 588176.0 | 6.4 |
| 588262.0 | 5.7 |
| 588601.0 | 6.0 |
| 588752.0 | 6.1 |
| 588983.0 | 6.5 |
| 589452.0 | 5.6 |
| 589512.0 | 7.4 |
| 590421.0 | 5.7 |
| 590892.0 | 6.9 |
| 591173.0 | 7.1 |
| 592392.0 | 7.0 |
| 592547.0 | 6.8 |
| 593894.0 | 6.0 |
| 593930.0 | 6.2 |
| 594675.0 | 5.9 |
| 595152.0 | 5.8 |
| 595509.0 | 6.2 |
| 596592.0 | 6.1 |
| 597473.0 | 6.5 |
| 598620.0 | 5.6 |
| 600052.0 | 5.6 |
| 600194.0 | 6.1 |
| 606023.0 | 6.1 |
| 606116.0 | 6.9 |
| 608112.0 | 5.5 |
| 609372.0 | 7.1 |
| 610172.0 | 6.0 |
| 610544.0 | 7.3 |
| 610632.0 | 6.9 |
| 611971.0 | 5.8 |
| 612072.0 | 6.8 |
| 612825.0 | 6.1 |
| 612976.0 | 7.0 |
| 613632.0 | 6.3 |
| 613992.0 | 6.2 |
| 614152.0 | 7.3 |
| 614219.0 | 6.5 |
| 616172.0 | 6.5 |
| 616752.0 | 6.6 |
| 617394.0 | 6.6 |
| 617709.0 | 5.5 |
| 620952.0 | 6.5 |
| 623772.0 | 6.0 |
| 627327.0 | 7.2 |
| 627398.0 | 6.9 |
| 627584.0 | 6.6 |
| 627640.0 | 6.7 |
| 627844.0 | 6.4 |
| 628152.0 | 7.3 |
| 628665.0 | 6.2 |
| 628822.0 | 5.6 |
| 629022.0 | 7.1 |
| 629031.0 | 6.9 |
| 629472.0 | 7.3 |
| 629532.0 | 6.8 |
| 629772.0 | 6.1 |
| 630690.0 | 5.9 |
| 631152.0 | 6.6 |
| 631380.0 | 5.9 |
| 632412.0 | 6.2 |
| 633852.0 | 6.3 |
| 634152.0 | 7.1 |
| 640453.0 | 5.8 |
| 645016.0 | 6.1 |
| 645372.0 | 6.8 |
| 646632.0 | 5.7 |
| 646992.0 | 7.5 |
| 647701.0 | 6.0 |
| 648072.0 | 6.8 |
| 648432.0 | 7.4 |
| 648492.0 | 5.8 |
| 649146.0 | 7.3 |
| 649692.0 | 7.3 |
| 650403.0 | 6.0 |
| 651372.0 | 5.6 |
| 651504.0 | 5.8 |
| 651751.0 | 6.7 |
| 652008.0 | 5.9 |
| 653279.0 | 5.9 |
| 653703.0 | 6.4 |
| 653832.0 | 5.6 |
| 655512.0 | 5.9 |
| 655919.0 | 5.6 |
| 656523.0 | 7.4 |
| 657072.0 | 6.8 |
| 659137.0 | 6.8 |
| 659344.0 | 5.6 |
| 665194.0 | 6.8 |
| 666912.0 | 6.2 |
| 667212.0 | 6.0 |
| 667766.0 | 6.1 |
| 667886.0 | 6.8 |
| 668147.0 | 6.5 |
| 668292.0 | 5.6 |
| 668532.0 | 5.5 |
| 669296.0 | 6.6 |
| 671172.0 | 7.4 |
| 671954.0 | 6.8 |
| 673371.0 | 5.6 |
| 673992.0 | 7.3 |
| 674412.0 | 7.4 |
| 677112.0 | 6.9 |
| 679718.0 | 6.0 |
| 680335.0 | 6.7 |
| 684372.0 | 7.5 |
| 686272.0 | 6.7 |
| 686900.0 | 6.8 |
| 687192.0 | 6.8 |
| 687312.0 | 7.5 |
| 687372.0 | 5.8 |
| 689534.0 | 6.6 |
| 690132.0 | 5.8 |
| 691272.0 | 5.7 |
| 691452.0 | 6.1 |
| 691512.0 | 5.8 |
| 692160.0 | 5.5 |
| 692568.0 | 6.8 |
| 697212.0 | 6.8 |
| 700548.0 | 5.9 |
| 702972.0 | 6.8 |
| 703092.0 | 7.3 |
| 703543.0 | 6.6 |
| 705014.0 | 6.3 |
| 705732.0 | 6.4 |
| 706973.0 | 7.4 |
| 707102.0 | 6.8 |
| 707172.0 | 5.7 |
| 708536.0 | 5.6 |
| 708732.0 | 6.0 |
| 708852.0 | 5.9 |
| 709516.0 | 7.0 |
| 709704.0 | 6.8 |
| 710930.0 | 6.8 |
| 712290.0 | 6.0 |
| 717180.0 | 7.2 |
| 722145.0 | 6.7 |
| 725032.0 | 5.7 |
| 726192.0 | 6.8 |
| 726687.0 | 7.2 |
| 727163.0 | 6.9 |
| 728892.0 | 5.9 |
| 729404.0 | 7.0 |
| 729951.0 | 6.6 |
| 730020.0 | 6.4 |
| 733392.0 | 6.2 |
| 735412.0 | 7.2 |
| 739032.0 | 6.7 |
| 740811.0 | 7.4 |
| 744289.0 | 7.1 |
| 744440.0 | 6.1 |
| 745726.0 | 6.4 |
| 745834.0 | 6.2 |
| 746540.0 | 6.7 |
| 746882.0 | 6.0 |
| 747132.0 | 5.7 |
| 747160.0 | 6.5 |
| 748240.0 | 6.6 |
| 748324.0 | 6.9 |
| 748760.0 | 5.7 |
| 750372.0 | 6.1 |
| 750492.0 | 7.1 |
| 751193.0 | 6.1 |
| 756049.0 | 6.6 |
| 756192.0 | 5.8 |
| 757023.0 | 6.1 |
| 757288.0 | 5.8 |
| 757872.0 | 6.2 |
| 763955.0 | 6.9 |
| 764268.0 | 7.5 |
| 764462.0 | 6.5 |
| 764892.0 | 7.2 |
| 765072.0 | 7.1 |
| 765132.0 | 7.5 |
| 765764.0 | 6.5 |
| 765990.0 | 6.1 |
| 766078.0 | 5.5 |
| 766152.0 | 7.1 |
| 766774.0 | 5.8 |
| 767160.0 | 7.5 |
| 767284.0 | 5.6 |
| 768320.0 | 6.3 |
| 768953.0 | 6.1 |
| 775032.0 | 5.8 |
| 776652.0 | 5.7 |
| 776823.0 | 5.8 |
| 780316.0 | 5.9 |
| 781032.0 | 5.7 |
| 781628.0 | 6.3 |
| 781949.0 | 7.1 |
| 782052.0 | 5.5 |
| 782112.0 | 6.6 |
| 782292.0 | 7.1 |
| 782589.0 | 6.8 |
| 784036.0 | 6.9 |
| 784243.0 | 6.6 |
| 784491.0 | 6.3 |
| 784872.0 | 6.9 |
| 785292.0 | 7.2 |
| 785440.0 | 7.1 |
| 785704.0 | 6.5 |
| 786000.0 | 7.2 |
| 786139.0 | 6.4 |
| 787520.0 | 5.6 |
| 787620.0 | 5.6 |
| 790245.0 | 6.9 |
| 791471.0 | 6.6 |
| 794952.0 | 5.8 |
| 795252.0 | 5.8 |
| 803652.0 | 5.6 |
| 804666.0 | 6.4 |
| 805212.0 | 6.0 |
| 805392.0 | 7.5 |
| 805512.0 | 7.3 |
| 806017.0 | 5.9 |
| 806092.0 | 6.3 |
| 806418.0 | 6.6 |
| 806712.0 | 7.0 |
| 806832.0 | 7.1 |
| 807972.0 | 6.4 |
| 809352.0 | 5.6 |
| 810115.0 | 6.8 |
| 810402.0 | 7.2 |
| 814781.0 | 7.2 |
| 815412.0 | 6.2 |
| 815669.0 | 6.0 |
| 816170.0 | 6.3 |
| 816472.0 | 6.2 |
| 817245.0 | 5.7 |
| 819149.0 | 6.2 |
| 822003.0 | 7.1 |
| 822612.0 | 6.9 |
| 823000.0 | 5.6 |
| 823045.0 | 7.2 |
| 823119.0 | 6.9 |
| 823184.0 | 6.6 |
| 823307.0 | 6.7 |
| 823572.0 | 7.3 |
| 823624.0 | 7.4 |
| 823640.0 | 6.0 |
| 823701.0 | 6.3 |
| 823752.0 | 6.2 |
| 823932.0 | 6.3 |
| 824357.0 | 6.3 |
| 824391.0 | 6.5 |
| 824599.0 | 5.8 |
| 824702.0 | 7.1 |
| 824732.0 | 7.2 |
| 825094.0 | 7.2 |
| 825107.0 | 6.5 |
| 825746.0 | 7.5 |
| 825816.0 | 6.5 |
| 826248.0 | 7.4 |
| 826289.0 | 7.1 |
| 826445.0 | 6.9 |
| 826692.0 | 7.1 |
| 826872.0 | 6.4 |
| 826992.0 | 6.8 |
| 827191.0 | 6.5 |
| 827302.0 | 7.0 |
| 827482.0 | 7.2 |
| 827522.0 | 5.8 |
| 827668.0 | 5.8 |
| 828252.0 | 5.6 |
| 828897.0 | 6.4 |
| 829202.0 | 6.5 |
| 829752.0 | 6.4 |
| 829872.0 | 5.7 |
| 830823.0 | 5.9 |
| 831252.0 | 6.3 |
| 831629.0 | 6.4 |
| 831807.0 | 5.6 |
| 832108.0 | 6.0 |
| 832512.0 | 6.6 |
| 833346.0 | 5.8 |
| 834192.0 | 6.2 |
| 836283.0 | 6.4 |
| 836712.0 | 5.9 |
| 837192.0 | 6.3 |
| 837363.0 | 7.0 |
| 841332.0 | 5.9 |
| 841392.0 | 6.3 |
| 841452.0 | 5.9 |
| 842892.0 | 7.0 |
| 843086.0 | 7.4 |
| 843367.0 | 5.6 |
| 843500.0 | 6.3 |
| 843790.0 | 5.6 |
| 843882.0 | 6.6 |
| 844272.0 | 6.4 |
| 844392.0 | 6.8 |
| 844633.0 | 6.8 |
| 844732.0 | 6.5 |
| 844860.0 | 6.9 |
| 845198.0 | 5.6 |
| 845412.0 | 7.2 |
| 846015.0 | 5.5 |
| 846852.0 | 6.7 |
| 846972.0 | 5.5 |
| 847032.0 | 6.3 |
| 847394.0 | 6.1 |
| 847495.0 | 7.1 |
| 847512.0 | 7.4 |
| 848134.0 | 5.9 |
| 848160.0 | 6.8 |
| 848592.0 | 6.2 |
| 848652.0 | 6.7 |
| 848712.0 | 6.5 |
| 848897.0 | 6.1 |
| 849275.0 | 5.8 |
| 849553.0 | 6.0 |
| 849644.0 | 5.9 |
| 849852.0 | 6.7 |
| 850032.0 | 6.9 |
| 850152.0 | 6.7 |
| 850435.0 | 6.2 |
| 850671.0 | 6.0 |
| 850820.0 | 6.8 |
| 850966.0 | 7.0 |
| 851721.0 | 6.5 |
| 852336.0 | 6.9 |
| 852540.0 | 5.8 |
| 854112.0 | 6.8 |
| 854172.0 | 6.5 |
| 854864.0 | 6.7 |
| 855068.0 | 5.8 |
| 855552.0 | 6.5 |
| 856992.0 | 6.5 |
| 857692.0 | 6.8 |
| 857846.0 | 5.5 |
| 860052.0 | 7.4 |
| 860112.0 | 7.2 |
| 860172.0 | 7.0 |
| 860716.0 | 6.2 |
| 861192.0 | 6.6 |
| 861432.0 | 7.3 |
| 861803.0 | 6.9 |
| 862333.0 | 6.4 |
| 862632.0 | 6.7 |
| 862752.0 | 6.2 |
| 862992.0 | 5.6 |
| 863052.0 | 7.5 |
| 863233.0 | 6.4 |
| 863328.0 | 7.2 |
| 863521.0 | 5.8 |
| 863652.0 | 6.5 |
| 863810.0 | 7.3 |
| 863927.0 | 5.9 |
| 864192.0 | 7.1 |
| 864252.0 | 6.3 |
| 864432.0 | 6.6 |
| 864922.0 | 6.5 |
| 865052.0 | 6.3 |
| 865494.0 | 6.9 |
| 865812.0 | 7.5 |
| 865872.0 | 7.2 |
| 865932.0 | 5.5 |
| 865992.0 | 6.0 |
| 866074.0 | 6.4 |
| 866228.0 | 7.3 |
| 866440.0 | 6.5 |
| 866617.0 | 6.0 |
| 866671.0 | 6.5 |
| 866700.0 | 5.7 |
| 866823.0 | 6.6 |
| 866882.0 | 5.6 |
| 866952.0 | 6.3 |
| 867012.0 | 5.8 |
| 867312.0 | 6.2 |
| 867432.0 | 6.3 |
| 867586.0 | 7.4 |
| 867701.0 | 7.1 |
| 867714.0 | 7.1 |
| 867761.0 | 6.5 |
| 867842.0 | 6.4 |
| 867935.0 | 7.4 |
| 867945.0 | 7.0 |
| 867957.0 | 6.5 |
| 868075.0 | 5.7 |
| 868217.0 | 6.5 |
| 868251.0 | 6.9 |
| 868316.0 | 7.0 |
| 868342.0 | 6.2 |
| 868392.0 | 7.1 |
| 868752.0 | 7.2 |
| 868812.0 | 6.5 |
| 869224.0 | 7.2 |
| 869458.0 | 6.3 |
| 869606.0 | 6.7 |
| 869892.0 | 6.1 |
| 870012.0 | 6.0 |
| 870132.0 | 7.3 |
| 870252.0 | 5.8 |
| 870394.0 | 7.1 |
| 870483.0 | 7.0 |
| 870878.0 | 6.7 |
| 871142.0 | 5.6 |
| 871512.0 | 6.9 |
| 871692.0 | 6.1 |
| 871881.0 | 6.1 |
| 872361.0 | 5.7 |
| 872679.0 | 6.1 |
| 872712.0 | 5.6 |
| 874392.0 | 7.1 |
| 875263.0 | 7.5 |
| 875349.0 | 5.6 |
| 875401.0 | 7.3 |
| 875466.0 | 6.0 |
| 877332.0 | 7.2 |
| 877971.0 | 6.4 |
| 878532.0 | 5.9 |
| 878592.0 | 6.0 |
| 880332.0 | 6.3 |
| 880722.0 | 6.1 |
| 880979.0 | 6.2 |
| 881344.0 | 6.1 |
| 881652.0 | 7.5 |
| 881832.0 | 6.4 |
| 881947.0 | 6.7 |
| 881971.0 | 6.7 |
| 882040.0 | 6.7 |
| 882288.0 | 7.4 |
| 882391.0 | 7.4 |
| 882416.0 | 7.0 |
| 882504.0 | 5.9 |
| 882741.0 | 7.0 |
| 882852.0 | 7.4 |
| 882912.0 | 7.3 |
| 883032.0 | 6.5 |
| 883092.0 | 7.3 |
| 883212.0 | 5.7 |
| 883272.0 | 6.3 |
| 883636.0 | 5.5 |
| 883678.0 | 5.7 |
| 883692.0 | 5.8 |
| 883941.0 | 7.5 |
| 884096.0 | 6.2 |
| 884152.0 | 6.2 |
| 884163.0 | 6.3 |
| 884232.0 | 7.3 |
| 884592.0 | 7.1 |
| 884812.0 | 5.8 |
| 884939.0 | 7.0 |
| 884972.0 | 6.2 |
| 885062.0 | 6.4 |
| 885178.0 | 6.6 |
| 885214.0 | 6.0 |
| 885382.0 | 5.8 |
| 885484.0 | 6.5 |
| 885527.0 | 7.0 |
| 885572.0 | 6.9 |
| 885672.0 | 7.3 |
| 885732.0 | 5.6 |
| 885972.0 | 6.1 |
| 886152.0 | 6.8 |
| 886366.0 | 7.3 |
| 886414.0 | 6.2 |
| 887232.0 | 6.1 |
| 887592.0 | 5.5 |
| 888091.0 | 7.2 |
| 888379.0 | 6.2 |
| 889213.0 | 6.0 |
| 889489.0 | 5.6 |
| 890292.0 | 6.9 |
| 891030.0 | 7.1 |
| 891067.0 | 6.4 |
| 891672.0 | 6.9 |
| 891912.0 | 5.5 |
| 891977.0 | 5.6 |
| 893112.0 | 7.2 |
| 893172.0 | 5.9 |
| 893292.0 | 7.0 |
| 893352.0 | 6.9 |
| 893509.0 | 6.3 |
| 893541.0 | 7.3 |
| 893727.0 | 6.9 |
| 894295.0 | 7.1 |
| 894492.0 | 7.1 |
| 894552.0 | 6.4 |
| 894792.0 | 6.4 |
| 895046.0 | 5.8 |
| 895280.0 | 5.6 |
| 895573.0 | 5.6 |
| 896649.0 | 6.2 |
| 897552.0 | 5.7 |
| 900859.0 | 7.2 |
| 901038.0 | 6.8 |
| 901346.0 | 6.2 |
| 901572.0 | 7.0 |
| 901632.0 | 7.2 |
| 901692.0 | 7.0 |
| 901812.0 | 7.3 |
| 902074.0 | 7.1 |
| 902194.0 | 5.6 |
| 902229.0 | 6.4 |
| 902274.0 | 7.1 |
| 902372.0 | 6.8 |
| 902386.0 | 7.0 |
| 902442.0 | 7.2 |
| 902619.0 | 6.0 |
| 902716.0 | 6.0 |
| 902769.0 | 6.9 |
| 902840.0 | 7.4 |
| 902854.0 | 6.3 |
| 902952.0 | 6.5 |
| 903072.0 | 5.9 |
| 903192.0 | 7.4 |
| 903312.0 | 6.5 |
| 903372.0 | 5.7 |
| 903432.0 | 7.3 |
| 903715.0 | 6.6 |
| 903780.0 | 7.1 |
| 903802.0 | 6.0 |
| 904016.0 | 7.2 |
| 904132.0 | 6.9 |
| 904152.0 | 6.9 |
| 904632.0 | 7.2 |
| 904692.0 | 6.9 |
| 904752.0 | 6.8 |
| 905754.0 | 5.6 |
| 906192.0 | 6.5 |
| 906252.0 | 5.8 |
| 908110.0 | 6.2 |
| 908577.0 | 7.2 |
| 908712.0 | 5.9 |
| 909012.0 | 6.0 |
| 909072.0 | 6.5 |
| 909399.0 | 6.7 |
| 909854.0 | 5.6 |
| 910094.0 | 7.0 |
| 912623.0 | 5.9 |
| 918721.0 | 6.0 |
| 919373.0 | 7.0 |
| 919379.0 | 6.8 |
| 919467.0 | 6.3 |
| 919639.0 | 6.5 |
| 920292.0 | 6.9 |
| 920788.0 | 5.9 |
| 920830.0 | 6.0 |
| 921053.0 | 6.6 |
| 921214.0 | 6.9 |
| 921972.0 | 6.1 |
| 922092.0 | 6.1 |
| 922152.0 | 6.6 |
| 922408.0 | 6.7 |
| 922754.0 | 7.0 |

922914.0 7.2
923172.0 6.6
923232.0 7.0
923532.0 5.9
923592.0 6.1
923738.0 6.5
923891.0 6.8
923984.0 5.6
924025.0 5.8
924336.0 5.8
924382.0 7.1
924552.0 6.9
924852.0 6.2
924972.0 6.3
925191.0 5.6
925478.0 6.3
925516.0 6.6
925649.0 6.3
925718.0 6.0
925778.0 6.1
926840.0 7.0
926881.0 5.9
927014.0 7.2
927175.0 5.9
927350.0 5.8
927393.0 6.7
927492.0 5.5
927552.0 6.6
927672.0 6.1
927912.0 7.5
929052.0 7.5
929112.0 5.6
929586.0 7.0
929882.0 7.2
931432.0 5.7
931651.0 6.8
932853.0 5.8
932916.0 6.9
933109.0 6.2
933252.0 7.2
934992.0 6.9
935244.0 6.6
935480.0 6.4
935535.0 6.1
935722.0 7.4
937005.0 6.6
937089.0 6.4
937342.0 6.6
937512.0 7.3
937932.0 5.9
937992.0 6.6
938278.0 5.9
938302.0 5.8
938399.0 5.8
939312.0 6.0
940181.0 6.9
940295.0 6.0
940327.0 7.3
941138.0 7.1
941293.0 6.0
941320.0 6.4
941345.0 7.2
941428.0 6.4
941498.0 6.4
941542.0 6.6
941732.0 7.2
941892.0 6.4
941952.0 6.2
942312.0 6.6
942400.0 6.7
943242.0 5.7
943632.0 6.6
943752.0 7.1
944395.0 6.5
944468.0 7.2
944507.0 6.5
945012.0 6.0
945132.0 7.4
946392.0 6.2
946572.0 6.0
946972.0 6.0
947010.0 7.2
947070.0 7.4
947376.0 6.3
947832.0 7.0
948072.0 6.2
948887.0 6.1
950317.0 7.1
950592.0 5.8
950772.0 6.5
952901.0 7.1
954708.0 7.1
959112.0 7.3
959292.0 6.2
959352.0 6.5
959472.0 6.6
959532.0 7.2
959592.0 6.2
959694.0 6.1
959876.0 5.8
959925.0 6.2
960020.0 6.8
960207.0 6.8
960242.0 6.3
960323.0 5.6
960552.0 7.1
960672.0 6.5
960792.0 7.1
960852.0 6.6
960912.0 6.4
960972.0 7.4
961032.0 5.8
961154.0 7.5
961315.0 6.8
961442.0 5.9
961477.0 5.6
961544.0 7.5
961708.0 6.2
961879.0 7.4
961970.0 6.0
962052.0 6.0
962172.0 6.8
962352.0 6.5
962412.0 6.0
962739.0 5.7
962807.0 7.2
962962.0 6.4
962975.0 6.5
963147.0 6.8
963405.0 5.9
963432.0 6.2
963552.0 7.2
963612.0 6.2
963672.0 7.2
963792.0 7.0
963912.0 5.8
964158.0 6.6
964259.0 6.8
964431.0 5.9
964574.0 5.6
964818.0 5.9
964872.0 5.8
965466.0 6.2
965527.0 6.1
965551.0 7.1
965797.0 7.0
965872.0 6.4
966167.0 5.5
966492.0 5.7
966552.0 6.6
967234.0 5.9
968304.0 5.6
968451.0 5.6
969145.0 6.1
969312.0 6.7
970692.0 6.7
970812.0 7.0
975252.0 6.5
975372.0 7.4
976150.0 6.0
976367.0 5.7
976872.0 5.9
977292.0 6.1
977314.0 6.0
977480.0 6.7
977832.0 6.7
978192.0 7.3
978420.0 7.4
978476.0 7.4
978617.0 7.0
978862.0 7.4
978962.0 6.5
979047.0 6.9
979139.0 6.2
979202.0 7.2
979632.0 5.9
979752.0 6.8
979907.0 7.5
980290.0 6.2
980332.0 6.2
980367.0 7.5
980420.0 7.4
980476.0 6.5
980522.0 5.5
980566.0 7.1
980643.0 6.2
980697.0 6.4
980712.0 6.6
980772.0 6.1
980892.0 7.5
981072.0 6.7
981132.0 6.0
981269.0 7.0
982113.0 6.2
982150.0 5.8
982272.0 7.3
982512.0 6.2
982632.0 6.0
982933.0 6.3
982967.0 6.3
983244.0 7.5
983592.0 6.6
983652.0 6.1
983772.0 5.8
983952.0 7.4
984072.0 6.3
984174.0 5.6
984245.0 6.9
984345.0 7.2
984690.0 5.6
984708.0 6.7
984812.0 5.9
985032.0 6.1
985092.0 5.8
985332.0 7.3
986158.0 6.1
986202.0 6.0
986283.0 6.4
986472.0 6.7
987155.0 6.7
987784.0 6.9
987820.0 5.7
989153.0 6.8
995592.0 6.6
998412.0 7.3
998575.0 7.3
998692.0 7.5
999262.0 6.4
999312.0 5.8
999852.0 6.3
1000110.0 5.6
1000230.0 5.9
1000289.0 6.4
1000602.0 6.9
1000932.0 6.7
1001112.0 6.4
1001172.0 7.4
1001441.0 7.0
1001608.0 5.7
1001651.0 6.7
1001719.0 7.3
1002250.0 6.4
1002312.0 7.3
1002432.0 6.6
1002552.0 6.7
1002672.0 6.3
1003187.0 6.5
1003509.0 6.4
1003587.0 6.7
1003714.0 5.8
1003729.0 6.2
1004112.0 5.9
1004172.0 6.0
1004232.0 7.4
1004322.0 6.9
1004357.0 6.5
1004470.0 6.6
1004502.0 6.6
1004567.0 6.4
1005040.0 6.1
1005766.0 6.9
1005832.0 7.2
1007052.0 5.7
1007112.0 6.5
1008192.0 7.1
1008252.0 7.1
1008954.0 6.5
1009509.0 6.4
1009572.0 6.8
1009752.0 5.7
1010159.0 5.8
1010594.0 5.8
1011192.0 6.3
1011372.0 6.7
1012204.0 5.9
1012300.0 6.2
1012752.0 5.5
1012960.0 7.4
1013052.0 7.5
1013519.0 6.3
1013585.0 6.4
1018016.0 6.4
1018110.0 6.2
1018152.0 7.3
1018392.0 5.9
1018452.0 6.7
1018890.0 5.7
1019093.0 7.0
1019154.0 5.7
1019238.0 6.4
1019269.0 5.7
1019445.0 7.4
1019544.0 6.1
1019772.0 6.9
1019832.0 5.6
1019892.0 5.8
1019952.0 7.4
1020187.0 6.6
1020544.0 5.6
1020730.0 7.0
1020918.0 6.0
1021092.0 6.2
1021392.0 6.2
1021613.0 6.6
1021632.0 6.6
1021749.0 7.2
1021762.0 6.6
1021950.0 7.1
1022105.0 5.6
1022114.0 7.0
1022532.0 7.3
1022712.0 6.8
1022772.0 6.0
1022832.0 7.3
1023056.0 6.9
1023125.0 6.9
1023150.0 5.5
1023733.0 6.7
1023872.0 6.6
1023882.0 7.4
1024032.0 5.5
1024092.0 6.7
1024332.0 6.1
1024830.0 6.7
1025059.0 5.8
1025094.0 7.1
1025254.0 7.2
1025592.0 6.5
1026316.0 5.6
1026330.0 6.2
1026394.0 6.4
1026664.0 7.2
1026721.0 5.6
1027539.0 6.2
1027669.0 7.1
1028132.0 6.6
1029146.0 7.4
1036932.0 6.6
1038372.0 6.1
1038492.0 6.6
1039325.0 6.6
1039444.0 7.1
1039504.0 7.2
1039752.0 6.1
1039932.0 6.0
1039992.0 6.7
1040172.0 6.5
1040424.0 5.6
1040624.0 6.1
1040670.0 6.9
1040975.0 6.9
1041067.0 6.7
1041612.0 5.6
1041892.0 6.3
1042031.0 5.7
1042134.0 5.7
1042359.0 6.3
1042390.0 6.9
1042532.0 6.6
1042604.0 6.4
1042812.0 6.5
1043112.0 7.1
1043424.0 6.6
1043520.0 5.8
1043695.0 6.9
1043707.0 5.8
1044192.0 6.4
1045386.0 5.9
1045512.0 6.1
1047012.0 7.1
1047542.0 6.5
1048812.0 7.4
1051999.0 6.6
1055712.0 7.3
1055772.0 6.9
1056291.0 6.5
1056624.0 6.9
1058216.0 5.7
1059134.0 6.9
1059176.0 5.6
1059244.0 7.0
1060542.0 5.5
1061207.0 6.5
1061472.0 6.2
1061652.0 5.9
1061712.0 6.9
1062100.0 5.6
1062177.0 6.2
1062220.0 6.7
1062295.0 7.4
1062792.0 7.1
1062912.0 6.2
1063671.0 6.7
1064592.0 6.9
1064712.0 6.3
1065092.0 6.7
1066392.0 6.9
1066501.0 6.1
1067352.0 7.3
1067412.0 6.8
1067472.0 7.1
1067692.0 6.6
1067719.0 6.7
1069574.0 6.6
1070560.0 5.9
1075275.0 6.5
1077432.0 6.7
1078812.0 5.9
1078932.0 7.2
1079382.0 6.9
1080060.0 7.4
1080132.0 7.0
1080432.0 5.9
1080661.0 6.4
1080937.0 5.9
1081072.0 5.5
1081100.0 6.0
1081175.0 7.1
1081343.0 6.8
1081872.0 6.3
1081992.0 7.3
1082118.0 5.5
1082132.0 7.4
1082306.0 5.9
1082364.0 5.8
1082580.0 7.1
1082707.0 6.2
1082794.0 6.2
1082803.0 6.0
1082868.0 5.9
1082944.0 6.0
1082952.0 6.6
1083525.0 6.5
1083631.0 6.8
1083645.0 7.0
1083700.0 6.3
1083805.0 7.1
1083856.0 5.8
1084157.0 5.8
1084386.0 5.9
1084752.0 7.2
1085083.0 6.9
1085582.0 6.7
1085638.0 6.0
1085806.0 6.5
1085832.0 6.2
1087228.0 6.1
1088217.0 7.3
1088247.0 6.5
1088571.0 6.6
1090272.0 7.4
1092456.0 7.2
1094228.0 6.0
1094772.0 7.3
1095027.0 5.6
1095770.0 7.1
1095972.0 6.9
1097084.0 7.3
1097352.0 6.4
1097412.0 6.3
1097532.0 7.4
1097712.0 6.1
1097772.0 6.1
1097832.0 6.4
1097900.0 7.3
1097951.0 6.2
1098032.0 6.8
1098070.0 6.1
1098189.0 5.7
1098217.0 6.1
1098279.0 6.6
1098322.0 7.0
1098499.0 7.5
1098602.0 5.7
1098645.0 5.8
1098704.0 7.0
1098710.0 5.7
1098792.0 7.3
1098852.0 6.6
1099445.0 6.3
1099613.0 6.0
1099650.0 7.0
1099832.0 6.0
1100141.0 6.6
1100352.0 6.8
1100910.0 7.4
1101202.0 5.9
1101563.0 6.8
1101633.0 5.8
1101972.0 7.3
1102432.0 6.5
1102512.0 7.1
1102609.0 7.5
1103059.0 7.3
1103352.0 6.2
1103412.0 5.5
1103472.0 5.6
1103592.0 6.6
1103682.0 6.7
1103723.0 5.7
1103964.0 7.0
1104178.0 5.9
1104234.0 6.1
1104271.0 6.0
1104339.0 6.3
1104444.0 5.7
1104792.0 5.7
1105032.0 7.4
1105255.0 5.9
1105760.0 5.7
1105869.0 6.2
1109964.0 6.3
1110792.0 6.1
1111244.0 7.1
1111752.0 6.4
1112914.0 5.8
1113169.0 6.0
1114632.0 5.9
1116990.0 6.1
1118206.0 6.7
1119312.0 7.5
1119867.0 6.2
1120308.0 7.2
1121348.0 7.2
1121548.0 7.4
1121606.0 6.6
1121792.0 7.4
1122012.0 7.3
1122252.0 5.9
1122312.0 6.7
1122404.0 6.2
1122415.0 6.2
1122624.0 5.7
1122653.0 5.9
1122831.0 6.0
1122870.0 7.2
1123045.0 6.4
1123332.0 7.1
1123572.0 5.6
1124454.0 6.1
1125330.0 7.4
1125450.0 5.7
1126152.0 5.7
1126700.0 5.8
1127014.0 6.4
1127178.0 6.4
1127205.0 6.6
1128364.0 5.6
1128500.0 5.5
1129032.0 6.5
1129212.0 6.4
1129272.0 7.5
1130164.0 6.9
1130772.0 6.6
1130952.0 7.4
1132902.0 5.5
1133712.0 6.5
1135092.0 5.8
1135272.0 5.8
1136156.0 5.7
1136220.0 6.3
1136292.0 7.3
1136652.0 5.8
1136876.0 7.3
1136932.0 5.8
1136972.0 7.5
1136995.0 6.6
1137039.0 6.3
1137105.0 7.3
1137172.0 5.5
1137222.0 6.0
1137382.0 6.0
1137474.0 5.6
1137496.0 7.0
1137550.0 6.4
1137597.0 6.9
1137672.0 7.0
1137732.0 6.2
1137912.0 6.1
1137972.0 6.8
1138032.0 5.9
1138092.0 6.4
1138249.0 6.0
1138260.0 6.7
1138307.0 6.0
1138354.0 7.1
1138422.0 7.4
1138628.0 7.3
1138641.0 6.8
1138696.0 6.9
1138732.0 6.8
1138814.0 7.4
1138842.0 6.3
1138870.0 6.5
1138890.0 6.0
1138934.0 7.5
1139030.0 7.1
1139112.0 7.4
1139172.0 6.2
1139232.0 6.4
1139292.0 5.9
1139352.0 5.6
1139472.0 6.0
1139532.0 6.9
1139694.0 5.6
1139730.0 6.2
1139970.0 6.9
1140000.0 7.3
1140040.0 6.2
1140218.0 6.5
1140282.0 7.2
1140342.0 7.0
1140432.0 6.7
1140470.0 5.6
1140507.0 6.2
1140672.0 6.1
1141448.0 6.1
1141724.0 5.9
1141982.0 6.3
1142790.0 5.8
1142822.0 7.4
1143137.0 5.7
1143492.0 7.1
1143732.0 6.7
1143852.0 7.0
1144306.0 6.8
1144334.0 6.3
1144374.0 7.4
1144471.0 6.6
1144564.0 7.2
1144605.0 7.2
1144778.0 6.4
1144992.0 5.7
1145292.0 7.3
1145694.0 6.8
1145970.0 5.7
1146286.0 6.8
1146312.0 5.8
1146432.0 5.9
1146552.0 6.1
1146612.0 5.8
1146792.0 7.5
1146887.0 6.5
1147040.0 7.1
1147532.0 7.2
1147752.0 6.6
1148052.0 6.6
1148299.0 6.0
1148539.0 7.0
1148732.0 7.0
1148932.0 7.3
1149110.0 6.4
1149552.0 6.2
1149897.0 7.4
1150077.0 6.7
1150350.0 6.7
1150533.0 7.4
1150992.0 6.3
1153414.0 6.5
1153752.0 5.7
1154166.0 6.5
1154360.0 6.8
1154450.0 5.9
1154474.0 6.9
1154556.0 5.7
1154810.0 6.7
1155072.0 6.3
1156392.0 6.1
1156452.0 7.0
1156960.0 5.9
1157759.0 5.8
1157826.0 5.9
1157952.0 5.8
1158802.0 7.1
1158930.0 6.2
1159043.0 6.9
1159272.0 6.8
1159632.0 7.3
1159890.0 5.7
1161867.0 6.7
1162332.0 7.3
1162452.0 6.9
1162766.0 7.2
1162942.0 6.0
1163120.0 6.8
1163952.0 7.5
1164638.0 7.0
1164672.0 6.0
1165212.0 6.5
1165272.0 6.9
1165332.0 6.2
1165392.0 6.0
1165452.0 6.9
1165512.0 6.7
1165685.0 7.4
1165794.0 7.1
1166079.0 5.9
1166333.0 7.2
1166652.0 6.4
1166772.0 5.6
1166832.0 5.9
1167356.0 6.0
1167541.0 6.1
1167736.0 6.7
1168092.0 7.1
1168152.0 6.0
1168272.0 6.4
1168332.0 7.3
1168662.0 7.1
1168696.0 6.4
1168745.0 5.7
1168805.0 5.7
1169412.0 6.4
1169772.0 7.3
1169832.0 6.8
1169915.0 5.7
1170150.0 6.6
1170285.0 7.3
1170391.0 6.7
1170482.0 5.8
1170567.0 5.8
1171212.0 7.1
1171616.0 6.5
1171662.0 5.7
1171901.0 5.7
1171949.0 6.8
1172292.0 6.0
1172412.0 7.0
1172592.0 6.9
1172811.0 7.1
1173510.0 5.9
1174152.0 5.8
1174414.0 6.8
1174418.0 6.5
1174597.0 6.2
1174669.0 7.1
1174800.0 6.9
1174819.0 6.6
1174913.0 7.2
1175112.0 6.2
1175232.0 6.1
1175292.0 5.9
1175352.0 7.1
1175412.0 7.3
1175532.0 6.0
1175845.0 5.8
1175852.0 6.2
1176030.0 6.9
1176502.0 7.2
1176552.0 6.9
1176612.0 5.9
1176672.0 5.5
1176732.0 5.6
1176852.0 6.5
1176972.0 6.6
1177032.0 6.4
1177094.0 6.2
1177205.0 6.2
1177313.0 7.4
1177366.0 6.8
1177617.0 6.4
1177638.0 7.0
1177699.0 5.9
1177823.0 6.9
1177886.0 6.1
1177960.0 6.6
1177971.0 6.6
1177992.0 7.2
1178052.0 6.7
1178232.0 7.2
1178292.0 5.8
1178352.0 5.7
1178600.0 5.8
1178679.0 5.7
1178730.0 6.3
1178868.0 5.5
1178903.0 6.8
1178918.0 7.4
1178962.0 5.6
1179050.0 5.9
1179100.0 5.7
1179294.0 6.0
1179415.0 7.0
1179612.0 6.2
1179852.0 6.7
1179912.0 7.0
1180021.0 5.8
1180062.0 5.7
1180103.0 7.2
1180147.0 7.4
1180163.0 7.2
1180210.0 6.3
1180245.0 7.4
1180397.0 6.5
1180417.0 6.4
1180442.0 5.8
1180482.0 6.9
1180521.0 7.5
1180697.0 5.7
1180820.0 6.0
1180860.0 6.2
1180872.0 6.5
1180932.0 6.9
1181292.0 6.4
1181491.0 6.2
1181512.0 7.5
1181685.0 5.6
1182176.0 5.7
1182205.0 6.6
1182432.0 6.7
1182552.0 5.5
1182672.0 7.2
1182732.0 5.7
1183039.0 5.6
1183199.0 7.0
1183212.0 6.3
1183349.0 7.4
1183707.0 6.4
1183752.0 5.9
1183872.0 7.4
1184292.0 6.4
1184410.0 6.1
1184482.0 6.8
1184767.0 6.8
1185192.0 6.7
1185372.0 6.2
1185552.0 5.6
1185672.0 6.0
1185848.0 7.1
1185930.0 5.8
1186136.0 6.0
1186311.0 7.1
1186405.0 6.3
1186632.0 6.8
1186692.0 5.6
1186812.0 6.2
1186932.0 5.6
1187112.0 5.5
1187362.0 5.7
1187457.0 6.8
1187662.0 7.1
1188072.0 5.9
1188192.0 7.4
1188837.0 7.1
1188930.0 6.5
1189512.0 5.9
1189812.0 6.6
1191012.0 6.2
1191192.0 5.9
1191372.0 6.5
1191731.0 7.1
1191905.0 6.7
1192095.0 6.3
1192392.0 6.0
1192512.0 7.4
1193067.0 7.3
1193122.0 5.8
1193470.0 6.3
1193507.0 6.7
1193832.0 5.7
1194252.0 6.8
1194778.0 6.8
1194838.0 6.5
1194867.0 7.4
1194913.0 6.9
1194925.0 5.5
1195392.0 7.5
1195512.0 6.8
1195632.0 7.0
1195752.0 6.1
1195876.0 7.2
1196020.0 6.4
1196187.0 5.8
1196610.0 6.7
1196647.0 6.0
1196712.0 5.8
1196772.0 6.3
1196832.0 6.2
1197810.0 6.8
1198038.0 7.3
1198116.0 5.8
1198152.0 6.9
1198272.0 7.4
1198512.0 6.2
1198572.0 6.0
1198632.0 7.0
1198906.0 5.6
1199056.0 6.8
1199273.0 6.8
1199445.0 6.7
1199832.0 6.7
1199892.0 7.2
1200367.0 6.1
1200712.0 6.0
1200743.0 7.3
1200900.0 7.0
1200925.0 7.5
1201032.0 6.3
1201152.0 6.4
1201212.0 5.7
1201512.0 6.8
1201717.0 6.8
1201733.0 7.0
1201825.0 6.0
1201911.0 6.6
1202532.0 6.2
1203912.0 6.2
1204651.0 5.9
1205472.0 6.8
1205652.0 6.3
1205772.0 7.4
1206249.0 6.6
1206479.0 7.1
1206625.0 6.7
1206762.0 6.1
1207535.0 5.6
1208177.0 6.8
1208532.0 7.3
1208712.0 5.6
1209097.0 6.5
1209164.0 7.3
1209417.0 5.6
1209672.0 5.9
1209912.0 6.6
1209972.0 6.1
1210092.0 6.9
1210222.0 7.2
1210322.0 6.9
1210432.0 7.4
1211025.0 5.6
1211112.0 5.6
1211232.0 5.7
1211704.0 5.8
1211772.0 7.2
1211784.0 6.7
1212028.0 5.8
1212094.0 5.8
1212145.0 6.6
1212214.0 5.7
1212267.0 6.6
1212332.0 5.6
1212405.0 7.3
1212552.0 6.3
1212672.0 7.2
1212792.0 6.6
1212852.0 5.5
1212972.0 6.7
1213107.0 6.4
1213172.0 7.4
1213195.0 6.4
1213253.0 6.8
1213317.0 6.6
1213382.0 6.5
1213447.0 7.0
1213498.0 6.4
1213574.0 7.0
1213594.0 7.2
1213689.0 6.7
1213830.0 5.7
1213836.0 7.1
1213958.0 6.9
1214172.0 7.3
1214232.0 7.1
1214292.0 5.9
1214412.0 7.1
1214685.0 6.7
1214713.0 5.6
1215059.0 5.9
1215099.0 5.9
1215184.0 6.9
1215207.0 6.1
1215291.0 5.7
1215552.0 6.5
1215672.0 6.5
1215732.0 7.4
1215792.0 7.0
1216097.0 6.3
1216122.0 6.5
1216543.0 7.1
1216742.0 7.2
1216790.0 6.3
1216809.0 7.3
1217112.0 5.8
1217172.0 7.0
1217352.0 6.3
1217850.0 7.3
1217931.0 6.1
1218024.0 6.8
1218099.0 6.5
1218132.0 5.7
1218152.0 5.6
1218189.0 6.4
1218261.0 7.4
1218295.0 6.4

1218312.0 6.6
1218372.0 7.4
1218432.0 7.1
1218492.0 6.6
1218552.0 7.5
1218612.0 6.9
1218672.0 5.6
1218732.0 6.7
1218867.0 6.5
1218929.0 6.6
1218992.0 5.6
1219030.0 6.8
1219084.0 6.5
1219107.0 7.4
1219189.0 7.3
1219253.0 6.5
1219282.0 7.4
1219329.0 6.2
1219405.0 5.7
1219477.0 6.7
1219744.0 5.9
1220112.0 5.9
1220312.0 7.3
1220665.0 7.3
1220682.0 7.0
1220810.0 6.6
1220955.0 6.1
1221005.0 5.7
1221040.0 6.7
1221127.0 6.9
1221192.0 6.1
1221432.0 7.1
1221672.0 6.7
1222009.0 6.0
1222080.0 5.8
1222197.0 5.7
1222692.0 7.1
1222872.0 7.2
1222932.0 6.2
1223300.0 7.3
1223441.0 6.7
1223850.0 7.2
1223949.0 6.8
1223998.0 7.3
1224252.0 7.5
1224372.0 6.2
1224552.0 6.6
1224657.0 6.2
1225014.0 6.4
1225692.0 5.9
1225812.0 7.1
1225872.0 6.8
1225932.0 5.6
1226207.0 6.0
1226326.0 6.9
1226397.0 7.5
1228752.0 5.6
1229164.0 7.5
1229597.0 6.3
1231272.0 6.8
1231916.0 6.2
1231972.0 6.9
1232282.0 5.9
1232952.0 6.7
1233132.0 6.5
1233192.0 6.5
1233277.0 6.7
1233477.0 5.8
1233683.0 5.9
1233747.0 6.0
1233843.0 7.1
1233892.0 6.0
1234392.0 7.1
1234572.0 7.2
1235045.0 6.9
1235167.0 5.6
1235193.0 6.8
1235592.0 5.9
1235652.0 6.4
1235832.0 7.0
1235952.0 7.2
1236012.0 6.0
1236344.0 6.8
1236413.0 6.4
1236488.0 7.4
1236763.0 6.9
1236777.0 6.0
1237152.0 7.0
1237272.0 6.3
1237332.0 6.0
1237593.0 5.7
1237903.0 5.9
1238032.0 7.0
1238324.0 6.5
1238400.0 6.9
1238832.0 7.0
1239249.0 7.3
1239277.0 6.4
1239369.0 6.2
1239734.0 6.9
1239912.0 7.5
1239972.0 7.3
1240212.0 6.8
1241174.0 6.6
1241252.0 5.9
1241337.0 5.6
1241352.0 6.4
1241772.0 5.9
1241832.0 5.9
1242005.0 6.4
1243212.0 6.5
1244043.0 6.7
1244836.0 7.1
1245029.0 5.9
1245099.0 5.8
1245732.0 5.8
1246152.0 6.2
1246310.0 5.7
1249672.0 6.6
1250232.0 7.2
1250412.0 7.5
1251394.0 6.0
1251552.0 6.1
1251732.0 6.4
1251852.0 6.0
1252119.0 6.7
1252450.0 6.0
1252472.0 6.7
1252490.0 6.9
1252521.0 7.5
1252561.0 7.1
1252932.0 6.1
1253232.0 6.0
1253352.0 6.5
1253448.0 6.1
1253478.0 6.4
1253488.0 5.5
1253527.0 5.7
1253657.0 6.8
1253702.0 7.1
1253761.0 5.8
1254023.0 7.3
1254032.0 6.5
1254162.0 7.1
1254212.0 5.8
1254256.0 5.7
1254274.0 6.4
1254281.0 5.9
1254372.0 6.1
1254432.0 6.2
1254552.0 7.4
1254732.0 6.7
1254873.0 5.6
1254923.0 6.0
1255024.0 6.4
1255122.0 7.4
1255197.0 7.1
1255249.0 6.3
1255326.0 7.2
1255812.0 7.1
1256172.0 6.8
1256232.0 6.3
1256292.0 7.4
1256403.0 6.3
1256660.0 6.8
1256703.0 5.9
1256720.0 7.2
1256747.0 6.8
1256763.0 6.0
1256910.0 6.9
1257104.0 6.6
1257192.0 5.7
1257492.0 6.8
1257552.0 5.5
1257672.0 6.0
1257823.0 7.0
1257903.0 6.4
1257953.0 5.6
1257987.0 6.2
1258311.0 6.0
1258373.0 6.9
1258437.0 7.1
1258692.0 6.0
1258812.0 7.0
1258992.0 6.9
1259112.0 7.4
1259234.0 6.7
1259283.0 6.4
1259308.0 6.8
1259333.0 6.9
1259390.0 7.2
1259404.0 6.4
1259442.0 6.1
1259461.0 5.9
1259542.0 7.3
1259584.0 7.5
1259642.0 6.5
1259804.0 6.4
1259877.0 6.8
1259893.0 6.8
1260002.0 7.2
1260252.0 7.2
1260372.0 7.5
1260432.0 6.9
1260552.0 7.3
1260808.0 6.4
1261333.0 6.3
1261505.0 7.2
1261992.0 7.2
1262240.0 6.4
1262344.0 7.5
1262380.0 7.4
1262457.0 6.7
1262590.0 6.3
1262890.0 6.1
1262925.0 7.3
1262952.0 6.5
1263012.0 5.9
1263192.0 5.6
1263432.0 7.2
1263777.0 6.1
1263807.0 7.3
1264353.0 5.7
1264752.0 6.8
1265454.0 7.5
1265670.0 5.6
1266457.0 5.6
1268952.0 7.0
1271284.0 6.5
1272930.0 7.5
1273032.0 6.4
1273332.0 6.3
1273452.0 5.7
1273619.0 6.6
1273857.0 6.6
1274047.0 6.2
1274532.0 7.5
1274592.0 6.2
1274652.0 5.9
1274712.0 7.0
1275398.0 6.6
1275443.0 6.9
1275613.0 6.4
1276332.0 6.8
1276474.0 7.2
1276723.0 5.7
1277412.0 6.6
1278180.0 5.8
1278370.0 6.4
1279032.0 7.5
1279092.0 5.9
1279152.0 5.9
1279349.0 6.3
1279968.0 6.1
1280084.0 7.2
1280146.0 7.2
1280352.0 6.8
1280792.0 6.1
1280853.0 6.9
1281205.0 6.3
1282483.0 6.3
1284301.0 5.7
1285961.0 7.2
1285992.0 5.7
1286412.0 7.0
1287342.0 6.9
1287672.0 7.1
1287732.0 6.1
1288232.0 7.5
1290552.0 7.0
1292172.0 6.3
1292707.0 5.6
1292812.0 6.3
1292937.0 7.5
1293068.0 7.4
1293252.0 6.0
1293312.0 6.4
1293372.0 5.7
1293432.0 6.3
1293492.0 7.4
1293612.0 6.2
1293672.0 5.7
1293874.0 7.2
1294061.0 6.2
1294282.0 6.6
1294308.0 6.0
1294393.0 7.3
1294577.0 5.9
1294692.0 7.5
1294812.0 6.6
1294872.0 5.6
1295052.0 6.7
1295112.0 7.0
1295305.0 6.8
1295315.0 6.9
1295396.0 6.1
1295570.0 7.1
1295725.0 5.7
1295820.0 7.0
1295869.0 6.1
1296252.0 6.5
1296312.0 6.3
1296837.0 6.9
1296902.0 5.9
1297041.0 6.2
1297140.0 6.7
1297512.0 6.5
1297572.0 6.8
1297692.0 7.2
1298105.0 5.6
1298217.0 5.8
1298587.0 6.3
1299012.0 5.6
1299072.0 6.2
1299132.0 5.9
1299657.0 6.7
1300392.0 7.5
1300572.0 7.1
1300632.0 6.8
1300692.0 7.2
1302615.0 6.0
1302640.0 5.6
1302940.0 6.4
1303008.0 6.5
1303088.0 6.8
1303572.0 6.8
1305132.0 7.3
1306512.0 5.6
1307058.0 7.3
1308287.0 7.3
1309789.0 5.6
1311896.0 6.3
1312682.0 5.5
1314287.0 6.0
1315152.0 6.4
1315365.0 5.7
1315692.0 7.1
1315884.0 5.6
1315929.0 7.0
1316292.0 6.2
1316532.0 6.2
1317232.0 6.4
1317852.0 5.7
1318412.0 5.9
1318679.0 6.2
1319076.0 7.4
1319172.0 6.3
1320464.0 5.6
1321190.0 7.2
1321362.0 6.7
1321474.0 7.1
1321552.0 6.7
1321674.0 7.0
1321932.0 5.5
1321962.0 6.8
1322552.0 7.3
1322824.0 6.1
1323080.0 7.5
1324082.0 6.5
1325052.0 7.4
1325172.0 5.7
1326312.0 5.8
1326857.0 6.4
1329552.0 6.4
1330459.0 6.6
1333217.0 6.7
1333360.0 7.2
1333387.0 7.4
1333442.0 6.7
1333504.0 7.1
1333572.0 6.8
1333692.0 6.6
1334111.0 6.8
1334139.0 7.2
1334337.0 6.0
1334522.0 6.9
1334559.0 7.2
1334902.0 7.0
1334922.0 6.2
1334942.0 6.0
1334952.0 5.6
1335432.0 7.5
1335512.0 7.3
1335587.0 7.1
1335714.0 7.3
1335899.0 6.4
1335912.0 6.6
1336122.0 5.7
1336241.0 5.6
1336572.0 7.1
1336632.0 5.7
1337742.0 5.7
1338072.0 5.7
1338702.0 7.2
1338747.0 7.4
1338816.0 6.7
1338858.0 5.9
1339452.0 6.4
1339830.0 6.8
1340496.0 7.1
1341577.0 6.5
1341758.0 6.9
1342005.0 7.2
1342272.0 7.1
1343464.0 7.1
1343772.0 7.1
1344481.0 6.7
1344508.0 7.0
1347285.0 7.3
1347403.0 5.5
1351272.0 7.0
1351422.0 6.7
1351790.0 7.0
1352042.0 5.9
1352087.0 6.3
1352172.0 6.9
1352532.0 6.8
1352652.0 6.8
1352712.0 6.8
1352967.0 6.8
1353002.0 7.0
1353399.0 6.0
1353672.0 5.6
1353852.0 6.2
1354320.0 7.3
1354569.0 6.2
1354615.0 6.9
1354628.0 6.2
1354652.0 7.4
1354760.0 6.0
1356088.0 6.1
1356224.0 6.0
1356792.0 7.3
1357240.0 6.0
1357689.0 6.3
1357901.0 7.0
1358172.0 5.7
1358232.0 6.6
1358914.0 6.6
1359672.0 6.9
1360821.0 7.0
1360872.0 7.0
1362252.0 6.5
1362792.0 6.0
1363501.0 5.9
1364232.0 6.5
1366692.0 6.7
1369376.0 6.3
1370934.0 7.5
1371252.0 6.9
1371957.0 6.8
1371980.0 7.2
1373144.0 6.1
1373770.0 5.7
1373832.0 6.0
1373892.0 6.5
1373952.0 7.3
1374390.0 7.0
1374415.0 7.3
1374535.0 7.1
1375512.0 7.3
1375572.0 6.8
1375632.0 6.6
1375752.0 7.5
1375913.0 6.4
1376374.0 6.3
1376487.0 6.2
1376527.0 5.8
1376952.0 7.1
1377510.0 6.9
1377687.0 5.5
1377914.0 5.5
1378272.0 6.7
1378452.0 6.2
1378512.0 5.7
1378706.0 7.0
1378815.0 6.5
1379652.0 7.4
1380301.0 7.1
1380694.0 6.7
1380949.0 7.4
1381212.0 7.5
1381452.0 6.9
1381633.0 6.0
1381759.0 6.4
1382253.0 6.5
1382532.0 7.0
1382712.0 5.5
1383804.0 5.7
1383912.0 6.4
1388292.0 7.3
1388592.0 6.6
1389010.0 6.4
1389216.0 6.3
1389732.0 6.9
1389852.0 7.1
1390032.0 5.7
1390092.0 6.2
1390933.0 6.6
1391109.0 6.0
1391172.0 5.6
1391352.0 6.1
1391412.0 5.6
1391472.0 6.8
1391592.0 6.5
1391817.0 5.5
1391982.0 5.9
1392009.0 6.1
1392076.0 6.1
1392173.0 7.0
1392179.0 6.2
1392250.0 5.9
1392552.0 6.5
1392612.0 6.9
1392732.0 6.9
1392972.0 7.2
1393110.0 6.6
1393153.0 7.4
1393383.0 7.3
1393387.0 6.7
1393949.0 6.3
1394472.0 6.8
1395259.0 6.8
1395343.0 7.0
1395552.0 5.8
1395672.0 7.4
1396643.0 7.3
1397112.0 6.1
1397533.0 7.1
1399317.0 6.9
1400798.0 5.7
1402263.0 7.4
1402412.0 6.7
1403112.0 6.8
1404252.0 6.6
1404708.0 5.9
1404763.0 7.3
1405632.0 5.7
1406690.0 6.0
1408452.0 7.1
1409010.0 5.6
1409089.0 6.7
1410312.0 7.5
1410482.0 6.8
1411189.0 5.6
1411692.0 6.7
1411886.0 5.8
1412644.0 6.1
1412712.0 6.6
1412772.0 7.5
1412952.0 6.4
1413012.0 6.4
1413072.0 6.3
1413433.0 6.6
1413788.0 7.1
1414011.0 5.7
1414272.0 6.4
1414452.0 6.6
1414572.0 6.6
1414632.0 6.5
1415378.0 5.9
1416235.0 7.5
1416624.0 6.0
1416719.0 7.4
1417852.0 5.7
1418413.0 5.7
1418472.0 6.4
1418592.0 6.9
1419603.0 6.1
1420332.0 7.0
1420513.0 6.0
1420689.0 6.7
1421310.0 6.1
1421412.0 6.4
1421532.0 5.6
1421652.0 6.4
1422160.0 7.2
1422371.0 7.1
1422580.0 7.0
1423032.0 5.7
1423413.0 5.9
1423510.0 6.3
1423748.0 7.3
1423820.0 7.3
1424070.0 5.5
1424103.0 6.9
1424472.0 5.5
1424592.0 6.3
1424652.0 6.7
1424787.0 6.9
1424810.0 7.0
1424870.0 6.2
1425501.0 6.3
1425792.0 6.0
1427472.0 6.2
1427661.0 6.4
1428106.0 7.0
1428912.0 5.8
1429032.0 6.6
1429467.0 6.9
1429562.0 6.0
1429573.0 6.8
1429742.0 6.9
1429767.0 7.2
1430172.0 5.9
1430232.0 6.3
1430352.0 6.6
1431267.0 6.2
1431288.0 5.7
1431395.0 7.5
1431912.0 6.0
1432665.0 7.4
1432795.0 5.6
1432932.0 6.3
1433052.0 6.8
1433569.0 6.7
1433599.0 5.5
1433628.0 6.1
1433661.0 7.0
1433676.0 6.3
1433774.0 5.8
1433995.0 6.2
1434432.0 6.7
1434492.0 7.3
1434899.0 6.8
1435078.0 7.2
1435342.0 7.2
1435482.0 7.2
1435496.0 7.3
1435752.0 7.1
1435872.0 6.2
1435932.0 6.3
1436172.0 6.4
1436232.0 7.0
1436309.0 6.2
1436374.0 5.7
1436385.0 7.2
1436732.0 5.8
1436843.0 6.9
1436922.0 6.6
1437192.0 6.5
1437252.0 7.3
1437312.0 5.6
1437432.0 7.1
1437492.0 6.1
1437552.0 6.8
1437612.0 7.4
1437779.0 7.3
1437801.0 6.7
1437870.0 5.7
1437914.0 6.0
1438050.0 6.6
1438106.0 5.9
1438348.0 7.4
1438444.0 6.3
1438531.0 6.6
1438620.0 5.8
1438632.0 5.9
1438692.0 5.9
1439112.0 6.0
1439204.0 6.1
1439239.0 5.6
1439490.0 7.0
1439808.0 6.0
1442374.0 7.3
1442597.0 7.1
1443432.0 7.2
1443968.0 6.9
1443984.0 5.6
1444039.0 6.0
1444130.0 6.7
1444382.0 7.3
1444452.0 5.9
1444512.0 5.8
1444572.0 5.7
1444632.0 7.4
1444752.0 6.2
1444872.0 7.5
1444943.0 5.8
1445067.0 6.7
1445097.0 6.4
1445112.0 7.5
1445134.0 6.7
1445180.0 7.1
1445332.0 6.2
1445363.0 6.1
1445527.0 6.0
1445617.0 6.3
1445660.0 6.6
1445692.0 7.5
1445733.0 5.6
1445832.0 6.7
1445892.0 6.8
1446012.0 7.5
1446252.0 6.7
1446582.0 6.8
1446704.0 6.2
1446743.0 5.5
1446818.0 6.8
1446843.0 7.3
1446872.0 7.2
1446960.0 6.9
1446980.0 6.4
1447020.0 7.1
1447092.0 5.9
1447147.0 6.9
1447175.0 7.4
1447272.0 7.4
1447332.0 5.5
1447512.0 7.4
1447752.0 6.6
1447817.0 6.6
1447902.0 6.7
1447922.0 7.4
1448002.0 6.7
1448035.0 5.6
1448140.0 7.2
1448247.0 5.7
1448277.0 7.4
1448360.0 7.0
1448482.0 6.7
1448520.0 6.8
1448534.0 7.0
1448619.0 5.7
1448712.0 6.6
1448772.0 6.4
1448832.0 7.2
1448892.0 7.2
1449012.0 7.4
1449132.0 7.5
1449192.0 6.2
1449291.0 7.0
1449320.0 6.7
1449482.0 6.4
1449639.0 6.9
1449716.0 6.1
1449742.0 7.2
1449777.0 7.1
1449827.0 6.8
1449951.0 5.8
1450017.0 5.8
1450332.0 6.2
1450512.0 7.4
1450572.0 7.5
1450939.0 6.7
1450989.0 7.2
1451016.0 6.7
1451078.0 6.1
1451139.0 5.6
1451219.0 6.0
1451302.0 6.1
1451345.0 6.3
1451430.0 6.3
1451512.0 7.4
1451892.0 5.6
1452012.0 6.9
1452158.0 6.6
1452194.0 7.3
1452240.0 7.0
1452440.0 7.1
1452620.0 7.5
1452844.0 6.0
1453152.0 6.4
1453626.0 6.7
1454034.0 6.0
1454152.0 7.1
1454274.0 6.1
1454308.0 6.1
1454411.0 5.6
1454532.0 6.4
1454772.0 5.6
1454832.0 6.5
1455094.0 7.4
1455356.0 7.0
1455744.0 6.0
1455753.0 6.6
1456492.0 6.2
1456528.0 5.6
1456708.0 7.4
1456894.0 7.0
1456922.0 7.1
1457279.0 6.1
1457352.0 6.4
1457712.0 7.3
1457772.0 5.8
1457954.0 6.1
1457981.0 5.9
1458049.0 6.3
1458057.0 6.6
1458143.0 7.3
1458286.0 6.5
1458415.0 6.8
1458588.0 5.8
1458852.0 6.3
1458912.0 6.0
1459032.0 6.2
1459212.0 6.4
1459272.0 6.2
1459457.0 5.7
1459690.0 5.5
1459902.0 7.2
1459958.0 7.1
1460352.0 7.1
1460412.0 7.2
1460472.0 5.9
1460652.0 5.8
1460712.0 5.6
1460790.0 5.9
1460863.0 6.6
1460921.0 7.5
1460960.0 5.9
1460994.0 7.0
1461085.0 6.8
1461311.0 7.1
1461452.0 7.2
1461482.0 5.7
1461620.0 5.6
1461648.0 6.9
1461852.0 5.6
1462032.0 7.2
1462152.0 6.5
1462319.0 6.1
1462355.0 6.2
1462423.0 5.5
1462596.0 6.4
1462745.0 7.2
1462882.0 6.4
1462935.0 6.6
1463093.0 7.1
1463172.0 6.6
1463232.0 6.3
1463292.0 6.4
1463412.0 7.3
1463742.0 7.4
1463786.0 7.0
1463814.0 7.4
1463876.0 6.9
1463903.0 7.1
1464048.0 6.4
1464170.0 7.3
1464320.0 6.0
1464367.0 6.5
1464417.0 7.2
1464462.0 5.5
1464538.0 7.2
1464852.0 6.4
1464912.0 7.1
1464972.0 6.2
1465234.0 5.8
1465378.0 5.7
1465531.0 5.5
1465690.0 7.5
1465798.0 5.9
1465854.0 6.1
1465883.0 7.2
1465992.0 6.1
1466232.0 7.2
1466292.0 6.8
1466352.0 7.4
1466612.0 5.8
1466643.0 6.1
1466713.0 6.2
1466896.0 6.0
1467240.0 6.5
1467432.0 6.4
1467492.0 7.4
1467672.0 6.5
1467732.0 6.4
1467792.0 6.4
1467852.0 7.2
1467912.0 7.5
1468018.0 7.5
1468092.0 6.8
1468217.0 5.6
1468305.0 7.4
1468332.0 6.9
1468434.0 5.5
1468492.0 5.6
1468596.0 5.8
1468673.0 6.0
1468701.0 5.9
1468792.0 6.2
1468833.0 7.1
1468992.0 6.8
1469052.0 5.9
1469112.0 6.2
1469172.0 6.1
1469232.0 7.3
1469554.0 6.1
1469734.0 5.5
1469824.0 6.4
1469835.0 7.1
1469894.0 7.4
1470142.0 7.1
1470312.0 5.5
1470432.0 5.7
1470873.0 6.9
1470984.0 6.4
1471099.0 6.5
1471204.0 6.6
1471254.0 5.8
1471331.0 6.0
1471425.0 6.4
1471560.0 7.0
1471570.0 6.4
1471658.0 6.3
1471722.0 6.3
1472172.0 6.2
1472232.0 5.7
1472367.0 6.1
1472387.0 6.7
1472446.0 5.7
1472596.0 6.3
1472733.0 7.4
1472778.0 6.3
1472784.0 5.7
1472822.0 6.6
1472904.0 7.4
1473135.0 5.5
1473155.0 6.0
1473192.0 5.8
1473252.0 5.8
1473312.0 6.3
1473552.0 5.6
1473612.0 5.8
1473672.0 7.4
1473759.0 5.9
1473775.0 5.7
1473804.0 5.5
1473850.0 6.1
1473982.0 5.6
1474097.0 6.7
1474135.0 7.1
1474215.0 6.3
1474344.0 6.0
1474506.0 5.6
1474565.0 6.5
1474614.0 7.3
1474632.0 7.0
1474752.0 6.3
1475507.0 5.6
1475524.0 6.6
1475712.0 7.5
1475952.0 7.3
1475984.0 5.7
1476072.0 5.5
1476132.0 6.0
1476192.0 5.8
1476312.0 6.6
1476432.0 5.7
1476727.0 5.9
1477239.0 6.3
1477252.0 7.4
1477264.0 6.7
1477512.0 6.2
1477572.0 6.0
1477692.0 5.5
1477752.0 6.7
1477872.0 6.8
1477932.0 5.9
1478206.0 5.6
1478422.0 7.2
1478712.0 6.4
1478735.0 5.7
1479252.0 7.0
1479432.0 5.6
1479716.0 7.4
1480002.0 5.8
1480632.0 7.5
1481327.0 7.3
1481358.0 7.3
1481761.0 6.9
1481952.0 7.1
1482252.0 6.6
1482634.0 6.3
1482769.0 6.6
1482872.0 7.2
1482974.0 6.9
1483039.0 7.1
1483165.0 5.8
1483272.0 7.3
1483752.0 6.7
1483889.0 6.1
1483917.0 6.5
1484073.0 6.8
1484219.0 6.0
1484272.0 7.1
1484285.0 6.6
1484342.0 5.8
1484438.0 7.5
1484528.0 7.1
1484612.0 7.2
1484637.0 6.7
1484712.0 6.1

1484772.0 6.7
1484832.0 7.4
1484952.0 6.5
1485192.0 5.8
1485467.0 6.1
1485507.0 6.6
1485708.0 6.0
1485783.0 5.7
1485972.0 5.9
1486002.0 6.2
1486103.0 6.9
1486212.0 6.0
1486332.0 7.0
1486392.0 5.8
1486512.0 6.6
1486572.0 6.5
1486929.0 6.4
1486937.0 5.6
1487064.0 5.9
1487087.0 6.2
1487201.0 6.0
1487318.0 5.5
1487342.0 7.0
1487488.0 5.6
1487592.0 6.9
1487712.0 6.5
1487772.0 6.8
1487892.0 6.8
1488012.0 6.9
1488072.0 5.9
1488174.0 7.3
1488209.0 5.7
1488234.0 7.4
1488264.0 6.8
1488272.0 5.7
1488315.0 5.9
1488360.0 6.6
1488377.0 7.4
1488432.0 6.6
1488500.0 5.9
1488522.0 7.1
1488537.0 6.7
1488604.0 6.0
1488623.0 7.4
1488648.0 5.6
1488853.0 5.9
1488901.0 6.9
1488967.0 6.1
1488989.0 5.7
1488995.0 6.0
1489032.0 6.0
1489152.0 6.1
1489212.0 7.3
1489272.0 6.2
1489332.0 7.3
1489452.0 5.6
1489580.0 7.4
1489627.0 6.0
1489724.0 6.6
1489743.0 6.9
1490080.0 6.8
1490109.0 6.7
1490129.0 6.0
1490215.0 6.0
1490245.0 6.9
1490342.0 5.5
1490456.0 7.2
1490532.0 6.1
1490592.0 7.0
1490892.0 6.4
1490952.0 7.1
1491375.0 7.0
1491388.0 6.9
1491534.0 5.6
1491550.0 7.4
1491684.0 7.4
1491753.0 5.8
1491781.0 5.9
1491815.0 6.5
1491841.0 5.9
1491874.0 7.4
1491912.0 6.7
1492032.0 6.4
1492152.0 5.8
1492392.0 6.3
1492456.0 6.0
1492500.0 6.1
1492531.0 6.6
1492607.0 6.9
1492649.0 6.5
1492670.0 5.8
1492723.0 5.8
1492904.0 6.4
1492982.0 6.3
1493095.0 7.4
1493230.0 6.7
1493300.0 6.5
1493352.0 6.7
1493412.0 6.6
1493592.0 6.9
1493772.0 7.2
1493832.0 5.7
1494032.0 7.1
1494117.0 7.2
1494142.0 7.2
1494194.0 6.2
1494912.0 7.2
1494972.0 6.6
1495212.0 6.6
1495272.0 6.4
1495420.0 5.9
1495952.0 6.7
1495982.0 6.7
1496141.0 7.1
1496712.0 5.9
1496796.0 5.6
1496808.0 5.9
1496825.0 5.6
1497215.0 5.7
1497512.0 5.9
1497792.0 5.8
1497852.0 5.9
1498092.0 7.0
1498339.0 7.3
1498477.0 6.8
1498909.0 7.4
1499412.0 6.5
1499472.0 5.8
1499652.0 6.6
1499719.0 7.3
1499757.0 7.3
1499855.0 5.9
1500150.0 6.1
1500172.0 5.6
1500392.0 6.1
1500421.0 6.6
1500492.0 6.3
1500537.0 6.2
1500612.0 6.3
1500672.0 6.8
1500972.0 5.5
1501032.0 7.5
1501182.0 5.5
1501472.0 6.5
1501565.0 5.9
1501658.0 6.2
1501724.0 6.9
1501874.0 6.5
1502232.0 5.7
1502292.0 5.7
1502816.0 7.0
1502909.0 6.4
1503142.0 7.0
1503172.0 7.4
1503256.0 7.0
1503432.0 5.8
1503612.0 6.5
1503672.0 6.6
1503732.0 6.8
1503792.0 6.6
1504164.0 6.9
1504197.0 6.8
1504284.0 6.7
1504351.0 6.5
1504403.0 6.1
1504459.0 6.0
1504592.0 6.1
1504625.0 5.7
1504678.0 6.2
1504710.0 5.7
1504872.0 5.6
1504932.0 6.9
1505172.0 5.6
1505292.0 5.7
1505352.0 7.0
1505594.0 7.4
1505665.0 5.8
1506040.0 6.5
1506191.0 7.1
1506672.0 5.7
1506922.0 7.2
1506949.0 7.3
1506959.0 7.1
1506996.0 6.5
1507062.0 6.3
1507231.0 5.8
1507255.0 7.2
1507306.0 6.8
1507471.0 7.4
1507545.0 6.8
1507595.0 6.7
1507717.0 7.4
1507740.0 6.9
1507812.0 6.4
1507872.0 6.6
1507932.0 5.9
1507992.0 7.0
1508112.0 6.6
1508172.0 7.3
1508232.0 6.6
1508312.0 6.4
1508345.0 6.8
1508407.0 6.0
1508451.0 6.4
1508497.0 6.6
1508509.0 6.0
1508547.0 6.2
1508572.0 5.9
1508615.0 6.7
1508622.0 6.0
1508704.0 6.9
1508752.0 6.4
1508886.0 7.0
1508912.0 7.5
1508937.0 6.8
1509042.0 6.5
1509095.0 7.2
1509130.0 7.4
1509252.0 5.5
1509312.0 6.9
1509372.0 6.8
1509432.0 7.4
1509492.0 6.8
1509552.0 6.7
1509612.0 5.6
1509672.0 6.2
1509740.0 7.4
1509916.0 5.6
1509950.0 5.9
1509989.0 6.5
1510074.0 7.0
1510119.0 7.1
1510172.0 5.7
1510544.0 7.5
1510692.0 7.3
1510752.0 5.8
1510812.0 6.0
1510872.0 7.2
1510992.0 6.0
1511310.0 5.6
1511325.0 6.5
1512020.0 6.2
1512372.0 6.8
1512492.0 5.7
1512552.0 7.5
1512907.0 5.6
1513191.0 6.1
1513440.0 7.2
1513572.0 6.0
1513752.0 6.0
1513812.0 5.7
1513872.0 6.1
1513932.0 5.8
1514060.0 7.2
1514172.0 6.2
1514200.0 6.0
1514304.0 7.1
1514361.0 6.9
1514487.0 6.4
1514507.0 5.7
1514715.0 6.2
1514762.0 6.7
1514824.0 6.3
1514946.0 6.9
1515192.0 6.1
1515252.0 6.7
1515842.0 6.1
1516008.0 6.6
1516139.0 5.5
1516239.0 6.5
1516572.0 6.2
1517408.0 6.6
1517432.0 6.6
1518012.0 6.3
1518192.0 5.7
1518493.0 5.9
1518590.0 6.6
1518873.0 5.7
1518992.0 7.4
1519692.0 7.1
1519886.0 5.7
1520183.0 7.4
1521192.0 6.0
1523405.0 6.7
1523712.0 7.4
1524642.0 5.9
1525092.0 6.4
1525624.0 6.1
1525631.0 6.4
1525755.0 7.2
1525904.0 6.3
1526095.0 5.9
1526177.0 7.4
1526239.0 6.2
1526389.0 6.4
1526472.0 6.1
1526532.0 7.4
1526652.0 5.7
1526712.0 6.0
1526772.0 6.8
1526892.0 6.9
1526952.0 6.5
1527061.0 6.7
1527093.0 7.1
1527281.0 5.8
1527377.0 6.6
1527399.0 5.5
1527481.0 7.4
1527501.0 5.6
1527563.0 6.0
1527594.0 7.2
1527658.0 6.0
1527704.0 6.1
1527790.0 5.9
1527854.0 5.8
1527893.0 6.6
1527912.0 6.0
1527972.0 5.8
1528032.0 6.2
1528272.0 6.9
1528452.0 6.7
1528479.0 6.2
1528567.0 5.7
1528700.0 6.8
1528725.0 7.3
1528777.0 6.0
1529114.0 6.1
1529161.0 6.4
1529352.0 5.5
1529472.0 5.6
1529592.0 6.6
1529772.0 6.8
1529832.0 5.8
1530188.0 6.8
1530242.0 6.8
1530259.0 6.9
1530314.0 6.5
1530321.0 5.8
1530562.0 6.5
1530631.0 5.9
1530687.0 7.2
1530852.0 6.3
1530912.0 6.7
1531032.0 6.1
1531272.0 6.9
1531341.0 6.5
1531430.0 7.3
1531517.0 6.5
1531602.0 7.0
1531686.0 6.9
1531703.0 6.8
1531755.0 6.3
1531872.0 7.0
1531923.0 6.8
1531993.0 6.7
1532292.0 6.0
1532352.0 6.9
1532472.0 6.7
1532532.0 6.1
1532592.0 5.7
1532712.0 7.0
1532991.0 6.3
1533029.0 7.4
1533162.0 6.6
1533335.0 6.0
1533348.0 6.2
1533446.0 5.8
1533488.0 6.4
1533500.0 7.2
1533525.0 5.7
1533591.0 6.1
1533597.0 5.7
1533627.0 7.2
1533672.0 7.0
1533852.0 7.2
1533912.0 6.3
1533972.0 6.4
1534092.0 5.6
1534152.0 7.2
1534219.0 5.9
1534335.0 7.0
1534417.0 7.5
1534504.0 5.7
1534568.0 7.4
1534671.0 6.1
1534683.0 7.0
1534812.0 7.3
1534867.0 7.1
1534996.0 6.5
1535030.0 7.1
1535064.0 6.7
1535112.0 6.5
1535172.0 6.2
1535232.0 6.9
1535532.0 6.5
1535827.0 7.4
1535835.0 6.1
1535874.0 7.2
1535920.0 5.9
1536004.0 7.4
1536042.0 7.5
1536092.0 5.9
1536125.0 6.9
1536147.0 7.4
1536792.0 5.7
1536852.0 7.5
1536912.0 5.7
1537260.0 6.3
1537306.0 6.8
1537435.0 5.7
1537681.0 6.5
1537835.0 6.7
1538877.0 6.3
1538900.0 6.9
1539120.0 6.2
1539145.0 7.3
1539279.0 7.1
1540587.0 5.7
1540932.0 6.2
1541352.0 6.0
1541752.0 6.4
1542058.0 6.2
1542261.0 6.3
1542372.0 5.5
1542492.0 6.0
1542959.0 7.4
1543342.0 6.3
1543752.0 6.2
1543812.0 5.7
1543872.0 6.8
1543992.0 7.0
1544112.0 7.1
1544172.0 6.7
1544545.0 7.3
1544637.0 7.4
1544802.0 5.6
1544823.0 6.5
1544946.0 6.8
1545012.0 6.8
1545160.0 6.0
1545372.0 5.5
1545492.0 7.2
1545552.0 6.0
1545778.0 6.8
1545897.0 6.0
1546030.0 7.5
1546197.0 5.8
1546568.0 5.8
1546692.0 6.3
1546932.0 6.2
1547112.0 6.3
1547272.0 6.0
1547433.0 5.9
1547509.0 7.1
1547519.0 5.6
1547635.0 5.8
1547767.0 7.3
1547797.0 6.3
1547835.0 6.7
1547862.0 7.3
1547899.0 7.0
1548000.0 6.8
1548072.0 5.8
1548132.0 7.3
1548192.0 6.5
1548252.0 6.8
1548432.0 5.7
1548678.0 6.5
1548871.0 6.2
1548901.0 6.5
1548966.0 5.6
1548989.0 5.6
1549066.0 7.1
1549125.0 5.9
1549192.0 6.2
1549399.0 7.0
1549472.0 6.0
1549499.0 7.4
1549572.0 6.7
1549632.0 6.6
1549812.0 6.7
1549872.0 5.6
1549932.0 6.1
1549992.0 6.1
1550239.0 7.5
1550314.0 6.7
1550407.0 6.8
1550544.0 7.3
1550774.0 5.9
1550792.0 6.3
1550946.0 6.0
1550952.0 5.8
1551132.0 5.8
1551192.0 5.9
1551252.0 5.7
1551372.0 5.8
1551564.0 7.0
1551717.0 7.1
1551950.0 7.0
1551974.0 7.1
1552076.0 7.0
1552217.0 6.9
1552247.0 6.3
1552632.0 7.1
1553122.0 5.8
1553257.0 5.6
1553337.0 6.7
1553401.0 5.7
1553427.0 6.7
1553599.0 6.0
1553629.0 7.1
1553952.0 5.5
1554132.0 7.1
1554192.0 7.3
1554420.0 5.8
1554627.0 5.6
1554884.0 6.1
1554991.0 6.3
1555095.0 6.6
1555392.0 5.6
1555857.0 6.5
1556038.0 6.8
1556065.0 6.9
1556114.0 7.3
1556222.0 6.5
1556366.0 6.4
1556707.0 6.7
1556952.0 7.0
1557335.0 6.0
1557461.0 6.5
1557552.0 5.7
1557912.0 6.4
1558043.0 5.9
1558212.0 5.8
1559044.0 6.5
1559064.0 7.4
1559357.0 7.2
1559832.0 6.2
1560860.0 6.8
1561634.0 7.1
1562592.0 5.9
1564212.0 7.0
1564604.0 6.0
1565412.0 5.6
1565772.0 5.6
1566972.0 7.3
1567092.0 7.1
1567152.0 6.9
1567272.0 7.5
1567424.0 6.5
1567472.0 5.9
1567504.0 6.7
1567592.0 6.9
1567720.0 7.2
1567972.0 7.3
1568047.0 6.9
1568087.0 6.8
1568141.0 6.6
1568163.0 7.0
1568185.0 5.9
1568217.0 6.4
1568292.0 5.5
1568352.0 6.5
1568472.0 7.3
1568782.0 6.2
1568817.0 6.4
1568840.0 5.9
1568863.0 6.4
1568936.0 7.2
1568974.0 5.5
1569054.0 7.2
1569058.0 6.4
1569093.0 6.1
1569150.0 6.7
1569181.0 6.6
1569263.0 7.3
1569274.0 5.6
1569296.0 6.3
1569304.0 7.2
1569333.0 6.0
1569437.0 5.7
1569577.0 7.5
1569672.0 7.2
1569852.0 6.7
1569912.0 7.1
1570032.0 6.7
1570152.0 6.2
1570257.0 6.8
1570344.0 6.7
1570452.0 5.6
1570556.0 6.8
1570590.0 5.6
1570704.0 7.3
1570712.0 7.0
1570734.0 6.1
1570855.0 7.3
1570974.0 6.8
1570999.0 6.9
1571112.0 6.9
1571292.0 6.8
1571472.0 5.7
1571532.0 6.3
1571592.0 6.2
1571800.0 6.7
1571822.0 7.0
1571950.0 7.2
1572086.0 5.9
1572211.0 6.0
1572232.0 6.6
1572253.0 6.8
1572272.0 6.2
1572308.0 6.7
1572387.0 6.0
1572547.0 6.0
1572612.0 6.0
1572672.0 6.6
1572732.0 7.0
1572972.0 7.2
1573032.0 7.5
1573288.0 7.1
1573371.0 7.4
1573410.0 5.6
1573432.0 6.0
1573560.0 7.0
1573587.0 7.0
1573723.0 6.9
1573765.0 6.6
1573770.0 6.6
1573977.0 6.9
1573992.0 6.8
1574172.0 5.6
1574232.0 6.0
1574472.0 5.9
1574602.0 6.9
1574650.0 7.0
1574744.0 5.9
1574812.0 6.4
1574872.0 6.5
1574918.0 5.6
1575231.0 5.6
1575257.0 6.8
1575357.0 6.0
1575374.0 5.5
1575432.0 6.9
1575552.0 6.6
1576031.0 7.0
1576230.0 6.6
1576232.0 6.8
1576330.0 6.7
1576379.0 6.0
1576384.0 6.4
1576453.0 6.1
1576540.0 6.6
1576577.0 6.5
1576621.0 6.5
1576678.0 5.7
1576697.0 6.1
1576707.0 7.4
1576788.0 7.3
1576825.0 5.7
1576849.0 5.6
1576872.0 6.4
1576932.0 6.6
1576992.0 5.7
1577112.0 5.8
1577292.0 5.9
1577466.0 6.5
1577485.0 6.8
1577690.0 7.4
1577815.0 5.8
1577830.0 5.7
1577906.0 6.1
1578945.0 5.6
1579046.0 6.8
1579134.0 5.6
1580504.0 6.9
1580597.0 5.7
1581192.0 6.4
1581252.0 5.6
1583970.0 5.7
1584372.0 6.7
1584931.0 6.1
1584984.0 7.1
1585156.0 6.4
1585381.0 5.7
1585632.0 6.8
1585752.0 6.4
1585812.0 7.4
1586245.0 5.8
1586308.0 6.1
1586607.0 6.1
1586623.0 6.1
1586641.0 7.2
1587072.0 6.1
1588169.0 6.0
1588244.0 6.4
1588325.0 6.5
1588334.0 5.6
1588452.0 5.7
1588572.0 6.2
1588812.0 6.7
1589099.0 7.1
1589143.0 7.3
1589152.0 7.1
1589230.0 7.1
1589315.0 6.1
1589390.0 5.6
1589558.0 7.1
1589832.0 6.4
1590072.0 6.1
1590132.0 6.5
1590192.0 7.0
1590312.0 7.4
1590383.0 6.8
1590442.0 6.1
1590549.0 5.8
1590638.0 6.6
1590784.0 7.0
1590801.0 5.7
1590856.0 7.4
1590931.0 6.7
1591062.0 7.3
1591127.0 6.4
1591812.0 5.7
1592079.0 7.0
1592132.0 6.4
1592202.0 6.7
1592332.0 7.0
1592574.0 6.7
1592712.0 6.5
1592832.0 5.6
1592952.0 5.6
1593012.0 7.3
1593321.0 6.3
1593333.0 6.5
1593771.0 6.7
1593815.0 7.3
1593838.0 5.6
1593942.0 6.4
1593998.0 5.6
1594090.0 7.5
1594332.0 6.2
1594779.0 6.7
1594857.0 7.5
1594935.0 5.8
1595179.0 5.6
1595237.0 6.2
1595275.0 5.8
1595382.0 6.3
1595453.0 7.3
1595652.0 7.3
1595892.0 7.3
1596012.0 5.5
1596180.0 6.3
1596269.0 7.1
1596547.0 6.6
1597092.0 6.1
1597272.0 5.8
1597392.0 7.4
1597662.0 6.6
1597679.0 7.2
1598100.0 6.2
1598302.0 5.7
1598352.0 7.1
1598451.0 5.6
1598592.0 5.6
1598652.0 6.5
1599036.0 6.4
1599452.0 7.2
1599640.0 7.3
1600272.0 5.7
1601002.0 6.0
1601472.0 6.2
1602295.0 5.6
1602560.0 7.2
1602972.0 5.8
1603649.0 6.5
1604073.0 5.7
1604185.0 6.0
1604221.0 7.2
1604352.0 6.5
1604592.0 7.4
1604712.0 6.9
1605113.0 7.2
1605571.0 6.6
1605972.0 6.1
1606032.0 6.3
1606361.0 6.3
1606415.0 6.5
1606672.0 7.3
1606712.0 7.3
1606834.0 7.5
1606856.0 6.8
1606907.0 7.4
1606989.0 6.5
1607059.0 7.4
1607079.0 5.6
1607112.0 5.7
1607172.0 5.9
1607232.0 7.3
1607292.0 5.9
1607352.0 5.9
1607412.0 5.6
1607472.0 6.6
1607657.0 6.8
1607715.0 6.1
1607730.0 7.0
1607811.0 7.4
1607837.0 6.9
1607867.0 7.2
1607918.0 7.5
1607950.0 5.9
1607968.0 7.1
1607987.0 6.1
1607990.0 7.2
1608224.0 6.9
1608477.0 6.7
1608552.0 6.9
1608672.0 7.1
1608732.0 6.4
1608792.0 6.5
1609220.0 6.7
1609272.0 6.6
1609340.0 6.3
1609460.0 5.5
1609482.0 7.5
1609817.0 7.3
1610352.0 7.0
1610970.0 5.9
1611024.0 6.2
1611136.0 7.4
1611250.0 5.9
1611282.0 7.1
1611432.0 6.5
1611552.0 7.1
1611612.0 6.6
1611672.0 5.8
1611732.0 7.1
1611792.0 7.5
1611852.0 7.3
1611987.0 5.9
1612109.0 6.9
1612135.0 6.1
1612181.0 5.7
1612314.0 7.2
1612355.0 7.1
1612390.0 6.1
1612521.0 7.2
1612635.0 6.9
1612779.0 6.4
1612872.0 6.9
1612932.0 7.3
1612992.0 6.3
1613112.0 7.0
1613232.0 7.0
1613292.0 7.1
1613352.0 7.4
1613429.0 5.6
1613475.0 6.0
1613728.0 5.8
1613735.0 7.1
1613760.0 5.6
1613777.0 7.3
1613828.0 6.9
1613898.0 7.4
1614054.0 5.7
1614065.0 6.6
1614157.0 5.8
1614172.0 7.2
1614293.0 6.0
1614312.0 7.0
1614372.0 6.5
1614492.0 6.7
1614552.0 5.9
1614732.0 7.1
1614792.0 6.6
1614865.0 5.5
1614902.0 6.8
1614930.0 7.0
1614959.0 6.2
1615080.0 6.5
1615161.0 6.1
1615228.0 7.5
1615292.0 5.7
1615311.0 6.1
1615319.0 7.5
1615390.0 7.2
1615467.0 7.0
1615540.0 6.8
1615554.0 6.0
1615627.0 6.9
1615731.0 6.0
1615752.0 7.4
1615872.0 6.6
1615932.0 5.6
1615992.0 7.0
1616112.0 6.3
1616172.0 5.9
1616232.0 6.9
1616321.0 5.9
1616537.0 6.6
1616547.0 6.1
1616608.0 7.4
1616643.0 5.9
1616725.0 6.0
1616734.0 6.3
1616810.0 7.3
1616859.0 7.2
1616881.0 6.7
1616888.0 6.3
1616897.0 5.8
1617111.0 7.4
1617192.0 6.5
1617252.0 6.2
1617372.0 7.3
1617432.0 6.7
1617492.0 6.6
1617612.0 6.7
1617732.0 6.9
1617852.0 6.2
1617969.0 7.1
1618084.0 7.1
1618211.0 6.8
1618340.0 6.6
1618478.0 6.7
1618625.0 7.4
1618752.0 6.2
1618812.0 6.3
1618932.0 6.6
1619052.0 6.8
1619112.0 6.3
1619281.0 6.9
1619665.0 6.7
1619877.0 6.7
1620072.0 6.9
1620312.0 6.7
1620372.0 6.6
1620432.0 5.6
1620675.0 5.9
1621362.0 7.2
1621387.0 6.4
1622274.0 7.2
1623132.0 6.4
1623252.0 5.6
1623602.0 5.9
1623882.0 6.3
1623955.0 7.2
1624007.0 5.7
1624632.0 7.4
1624812.0 5.7
1625062.0 6.9
1625115.0 7.1
1626192.0 7.3
1626378.0 7.3
1627141.0 6.1
1627252.0 6.7
1627392.0 6.9
1627888.0 6.6
1627960.0 6.7
1628077.0 6.0
1628134.0 7.3
1628178.0 6.8
1628217.0 6.0
1628243.0 7.2
1628304.0 6.1
1628398.0 5.7
1628626.0 6.6
1628659.0 7.4
1629012.0 7.5
1629430.0 6.3
1629542.0 6.5
1630045.0 6.7
1630212.0 6.9
1630743.0 7.0
1631104.0 6.3
1631352.0 5.9
1631528.0 7.2
1631652.0 6.7
1631772.0 6.5
1632072.0 6.1
1632992.0 7.0
1633272.0 6.7
1633512.0 7.4
1633608.0 7.0
1634278.0 6.7
1634362.0 7.3
1634652.0 6.3
1634712.0 5.6
1635209.0 6.6
1635285.0 6.8
1635668.0 6.6
1635886.0 6.4
1635912.0 7.2
1636572.0 6.4

1637172.0 7.4
1637226.0 7.3
1637954.0 6.7
1638057.0 6.7
1639152.0 5.7
1639486.0 7.5
1639508.0 7.4
1639598.0 6.8
1639999.0 6.9
1640155.0 7.0
1640292.0 7.2
1640592.0 6.7
1640972.0 6.8
1642032.0 6.8
1642252.0 7.1
1643014.0 7.4
1643733.0 6.0
1644612.0 6.0
1644792.0 7.4
1644852.0 6.7
1645196.0 5.9
1645279.0 5.7
1645652.0 7.0
1645992.0 6.4
1646052.0 6.9
1646112.0 6.0
1646292.0 6.6
1646352.0 7.1
1646472.0 7.3
1646580.0 6.7
1646607.0 6.0
1646632.0 5.7
1646763.0 5.9
1647432.0 6.6
1647612.0 6.3
1648133.0 5.7
1648455.0 7.3
1648681.0 5.5
1648769.0 6.4
1649052.0 6.6
1649519.0 5.9
1650063.0 5.7
1650672.0 7.3
1651056.0 6.8
1651269.0 7.4
1652052.0 6.7
1652112.0 5.7
1652435.0 6.7
1652545.0 6.7
1652570.0 6.7
1652971.0 7.1
1653069.0 6.0
1653104.0 7.0
1653192.0 6.4
1653252.0 6.6
1653372.0 5.8
1653432.0 6.8
1653552.0 6.6
1653672.0 7.0
1653757.0 7.2
1653805.0 7.4
1653857.0 5.6
1654032.0 6.5
1654051.0 6.3
1654221.0 6.6
1654309.0 6.5
1654400.0 5.8
1654531.0 7.0
1654589.0 5.7
1654627.0 6.6
1654632.0 7.4
1654752.0 6.4
1654872.0 7.4
1654992.0 6.2
1655172.0 5.6
1655282.0 7.1
1655533.0 6.9
1655600.0 6.7
1655667.0 6.5
1655813.0 6.8
1655865.0 7.0
1656072.0 6.9
1656132.0 6.5
1656192.0 5.9
1656252.0 5.8
1656432.0 6.2
1656492.0 7.1
1656552.0 7.2
1656614.0 5.6
1656691.0 5.9
1656758.0 5.8
1656821.0 5.5
1657015.0 5.6
1657219.0 7.0
1657301.0 5.6
1657372.0 5.5
1657572.0 6.0
1657992.0 6.0
1658062.0 7.4
1658318.0 5.7
1658395.0 6.2
1658416.0 6.3
1658564.0 5.9
1658694.0 6.5
1658711.0 6.6
1658774.0 7.5
1658952.0 6.1
1659012.0 7.4
1659192.0 6.5
1659252.0 6.7
1659499.0 7.4
1660262.0 6.4
1660359.0 7.2
1660452.0 6.6
1660572.0 5.6
1660692.0 5.7
1660812.0 6.0
1660872.0 6.2
1661014.0 7.3
1661022.0 5.6
1661172.0 6.9
1661214.0 6.7
1661334.0 7.0
1661453.0 5.9
1661463.0 5.8
1661506.0 7.1
1661595.0 5.6
1661687.0 7.0
1661762.0 6.5
1661832.0 5.9
1662072.0 7.3
1662192.0 5.8
1662612.0 5.7
1662825.0 5.7
1662858.0 7.0
1663202.0 7.1
1663272.0 5.6
1664215.0 5.8
1664712.0 6.4
1664832.0 6.9
1665192.0 7.3
1665616.0 5.6
1665966.0 6.3
1666087.0 7.1
1666152.0 6.9
1666212.0 6.0
1666452.0 7.3
1666572.0 7.1
1666849.0 6.6
1666966.0 6.6
1667832.0 6.6
1668012.0 5.9
1668408.0 5.9
1668591.0 6.9
1668638.0 7.0
1668812.0 7.0
1669332.0 7.1
1669747.0 5.5
1669868.0 5.6
1670472.0 6.0
1670592.0 7.2
1670652.0 5.6
1671064.0 6.3
1671121.0 6.0
1671462.0 5.8
1671546.0 6.8
1671912.0 5.6
1672332.0 5.5
1672392.0 5.8
1672627.0 6.4
1672674.0 6.7
1672915.0 7.1
1673107.0 6.9
1673192.0 7.4
1674317.0 6.4
1674780.0 6.5
1675092.0 6.5
1675152.0 7.2
1675399.0 6.0
1675763.0 6.8
1676950.0 5.6
1677477.0 7.1
1678292.0 5.7
1678636.0 7.2
1679232.0 7.3
1680044.0 7.0
1681414.0 5.6
1681514.0 7.5
1682172.0 5.9
1682472.0 5.7
1682687.0 6.4
1682794.0 6.0
1682857.0 6.3
1683032.0 6.7
1683127.0 5.9
1683287.0 7.3
1683330.0 5.8
1683612.0 5.5
1683792.0 5.6
1684233.0 6.5
1684300.0 7.1
1684326.0 6.1
1684800.0 5.8
1684932.0 7.4
1685052.0 7.2
1685232.0 6.0
1685502.0 6.8
1685650.0 5.6
1685789.0 6.9
1685870.0 6.8
1686094.0 6.5
1686152.0 5.5
1686202.0 6.2
1686260.0 6.5
1686732.0 6.1
1686792.0 6.5
1687431.0 6.1
1687752.0 7.0
1687992.0 7.4
1688052.0 7.2
1688112.0 5.8
1688232.0 6.2
1688450.0 6.4
1688484.0 6.0
1689005.0 7.0
1689164.0 6.5
1689192.0 7.2
1689672.0 6.1
1689737.0 6.0
1689817.0 6.3
1689887.0 5.8
1690078.0 7.5
1690120.0 6.9
1690165.0 7.1
1690692.0 5.7
1691445.0 6.2
1691528.0 5.7
1692649.0 7.0
1692739.0 7.0
1692820.0 6.0
1693331.0 6.1
1693377.0 5.9
1693632.0 6.3
1693812.0 6.8
1693872.0 7.4
1693992.0 5.7
1694080.0 7.0
1694133.0 6.8
1694265.0 6.5
1694350.0 7.2
1694415.0 6.4
1694794.0 7.0
1695132.0 6.8
1695694.0 6.3
1696320.0 7.0
1696392.0 5.9
1696452.0 5.7
1696632.0 5.6
1696812.0 6.3
1696942.0 6.6
1697003.0 6.8
1697220.0 6.9
1697279.0 6.6
1697562.0 6.8
1697892.0 5.6
1698012.0 6.5
1698132.0 5.7
1698312.0 6.7
1698625.0 7.4
1698654.0 6.2
1699096.0 7.5
1699127.0 6.1
1699151.0 6.9
1699213.0 6.9
1699230.0 6.3
1699452.0 7.0
1699512.0 6.1
1699692.0 5.9
1699825.0 7.5
1699886.0 7.4
1699907.0 6.4
1699953.0 6.9
1699984.0 6.8
1700035.0 6.0
1700092.0 6.1
1700144.0 6.5
1700191.0 6.3
1700344.0 6.8
1700351.0 6.7
1700361.0 6.9
1700448.0 5.8
1700465.0 6.6
1700519.0 5.9
1700647.0 7.2
1700693.0 7.4
1700712.0 6.1
1700832.0 5.5
1700952.0 7.5
1701072.0 6.8
1701132.0 5.9
1701192.0 5.6
1701267.0 5.6
1701321.0 6.1
1701328.0 7.4
1701381.0 5.6
1701390.0 6.8
1701420.0 6.1
1701442.0 7.4
1701505.0 6.5
1701661.0 5.8
1701790.0 5.7
1701834.0 6.0
1701938.0 6.3
1702212.0 7.0
1702332.0 6.5
1702512.0 5.7
1702572.0 6.2
1702751.0 7.0
1702883.0 5.9
1702902.0 6.6
1702956.0 6.8
1703022.0 6.9
1703146.0 6.7
1703213.0 7.4
1703274.0 5.8
1703391.0 7.2
1703402.0 6.8
1703411.0 6.8
1703455.0 5.9
1703523.0 6.8
1703555.0 6.9
1703568.0 7.3
1703592.0 5.9
1703652.0 7.5
1703712.0 7.0
1703772.0 5.7
1703832.0 7.3
1703892.0 7.5
1703952.0 7.2
1704012.0 5.8
1704180.0 7.5
1704224.0 7.0
1704242.0 6.0
1704250.0 5.7
1704260.0 6.7
1704296.0 5.9
1704402.0 5.6
1704453.0 7.5
1704606.0 7.1
1704661.0 6.0
1704704.0 7.4
1704715.0 7.4
1704757.0 6.0
1704790.0 7.2
1704820.0 6.8
1704836.0 6.3
1704942.0 6.4
1705025.0 6.0
1705152.0 7.3
1705212.0 6.0
1705272.0 5.9
1705452.0 7.1
1705512.0 6.5
1705768.0 6.6
1706532.0 6.8
1706772.0 6.8
1707236.0 6.1
1707320.0 6.0
1707912.0 6.6
1708032.0 7.1
1708479.0 6.0
1709197.0 5.9
1709242.0 6.7
1709292.0 6.2
1709592.0 5.8
1709772.0 5.6
1709917.0 7.2
1709965.0 7.4
1710342.0 5.7
1710489.0 6.2
1710792.0 7.4
1711092.0 6.0
1711566.0 7.5
1712292.0 7.2
1712412.0 6.1
1712532.0 6.3
1712652.0 7.4
1712853.0 6.2
1713171.0 6.7
1713223.0 5.5
1713311.0 6.2
1713732.0 6.2
1713792.0 6.7
1713912.0 6.1
1714032.0 7.2
1714152.0 6.8
1714347.0 6.5
1714467.0 6.0
1714577.0 5.7
1714989.0 7.0
1715097.0 7.4
1715172.0 6.3
1715472.0 7.0
1715739.0 7.2
1715825.0 7.4
1715980.0 6.8
1716540.0 7.1
1716612.0 7.1
1716672.0 6.2
1716732.0 5.6
1716972.0 6.6
1717328.0 6.5
1717355.0 6.0
1717992.0 5.8
1718052.0 7.2
1718472.0 5.9
1718781.0 7.5
1718929.0 6.8
1719047.0 6.9
1719386.0 7.2
1719552.0 7.2
1719852.0 6.8
1720154.0 5.9
1720195.0 5.7
1720281.0 7.1
1720484.0 5.5
1720750.0 7.2
1720872.0 6.8
1721232.0 6.3
1721352.0 6.4
1721739.0 6.2
1721806.0 6.4
1722022.0 7.3
1722142.0 6.9
1722170.0 5.7
1722279.0 6.4
1722300.0 6.7
1722672.0 6.9
1722907.0 5.5
1723122.0 6.6
1723440.0 7.5
1723472.0 5.8
1723497.0 5.9
1723577.0 5.6
1723584.0 6.0
1723608.0 6.7
1723691.0 7.1
1723735.0 6.7
1723752.0 6.8
1723812.0 6.5
1723932.0 5.6
1724052.0 6.5
1724172.0 6.1
1724232.0 5.5
1724431.0 5.8
1724464.0 6.1
1724509.0 6.5
1724539.0 7.2
1724596.0 7.3
1724672.0 6.0
1724690.0 7.1
1724782.0 6.4
1724854.0 6.9
1724958.0 5.8
1725022.0 5.8
1725080.0 6.9
1725125.0 6.9
1725192.0 7.1
1725312.0 5.7
1725432.0 6.7
1725612.0 6.8
1725948.0 5.7
1725988.0 5.7
1726110.0 6.0
1726139.0 7.3
1726216.0 5.6
1726325.0 5.7
1726386.0 5.8
1726430.0 6.8
1726872.0 5.6
1727112.0 6.1
1727414.0 6.2
1727485.0 6.4
1727520.0 6.2
1727647.0 6.7
1727758.0 6.6
1727788.0 7.4
1727878.0 6.1
1727982.0 6.5
1728072.0 6.0
1728132.0 6.2
1728252.0 6.8
1728312.0 6.0
1728372.0 5.5
1728432.0 7.3
1728492.0 5.7
1728552.0 7.0
1728686.0 7.2
1729000.0 7.2
1729112.0 6.4
1729132.0 7.0
1729182.0 5.8
1729197.0 5.9
1729301.0 5.9
1729692.0 5.7
1729752.0 5.9
1729872.0 7.1
1729932.0 6.0
1729992.0 6.0
1730108.0 5.9
1730160.0 6.2
1730190.0 5.8
1730242.0 6.1
1730468.0 6.8
1730500.0 6.5
1730544.0 6.0
1730566.0 7.3
1730760.0 6.5
1730844.0 6.6
1730952.0 6.2
1731012.0 6.7
1731132.0 6.6
1731252.0 6.2
1731312.0 7.4
1731432.0 6.8
1731510.0 6.8
1731638.0 5.7
1731837.0 6.6
1732124.0 7.3
1732156.0 6.0
1732392.0 6.4
1732452.0 6.1
1732572.0 7.2
1732752.0 6.6
1732872.0 6.0
1732996.0 6.0
1733226.0 7.3
1733544.0 7.4
1733600.0 6.1
1733637.0 7.3
1733724.0 6.2
1733767.0 5.7
1733892.0 5.9
1733952.0 5.9
1734012.0 6.1
1734072.0 6.7
1734252.0 6.1
1734312.0 7.2
1734418.0 7.3
1734577.0 5.9
1734682.0 7.3
1734781.0 7.2
1734818.0 7.5
1734979.0 7.4
1735332.0 6.5
1735392.0 7.0
1735452.0 6.6
1735512.0 6.8
1735905.0 5.5
1736510.0 5.5
1736612.0 7.2
1736772.0 6.1
1737279.0 6.5
1737613.0 7.5
1737648.0 7.2
1737882.0 5.5
1737954.0 6.7
1738051.0 7.2
1738152.0 6.0
1738272.0 6.0
1738392.0 6.6
1738512.0 7.1
1738572.0 6.9
1738632.0 5.8
1738801.0 6.6
1738932.0 6.5
1739050.0 6.8
1739100.0 5.9
1739214.0 7.3
1739298.0 7.3
1739386.0 6.7
1739449.0 7.5
1739555.0 6.9
1739652.0 5.8
1739772.0 5.9
1739832.0 7.1
1739892.0 7.5
1739952.0 6.7
1740217.0 6.9
1740356.0 6.5
1740414.0 6.9
1740466.0 7.0
1740526.0 7.2
1740600.0 7.0
1740640.0 6.8
1740733.0 7.1
1740910.0 6.3
1741030.0 6.3
1741032.0 5.5
1741152.0 6.2
1741212.0 7.1
1741332.0 6.3
1741607.0 6.0
1741682.0 6.1
1741696.0 7.2
1741707.0 7.0
1741809.0 6.9
1741849.0 5.8
1741859.0 6.2
1741974.0 6.5
1742096.0 6.8
1742152.0 5.9
1742225.0 6.1
1742286.0 6.1
1742310.0 7.0
1742352.0 7.4
1742386.0 5.9
1742396.0 6.9
1742452.0 5.8
1742472.0 6.8
1742532.0 6.7
1742652.0 6.8
1742712.0 6.6
1742772.0 7.3
1742832.0 6.0
1742892.0 6.4
1742952.0 6.4
1743024.0 6.1
1743198.0 7.2
1743315.0 5.5
1743394.0 6.9
1743480.0 7.3
1743675.0 6.4
1743912.0 6.6
1744272.0 7.1
1744332.0 6.2
1744392.0 7.3
1744775.0 7.3
1745070.0 6.7
1745291.0 7.3
1745472.0 7.2
1745592.0 6.6
1746378.0 6.5
1748212.0 6.2
1748412.0 6.7
1749052.0 7.4
1749422.0 5.8
1749644.0 6.7
1749972.0 6.9
1750289.0 6.8
1750859.0 7.2
1751018.0 5.6
1751112.0 6.3
1751172.0 7.3
1752060.0 6.9
1752518.0 7.4
1753322.0 6.3
1753794.0 6.3
1753890.0 6.3
1755283.0 6.1
1755326.0 6.5
1755416.0 7.3
1756431.0 6.7
1756638.0 5.8
1756823.0 6.3
1756872.0 6.7
1758144.0 6.5
1758179.0 6.9
1759193.0 5.9
1759245.0 7.5
1760232.0 7.2
1762632.0 6.9
1762812.0 6.2
1762872.0 6.7
1762932.0 6.6
1762992.0 7.1
1763175.0 6.2
1763209.0 6.4
1763245.0 5.6
1763503.0 7.4
1763677.0 6.3
1764034.0 6.3
1764132.0 6.6
1764192.0 7.2
1764312.0 5.7
1764372.0 7.5
1764432.0 5.9
1764492.0 6.5
1764552.0 7.1
1764907.0 6.9
1765028.0 6.8
1765123.0 6.0
1765289.0 5.8
1765346.0 5.6
1765441.0 6.2
1765692.0 6.6
1765752.0 7.2
1765812.0 6.7
1765992.0 5.9
1766091.0 7.1
1766455.0 6.8
1766659.0 7.1
1767012.0 6.3
1767072.0 6.3
1767312.0 5.9
1767372.0 7.1
1767432.0 7.1
1767602.0 7.4
1767701.0 5.6
1767841.0 6.0
1767854.0 6.4
1768282.0 6.2
1768315.0 7.5
1768872.0 7.1
1769273.0 5.8
1769395.0 6.6
1769488.0 6.7
1769532.0 6.5
1769587.0 7.1
1769607.0 7.5
1769952.0 6.0
1770012.0 7.0
1770072.0 5.8
1770132.0 5.9
1770477.0 7.3
1770610.0 6.7
1770972.0 6.4
1771142.0 7.5
1771272.0 6.0
1771332.0 6.2
1771392.0 7.1
1771512.0 7.2
1771632.0 7.1
1772041.0 7.2
1772162.0 7.2
1772415.0 6.6
1772952.0 6.7
1773012.0 7.3
1773314.0 5.6
1773387.0 6.5
1773720.0 6.6
1773732.0 7.2
1774047.0 7.0
1774119.0 6.5
1774152.0 7.5
1774452.0 6.5
1774816.0 5.6
1775074.0 6.9
1775554.0 6.3
1775652.0 5.6
1776559.0 5.6
1777212.0 6.8
1777607.0 6.6
1778042.0 7.3
1778379.0 5.7
1778411.0 6.4
1778832.0 6.7
1778892.0 5.9
1779395.0 6.6
1779578.0 6.9
1779632.0 6.8
1779672.0 6.8
1780272.0 6.0
1780766.0 5.7
1780781.0 5.5
1780811.0 6.7
1780897.0 6.6
1781032.0 7.2
1781204.0 6.7
1781532.0 6.5
1781592.0 6.7
1781712.0 5.7
1781955.0 5.9
1782130.0 5.9
1782327.0 7.3
1782525.0 7.2
1782610.0 6.3
1782731.0 6.1
1782737.0 6.2
1782852.0 7.3
1782912.0 7.1
1782972.0 7.1
1783032.0 5.5
1783212.0 7.1
1783380.0 7.4
1783446.0 6.5
1783518.0 5.8
1783716.0 5.6
1783908.0 7.4
1784195.0 7.0
1784810.0 6.3
1784957.0 6.3
1786921.0 5.7
1787671.0 6.9
1788261.0 6.8
1788275.0 7.2
1788392.0 6.8
1788502.0 5.7
1788612.0 6.8
1788732.0 7.2
1788912.0 7.4
1788972.0 5.6
1789032.0 6.6
1789100.0 6.8
1789277.0 6.1
1789315.0 6.3
1789482.0 7.1
1789583.0 6.6
1789683.0 6.1
1789769.0 7.1
1790052.0 6.1
1790172.0 6.2
1790352.0 6.4
1791100.0 6.9
1791287.0 6.3
1791552.0 6.8
1792012.0 7.4
1793112.0 5.6
1793172.0 5.5
1793451.0 5.6
1793742.0 6.9
1793824.0 7.2
1793917.0 5.9
1795049.0 7.4
1795160.0 7.1
1797252.0 5.7
1798992.0 6.6
1799422.0 5.7
1800372.0 6.6
1800432.0 7.3
1800640.0 7.2
1800730.0 6.9
1800789.0 5.5
1800816.0 7.2
1801009.0 6.1
1801152.0 6.7
1801267.0 6.0
1801433.0 6.8
1801692.0 7.4
1801932.0 6.9
1802160.0 7.0
1802331.0 6.6
1802442.0 6.1
1802538.0 6.8
1802578.0 6.0
1802689.0 6.7
1802797.0 5.5
1802862.0 6.6
1802910.0 6.7
1803012.0 5.7
1803192.0 6.1
1803252.0 7.3
1803312.0 6.7
1803372.0 7.5
1803432.0 5.9
1803508.0 7.3
1803512.0 6.7
1803584.0 6.0
1803620.0 5.9
1803634.0 6.2
1803714.0 6.5
1803800.0 6.5
1803870.0 5.6
1803892.0 5.9
1803907.0 6.1
1804009.0 6.2
1804041.0 7.0
1804060.0 6.7
1804114.0 6.6
1804132.0 6.9
1804234.0 6.2
1804254.0 5.6
1804512.0 7.0
1804572.0 5.5
1804632.0 5.5
1804752.0 5.9
1804812.0 6.8
1805025.0 6.9
1805095.0 6.3
1805207.0 5.9
1805324.0 5.8
1805400.0 6.5
1805648.0 6.8
1805721.0 7.5
1805754.0 5.8
1805832.0 6.7
1805892.0 5.9
1806192.0 5.7
1806410.0 5.7
1806495.0 5.9
1806720.0 5.9
1806779.0 7.4
1806868.0 6.6
1806917.0 6.6
1807240.0 5.7
1807272.0 6.4
1807392.0 6.6
1807452.0 6.0
1807572.0 6.9
1807632.0 6.4
1807692.0 7.5
1807867.0 6.1
1808122.0 7.5
1808522.0 7.4
1808618.0 6.5
1808647.0 6.0
1808772.0 6.4
1809072.0 6.5
1809132.0 6.4
1809192.0 5.9
1809294.0 6.0
1809360.0 5.8
1809957.0 6.4
1809985.0 5.6
1810024.0 7.4
1810392.0 6.9
1810452.0 5.6
1810572.0 7.0
1810937.0 6.7
1810994.0 6.7
1811073.0 5.7
1811217.0 5.5
1811290.0 5.7
1811364.0 6.7
1811832.0 6.7
1812072.0 5.9
1812287.0 5.6
1812565.0 7.2
1813002.0 6.2
1813092.0 7.2
1813332.0 7.0
1813512.0 5.9
1813662.0 6.0
1814652.0 5.8
1815874.0 6.3
1815912.0 6.3
1816152.0 6.3
1816332.0 6.7
1817964.0 6.8
1818199.0 6.6
1818387.0 6.8
1818560.0 7.2
1818683.0 6.3
1819152.0 6.8
1819212.0 5.6
1819272.0 7.1
1819424.0 5.7
1819499.0 5.8
1819734.0 6.5
1819952.0 5.5
1820178.0 6.6
1820232.0 6.8
1820352.0 7.2
1820472.0 7.2
1820838.0 6.9
1821386.0 7.3
1821441.0 6.0
1821672.0 5.8
1821792.0 6.1
1821912.0 5.8
1822092.0 6.9
1822266.0 6.2
1822399.0 5.6
1822458.0 6.3
1822681.0 5.5
1822731.0 6.2
1822802.0 6.4
1822961.0 5.8
1823112.0 5.6
1823292.0 6.9
1823472.0 6.7
1823532.0 6.1
1823700.0 6.2
1823903.0 6.1
1824002.0 6.6
1824129.0 7.3
1824180.0 6.3
1824332.0 6.5
1824380.0 5.7
1824552.0 7.2
1824612.0 7.3
1824672.0 7.0
1825032.0 6.7
1825184.0 6.3
1825645.0 6.9
1825992.0 5.8
1826052.0 6.6
1826292.0 6.4
1826352.0 7.4
1826412.0 6.4
1826472.0 5.8
1826581.0 6.3
1826614.0 6.1
1826767.0 5.9

1826939.0 7.3
1828467.0 6.5
1829436.0 6.6
1829753.0 6.2
1829796.0 6.5
1829989.0 6.2
1830033.0 6.4
1830183.0 7.1
1830256.0 7.5
1830312.0 6.3
1830372.0 6.3
1830612.0 6.0
1830672.0 6.4
1830910.0 6.0
1830945.0 5.9
1831102.0 6.8
1831204.0 5.9
1831299.0 7.5
1831613.0 5.8
1831745.0 7.2
1831932.0 6.3
1832578.0 6.0
1833432.0 6.4
1833800.0 6.1
1834000.0 6.7
1834055.0 5.6
1834752.0 6.2
1834932.0 7.3
1835617.0 6.4
1835627.0 6.4
1839132.0 6.6
1839312.0 6.8
1839498.0 6.4
1840212.0 7.4
1840572.0 5.7
1840632.0 7.3
1841031.0 7.4
1841196.0 7.2
1841639.0 6.9
1841689.0 6.4
1841952.0 7.3
1842012.0 6.4
1842597.0 7.1
1842645.0 5.9
1843632.0 6.3
1843752.0 5.8
1844016.0 6.3
1844289.0 6.1
1844772.0 7.3
1844832.0 7.2
1844892.0 6.3
1845262.0 6.6
1845621.0 6.7
1845898.0 5.5
1846272.0 6.0
1846392.0 7.5
1846512.0 6.8
1846780.0 6.1
1847079.0 6.0
1847201.0 5.6
1847400.0 5.6
1847580.0 6.0
1848363.0 5.9
1848429.0 6.0
1848671.0 5.9
1848769.0 6.8
1849092.0 6.5
1849801.0 6.0
1849965.0 6.9
1850021.0 6.7
1850092.0 7.0
1850232.0 6.5
1850467.0 7.2
1851024.0 7.0
1851176.0 6.1
1851202.0 6.3
1851345.0 5.5
1851498.0 6.6
1851863.0 6.6
1851910.0 6.5
1853472.0 5.7
1853592.0 6.4
1853712.0 6.9
1854051.0 6.2
1854264.0 7.4
1854912.0 6.5
1855092.0 7.2
1855212.0 5.9
1855517.0 6.5
1855529.0 6.4
1856850.0 6.8
1858985.0 5.7
1859172.0 6.2
1859868.0 6.6
1859961.0 5.8
1860016.0 6.3
1860141.0 6.5
1860167.0 6.6
1860304.0 7.2
1860409.0 7.4
1860532.0 5.6
1860552.0 7.3
1860672.0 5.7
1860792.0 7.3
1861032.0 6.1
1861149.0 5.6
1861192.0 6.2
1861328.0 5.6
1861473.0 7.3
1861667.0 7.2
1861762.0 7.4
1861800.0 5.9
1861846.0 7.2
1862412.0 5.6
1862551.0 7.2
1862675.0 7.4
1862939.0 7.4
1863232.0 6.6
1863299.0 6.4
1863792.0 5.6
1863992.0 6.7
1864168.0 6.8
1864179.0 7.2
1864251.0 6.0
1864329.0 7.3
1864452.0 5.7
1864580.0 6.8
1864643.0 6.4
1864712.0 6.6
1864813.0 6.7
1864847.0 7.3
1864992.0 5.9
1865112.0 5.8
1865292.0 6.6
1865352.0 6.2
1865444.0 6.4
1865541.0 6.9
1865598.0 6.5
1865738.0 6.3
1865760.0 6.4
1866113.0 6.6
1866303.0 5.9
1866372.0 6.1
1866432.0 6.6
1866492.0 6.9
1866612.0 6.9
1866732.0 6.0
1866996.0 6.9
1867212.0 6.3
1867268.0 6.0
1867546.0 7.3
1867568.0 6.2
1867812.0 6.0
1867872.0 5.9
1867932.0 6.1
1867992.0 5.7
1868052.0 6.9
1868112.0 7.2
1868172.0 7.0
1868232.0 6.2
1868346.0 7.1
1868471.0 5.8
1868530.0 7.3
1868576.0 5.7
1868605.0 6.7
1868657.0 7.0
1868700.0 7.1
1868748.0 7.1
1869040.0 7.5
1869046.0 7.3
1869076.0 7.5
1869089.0 6.3
1869252.0 5.8
1869372.0 5.9
1869492.0 6.7
1869552.0 5.6
1869746.0 6.8
1869957.0 6.8
1869974.0 6.6
1870192.0 5.6
1870197.0 6.5
1870309.0 7.2
1870339.0 5.6
1870372.0 6.6
1870471.0 6.0
1871221.0 6.2
1871570.0 6.5
1872192.0 6.2
1872252.0 6.7
1872312.0 6.8
1872372.0 6.5
1872432.0 6.0
1872492.0 6.0
1872552.0 7.4
1872734.0 7.1
1872739.0 6.8
1872819.0 7.1
1872887.0 5.8
1873049.0 7.2
1873220.0 6.4
1873380.0 6.8
1873398.0 7.4
1873419.0 6.3
1873812.0 5.8
1873992.0 5.6
1874082.0 6.8
1874226.0 5.9
1874241.0 5.7
1877060.0 6.3
1878132.0 6.2
1878192.0 6.3
1879572.0 6.6
1880772.0 5.5
1880892.0 5.7
1881072.0 6.7
1881132.0 6.9
1881427.0 7.0
1882332.0 5.9
1882572.0 6.0
1882632.0 6.1
1883302.0 7.3
1883464.0 6.9
1883536.0 6.9
1883892.0 6.6
1884322.0 6.2
1884343.0 6.3
1885152.0 7.2
1885332.0 7.2
1885512.0 6.5
1885812.0 5.8
1885945.0 6.1
1886340.0 6.2
1886472.0 6.2
1886592.0 6.0
1886652.0 6.1
1887125.0 6.3
1887483.0 6.8
1888152.0 6.6
1888272.0 7.3
1889652.0 6.0
1889922.0 7.2
1890545.0 6.4
1890852.0 7.0
1894511.0 6.5
1899036.0 6.3
1899081.0 7.1
1899171.0 6.6
1900142.0 6.9
1900189.0 7.1
1900490.0 6.8
1900590.0 7.4
1903077.0 7.1
1903682.0 6.1
1903932.0 7.4
1903992.0 6.5
1904172.0 6.5
1904655.0 6.6
1905008.0 6.7
1905192.0 7.0
1905612.0 5.7
1906584.0 6.9
1906692.0 5.5
1906812.0 6.2
1906932.0 6.7
1907052.0 6.5
1907112.0 5.7
1907264.0 7.2
1907396.0 5.5
1907536.0 7.3
1907680.0 5.7
1907864.0 5.7
1908014.0 7.0
1908132.0 6.5
1908252.0 6.9
1908492.0 6.3
1908552.0 5.8
1908676.0 6.3
1909090.0 6.2
1909142.0 7.2
1909195.0 7.1
1909318.0 5.5
1909413.0 6.0
1909512.0 5.9
1909572.0 7.4
1909632.0 5.7
1909812.0 7.4
1909872.0 6.5
1910127.0 7.2
1910345.0 6.3
1910422.0 6.4
1910458.0 5.8
1910544.0 6.2
1910569.0 6.4
1910661.0 6.2
1910788.0 7.2
1910803.0 5.7
1910914.0 7.3
1911012.0 6.0
1911132.0 7.4
1911252.0 6.0
1911312.0 6.3
1911630.0 7.2
1911790.0 7.0
1911837.0 5.6
1911955.0 6.0
1912054.0 6.1
1912089.0 7.3
1912192.0 5.9
1912287.0 5.8
1912512.0 6.5
1912812.0 6.6
1912872.0 6.1
1913130.0 5.7
1913262.0 6.4
1913952.0 6.5
1914427.0 6.4
1914545.0 6.7
1915009.0 6.2
1915140.0 6.7
1915692.0 7.0
1915905.0 7.4
1917254.0 7.3
1918572.0 7.0
1918833.0 7.2
1919115.0 7.2
1919280.0 5.5
1919324.0 6.2
1919712.0 5.7
1919772.0 5.8
1920072.0 6.7
1920909.0 6.8
1921332.0 6.4
1922652.0 6.0
1922952.0 5.7
1923972.0 6.7
1924092.0 6.3
1924152.0 6.9
1924502.0 5.8
1924530.0 5.8
1924615.0 6.6
1924712.0 7.4
1924846.0 6.2
1924922.0 6.8
1925000.0 6.9
1925057.0 5.9
1925155.0 5.5
1925224.0 7.5
1925275.0 7.2
1925352.0 6.4
1925412.0 6.1
1925592.0 7.4
1925652.0 7.1
1925899.0 6.5
1925984.0 6.5
1926079.0 7.1
1926601.0 7.4
1926728.0 6.4
1926972.0 5.7
1927032.0 6.2
1927152.0 6.1
1927212.0 5.6
1927373.0 6.1
1927555.0 6.8
1927652.0 5.9
1927753.0 7.0
1928152.0 7.5
1928232.0 6.0
1928532.0 5.9
1928652.0 6.8
1928776.0 7.3
1928784.0 7.5
1928840.0 7.4
1928854.0 5.8
1929633.0 6.8
1929658.0 7.2
1929672.0 6.6
1929792.0 7.3
1929852.0 5.7
1930385.0 6.8
1930848.0 6.5
1931051.0 6.3
1931992.0 6.8
1933274.0 7.3
1933306.0 5.9
1933409.0 5.6
1933622.0 7.3
1933679.0 6.6
1933964.0 6.2
1934292.0 6.8
1935672.0 5.6
1935912.0 6.6
1936412.0 7.3
1936572.0 6.6
1936582.0 6.2
1936704.0 6.5
1937292.0 7.4
1937532.0 7.0
1937552.0 7.3
1937591.0 6.3
1937715.0 7.0
1937942.0 5.7
1938046.0 7.0
1938252.0 7.4
1938372.0 6.9
1938432.0 6.2
1938492.0 6.1
1938552.0 7.4
1938612.0 6.0
1938672.0 6.3
1938792.0 6.5
1938977.0 5.8
1939365.0 6.8
1939872.0 6.6
1940932.0 5.8
1942192.0 6.9
1942369.0 7.4
1942632.0 6.7
1942992.0 5.6
1943182.0 6.3
1943204.0 5.9
1943383.0 6.7
1943714.0 5.7
1944192.0 6.7
1944372.0 6.4
1944733.0 5.8
1951101.0 6.5
1951146.0 5.9
1952832.0 6.6
1959182.0 6.8
1959293.0 6.8
1959358.0 6.1
1960032.0 7.3
1960212.0 5.8
1960480.0 5.7
1962162.0 7.4
1962384.0 6.0
1962520.0 5.6
1963272.0 7.2
1965852.0 6.0
1966092.0 6.7
1966152.0 5.5
1966352.0 5.5
1966365.0 6.9
1966481.0 7.4
1966879.0 7.1
1967232.0 6.6
1967412.0 6.8
1967472.0 6.3
1968000.0 7.3
1968042.0 6.2
1968672.0 7.1
1968732.0 6.2
1968792.0 5.9
1968912.0 6.8
1968972.0 7.0
1969149.0 6.6
1969262.0 5.7
1969359.0 5.6
1969380.0 5.9
1969732.0 6.7
1969782.0 6.2
1969912.0 7.2
1970052.0 6.7
1970659.0 6.3
1970776.0 7.3
1970804.0 6.5
1970832.0 6.4
1971364.0 7.0
1971672.0 6.4
1971792.0 7.0
1972043.0 6.8
1972331.0 7.1
1972450.0 6.3
1972728.0 7.3
1972776.0 5.9
1972932.0 7.4
1973112.0 6.0
1973172.0 6.9
1973232.0 6.5
1973647.0 6.0
1973680.0 5.9
1973795.0 5.9
1973841.0 6.1
1973984.0 6.2
1974123.0 6.2
1974312.0 7.4
1974732.0 5.9
1975078.0 5.7
1975172.0 7.4
1975245.0 6.0
1975686.0 7.3
1975932.0 7.2
1976112.0 5.5
1976416.0 6.1
1976656.0 5.9
1976916.0 5.6
1977063.0 6.2
1977612.0 6.0
1977759.0 7.1
1977940.0 6.7
1978059.0 7.1
1978442.0 6.9
1978522.0 5.8
1978579.0 5.6
1978632.0 6.8
1978812.0 6.4
1978992.0 6.8
1979112.0 6.1
1979327.0 6.6
1979454.0 6.3
1979516.0 6.9
1979718.0 6.0
1979828.0 7.4
1979876.0 5.5
1980062.0 6.4
1980072.0 5.7
1980132.0 6.3
1980192.0 6.3
1980552.0 6.2
1981812.0 6.4
1981932.0 6.6
1982626.0 5.6
1982720.0 6.4
1983012.0 5.7
1984452.0 6.9
1985479.0 5.7
1986392.0 6.1
1986702.0 7.0
1987055.0 7.2
1993032.0 7.4
1993212.0 5.7
1993452.0 6.4
1993512.0 5.9
1993754.0 5.5
1993961.0 7.4
1993973.0 7.1
1994106.0 6.2
1994205.0 6.8
1994442.0 6.6
1994652.0 7.4
1994952.0 6.1
1995265.0 7.4
1995497.0 7.4
1995582.0 6.2
1995612.0 6.9
1995692.0 5.8
1995750.0 5.8
1995882.0 7.0
1995912.0 6.3
1995972.0 5.6
1996152.0 7.3
1996272.0 6.9
1996332.0 5.7
1996392.0 7.2
1996453.0 5.6
1996597.0 6.0
1996750.0 6.7
1996801.0 6.3
1997139.0 5.7
1997224.0 7.1
1997255.0 5.6
1997412.0 5.8
1997592.0 7.2
1997832.0 6.5
1997949.0 5.8
1998150.0 6.0
1998217.0 6.5
1998268.0 7.4
1998432.0 5.6
1998627.0 6.0
1998912.0 7.1
1998972.0 7.2
1999434.0 7.5
2000023.0 6.2
2000170.0 7.2
2000472.0 6.3
2000532.0 5.7
2000652.0 7.4
2000932.0 5.7
2000984.0 5.9
2001322.0 6.8
2001972.0 7.1
2002545.0 6.9
2002932.0 6.9
2002984.0 6.5
2003110.0 6.2
2003532.0 6.5
2004003.0 5.6
2005824.0 5.7
2007912.0 5.7
2008752.0 6.7
2009052.0 7.2
2009352.0 6.9
2009604.0 7.3
2009622.0 5.9
2009717.0 7.2
2009865.0 6.0
2009995.0 5.9
2010372.0 5.5
2010552.0 7.0
2010882.0 6.6
2011170.0 5.8
2011812.0 6.6
2011932.0 6.4
2012112.0 7.4
2013023.0 6.5
2013735.0 7.2
2014107.0 5.9
2014136.0 6.1
2014396.0 6.7
2014425.0 5.7
2014584.0 6.6
2014632.0 5.8
2014692.0 7.4
2014872.0 5.6
2015112.0 5.6
2015365.0 5.6
2015628.0 6.5
2015708.0 5.6
2015826.0 6.5
2015982.0 6.1
2016132.0 7.1
2016192.0 6.2
2016252.0 6.6
2016312.0 7.2
2016745.0 6.9
2016762.0 7.0
2016805.0 6.6
2017017.0 6.0
2017044.0 6.8
2017124.0 6.5
2017149.0 6.6
2017482.0 6.4
2017512.0 7.1
2017632.0 6.2
2018072.0 6.9
2018311.0 5.8
2019192.0 5.5
2019312.0 7.3
2019514.0 6.5
2020304.0 6.0
2020372.0 6.5
2021014.0 6.9
2021832.0 6.1
2021952.0 5.6
2022012.0 6.1
2022072.0 7.2
2022470.0 6.0
2023272.0 6.4
2025012.0 6.2
2025541.0 7.0
2026392.0 6.1
2026762.0 5.9
2028782.0 6.4
2030052.0 5.9
2030532.0 5.9
2030652.0 6.0
2032624.0 5.7
2033247.0 7.3
2037450.0 6.5
2037972.0 7.1
2040732.0 7.5
2040912.0 6.0
2041760.0 6.8
2042112.0 7.3
2042412.0 7.1
2042739.0 5.9
2043432.0 5.5
2044807.0 5.7
2045494.0 6.5
2046082.0 7.2
2047167.0 6.7
2047565.0 7.1
2047703.0 6.2
2048917.0 6.2
2049778.0 5.5
2051316.0 5.6
2051428.0 5.9
2051662.0 7.5
2051776.0 7.0
2051956.0 7.3
2052192.0 6.5
2052752.0 5.7
2053512.0 6.3
2053992.0 6.8
2054795.0 6.2
2054852.0 6.8
2054952.0 6.5
2055012.0 6.0
2055072.0 6.8
2055132.0 7.1
2055252.0 6.3
2055432.0 7.1
2055551.0 7.2
2055782.0 7.2
2055947.0 6.2
2056095.0 7.5
2056213.0 6.4
2056240.0 5.6
2056260.0 7.3
2056281.0 6.4
2056329.0 7.1
2056392.0 6.0
2056452.0 6.2
2056632.0 6.8
2056932.0 7.2
2056957.0 7.0
2057037.0 6.1
2057130.0 7.0
2057247.0 5.6
2057341.0 6.2
2057349.0 6.5
2057393.0 5.8
2057583.0 6.9
2057676.0 5.7
2057722.0 6.3
2057745.0 5.7
2058072.0 6.3
2058192.0 6.8
2058433.0 6.7
2059022.0 7.2
2059045.0 5.6
2059203.0 6.6
2059230.0 6.2
2059272.0 5.8
2059332.0 5.7
2059452.0 6.6
2059512.0 7.4
2059692.0 7.3
2059964.0 7.5
2060577.0 6.0
2060772.0 6.7
2061556.0 6.7
2061912.0 6.0
2062007.0 7.4
2062152.0 6.1
2062272.0 7.3
2062993.0 7.1
2063588.0 7.4
2063592.0 6.4
2063652.0 6.0
2063772.0 7.0
2064151.0 6.8
2064230.0 5.6
2064494.0 5.7
2065092.0 5.7
2065452.0 5.5
2065512.0 7.1
2065802.0 6.1
2065873.0 7.2
2066592.0 5.6
2066712.0 6.1
2066772.0 6.3
2067809.0 6.1
2068681.0 7.1
2069472.0 6.2
2069712.0 6.5
2070202.0 6.8
2071032.0 7.0
2071092.0 7.3
2071494.0 6.9
2071726.0 5.5
2075052.0 6.9
2075112.0 7.5
2075172.0 7.2
2075697.0 7.2
2076102.0 6.0
2076672.0 6.1
2077153.0 6.0
2077304.0 6.9
2077479.0 6.4
2077706.0 6.3
2077830.0 6.9
2077916.0 5.7
2078052.0 7.2
2078352.0 5.9
2078412.0 6.3
2078472.0 6.3
2078970.0 6.0
2079197.0 7.4
2079404.0 6.3
2079492.0 6.2
2079612.0 6.7
2079732.0 6.2
2079792.0 5.6
2079852.0 7.2
2080066.0 7.4
2080124.0 6.4
2080360.0 6.2
2080647.0 6.0
2080684.0 7.3
2080780.0 5.7
2080992.0 5.8
2081112.0 5.6
2081172.0 6.3
2081232.0 7.4
2081352.0 6.3
2081480.0 6.1
2081552.0 7.2
2081617.0 7.5
2081688.0 5.8
2081900.0 7.1
2082612.0 5.7
2082857.0 5.9
2083196.0 6.7
2083932.0 5.5
2084788.0 7.3
2084872.0 6.9
2085552.0 6.6
2085907.0 6.2
2086449.0 6.1
2086632.0 7.4
2086752.0 6.1
2087052.0 5.8
2087932.0 7.0
2088702.0 5.8
2088737.0 7.3
2088863.0 5.9
2089752.0 6.0
2090292.0 7.4
2091312.0 6.7
2091432.0 6.9
2092034.0 6.1
2093087.0 7.1
2093435.0 5.6
2093446.0 5.8
2093609.0 7.0
2093832.0 6.7
2094312.0 6.5
2094380.0 6.6
2094629.0 5.6
2094642.0 7.5
2095019.0 7.3
2095156.0 6.5
2095200.0 7.3
2095512.0 5.8
2095632.0 7.0
2095692.0 6.8
2095752.0 7.2
2095972.0 6.2
2096337.0 6.1
2096354.0 7.5
2096522.0 6.7
2096772.0 6.0
2096892.0 6.8
2097535.0 7.0
2097550.0 5.5
2097642.0 6.1
2097967.0 7.1
2098332.0 5.7
2098512.0 6.4
2099033.0 7.4
2099554.0 7.3
2099564.0 6.4
2099592.0 6.7
2099652.0 7.3
2099892.0 6.5
2099952.0 7.5
2100483.0 5.5
2101392.0 6.0
2101452.0 5.9
2101585.0 6.6
2101820.0 7.0
2102233.0 6.0
2102592.0 7.1
2102952.0 5.5
2104754.0 5.7
2104772.0 5.6
2106792.0 5.5
2107032.0 6.2
2107272.0 5.8
2107350.0 5.9
2107448.0 7.2
2107515.0 6.4
2108068.0 6.8
2108089.0 6.6
2108412.0 7.1
2108532.0 5.7
2108592.0 6.6
2108933.0 6.8
2109114.0 6.1
2109162.0 6.4
2109792.0 5.7
2109852.0 6.1
2109972.0 6.5
2110032.0 7.3
2110294.0 5.7
2110422.0 7.1
2110500.0 5.8
2110885.0 6.7
2111232.0 5.9
2113967.0 6.5
2113992.0 6.6
2114472.0 6.8
2114544.0 6.2
2115306.0 6.4
2115552.0 6.8
2115672.0 5.9
2115912.0 7.4
2116200.0 7.3
2116494.0 6.4
2116517.0 6.6
2116537.0 6.5
2116687.0 5.8
2116809.0 5.7
2117705.0 7.0
2117817.0 6.8
2118055.0 6.4
2118127.0 6.5
2118173.0 7.0
2118264.0 6.0
2119597.0 5.8
2119752.0 7.1
2119992.0 6.1
2120455.0 6.3
2120802.0 7.4
2121552.0 6.5
2122299.0 6.3
2122490.0 7.5
2122500.0 7.1
2122623.0 6.9
2122632.0 6.6
2123300.0 7.3
2123617.0 7.3
2123995.0 5.8
2124739.0 5.8
2124777.0 7.2
2124993.0 5.9
2125752.0 5.7
2126107.0 6.4
2126123.0 7.1
2126377.0 6.5
2127132.0 7.3
2127252.0 7.1
2128072.0 6.1
2128752.0 6.4
2129159.0 6.1
2129560.0 6.3
2129832.0 6.3
2130012.0 5.8
2130252.0 6.1
2131572.0 6.1
2131902.0 5.7
2132243.0 5.8
2132410.0 6.6
2132772.0 7.2
2134120.0 6.4
2134332.0 6.4
2135304.0 6.3
2135652.0 6.1
2135772.0 7.4
2135832.0 7.3
2135952.0 6.0
2136136.0 5.5
2136862.0 6.9
2139542.0 6.2
2140272.0 6.8
2141201.0 6.8
2141532.0 5.6
2142030.0 7.4
2142070.0 5.7
2142107.0 6.7
2146705.0 7.2
2148547.0 7.1
2148852.0 7.2
2149992.0 6.3
2150954.0 7.2
2151234.0 6.3
2151374.0 7.0
2152761.0 7.1
2152992.0 7.3
2153172.0 7.1
2153292.0 6.1
2153452.0 6.5
2153557.0 5.8
2153909.0 6.4
2154290.0 6.6
2154312.0 6.3
2154432.0 6.5
2154492.0 6.8
2154672.0 6.7
2154732.0 6.9
2154877.0 7.2
2155349.0 7.2
2155660.0 6.0
2156627.0 7.5
2156652.0 7.3
2157192.0 6.2
2157492.0 6.0
2157552.0 7.1
2157672.0 6.5
2158380.0 7.0
2158422.0 5.8
2158450.0 6.4

2159209.0 5.5
2159527.0 6.3
2159880.0 6.1
2159955.0 5.8
2160132.0 6.4
2161572.0 6.5
2161992.0 6.1
2162776.0 6.8
2162898.0 5.9
2162919.0 5.9
2162939.0 5.7
2163012.0 5.8
2163072.0 6.1
2163252.0 5.7
2163312.0 7.1
2163497.0 7.5
2163566.0 5.7
2163640.0 6.6
2163706.0 6.9
2163783.0 6.9
2163932.0 5.9
2164082.0 6.2
2164168.0 6.5
2164309.0 5.9
2164392.0 7.1
2164452.0 7.1
2164692.0 7.4
2164872.0 6.4
2165032.0 6.4
2165362.0 7.1
2165486.0 5.9
2165570.0 6.1
2165581.0 5.7
2165592.0 5.7
2165722.0 7.1
2165892.0 7.5
2166072.0 6.7
2166132.0 7.2
2166252.0 7.2
2166312.0 6.5
2166431.0 6.3
2166530.0 7.1
2166544.0 7.2
2166926.0 7.3
2167045.0 6.7
2167151.0 6.4
2167192.0 6.0
2167272.0 7.0
2167332.0 6.4
2167512.0 7.2
2167572.0 6.8
2168454.0 6.4
2168602.0 6.9
2168832.0 6.9
2169012.0 6.7
2169340.0 5.8
2169385.0 6.8
2169589.0 6.6
2169670.0 7.2
2169720.0 5.5
2169833.0 6.6
2170033.0 7.0
2170212.0 5.8
2170332.0 5.8
2170703.0 6.0
2170899.0 6.0
2171124.0 7.4
2171140.0 6.0
2171371.0 5.9
2171652.0 6.8
2172072.0 6.4
2172332.0 5.6
2172589.0 5.8
2172724.0 7.2
2173032.0 6.5
2173609.0 6.6
2173661.0 5.6
2173708.0 7.3
2173717.0 6.9
2173881.0 7.4
2173943.0 6.0
2174276.0 5.7
2174532.0 5.8
2175182.0 6.7
2175360.0 6.3
2175492.0 6.5
2175686.0 5.7
2175902.0 5.9
2175912.0 6.3
2177320.0 5.7
2177934.0 6.8
2178555.0 6.1
2178634.0 6.1
2178649.0 7.4
2179032.0 5.9
2179575.0 7.3
2180160.0 6.1
2180352.0 6.8
2180592.0 7.2
2181732.0 6.4
2182458.0 6.0
2184792.0 6.9
2185556.0 6.7
2191317.0 7.1
2193914.0 5.5
2193999.0 6.7
2194026.0 6.3
2194277.0 6.4
2194364.0 6.2
2194510.0 6.0
2194812.0 5.5
2195289.0 6.9
2195349.0 7.4
2196252.0 5.7
2196635.0 6.6
2196738.0 5.6
2196882.0 6.1
2196892.0 5.7
2197127.0 7.2
2197172.0 7.1
2197179.0 6.0
2197238.0 5.6
2197363.0 7.2
2197632.0 7.0
2198147.0 6.5
2198420.0 7.0
2198803.0 5.9
2198825.0 5.8
2198878.0 6.7
2198952.0 7.0
2199372.0 6.9
2199713.0 6.6
2200262.0 6.2
2200364.0 5.6
2202397.0 5.6
2202428.0 6.6
2202830.0 7.0
2203332.0 5.9
2203572.0 6.0
2203692.0 6.2
2203752.0 6.3
2203856.0 6.0
2203921.0 5.8
2203972.0 6.9
2204037.0 6.1
2204158.0 6.6
2204285.0 7.0
2204396.0 6.4
2204409.0 5.7
2204464.0 7.2
2204475.0 7.1
2204712.0 7.4
2204772.0 6.1
2204832.0 7.2
2204952.0 5.7
2205132.0 6.0
2205192.0 6.7
2205320.0 7.2
2205354.0 6.2
2205520.0 7.0
2205590.0 7.5
2205694.0 6.7
2205812.0 5.6
2205882.0 6.9
2206002.0 7.0
2206084.0 5.9
2206272.0 6.9
2206452.0 6.3
2206512.0 7.4
2206632.0 6.4
2206802.0 7.1
2206852.0 6.7
2206929.0 6.9
2206987.0 6.1
2207075.0 5.6
2207361.0 7.1
2207590.0 7.0
2207652.0 5.8
2207712.0 7.4
2207772.0 7.5
2207892.0 6.8
2207952.0 6.6
2208012.0 6.4
2208072.0 5.9
2208242.0 6.5
2208288.0 6.9
2208322.0 6.5
2208464.0 6.0
2208527.0 5.5
2208559.0 6.4
2208715.0 6.2
2208842.0 6.4
2208877.0 6.1
2208995.0 7.4
2209272.0 6.2
2209452.0 6.8
2209665.0 6.3
2209686.0 7.2
2209858.0 7.2
2209897.0 6.6
2209994.0 7.2
2210110.0 6.4
2210153.0 7.1
2210351.0 6.0
2210532.0 7.5
2211050.0 5.8
2211088.0 5.6
2211156.0 6.7
2211364.0 6.9
2211446.0 5.6
2211473.0 6.8
2211485.0 6.7
2211516.0 6.5
2211586.0 5.9
2211670.0 6.1
2211697.0 5.6
2211717.0 6.1
2211745.0 6.4
2211783.0 5.8
2211805.0 6.1
2211857.0 6.5
2211912.0 6.1
2212032.0 6.2
2212092.0 6.2
2212152.0 7.2
2212212.0 6.4
2212332.0 6.2
2212473.0 7.3
2212630.0 6.1
2212641.0 7.3
2212811.0 6.9
2212992.0 6.8
2213162.0 7.1
2213258.0 6.7
2213284.0 7.3
2213349.0 6.2
2213352.0 6.3
2213472.0 7.0
2213712.0 6.3
2213832.0 7.5
2213905.0 6.1
2215272.0 6.5
2215687.0 6.6
2216900.0 6.5
2218152.0 6.8
2218314.0 6.6
2218357.0 5.6
2218760.0 6.0
2219078.0 7.5
2219893.0 5.7
2220023.0 6.6
2220672.0 6.9
2221202.0 5.7
2222052.0 6.3
2223314.0 6.1
2223357.0 6.4
2223399.0 6.3
2223852.0 7.5
2225773.0 6.6
2227689.0 5.8
2227812.0 6.7
2228052.0 7.3
2229492.0 6.5
2230465.0 6.7
2231610.0 6.9
2232132.0 6.4
2232252.0 6.8
2233243.0 6.5
2233572.0 6.7
2233752.0 5.7
2234327.0 5.9
2234571.0 7.2
2234662.0 6.6
2236812.0 6.7
2237832.0 6.5
2238192.0 5.8
2239098.0 6.4
2239194.0 7.4
2239242.0 7.3
2239392.0 6.8
2240352.0 6.9
2243712.0 7.0
2245392.0 6.6
2245629.0 6.4
2245665.0 6.9
2246712.0 6.7
2247121.0 6.4
2247783.0 6.6
2248461.0 5.7
2250650.0 7.2
2250792.0 7.5
2250912.0 6.9
2251092.0 7.3
2251383.0 6.8
2251662.0 6.6
2251703.0 6.4
2251912.0 6.1
2252012.0 5.6
2252080.0 7.4
2252352.0 6.6
2252412.0 6.5
2252960.0 5.6
2253012.0 7.2
2253172.0 5.6
2253197.0 7.5
2253390.0 6.1
2253672.0 7.2
2254889.0 7.4
2254987.0 5.7
2255030.0 7.1
2257192.0 5.7
2257260.0 5.8
2257802.0 6.8
2257809.0 6.8
2258112.0 6.6
2258172.0 6.6
2264112.0 6.6
2271840.0 6.1
2272079.0 6.5
2272342.0 6.5
2272752.0 6.2
2272812.0 6.4
2274072.0 5.8
2274192.0 5.8
2274664.0 7.4
2274760.0 7.5
2275992.0 6.0
2276193.0 7.0
2276772.0 5.8
2276892.0 7.0
2276952.0 6.4
2277012.0 6.5
2277530.0 6.2
2278452.0 5.7
2278751.0 6.6
2279237.0 7.3
2283737.0 6.5
2284546.0 6.4
2284693.0 7.5
2284718.0 5.7
2285352.0 6.1
2286553.0 6.4
2286648.0 7.3
2286724.0 6.9
2286767.0 6.6
2287032.0 7.5
2287391.0 7.4
2287497.0 5.9
2287552.0 6.6
2287800.0 5.6
2288232.0 5.7
2288824.0 7.0
2289393.0 6.4
2289732.0 7.3
2289792.0 6.7
2290032.0 6.4
2290224.0 7.0
2290316.0 6.3
2290372.0 7.0
2290388.0 6.8
2290413.0 6.4
2290525.0 7.4
2290690.0 5.8
2290874.0 5.8
2290909.0 5.9
2291112.0 7.1
2291232.0 5.8
2291696.0 7.0
2291724.0 6.7
2291752.0 6.8
2291779.0 6.5
2291982.0 6.2
2292017.0 6.9
2292136.0 6.1
2292238.0 6.5
2292251.0 5.7
2292404.0 7.0
2292612.0 5.7
2292672.0 7.3
2292792.0 6.6
2292852.0 7.4
2292972.0 5.7
2295113.0 6.0
2295161.0 5.6
2295320.0 7.3
2296047.0 5.8
2296992.0 6.1
2297352.0 6.8
2297481.0 6.6
2297517.0 6.3
2297535.0 6.5
2297606.0 6.6
2297705.0 5.6
2297789.0 6.5
2298053.0 5.8
2298236.0 6.6
2302632.0 7.1
2305812.0 7.2
2307372.0 5.6
2307580.0 6.3
2309344.0 6.8
2310072.0 6.3
2310469.0 5.9
2310552.0 6.8
2311268.0 6.3
2311272.0 7.5
2311332.0 6.6
2311392.0 7.3
2311512.0 5.9
2311632.0 6.8
2311849.0 7.3
2312237.0 6.8
2312340.0 7.1
2312483.0 6.3
2312772.0 6.9
2313132.0 7.2
2313264.0 7.1
2313325.0 6.1
2313827.0 6.5
2313970.0 6.6
2313985.0 6.5
2314092.0 7.4
2314332.0 6.6
2314715.0 5.8
2314755.0 5.5
2315172.0 6.7
2315280.0 7.4
2315592.0 6.5
2315772.0 5.9
2316072.0 6.9
2316472.0 6.3
2316485.0 6.4
2316622.0 6.4
2316722.0 7.3
2317092.0 7.0
2317332.0 5.9
2317452.0 6.2
2317670.0 6.1
2318652.0 6.8
2319328.0 5.9
2321276.0 6.7
2321712.0 6.1
2322702.0 6.0
2322752.0 5.6
2323704.0 5.6
2323715.0 6.0
2323720.0 6.3
2324127.0 6.8
2324412.0 6.3
2325050.0 6.7
2325080.0 6.8
2325467.0 5.9
2325672.0 5.7
2326533.0 7.5
2327412.0 7.3
2327744.0 5.6
2327753.0 6.5
2328143.0 7.2
2328200.0 6.4
2328341.0 5.7
2328374.0 6.7
2328426.0 6.4
2328672.0 6.2
2329339.0 6.4
2330112.0 5.9
2330292.0 6.9
2330646.0 5.6
2330865.0 6.4
2331492.0 6.1
2332056.0 6.3
2332140.0 7.2
2332222.0 5.7
2332376.0 6.4
2332576.0 6.6
2332693.0 6.0
2332707.0 5.5
2332740.0 6.8
2332824.0 6.5
2332872.0 6.3
2332992.0 6.3
2333052.0 5.5
2333172.0 6.6
2333505.0 6.7
2333516.0 6.8
2333577.0 5.8
2333692.0 6.5
2333883.0 7.1
2333949.0 7.0
2334102.0 7.0
2334372.0 6.7
2334552.0 7.0
2334792.0 6.3
2335066.0 5.8
2335086.0 5.6
2335136.0 6.9
2335465.0 7.5
2335736.0 6.1
2335812.0 6.3
2335992.0 6.8
2336112.0 5.9
2336172.0 7.2
2336329.0 7.2
2336372.0 6.2
2336640.0 5.5
2336700.0 6.1
2337612.0 7.1
2337786.0 6.5
2338054.0 6.0
2338364.0 5.8
2338456.0 5.6
2338525.0 5.8
2342425.0 7.0
2342522.0 6.2
2343192.0 7.5
2343252.0 6.2
2344053.0 6.7
2344297.0 6.1
2344692.0 5.9
2346525.0 7.4
2346958.0 6.3
2347997.0 5.9
2348002.0 6.6
2349780.0 6.2
2353512.0 6.6
2353702.0 7.4
2354421.0 6.2
2355115.0 7.0
2355224.0 5.6
2355555.0 6.0
2356212.0 5.9
2356788.0 6.5
2357948.0 6.9
2359152.0 6.0
2359212.0 6.0
2359362.0 5.7
2359422.0 7.3
2360977.0 7.1
2361534.0 7.5
2363232.0 5.7
2364487.0 5.7
2364612.0 7.3
2366785.0 6.0
2367365.0 5.6
2368099.0 6.0
2368216.0 5.6
2368242.0 5.6
2368869.0 6.5
2369352.0 6.7
2369819.0 5.5
2370247.0 7.1
2371477.0 5.7
2371699.0 6.5
2371872.0 5.9
2372052.0 6.0
2372172.0 6.6
2372232.0 6.7
2372361.0 6.7
2372434.0 7.3
2372497.0 6.1
2372540.0 6.2
2372688.0 7.0
2372904.0 6.3
2373182.0 7.5
2373492.0 5.7
2374574.0 7.1
2375982.0 6.1
2376252.0 5.8
2376492.0 5.6
2376959.0 5.9
2377632.0 6.8
2378952.0 5.5
2379577.0 6.0
2381245.0 6.6
2383064.0 6.6
2383882.0 6.3
2385542.0 6.8
2386212.0 6.9
2387121.0 7.3
2387185.0 5.6
2387712.0 5.7
2387832.0 6.1
2387952.0 6.2
2388527.0 6.0
2388833.0 6.5
2389032.0 5.9
2389152.0 7.4
2389452.0 7.3
2389815.0 6.3
2390832.0 6.3
2390952.0 5.6
2391912.0 6.1
2392032.0 7.2
2392332.0 6.5
2392392.0 7.5
2392647.0 6.5
2392802.0 5.6
2392830.0 7.3
2392912.0 6.2
2392960.0 6.8
2393352.0 6.7
2393412.0 6.4
2393712.0 7.2
2393896.0 7.5
2394041.0 6.2
2394138.0 5.7
2394413.0 7.2
2394518.0 7.4
2394587.0 7.3
2394702.0 5.9
2394762.0 5.6
2394852.0 7.2
2394972.0 5.5
2395032.0 6.5
2395366.0 5.6
2395681.0 7.0
2395786.0 7.4
2396007.0 5.5
2396126.0 6.8
2396224.0 7.2
2396232.0 6.7
2396292.0 6.9
2396352.0 5.9
2396592.0 5.8
2396652.0 6.9
2396712.0 6.7
2396868.0 6.3
2396966.0 6.2
2397167.0 5.6
2397317.0 6.9
2397387.0 7.4
2397553.0 7.3
2397615.0 6.4
2397672.0 7.1
2397912.0 6.2
2398092.0 7.1
2398626.0 6.0
2398653.0 6.3
2399112.0 6.8
2399412.0 6.7
2399472.0 7.1
2400475.0 7.5
2400792.0 5.7
2401125.0 7.1
2402052.0 5.9
2402232.0 7.2
2402753.0 6.9
2403155.0 6.4
2403270.0 6.9
2403552.0 6.9
2406089.0 7.5
2406970.0 6.2
2407056.0 5.5
2407588.0 6.7
2407812.0 5.7
2408112.0 5.6
2409192.0 7.4
2409252.0 7.0
2409492.0 6.2
2409847.0 5.5
2409907.0 6.0
2409977.0 7.0
2410692.0 7.1
2410752.0 5.8
2410932.0 6.8
2410992.0 5.8
2411052.0 6.6
2411307.0 6.1
2411347.0 6.3
2411541.0 5.8
2411562.0 5.6
2411975.0 7.0
2412072.0 6.4
2412719.0 6.3
2412980.0 6.4
2413080.0 6.8
2413512.0 7.0
2414616.0 6.6
2414647.0 6.4
2415012.0 6.3
2415132.0 6.8
2415252.0 7.5
2415312.0 7.2
2415512.0 6.6
2415892.0 6.3
2415987.0 6.6
2416037.0 5.8
2417437.0 6.7
2419003.0 7.4
2419042.0 6.2
2419632.0 7.2
2419752.0 7.0
2419895.0 6.4
2420174.0 7.4
2420299.0 7.1
2420477.0 6.2
2420712.0 5.8
2420772.0 6.5
2420832.0 6.8
2420952.0 7.1
2421132.0 6.6
2421480.0 7.2
2421579.0 7.3
2421924.0 5.5
2422009.0 6.4
2422087.0 6.6
2422212.0 5.7
2422392.0 6.6
2422452.0 6.8
2422572.0 7.2
2422766.0 6.3
2422891.0 7.1
2423035.0 5.9
2423066.0 5.9
2423509.0 5.9
2423772.0 5.7
2425092.0 7.3
2425212.0 6.4
2425651.0 7.0
2425691.0 5.9
2425803.0 7.0
2426472.0 6.3
2426532.0 5.7
2426592.0 7.0
2426892.0 5.7
2427028.0 6.7
2427466.0 6.4
2427516.0 6.3
2427912.0 6.2
2428092.0 6.9
2428814.0 7.2
2428897.0 6.5
2429124.0 6.4
2430193.0 6.3
2430792.0 6.9
2430972.0 6.8
2431092.0 6.3
2431212.0 6.1
2431395.0 6.3
2432352.0 7.1
2433574.0 7.2
2433579.0 5.6
2433912.0 7.0
2434317.0 7.1
2434500.0 5.6
2434808.0 6.1
2434882.0 7.1
2435472.0 6.4
2435592.0 5.5
2435681.0 7.0
2435918.0 5.8
2436038.0 7.2
2436174.0 7.2
2436382.0 6.4
2437032.0 7.1
2437288.0 7.1
2437464.0 6.1
2437602.0 6.2
2437860.0 6.2
2437982.0 7.5
2438172.0 6.1
2438232.0 7.3
2438292.0 6.3
2438753.0 6.0
2438799.0 7.5
2438989.0 7.1
2440628.0 6.9
2440847.0 7.1
2441052.0 7.4
2442099.0 6.5
2442168.0 7.3
2442252.0 6.2
2443128.0 7.4
2444770.0 7.1
2445492.0 7.0
2446632.0 6.4
2446752.0 6.4
2448732.0 6.3
2448881.0 7.0
2449132.0 7.2
2449632.0 7.4
2449752.0 5.7
2449992.0 6.1
2451252.0 6.2
2451432.0 6.0
2451524.0 6.2
2454012.0 7.4
2454252.0 7.3
2456399.0 7.1
2458452.0 6.5
2466648.0 6.2
2466852.0 6.8
2468010.0 7.1
2469043.0 5.9
2469407.0 6.6
2469517.0 7.5
2469732.0 6.9
2469792.0 6.7
2469852.0 7.3
2470538.0 6.4
2470922.0 7.3
2471057.0 5.6
2472672.0 5.6
2473530.0 6.2
2473565.0 6.2
2473768.0 6.2
2475612.0 6.2
2475912.0 7.1
2476208.0 5.8
2476357.0 6.0
2476589.0 7.5
2476801.0 6.6
2477352.0 6.8
2477491.0 5.7
2477570.0 5.8
2477885.0 5.6
2478050.0 7.2
2478312.0 6.0
2478612.0 7.3
2478732.0 6.4
2479189.0 5.7
2480572.0 5.7
2480725.0 6.3
2480770.0 5.6
2480905.0 5.5
2480950.0 6.7
2481033.0 7.5
2481192.0 5.8
2481312.0 6.1
2482872.0 5.9
2482992.0 6.7
2483564.0 6.0
2483829.0 6.4
2484054.0 6.9
2484831.0 6.0
2484952.0 6.4
2485372.0 7.1
2485436.0 5.8
2485484.0 5.9
2485752.0 7.0
2485932.0 7.2
2485992.0 6.9
2486461.0 7.3
2486577.0 7.1
2486639.0 7.1
2486740.0 7.3
2486876.0 7.5
2486952.0 6.1
2487252.0 6.2
2487608.0 6.2
2487682.0 6.3
2488123.0 6.1
2488194.0 7.2
2488452.0 5.6
2488512.0 5.6
2488692.0 6.3
2488872.0 5.6
2489216.0 6.2
2489275.0 6.4
2489491.0 6.6
2492688.0 6.5
2492892.0 7.2
2493452.0 7.0
2493769.0 6.0
2494452.0 5.9
2494632.0 6.4
2496226.0 5.9
2496288.0 7.2
2503622.0 6.8
2506944.0 5.8
2508543.0 6.9
2508672.0 6.3
2509182.0 6.2
2509365.0 5.6
2509974.0 6.4
2510112.0 6.1
2510633.0 6.0
2511121.0 6.6
2511432.0 7.4
2511492.0 5.8
2511912.0 6.8
2512005.0 6.3
2512570.0 7.5
2513172.0 5.7
2514612.0 6.8
2514905.0 7.3
2515050.0 6.9
2515305.0 6.6
2515338.0 6.2
2515992.0 6.2
2516052.0 7.3
2518932.0 5.8
2519112.0 6.5
2519379.0 5.5
2519850.0 7.0
2520132.0 6.4
2520192.0 5.5
2520552.0 7.4
2520665.0 6.4
2520721.0 6.8
2521118.0 6.5
2521126.0 7.3
2521512.0 6.7
2521932.0 7.0
2522221.0 7.4
2522503.0 5.5
2522663.0 5.8
2522952.0 7.0
2523012.0 6.7
2523685.0 5.5
2524115.0 5.5
2524216.0 6.7
2524340.0 7.1
2524392.0 6.2
2524632.0 7.5
2524692.0 6.0
2524812.0 7.2
2525030.0 6.1
2525175.0 5.5
2525223.0 7.0
2525450.0 7.1
2525559.0 5.5
2526132.0 7.2
2526393.0 6.0
2526466.0 7.4
2526553.0 7.1
2526573.0 7.1
2526700.0 5.9
2526731.0 6.6
2527067.0 5.9
2527135.0 6.3
2527392.0 6.1
2527512.0 5.8
2528116.0 6.9
2528134.0 6.0
2528225.0 5.7
2528459.0 7.0
2528609.0 6.6
2528952.0 7.5
2529012.0 6.6
2529436.0 6.8
2529480.0 5.7
2530016.0 7.4
2530057.0 6.4
2530152.0 6.3
2530512.0 6.0
2530572.0 6.1
2530632.0 6.9
2531712.0 6.1
2531832.0 6.2
2532012.0 6.4
2532072.0 7.5
2532381.0 5.8
2532476.0 6.8
2532502.0 6.2
2532749.0 5.8
2532813.0 6.1
2533003.0 6.0
2533092.0 6.3
2533332.0 6.8
2533392.0 6.5
2533452.0 5.9
2533512.0 7.0
2533900.0 7.5
2534472.0 6.9
2535523.0 6.9
2535912.0 6.3
2537532.0 5.7
2538642.0 6.4
2539272.0 6.0
2542857.0 6.7
2543352.0 5.5
2544069.0 7.3
2545200.0 6.6
2545362.0 6.1
2545561.0 7.0
2546232.0 5.7
2547322.0 7.3
2547432.0 6.0
2547852.0 6.1
2548592.0 7.3
2550189.0 5.6
2551176.0 5.7
2552519.0 6.2
2552742.0 7.1
2553846.0 5.7
2554007.0 5.5
2554752.0 7.4
2561319.0 6.9
2562673.0 6.8
2563632.0 6.7
2564565.0 6.4
2565132.0 6.1
2565674.0 6.2
2566893.0 5.9

2567184.0 6.0
2567449.0 7.0
2567553.0 5.7
2567652.0 7.0
2568072.0 6.8
2568230.0 7.0
2568310.0 5.5
2568340.0 7.2
2568495.0 7.3
2569092.0 6.2
2569332.0 7.4
2569854.0 7.5
2570200.0 6.7
2570592.0 6.8
2571681.0 6.9
2571972.0 6.8
2572872.0 7.2
2572944.0 6.9
2573070.0 6.1
2573970.0 6.4
2574069.0 6.1
2574912.0 6.9
2579352.0 7.4
2581399.0 6.6
2581992.0 6.0
2582412.0 7.3
2582840.0 7.0
2584092.0 7.3
2584400.0 5.8
2584509.0 6.1
2584580.0 5.9
2584720.0 6.4
2584868.0 6.1
2584872.0 7.0
2584932.0 7.1
2585052.0 6.9
2585112.0 6.7
2585232.0 6.1
2585292.0 5.6
2585352.0 7.0
2585588.0 6.7
2585620.0 7.3
2585641.0 6.0
2585867.0 6.3
2585947.0 7.1
2585997.0 7.2
2586024.0 6.7
2587408.0 6.1
2587520.0 6.6
2588112.0 6.5
2588232.0 7.0
2589252.0 7.1
2589432.0 5.7
2590142.0 6.2
2590206.0 5.5
2590632.0 5.7
2592707.0 7.1
2592725.0 5.7
2594680.0 7.3
2595252.0 6.1
2599332.0 5.7
2599392.0 5.6
2601192.0 6.4
2601535.0 5.9
2603584.0 6.6
2604288.0 6.9
2604474.0 6.1
2604610.0 6.4
2606197.0 6.1
2606712.0 5.6
2607381.0 6.0
2607572.0 6.4
2608152.0 6.2
2608212.0 5.5
2608272.0 6.7
2608780.0 5.5
2608982.0 7.4
2610004.0 6.4
2610047.0 7.2
2610367.0 6.5
2610490.0 7.1
2611546.0 6.9
2612074.0 6.1
2612209.0 6.0
2612472.0 7.4
2612969.0 6.7
2613479.0 7.3
2613792.0 7.0
2614032.0 6.3
2614504.0 5.8
2614732.0 6.3
2615472.0 6.6
2616972.0 6.5
2617107.0 6.3
2617137.0 7.1
2618472.0 7.0
2620525.0 6.4
2621854.0 5.8
2622552.0 6.9
2622672.0 5.6
2622732.0 5.8
2622792.0 6.3
2622868.0 7.5
2623377.0 6.5
2623538.0 6.9
2623872.0 6.6
2623992.0 5.7
2624112.0 5.9
2625492.0 6.7
2626872.0 7.5
2626992.0 6.4
2627299.0 7.1
2627737.0 6.7
2629632.0 6.5
2630556.0 6.8
2630824.0 7.3
2631192.0 7.2
2631252.0 7.3
2632452.0 7.3
2632872.0 6.6
2634012.0 7.2
2635332.0 5.7
2636003.0 6.0
2636325.0 6.9
2637192.0 6.9
2637438.0 6.4
2639058.0 6.7
2640914.0 5.7
2640959.0 7.3
2641512.0 7.4
2641607.0 6.1
2641993.0 6.2
2643104.0 6.2
2643414.0 7.1
2643844.0 5.8
2644032.0 7.1
2644092.0 6.8
2644212.0 6.4
2644332.0 6.3
2644722.0 5.9
2644882.0 6.0
2645160.0 6.1
2645412.0 5.9
2645592.0 6.9
2645652.0 5.5
2646261.0 6.5
2646302.0 7.4
2646320.0 5.6
2646436.0 6.5
2646586.0 6.2
2647152.0 6.3
2647362.0 7.3
2647480.0 7.3
2647563.0 5.8
2647989.0 5.8
2648072.0 7.4
2648098.0 7.1
2648472.0 7.1
2648917.0 6.9
2649004.0 6.9
2649732.0 7.2
2650940.0 5.8
2651532.0 7.5
2651938.0 6.0
2652203.0 6.6
2652436.0 6.7
2653095.0 7.4
2654292.0 6.4
2654352.0 5.5
2654412.0 6.3
2654880.0 5.6
2655992.0 5.7
2656857.0 7.4
2660052.0 6.4
2660442.0 6.8
2660574.0 6.7
2660706.0 7.1
2660849.0 6.4
2661252.0 5.6
2661907.0 7.2
2663491.0 6.8
2664192.0 6.1
2664767.0 5.9
2664859.0 7.1
2664914.0 5.6
2665222.0 5.5
2665362.0 7.3
2665692.0 6.3
2666164.0 7.5
2666207.0 6.8
2666430.0 6.9
2666540.0 5.6
2666560.0 6.1
2667132.0 7.1
2669492.0 7.5
2670748.0 7.0
2670793.0 6.4
2670832.0 6.9
2671512.0 7.3
2671979.0 7.3
2672023.0 6.4
2672402.0 5.6
2672544.0 7.1
2672772.0 7.1
2672832.0 6.8
2673012.0 6.7
2673072.0 6.4
2673273.0 6.6
2673777.0 5.8
2674572.0 6.6
2674806.0 6.4
2675194.0 6.4
2675560.0 7.0
2676072.0 5.5
2676384.0 6.4
2677092.0 6.4
2677152.0 6.8
2677392.0 5.7
2677814.0 7.4
2678592.0 6.9
2680092.0 6.2
2680489.0 5.9
2680733.0 6.6
2680988.0 6.9
2683092.0 7.5
2683212.0 5.6
2684077.0 6.0
2684177.0 5.8
2684712.0 7.4
2684883.0 5.5
2685076.0 5.8
2685165.0 6.4
2685229.0 6.9
2685248.0 5.6
2685732.0 6.0
2685852.0 6.3
2685972.0 6.5
2686464.0 6.5
2686690.0 7.2
2686876.0 7.2
2686900.0 7.5
2687111.0 6.6
2687352.0 7.1
2687472.0 5.6
2688672.0 6.5
2688792.0 6.8
2689901.0 7.0
2690112.0 7.1
2691091.0 7.2
2691552.0 6.9
2692420.0 5.6
2692474.0 6.9
2692556.0 7.3
2693052.0 7.3
2693693.0 6.4
2694182.0 5.7
2694312.0 6.7
2694792.0 5.8
2695111.0 7.2
2698872.0 5.5
2700492.0 7.5
2700834.0 6.0
2700985.0 6.1
2701162.0 7.3
2701209.0 6.8
2701512.0 6.2
2701572.0 6.9
2701632.0 5.7
2701692.0 5.6
2701872.0 6.8
2702052.0 5.5
2702150.0 5.6
2702227.0 5.9
2702530.0 5.8
2702577.0 6.5
2702940.0 5.9
2704000.0 5.5
2704126.0 6.2
2704512.0 6.2
2705009.0 6.4
2705202.0 7.5
2705316.0 6.4
2706012.0 6.1
2706509.0 7.3
2707980.0 7.4
2708550.0 7.1
2708639.0 6.8
2712644.0 5.9
2712892.0 7.2
2715785.0 5.8
2716092.0 6.2
2716512.0 7.5
2717352.0 6.8
2717832.0 7.0
2719032.0 7.2
2719643.0 6.0
2720210.0 6.9
2721450.0 7.3
2722498.0 6.6
2722590.0 7.1
2722943.0 6.0
2723112.0 6.0
2723232.0 7.4
2725032.0 7.3
2725163.0 6.9
2725744.0 7.3
2726800.0 6.4
2727492.0 6.3
2730854.0 7.1
2731237.0 6.3
2732384.0 6.1
2732408.0 6.4
2732486.0 6.3
2734752.0 7.3
2735112.0 5.9
2736950.0 7.1
2738729.0 7.2
2739924.0 6.4
2739962.0 6.6
2740392.0 6.3
2740572.0 5.9
2740752.0 6.4
2740954.0 7.2
2741304.0 7.2
2741686.0 6.7
2742192.0 6.4
2742312.0 6.0
2742379.0 6.3
2743157.0 5.7
2743268.0 6.4
2743946.0 7.1
2744625.0 6.7
2745012.0 6.4
2745521.0 7.3
2745751.0 5.6
2745840.0 7.0
2747196.0 7.3
2747892.0 5.6
2748072.0 5.6
2749272.0 5.9
2749512.0 5.8
2750372.0 5.7
2750394.0 6.8
2751674.0 6.0
2751746.0 6.0
2751912.0 7.1
2752212.0 7.0
2752332.0 6.7
2753057.0 7.3
2753278.0 6.5
2753350.0 6.5
2754912.0 6.9
2757619.0 7.4
2757792.0 6.0
2758152.0 6.6
2758602.0 6.1
2758771.0 6.7
2759033.0 6.7
2759592.0 6.1
2759844.0 6.8
2760075.0 5.8
2761361.0 6.7
2761440.0 7.2
2761504.0 7.5
2761557.0 5.7
2761992.0 6.0
2762232.0 6.5
2762352.0 6.3
2763612.0 7.4
2764055.0 6.3
2764082.0 7.4
2764447.0 5.8
2764992.0 6.9
2765112.0 6.8
2765352.0 6.5
2765607.0 7.4
2766672.0 6.3
2769038.0 5.7
2769904.0 5.9
2770115.0 6.5
2770578.0 6.8
2770992.0 6.6
2778312.0 5.9
2779122.0 6.0
2779332.0 6.4
2779452.0 5.6
2779572.0 7.1
2779692.0 6.7
2779910.0 6.7
2780159.0 7.4
2780210.0 7.0
2780511.0 5.9
2780712.0 6.1
2780832.0 7.1
2780892.0 6.2
2781386.0 7.4
2781565.0 6.9
2782332.0 7.4
2782714.0 7.1
2782825.0 7.3
2782880.0 6.4
2782979.0 7.1
2783652.0 7.3
2783712.0 6.2
2785452.0 6.7
2785626.0 7.5
2786066.0 6.1
2786343.0 6.8
2791385.0 7.5
2791968.0 7.5
2792352.0 6.1
2793151.0 7.0
2793414.0 7.1
2793596.0 5.8
2793972.0 6.6
2794152.0 5.7
2794508.0 6.4
2796464.0 5.6
2800449.0 6.8
2800548.0 6.1
2801352.0 6.5
2803639.0 6.3
2804172.0 5.6
2804375.0 6.7
2804454.0 6.2
2806219.0 6.7
2808252.0 6.1
2808995.0 7.2
2809104.0 6.0
2812384.0 7.3
2816210.0 7.0
2816712.0 6.4
2817539.0 6.7
2818958.0 6.7
2819772.0 7.1
2820187.0 7.1
2820745.0 5.6
2820779.0 5.8
2820856.0 6.1
2821032.0 6.5
2821092.0 5.8
2821272.0 6.6
2821452.0 6.7
2821838.0 7.5
2822022.0 7.0
2822363.0 6.1
2822398.0 6.3
2822772.0 6.0
2822892.0 6.1
2823342.0 5.8
2823405.0 6.6
2823617.0 6.9
2823653.0 7.1
2823802.0 5.6
2825010.0 5.6
2825146.0 5.9
2825177.0 7.0
2825412.0 6.8
2825832.0 6.5
2826486.0 6.3
2827152.0 6.3
2828592.0 7.2
2829732.0 5.6
2829972.0 7.4
2830032.0 6.9
2831106.0 6.4
2831592.0 6.0
2833218.0 7.0
2835981.0 6.4
2839812.0 6.8
2839932.0 7.3
2840232.0 7.1
2841252.0 5.9
2841789.0 6.1
2842752.0 6.1
2843292.0 7.3
2845812.0 5.7
2847192.0 6.4
2847372.0 6.2
2848217.0 5.9
2848452.0 7.1
2848632.0 6.0
2850012.0 6.3
2851001.0 5.9
2857032.0 6.3
2858772.0 6.0
2860720.0 6.2
2862113.0 6.6
2862604.0 6.4
2862702.0 5.6
2864970.0 5.7
2866092.0 5.6
2866630.0 6.1
2866906.0 7.3
2868090.0 7.1
2868338.0 6.3
2869167.0 6.5
2869619.0 6.8
2869712.0 6.1
2870112.0 6.0
2870412.0 6.5
2870809.0 6.3
2871400.0 6.8
2872172.0 5.8
2877017.0 5.5
2877672.0 7.5
2881366.0 6.9
2881512.0 7.5
2881752.0 6.9
2882160.0 6.8
2882168.0 7.3
2882360.0 5.6
2882525.0 6.9
2882846.0 5.6
2882916.0 6.9
2883192.0 7.4
2883312.0 6.2
2883432.0 6.5
2883680.0 6.3
2884268.0 7.2
2884452.0 6.2
2885219.0 7.2
2885892.0 6.5
2885952.0 5.7
2886252.0 5.6
2886312.0 6.0
2887272.0 6.1
2887392.0 7.0
2887632.0 5.8
2887857.0 5.6
2887878.0 6.2
2888224.0 6.6
2888247.0 6.6
2888772.0 6.7
2889072.0 7.1
2889344.0 6.7
2889673.0 6.6
2890152.0 7.2
2890726.0 7.2
2891056.0 5.6
2891307.0 6.3
2891652.0 7.2
2892012.0 6.3
2894532.0 7.4
2896152.0 5.9
2896983.0 6.3
2897240.0 5.6
2897302.0 6.4
2897772.0 5.8
2898676.0 7.3
2899396.0 5.6
2901384.0 6.5
2901912.0 6.7
2902379.0 6.4
2905406.0 5.9
2906782.0 6.1
2907912.0 5.6
2908000.0 7.0
2908671.0 7.5
2908755.0 6.1
2908872.0 7.2
2908932.0 7.4
2909424.0 7.2
2909755.0 5.5
2909927.0 7.3
2910169.0 7.2
2910312.0 5.8
2911164.0 6.3
2912542.0 5.9
2913192.0 5.7
2914566.0 6.8
2915589.0 5.6
2915909.0 7.0
2915923.0 5.6
2916851.0 6.5
2917337.0 6.7
2917872.0 6.5
2919372.0 6.6
2919432.0 7.0
2919557.0 7.1
2920452.0 6.4
2920512.0 6.9
2920949.0 7.4
2921558.0 6.5
2921824.0 6.1
2922445.0 6.6
2922501.0 7.4
2922539.0 6.5
2922645.0 7.1
2922792.0 6.6
2922959.0 7.4
2923271.0 7.3
2923752.0 5.5
2924333.0 5.8
2924712.0 7.2
2925132.0 7.0
2925567.0 6.1
2926152.0 5.9
2926452.0 6.5
2926949.0 5.6
2928769.0 6.4
2929650.0 7.3
2929954.0 6.7
2930352.0 5.8
2930532.0 7.0
2930892.0 6.5
2931047.0 6.4
2931690.0 7.1
2931818.0 6.0
2932332.0 6.2
2934027.0 6.9
2934228.0 7.0
2934320.0 6.8
2935032.0 6.5
2936232.0 7.5
2936352.0 6.2
2936532.0 5.9
2936652.0 6.9
2936837.0 7.1
2936967.0 7.4
2937657.0 7.1
2937672.0 7.0
2938562.0 6.5
2940273.0 6.8
2943492.0 6.0
2945112.0 7.1
2945172.0 6.6
2949010.0 6.8
2949612.0 7.3
2949834.0 7.1
2951220.0 6.9
2953872.0 5.5
2955312.0 5.6
2955372.0 6.9
2955872.0 6.6
2955882.0 6.6
2957044.0 6.1
2957718.0 6.9
2958012.0 6.0
2958192.0 5.9
2958746.0 7.1
2961072.0 6.9
2961442.0 7.3
2962332.0 6.0
2962452.0 7.1
2962826.0 5.9
2963284.0 6.2
2963515.0 6.6
2964137.0 6.4
2964306.0 6.9
2964558.0 5.5
2964697.0 5.9
2964799.0 7.3
2968890.0 6.0
2969135.0 6.3
2970912.0 6.9
2971212.0 6.5
2972005.0 5.9
2973672.0 5.8
2974425.0 6.1
2976379.0 5.8
2976399.0 6.3
2977032.0 6.9
2977283.0 7.1
2977620.0 6.0
2978052.0 7.3
2978847.0 6.1
2979552.0 6.9
2985192.0 6.1
2985552.0 6.2
2989632.0 7.5
2989692.0 6.1
2992223.0 6.8
2993450.0 6.1
2993690.0 5.6
2993751.0 6.0
2996629.0 6.5
2997615.0 6.1
2997667.0 6.3
2998135.0 5.8
2998152.0 5.7
2998392.0 6.4
2998512.0 6.8
2998632.0 7.0
2999249.0 7.4
2999429.0 5.8
2999772.0 6.7
3001212.0 6.3
3001272.0 6.7
3002532.0 6.7
3002592.0 5.8
3004032.0 6.8
3004588.0 5.9
3005268.0 6.4
3005333.0 5.8
3005592.0 7.5
3006076.0 7.0
3007032.0 7.1
3008292.0 6.4
3010923.0 5.9
3010982.0 7.1
3013840.0 5.7
3013992.0 5.7
3014620.0 6.1
3015032.0 7.4
3015224.0 5.8
3015852.0 6.8
3015912.0 6.8
3016087.0 6.1
3016516.0 6.1
3016625.0 5.7
3017112.0 7.0
3017292.0 7.0
3018792.0 6.2
3018901.0 7.2
3019112.0 6.7
3019142.0 5.6
3020172.0 7.4
3021160.0 6.8
3021252.0 5.9
3021372.0 6.5
3021552.0 5.7
3021970.0 5.5
3024874.0 5.7
3025063.0 7.1
3029100.0 7.2
3033995.0 5.6
3034272.0 6.3
3035077.0 7.3
3035497.0 7.3
3035592.0 6.9
3035892.0 5.7
3036290.0 7.2
3036488.0 6.4
3036817.0 6.2
3036982.0 5.7
3037642.0 6.0
3037702.0 7.3
3037742.0 7.0
3037862.0 6.2
3038398.0 7.4
3038532.0 7.0
3038652.0 5.6
3038712.0 7.5
3039185.0 6.2
3039527.0 6.2
3040152.0 6.1
3040748.0 6.1
3041081.0 7.2
3041197.0 5.9
3041712.0 6.4
3042664.0 6.1
3042792.0 5.6
3043754.0 5.6
3045313.0 7.2
3046152.0 6.0
3050172.0 6.7
3054131.0 7.3
3054219.0 7.0
3054956.0 7.2
3055300.0 6.2
3055307.0 6.6
3055624.0 6.9
3056052.0 5.6
3057672.0 5.9
3058087.0 7.3
3058452.0 6.2
3059768.0 6.8
3059848.0 7.4
3060739.0 6.2
3060852.0 6.0
3060900.0 7.1
3061512.0 7.4
3065527.0 6.8
3065744.0 6.5
3067572.0 6.4
3068050.0 6.5
3074044.0 7.0
3075460.0 7.3
3075972.0 5.6
3076032.0 5.5
3076392.0 6.8
3076850.0 5.7
3077472.0 6.5
3078527.0 6.7
3078792.0 7.4
3079032.0 7.4
3079092.0 5.9
3079272.0 6.8
3079438.0 7.2
3079454.0 6.1
3079905.0 7.1
3080472.0 6.6
3081076.0 6.9
3081214.0 6.9
3081792.0 7.1
3082460.0 6.8
3082747.0 6.7
3083592.0 6.5
3085111.0 5.7
3085990.0 6.9
3087094.0 5.7
3088402.0 6.3
3091264.0 6.0
3093097.0 6.0
3093312.0 7.0
3093432.0 6.0
3093803.0 6.0
3093845.0 6.6
3094568.0 5.7
3094872.0 6.8
3095307.0 7.3
3095922.0 6.9
3098730.0 7.3
3099252.0 5.6
3099432.0 6.8
3100692.0 5.8
3102422.0 6.7
3104532.0 7.4
3105012.0 6.7
3106272.0 7.4
3106972.0 6.4
3114852.0 7.1
3115032.0 6.0
3115973.0 7.3
3117395.0 5.7
3118255.0 6.3
3118525.0 6.8
3118604.0 6.2
3118679.0 6.0
3118901.0 7.1
3118949.0 5.6
3119232.0 6.9
3119472.0 6.2
3120388.0 6.0
3120414.0 6.0
3120552.0 7.2
3120732.0 7.1
3121032.0 6.1
3121226.0 5.5
3122292.0 7.0
3122472.0 5.8
3123037.0 6.2
3123492.0 6.5
3123732.0 6.8
3124216.0 5.6
3124736.0 6.5
3125052.0 7.0
3129312.0 5.6
3130383.0 6.6
3130600.0 6.0
3130632.0 6.1
3131582.0 7.2
3132072.0 6.4
3132432.0 7.0
3132552.0 5.6
3132632.0 5.8
3132849.0 7.3
3134205.0 7.2
3134488.0 7.1
3134909.0 7.0
3134952.0 6.9
3136392.0 5.6
3137892.0 5.7
3138689.0 5.9
3144442.0 6.2
3155532.0 7.1
3155738.0 7.1
3156612.0 7.4
3157112.0 7.1
3157421.0 6.8
3157764.0 7.4
3158292.0 6.6
3158352.0 7.1
3158561.0 6.1
3159236.0 6.6
3159792.0 6.1
3159912.0 7.2
3160188.0 7.2
3160252.0 6.9
3160582.0 6.1
3161112.0 7.4
3161426.0 7.2
3161826.0 6.4
3161870.0 7.0
3162311.0 6.0
3162492.0 5.9
3162905.0 7.3
3164649.0 5.8
3164842.0 6.2
3165192.0 6.9
3165252.0 5.9
3165672.0 6.9
3166599.0 6.2
3168730.0 6.8
3169230.0 7.3
3169692.0 7.2
3170167.0 7.3
3171432.0 7.5
3171676.0 6.1
3171821.0 5.9
3171967.0 7.3
3172036.0 7.2
3172263.0 6.1
3172452.0 6.0
3173038.0 6.4
3173215.0 6.8
3173892.0 7.5
3174980.0 6.8
3175168.0 6.3
3175752.0 5.7
3176437.0 6.6
3176680.0 7.4
3177561.0 7.2
3177624.0 6.6
3177746.0 7.2
3177934.0 6.4
3178272.0 7.1
3178864.0 7.5
3180338.0 6.7
3180421.0 6.5
3187559.0 6.2
3189253.0 5.8
3190562.0 5.7
3193092.0 6.9
3194706.0 6.7
3195229.0 7.4
3195424.0 7.3
3196025.0 6.5
3196407.0 6.2
3196932.0 7.4
3197352.0 5.7
3197487.0 5.7
3198066.0 6.3
3198492.0 5.8
3198672.0 7.4
3198792.0 5.7
3198962.0 6.4
3199272.0 7.5
3199932.0 6.3
3200112.0 5.8
3200306.0 7.5
3200410.0 5.7
3201552.0 5.8
3202396.0 7.3
3202812.0 5.6
3203218.0 6.0
3203430.0 5.6
3203958.0 7.4
3205192.0 6.1
3205512.0 6.6
3205572.0 7.1
3205632.0 6.4
3205752.0 5.8
3206422.0 7.2
3206492.0 6.6
3206578.0 5.7
3207132.0 7.0
3207539.0 6.6
3209348.0 5.7
3209385.0 6.9
3209452.0 5.8
3209590.0 5.7
3210192.0 6.1
3210674.0 6.3
3210696.0 5.6
3211123.0 7.4
3212162.0 7.3
3212276.0 7.3
3212712.0 6.7
3212832.0 5.8
3213192.0 6.8
3213539.0 6.4
3213656.0 7.0
3213693.0 7.4
3214771.0 7.1
3214826.0 7.0
3215216.0 5.9
3215257.0 6.6
3215712.0 7.4
3216072.0 6.7
3217032.0 7.0
3217706.0 7.3
3218124.0 7.1
3218314.0 5.7
3220462.0 6.9
3221200.0 5.9
3221352.0 6.3
3222235.0 6.8
3223759.0 5.7
3223810.0 7.0
3225275.0 6.4
3225912.0 5.7
3227767.0 7.4

3232160.0 6.2
3233232.0 6.6
3233352.0 6.6
3233693.0 5.7
3234168.0 6.9
3234213.0 6.1
3234792.0 6.1
3234936.0 6.1
3235042.0 6.9
3235227.0 6.8
3235514.0 7.4
3238042.0 7.4
3238378.0 7.1
3238409.0 6.1
3239052.0 6.2
3240699.0 5.7
3240949.0 6.6
3241692.0 6.3
3242082.0 7.3
3242505.0 7.2
3242705.0 7.5
3243432.0 5.9
3243822.0 6.4
3244692.0 5.8
3246012.0 6.7
3246312.0 6.1
3246800.0 6.1
3247130.0 7.3
3248024.0 6.5
3248354.0 6.1
3248629.0 6.4
3249012.0 5.8
3249072.0 5.5
3249974.0 6.9
3250272.0 7.2
3250392.0 7.3
3250452.0 5.6
3250572.0 7.3
3250632.0 6.8
3250712.0 7.0
3250844.0 6.9
3251017.0 6.2
3251034.0 6.6
3251488.0 5.6
3251505.0 5.7
3251652.0 7.2
3252248.0 6.2
3252418.0 6.5
3252649.0 5.7
3252915.0 7.3
3253392.0 5.9
3253650.0 7.4
3253824.0 7.0
3254090.0 6.9
3254170.0 5.6
3254772.0 5.6
3255207.0 7.3
3255864.0 6.1
3256392.0 5.5
3257532.0 6.3
3260232.0 7.1
3261792.0 7.0
3267732.0 5.5
3268413.0 6.2
3269848.0 6.4
3270372.0 7.1
3270552.0 7.5
3270732.0 6.8
3271229.0 5.6
3271447.0 7.0
3272172.0 6.4
3272361.0 7.0
3272786.0 6.5
3273252.0 7.4
3273432.0 6.5
3273492.0 6.6
3273672.0 6.7
3274044.0 7.3
3274054.0 6.5
3274354.0 5.7
3274413.0 5.9
3274992.0 5.5
3275342.0 6.7
3275523.0 6.5
3275894.0 5.7
3276025.0 7.1
3276039.0 6.9
3276492.0 7.3
3277046.0 7.4
3278494.0 6.3
3278591.0 5.7
3278659.0 6.6
3279432.0 7.1
3280993.0 7.1
3284712.0 5.6
3286572.0 6.9
3286862.0 6.3
3287712.0 6.5
3287772.0 7.0
3287892.0 5.9
3288661.0 7.1
3288726.0 6.9
3289092.0 6.2
3289332.0 5.8
3289392.0 5.7
3289452.0 6.5
3289512.0 6.6
3289984.0 5.9
3290019.0 6.1
3290042.0 6.4
3290049.0 6.5
3290142.0 6.4
3290652.0 6.5
3290772.0 7.3
3290832.0 6.1
3290892.0 6.9
3290952.0 7.2
3291016.0 6.7
3291141.0 6.8
3291154.0 6.1
3291553.0 6.3
3291668.0 7.3
3291800.0 6.6
3292032.0 6.7
3292332.0 6.9
3292758.0 6.4
3292929.0 7.4
3293034.0 5.7
3293412.0 7.2
3293532.0 6.1
3293903.0 6.6
3294228.0 7.5
3294430.0 6.3
3295152.0 6.2
3295212.0 5.8
3295661.0 6.9
3295832.0 6.4
3296167.0 5.7
3296292.0 6.5
3297096.0 6.4
3297502.0 6.6
3297792.0 6.4
3299532.0 7.4
3300467.0 6.0
3301468.0 7.4
3304446.0 7.4
3305232.0 6.3
3305824.0 5.9
3310410.0 6.1
3310692.0 7.0
3310872.0 5.8
3311112.0 6.7
3312252.0 6.9
3312666.0 6.6
3312748.0 6.4
3313087.0 7.4
3313572.0 7.2
3313632.0 6.4
3313692.0 6.2
3314308.0 7.1
3314687.0 6.8
3314881.0 5.9
3315192.0 7.3
3315252.0 6.7
3315312.0 6.0
3315432.0 5.7
3320079.0 5.7
3322070.0 6.3
3322741.0 6.4
3322814.0 6.2
3322858.0 6.3
3325152.0 6.8
3327550.0 7.4
3327812.0 5.5
3327972.0 7.2
3328152.0 6.2
3328332.0 6.9
3328643.0 6.4
3328862.0 5.7
3328902.0 7.1
3329652.0 6.5
3330054.0 7.0
3330100.0 7.0
3330226.0 7.4
3330326.0 7.0
3330364.0 6.3
3330438.0 6.2
3330509.0 7.1
3331092.0 5.5
3331506.0 7.3
3331661.0 5.6
3331839.0 7.0
3332204.0 5.8
3332292.0 6.1
3332652.0 5.8
3332950.0 6.1
3332979.0 7.5
3333262.0 6.8
3333449.0 6.7
3333672.0 6.2
3333972.0 5.5
3334092.0 5.8
3334422.0 5.8
3334875.0 6.4
3335112.0 6.6
3336017.0 6.0
3336174.0 7.5
3336852.0 7.2
3337626.0 5.9
3338232.0 7.0
3338957.0 5.5
3339552.0 6.1
3341232.0 6.0
3341692.0 5.6
3341790.0 6.9
3342792.0 7.4
3344774.0 5.9
3348647.0 7.4
3348975.0 7.3
3349049.0 5.7
3349572.0 6.2
3349932.0 6.7
3350123.0 6.8
3350207.0 5.6
3350412.0 7.3
3351915.0 6.1
3353010.0 7.3
3353253.0 6.3
3354505.0 5.8
3354764.0 5.9
3355572.0 6.9
3355632.0 7.1
3356130.0 6.6
3356705.0 6.1
3356832.0 7.5
3356892.0 6.6
3357132.0 7.2
3357434.0 6.0
3357668.0 6.6
3358855.0 7.1
3359712.0 6.1
3360362.0 7.1
3360595.0 6.2
3360809.0 5.8
3361092.0 6.3
3362072.0 6.5
3362532.0 6.5
3362832.0 7.5
3363012.0 5.9
3363336.0 6.8
3363893.0 7.1
3364888.0 7.1
3364898.0 7.5
3364978.0 7.2
3365232.0 7.0
3366332.0 6.8
3368532.0 6.9
3368712.0 6.0
3368876.0 6.2
3371084.0 6.4
3371172.0 6.8
3372107.0 6.0
3372386.0 7.0
3372792.0 5.6
3373969.0 6.2
3374617.0 7.0
3374698.0 5.9
3377441.0 6.8
3378432.0 6.2
3378880.0 6.9
3379320.0 6.9
3380353.0 6.8
3383265.0 6.9
3383617.0 7.3
3384492.0 6.4
3384692.0 6.4
3384885.0 7.2
3384958.0 5.6
3387812.0 6.4
3387939.0 6.4
3388310.0 6.3
3389137.0 7.4
3390012.0 6.1
3390502.0 6.9
3390992.0 6.0
3391087.0 6.9
3391114.0 6.0
3391272.0 5.5
3392017.0 7.3
3392633.0 6.8
3392772.0 6.2
3393589.0 5.7
3393872.0 7.3
3394699.0 6.3
3395652.0 7.2
3395832.0 5.6
3395952.0 5.8
3396012.0 6.2
3396143.0 7.1
3396256.0 7.0
3396294.0 7.3
3396598.0 6.8
3396617.0 7.3
3396986.0 6.4
3398652.0 6.8
3399225.0 7.2
3400092.0 6.7
3402792.0 6.2
3406595.0 7.4
3407352.0 6.5
3407592.0 6.1
3407687.0 6.7
3408486.0 7.3
3408549.0 6.9
3409992.0 7.2
3410292.0 6.0
3410727.0 6.4
3412046.0 7.3
3412932.0 7.2
3413352.0 6.1
3415723.0 6.3
3415872.0 7.0
3417192.0 6.9
3417813.0 6.5
3418297.0 6.0
3418752.0 7.2
3419289.0 5.6
3419619.0 5.6
3420905.0 7.0
3424392.0 6.7
3424812.0 7.5
3425832.0 5.6
3426423.0 6.1
3427572.0 6.3
3427865.0 5.8
3428952.0 7.0
3429305.0 6.4
3429617.0 5.6
3429702.0 5.9
3429707.0 7.2
3429978.0 5.8
3430212.0 6.4
3430512.0 5.5
3430572.0 7.3
3431309.0 6.0
3431464.0 7.4
3431537.0 6.4
3431591.0 7.1
3431952.0 7.0
3432532.0 6.5
3432726.0 6.5
3433212.0 6.6
3433904.0 7.4
3434325.0 5.6
3434532.0 6.7
3435086.0 6.1
3437079.0 6.0
3437652.0 7.0
3437712.0 7.0
3438792.0 6.5
3438912.0 7.4
3439960.0 5.6
3440230.0 6.1
3440532.0 6.7
3443532.0 6.7
3446112.0 6.6
3446292.0 7.3
3446645.0 6.6
3446679.0 6.6
3446703.0 7.4
3446930.0 5.6
3446932.0 6.4
3446947.0 7.3
3446977.0 7.2
3447055.0 6.8
3447236.0 6.9
3447338.0 7.1
3447432.0 6.9
3447732.0 6.7
3448419.0 5.9
3448534.0 7.4
3448796.0 7.0
3448932.0 6.7
3449052.0 6.0
3449172.0 6.5
3449449.0 5.7
3449623.0 6.4
3449807.0 5.8
3450492.0 7.4
3450884.0 6.9
3451495.0 6.3
3452384.0 6.9
3452475.0 6.3
3452818.0 6.6
3453492.0 5.6
3457572.0 7.5
3457932.0 6.8
3458682.0 7.3
3458716.0 5.7
3460392.0 6.1
3461127.0 6.1
3464440.0 6.2
3465817.0 6.2
3466125.0 6.4
3466152.0 5.8
3466332.0 7.1
3466632.0 5.6
3466764.0 7.4
3466894.0 5.8
3467113.0 6.7
3467178.0 6.7
3467322.0 6.1
3467404.0 7.1
3467592.0 6.2
3467712.0 5.5
3467832.0 7.3
3468850.0 7.1
3469032.0 5.6
3469152.0 6.5
3469392.0 5.8
3469512.0 5.6
3469644.0 5.7
3470386.0 7.0
3470532.0 5.6
3470772.0 6.3
3471045.0 6.5
3471105.0 6.5
3471114.0 6.7
3471355.0 6.3
3471422.0 5.7
3471655.0 6.2
3471777.0 7.0
3471904.0 6.5
3472721.0 5.8
3472899.0 5.9
3473171.0 7.1
3473293.0 5.7
3473652.0 5.5
3473832.0 6.3
3473908.0 5.7
3474243.0 6.6
3474562.0 7.0
3474972.0 6.9
3475032.0 6.7
3475690.0 7.3
3475892.0 5.5
3476015.0 7.3
3476122.0 7.1
3476151.0 7.1
3476352.0 6.9
3476850.0 5.9
3476873.0 7.1
3477056.0 6.0
3477158.0 6.5
3477792.0 6.4
3477912.0 6.7
3478032.0 7.0
3478763.0 7.4
3478987.0 7.2
3479089.0 6.7
3479112.0 6.0
3479352.0 5.5
3479992.0 6.4
3480552.0 6.1
3480792.0 6.1
3481102.0 6.9
3481289.0 6.1
3481327.0 5.5
3481802.0 6.4
3481972.0 6.8
3481992.0 7.4
3482112.0 5.5
3482292.0 6.6
3482352.0 6.5
3482804.0 7.1
3482930.0 7.0
3483343.0 6.8
3483352.0 6.2
3483432.0 5.6
3483612.0 6.8
3483792.0 6.9
3483993.0 5.8
3484050.0 6.1
3484172.0 6.8
3484437.0 7.4
3484638.0 6.3
3484764.0 6.6
3485232.0 6.2
3485352.0 5.9
3485479.0 5.7
3485647.0 6.7
3485702.0 5.5
3486162.0 6.7
3486312.0 6.7
3486612.0 6.8
3487006.0 6.8
3487069.0 6.5
3487109.0 6.8
3487262.0 6.6
3487343.0 6.4
3487453.0 7.1
3487697.0 6.6
3487752.0 5.8
3487872.0 7.2
3487992.0 6.4
3488172.0 5.8
3488343.0 7.5
3488443.0 6.8
3488577.0 7.2
3488751.0 5.7
3488904.0 7.2
3488983.0 7.0
3489192.0 6.0
3489312.0 7.2
3489612.0 6.4
3489912.0 6.5
3490255.0 5.5
3490453.0 5.7
3491302.0 6.1
3492252.0 6.0
3494064.0 6.8
3498488.0 5.6
3500242.0 6.0
3502272.0 6.9
3504252.0 6.1
3504293.0 5.8
3505512.0 6.4
3505745.0 5.7
3506712.0 5.9
3507635.0 7.1
3507846.0 6.5
3508092.0 6.7
3508272.0 5.7
3508332.0 7.1
3508784.0 6.1
3509068.0 6.7
3509330.0 5.6
3509925.0 6.1
3510470.0 6.9
3510581.0 6.7
3510704.0 5.6
3511587.0 6.1
3512292.0 5.5
3512472.0 6.9
3512532.0 6.1
3512592.0 6.3
3512974.0 7.2
3513103.0 7.4
3513543.0 6.6
3514316.0 6.9
3514494.0 7.5
3514735.0 7.1
3515854.0 6.9
3516166.0 6.4
3516257.0 7.0
3516405.0 6.1
3516552.0 6.2
3519792.0 6.1
3520631.0 7.2
3523117.0 6.9
3523380.0 7.2
3523493.0 7.4
3523752.0 7.4
3523812.0 7.2
3524052.0 7.5
3524232.0 6.2
3524731.0 6.4
3524802.0 6.6
3524902.0 6.1
3525192.0 6.7
3525252.0 7.4
3525312.0 6.7
3525552.0 6.6
3525740.0 7.4
3525816.0 6.0
3525827.0 5.9
3525903.0 5.7
3526089.0 7.4
3526097.0 6.5
3526132.0 5.7
3526503.0 6.4
3526517.0 6.2
3526812.0 6.7
3526992.0 6.8
3527112.0 6.9
3527229.0 6.3
3527390.0 6.3
3527401.0 5.9
3527429.0 7.4
3527653.0 5.7
3527694.0 6.3
3527768.0 6.4
3527897.0 7.5
3527959.0 7.2
3528033.0 5.6
3528072.0 6.4
3528312.0 6.7
3528492.0 7.0
3528648.0 7.4
3528734.0 5.6
3528976.0 5.6
3529174.0 5.7
3529216.0 7.3
3529572.0 6.5
3529932.0 6.2
3530078.0 5.6
3530120.0 6.9
3530374.0 5.5
3530493.0 5.8
3530506.0 5.8
3530657.0 5.9
3530709.0 6.1
3531072.0 6.5
3531252.0 6.6
3531666.0 7.1
3531734.0 5.7
3531856.0 6.7
3532330.0 5.7
3532392.0 5.8
3534132.0 5.5
3537012.0 6.7
3537072.0 7.0
3537498.0 6.0
3539482.0 5.8
3543710.0 5.5
3544032.0 6.5
3544392.0 6.2
3544507.0 6.0
3545352.0 5.9
3545412.0 5.6
3545652.0 7.3
3545832.0 6.2
3546412.0 5.5
3546584.0 7.1
3546657.0 7.0
3546912.0 7.2
3547404.0 5.8
3547597.0 7.0
3547944.0 6.0
3548352.0 7.2
3548532.0 5.6
3548939.0 5.8
3548955.0 6.3
3549014.0 6.2
3549044.0 6.4
3549175.0 6.6
3549190.0 7.3
3549236.0 7.4
3549732.0 7.3
3549912.0 6.8
3550376.0 6.0
3550812.0 6.6
3551412.0 6.5
3552457.0 5.8
3552478.0 7.1
3552494.0 5.9
3552612.0 5.7
3552852.0 5.6
3553514.0 5.6
3553992.0 6.3
3554607.0 6.8
3555279.0 6.4
3555409.0 6.7
3556255.0 7.3
3556514.0 6.6
3557352.0 6.5
3558119.0 6.8
3558792.0 6.2
3560232.0 6.8
3560725.0 7.1
3560967.0 7.1
3561139.0 7.4
3561312.0 5.7
3561672.0 6.1
3562632.0 6.1
3562692.0 6.7
3563112.0 5.7
3563361.0 6.5
3563775.0 5.9
3563787.0 6.3
3564072.0 7.2
3564192.0 7.3
3564252.0 5.9
3564492.0 7.3
3564767.0 6.4
3564991.0 7.1
3565407.0 5.8
3565444.0 6.6
3565572.0 7.0
3565692.0 6.0
3565812.0 7.1
3565872.0 6.1
3566084.0 6.6
3566154.0 7.0
3566251.0 5.8
3566396.0 5.7
3566519.0 5.9
3566534.0 6.3
3566670.0 5.7
3567012.0 6.4
3567709.0 6.5
3567723.0 5.6
3567935.0 6.7
3568307.0 5.9
3568572.0 7.2
3568752.0 6.5
3568812.0 7.0
3569166.0 7.2
3569353.0 5.7
3569678.0 7.0
3569699.0 5.7
3569832.0 7.0
3570132.0 5.8
3570739.0 6.7
3571195.0 7.2
3571572.0 5.5
3571692.0 6.6
3571863.0 6.5
3572305.0 7.2
3572952.0 7.2
3573072.0 6.4
3573450.0 7.1
3574512.0 6.4
3576072.0 7.1
3576288.0 7.2
3580212.0 7.1
3582912.0 5.9
3583909.0 7.4
3584292.0 7.3
3585100.0 6.8
3585184.0 6.7
3585599.0 7.0
3585732.0 5.6
3585792.0 6.9
3585852.0 6.1
3586375.0 6.6
3586436.0 5.8
3586584.0 6.2
3586620.0 7.2
3586710.0 5.7
3587718.0 6.2
3587928.0 6.8
3587946.0 6.9
3588046.0 7.4
3588113.0 7.0
3588246.0 7.1
3588309.0 7.1
3588432.0 7.3
3588672.0 7.3
3588792.0 6.8
3588972.0 6.7
3589032.0 6.4
3589246.0 6.7
3589355.0 5.7
3589411.0 7.4
3589482.0 5.8
3589530.0 5.5
3589714.0 7.0
3589933.0 6.5
3589992.0 7.0
3590052.0 7.3
3590472.0 5.7
3590902.0 7.4
3591053.0 6.6
3591379.0 6.5
3591492.0 7.3
3592112.0 7.3
3592153.0 5.7
3592170.0 6.7
3592315.0 6.9
3593232.0 6.5
3593352.0 5.7
3594148.0 7.1
3594279.0 7.0
3594312.0 6.8
3594732.0 6.1
3595494.0 6.5
3595517.0 6.6
3595662.0 6.8
3596503.0 5.7
3596911.0 7.2
3596933.0 6.9
3597113.0 7.1
3597143.0 6.3
3597192.0 6.6
3597312.0 6.7
3598040.0 6.3
3598059.0 7.4
3598425.0 6.2
3598632.0 5.8
3601445.0 6.1
3602500.0 5.9
3602790.0 6.2
3602918.0 6.0
3603799.0 6.7
3603897.0 5.6
3604077.0 6.7
3604162.0 6.6
3604304.0 7.2
3604383.0 5.8
3604632.0 7.2
3604812.0 6.1
3604983.0 6.3
3605027.0 6.3
3605222.0 7.1
3605267.0 7.4
3605425.0 7.2
3605529.0 6.5
3605598.0 7.5
3605662.0 7.1
3605892.0 6.9
3606427.0 7.0
3606462.0 6.3
3606916.0 7.2
3607059.0 7.3
3607272.0 7.5
3607332.0 6.4
3607632.0 5.6
3608185.0 6.3
3608712.0 6.7
3608892.0 6.1
3609192.0 7.3
3609686.0 7.3
3609810.0 6.3
3610106.0 7.0
3610120.0 7.0
3610332.0 6.2
3610452.0 7.1
3610512.0 7.4
3610632.0 5.8
3611712.0 7.3
3612012.0 5.7
3612376.0 6.8
3613032.0 7.3
3613332.0 6.6
3613452.0 6.6
3613646.0 6.7
3613829.0 7.0
3614712.0 6.9
3614892.0 5.5
3614952.0 6.8
3615018.0 7.0
3615318.0 6.8
3616480.0 7.5
3616619.0 7.4
3617652.0 6.3
3619429.0 6.1
3621320.0 7.4
3622384.0 6.7
3622534.0 6.7
3622839.0 5.6
3622850.0 5.7
3623172.0 6.4
3623532.0 5.9
3623825.0 6.0
3623973.0 6.3
3624282.0 5.7
3624612.0 6.5
3625032.0 5.6
3625662.0 6.5
3626112.0 6.3
3626232.0 6.2
3626764.0 6.2
3627228.0 5.6
3627552.0 6.4
3628283.0 6.1
3628306.0 7.4
3628329.0 7.0
3628369.0 5.8
3628406.0 6.4
3628484.0 5.6
3628872.0 7.4
3629052.0 7.2
3629232.0 5.8
3629737.0 6.8
3629807.0 6.3
3630076.0 6.3
3630118.0 7.4
3630218.0 7.1
3630262.0 6.7
3630312.0 7.2
3630612.0 5.6
3631256.0 5.8
3631425.0 6.5
3631697.0 6.9
3631872.0 7.0
3632052.0 6.4
3632232.0 5.7
3632436.0 6.2
3632524.0 6.4
3633038.0 6.2
3633252.0 5.6
3633372.0 6.2
3633790.0 7.3
3634042.0 5.6
3634090.0 6.1
3634106.0 7.4
3634342.0 6.5
3634569.0 7.5
3634932.0 6.3
3635675.0 6.9
3636034.0 6.2
3638167.0 6.0
3639012.0 6.3
3639997.0 7.2
3642252.0 5.9
3642598.0 6.2
3642625.0 6.7
3642664.0 6.8
3642679.0 6.5
3642704.0 6.3
3642777.0 6.0
3643086.0 6.2
3643272.0 6.5
3643392.0 6.1
3643632.0 6.9
3643752.0 6.4
3643867.0 7.4
3644094.0 6.9
3644131.0 5.7
3644145.0 5.8
3644172.0 7.1
3644395.0 6.4
3644407.0 7.3
3644502.0 7.2
3644533.0 7.3
3644555.0 7.0
3644599.0 6.8
3644952.0 7.4
3645192.0 5.9
3645307.0 6.5
3645724.0 6.2
3645865.0 6.3
3645921.0 7.2
3646332.0 6.1
3646512.0 6.6
3646632.0 5.5
3647100.0 6.1
3647379.0 7.0
3647413.0 7.1
3647485.0 5.7
3648442.0 7.1
3648611.0 6.2
3648638.0 7.1
3648715.0 5.9
3648889.0 7.1
3650202.0 5.9
3650422.0 7.0
3652272.0 6.2
3653178.0 5.7
3653352.0 7.3
3653472.0 6.1
3653772.0 6.8
3654177.0 6.0
3654195.0 7.4
3654619.0 6.0
3655032.0 6.4
3655387.0 6.4
3658870.0 7.2
3659352.0 5.9
3660481.0 7.0
3661975.0 5.7
3662472.0 6.6
3662613.0 6.2
3662728.0 6.6
3662956.0 5.7
3663366.0 6.9
3664215.0 6.5
3664327.0 6.1
3664362.0 5.9
3664629.0 7.2
3665112.0 6.3
3665172.0 5.8
3665537.0 6.0
3666104.0 6.0
3666235.0 7.4
3667353.0 6.7
3667932.0 6.8
3668112.0 5.5
3668511.0 5.7
3669108.0 6.9
3669432.0 6.4
3669492.0 6.5
3669612.0 6.7
3669745.0 5.8
3669900.0 6.2
3670512.0 6.7
3670932.0 7.5
3670992.0 5.8
3671537.0 7.4
3671609.0 6.0
3671886.0 5.5
3671964.0 6.1
3672739.0 6.2

3673057.0 5.8
3673201.0 7.4
3673417.0 6.7
3675012.0 6.0
3675132.0 6.7
3675372.0 7.1
3675432.0 7.1
3675977.0 5.6
3676452.0 6.0
3676512.0 5.6
3677258.0 5.8
3677398.0 6.2
3677463.0 5.9
3678132.0 7.3
3680484.0 5.9
3680772.0 5.9
3680952.0 6.4
3681384.0 7.3
3681397.0 7.0
3681473.0 6.4
3681508.0 7.2
3681600.0 5.5
3681617.0 5.9
3681726.0 6.0
3682332.0 6.4
3682452.0 7.3
3682572.0 6.1
3682632.0 7.1
3682845.0 5.7
3682904.0 5.5
3683019.0 7.2
3683036.0 7.2
3683054.0 6.7
3683094.0 6.0
3683128.0 5.8
3683144.0 5.6
3683183.0 7.2
3683227.0 6.9
3683247.0 5.8
3683265.0 7.3
3683338.0 6.7
3683489.0 7.0
3683592.0 7.4
3683712.0 6.7
3683832.0 7.5
3683892.0 5.8
3683952.0 7.4
3684012.0 5.7
3684216.0 6.8
3684327.0 5.5
3684356.0 6.1
3684472.0 6.5
3684707.0 6.9
3684892.0 6.7
3684924.0 7.3
3684958.0 6.4
3685092.0 6.6
3685152.0 6.7
3685212.0 6.8
3685608.0 5.6
3685959.0 7.5
3686407.0 7.3
3687036.0 7.1
3687128.0 6.5
3687172.0 5.8
3687470.0 6.6
3687618.0 5.5
3688092.0 5.7
3688651.0 6.9
3689532.0 7.4
3690248.0 6.1
3691780.0 7.3
3692652.0 7.1
3692780.0 6.0
3692860.0 5.9
3693590.0 5.5
3693732.0 7.3
3695008.0 6.3
3695232.0 6.3
3695472.0 6.0
3695837.0 5.6
3696162.0 6.4
3696525.0 7.1
3697297.0 6.8
3700195.0 7.1
3700337.0 6.1
3700932.0 6.0
3701875.0 7.5
3701979.0 5.8
3702074.0 6.6
3702672.0 7.3
3702732.0 6.1
3702862.0 7.2
3703205.0 6.2
3703399.0 7.3
3703598.0 7.3
3703690.0 6.6
3705998.0 6.3
3706044.0 6.7
3706203.0 7.0
3706247.0 5.8
3706632.0 7.1
3706812.0 6.1
3707705.0 5.7
3708072.0 6.2
3708647.0 6.4
3709410.0 6.7
3709418.0 6.8
3709512.0 5.5
3710739.0 6.8
3710909.0 5.9
3710952.0 6.5
3712288.0 5.8
3712392.0 6.6
3712572.0 5.6
3713105.0 6.2
3713760.0 6.4
3714930.0 5.6
3715512.0 5.7
3715572.0 5.9
3716447.0 6.9
3716712.0 5.5
3717847.0 6.6
3718804.0 5.9
3719053.0 5.6
3719419.0 6.4
3719832.0 6.3
3720242.0 6.6
3720314.0 6.6
3720435.0 5.6
3720590.0 6.0
3721452.0 6.7
3721653.0 7.2
3721838.0 5.6
3721953.0 5.9
3722048.0 7.4
3722163.0 6.0
3722267.0 5.8
3722384.0 5.7
3722472.0 6.4
3722532.0 6.2
3722712.0 5.7
3722772.0 6.3
3722892.0 6.8
3722952.0 6.1
3723176.0 6.5
3723282.0 7.2
3723307.0 6.2
3723526.0 6.3
3723688.0 6.5
3723703.0 5.8
3723972.0 6.6
3724092.0 7.1
3724570.0 7.1
3724786.0 5.5
3724889.0 6.9
3725068.0 7.0
3725089.0 6.9
3725472.0 5.7
3725532.0 6.7
3725592.0 6.7
3726117.0 5.7
3726280.0 6.2
3726451.0 7.3
3726587.0 6.0
3726852.0 7.1
3727032.0 5.6
3727092.0 6.7
3727651.0 5.7
3727826.0 6.6
3727842.0 6.7
3728944.0 5.7
3729214.0 6.9
3729972.0 6.7
3730648.0 7.1
3730908.0 5.7
3731592.0 6.4
3732732.0 6.4
3733635.0 7.0
3735302.0 6.4
3737475.0 7.0
3737999.0 6.8
3738612.0 5.9
3738860.0 5.7
3738861.0 7.1
3739751.0 6.1
3739932.0 7.4
3740172.0 6.4
3740698.0 7.2
3741552.0 7.5
3741672.0 6.3
3741753.0 7.0
3742173.0 6.5
3742494.0 5.8
3742569.0 5.8
3742620.0 6.8
3742692.0 5.5
3742872.0 6.0
3743547.0 6.0
3744132.0 7.0
3745449.0 5.9
3745512.0 5.7
3745572.0 7.1
3745692.0 7.1
3745812.0 6.1
3748083.0 7.1
3748572.0 7.5
3749552.0 7.1
3750132.0 7.4
3750502.0 6.8
3750577.0 6.3
3751512.0 5.7
3752127.0 6.7
3754143.0 6.2
3754512.0 6.4
3754572.0 5.7
3756473.0 6.1
3757092.0 6.5
3757332.0 5.8
3757795.0 6.8
3757906.0 7.1
3758008.0 6.3
3758083.0 5.9
3758114.0 6.7
3758140.0 7.2
3758224.0 7.4
3758532.0 7.1
3758952.0 6.2
3759077.0 6.1
3759146.0 5.9
3759339.0 7.2
3759426.0 6.8
3759714.0 5.6
3759972.0 6.0
3760152.0 7.3
3760212.0 7.3
3760478.0 5.6
3760597.0 7.3
3760640.0 6.0
3760665.0 7.4
3760768.0 7.4
3761128.0 6.4
3761160.0 6.1
3761179.0 6.4
3761310.0 7.2
3761352.0 7.4
3761472.0 6.8
3761532.0 5.9
3761652.0 6.6
3761712.0 6.1
3761772.0 6.6
3761832.0 7.3
3761945.0 7.0
3762112.0 6.8
3762153.0 6.1
3762270.0 7.5
3762352.0 7.3
3762448.0 6.2
3762510.0 6.5
3762589.0 5.6
3762722.0 6.4
3762739.0 7.0
3762747.0 6.3
3762792.0 6.4
3762912.0 6.2
3762972.0 7.2
3763032.0 5.8
3763272.0 6.5
3763398.0 7.5
3763422.0 6.4
3763748.0 6.7
3763915.0 6.0
3764352.0 5.7
3764652.0 5.7
3765159.0 5.7
3765289.0 6.2
3765577.0 7.0
3765792.0 5.7
3765972.0 7.3
3766092.0 6.8
3766842.0 6.1
3767172.0 6.8
3767412.0 7.2
3767665.0 6.1
3767759.0 5.7
3768309.0 7.3
3768330.0 6.2
3768445.0 7.1
3768552.0 6.7
3769890.0 6.4
3771308.0 6.3
3771432.0 7.2
3771672.0 6.0
3777672.0 6.8
3778812.0 6.1
3778872.0 7.3
3779052.0 6.8
3779363.0 5.9
3779700.0 7.4
3780132.0 5.7
3780492.0 7.4
3780789.0 6.6
3781192.0 6.0
3781254.0 6.5
3781317.0 6.3
3781375.0 6.1
3781454.0 6.1
3781512.0 6.2
3781872.0 6.6
3782426.0 7.5
3783432.0 5.7
3784177.0 6.7
3784290.0 6.6
3784392.0 7.2
3784452.0 6.1
3785017.0 6.6
3785134.0 5.6
3785341.0 5.6
3785398.0 6.1
3786192.0 6.2
3786727.0 6.9
3786772.0 5.9
3795239.0 6.6
3797532.0 6.4
3797592.0 5.9
3797832.0 5.5
3797892.0 5.5
3798557.0 5.7
3798912.0 6.3
3798972.0 6.7
3799092.0 7.1
3799425.0 6.2
3799548.0 7.1
3799657.0 6.2
3799701.0 7.0
3799769.0 6.8
3799848.0 6.3
3799919.0 7.3
3799942.0 7.4
3800056.0 5.9
3800072.0 6.7
3800232.0 5.7
3800532.0 6.4
3800652.0 7.0
3800779.0 7.0
3800793.0 6.3
3800813.0 6.2
3801069.0 6.4
3801142.0 5.9
3801245.0 6.5
3801448.0 5.8
3801498.0 5.8
3801562.0 7.2
3801732.0 5.8
3801972.0 6.9
3802032.0 6.9
3802092.0 7.5
3802243.0 6.4
3802380.0 6.1
3802474.0 6.4
3802530.0 6.0
3802788.0 6.5
3802949.0 7.3
3803104.0 6.6
3803112.0 5.9
3803292.0 7.1
3803412.0 6.3
3804155.0 5.6
3804271.0 7.2
3804344.0 7.0
3804552.0 7.4
3804612.0 6.8
3805032.0 5.5
3805159.0 6.9
3805370.0 6.0
3805585.0 6.6
3805672.0 6.8
3805789.0 6.7
3805992.0 6.8
3806352.0 7.0
3807075.0 7.4
3807404.0 6.4
3807552.0 6.3
3808103.0 6.6
3808992.0 5.6
3811167.0 6.3
3811428.0 5.5
3811659.0 6.1
3811752.0 6.7
3812628.0 7.4
3812702.0 6.6
3815877.0 6.5
3816432.0 5.7
3816552.0 5.8
3816669.0 7.5
3817287.0 7.5
3817752.0 7.4
3818375.0 7.1
3819132.0 6.2
3819724.0 5.6
3819774.0 6.5
3820089.0 5.7
3820289.0 5.8
3821089.0 6.4
3821469.0 5.8
3821555.0 5.9
3821678.0 6.0
3821775.0 6.0
3821832.0 7.1
3821952.0 6.4
3822072.0 7.5
3822132.0 7.0
3822252.0 5.8
3822761.0 6.1
3822990.0 5.8
3823011.0 5.7
3823025.0 5.6
3823178.0 5.8
3823272.0 6.9
3823332.0 6.3
3823452.0 6.2
3823692.0 6.6
3823997.0 6.5
3824682.0 7.5
3825012.0 5.7
3825192.0 6.4
3825270.0 6.1
3825506.0 7.2
3825783.0 6.3
3826733.0 6.8
3826871.0 6.1
3827832.0 6.4
3828236.0 6.5
3829092.0 5.6
3829392.0 6.5
3829762.0 6.8
3834149.0 6.4
3836098.0 6.7
3837732.0 5.9
3837972.0 6.4
3838032.0 6.3
3838657.0 6.9
3838692.0 7.0
3838755.0 6.7
3838763.0 7.0
3839172.0 6.6
3839292.0 7.3
3839352.0 6.8
3839412.0 6.7
3839472.0 6.3
3839532.0 6.0
3839592.0 5.9
3839757.0 7.4
3839846.0 6.1
3839952.0 7.3
3840044.0 5.7
3840147.0 7.0
3840174.0 7.0
3840325.0 5.8
3840452.0 7.5
3840515.0 5.8
3840732.0 6.8
3840852.0 6.2
3840912.0 7.3
3840972.0 6.9
3841032.0 6.2
3841142.0 7.1
3841457.0 6.4
3841586.0 7.4
3841764.0 6.7
3841992.0 5.7
3842172.0 6.6
3842232.0 7.2
3842292.0 6.4
3842412.0 7.5
3842754.0 5.9
3843328.0 6.0
3843392.0 7.1
3843552.0 6.1
3843612.0 5.8
3843732.0 6.2
3843792.0 5.6
3844012.0 6.9
3844332.0 6.7
3844537.0 6.6
3845232.0 6.2
3845352.0 5.5
3845477.0 6.8
3845644.0 5.8
3845809.0 7.5
3845913.0 6.1
3846082.0 6.6
3846172.0 6.5
3846285.0 7.0
3846492.0 7.4
3847812.0 6.8
3849801.0 5.9
3850586.0 5.9
3851865.0 7.1
3852132.0 6.6
3852252.0 7.0
3854872.0 6.6
3854930.0 6.2
3856010.0 6.1
3856872.0 7.0
3857002.0 6.3
3857055.0 6.1
3857832.0 5.8
3857892.0 7.2
3858584.0 6.4
3858606.0 6.9
3858656.0 6.6
3858696.0 6.1
3858719.0 6.0
3858746.0 7.0
3858798.0 5.9
3859006.0 6.7
3859042.0 6.0
3859217.0 6.2
3859272.0 5.6
3859332.0 5.5
3859752.0 5.6
3859874.0 6.6
3860186.0 6.2
3860712.0 7.5
3860832.0 6.5
3860952.0 6.0
3862918.0 7.0
3863712.0 7.1
3863892.0 6.6
3864012.0 6.5
3864194.0 6.1
3864207.0 6.6
3864766.0 6.3
3864970.0 5.8
3864999.0 5.5
3865017.0 6.7
3865032.0 7.0
3865212.0 6.3
3865392.0 6.5
3865452.0 7.5
3865782.0 6.8
3865838.0 6.2
3866093.0 7.0
3866268.0 6.3
3866283.0 7.4
3866433.0 7.0
3866472.0 6.7
3866532.0 7.0
3866952.0 6.1
3867580.0 7.0
3868392.0 6.0
3868482.0 7.1
3869198.0 6.2
3869352.0 5.6
3869832.0 7.3
3869984.0 5.9
3870792.0 6.3
3870912.0 7.0
3871662.0 5.6
3871960.0 5.9
3872799.0 6.3
3873634.0 7.2
3874984.0 5.6
3875922.0 7.1
3877910.0 6.8
3878172.0 6.4
3878292.0 7.1
3879092.0 6.1
3879139.0 6.1
3879160.0 5.6
3879432.0 5.5
3879792.0 6.6
3880115.0 6.3
3880335.0 5.8
3880538.0 7.1
3880700.0 6.7
3880932.0 7.4
3881052.0 6.1
3881112.0 6.0
3881172.0 5.5
3881232.0 6.9
3881352.0 5.6
3881780.0 6.3
3881810.0 5.6
3881939.0 6.6
3882161.0 5.6
3882298.0 6.7
3882492.0 7.0
3882552.0 6.1
3882732.0 6.6
3882792.0 6.6
3883044.0 6.0
3883134.0 6.5
3883199.0 6.0
3883528.0 5.8
3884490.0 7.0
3885042.0 6.6
3886274.0 6.1
3886397.0 6.4
3886872.0 7.0
3886992.0 5.8
3887230.0 6.2
3887449.0 6.2
3887669.0 5.7
3888312.0 6.5
3888432.0 6.7
3889203.0 5.8
3893697.0 5.6
3895512.0 6.3
3895632.0 6.4
3896439.0 5.8
3896772.0 6.7
3896952.0 7.3
3897012.0 6.6
3897488.0 5.7
3897899.0 6.7
3898933.0 5.5
3899014.0 6.9
3899131.0 5.6
3899277.0 5.6
3899350.0 7.1
3899532.0 6.5
3899712.0 6.6
3900162.0 6.8
3900311.0 6.8
3900370.0 6.4
3900548.0 6.4
3900591.0 6.7
3900700.0 7.1
3900746.0 5.8
3900776.0 6.2
3901092.0 6.4
3901152.0 5.7
3901605.0 5.6
3901823.0 7.2
3902952.0 6.0
3903152.0 7.2
3903519.0 6.2
3903794.0 5.8
3903816.0 6.3
3903859.0 5.7
3905309.0 7.4
3905532.0 5.6
3905592.0 7.1
3905652.0 7.1
3906043.0 7.0
3906140.0 5.8
3906234.0 6.7
3906319.0 6.1
3906609.0 5.8
3906665.0 5.7
3906702.0 5.8
3906771.0 6.4
3906852.0 5.7
3906912.0 6.2
3906972.0 7.1
3907272.0 6.2
3907451.0 6.3
3907466.0 6.1
3907530.0 7.3
3908037.0 7.3
3908292.0 6.8
3909092.0 6.7
3909912.0 6.7
3910092.0 6.6
3910264.0 5.5
3910444.0 7.3
3911412.0 6.3
3912014.0 7.4
3912552.0 7.0
3913847.0 7.1
3914642.0 6.1
3914835.0 6.6
3916015.0 6.6
3916165.0 7.0
3916424.0 6.7
3916607.0 7.3
3917112.0 7.0
3917172.0 7.1
3917232.0 6.1
3917292.0 7.5
3917352.0 7.5
3917484.0 7.4
3917534.0 6.3
3917625.0 5.9
3917666.0 6.4
3918219.0 6.4
3918296.0 6.2
3918895.0 5.6
3918920.0 7.3
3919652.0 6.8
3919682.0 7.2
3920112.0 5.9
3920735.0 7.2
3920886.0 5.7
3920934.0 6.8
3921372.0 6.6
3921492.0 6.8
3921552.0 6.7
3921944.0 5.7
3921957.0 6.2
3922333.0 6.9
3922342.0 6.9
3922692.0 7.1
3923747.0 6.8
3923822.0 7.3
3924072.0 6.5
3924432.0 6.8
3924492.0 5.8
3924552.0 5.7
3926555.0 7.1
3926623.0 7.3
3933012.0 7.0
3933907.0 7.3
3934512.0 6.3
3935592.0 7.2
3935772.0 6.8
3936584.0 6.0
3936688.0 6.6
3936975.0 6.3
3937212.0 7.3
3937392.0 6.9
3937512.0 6.5
3937809.0 5.9
3938532.0 6.7
3938712.0 7.2
3938952.0 5.5
3939276.0 7.2
3939364.0 6.2
3939455.0 5.9
3939665.0 6.7
3939871.0 6.3
3940392.0 6.9
3940615.0 7.0
3940643.0 6.0
3940731.0 7.1
3941155.0 6.5
3941301.0 6.1
3941412.0 6.2
3941592.0 6.2
3942296.0 7.1
3942431.0 6.3
3942551.0 7.0
3942852.0 6.5
3943866.0 7.4
3944652.0 6.4
3944712.0 7.4
3945002.0 5.5
3946364.0 6.7
3946533.0 7.3
3947412.0 5.8
3947592.0 6.7
3947744.0 7.3
3947904.0 7.2
3948075.0 5.7
3948469.0 6.0
3948672.0 6.6
3948852.0 6.5
3949130.0 6.6
3950581.0 6.7
3950886.0 6.9
3952290.0 7.2
3952438.0 6.9
3952647.0 7.0
3952692.0 7.1
3954672.0 7.1
3954792.0 7.3
3955424.0 7.0
3955552.0 7.4
3955932.0 5.9
3956112.0 6.7
3956358.0 6.1
3956572.0 7.0
3957017.0 6.2
3957672.0 6.7
3958135.0 6.3
3958500.0 7.3
3958556.0 6.6
3958752.0 6.2
3958932.0 7.4
3958992.0 6.4
3959644.0 6.1
3959678.0 7.1
3959721.0 6.5
3959872.0 6.1
3960192.0 5.8
3960667.0 7.3
3960897.0 5.7
3961028.0 6.0
3962617.0 6.9
3963942.0 7.5
3966192.0 5.6
3967070.0 5.9
3967122.0 6.6
3970272.0 5.7
3970739.0 7.1
3972687.0 6.2
3974712.0 7.3
3974832.0 6.7
3975097.0 6.9
3975254.0 6.8
3975287.0 5.6
3975407.0 5.7
3975804.0 6.4
3975912.0 6.3
3975972.0 6.4
3976272.0 5.5
3976332.0 6.2
3976472.0 6.6
3976497.0 5.6
3976666.0 5.7
3976797.0 7.2
3976911.0 7.5
3977002.0 7.3
3977412.0 6.3
3977532.0 7.3
3978012.0 5.6
3978099.0 7.2
3978504.0 6.6
3978560.0 6.2
3978852.0 6.2
3978912.0 7.2
3979032.0 6.1
3979391.0 6.3
3979447.0 6.8
3979708.0 6.1
3979807.0 6.5
3980292.0 7.4
3981574.0 7.1
3983719.0 6.1
3985165.0 6.6
3985547.0 6.5
3987261.0 6.7
3987732.0 7.2
3990552.0 5.6
3991170.0 7.2
3991191.0 6.0
3991560.0 7.1
3991638.0 6.7
3991872.0 5.6
3991932.0 6.3
3992052.0 5.7
3992112.0 6.8
3992172.0 7.3
3992403.0 5.6
3992432.0 6.8
3992771.0 5.6
3993552.0 5.8
3993957.0 6.2
3994135.0 6.5
3994872.0 6.8
3995112.0 6.9
3995262.0 6.3
3995313.0 5.7
3995382.0 7.3
3995524.0 6.1
3995719.0 7.2
3995858.0 6.5
3995915.0 7.0
3996192.0 6.2
3996372.0 7.1
3996492.0 5.7
3996552.0 7.4
3996628.0 6.2
3996669.0 7.2
3996704.0 7.5
3996714.0 6.1
3996748.0 5.9
3996863.0 6.1
3996990.0 5.7
3997185.0 5.8
3997325.0 7.4
3997432.0 6.0
3997470.0 7.0
3997872.0 6.8
3997932.0 6.0
3998070.0 6.0
3998093.0 6.7
3998360.0 5.7
3998367.0 6.4
3998497.0 6.1
3998534.0 7.3
3998740.0 7.4
3998805.0 7.1
3999903.0 7.5
4000198.0 7.2
4000323.0 6.3
4000452.0 5.6
4001424.0 5.9
4001600.0 6.6
4001627.0 6.3
4003572.0 6.7
4013592.0 6.9
4013912.0 5.7
4014717.0 7.4
4015152.0 5.5
4015704.0 6.0
4015763.0 6.7
4015949.0 5.9
4016027.0 5.9
4016232.0 6.2
4016292.0 7.5
4017182.0 6.6
4017261.0 6.8
4017612.0 6.9
4017732.0 5.8
4017912.0 6.7
4018032.0 6.7
4018092.0 5.7
4018291.0 5.5
4018378.0 6.1
4018427.0 6.4
4018513.0 6.8
4018672.0 6.5
4018905.0 6.3
4020852.0 6.3
4024601.0 6.7
4025052.0 6.7
4027618.0 7.0
4027732.0 7.1
4028657.0 7.1
4029151.0 6.0
4029312.0 7.2
4030387.0 6.2
4030475.0 5.6
4030490.0 5.6
4030565.0 7.4
4030622.0 7.3
4030872.0 7.2
4031237.0 6.9
4031705.0 6.8
4032132.0 6.5
4032192.0 6.4
4032492.0 6.2
4032552.0 5.7
4032834.0 7.4
4032947.0 6.5
4032968.0 6.8
4032989.0 6.6
4033195.0 7.3
4033260.0 5.6
4033512.0 6.2
4033692.0 5.6
4033752.0 5.6
4033812.0 6.3
4033992.0 5.6
4034189.0 7.2
4034259.0 5.8
4034269.0 6.4
4034389.0 7.5
4034624.0 5.9
4034633.0 5.8
4034737.0 6.1
4034829.0 6.3
4034868.0 7.3
4034918.0 5.7
4034927.0 6.2
4034952.0 6.4
4035072.0 5.5
4035192.0 7.1
4035252.0 6.6
4035312.0 5.9
4035432.0 7.4
4035632.0 7.4
4035680.0 6.9
4035691.0 6.1
4035799.0 5.5
4035864.0 6.7
4035952.0 6.0
4035977.0 6.7
4036005.0 6.0
4036113.0 6.2
4036149.0 6.8
4036254.0 5.5
4036260.0 7.0
4036335.0 7.3
4036392.0 6.3
4036452.0 7.3
4036572.0 6.6
4036692.0 6.2
4036752.0 6.6
4036984.0 5.8
4037007.0 6.6
4037044.0 6.4
4037173.0 5.9
4037278.0 6.7
4037445.0 5.9
4037510.0 6.6
4037532.0 6.5
4037832.0 7.4
4037952.0 6.3
4038132.0 5.6
4038580.0 6.9

4038808.0 5.8
4039017.0 7.0
4039182.0 6.9
4039332.0 6.1
4039632.0 6.8
4039844.0 6.5
4040039.0 6.2
4040152.0 5.9
4040531.0 5.6
4040772.0 5.5
4041072.0 7.5
4041257.0 7.0
4041839.0 6.5
4041884.0 6.3
4042332.0 7.4
4042452.0 6.0
4043104.0 6.4
4043442.0 6.2
4044072.0 7.4
4044602.0 7.0
4045669.0 6.3
4045790.0 7.1
4045806.0 6.5
4046319.0 6.0
4047972.0 7.0
4048032.0 6.5
4048756.0 5.8
4050792.0 7.1
4052980.0 7.4
4053136.0 6.2
4053347.0 6.9
4053353.0 6.3
4053504.0 7.3
4053672.0 6.4
4053792.0 6.8
4054032.0 7.5
4054092.0 6.7
4054398.0 7.2
4054500.0 6.5
4054524.0 6.0
4054572.0 5.7
4054625.0 5.7
4054725.0 7.4
4054739.0 5.7
4054826.0 7.1
4055046.0 6.4
4055081.0 7.1
4055097.0 6.9
4055172.0 5.5
4055767.0 6.3
4056013.0 6.5
4056507.0 6.0
4056792.0 7.3
4056912.0 6.5
4057032.0 7.1
4057121.0 6.5
4057158.0 7.1
4057694.0 5.6
4057716.0 7.2
4057864.0 6.6
4058112.0 6.4
4058232.0 7.3
4058412.0 6.0
4058472.0 6.5
4058565.0 7.4
4058760.0 6.8
4058897.0 6.4
4058964.0 7.0
4059612.0 5.7
4059732.0 6.9
4059792.0 5.9
4060400.0 6.7
4060480.0 6.4
4060605.0 6.1
4060622.0 7.4
4060872.0 6.3
4061112.0 7.2
4061292.0 6.7
4061513.0 5.7
4061739.0 6.4
4062492.0 6.2
4062969.0 5.7
4063007.0 5.6
4063537.0 6.3
4064786.0 6.0
4066632.0 6.5
4066932.0 6.9
4067525.0 6.4
4067543.0 6.4
4067598.0 5.6
4068719.0 7.2
4069692.0 6.1
4070198.0 7.0
4070297.0 7.3
4070590.0 6.3
4070826.0 6.5
4070873.0 6.4
4070943.0 6.1
4071505.0 7.4
4071658.0 6.9
4071759.0 6.3
4071848.0 6.1
4072013.0 5.7
4072192.0 6.8
4072240.0 7.1
4072269.0 6.8
4072344.0 6.1
4072374.0 6.5
4072452.0 6.4
4072512.0 6.7
4072632.0 7.4
4072692.0 7.1
4072752.0 6.2
4072872.0 6.9
4073013.0 6.2
4073052.0 6.6
4073079.0 6.4
4073083.0 6.5
4073145.0 7.1
4073152.0 6.7
4073273.0 7.3
4073338.0 6.6
4073392.0 6.2
4073506.0 6.7
4073575.0 5.8
4073752.0 5.8
4073808.0 7.0
4073832.0 5.8
4073892.0 5.8
4073952.0 7.4
4074012.0 6.5
4074312.0 6.8
4074382.0 6.2
4074409.0 5.9
4074449.0 5.7
4074496.0 6.0
4074555.0 7.3
4074584.0 6.9
4074677.0 6.0
4074798.0 6.0
4074855.0 7.3
4074944.0 7.3
4075392.0 7.2
4075452.0 7.0
4075572.0 5.6
4076173.0 7.0
4076409.0 6.3
4076589.0 6.2
4076670.0 7.2
4076678.0 5.6
4076697.0 6.1
4076712.0 6.9
4076772.0 6.6
4076832.0 6.5
4076892.0 6.0
4077072.0 7.0
4077132.0 6.5
4077192.0 7.0
4077427.0 6.3
4077448.0 6.5
4077461.0 5.7
4077685.0 6.1
4077709.0 6.8
4077772.0 5.7
4077874.0 6.7
4077881.0 5.8
4077886.0 6.0
4078010.0 6.6
4078023.0 6.7
4078128.0 5.7
4078152.0 6.5
4078212.0 6.8
4078332.0 6.3
4078512.0 6.2
4078572.0 7.0
4078774.0 7.5
4078979.0 6.1
4079154.0 6.6
4079178.0 6.9
4079345.0 5.9
4079513.0 5.9
4079652.0 6.0
4079712.0 6.5
4079832.0 6.0
4079892.0 5.7
4079952.0 5.7
4080175.0 6.9
4080196.0 6.1
4080841.0 6.1
4080975.0 6.7
4081332.0 6.8
4081512.0 6.4
4081625.0 7.0
4081668.0 6.6
4081887.0 6.1
4081975.0 7.1
4082221.0 6.7
4082376.0 5.9
4082532.0 6.4
4082652.0 6.0
4082712.0 5.8
4082832.0 6.1
4082892.0 6.0
4082952.0 6.5
4083046.0 7.3
4083194.0 5.9
4083274.0 6.4
4083439.0 7.4
4083470.0 6.7
4083554.0 7.2
4083873.0 6.5
4083972.0 6.3
4084392.0 5.9
4086852.0 7.0
4086912.0 7.2
4087272.0 7.1
4089792.0 5.6
4090309.0 6.6
4090340.0 5.8
4090864.0 7.3
4091085.0 5.8
4091106.0 7.0
4091232.0 5.7
4092045.0 5.6
4092404.0 6.2
4092469.0 6.9
4092672.0 6.1
4093182.0 7.4
4093216.0 6.2
4093295.0 5.5
4093992.0 7.3
4094352.0 6.2
4094472.0 7.4
4094557.0 6.3
4094571.0 6.2
4094699.0 7.5
4094713.0 5.6
4094744.0 6.5
4095040.0 6.6
4095240.0 6.6
4095309.0 6.0
4095995.0 6.7
4096225.0 6.4
4096269.0 5.9
4096537.0 6.7
4097172.0 7.3
4097484.0 6.3
4097582.0 7.3
4097604.0 6.5
4097765.0 6.5
4097823.0 5.5
4097895.0 5.7
4098025.0 7.1
4098190.0 7.2
4098372.0 7.4
4098432.0 5.6
4098492.0 5.7
4098552.0 5.8
4098792.0 6.0
4098886.0 6.7
4099246.0 5.8
4099489.0 6.3
4099696.0 7.0
4099752.0 6.4
4099812.0 5.9
4099932.0 6.5
4100052.0 6.5
4100172.0 6.3
4100347.0 6.2
4100413.0 6.9
4100450.0 7.1
4100497.0 6.5
4100729.0 6.9
4100766.0 6.4
4100834.0 5.7
4100849.0 7.0
4101014.0 6.9
4101372.0 6.0
4101672.0 6.3
4101733.0 5.9
4101848.0 6.9
4102066.0 7.5
4102176.0 6.7
4102380.0 6.3
4102433.0 6.6
4102496.0 6.6
4102992.0 5.8
4103581.0 7.2
4103609.0 6.5
4103733.0 6.3
4103967.0 6.2
4104037.0 6.7
4104072.0 6.6
4104372.0 5.5
4104492.0 7.2
4104699.0 6.1
4106400.0 7.3
4106512.0 5.7
4107132.0 5.5
4108872.0 7.1
4109122.0 7.1
4109150.0 7.1
4109337.0 7.2
4109892.0 6.2
4110489.0 6.5
4111198.0 7.5
4111632.0 6.3
4111908.0 5.8
4111950.0 7.3
4111980.0 6.4
4112064.0 7.4
4112092.0 7.3
4112183.0 7.2
4112227.0 7.3
4112374.0 5.8
4112430.0 5.8
4112619.0 7.2
4112642.0 6.7
4112679.0 6.9
4112772.0 6.3
4112892.0 6.9
4112952.0 7.4
4113346.0 7.4
4113379.0 5.8
4113502.0 6.2
4114082.0 6.1
4114138.0 6.9
4114152.0 6.1
4114272.0 6.4
4114392.0 7.0
4114452.0 6.7
4114512.0 6.0
4114935.0 6.7
4115106.0 6.4
4115425.0 6.4
4115499.0 7.1
4115529.0 7.2
4115952.0 5.9
4116258.0 6.7
4116386.0 7.5
4116655.0 6.7
4116932.0 7.3
4116940.0 6.5
4117611.0 5.7
4117692.0 6.0
4117715.0 5.9
4117969.0 5.7
4118269.0 5.5
4118436.0 7.2
4118472.0 6.8
4118892.0 6.5
4119108.0 5.7
4119341.0 6.8
4120032.0 5.6
4120152.0 6.6
4120272.0 6.9
4120392.0 6.6
4120502.0 6.1
4120703.0 7.1
4121532.0 5.7
4121924.0 6.9
4121985.0 6.0
4122792.0 6.0
4123936.0 6.6
4123998.0 5.7
4124292.0 7.1
4124896.0 7.5
4127862.0 6.0
4128549.0 6.7
4128552.0 5.5
4128972.0 5.6
4129516.0 6.0
4130847.0 6.3
4131193.0 7.4
4131367.0 5.5
4132187.0 6.4
4132693.0 7.3
4132932.0 5.6
4133594.0 5.6
4133633.0 5.8
4133779.0 7.1
4133962.0 6.4
4134027.0 6.9
4134060.0 6.3
4134372.0 5.9
4134552.0 6.4
4135470.0 5.6
4135589.0 6.4
4136052.0 6.2
4136172.0 5.8
4136397.0 6.1
4136713.0 7.3
4137612.0 6.9
4137672.0 5.8
4137755.0 6.0
4137868.0 5.7
4137885.0 6.7
4138255.0 6.3
4138275.0 7.2
4138430.0 7.3
4138506.0 6.5
4138692.0 6.3
4138872.0 5.9
4138932.0 6.6
4138992.0 6.6
4139112.0 7.0
4139251.0 6.9
4139366.0 6.9
4139397.0 7.4
4139462.0 5.5
4139674.0 5.5
4139783.0 6.2
4140192.0 6.0
4140252.0 7.4
4140492.0 6.1
4140742.0 6.8
4140774.0 5.9
4140825.0 6.2
4140832.0 7.2
4141015.0 6.3
4141085.0 7.4
4141199.0 5.9
4141326.0 6.3
4141476.0 6.2
4141752.0 6.3
4141872.0 5.9
4141992.0 6.3
4142191.0 7.3
4142287.0 7.5
4142362.0 6.4
4142654.0 5.6
4142782.0 6.5
4142799.0 5.9
4142884.0 6.1
4142952.0 5.7
4143132.0 6.3
4143432.0 6.4
4143604.0 7.1
4143653.0 5.7
4143689.0 5.9
4143907.0 5.9
4144019.0 6.2
4144247.0 6.7
4144752.0 6.2
4147272.0 6.9
4147512.0 7.3
4148499.0 5.8
4148657.0 6.9
4149667.0 6.0
4150056.0 5.9
4150272.0 6.4
4150392.0 7.0
4150696.0 5.9
4150986.0 7.1
4151264.0 7.3
4151297.0 6.5
4151577.0 5.5
4151592.0 5.7
4151892.0 6.2
4152012.0 6.4
4152072.0 6.7
4152159.0 6.6
4152177.0 6.1
4152476.0 6.6
4152552.0 7.2
4152677.0 7.4
4152806.0 6.6
4152920.0 7.2
4152960.0 6.2
4153032.0 6.6
4153272.0 5.8
4153332.0 7.1
4153392.0 7.2
4153452.0 6.0
4153657.0 7.0
4154104.0 6.4
4154187.0 5.9
4154532.0 7.0
4154592.0 6.8
4155377.0 5.8
4155434.0 7.1
4155519.0 7.1
4155912.0 6.2
4156332.0 6.3
4156741.0 6.4
4157652.0 7.4
4158197.0 7.1
4158705.0 7.2
4158792.0 5.6
4158912.0 5.8
4159337.0 6.2
4159366.0 6.5
4159739.0 7.2
4160232.0 6.7
4160292.0 5.9
4160352.0 6.3
4160412.0 7.3
4160472.0 5.8
4160532.0 6.2
4160592.0 6.3
4162373.0 7.1
4163078.0 5.7
4163532.0 6.4
4163592.0 5.8
4163764.0 5.8
4163914.0 7.0
4168639.0 7.4
4168845.0 6.9
4169672.0 7.5
4169877.0 5.9
4169904.0 6.6
4170612.0 6.8
4170910.0 6.0
4171161.0 5.9
4171508.0 6.7
4171585.0 6.4
4171657.0 6.8
4172052.0 6.9
4172172.0 7.1
4172232.0 5.6
4172859.0 5.9
4172933.0 7.5
4173372.0 6.6
4173492.0 6.8
4174111.0 6.0
4174312.0 6.3
4174752.0 7.2
4175730.0 6.4
4175973.0 6.7
4176252.0 7.5
4176492.0 5.7
4176859.0 7.2
4177418.0 5.7
4177632.0 6.5
4177752.0 6.9
4178743.0 5.8
4179252.0 7.5
4180033.0 7.3
4180138.0 6.1
4180392.0 6.4
4180872.0 5.8
4181050.0 6.3
4181553.0 5.9
4181635.0 5.9
4182456.0 5.6
4182543.0 7.0
4182828.0 7.0
4183139.0 6.7
4183392.0 6.7
4183512.0 6.9
4184031.0 5.7
4184048.0 6.6
4184078.0 6.5
4184399.0 7.4
4184516.0 6.1
4184689.0 7.2
4184832.0 6.7
4184952.0 7.5
4185497.0 6.6
4186452.0 5.9
4186703.0 6.3
4186824.0 5.7
4187029.0 7.3
4187224.0 5.8
4187279.0 6.5
4187422.0 6.0
4187447.0 6.2
4187529.0 7.4
4187772.0 7.4
4187832.0 5.7
4187952.0 6.5
4188467.0 7.2
4188868.0 6.8
4189588.0 6.2
4189875.0 6.9
4189967.0 5.7
4189984.0 6.8
4190017.0 6.7
4190040.0 7.0
4190062.0 6.1
4190130.0 6.8
4190532.0 5.5
4191099.0 6.8
4191135.0 6.0
4191446.0 6.1
4191514.0 7.1
4191753.0 6.8
4191892.0 5.5
4191912.0 7.2
4192032.0 6.1
4192272.0 7.0
4192332.0 6.8
4192392.0 7.4
4192493.0 6.4
4192527.0 6.1
4192619.0 5.5
4192805.0 7.3
4192925.0 6.0
4192980.0 5.6
4193092.0 5.6
4193205.0 5.9
4193239.0 6.1
4193652.0 6.1
4193914.0 5.8
4194040.0 5.9
4194485.0 5.7
4194500.0 5.8
4194539.0 6.7
4194567.0 5.8
4194622.0 7.5
4194662.0 7.3
4194715.0 5.7
4194742.0 6.3
4194912.0 5.6
4194972.0 6.0
4195032.0 7.2
4195212.0 5.5
4195373.0 7.4
4195406.0 7.0
4195431.0 7.1
4195450.0 7.4
4195650.0 6.4
4195858.0 7.1
4195883.0 5.9
4195968.0 6.0
4196016.0 5.8
4196097.0 7.5
4196352.0 6.2
4196652.0 5.6
4196712.0 6.0
4197200.0 6.3
4197557.0 7.4
4197596.0 6.8
4198032.0 7.5
4198535.0 7.1
4199112.0 6.9
4199172.0 5.8
4199232.0 6.2
4199532.0 6.8
4199807.0 7.0
4199829.0 6.1
4199899.0 7.1
4200535.0 6.1
4200792.0 7.3
4201207.0 5.7
4201598.0 5.6
4205928.0 5.6
4206060.0 5.6
4207338.0 7.2
4207484.0 7.2
4207752.0 6.5
4208301.0 5.6
4208945.0 6.9
4209009.0 6.6
4209182.0 6.9
4211452.0 5.5
4211782.0 6.8
4211843.0 6.6
4211899.0 5.6
4212844.0 6.2
4212882.0 6.5
4213119.0 6.6
4213217.0 7.4
4215192.0 6.7
4215788.0 6.3
4218012.0 7.5
4219392.0 5.7
4219572.0 6.0
4220712.0 5.9
4222452.0 6.8
4222572.0 6.7
4222632.0 7.1
4222744.0 7.5
4223772.0 5.6
4223952.0 5.7
4224337.0 6.6
4224439.0 5.9
4224973.0 7.2
4225512.0 6.1
4226048.0 6.3
4226207.0 6.3
4226305.0 7.3
4226333.0 5.7
4226357.0 6.8
4227040.0 6.4
4228092.0 6.0
4228212.0 7.1
4228332.0 7.0
4228864.0 7.0
4228913.0 6.8
4229192.0 6.9
4229232.0 5.7
4229315.0 7.3
4229352.0 6.3
4229592.0 6.8
4229937.0 6.6
4230134.0 5.8
4230192.0 5.9
4230464.0 6.8
4230525.0 6.5
4230580.0 6.9
4230617.0 5.5
4230679.0 6.3
4230780.0 5.6
4230972.0 7.5
4231032.0 7.0
4231092.0 7.4
4232064.0 7.0
4232232.0 7.3
4232889.0 6.5
4232920.0 6.2
4233191.0 7.2
4233237.0 7.3
4233259.0 6.6
4233325.0 7.0
4233452.0 5.5
4233487.0 6.0
4233732.0 7.2
4233912.0 7.0
4234032.0 6.7
4234542.0 6.0
4234622.0 7.4
4234682.0 5.9
4234980.0 6.2
4235019.0 5.8
4235352.0 5.6
4235412.0 6.2
4235532.0 6.9
4235944.0 5.8
4235961.0 6.3
4236480.0 6.2
4236520.0 7.5
4236552.0 5.7
4236792.0 6.7
4237484.0 6.0
4237704.0 5.7
4237846.0 5.8
4237992.0 6.0
4239120.0 6.1
4240169.0 6.4
4240216.0 6.3
4242312.0 5.7
4248132.0 6.1
4248192.0 6.8
4248372.0 5.7
4248432.0 6.0
4248492.0 6.4
4248687.0 6.3
4249012.0 7.2
4249027.0 6.9
4249077.0 7.5
4249482.0 6.5
4249512.0 7.3
4249572.0 7.1
4249752.0 7.1
4250057.0 7.2
4250120.0 7.0
4250622.0 6.1
4250849.0 6.9
4251372.0 6.1
4251801.0 6.9
4252132.0 6.2
4252167.0 7.0
4252392.0 6.2
4252632.0 6.0
4252752.0 5.8
4253139.0 6.0
4253339.0 6.7
4254012.0 5.6
4254732.0 6.7
4258392.0 6.5
4259452.0 5.6
4260169.0 6.7
4262832.0 6.5
4262952.0 7.1
4263255.0 5.9
4266094.0 7.2
4266270.0 6.6
4266280.0 6.8
4266296.0 6.8
4266582.0 6.3
4267092.0 6.6
4267152.0 6.6
4267212.0 6.4
4267272.0 7.1
4267369.0 7.1
4267415.0 5.5
4267439.0 7.2
4267602.0 5.8
4267677.0 7.4
4267719.0 6.3
4267735.0 7.0
4267764.0 7.1
4268073.0 6.4
4268191.0 5.8
4268208.0 6.8
4268232.0 6.6
4268292.0 5.9
4268532.0 7.5
4268712.0 5.7
4268782.0 6.8
4269208.0 7.5
4269341.0 5.8
4269394.0 6.3
4269418.0 6.5
4269481.0 6.4
4269547.0 7.4
4269573.0 7.4
4269589.0 6.1
4269619.0 6.2
4269732.0 6.1
4269912.0 6.7
4269972.0 5.7
4270032.0 6.0
4270092.0 7.5
4270152.0 5.8
4270229.0 6.9
4270258.0 6.8
4270415.0 7.0
4270529.0 7.5
4270908.0 6.7
4271110.0 5.6
4271292.0 6.0
4271472.0 6.7
4271908.0 7.4
4272015.0 6.2
4272098.0 6.8
4272168.0 6.6
4272235.0 6.6
4272488.0 6.5
4272672.0 5.8
4272792.0 7.0
4272852.0 5.7
4272912.0 6.9
4273032.0 6.5
4273189.0 6.6
4273271.0 7.4
4273684.0 6.7
4273879.0 6.5
4274352.0 6.2
4274412.0 6.5
4274915.0 6.7
4275045.0 6.8
4275158.0 6.6
4276402.0 6.5
4276525.0 7.4
4276868.0 7.2
4277232.0 5.9
4278019.0 7.0
4278138.0 6.3
4278164.0 6.6
4278312.0 6.9
4278372.0 7.1
4278552.0 6.6
4278612.0 6.8
4279992.0 6.4
4286290.0 6.8
4287839.0 6.2
4287962.0 6.4
4288367.0 6.3
4288512.0 6.9
4289100.0 5.9
4289237.0 7.3
4289515.0 6.8
4289892.0 6.2
4290495.0 6.8
4291632.0 6.4
4291752.0 6.7
4292376.0 6.2
4292650.0 6.4
4293291.0 6.9
4294122.0 6.0
4294212.0 7.3
4294572.0 6.1
4296625.0 6.2
4297512.0 7.0

Table A-1: 0-reduced time series of 10815 moonquakes of unspecified magnitudes that occurred from 1969–1977, after removal of 2243 unnatural events. Data from the Nakamura's *Levent.1008* Moonquakes Catalog (2008 update). Seismic magnitudes random.

| T(min) | Mag' |
|---|---|
| 0.0 | 5.7 |
| 841.0 | 6.7 |
| 39252.0 | 7.0 |
| 40362.0 | 5.7 |
| 171792.0 | 6.8 |
| 210283.0 | 5.9 |
| 213372.0 | 6.4 |
| 218606.0 | 5.5 |
| 254367.0 | 5.8 |
| 255132.0 | 7.1 |
| 255662.0 | 6.6 |
| 256056.0 | 6.8 |
| 256812.0 | 6.2 |
| 257238.0 | 6.9 |
| 259009.0 | 6.3 |
| 259392.0 | 6.6 |
| 260772.0 | 6.7 |
| 295241.0 | 6.2 |
| 295392.0 | 5.8 |
| 295632.0 | 6.3 |
| 296090.0 | 6.2 |
| 296142.0 | 5.8 |
| 296680.0 | 5.7 |
| 296832.0 | 6.5 |
| 297132.0 | 6.5 |
| 297192.0 | 5.8 |
| 297647.0 | 7.5 |
| 298392.0 | 6.2 |
| 298717.0 | 6.3 |
| 298837.0 | 5.6 |
| 298907.0 | 6.4 |
| 299064.0 | 7.1 |
| 299189.0 | 7.2 |
| 299243.0 | 5.9 |
| 299476.0 | 5.6 |
| 300240.0 | 6.2 |
| 300663.0 | 7.4 |
| 302712.0 | 6.0 |
| 338712.0 | 5.8 |
| 339362.0 | 7.1 |
| 340152.0 | 6.5 |
| 341592.0 | 6.9 |
| 342462.0 | 7.1 |
| 342972.0 | 7.3 |
| 345076.0 | 7.5 |
| 345612.0 | 7.1 |
| 347671.0 | 5.9 |
| 380207.0 | 6.6 |
| 381972.0 | 6.2 |
| 384267.0 | 7.2 |
| 386734.0 | 6.3 |
| 389800.0 | 7.4 |
| 390123.0 | 6.3 |
| 427244.0 | 7.3 |
| 430573.0 | 6.8 |
| 430872.0 | 7.5 |
| 430992.0 | 5.9 |
| 431360.0 | 5.8 |
| 431519.0 | 5.7 |
| 431844.0 | 6.5 |
| 432020.0 | 6.6 |
| 432252.0 | 6.9 |
| 432759.0 | 6.6 |
| 432794.0 | 6.3 |
| 433404.0 | 7.0 |
| 465252.0 | 5.5 |
| 467342.0 | 6.6 |
| 467998.0 | 6.5 |
| 469812.0 | 5.6 |
| 470377.0 | 6.3 |
| 470739.0 | 6.1 |
| 470922.0 | 7.5 |
| 471918.0 | 6.0 |
| 472003.0 | 6.8 |
| 472115.0 | 7.0 |
| 472314.0 | 6.6 |
| 472512.0 | 6.1 |
| 473230.0 | 6.3 |
| 473308.0 | 6.2 |
| 473734.0 | 7.5 |
| 473952.0 | 6.1 |
| 474490.0 | 7.2 |
| 507072.0 | 7.3 |
| 507312.0 | 5.6 |
| 507738.0 | 5.7 |
| 508187.0 | 6.1 |
| 508512.0 | 5.8 |
| 508752.0 | 7.1 |
| 508812.0 | 5.7 |
| 509557.0 | 7.5 |
| 509706.0 | 6.4 |
| 509832.0 | 6.6 |
| 510072.0 | 6.8 |
| 511245.0 | 7.2 |
| 511692.0 | 6.9 |
| 513594.0 | 5.8 |
| 514452.0 | 6.5 |
| 514769.0 | 7.2 |
| 514796.0 | 6.5 |
| 516842.0 | 6.8 |
| 550272.0 | 5.8 |
| 551198.0 | 6.3 |
| 551552.0 | 6.5 |
| 552597.0 | 7.4 |
| 552930.0 | 5.7 |
| 553917.0 | 6.7 |
| 554772.0 | 7.4 |
| 555175.0 | 7.3 |
| 555469.0 | 6.5 |
| 558730.0 | 7.3 |
| 559272.0 | 7.1 |
| 559530.0 | 6.8 |
| 592392.0 | 5.8 |
| 592547.0 | 7.1 |
| 593894.0 | 7.2 |
| 593930.0 | 6.2 |
| 594675.0 | 6.5 |
| 595152.0 | 7.3 |
| 595509.0 | 7.2 |
| 596592.0 | 6.0 |
| 597473.0 | 7.4 |
| 598620.0 | 6.1 |
| 600052.0 | 7.2 |
| 600194.0 | 7.4 |
| 633852.0 | 5.9 |
| 634152.0 | 6.5 |
| 640453.0 | 6.9 |
| 677112.0 | 5.6 |
| 679718.0 | 5.5 |
| 680335.0 | 6.9 |
| 684372.0 | 6.0 |
| 686272.0 | 6.9 |
| 722145.0 | 7.0 |
| 725032.0 | 6.9 |
| 726192.0 | 6.9 |
| 726687.0 | 6.7 |
| 727163.0 | 6.2 |
| 763955.0 | 6.2 |
| 764268.0 | 5.9 |
| 764462.0 | 7.4 |
| 764892.0 | 6.1 |
| 765072.0 | 5.7 |
| 765132.0 | 7.0 |
| 765764.0 | 7.1 |
| 765990.0 | 6.8 |
| 766078.0 | 7.4 |
| 766152.0 | 7.2 |
| 766774.0 | 6.1 |
| 767160.0 | 5.9 |
| 767284.0 | 7.3 |
| 768320.0 | 5.8 |
| 768953.0 | 6.8 |
| 805212.0 | 5.8 |
| 805392.0 | 6.4 |
| 805512.0 | 6.9 |
| 806017.0 | 6.6 |
| 806092.0 | 7.2 |
| 806418.0 | 6.5 |
| 806712.0 | 6.0 |
| 806832.0 | 6.5 |
| 807972.0 | 6.3 |
| 809352.0 | 7.1 |
| 810115.0 | 6.2 |
| 810402.0 | 6.5 |
| 814781.0 | 5.7 |
| 848592.0 | 7.2 |
| 848652.0 | 6.9 |
| 848712.0 | 6.0 |
| 848897.0 | 6.8 |
| 849275.0 | 6.2 |
| 849553.0 | 5.7 |
| 849644.0 | 6.3 |
| 849852.0 | 7.0 |
| 850032.0 | 6.7 |
| 850152.0 | 7.2 |
| 850435.0 | 7.2 |
| 850671.0 | 6.1 |
| 850820.0 | 6.2 |
| 850966.0 | 6.7 |
| 851721.0 | 6.4 |
| 852336.0 | 7.0 |
| 852540.0 | 6.8 |
| 854112.0 | 7.1 |
| 854172.0 | 5.8 |
| 854864.0 | 6.8 |
| 855068.0 | 7.2 |
| 855552.0 | 6.5 |
| 856992.0 | 7.5 |
| 857692.0 | 6.3 |
| 857846.0 | 7.4 |
| 890292.0 | 5.8 |
| 891030.0 | 6.3 |
| 891067.0 | 5.8 |
| 891672.0 | 7.2 |
| 891912.0 | 6.3 |
| 891977.0 | 5.6 |
| 893112.0 | 6.8 |
| 893172.0 | 7.4 |
| 893292.0 | 7.0 |
| 893352.0 | 6.3 |
| 893509.0 | 7.5 |
| 893541.0 | 6.9 |
| 893727.0 | 5.8 |
| 894295.0 | 6.3 |
| 894492.0 | 6.8 |
| 894552.0 | 5.5 |
| 894792.0 | 6.7 |
| 895046.0 | 5.7 |
| 895280.0 | 7.4 |
| 895573.0 | 6.0 |
| 896649.0 | 7.5 |
| 897552.0 | 5.9 |
| 933252.0 | 6.8 |
| 934992.0 | 6.0 |
| 935244.0 | 7.1 |
| 935480.0 | 5.8 |
| 935535.0 | 5.8 |
| 935722.0 | 6.8 |
| 937005.0 | 6.5 |
| 937089.0 | 5.7 |
| 937342.0 | 6.8 |
| 937512.0 | 5.8 |
| 937932.0 | 7.1 |
| 937992.0 | 6.1 |
| 938278.0 | 7.3 |
| 938302.0 | 6.5 |
| 938399.0 | 6.0 |
| 939312.0 | 6.8 |
| 940181.0 | 5.7 |
| 940295.0 | 6.1 |
| 940327.0 | 5.8 |
| 941138.0 | 7.1 |
| 941293.0 | 6.2 |
| 941320.0 | 7.4 |
| 941345.0 | 6.8 |
| 941428.0 | 5.6 |
| 941498.0 | 6.3 |
| 941542.0 | 5.8 |
| 941732.0 | 7.4 |
| 941892.0 | 5.9 |
| 941952.0 | 6.1 |
| 942312.0 | 5.5 |
| 942400.0 | 7.1 |
| 976367.0 | 6.4 |
| 976872.0 | 6.7 |
| 977292.0 | 5.8 |
| 977314.0 | 7.1 |
| 977480.0 | 6.2 |
| 977832.0 | 6.3 |
| 978192.0 | 6.5 |
| 978420.0 | 6.1 |
| 978476.0 | 7.1 |
| 978617.0 | 6.7 |
| 978862.0 | 6.3 |
| 978962.0 | 6.5 |
| 979047.0 | 6.2 |
| 979139.0 | 7.0 |
| 979202.0 | 7.1 |
| 979632.0 | 7.2 |
| 979752.0 | 6.6 |
| 979907.0 | 6.1 |
| 980290.0 | 7.3 |
| 980332.0 | 7.2 |
| 980367.0 | 5.7 |
| 980420.0 | 7.3 |
| 980476.0 | 6.2 |
| 980522.0 | 6.2 |
| 980566.0 | 6.9 |
| 980643.0 | 5.9 |
| 980697.0 | 6.9 |
| 980712.0 | 7.0 |
| 980772.0 | 7.1 |
| 980892.0 | 6.3 |
| 981072.0 | 6.2 |
| 981132.0 | 6.1 |
| 981269.0 | 6.4 |
| 982113.0 | 6.7 |
| 982150.0 | 6.3 |
| 982272.0 | 6.2 |
| 982512.0 | 5.6 |
| 982632.0 | 6.8 |
| 982933.0 | 6.9 |
| 982967.0 | 5.7 |
| 983244.0 | 5.8 |
| 983592.0 | 6.4 |
| 983652.0 | 6.5 |
| 983772.0 | 6.6 |
| 983952.0 | 6.0 |
| 984072.0 | 6.9 |
| 984174.0 | 7.3 |
| 984245.0 | 7.2 |
| 984345.0 | 5.8 |
| 984690.0 | 7.3 |
| 984708.0 | 6.1 |
| 984812.0 | 7.1 |
| 985032.0 | 6.3 |
| 985092.0 | 7.0 |
| 985332.0 | 6.2 |
| 986158.0 | 5.8 |
| 986202.0 | 6.2 |
| 986283.0 | 6.8 |
| 1018110.0 | 5.6 |
| 1018152.0 | 6.7 |
| 1018392.0 | 5.8 |
| 1018452.0 | 7.3 |
| 1018890.0 | 5.7 |
| 1019093.0 | 6.8 |
| 1019154.0 | 6.3 |
| 1019238.0 | 7.3 |
| 1019269.0 | 6.5 |
| 1019445.0 | 7.2 |
| 1019544.0 | 6.8 |
| 1019772.0 | 7.4 |
| 1019832.0 | 5.9 |
| 1019892.0 | 7.2 |
| 1019952.0 | 7.3 |
| 1020187.0 | 7.3 |
| 1020544.0 | 7.5 |
| 1020730.0 | 6.6 |
| 1020918.0 | 5.5 |
| 1021092.0 | 5.8 |
| 1021392.0 | 5.7 |
| 1021613.0 | 5.9 |
| 1021632.0 | 5.7 |
| 1021749.0 | 7.4 |
| 1021762.0 | 6.0 |
| 1021950.0 | 7.3 |
| 1022105.0 | 6.3 |
| 1022114.0 | 6.0 |
| 1022532.0 | 6.1 |
| 1022712.0 | 5.7 |
| 1022772.0 | 7.3 |
| 1022832.0 | 5.9 |
| 1023056.0 | 7.2 |
| 1023125.0 | 6.1 |
| 1023150.0 | 7.4 |
| 1023733.0 | 6.4 |
| 1023872.0 | 6.5 |
| 1023882.0 | 7.0 |
| 1024032.0 | 5.7 |
| 1024092.0 | 6.3 |
| 1024332.0 | 6.2 |
| 1024830.0 | 6.7 |
| 1025059.0 | 5.8 |
| 1025094.0 | 7.1 |
| 1025254.0 | 6.5 |
| 1025592.0 | 6.0 |
| 1026316.0 | 6.4 |
| 1026330.0 | 7.2 |
| 1026394.0 | 6.4 |
| 1026664.0 | 7.0 |
| 1026721.0 | 7.2 |
| 1027539.0 | 6.6 |
| 1027669.0 | 7.4 |
| 1028132.0 | 7.4 |
| 1060542.0 | 6.8 |
| 1061207.0 | 6.2 |
| 1061472.0 | 6.5 |
| 1061652.0 | 5.8 |
| 1061712.0 | 5.7 |
| 1062100.0 | 6.6 |
| 1062177.0 | 6.0 |
| 1062220.0 | 7.2 |
| 1062295.0 | 7.0 |
| 1062792.0 | 6.4 |
| 1062912.0 | 6.6 |
| 1063671.0 | 6.6 |
| 1064592.0 | 7.4 |
| 1064712.0 | 5.6 |
| 1065092.0 | 7.4 |
| 1066392.0 | 6.5 |
| 1066501.0 | 6.8 |
| 1067352.0 | 7.2 |
| 1067412.0 | 6.0 |
| 1067472.0 | 5.6 |
| 1067692.0 | 6.3 |
| 1067719.0 | 7.3 |
| 1069574.0 | 5.9 |
| 1103059.0 | 6.5 |
| 1103352.0 | 7.2 |
| 1103412.0 | 7.2 |
| 1103472.0 | 6.0 |
| 1103592.0 | 5.8 |
| 1103682.0 | 6.9 |
| 1103723.0 | 7.3 |
| 1103964.0 | 6.6 |
| 1104178.0 | 7.1 |
| 1104234.0 | 6.9 |
| 1104271.0 | 5.8 |
| 1104339.0 | 7.4 |
| 1104444.0 | 6.1 |
| 1104792.0 | 6.9 |
| 1105032.0 | 6.8 |
| 1105255.0 | 7.2 |
| 1105760.0 | 6.4 |
| 1105869.0 | 6.2 |
| 1109964.0 | 6.0 |
| 1110792.0 | 5.6 |
| 1111244.0 | 6.8 |
| 1111752.0 | 6.3 |
| 1112914.0 | 7.0 |
| 1144992.0 | 6.0 |
| 1145292.0 | 6.6 |
| 1145694.0 | 6.3 |
| 1145970.0 | 5.5 |
| 1146286.0 | 7.4 |
| 1146312.0 | 5.5 |
| 1146432.0 | 5.6 |
| 1146552.0 | 6.0 |
| 1146612.0 | 6.6 |
| 1146792.0 | 7.0 |
| 1146887.0 | 6.6 |
| 1147040.0 | 5.6 |
| 1147532.0 | 5.6 |
| 1147752.0 | 6.5 |
| 1148052.0 | 7.1 |
| 1148299.0 | 6.9 |
| 1148539.0 | 5.7 |
| 1148732.0 | 6.4 |
| 1148932.0 | 7.0 |
| 1149110.0 | 6.0 |
| 1149552.0 | 6.6 |
| 1149897.0 | 5.8 |
| 1150077.0 | 5.7 |
| 1150350.0 | 7.4 |
| 1150533.0 | 5.8 |
| 1150992.0 | 7.1 |
| 1153414.0 | 7.4 |
| 1153752.0 | 7.0 |
| 1154166.0 | 6.2 |
| 1154360.0 | 6.3 |
| 1154450.0 | 7.3 |
| 1154474.0 | 6.4 |
| 1154556.0 | 5.8 |
| 1154810.0 | 7.0 |
| 1186632.0 | 7.3 |
| 1186692.0 | 6.8 |
| 1186812.0 | 6.0 |
| 1186932.0 | 5.7 |
| 1187112.0 | 6.8 |
| 1187362.0 | 6.3 |
| 1187457.0 | 5.8 |
| 1187662.0 | 6.6 |
| 1188072.0 | 6.8 |
| 1188192.0 | 7.1 |
| 1188837.0 | 6.6 |
| 1188930.0 | 6.8 |
| 1189512.0 | 5.8 |
| 1189812.0 | 6.8 |
| 1191012.0 | 5.8 |
| 1191192.0 | 5.7 |
| 1191372.0 | 7.2 |
| 1191731.0 | 5.7 |
| 1191905.0 | 7.0 |
| 1192095.0 | 6.9 |
| 1192392.0 | 6.9 |
| 1192512.0 | 7.1 |
| 1193067.0 | 7.3 |
| 1193122.0 | 5.8 |
| 1193470.0 | 7.0 |
| 1193507.0 | 7.4 |
| 1193832.0 | 6.4 |
| 1194252.0 | 6.4 |
| 1194778.0 | 5.7 |
| 1194838.0 | 6.5 |
| 1194867.0 | 6.9 |
| 1194913.0 | 5.7 |
| 1194925.0 | 6.7 |
| 1195392.0 | 5.8 |
| 1195512.0 | 7.0 |
| 1195632.0 | 6.9 |
| 1195752.0 | 6.4 |
| 1195876.0 | 5.5 |
| 1196020.0 | 6.0 |
| 1196187.0 | 5.6 |
| 1196610.0 | 6.7 |
| 1196647.0 | 5.8 |
| 1231272.0 | 5.9 |
| 1231916.0 | 6.3 |
| 1231972.0 | 6.5 |
| 1232282.0 | 5.8 |
| 1232952.0 | 6.3 |
| 1233132.0 | 6.6 |
| 1233192.0 | 6.5 |
| 1233277.0 | 6.5 |
| 1233477.0 | 7.0 |
| 1233683.0 | 6.0 |
| 1233747.0 | 7.4 |
| 1233843.0 | 6.6 |
| 1233892.0 | 6.1 |
| 1234392.0 | 6.8 |
| 1234572.0 | 6.6 |
| 1235045.0 | 5.9 |
| 1235167.0 | 6.9 |
| 1235193.0 | 7.1 |
| 1235592.0 | 6.6 |
| 1235652.0 | 6.2 |
| 1235832.0 | 6.5 |
| 1235952.0 | 7.0 |
| 1236012.0 | 5.8 |
| 1236344.0 | 6.5 |
| 1236413.0 | 6.7 |
| 1236488.0 | 6.8 |
| 1236763.0 | 6.1 |
| 1236777.0 | 5.7 |
| 1237152.0 | 6.5 |
| 1237272.0 | 7.1 |
| 1237332.0 | 7.3 |
| 1237593.0 | 6.2 |
| 1237903.0 | 5.8 |
| 1238032.0 | 6.1 |
| 1238324.0 | 5.8 |
| 1238400.0 | 7.4 |
| 1238832.0 | 5.6 |
| 1239249.0 | 6.2 |
| 1239277.0 | 5.9 |
| 1239369.0 | 6.9 |
| 1239734.0 | 6.4 |
| 1272930.0 | 6.2 |
| 1273032.0 | 5.6 |
| 1273332.0 | 5.7 |
| 1273452.0 | 7.1 |
| 1273619.0 | 6.1 |
| 1273857.0 | 7.4 |
| 1274047.0 | 7.4 |
| 1274532.0 | 6.6 |
| 1274592.0 | 6.0 |
| 1274652.0 | 5.6 |
| 1274712.0 | 7.2 |
| 1275398.0 | 6.6 |
| 1275443.0 | 5.5 |
| 1275613.0 | 6.7 |
| 1276332.0 | 6.5 |
| 1276474.0 | 6.8 |
| 1276723.0 | 6.8 |
| 1277412.0 | 6.3 |
| 1278180.0 | 6.1 |
| 1278370.0 | 7.1 |
| 1279032.0 | 5.8 |
| 1279092.0 | 6.9 |
| 1279152.0 | 6.5 |
| 1279349.0 | 5.5 |
| 1279968.0 | 6.7 |
| 1280084.0 | 5.9 |
| 1280146.0 | 6.8 |
| 1280352.0 | 5.5 |
| 1280792.0 | 6.0 |
| 1280853.0 | 7.4 |
| 1281205.0 | 7.5 |
| 1315152.0 | 7.2 |
| 1315365.0 | 6.4 |
| 1315692.0 | 5.6 |
| 1315884.0 | 6.3 |
| 1315929.0 | 5.7 |
| 1316292.0 | 7.0 |
| 1316532.0 | 5.7 |
| 1317232.0 | 6.0 |
| 1317852.0 | 5.6 |
| 1318412.0 | 6.6 |
| 1318679.0 | 7.0 |
| 1319076.0 | 6.2 |
| 1319172.0 | 6.8 |
| 1320464.0 | 6.9 |
| 1321190.0 | 6.5 |
| 1321362.0 | 6.6 |
| 1321474.0 | 6.4 |
| 1321552.0 | 7.2 |
| 1321674.0 | 7.4 |
| 1321932.0 | 5.5 |
| 1321962.0 | 6.2 |
| 1322552.0 | 6.3 |
| 1322824.0 | 5.6 |
| 1323080.0 | 5.7 |
| 1324082.0 | 6.6 |
| 1358172.0 | 6.9 |
| 1358232.0 | 5.6 |
| 1358914.0 | 6.1 |
| 1359672.0 | 5.5 |
| 1360821.0 | 7.4 |
| 1360872.0 | 7.3 |
| 1362252.0 | 6.2 |
| 1362792.0 | 6.8 |
| 1363501.0 | 6.6 |
| 1364232.0 | 7.4 |
| 1366692.0 | 6.6 |
| 1400798.0 | 7.1 |
| 1402263.0 | 6.9 |
| 1402412.0 | 6.0 |
| 1403112.0 | 6.8 |
| 1404252.0 | 5.8 |
| 1404708.0 | 5.7 |
| 1404763.0 | 6.1 |
| 1405632.0 | 6.9 |
| 1406690.0 | 6.7 |
| 1408452.0 | 6.0 |
| 1409010.0 | 5.5 |
| 1409089.0 | 6.1 |
| 1443432.0 | 6.3 |
| 1443968.0 | 6.9 |
| 1443984.0 | 6.6 |
| 1444039.0 | 6.9 |
| 1444130.0 | 6.5 |
| 1444382.0 | 5.8 |
| 1444452.0 | 6.0 |
| 1444512.0 | 6.7 |
| 1444572.0 | 7.1 |
| 1444632.0 | 5.5 |
| 1444752.0 | 5.9 |
| 1444872.0 | 6.2 |
| 1444943.0 | 6.4 |
| 1445067.0 | 7.0 |
| 1445097.0 | 6.2 |
| 1445112.0 | 7.0 |
| 1445134.0 | 7.0 |
| 1445180.0 | 5.6 |
| 1445332.0 | 5.9 |
| 1445363.0 | 7.2 |
| 1445527.0 | 6.2 |
| 1445617.0 | 6.4 |
| 1445660.0 | 6.3 |
| 1445692.0 | 6.8 |
| 1445733.0 | 6.9 |
| 1445832.0 | 6.2 |
| 1445892.0 | 7.2 |
| 1446012.0 | 7.0 |
| 1446252.0 | 6.0 |
| 1446582.0 | 7.4 |
| 1446704.0 | 7.1 |
| 1446743.0 | 6.1 |
| 1446818.0 | 6.7 |
| 1446843.0 | 7.0 |
| 1446872.0 | 7.1 |
| 1446960.0 | 6.4 |
| 1446980.0 | 7.3 |
| 1447020.0 | 6.4 |
| 1447092.0 | 6.7 |
| 1447147.0 | 6.9 |
| 1447175.0 | 6.4 |
| 1447272.0 | 6.7 |
| 1447332.0 | 7.3 |
| 1447512.0 | 5.7 |
| 1447752.0 | 5.7 |
| 1447817.0 | 5.7 |
| 1447902.0 | 6.2 |
| 1447922.0 | 7.0 |
| 1448002.0 | 5.7 |
| 1448035.0 | 7.2 |
| 1448140.0 | 5.8 |
| 1448247.0 | 6.8 |
| 1448277.0 | 7.0 |
| 1448360.0 | 6.9 |
| 1448482.0 | 7.1 |
| 1448520.0 | 6.3 |
| 1448534.0 | 6.3 |
| 1448619.0 | 6.8 |
| 1448712.0 | 5.8 |
| 1448772.0 | 7.4 |
| 1448832.0 | 6.7 |
| 1448892.0 | 5.7 |
| 1449012.0 | 5.6 |
| 1449132.0 | 5.9 |
| 1449192.0 | 6.3 |
| 1449291.0 | 6.4 |
| 1449320.0 | 7.1 |
| 1449482.0 | 6.4 |
| 1449639.0 | 6.6 |
| 1449716.0 | 7.0 |
| 1449742.0 | 7.1 |
| 1449777.0 | 6.4 |
| 1449827.0 | 6.9 |
| 1449951.0 | 6.4 |
| 1450017.0 | 6.0 |
| 1450332.0 | 7.5 |
| 1450512.0 | 6.2 |
| 1450572.0 | 7.4 |
| 1450939.0 | 6.0 |
| 1450989.0 | 7.0 |
| 1451016.0 | 5.6 |
| 1451078.0 | 5.7 |
| 1451139.0 | 5.6 |
| 1451219.0 | 7.5 |
| 1451302.0 | 6.6 |
| 1451345.0 | 7.1 |
| 1451430.0 | 6.1 |
| 1451512.0 | 7.4 |
| 1451892.0 | 7.3 |
| 1452012.0 | 6.8 |
| 1452158.0 | 5.7 |
| 1452194.0 | 6.2 |
| 1452240.0 | 5.6 |
| 1452440.0 | 6.9 |
| 1452620.0 | 6.1 |
| 1452844.0 | 6.6 |
| 1486103.0 | 7.4 |
| 1486212.0 | 6.1 |
| 1486332.0 | 5.8 |
| 1486392.0 | 7.2 |
| 1486512.0 | 5.9 |
| 1486572.0 | 6.0 |
| 1486929.0 | 5.9 |
| 1486937.0 | 6.0 |
| 1487064.0 | 7.3 |
| 1487087.0 | 6.2 |
| 1487201.0 | 6.8 |
| 1487318.0 | 6.9 |
| 1487342.0 | 6.8 |
| 1487488.0 | 7.0 |
| 1487592.0 | 5.8 |
| 1487712.0 | 7.0 |
| 1487772.0 | 5.9 |
| 1487892.0 | 7.4 |
| 1488012.0 | 6.5 |
| 1488072.0 | 6.8 |
| 1488174.0 | 7.4 |
| 1488209.0 | 7.2 |
| 1488234.0 | 6.5 |
| 1488264.0 | 5.5 |
| 1488272.0 | 7.3 |
| 1488315.0 | 6.9 |
| 1488360.0 | 5.9 |
| 1488377.0 | 6.1 |
| 1488432.0 | 6.9 |
| 1488500.0 | 6.9 |
| 1488522.0 | 6.2 |
| 1488537.0 | 5.6 |
| 1488604.0 | 6.2 |
| 1488623.0 | 7.3 |
| 1488648.0 | 5.9 |
| 1488853.0 | 6.5 |
| 1488901.0 | 6.4 |
| 1488967.0 | 6.2 |
| 1488989.0 | 6.6 |
| 1488995.0 | 5.9 |
| 1489032.0 | 7.5 |
| 1489152.0 | 6.6 |
| 1489212.0 | 7.5 |
| 1489272.0 | 7.3 |
| 1489332.0 | 7.3 |
| 1489452.0 | 6.8 |
| 1489580.0 | 6.3 |
| 1489627.0 | 6.0 |
| 1489724.0 | 7.0 |
| 1489743.0 | 6.8 |
| 1490080.0 | 5.8 |
| 1490109.0 | 6.2 |
| 1490129.0 | 5.7 |
| 1490215.0 | 5.9 |
| 1490245.0 | 6.3 |
| 1490342.0 | 6.6 |
| 1490456.0 | 6.5 |
| 1490532.0 | 5.9 |
| 1490592.0 | 6.5 |
| 1490892.0 | 5.6 |
| 1490952.0 | 6.6 |
| 1491375.0 | 7.2 |
| 1491388.0 | 5.8 |
| 1491534.0 | 5.9 |
| 1491550.0 | 6.2 |
| 1491684.0 | 6.0 |
| 1491753.0 | 6.6 |
| 1491781.0 | 5.8 |
| 1491815.0 | 6.0 |
| 1491841.0 | 5.5 |
| 1491874.0 | 5.7 |
| 1491912.0 | 6.5 |
| 1492032.0 | 6.9 |
| 1492152.0 | 7.3 |
| 1492392.0 | 6.1 |
| 1492456.0 | 6.9 |
| 1492500.0 | 5.6 |
| 1492531.0 | 5.8 |
| 1492607.0 | 6.2 |
| 1492649.0 | 7.2 |
| 1492670.0 | 6.7 |
| 1492723.0 | 6.3 |
| 1492904.0 | 6.4 |
| 1492982.0 | 6.6 |
| 1493095.0 | 7.3 |
| 1493230.0 | 7.0 |
| 1493300.0 | 6.8 |
| 1493352.0 | 6.0 |
| 1493412.0 | 5.9 |
| 1493592.0 | 6.9 |
| 1493772.0 | 5.7 |
| 1493832.0 | 6.6 |
| 1494032.0 | 6.8 |
| 1494117.0 | 5.7 |
| 1494142.0 | 6.9 |
| 1494194.0 | 6.6 |
| 1494912.0 | 5.9 |
| 1494972.0 | 6.6 |
| 1495212.0 | 6.3 |
| 1495272.0 | 6.5 |
| 1495420.0 | 6.9 |
| 1495952.0 | 6.7 |
| 1495982.0 | 6.1 |
| 1496141.0 | 6.2 |
| 1527854.0 | 7.4 |
| 1527893.0 | 7.2 |
| 1527912.0 | 6.4 |
| 1527972.0 | 6.3 |
| 1528032.0 | 7.1 |
| 1528272.0 | 6.2 |
| 1528452.0 | 6.7 |
| 1528479.0 | 6.3 |
| 1528567.0 | 6.8 |
| 1528700.0 | 7.3 |
| 1528725.0 | 6.1 |
| 1528777.0 | 7.4 |
| 1529114.0 | 6.0 |
| 1529161.0 | 6.2 |
| 1529352.0 | 6.9 |
| 1529472.0 | 6.4 |
| 1529592.0 | 7.3 |
| 1529772.0 | 6.5 |
| 1529832.0 | 5.7 |
| 1530188.0 | 6.4 |
| 1530242.0 | 7.1 |
| 1530259.0 | 6.1 |
| 1530314.0 | 6.0 |
| 1530321.0 | 5.6 |
| 1530562.0 | 7.2 |
| 1530631.0 | 6.6 |
| 1530687.0 | 5.7 |
| 1530852.0 | 6.2 |
| 1530912.0 | 6.5 |
| 1531032.0 | 7.1 |
| 1531272.0 | 6.1 |
| 1531341.0 | 5.6 |
| 1531430.0 | 6.1 |
| 1531517.0 | 7.1 |
| 1531602.0 | 7.3 |
| 1531686.0 | 5.7 |
| 1531703.0 | 7.0 |
| 1531755.0 | 5.8 |
| 1531872.0 | 6.4 |
| 1531923.0 | 7.4 |
| 1531993.0 | 6.0 |
| 1532292.0 | 6.4 |
| 1532352.0 | 5.9 |
| 1532472.0 | 6.6 |
| 1532532.0 | 7.3 |
| 1532592.0 | 7.1 |
| 1532712.0 | 5.9 |
| 1532991.0 | 6.9 |
| 1533029.0 | 7.1 |
| 1533162.0 | 6.5 |
| 1533335.0 | 5.9 |
| 1533348.0 | 6.8 |
| 1533446.0 | 7.4 |
| 1533488.0 | 6.6 |
| 1533500.0 | 5.8 |
| 1533525.0 | 7.3 |
| 1533591.0 | 6.6 |
| 1533597.0 | 6.2 |
| 1533627.0 | 6.7 |
| 1533672.0 | 6.0 |
| 1533852.0 | 6.1 |
| 1533912.0 | 6.9 |
| 1533972.0 | 6.7 |
| 1534092.0 | 7.0 |
| 1534152.0 | 6.8 |
| 1534219.0 | 6.5 |
| 1534335.0 | 7.1 |
| 1534417.0 | 5.7 |
| 1534504.0 | 6.2 |
| 1534568.0 | 7.0 |
| 1534671.0 | 5.6 |
| 1534683.0 | 6.1 |
| 1534812.0 | 7.0 |
| 1534867.0 | 6.4 |
| 1534996.0 | 7.4 |
| 1535030.0 | 6.3 |
| 1535064.0 | 6.9 |
| 1535112.0 | 7.2 |
| 1535172.0 | 5.7 |
| 1535232.0 | 5.6 |
| 1535532.0 | 7.3 |
| 1535827.0 | 5.6 |
| 1535835.0 | 6.8 |
| 1535874.0 | 5.6 |
| 1535920.0 | 6.4 |
| 1536004.0 | 5.8 |
| 1536042.0 | 7.2 |
| 1536092.0 | 6.7 |
| 1536125.0 | 5.5 |
| 1536147.0 | 6.5 |
| 1536792.0 | 6.1 |
| 1536852.0 | 6.8 |
| 1536912.0 | 6.1 |
| 1537260.0 | 6.5 |
| 1537306.0 | 6.6 |
| 1537435.0 | 6.7 |
| 1537681.0 | 5.8 |
| 1537835.0 | 5.8 |
| 1571112.0 | 6.9 |
| 1571292.0 | 6.3 |
| 1571472.0 | 6.0 |
| 1571532.0 | 7.5 |
| 1571592.0 | 6.7 |
| 1571800.0 | 7.1 |
| 1571822.0 | 6.1 |
| 1571950.0 | 6.9 |
| 1572086.0 | 7.2 |
| 1572211.0 | 6.2 |
| 1572232.0 | 6.4 |
| 1572253.0 | 5.8 |
| 1572272.0 | 6.2 |
| 1572308.0 | 7.4 |
| 1572387.0 | 6.6 |
| 1572547.0 | 5.7 |
| 1572612.0 | 6.9 |
| 1572672.0 | 7.4 |
| 1572732.0 | 6.8 |
| 1572972.0 | 6.2 |
| 1573032.0 | 7.1 |
| 1573288.0 | 5.7 |
| 1573371.0 | 6.7 |
| 1573410.0 | 6.9 |
| 1573432.0 | 6.9 |
| 1573560.0 | 5.7 |
| 1573587.0 | 6.9 |
| 1573723.0 | 5.8 |
| 1573765.0 | 7.4 |
| 1573770.0 | 6.2 |
| 1573977.0 | 5.8 |
| 1573992.0 | 6.1 |
| 1574172.0 | 6.6 |
| 1574232.0 | 6.6 |
| 1574472.0 | 7.4 |
| 1574602.0 | 7.4 |
| 1574650.0 | 6.6 |
| 1574744.0 | 6.4 |
| 1574812.0 | 6.5 |
| 1574872.0 | 6.3 |
| 1574918.0 | 5.7 |
| 1575231.0 | 7.2 |
| 1575257.0 | 7.2 |
| 1575357.0 | 7.0 |
| 1575374.0 | 6.7 |
| 1575432.0 | 5.8 |
| 1575552.0 | 6.8 |
| 1576031.0 | 6.3 |
| 1576230.0 | 6.6 |
| 1576232.0 | 6.8 |
| 1576330.0 | 6.0 |
| 1576379.0 | 6.6 |
| 1576384.0 | 7.4 |
| 1576453.0 | 6.1 |
| 1576540.0 | 6.1 |
| 1576577.0 | 7.3 |
| 1576621.0 | 5.8 |
| 1576678.0 | 6.3 |
| 1576697.0 | 6.9 |
| 1576707.0 | 6.6 |
| 1576788.0 | 6.2 |
| 1576825.0 | 5.5 |
| 1576849.0 | 6.3 |
| 1576872.0 | 5.9 |
| 1576932.0 | 5.8 |
| 1576992.0 | 7.3 |
| 1577112.0 | 7.4 |
| 1577292.0 | 7.2 |
| 1577466.0 | 6.1 |
| 1577485.0 | 5.5 |
| 1577690.0 | 6.6 |
| 1577815.0 | 6.1 |
| 1577830.0 | 7.1 |
| 1577906.0 | 6.3 |
| 1578945.0 | 6.9 |
| 1579046.0 | 7.1 |
| 1579134.0 | 5.8 |
| 1580504.0 | 7.1 |
| 1580597.0 | 7.3 |
| 1612872.0 | 7.2 |
| 1612932.0 | 7.4 |
| 1612992.0 | 7.4 |
| 1613112.0 | 5.7 |
| 1613232.0 | 6.3 |
| 1613292.0 | 6.9 |
| 1613352.0 | 5.8 |
| 1613429.0 | 6.5 |
| 1613475.0 | 6.9 |
| 1613728.0 | 5.7 |
| 1613735.0 | 6.1 |
| 1613760.0 | 6.8 |

1613777.0 5.8
1613828.0 7.3
1613898.0 5.7
1614054.0 5.7
1614065.0 5.8
1614157.0 5.6
1614172.0 7.1
1614293.0 7.0
1614312.0 5.6
1614372.0 6.5
1614492.0 6.0
1614552.0 6.7
1614732.0 7.3
1614792.0 6.5
1614865.0 6.8
1614902.0 6.9
1614930.0 5.8
1614959.0 6.5
1615080.0 6.0
1615161.0 6.9
1615228.0 6.3
1615292.0 7.4
1615311.0 6.6
1615319.0 6.1
1615390.0 7.2
1615467.0 7.2
1615540.0 7.5
1615554.0 7.1
1615627.0 6.4
1615731.0 6.8
1615752.0 7.1
1615872.0 6.5
1615932.0 6.1
1615992.0 6.3
1616112.0 6.1
1616172.0 7.4
1616232.0 6.6
1616321.0 7.2
1616537.0 6.3
1616547.0 5.5
1616608.0 6.1
1616643.0 6.6
1616725.0 5.8
1616734.0 6.5
1616810.0 6.4
1616859.0 6.4
1616881.0 6.5
1616888.0 6.0
1616897.0 6.2
1617111.0 6.0
1617192.0 6.5
1617252.0 5.6
1617372.0 5.7
1617432.0 7.1
1617492.0 6.5
1617612.0 6.7
1617732.0 5.8
1617852.0 6.0
1617969.0 5.5
1618084.0 6.5
1618211.0 5.9
1618340.0 6.0
1618478.0 7.3
1618625.0 6.7
1618752.0 6.6
1618812.0 6.1
1618932.0 6.9
1619052.0 7.1
1619112.0 7.0
1619281.0 7.1
1619665.0 7.4
1619877.0 6.5
1620072.0 6.1
1620312.0 7.1
1620372.0 5.5
1620432.0 7.1
1620675.0 5.8
1621362.0 6.5
1621387.0 5.6
1622274.0 7.2
1656072.0 6.9
1656132.0 6.1
1656192.0 6.3
1656252.0 6.1
1656432.0 6.4
1656492.0 6.2
1656552.0 6.7
1656614.0 5.6
1656691.0 7.2
1656758.0 6.8
1656821.0 6.6
1657015.0 7.3
1657219.0 6.9
1657301.0 5.6
1657372.0 6.2
1657572.0 5.6
1657992.0 7.2
1658062.0 6.2
1658318.0 5.9
1658395.0 7.1
1658416.0 6.3
1658564.0 6.7
1658694.0 6.5
1658711.0 6.2
1658774.0 5.9
1658952.0 6.9
1659012.0 6.9
1659192.0 6.0
1659252.0 5.5
1659499.0 5.5
1660262.0 7.0
1660359.0 6.5
1660452.0 7.0
1660572.0 7.4
1660692.0 7.4
1660812.0 6.2
1660872.0 6.1
1661014.0 7.4
1661022.0 5.8
1661172.0 7.1
1661214.0 5.8
1661334.0 6.6
1661453.0 5.6
1661463.0 5.8
1661506.0 6.9
1661595.0 7.2
1661687.0 6.4
1661762.0 6.6
1661832.0 5.7
1662072.0 6.0
1662192.0 6.9
1662612.0 5.6
1662825.0 5.6
1662858.0 7.0
1663202.0 5.8
1663272.0 6.9
1664215.0 6.6
1664712.0 7.1
1664832.0 7.0
1665192.0 6.4
1665616.0 6.7
1665966.0 6.5
1666087.0 7.5
1697892.0 6.2
1698012.0 7.1
1698132.0 6.4
1698312.0 7.4
1698625.0 6.1
1698654.0 6.8
1699096.0 6.5
1699127.0 6.5
1699151.0 6.0
1699213.0 5.7
1699230.0 7.2
1699452.0 5.9
1699512.0 5.7
1699692.0 6.8
1699825.0 6.1
1699886.0 6.6
1699907.0 5.7
1699953.0 6.6
1699984.0 6.3
1700035.0 6.8
1700092.0 6.2
1700144.0 7.2
1700191.0 6.0
1700344.0 5.9
1700351.0 5.7
1700361.0 5.6
1700448.0 5.8
1700465.0 6.8
1700519.0 6.5
1700647.0 6.5
1700693.0 7.2
1700712.0 6.7
1700832.0 6.4
1700952.0 5.6
1701072.0 6.3
1701132.0 5.8
1701192.0 6.3
1701267.0 5.6
1701321.0 6.4
1701328.0 6.0
1701381.0 6.8
1701390.0 5.9
1701420.0 7.1
1701442.0 5.7
1701505.0 6.8
1701661.0 5.9
1701790.0 6.5
1701834.0 6.5
1701938.0 7.1
1702212.0 6.9
1702332.0 5.7
1702512.0 6.1
1702572.0 7.3
1702751.0 6.7
1702883.0 5.6
1702902.0 6.4
1702956.0 5.7
1703022.0 5.9
1703146.0 7.5
1703213.0 6.7
1703274.0 5.7
1703391.0 7.3
1703402.0 6.8
1703411.0 5.6
1703455.0 6.6
1703523.0 6.3
1703555.0 6.6
1703568.0 6.3
1703592.0 5.5
1703652.0 6.5
1703712.0 5.7
1703772.0 7.4
1703832.0 6.2
1703892.0 6.9
1703952.0 6.8
1704012.0 7.5
1704180.0 6.6
1704224.0 7.0
1704242.0 6.5
1704250.0 5.8
1704260.0 6.9
1704296.0 6.5
1704402.0 7.1
1704453.0 7.0
1704606.0 6.9
1704661.0 7.4
1704704.0 5.8
1704715.0 6.2
1704757.0 6.4
1704790.0 6.0
1704820.0 5.8
1704836.0 6.9
1704942.0 5.6
1705025.0 7.4
1705152.0 5.7
1705212.0 6.0
1705272.0 6.5
1705452.0 7.4
1705512.0 5.6
1705768.0 6.4
1706532.0 5.6
1706772.0 7.4
1707236.0 7.1
1707320.0 6.9
1739555.0 6.2
1739652.0 5.7
1739772.0 7.1
1739832.0 5.7
1739892.0 5.5
1739952.0 7.3
1740217.0 6.7
1740356.0 5.8
1740414.0 6.6
1740466.0 5.6
1740526.0 7.3
1740600.0 6.8
1740640.0 5.6
1740733.0 6.4
1740910.0 6.4
1741030.0 7.1
1741032.0 6.8
1741152.0 6.4
1741212.0 5.9
1741332.0 6.6
1741607.0 5.9
1741682.0 6.1
1741696.0 5.6
1741707.0 6.8
1741809.0 6.3
1741849.0 6.6
1741859.0 6.0
1741974.0 6.1
1742096.0 7.4
1742152.0 6.0
1742225.0 5.8
1742286.0 6.1
1742310.0 5.7
1742352.0 6.4
1742386.0 6.4
1742396.0 6.2
1742452.0 6.3
1742472.0 5.9
1742532.0 6.4
1742652.0 5.5
1742712.0 5.5
1742772.0 7.2
1742832.0 6.2
1742892.0 6.0
1742952.0 6.6
1743024.0 6.4
1743198.0 5.8
1743315.0 6.2
1743394.0 6.9
1743480.0 6.9
1743675.0 7.0
1743912.0 6.4
1744272.0 5.8
1744332.0 7.2
1744392.0 5.6
1744775.0 6.2
1745070.0 6.7
1745291.0 7.2
1745472.0 5.9
1745592.0 7.3
1746378.0 6.6
1748212.0 6.1
1748412.0 6.6
1749052.0 6.8
1749422.0 6.4
1782737.0 6.6
1782852.0 7.1
1782912.0 6.8
1782972.0 6.6
1783032.0 5.7
1783212.0 5.8
1783380.0 5.8
1783446.0 7.2
1783518.0 6.6
1783716.0 6.1
1783908.0 5.8
1784195.0 6.5
1784810.0 7.4
1784957.0 6.6
1786921.0 7.2
1787671.0 6.3
1788261.0 5.7
1788275.0 6.0
1788392.0 5.7
1788502.0 7.1
1788612.0 7.3
1788732.0 6.5
1788912.0 7.4
1788972.0 6.0
1789032.0 7.5
1789100.0 7.0
1789277.0 6.4
1789315.0 6.1
1789482.0 6.2
1789583.0 5.9
1789683.0 6.5
1789769.0 6.1
1790052.0 6.8
1790172.0 5.6
1790352.0 7.5
1791100.0 7.3
1791287.0 5.8
1791552.0 5.8
1792012.0 7.2
1823112.0 7.2
1823292.0 5.7
1823472.0 6.4
1823532.0 6.1
1823700.0 7.3
1823903.0 6.3
1824002.0 7.4
1824129.0 6.9
1824180.0 6.7
1824332.0 7.2
1824380.0 7.1
1824552.0 6.7
1824612.0 5.8
1824672.0 7.1
1825032.0 6.0
1825184.0 6.3
1825645.0 5.5
1825992.0 7.5
1826052.0 7.3
1826292.0 6.0
1826352.0 5.9
1826412.0 5.7
1826472.0 7.0
1826581.0 7.5
1826614.0 6.6
1826767.0 5.8
1826939.0 6.1
1828467.0 6.8
1829436.0 6.1
1829753.0 6.5
1829796.0 6.4
1829989.0 5.9
1830033.0 6.4
1830183.0 6.3
1830256.0 7.4
1830312.0 5.7
1830372.0 5.7
1830612.0 5.9
1830672.0 7.2
1830910.0 6.5
1830945.0 5.7
1831102.0 7.0
1831204.0 5.7
1831299.0 6.6
1831613.0 6.2
1831745.0 5.6
1831932.0 5.7
1832578.0 6.1
1866303.0 5.7
1866372.0 5.9
1866432.0 7.0
1866492.0 7.3
1866612.0 6.6
1866732.0 5.8
1866996.0 6.5
1867212.0 6.4
1867268.0 6.2
1867546.0 5.9
1867568.0 6.6
1867812.0 7.4
1867872.0 5.5
1867932.0 6.0
1867992.0 6.4
1868052.0 7.2
1868112.0 6.5
1868172.0 6.4
1868232.0 7.3
1868346.0 6.5
1868471.0 7.1
1868530.0 5.8
1868576.0 7.0
1868605.0 7.2
1868657.0 5.9
1868700.0 6.3
1868748.0 6.6
1869040.0 6.5
1869046.0 6.5
1869076.0 5.9
1869089.0 6.3
1869252.0 6.1
1869372.0 7.4
1869492.0 7.3
1869552.0 7.4
1869746.0 6.9
1869957.0 5.5
1869974.0 6.2
1870192.0 6.5
1870197.0 5.7
1870309.0 7.2
1870339.0 7.1
1870372.0 6.6
1870471.0 6.1
1871221.0 6.2
1871570.0 6.1
1872192.0 6.8
1872252.0 7.4
1872312.0 7.5
1872372.0 6.5
1872432.0 7.1
1872492.0 5.8
1872552.0 6.8
1872734.0 6.6
1872739.0 6.3
1872819.0 7.5
1872887.0 7.1
1873049.0 6.8
1873220.0 5.7
1873380.0 5.8
1873398.0 6.7
1873419.0 7.0
1873812.0 7.4
1873992.0 7.0
1874082.0 7.4
1874226.0 5.9
1874241.0 7.1
1908014.0 7.4
1908132.0 6.1
1908252.0 6.6
1908492.0 7.0
1908552.0 7.1
1908676.0 7.3
1909090.0 7.3
1909142.0 7.3
1909195.0 6.5
1909318.0 7.5
1909413.0 5.8
1909512.0 7.4
1909572.0 5.6
1909632.0 5.5
1909812.0 6.7
1909872.0 6.4
1910127.0 7.4
1910345.0 6.8
1910422.0 5.7
1910458.0 6.9
1910544.0 6.4
1910569.0 5.8
1910661.0 6.0
1910788.0 5.8
1910803.0 6.1
1910914.0 6.8
1911012.0 5.6
1911132.0 7.1
1911252.0 7.4
1911312.0 5.8
1911630.0 6.3
1911790.0 6.3
1911837.0 7.0
1911955.0 7.4
1912054.0 7.1
1912089.0 7.0
1912192.0 6.2
1912287.0 5.7
1912512.0 5.9
1912812.0 6.1
1912872.0 6.6
1913130.0 5.5
1913262.0 6.9
1913952.0 7.1
1914427.0 7.3
1914545.0 5.7
1915009.0 7.0
1915140.0 7.0
1915692.0 6.7
1915905.0 5.7
1917254.0 6.6
1952832.0 7.0
1959182.0 5.6
1959293.0 6.0
1959358.0 7.0
1960032.0 5.7
1960212.0 5.9
1960480.0 6.5
1994442.0 5.6
1994652.0 7.4
1994952.0 5.8
1995265.0 5.6
1995497.0 5.6
1995582.0 7.1
1995612.0 7.4
1995692.0 6.2
1995750.0 6.9
1995882.0 6.8
1995912.0 6.9
1995972.0 7.3
1996152.0 7.0
1996272.0 6.9
1996332.0 5.6
1996392.0 6.3
1996453.0 5.7
1996597.0 7.5
1996750.0 5.9
1996801.0 6.8
1997139.0 5.9
1997224.0 6.6
1997255.0 5.8
1997412.0 6.1
1997592.0 6.1
1997832.0 5.5
1997949.0 6.1
1998150.0 6.6
1998217.0 7.2
1998268.0 6.0
1998432.0 5.9
1998627.0 7.3
1998912.0 6.2
1998972.0 7.0
1999434.0 7.0
2000023.0 6.2
2000170.0 5.7
2000472.0 5.5
2000532.0 5.6
2000652.0 6.0
2000932.0 6.2
2000984.0 6.2
2001322.0 6.7
2001972.0 6.6
2002545.0 7.4
2002932.0 7.4
2002984.0 6.8
2003110.0 5.7
2003532.0 7.1
2004003.0 6.0
2037450.0 7.4
2037972.0 6.9
2040732.0 6.0
2040912.0 6.0
2041760.0 6.0
2042112.0 7.3
2042412.0 7.0
2042739.0 6.6
2043432.0 6.9
2044807.0 5.7
2045494.0 6.0
2046082.0 5.6
2079404.0 6.0
2079492.0 6.5
2079612.0 6.7
2079732.0 7.2
2079792.0 6.1
2079852.0 6.4
2080066.0 6.4
2080124.0 6.0
2080360.0 5.7
2080647.0 6.3
2080684.0 6.4
2080780.0 6.7
2080992.0 6.9
2081112.0 7.4
2081172.0 7.4
2081232.0 6.4
2081352.0 5.7
2081480.0 5.5
2081552.0 5.9
2081617.0 7.3
2081688.0 6.3
2081900.0 6.9
2082612.0 6.6
2082857.0 7.5
2083196.0 7.1
2083932.0 6.1
2084788.0 6.5
2084872.0 7.0
2085552.0 6.3
2085907.0 6.9
2086449.0 7.1
2086632.0 7.4
2086752.0 7.2
2087052.0 6.9
2087932.0 5.6
2088702.0 5.6
2088737.0 6.4
2088863.0 6.1
2122623.0 7.0
2122632.0 6.5
2123300.0 7.3
2123617.0 7.0
2123995.0 7.5
2124739.0 7.3
2124777.0 5.6
2124993.0 6.5
2125752.0 5.6
2126107.0 7.2
2126123.0 6.5
2126377.0 6.8
2127132.0 6.8
2127252.0 7.3
2128072.0 6.9
2128752.0 6.6
2129159.0 7.4
2129560.0 6.2
2129832.0 7.1
2130012.0 6.2
2130252.0 6.4
2131572.0 7.0
2131902.0 6.9
2132243.0 6.0
2132410.0 7.0
2164392.0 6.8
2164452.0 6.8
2164692.0 6.2
2164872.0 6.6
2165032.0 6.3
2165362.0 5.6
2165486.0 6.3
2165570.0 7.3
2165581.0 5.8
2165592.0 5.7
2165722.0 7.5
2165892.0 5.7
2166072.0 7.4
2166132.0 6.7
2166252.0 6.8
2166312.0 7.0
2166431.0 5.7
2166530.0 6.7
2166544.0 5.7
2166926.0 5.5
2167045.0 6.6
2167151.0 6.9
2167192.0 7.3
2167272.0 6.0
2167332.0 7.2
2167512.0 7.2
2167572.0 6.1
2168454.0 5.5
2168602.0 5.7
2168832.0 7.0
2169012.0 5.7
2169340.0 6.6
2169385.0 6.0
2169589.0 5.9
2169670.0 6.7
2169720.0 6.7
2169833.0 7.0
2170033.0 6.3
2170212.0 7.0
2170332.0 5.7
2170703.0 6.4
2170899.0 6.8
2171124.0 7.4
2171140.0 6.4
2171371.0 6.4
2171652.0 6.1
2172072.0 7.2
2172332.0 6.1
2172589.0 6.4
2172724.0 7.4
2173032.0 5.8
2173609.0 7.2
2173661.0 6.2
2173708.0 7.1
2173717.0 7.0
2173881.0 6.7
2173943.0 5.7
2174276.0 7.3
2207590.0 6.9
2207652.0 6.2
2207712.0 6.4
2207772.0 6.6
2207892.0 5.9
2207952.0 6.3
2208012.0 6.6
2208072.0 6.4
2208242.0 7.0
2208288.0 6.7
2208322.0 5.8
2208464.0 6.4
2208527.0 6.7
2208559.0 6.3
2208715.0 7.1
2208842.0 6.8
2208877.0 6.1
2208995.0 5.7
2209272.0 6.1
2209452.0 6.1
2209665.0 6.5
2209686.0 6.4
2209858.0 6.2
2209897.0 6.6
2209994.0 7.1
2210110.0 6.1
2210153.0 6.9
2210351.0 7.3
2210532.0 5.9
2211050.0 6.2
2211088.0 6.2
2211156.0 6.8
2211364.0 6.7
2211446.0 6.9
2211473.0 6.1
2211485.0 7.2
2211516.0 6.7
2211586.0 6.9
2211670.0 6.6
2211697.0 7.4
2211717.0 6.9
2211745.0 7.3
2211783.0 6.4
2211805.0 6.7
2211857.0 7.3
2211912.0 6.2
2212032.0 6.2
2212092.0 6.2
2212152.0 7.5
2212212.0 6.2
2212332.0 7.0
2212473.0 7.0
2212630.0 7.5
2212641.0 7.4
2212811.0 5.8
2212992.0 7.3
2213162.0 6.2
2213258.0 6.7
2213284.0 6.2
2213349.0 7.0
2213352.0 6.8
2213472.0 5.8
2213712.0 7.0
2213832.0 6.4
2213905.0 6.1
2215272.0 6.4
2215687.0 5.9
2216900.0 7.4
2250650.0 6.2
2250792.0 7.4
2250912.0 7.0
2251092.0 7.2
2251383.0 7.4
2251662.0 5.7
2251703.0 5.5
2251912.0 7.4
2252012.0 6.3
2252080.0 7.3
2252352.0 6.6
2252412.0 7.1
2252960.0 5.8
2253012.0 6.1
2253172.0 6.8
2253197.0 7.0
2253390.0 6.8
2253672.0 7.2
2254889.0 5.9
2254987.0 6.5
2255030.0 6.0
2257192.0 5.7
2257260.0 6.9
2257802.0 7.1
2257809.0 7.0
2258112.0 5.5
2258172.0 6.2
2292612.0 5.8
2292672.0 6.8
2292792.0 7.1
2292852.0 6.6
2292972.0 7.2
2295113.0 6.4
2295161.0 6.1
2295320.0 6.5
2296047.0 6.3
2296992.0 6.2
2297352.0 6.2
2297481.0 6.6
2297517.0 7.4
2297535.0 6.3
2297606.0 7.1
2297705.0 5.5
2297789.0 5.5
2298053.0 5.8
2298236.0 6.2
2312772.0 6.9
2313132.0 7.5
2313264.0 7.4
2313325.0 7.1
2313827.0 6.2
2313970.0 6.9
2313985.0 6.2
2314092.0 6.6
2314332.0 6.6
2314715.0 6.6
2314755.0 6.0
2315172.0 5.8
2315280.0 7.4
2315592.0 5.8
2315772.0 7.3
2316072.0 7.2
2316472.0 5.6
2316485.0 6.4
2316622.0 6.5
2316722.0 7.4
2317092.0 6.1
2317332.0 6.9
2317452.0 7.3
2317670.0 6.8
2318652.0 6.5
2319328.0 5.6
2321276.0 6.7
2321712.0 7.1
2322702.0 6.5
2334372.0 6.2
2334552.0 6.4
2334792.0 6.8
2335066.0 6.0
2335086.0 6.1
2335136.0 5.9
2335465.0 6.7
2335736.0 6.6
2335812.0 6.4
2335992.0 5.7
2336112.0 7.0
2336172.0 6.4
2336329.0 7.5
2336372.0 5.6
2336640.0 6.9
2336700.0 6.9
2337612.0 6.6
2337786.0 6.8
2338054.0 7.3
2338364.0 6.2
2338456.0 6.6
2338525.0 5.9
2342425.0 6.1
2342522.0 6.2
2343192.0 5.5
2343252.0 5.5
2344053.0 6.2
2344297.0 6.6
2376252.0 7.0
2376492.0 6.1
2376959.0 6.0
2377632.0 6.6
2378952.0 6.3
2379577.0 6.5
2381245.0 6.4
2383064.0 6.3
2383882.0 7.4
2385542.0 7.3
2416632.0 6.7
2419752.0 5.6
2419895.0 6.0
2420174.0 7.2
2420299.0 6.7
2420477.0 7.1
2420712.0 7.0
2420772.0 7.0
2420832.0 7.3
2420952.0 6.9
2421132.0 6.6
2421480.0 6.6
2421579.0 7.3
2421924.0 5.8
2422009.0 7.2
2422087.0 7.3
2422212.0 6.3
2422392.0 5.8
2422452.0 6.1
2422572.0 5.7
2422766.0 6.8
2422891.0 6.9
2423035.0 7.5
2423066.0 6.4
2423509.0 6.1
2423772.0 6.3
2425092.0 5.6
2425212.0 7.0
2425651.0 7.2
2425691.0 6.0
2425803.0 5.8
2426472.0 6.7
2426532.0 6.0
2426592.0 5.7
2426892.0 6.6
2427028.0 6.6
2427466.0 5.9
2427516.0 6.8
2427912.0 7.5
2428092.0 7.0
2428814.0 6.0
2428897.0 7.3
2429124.0 6.0
2466648.0 6.4
2466852.0 6.7
2468010.0 7.2
2469043.0 6.7
2469407.0 7.0
2469517.0 6.5
2469732.0 7.3
2469792.0 7.1
2469852.0 7.0
2470538.0 6.6
2470922.0 6.8
2506944.0 6.3
2508543.0 6.3
2508672.0 5.7
2509182.0 7.3
2509365.0 7.4
2509974.0 6.8
2510112.0 7.4
2510633.0 7.3
2511121.0 6.2
2511432.0 6.4
2511492.0 7.3
2511912.0 6.7
2512005.0 6.6
2512570.0 7.3
2513172.0 5.9
2546232.0 6.8
2547322.0 7.2
2547432.0 6.8
2547852.0 7.4
2548592.0 5.9
2550189.0 7.4
2551176.0 7.3
2552519.0 5.6
2552742.0 5.6
2553846.0 5.7
2554007.0 7.1
2554752.0 5.8
2589252.0 7.1
2589432.0 5.6
2590142.0 7.0
2590206.0 6.6
2590632.0 7.0
2592707.0 7.2
2592725.0 5.7
2594680.0 5.9
2595252.0 5.5
2632452.0 7.2
2632872.0 6.3
2634012.0 6.9
2635332.0 5.8
2636003.0 5.6
2636325.0 7.5
2637192.0 5.6
2637438.0 6.4
2639058.0 6.8
2640914.0 7.3
2640959.0 5.5
2641512.0 7.5
2641607.0 6.3
2641993.0 6.8
2674572.0 6.2
2674806.0 6.6
2675194.0 7.2
2675560.0 6.4
2676072.0 7.4
2676384.0 7.1
2677092.0 6.4
2677152.0 6.6
2677392.0 6.8

| | | | | | | |
|---|---|---|---|---|---|---|
| 2677814.0 7.0 | 3021252.0 6.8 | 3446932.0 6.0 | 3571692.0 7.1 | 3823452.0 5.6 | 3997185.0 6.6 | 4082221.0 5.7 |
| 2678592.0 6.9 | 3021372.0 7.0 | 3446947.0 6.6 | 3571863.0 6.4 | 3823692.0 6.7 | 3997325.0 6.7 | 4082376.0 5.5 |
| 2680092.0 7.5 | 3021552.0 6.0 | 3446977.0 7.5 | 3572305.0 5.8 | 3823997.0 6.2 | 3997432.0 6.6 | 4082532.0 7.1 |
| 2680489.0 6.5 | 3021970.0 7.2 | 3447055.0 7.3 | 3572952.0 6.3 | 3824682.0 6.2 | 3997470.0 6.8 | 4082652.0 7.1 |
| 2680733.0 6.2 | 3057672.0 5.7 | 3447236.0 6.7 | 3573072.0 6.0 | 3825012.0 6.4 | 3997872.0 7.3 | 4082712.0 6.8 |
| 2680988.0 6.2 | 3058087.0 6.3 | 3447338.0 6.7 | 3573450.0 5.9 | 3825192.0 5.8 | 3997932.0 6.7 | 4082832.0 7.4 |
| 2683092.0 5.9 | 3058452.0 6.4 | 3447432.0 6.4 | 3574512.0 7.2 | 3825270.0 7.0 | 3998070.0 6.0 | 4082892.0 7.4 |
| 2683212.0 5.8 | 3059768.0 6.0 | 3447732.0 7.2 | 3576072.0 5.6 | 3825506.0 5.9 | 3998093.0 6.9 | 4082952.0 6.2 |
| 2684077.0 5.8 | 3059848.0 7.1 | 3448419.0 6.5 | 3576288.0 6.9 | 3825783.0 6.6 | 3998360.0 5.6 | 4083046.0 7.5 |
| 2717352.0 6.3 | 3060739.0 5.8 | 3448534.0 7.2 | 3610106.0 5.8 | 3826733.0 6.7 | 3998367.0 7.3 | 4083194.0 6.8 |
| 2717832.0 5.8 | 3060852.0 6.3 | 3448796.0 5.8 | 3610120.0 6.6 | 3826871.0 7.1 | 3998497.0 6.2 | 4083274.0 6.6 |
| 2719032.0 5.6 | 3060900.0 6.9 | 3448932.0 6.7 | 3610332.0 6.9 | 3827832.0 6.9 | 3998534.0 7.0 | 4083439.0 6.4 |
| 2719643.0 6.4 | 3061512.0 5.8 | 3449052.0 6.4 | 3610452.0 6.5 | 3828236.0 7.5 | 3998740.0 6.4 | 4083470.0 6.6 |
| 2720210.0 6.5 | 3065527.0 6.2 | 3449172.0 6.4 | 3610512.0 6.6 | 3829092.0 7.1 | 3998805.0 6.9 | 4083554.0 5.6 |
| 2721450.0 5.8 | 3065744.0 5.8 | 3449449.0 5.8 | 3610632.0 7.3 | 3829392.0 5.8 | 3999903.0 7.4 | 4083873.0 6.3 |
| 2722498.0 6.3 | 3099252.0 5.8 | 3449623.0 6.6 | 3611712.0 6.7 | 3829762.0 6.6 | 4000198.0 7.0 | 4083972.0 5.8 |
| 2722590.0 7.1 | 3099432.0 6.9 | 3449807.0 7.3 | 3612012.0 7.4 | 3864999.0 6.9 | 4000323.0 7.4 | 4084392.0 6.2 |
| 2722943.0 6.1 | 3100692.0 6.8 | 3483432.0 7.2 | 3612376.0 6.3 | 3865017.0 5.8 | 4000452.0 5.5 | 4118436.0 6.4 |
| 2723112.0 6.2 | 3102422.0 6.9 | 3483612.0 5.9 | 3613032.0 6.6 | 3865032.0 6.9 | 4001424.0 7.5 | 4118472.0 5.8 |
| 2723232.0 5.8 | 3104532.0 7.1 | 3483792.0 7.0 | 3613332.0 6.8 | 3865212.0 7.0 | 4001600.0 6.5 | 4118892.0 7.5 |
| 2725032.0 6.8 | 3105012.0 6.8 | 3483993.0 5.7 | 3613452.0 5.7 | 3865392.0 7.0 | 4001627.0 6.6 | 4119108.0 7.1 |
| 2725163.0 6.5 | 3106272.0 6.6 | 3484050.0 5.6 | 3613646.0 6.4 | 3865452.0 5.7 | 4034918.0 5.9 | 4119341.0 7.3 |
| 2725744.0 6.9 | 3106972.0 6.3 | 3484172.0 7.5 | 3613829.0 6.5 | 3865782.0 6.9 | 4034927.0 5.9 | 4120032.0 6.4 |
| 2726800.0 6.2 | 3144442.0 7.0 | 3484437.0 5.5 | 3614712.0 6.3 | 3865838.0 7.4 | 4034952.0 7.5 | 4120152.0 7.0 |
| 2761361.0 6.9 | 3187559.0 6.7 | 3484638.0 7.4 | 3614892.0 5.7 | 3866093.0 6.5 | 4035072.0 6.9 | 4120272.0 6.7 |
| 2761440.0 7.0 | 3189253.0 6.5 | 3484764.0 6.1 | 3614952.0 7.1 | 3866268.0 6.6 | 4035192.0 6.2 | 4120392.0 5.8 |
| 2761504.0 7.3 | 3190562.0 5.8 | 3485232.0 7.4 | 3615018.0 6.9 | 3866283.0 6.1 | 4035252.0 6.7 | 4120502.0 5.9 |
| 2761557.0 5.9 | 3193092.0 6.3 | 3485352.0 7.3 | 3615318.0 7.2 | 3866433.0 6.0 | 4035312.0 6.7 | 4120703.0 6.2 |
| 2761992.0 5.9 | 3227767.0 5.8 | 3485479.0 7.3 | 3616480.0 5.6 | 3866472.0 5.8 | 4035432.0 6.9 | 4121532.0 5.5 |
| 2762232.0 6.9 | 3232160.0 7.2 | 3485647.0 6.0 | 3616619.0 5.5 | 3866532.0 6.2 | 4035632.0 6.9 | 4121924.0 7.4 |
| 2762352.0 6.2 | 3233232.0 7.0 | 3485702.0 6.8 | 3617652.0 7.0 | 3866952.0 7.2 | 4035680.0 6.8 | 4121985.0 7.0 |
| 2763612.0 6.2 | 3233352.0 6.8 | 3486162.0 5.5 | 3619429.0 7.5 | 3867580.0 6.8 | 4035691.0 7.3 | 4122792.0 6.2 |
| 2764055.0 6.2 | 3233693.0 6.5 | 3486312.0 6.4 | 3652272.0 6.6 | 3868392.0 7.0 | 4035799.0 5.8 | 4123936.0 7.4 |
| 2764082.0 6.9 | 3234168.0 7.1 | 3486612.0 6.0 | 3653178.0 5.8 | 3868482.0 5.9 | 4035864.0 5.6 | 4123998.0 5.6 |
| 2764447.0 6.2 | 3234213.0 7.3 | 3487006.0 5.9 | 3653352.0 6.1 | 3869198.0 7.0 | 4035952.0 6.7 | 4124292.0 7.3 |
| 2764992.0 7.4 | 3234792.0 7.2 | 3487069.0 6.2 | 3653472.0 7.4 | 3869352.0 5.7 | 4035977.0 6.4 | 4124896.0 7.4 |
| 2765112.0 6.8 | 3234936.0 6.6 | 3487109.0 6.0 | 3653772.0 6.9 | 3869832.0 6.9 | 4036005.0 6.6 | 4127862.0 6.2 |
| 2765352.0 7.1 | 3235042.0 6.5 | 3487262.0 6.2 | 3654177.0 6.7 | 3869984.0 6.3 | 4036113.0 6.5 | 4162373.0 6.2 |
| 2765607.0 7.3 | 3235227.0 6.9 | 3487343.0 7.2 | 3654195.0 7.1 | 3870792.0 5.9 | 4036149.0 5.5 | 4163078.0 7.4 |
| 2766672.0 6.2 | 3235514.0 6.5 | 3487453.0 6.4 | 3654619.0 7.1 | 3870912.0 7.2 | 4036254.0 6.9 | 4163532.0 6.6 |
| 2769038.0 6.9 | 3270372.0 6.0 | 3487697.0 6.2 | 3655032.0 7.3 | 3871662.0 6.2 | 4036260.0 6.5 | 4163592.0 6.9 |
| 2769904.0 5.8 | 3270552.0 7.0 | 3487752.0 7.4 | 3655387.0 6.4 | 3871960.0 7.2 | 4036335.0 6.0 | 4163764.0 7.1 |
| 2770115.0 5.7 | 3270732.0 6.0 | 3487872.0 7.0 | 3658870.0 5.9 | 3872799.0 6.0 | 4036392.0 7.2 | 4163914.0 5.7 |
| 2803639.0 6.1 | 3271229.0 6.9 | 3487992.0 7.2 | 3659352.0 5.6 | 3873634.0 7.4 | 4036452.0 7.2 | 4168639.0 6.7 |
| 2804172.0 6.3 | 3271447.0 6.2 | 3488172.0 6.7 | 3660481.0 6.8 | 3874984.0 6.1 | 4036572.0 5.9 | 4168845.0 5.6 |
| 2804375.0 7.0 | 3272172.0 5.8 | 3488343.0 6.3 | 3693732.0 5.9 | 3906771.0 7.3 | 4036692.0 6.8 | 4169672.0 6.1 |
| 2804454.0 5.8 | 3272361.0 5.6 | 3488443.0 6.4 | 3695008.0 7.1 | 3906852.0 6.0 | 4036752.0 7.0 | 4169877.0 5.7 |
| 2806219.0 7.3 | 3272786.0 7.3 | 3488577.0 5.7 | 3695232.0 6.7 | 3906912.0 6.2 | 4036984.0 7.4 | 4169904.0 6.8 |
| 2808252.0 6.8 | 3273252.0 7.2 | 3488751.0 6.6 | 3695472.0 6.5 | 3906972.0 5.9 | 4037007.0 7.1 | 4170612.0 6.0 |
| 2808995.0 6.5 | 3273432.0 5.7 | 3488904.0 6.5 | 3695837.0 6.8 | 3907272.0 5.8 | 4037044.0 5.7 | 4170910.0 7.0 |
| 2809104.0 6.3 | 3273492.0 5.8 | 3488983.0 6.8 | 3696162.0 6.5 | 3907451.0 6.1 | 4037173.0 7.3 | 4171161.0 7.3 |
| 2845812.0 6.9 | 3273672.0 7.4 | 3489192.0 6.0 | 3696525.0 6.9 | 3907466.0 6.4 | 4037278.0 5.6 | 4171508.0 6.3 |
| 2847192.0 7.1 | 3274044.0 5.8 | 3489312.0 5.9 | 3697297.0 6.3 | 3907530.0 7.1 | 4037445.0 7.0 | 4171585.0 7.3 |
| 2847372.0 6.3 | 3274054.0 5.9 | 3489612.0 6.3 | 3700195.0 6.2 | 3908037.0 7.3 | 4037510.0 7.1 | 4171657.0 5.6 |
| 2848217.0 6.4 | 3274354.0 7.3 | 3489912.0 6.4 | 3700337.0 6.2 | 3908292.0 5.5 | 4037532.0 6.9 | 4205928.0 6.7 |
| 2848452.0 6.4 | 3274413.0 7.1 | 3490255.0 5.7 | 3700932.0 7.3 | 3909092.0 5.7 | 4037832.0 6.7 | 4206060.0 5.8 |
| 2848632.0 5.5 | 3274992.0 5.7 | 3490453.0 6.1 | 3701875.0 7.3 | 3909912.0 5.6 | 4037952.0 6.5 | 4207338.0 6.5 |
| 2850012.0 6.9 | 3275342.0 6.6 | 3491302.0 6.1 | 3701979.0 6.2 | 3910092.0 7.0 | 4038132.0 7.5 | 4207484.0 6.0 |
| 2851001.0 7.3 | 3275523.0 7.1 | 3492252.0 6.0 | 3702074.0 7.0 | 3910264.0 7.3 | 4038580.0 5.6 | 4207752.0 7.3 |
| 2887272.0 6.8 | 3275894.0 6.9 | 3525192.0 7.1 | 3702672.0 6.0 | 3910444.0 6.1 | 4038808.0 7.4 | 4208301.0 5.6 |
| 2887392.0 7.4 | 3276025.0 5.7 | 3525252.0 6.2 | 3702732.0 5.6 | 3911412.0 5.5 | 4039017.0 5.6 | 4208945.0 7.3 |
| 2887632.0 6.0 | 3276039.0 7.3 | 3525312.0 5.8 | 3702862.0 6.3 | 3912014.0 6.9 | 4039182.0 7.1 | 4209009.0 5.8 |
| 2887857.0 5.6 | 3276492.0 5.9 | 3525552.0 6.3 | 3703205.0 6.6 | 3912552.0 7.1 | 4039332.0 5.6 | 4209182.0 6.9 |
| 2887878.0 6.5 | 3277046.0 5.8 | 3525740.0 6.6 | 3703399.0 7.2 | 3913847.0 5.8 | 4039632.0 7.4 | 4211452.0 6.7 |
| 2888224.0 7.4 | 3278494.0 7.2 | 3525816.0 7.0 | 3703598.0 5.6 | 3914642.0 5.8 | 4039844.0 7.2 | 4211782.0 5.7 |
| 2888247.0 5.6 | 3278591.0 6.6 | 3525827.0 7.1 | 3703690.0 6.1 | 3914835.0 7.3 | 4040039.0 7.4 | 4211843.0 6.2 |
| 2888772.0 6.8 | 3278659.0 6.9 | 3525903.0 6.3 | 3737475.0 6.4 | 3916015.0 5.7 | 4040152.0 7.1 | 4211899.0 7.2 |
| 2889072.0 6.5 | 3279432.0 6.9 | 3526089.0 6.6 | 3737999.0 6.7 | 3916165.0 5.5 | 4040531.0 6.7 | 4212844.0 7.3 |
| 2889344.0 6.7 | 3312252.0 5.6 | 3526097.0 6.9 | 3738612.0 7.2 | 3916424.0 5.7 | 4040772.0 5.6 | 4212882.0 7.5 |
| 2889673.0 7.1 | 3312666.0 7.4 | 3526132.0 7.1 | 3738860.0 5.6 | 3916607.0 6.4 | 4041072.0 6.2 | 4213119.0 6.9 |
| 2890152.0 7.0 | 3312748.0 6.2 | 3526503.0 6.5 | 3738861.0 7.3 | 3950581.0 6.3 | 4041257.0 6.9 | 4213217.0 6.3 |
| 2890726.0 6.5 | 3313087.0 6.1 | 3526517.0 6.4 | 3739751.0 6.4 | 3950886.0 6.3 | 4041839.0 6.9 | 4248132.0 6.4 |
| 2891056.0 7.2 | 3313572.0 5.9 | 3526812.0 5.6 | 3739932.0 6.2 | 3952290.0 7.5 | 4041884.0 6.6 | 4248192.0 7.2 |
| 2891307.0 7.4 | 3313632.0 7.4 | 3526992.0 7.4 | 3740172.0 6.3 | 3952438.0 6.8 | 4042332.0 6.3 | 4248372.0 7.4 |
| 2891652.0 5.6 | 3313692.0 6.6 | 3527112.0 7.3 | 3740698.0 7.2 | 3952647.0 5.8 | 4042452.0 5.8 | 4248432.0 6.6 |
| 2892012.0 5.7 | 3314308.0 7.4 | 3527229.0 5.8 | 3741552.0 6.3 | 3952692.0 6.1 | 4043104.0 6.2 | 4248492.0 7.3 |
| 2894532.0 6.0 | 3314687.0 6.5 | 3527390.0 6.5 | 3741672.0 6.0 | 3954672.0 7.0 | 4043442.0 6.4 | 4248687.0 7.2 |
| 2896152.0 5.9 | 3314881.0 6.5 | 3527401.0 5.8 | 3741753.0 6.8 | 3954792.0 7.1 | 4044072.0 6.8 | 4249012.0 7.5 |
| 2896983.0 7.2 | 3315192.0 5.5 | 3527429.0 6.1 | 3742173.0 6.8 | 3955424.0 6.3 | 4044602.0 6.3 | 4249027.0 5.5 |
| 2897240.0 7.2 | 3315252.0 6.2 | 3527653.0 6.9 | 3742494.0 7.5 | 3955552.0 6.7 | 4076670.0 6.5 | 4249077.0 5.8 |
| 2930532.0 6.7 | 3315312.0 6.3 | 3527694.0 5.8 | 3742569.0 6.3 | 3955932.0 5.5 | 4076678.0 6.8 | 4249482.0 6.7 |
| 2930892.0 5.7 | 3315432.0 6.7 | 3527768.0 7.5 | 3742620.0 6.6 | 3956112.0 6.0 | 4076697.0 7.3 | 4249512.0 7.4 |
| 2931047.0 5.9 | 3320079.0 5.6 | 3527897.0 5.7 | 3742692.0 6.3 | 3956358.0 6.3 | 4076712.0 6.8 | 4249572.0 6.7 |
| 2931690.0 6.3 | 3322070.0 5.8 | 3527959.0 7.2 | 3742872.0 7.4 | 3956572.0 5.8 | 4076772.0 6.5 | 4249752.0 7.3 |
| 2931818.0 6.7 | 3355572.0 6.5 | 3528033.0 6.1 | 3743547.0 6.6 | 3957017.0 6.5 | 4076832.0 6.9 | 4250057.0 5.7 |
| 2932332.0 7.1 | 3355632.0 5.7 | 3528072.0 6.8 | 3744132.0 6.6 | 3957672.0 5.6 | 4076892.0 6.2 | 4250120.0 5.7 |
| 2934027.0 7.4 | 3356130.0 5.8 | 3528312.0 6.0 | 3745449.0 6.4 | 3958135.0 6.1 | 4077072.0 6.8 | 4250622.0 6.8 |
| 2934228.0 7.1 | 3356705.0 6.7 | 3528492.0 5.9 | 3745512.0 5.8 | 3958500.0 6.0 | 4077132.0 6.5 | 4250849.0 7.5 |
| 2934320.0 6.4 | 3356832.0 7.2 | 3528648.0 6.4 | 3745572.0 6.7 | 3958556.0 5.5 | 4077192.0 7.5 | 4251372.0 5.7 |
| 2935032.0 7.1 | 3356892.0 6.4 | 3528734.0 6.1 | 3745692.0 5.8 | 3958752.0 5.6 | 4077427.0 6.2 | 4251801.0 7.1 |
| 2936232.0 7.4 | 3357132.0 6.7 | 3528976.0 6.3 | 3745812.0 5.7 | 3958932.0 7.3 | 4077448.0 6.8 | 4252132.0 7.0 |
| 2936352.0 7.0 | 3357434.0 6.8 | 3529174.0 7.1 | 3780132.0 5.6 | 3958992.0 6.6 | 4077461.0 6.9 | 4252167.0 6.5 |
| 2936532.0 6.0 | 3357668.0 7.2 | 3529216.0 7.3 | 3780492.0 6.0 | 3959644.0 6.9 | 4077685.0 7.4 | 4252392.0 6.6 |
| 2936652.0 7.2 | 3358855.0 5.5 | 3529572.0 5.7 | 3780789.0 6.2 | 3959678.0 7.3 | 4077709.0 5.9 | 4252632.0 7.2 |
| 2936837.0 6.9 | 3359712.0 6.7 | 3529932.0 6.4 | 3781192.0 6.9 | 3959721.0 6.3 | 4077772.0 6.2 | 4252752.0 7.4 |
| 2936967.0 7.4 | 3360362.0 7.1 | 3530078.0 6.1 | 3781254.0 6.3 | 3959872.0 6.1 | 4077874.0 6.4 | 4253139.0 6.8 |
| 2937657.0 6.5 | 3360595.0 6.3 | 3530120.0 5.8 | 3781317.0 5.9 | 3991872.0 5.7 | 4077881.0 6.5 | 4253339.0 6.4 |
| 2937672.0 6.4 | 3360809.0 6.8 | 3530374.0 5.8 | 3781375.0 6.6 | 3991932.0 6.6 | 4077886.0 6.0 | 4254012.0 6.7 |
| 2938562.0 5.6 | 3361092.0 6.7 | 3530493.0 6.1 | 3781454.0 6.5 | 3992052.0 6.1 | 4078010.0 6.7 | 4254732.0 6.2 |
| 2940273.0 7.3 | 3362072.0 6.0 | 3530506.0 6.4 | 3781512.0 5.6 | 3992112.0 5.9 | 4078023.0 7.2 | 4288367.0 7.2 |
| 2973672.0 5.8 | 3362532.0 6.9 | 3530657.0 7.2 | 3781872.0 5.9 | 3992172.0 6.9 | 4078128.0 7.5 | 4288512.0 6.1 |
| 2974425.0 6.7 | 3362832.0 5.7 | 3530709.0 6.8 | 3782426.0 7.5 | 3992403.0 5.9 | 4078152.0 5.8 | 4289100.0 7.5 |
| 2976379.0 5.8 | 3363012.0 6.0 | 3531072.0 6.2 | 3783432.0 7.2 | 3992432.0 6.8 | 4078212.0 6.2 | 4289237.0 7.5 |
| 2976399.0 6.9 | 3363336.0 6.8 | 3531252.0 5.8 | 3784177.0 5.7 | 3992771.0 6.3 | 4078332.0 5.6 | 4289515.0 7.0 |
| 2977032.0 7.2 | 3363893.0 7.0 | 3531666.0 6.5 | 3784290.0 6.1 | 3993552.0 6.1 | 4078512.0 6.7 | 4289892.0 6.3 |
| 2977283.0 6.1 | 3364888.0 5.6 | 3531734.0 7.3 | 3784392.0 7.3 | 3993957.0 6.5 | 4078572.0 7.4 | 4290495.0 6.2 |
| 2977620.0 6.9 | 3364898.0 7.3 | 3531856.0 6.4 | 3784452.0 6.0 | 3994135.0 6.8 | 4078774.0 7.0 | 4291632.0 5.9 |
| 2978052.0 6.2 | 3364978.0 6.5 | 3532330.0 5.9 | 3785017.0 7.4 | 3994872.0 6.2 | 4078979.0 5.9 | 4291752.0 6.7 |
| 2978847.0 6.4 | 3365232.0 7.2 | 3532392.0 6.5 | 3785134.0 6.4 | 3995112.0 5.8 | 4079154.0 6.8 | 4292376.0 6.3 |
| 2979552.0 7.2 | 3398652.0 6.0 | 3534132.0 7.3 | 3785341.0 5.6 | 3995262.0 6.5 | 4079178.0 5.6 | 4292650.0 6.3 |
| 3013992.0 6.7 | 3399225.0 6.5 | 3567012.0 7.4 | 3785398.0 5.6 | 3995313.0 7.1 | 4079345.0 6.3 | 4293291.0 7.5 |
| 3014620.0 5.8 | 3400092.0 6.6 | 3567709.0 5.8 | 3786192.0 7.3 | 3995382.0 6.3 | 4079513.0 7.3 | 4294122.0 6.3 |
| 3015032.0 5.7 | 3402792.0 7.3 | 3567723.0 6.2 | 3786727.0 5.8 | 3995524.0 5.5 | 4079652.0 6.1 | 4294212.0 5.7 |
| 3015224.0 6.1 | 3406595.0 6.0 | 3567935.0 6.1 | 3786772.0 6.9 | 3995719.0 6.7 | 4079712.0 6.4 | 4294572.0 7.2 |
| 3015852.0 7.3 | 3407352.0 7.3 | 3568307.0 6.0 | 3821775.0 5.5 | 3995858.0 6.9 | 4079832.0 6.2 | 4296625.0 7.4 |
| 3015912.0 7.4 | 3407592.0 7.2 | 3568572.0 7.1 | 3821832.0 6.7 | 3995915.0 5.7 | 4079892.0 7.3 | 4297512.0 5.9 |
| 3016087.0 5.9 | 3407687.0 6.7 | 3568752.0 6.6 | 3821952.0 6.3 | 3996192.0 7.2 | 4079952.0 6.2 | |
| 3016516.0 7.4 | 3408486.0 5.9 | 3568812.0 6.2 | 3822072.0 5.5 | 3996372.0 6.9 | 4080175.0 6.1 | |
| 3016625.0 6.5 | 3440230.0 6.3 | 3569166.0 7.5 | 3822132.0 6.9 | 3996492.0 7.2 | 4080196.0 5.9 | |
| 3017112.0 5.5 | 3440532.0 5.7 | 3569353.0 6.4 | 3822252.0 7.4 | 3996552.0 5.8 | 4080841.0 7.1 | |
| 3017292.0 7.1 | 3443532.0 6.8 | 3569678.0 5.5 | 3822761.0 6.0 | 3996628.0 7.1 | 4080975.0 6.8 | |
| 3018792.0 5.9 | 3446112.0 6.4 | 3569699.0 7.3 | 3822990.0 6.7 | 3996669.0 6.4 | 4081332.0 7.3 | |
| 3018901.0 6.3 | 3446292.0 6.6 | 3569832.0 6.0 | 3823011.0 7.0 | 3996704.0 5.8 | 4081512.0 6.5 | |
| 3019112.0 7.2 | 3446645.0 6.5 | 3570132.0 7.3 | 3823025.0 5.8 | 3996714.0 7.4 | 4081625.0 7.1 | |
| 3019142.0 6.7 | 3446679.0 5.8 | 3570739.0 7.2 | 3823178.0 7.5 | 3996748.0 7.1 | 4081668.0 7.5 | |
| 3020172.0 6.7 | 3446703.0 7.5 | 3571195.0 6.0 | 3823272.0 7.4 | 3996863.0 6.8 | 4081887.0 7.4 | |
| 3021160.0 5.5 | 3446930.0 6.1 | 3571572.0 7.3 | 3823332.0 6.9 | 3996990.0 7.3 | 4081975.0 7.4 | |

Table A-2: 0-reduced time series of 2865 moonquakes from Table A-1 that occurred during the Moon's traversals of the Earth's magnetotail, i.e., within three days from the full Moon.

| T(min) | Mag' |
|---|---|
| 37461.0 | 5.6 |
| 175872.0 | 6.8 |
| 177312.0 | 7.0 |
| 179112.0 | 6.6 |
| 179287.0 | 6.3 |
| 180192.0 | 6.7 |
| 180372.0 | 6.5 |
| 180720.0 | 7.4 |
| 181167.0 | 7.2 |
| 181512.0 | 6.9 |
| 181812.0 | 7.4 |
| 182104.0 | 5.7 |
| 182896.0 | 6.7 |
| 183012.0 | 5.8 |
| 188261.0 | 7.1 |
| 189885.0 | 5.5 |
| 192012.0 | 6.0 |
| 194115.0 | 5.6 |
| 195295.0 | 5.6 |
| 195704.0 | 5.7 |
| 195972.0 | 5.5 |
| 196092.0 | 7.2 |
| 196507.0 | 5.9 |
| 196997.0 | 6.7 |
| 197352.0 | 6.0 |
| 197772.0 | 7.5 |
| 198418.0 | 6.0 |
| 198912.0 | 6.5 |
| 199152.0 | 6.2 |
| 199629.0 | 5.8 |
| 199866.0 | 6.2 |
| 200081.0 | 6.4 |
| 200292.0 | 6.2 |
| 200412.0 | 6.9 |
| 201510.0 | 7.0 |
| 201596.0 | 5.8 |
| 202414.0 | 7.4 |
| 206112.0 | 7.4 |
| 219432.0 | 6.2 |
| 219712.0 | 5.9 |
| 220452.0 | 6.4 |
| 220752.0 | 5.9 |
| 220812.0 | 7.2 |
| 221019.0 | 6.2 |
| 221892.0 | 7.5 |
| 223127.0 | 5.6 |
| 226152.0 | 5.7 |
| 228810.0 | 7.2 |
| 229512.0 | 7.0 |
| 229657.0 | 6.6 |
| 229734.0 | 6.3 |
| 231092.0 | 7.4 |
| 231126.0 | 6.9 |
| 231375.0 | 5.9 |
| 231677.0 | 5.5 |
| 233472.0 | 5.9 |
| 234100.0 | 7.1 |
| 234610.0 | 6.2 |
| 234852.0 | 7.1 |
| 235272.0 | 6.0 |
| 235400.0 | 7.2 |
| 236221.0 | 5.7 |
| 236412.0 | 7.5 |
| 236941.0 | 6.2 |
| 237172.0 | 7.1 |
| 237672.0 | 7.0 |
| 237732.0 | 7.5 |
| 237972.0 | 5.7 |
| 238032.0 | 6.6 |
| 239292.0 | 7.4 |
| 239412.0 | 6.0 |
| 240438.0 | 6.1 |
| 240612.0 | 6.7 |
| 241967.0 | 5.9 |
| 242693.0 | 7.4 |
| 246492.0 | 6.8 |
| 262794.0 | 5.8 |
| 262979.0 | 6.3 |
| 266429.0 | 7.1 |
| 267430.0 | 5.9 |
| 268212.0 | 6.8 |
| 270702.0 | 5.8 |
| 270912.0 | 6.5 |
| 270972.0 | 7.2 |
| 272292.0 | 7.1 |
| 272831.0 | 6.3 |
| 273042.0 | 7.1 |
| 273092.0 | 7.1 |
| 273429.0 | 5.9 |
| 274727.0 | 6.3 |
| 274841.0 | 6.1 |
| 276181.0 | 5.9 |
| 276292.0 | 5.8 |
| 276732.0 | 6.5 |
| 276972.0 | 6.0 |
| 277364.0 | 7.4 |
| 277479.0 | 6.1 |
| 278412.0 | 6.5 |
| 279432.0 | 5.7 |
| 279552.0 | 6.4 |
| 280191.0 | 5.5 |
| 282732.0 | 6.4 |
| 284112.0 | 6.9 |
| 287866.0 | 6.1 |
| 292973.0 | 6.1 |
| 305652.0 | 6.6 |
| 307723.0 | 6.7 |
| 312160.0 | 7.3 |
| 312185.0 | 5.8 |
| 312552.0 | 6.0 |
| 312732.0 | 7.4 |
| 312972.0 | 6.6 |
| 313151.0 | 6.4 |
| 313413.0 | 5.7 |
| 313816.0 | 6.4 |
| 314412.0 | 6.6 |
| 314472.0 | 6.3 |
| 314740.0 | 7.1 |
| 314864.0 | 6.7 |
| 315612.0 | 6.9 |
| 317834.0 | 7.4 |
| 317962.0 | 6.7 |
| 318237.0 | 6.6 |
| 318312.0 | 6.0 |
| 318732.0 | 7.1 |
| 319546.0 | 7.2 |
| 320438.0 | 6.6 |
| 321067.0 | 5.9 |
| 326649.0 | 6.9 |
| 331752.0 | 6.5 |
| 332331.0 | 6.2 |
| 333648.0 | 7.4 |
| 334152.0 | 5.5 |
| 336012.0 | 7.3 |
| 338233.0 | 6.2 |
| 352447.0 | 5.9 |
| 352660.0 | 6.1 |
| 352732.0 | 7.3 |
| 353447.0 | 7.1 |
| 353529.0 | 7.1 |
| 353583.0 | 5.5 |
| 353872.0 | 5.9 |
| 354120.0 | 6.3 |
| 354219.0 | 5.5 |
| 354930.0 | 5.9 |
| 355251.0 | 5.7 |
| 355720.0 | 7.0 |
| 355752.0 | 5.6 |
| 355872.0 | 6.7 |
| 356052.0 | 5.8 |
| 356172.0 | 6.4 |
| 356232.0 | 7.0 |
| 356415.0 | 7.3 |
| 358155.0 | 6.5 |
| 359221.0 | 6.3 |
| 361265.0 | 5.8 |
| 362879.0 | 7.3 |
| 363562.0 | 5.8 |
| 369317.0 | 5.8 |
| 371530.0 | 6.5 |
| 371652.0 | 6.7 |
| 371892.0 | 6.5 |
| 371952.0 | 5.7 |
| 373854.0 | 6.5 |
| 373937.0 | 6.7 |
| 374242.0 | 6.2 |
| 374472.0 | 6.1 |
| 375054.0 | 6.4 |
| 376587.0 | 6.7 |
| 377532.0 | 6.6 |
| 377987.0 | 5.9 |
| 378792.0 | 6.3 |
| 379092.0 | 6.8 |
| 379212.0 | 6.2 |
| 379613.0 | 7.4 |
| 379840.0 | 5.5 |
| 391872.0 | 5.5 |
| 392112.0 | 5.5 |
| 392172.0 | 7.3 |
| 392301.0 | 6.8 |
| 392561.0 | 7.0 |
| 392604.0 | 5.9 |
| 392933.0 | 7.3 |
| 393372.0 | 7.2 |
| 393734.0 | 6.8 |
| 394427.0 | 6.1 |
| 394692.0 | 7.1 |
| 395220.0 | 6.7 |
| 395425.0 | 6.2 |
| 396252.0 | 5.5 |
| 397128.0 | 5.7 |
| 397572.0 | 6.6 |
| 397812.0 | 7.3 |
| 399072.0 | 5.8 |
| 399252.0 | 5.5 |
| 399372.0 | 6.7 |
| 402372.0 | 5.7 |
| 402462.0 | 6.1 |
| 405400.0 | 7.3 |
| 408248.0 | 5.7 |
| 411377.0 | 6.4 |
| 412876.0 | 7.0 |
| 413352.0 | 6.7 |
| 413712.0 | 6.3 |
| 413772.0 | 7.1 |
| 414912.0 | 6.4 |
| 415534.0 | 5.7 |
| 415700.0 | 7.2 |
| 416878.0 | 7.1 |
| 417012.0 | 7.2 |
| 417792.0 | 6.9 |
| 417912.0 | 5.9 |
| 418152.0 | 7.5 |
| 418902.0 | 6.7 |
| 419292.0 | 5.7 |
| 419412.0 | 5.8 |
| 419759.0 | 7.2 |
| 420852.0 | 6.8 |
| 421833.0 | 6.0 |
| 422552.0 | 6.6 |
| 434184.0 | 6.0 |
| 436512.0 | 5.7 |
| 436632.0 | 5.6 |
| 436812.0 | 6.5 |
| 452412.0 | 6.5 |
| 452833.0 | 6.3 |
| 452915.0 | 6.0 |
| 453259.0 | 7.0 |
| 454272.0 | 6.9 |
| 455675.0 | 7.0 |
| 456372.0 | 6.4 |
| 456912.0 | 5.6 |
| 457846.0 | 5.8 |
| 458412.0 | 6.2 |
| 459083.0 | 5.5 |
| 459552.0 | 6.3 |
| 459912.0 | 5.7 |
| 460027.0 | 5.6 |
| 461112.0 | 6.7 |
| 463727.0 | 7.3 |
| 475872.0 | 6.9 |
| 475933.0 | 6.3 |
| 476692.0 | 6.9 |
| 477977.0 | 6.1 |
| 485352.0 | 7.3 |
| 488791.0 | 6.4 |
| 489317.0 | 7.2 |
| 490032.0 | 6.6 |
| 491412.0 | 6.1 |
| 492024.0 | 6.1 |
| 493690.0 | 6.2 |
| 494627.0 | 6.7 |
| 494790.0 | 5.9 |
| 494850.0 | 7.1 |
| 497901.0 | 7.3 |
| 498492.0 | 6.6 |
| 498672.0 | 7.2 |
| 499284.0 | 6.6 |
| 499932.0 | 6.2 |
| 500869.0 | 7.0 |
| 502872.0 | 6.6 |
| 503412.0 | 7.0 |
| 504962.0 | 7.1 |
| 518712.0 | 5.7 |
| 520924.0 | 7.4 |
| 527112.0 | 6.7 |
| 527352.0 | 7.4 |
| 528462.0 | 6.2 |
| 528552.0 | 7.2 |
| 528732.0 | 7.1 |
| 529800.0 | 6.2 |
| 530472.0 | 6.0 |
| 530674.0 | 7.0 |
| 531381.0 | 6.0 |
| 532992.0 | 6.1 |
| 533172.0 | 6.0 |
| 534044.0 | 5.8 |
| 534053.0 | 6.6 |
| 534372.0 | 6.7 |
| 534552.0 | 7.4 |
| 534672.0 | 6.0 |
| 534792.0 | 5.9 |
| 536521.0 | 6.2 |
| 537372.0 | 6.7 |
| 538217.0 | 6.6 |
| 538752.0 | 7.4 |
| 538932.0 | 6.1 |
| 544160.0 | 6.9 |
| 544392.0 | 6.2 |
| 544632.0 | 6.6 |
| 544872.0 | 6.8 |
| 544950.0 | 6.9 |
| 545029.0 | 6.9 |
| 546012.0 | 6.7 |
| 546635.0 | 6.4 |
| 546870.0 | 7.3 |
| 547916.0 | 6.8 |
| 548306.0 | 7.2 |
| 548458.0 | 6.8 |
| 548482.0 | 5.9 |
| 548597.0 | 7.3 |
| 549290.0 | 6.8 |
| 549364.0 | 6.6 |
| 549700.0 | 7.4 |
| 549920.0 | 7.3 |
| 549950.0 | 6.6 |
| 560194.0 | 7.0 |
| 561912.0 | 6.6 |
| 563292.0 | 5.9 |
| 566639.0 | 5.7 |
| 567672.0 | 7.0 |
| 567912.0 | 7.3 |
| 568222.0 | 5.8 |
| 569772.0 | 7.0 |
| 570040.0 | 6.7 |
| 570552.0 | 5.8 |
| 570672.0 | 6.7 |
| 570732.0 | 6.9 |
| 571140.0 | 7.0 |
| 571932.0 | 6.8 |
| 572482.0 | 7.1 |
| 572592.0 | 6.9 |
| 572810.0 | 5.5 |
| 572987.0 | 7.2 |
| 573432.0 | 6.0 |
| 574144.0 | 6.7 |
| 574590.0 | 6.4 |
| 575052.0 | 6.5 |
| 575495.0 | 5.9 |
| 577333.0 | 7.2 |
| 579896.0 | 7.4 |
| 579965.0 | 7.0 |
| 582952.0 | 6.8 |
| 585322.0 | 6.7 |
| 586740.0 | 7.1 |
| 587204.0 | 5.7 |
| 588176.0 | 5.9 |
| 588262.0 | 6.6 |
| 588601.0 | 6.1 |
| 588752.0 | 6.2 |
| 588983.0 | 6.1 |
| 589452.0 | 7.4 |
| 589512.0 | 6.2 |
| 590421.0 | 7.0 |
| 590892.0 | 6.3 |
| 591173.0 | 6.7 |
| 606023.0 | 5.6 |
| 606116.0 | 6.6 |
| 608112.0 | 7.3 |
| 609372.0 | 5.6 |
| 610172.0 | 6.4 |
| 610544.0 | 7.1 |
| 610632.0 | 5.9 |
| 611971.0 | 6.6 |
| 612072.0 | 5.8 |
| 612825.0 | 5.9 |
| 612976.0 | 6.4 |
| 613632.0 | 7.0 |
| 613992.0 | 5.9 |
| 614152.0 | 7.4 |
| 614219.0 | 7.0 |
| 616172.0 | 6.4 |
| 616752.0 | 7.2 |
| 617394.0 | 7.2 |
| 617709.0 | 6.4 |
| 620952.0 | 6.9 |
| 623772.0 | 7.0 |
| 627327.0 | 6.7 |
| 627398.0 | 6.9 |
| 627584.0 | 5.8 |
| 627640.0 | 7.4 |
| 627844.0 | 5.7 |
| 628152.0 | 5.9 |
| 628665.0 | 7.4 |
| 628822.0 | 7.1 |
| 629022.0 | 7.2 |
| 629031.0 | 6.0 |
| 629472.0 | 6.3 |
| 629532.0 | 6.0 |
| 629772.0 | 6.6 |
| 630690.0 | 6.8 |
| 631152.0 | 7.3 |
| 631380.0 | 7.0 |
| 632412.0 | 7.1 |
| 645016.0 | 5.7 |
| 645372.0 | 7.3 |
| 646632.0 | 6.7 |
| 646992.0 | 6.9 |
| 647701.0 | 6.3 |
| 648072.0 | 6.2 |
| 648432.0 | 6.7 |
| 648492.0 | 6.5 |
| 649146.0 | 6.8 |
| 649692.0 | 6.5 |
| 650403.0 | 7.0 |
| 651372.0 | 5.9 |
| 651504.0 | 5.7 |
| 651751.0 | 5.7 |
| 652008.0 | 7.4 |
| 653279.0 | 7.2 |
| 653703.0 | 6.2 |
| 653832.0 | 6.2 |
| 655512.0 | 6.3 |
| 655919.0 | 6.0 |
| 656523.0 | 6.0 |
| 657072.0 | 5.8 |
| 659137.0 | 7.0 |
| 659344.0 | 5.7 |
| 665194.0 | 7.0 |
| 666912.0 | 7.0 |
| 667212.0 | 6.5 |
| 667766.0 | 6.5 |
| 667886.0 | 6.0 |
| 668147.0 | 5.7 |
| 668292.0 | 5.6 |
| 668532.0 | 7.1 |
| 669296.0 | 7.2 |
| 671172.0 | 5.9 |
| 671954.0 | 5.9 |
| 673371.0 | 5.7 |
| 673992.0 | 7.0 |
| 674412.0 | 7.0 |
| 686900.0 | 6.2 |
| 687192.0 | 6.1 |
| 687312.0 | 6.1 |
| 687372.0 | 6.4 |
| 689534.0 | 6.7 |
| 690132.0 | 6.2 |
| 691272.0 | 6.9 |
| 691452.0 | 5.5 |
| 691512.0 | 6.4 |
| 692160.0 | 7.2 |
| 692568.0 | 6.6 |
| 697212.0 | 6.8 |
| 700548.0 | 6.6 |
| 702972.0 | 7.2 |
| 703092.0 | 6.4 |
| 703543.0 | 7.0 |
| 705014.0 | 6.0 |
| 705732.0 | 6.8 |
| 706973.0 | 6.8 |
| 707102.0 | 6.5 |
| 707172.0 | 5.7 |
| 708536.0 | 6.5 |
| 708732.0 | 5.6 |
| 708852.0 | 7.0 |
| 709516.0 | 6.5 |
| 709704.0 | 6.9 |
| 710930.0 | 5.5 |
| 712290.0 | 6.3 |
| 717180.0 | 6.9 |
| 728892.0 | 6.1 |
| 729404.0 | 6.1 |
| 729951.0 | 6.7 |
| 730020.0 | 6.7 |
| 733392.0 | 7.4 |
| 735412.0 | 6.9 |
| 739032.0 | 7.2 |
| 740811.0 | 6.9 |
| 744289.0 | 7.5 |
| 744440.0 | 6.9 |
| 745726.0 | 6.2 |
| 745834.0 | 6.6 |
| 746540.0 | 6.2 |
| 746882.0 | 6.6 |
| 747132.0 | 6.8 |
| 747160.0 | 6.6 |
| 748240.0 | 6.3 |
| 748324.0 | 6.8 |
| 748760.0 | 7.5 |
| 750372.0 | 6.6 |
| 750492.0 | 6.1 |
| 751193.0 | 7.4 |
| 756049.0 | 6.4 |
| 756192.0 | 6.7 |
| 757023.0 | 7.3 |
| 757288.0 | 5.8 |
| 757872.0 | 6.8 |
| 775032.0 | 6.8 |
| 776652.0 | 7.2 |
| 776823.0 | 6.0 |
| 780316.0 | 5.8 |
| 781032.0 | 5.7 |
| 781628.0 | 5.7 |
| 781949.0 | 6.9 |
| 782052.0 | 6.3 |
| 782112.0 | 7.5 |
| 782292.0 | 6.6 |
| 782589.0 | 5.8 |
| 784036.0 | 6.2 |
| 784243.0 | 6.5 |
| 784491.0 | 5.8 |
| 784872.0 | 6.0 |
| 785292.0 | 6.9 |
| 785440.0 | 6.8 |
| 785704.0 | 7.5 |
| 786000.0 | 6.3 |
| 786139.0 | 5.5 |
| 787520.0 | 6.8 |
| 787620.0 | 5.7 |
| 790245.0 | 7.3 |
| 791471.0 | 6.1 |
| 794952.0 | 7.4 |
| 795252.0 | 6.3 |
| 803652.0 | 7.2 |
| 804666.0 | 6.5 |
| 815412.0 | 6.9 |
| 815669.0 | 5.8 |
| 816170.0 | 5.6 |
| 816472.0 | 6.8 |
| 817245.0 | 5.9 |
| 819149.0 | 6.9 |
| 822003.0 | 7.3 |
| 822612.0 | 6.8 |
| 823000.0 | 5.7 |
| 823045.0 | 7.4 |
| 823119.0 | 7.1 |
| 823184.0 | 7.3 |
| 823307.0 | 5.9 |
| 823572.0 | 5.6 |
| 823624.0 | 6.9 |
| 823640.0 | 5.7 |
| 823701.0 | 6.1 |
| 823752.0 | 6.3 |
| 823932.0 | 6.5 |
| 824357.0 | 7.5 |
| 824391.0 | 6.2 |
| 824599.0 | 5.6 |
| 824702.0 | 6.4 |
| 824732.0 | 7.3 |
| 825094.0 | 6.0 |
| 825107.0 | 7.0 |
| 825746.0 | 5.8 |
| 825816.0 | 5.7 |
| 826248.0 | 6.5 |
| 826289.0 | 7.4 |
| 826445.0 | 7.0 |
| 826692.0 | 6.2 |
| 826872.0 | 6.6 |
| 826992.0 | 6.7 |
| 827191.0 | 6.7 |
| 827302.0 | 5.8 |
| 827482.0 | 6.4 |
| 827522.0 | 7.5 |
| 827668.0 | 7.4 |
| 828252.0 | 6.7 |
| 828897.0 | 7.0 |
| 829202.0 | 5.6 |
| 829752.0 | 5.6 |
| 829872.0 | 6.5 |
| 830823.0 | 6.3 |
| 831252.0 | 7.0 |
| 831629.0 | 7.3 |
| 831807.0 | 5.6 |
| 832108.0 | 6.5 |
| 832512.0 | 6.7 |
| 833346.0 | 6.4 |
| 834192.0 | 7.2 |
| 836283.0 | 5.5 |
| 836712.0 | 6.7 |
| 837192.0 | 6.8 |
| 837363.0 | 7.1 |
| 841332.0 | 7.4 |
| 841392.0 | 5.6 |
| 841452.0 | 6.8 |
| 842892.0 | 6.3 |
| 843086.0 | 7.2 |
| 843367.0 | 6.1 |
| 843500.0 | 7.0 |
| 843790.0 | 6.1 |
| 843882.0 | 6.9 |
| 844272.0 | 7.4 |
| 844392.0 | 6.5 |
| 844633.0 | 6.2 |
| 844732.0 | 5.9 |
| 844860.0 | 6.3 |
| 845198.0 | 6.2 |
| 845412.0 | 5.8 |
| 846015.0 | 6.6 |
| 846852.0 | 7.4 |
| 846972.0 | 5.9 |
| 847032.0 | 6.0 |
| 847394.0 | 7.4 |
| 847495.0 | 5.6 |
| 847512.0 | 5.7 |
| 848134.0 | 5.9 |
| 848160.0 | 6.2 |
| 860052.0 | 6.5 |
| 860112.0 | 7.0 |
| 860172.0 | 7.1 |
| 860716.0 | 6.1 |
| 861192.0 | 6.7 |
| 861432.0 | 6.2 |
| 861803.0 | 6.8 |
| 862333.0 | 5.7 |
| 862632.0 | 7.3 |
| 862752.0 | 7.1 |
| 862992.0 | 7.3 |
| 863052.0 | 6.6 |
| 863233.0 | 7.2 |
| 863328.0 | 6.8 |
| 863521.0 | 6.3 |
| 863652.0 | 5.6 |
| 863810.0 | 6.1 |
| 863927.0 | 6.0 |
| 864192.0 | 6.5 |
| 864252.0 | 6.8 |
| 864432.0 | 6.7 |
| 864922.0 | 7.3 |
| 865052.0 | 7.2 |
| 865494.0 | 5.7 |
| 865812.0 | 6.4 |
| 865872.0 | 6.3 |
| 865932.0 | 5.7 |
| 865992.0 | 5.5 |
| 866074.0 | 6.6 |
| 866228.0 | 6.2 |
| 866440.0 | 5.9 |
| 866617.0 | 7.5 |
| 866671.0 | 6.1 |
| 866700.0 | 6.7 |
| 866823.0 | 5.5 |
| 866882.0 | 6.2 |
| 866952.0 | 7.3 |
| 867012.0 | 7.3 |
| 867312.0 | 7.4 |
| 867432.0 | 6.9 |
| 867586.0 | 6.8 |
| 867701.0 | 7.4 |
| 867714.0 | 6.9 |
| 867761.0 | 5.9 |
| 867842.0 | 6.3 |
| 867935.0 | 6.0 |
| 867945.0 | 5.9 |
| 867957.0 | 6.5 |
| 868075.0 | 6.8 |
| 868217.0 | 6.3 |
| 868251.0 | 7.5 |
| 868316.0 | 6.7 |
| 868342.0 | 5.9 |
| 868392.0 | 7.1 |
| 868752.0 | 6.8 |
| 868812.0 | 7.4 |
| 869224.0 | 7.5 |
| 869458.0 | 6.4 |
| 869606.0 | 5.8 |
| 869892.0 | 6.7 |
| 870012.0 | 5.9 |
| 870132.0 | 6.5 |
| 870252.0 | 6.3 |
| 870394.0 | 6.4 |
| 870483.0 | 6.7 |
| 870878.0 | 6.6 |
| 871142.0 | 5.5 |
| 871512.0 | 6.1 |
| 871692.0 | 5.5 |
| 871881.0 | 6.9 |
| 872361.0 | 5.6 |
| 872679.0 | 7.2 |
| 872712.0 | 5.9 |
| 874392.0 | 7.0 |
| 875263.0 | 6.5 |
| 875349.0 | 6.2 |
| 875401.0 | 7.4 |
| 875466.0 | 6.3 |
| 877332.0 | 5.6 |
| 877971.0 | 6.6 |
| 878532.0 | 6.1 |
| 878592.0 | 5.9 |
| 880332.0 | 6.1 |
| 880722.0 | 6.2 |
| 880979.0 | 6.3 |
| 881344.0 | 7.0 |
| 881652.0 | 6.6 |
| 881832.0 | 5.5 |
| 881947.0 | 6.2 |
| 881971.0 | 6.1 |
| 882040.0 | 7.0 |
| 882288.0 | 7.0 |
| 882391.0 | 6.6 |
| 882416.0 | 7.2 |
| 882504.0 | 6.4 |
| 882741.0 | 6.5 |
| 882852.0 | 7.1 |
| 882912.0 | 6.6 |
| 883032.0 | 5.8 |
| 883092.0 | 7.0 |
| 883212.0 | 6.8 |
| 883272.0 | 5.7 |
| 883636.0 | 5.7 |
| 883678.0 | 7.5 |
| 883692.0 | 6.4 |
| 883941.0 | 7.2 |
| 884096.0 | 6.3 |
| 884152.0 | 6.1 |
| 884163.0 | 6.7 |
| 884232.0 | 6.1 |
| 884592.0 | 6.7 |
| 884812.0 | 5.7 |
| 884939.0 | 5.6 |
| 884972.0 | 7.1 |
| 885062.0 | 6.9 |
| 885178.0 | 6.9 |
| 885214.0 | 6.2 |
| 885382.0 | 6.0 |
| 885484.0 | 7.1 |
| 885527.0 | 5.5 |
| 885572.0 | 6.9 |
| 885672.0 | 7.4 |
| 885732.0 | 5.8 |
| 885972.0 | 7.4 |
| 886152.0 | 7.5 |
| 886366.0 | 7.2 |
| 886414.0 | 6.5 |
| 887232.0 | 6.9 |
| 887592.0 | 6.6 |
| 888091.0 | 6.4 |
| 888379.0 | 5.6 |
| 889213.0 | 6.4 |
| 889489.0 | 6.6 |
| 900859.0 | 6.8 |
| 901038.0 | 6.8 |
| 901346.0 | 6.4 |
| 901572.0 | 7.4 |
| 901632.0 | 7.3 |
| 901692.0 | 5.7 |
| 901812.0 | 6.4 |
| 902074.0 | 6.0 |
| 902194.0 | 5.8 |
| 902229.0 | 6.9 |
| 902274.0 | 6.2 |
| 902372.0 | 7.2 |
| 902386.0 | 7.2 |
| 902442.0 | 6.8 |
| 902619.0 | 6.1 |
| 902716.0 | 6.6 |
| 902769.0 | 7.0 |
| 902840.0 | 5.9 |
| 902854.0 | 6.3 |
| 902952.0 | 6.4 |
| 903072.0 | 7.0 |
| 903192.0 | 6.1 |
| 903312.0 | 7.0 |
| 903372.0 | 6.8 |
| 903432.0 | 5.8 |
| 903715.0 | 6.7 |
| 903780.0 | 6.1 |
| 903802.0 | 6.0 |
| 904016.0 | 6.5 |
| 904132.0 | 6.9 |
| 904152.0 | 6.7 |
| 904632.0 | 5.9 |
| 904692.0 | 6.3 |
| 904752.0 | 6.0 |
| 905754.0 | 6.1 |
| 906192.0 | 7.3 |
| 906252.0 | 5.6 |
| 908110.0 | 5.9 |
| 908577.0 | 5.8 |
| 908712.0 | 7.2 |
| 909012.0 | 6.8 |
| 909072.0 | 7.5 |
| 909399.0 | 7.2 |
| 909854.0 | 7.1 |
| 910094.0 | 7.2 |
| 912623.0 | 6.3 |
| 918721.0 | 5.6 |
| 919373.0 | 6.2 |
| 919379.0 | 6.7 |
| 919467.0 | 7.0 |
| 919639.0 | 7.1 |
| 920292.0 | 6.6 |
| 920788.0 | 5.6 |
| 920830.0 | 6.4 |
| 921053.0 | 7.0 |
| 921214.0 | 6.2 |
| 921972.0 | 7.3 |
| 922092.0 | 6.3 |
| 922152.0 | 6.6 |
| 922408.0 | 6.4 |
| 922754.0 | 6.0 |
| 922914.0 | 6.1 |
| 923172.0 | 6.1 |
| 923232.0 | 7.4 |
| 923532.0 | 6.3 |
| 923592.0 | 6.8 |
| 923738.0 | 6.4 |
| 923891.0 | 7.1 |
| 923984.0 | 6.4 |
| 924025.0 | 7.1 |
| 924336.0 | 7.3 |
| 924382.0 | 7.4 |
| 924552.0 | 6.5 |
| 924852.0 | 7.3 |
| 924972.0 | 5.8 |
| 925191.0 | 6.9 |
| 925478.0 | 7.1 |
| 925516.0 | 5.7 |
| 925649.0 | 5.6 |
| 925718.0 | 7.0 |
| 925778.0 | 5.5 |
| 926840.0 | 7.2 |
| 926881.0 | 6.1 |
| 927014.0 | 6.8 |
| 927175.0 | 6.1 |
| 927350.0 | 7.5 |
| 927393.0 | 6.9 |
| 927492.0 | 6.5 |
| 927552.0 | 7.0 |
| 927672.0 | 6.9 |
| 927912.0 | 6.7 |
| 929052.0 | 6.6 |
| 929112.0 | 6.3 |
| 929586.0 | 6.3 |
| 929882.0 | 7.4 |
| 931432.0 | 6.6 |
| 931651.0 | 6.3 |
| 932853.0 | 6.7 |
| 932916.0 | 7.2 |
| 933109.0 | 7.4 |
| 943242.0 | 6.8 |
| 943632.0 | 6.8 |
| 943752.0 | 7.2 |
| 944395.0 | 7.3 |
| 944468.0 | 5.6 |
| 944507.0 | 6.9 |
| 945012.0 | 5.9 |
| 945132.0 | 5.8 |
| 946392.0 | 6.9 |
| 946572.0 | 6.6 |
| 946972.0 | 6.4 |
| 947010.0 | 5.5 |
| 947070.0 | 6.5 |
| 947376.0 | 5.6 |
| 947832.0 | 7.4 |
| 948072.0 | 7.1 |
| 948887.0 | 6.5 |
| 950317.0 | 6.1 |
| 950592.0 | 6.8 |
| 950772.0 | 6.4 |
| 952901.0 | 7.1 |
| 954708.0 | 6.4 |
| 959112.0 | 7.1 |
| 959292.0 | 6.2 |
| 959352.0 | 5.9 |
| 959472.0 | 6.5 |
| 959532.0 | 7.0 |
| 959592.0 | 6.7 |
| 959694.0 | 7.0 |
| 959876.0 | 6.0 |
| 959925.0 | 6.8 |
| 960020.0 | 7.2 |
| 960207.0 | 6.0 |
| 960242.0 | 6.7 |
| 960323.0 | 6.6 |
| 960552.0 | 5.5 |
| 960672.0 | 6.8 |
| 960792.0 | 7.1 |
| 960852.0 | 6.4 |
| 960912.0 | 7.3 |
| 960972.0 | 6.3 |
| 961032.0 | 6.9 |
| 961154.0 | 6.4 |
| 961315.0 | 6.8 |
| 961442.0 | 6.9 |
| 961477.0 | 6.9 |
| 961544.0 | 7.5 |
| 961708.0 | 5.7 |
| 961879.0 | 7.0 |
| 961970.0 | 6.3 |
| 962052.0 | 6.4 |
| 962172.0 | 6.6 |
| 962352.0 | 6.8 |
| 962412.0 | 6.6 |
| 962739.0 | 6.2 |
| 962807.0 | 7.0 |
| 962962.0 | 5.8 |
| 962975.0 | 7.2 |
| 963147.0 | 6.0 |
| 963405.0 | 7.1 |
| 963432.0 | 7.3 |
| 963552.0 | 7.1 |
| 963612.0 | 7.4 |
| 963672.0 | 6.1 |
| 963792.0 | 7.1 |
| 963912.0 | 6.1 |
| 964158.0 | 7.0 |
| 964259.0 | 6.8 |
| 964431.0 | 7.4 |
| 964574.0 | 5.9 |
| 964818.0 | 5.7 |
| 964872.0 | 7.0 |
| 965466.0 | 7.3 |
| 965527.0 | 5.6 |
| 965551.0 | 7.0 |
| 965797.0 | 5.9 |
| 965872.0 | 6.9 |
| 966167.0 | 5.7 |
| 966492.0 | 7.4 |
| 966552.0 | 7.3 |
| 967234.0 | 6.8 |
| 968304.0 | 6.2 |
| 968451.0 | 6.8 |
| 969145.0 | 5.8 |
| 969312.0 | 6.1 |
| 970692.0 | 6.2 |
| 970812.0 | 5.7 |
| 975252.0 | 6.9 |
| 975372.0 | 6.9 |
| 976150.0 | 6.4 |
| 986472.0 | 6.4 |
| 987155.0 | 7.3 |
| 987784.0 | 5.6 |
| 987820.0 | 5.5 |
| 989153.0 | 6.4 |
| 995592.0 | 7.1 |
| 998412.0 | 7.3 |
| 998575.0 | 7.4 |
| 998692.0 | 7.3 |
| 999262.0 | 6.2 |
| 999312.0 | 6.9 |
| 999852.0 | 7.4 |
| 1000110.0 | 6.2 |
| 1000230.0 | 6.9 |
| 1000289.0 | 5.9 |
| 1000602.0 | 6.6 |
| 1000932.0 | 6.3 |
| 1001112.0 | 6.2 |
| 1001172.0 | 7.2 |
| 1001441.0 | 6.1 |
| 1001608.0 | 6.2 |
| 1001651.0 | 7.2 |
| 1001719.0 | 6.8 |
| 1002250.0 | 6.9 |
| 1002312.0 | 7.4 |
| 1002432.0 | 6.8 |
| 1002552.0 | 6.6 |
| 1002672.0 | 5.9 |
| 1003187.0 | 7.5 |
| 1003509.0 | 6.3 |
| 1003587.0 | 7.3 |
| 1003714.0 | 6.1 |
| 1003729.0 | 6.6 |
| 1004112.0 | 6.5 |
| 1004172.0 | 6.9 |
| 1004232.0 | 5.5 |
| 1004322.0 | 6.3 |
| 1004357.0 | 6.4 |
| 1004470.0 | 7.2 |
| 1004502.0 | 7.5 |
| 1004567.0 | 5.9 |
| 1005040.0 | 6.4 |
| 1005766.0 | 6.4 |
| 1005832.0 | 5.9 |
| 1007052.0 | 5.9 |
| 1007112.0 | 6.5 |
| 1008192.0 | 5.9 |
| 1008252.0 | 7.3 |
| 1008954.0 | 5.6 |
| 1009509.0 | 7.5 |
| 1009572.0 | 6.0 |
| 1009752.0 | 7.1 |
| 1010159.0 | 6.3 |
| 1010594.0 | 7.5 |
| 1011192.0 | 6.3 |
| 1011372.0 | 5.6 |
| 1012204.0 | 6.8 |
| 1012300.0 | 6.2 |
| 1012752.0 | 6.0 |
| 1012960.0 | 6.2 |
| 1013052.0 | 6.6 |
| 1013519.0 | 7.4 |
| 1013585.0 | 6.3 |
| 1018016.0 | 7.1 |
| 1029146.0 | 6.7 |
| 1036932.0 | 6.3 |
| 1038372.0 | 7.3 |
| 1038492.0 | 6.3 |
| 1039325.0 | 7.3 |
| 1039444.0 | 7.5 |
| 1039504.0 | 6.1 |
| 1039752.0 | 6.6 |
| 1039932.0 | 7.4 |
| 1039992.0 | 7.5 |
| 1040172.0 | 7.0 |
| 1040424.0 | 5.6 |

1040624.0 7.3
1040670.0 5.9
1040975.0 6.4
1041067.0 7.4
1041612.0 6.0
1041892.0 5.9
1042031.0 6.9
1042134.0 7.5
1042359.0 5.7
1042390.0 7.1
1042532.0 6.5
1042604.0 7.5
1042812.0 7.2
1043112.0 7.2
1043424.0 5.7
1043520.0 7.3
1043695.0 6.9
1043707.0 6.9
1044192.0 6.3
1045386.0 6.2
1045512.0 7.2
1047012.0 6.5
1047542.0 5.9
1048812.0 5.9
1051999.0 6.2
1055712.0 7.0
1055772.0 6.7
1056291.0 6.0
1056624.0 5.7
1058216.0 7.4
1059134.0 7.2
1059176.0 6.5
1059244.0 7.3
1070560.0 7.0
1075275.0 6.0
1077432.0 6.9
1078812.0 5.9
1078932.0 7.4
1079382.0 5.8
1080060.0 6.7
1080132.0 6.9
1080432.0 7.4
1080661.0 6.9
1080937.0 5.7
1081072.0 7.1
1081100.0 6.5
1081175.0 6.9
1081343.0 7.1
1081872.0 6.2
1081992.0 6.0
1082118.0 6.7
1082132.0 6.9
1082306.0 6.8
1082364.0 6.8
1082580.0 6.0
1082707.0 7.0
1082794.0 6.6
1082803.0 5.5
1082868.0 6.8
1082944.0 6.0
1082952.0 7.4
1083525.0 6.6
1083631.0 6.0
1083645.0 7.5
1083700.0 5.9
1083805.0 7.2
1083856.0 6.0
1084157.0 6.6
1084386.0 6.5
1084752.0 5.6
1085083.0 6.6
1085582.0 5.9
1085638.0 6.4
1085806.0 6.3
1085832.0 7.1
1087228.0 7.1
1088217.0 6.7
1088247.0 7.4
1088571.0 5.8
1090272.0 5.7
1092456.0 6.8
1094228.0 6.8
1094772.0 6.9
1095027.0 7.2
1095770.0 5.6
1095972.0 7.3
1097084.0 6.6
1097352.0 5.6
1097412.0 5.6
1097532.0 7.2
1097712.0 5.9
1097772.0 7.0
1097832.0 5.6
1097900.0 5.5
1097951.0 6.7
1098032.0 5.7
1098070.0 6.1
1098189.0 6.7
1098217.0 6.3
1098279.0 6.7
1098322.0 6.4
1098499.0 5.7
1098602.0 6.5
1098645.0 6.9
1098704.0 7.4
1098710.0 6.4
1098792.0 6.5
1098852.0 7.4
1099445.0 7.5
1099613.0 6.4
1099650.0 7.1
1099832.0 5.8
1100141.0 5.9
1100352.0 7.3
1100910.0 6.4
1101202.0 6.7
1101563.0 7.2
1101633.0 6.4
1101972.0 6.8
1102432.0 6.5
1102512.0 5.8
1102609.0 5.7
1113169.0 6.1
1114632.0 7.0
1116990.0 5.9
1118206.0 6.9
1119312.0 7.3
1119867.0 5.6
1120308.0 5.5
1121348.0 7.4
1121548.0 6.7
1121606.0 6.4
1121792.0 6.1
1122012.0 6.9
1122252.0 6.9
1122312.0 5.8
1122404.0 5.8
1122415.0 6.3
1122624.0 5.6
1122653.0 6.1
1122831.0 5.9
1122870.0 6.6
1123045.0 6.1
1123332.0 6.4
1123572.0 6.9
1124454.0 6.4
1125330.0 7.4
1125450.0 7.1
1126152.0 7.0
1126700.0 6.4
1127014.0 6.7
1127178.0 7.0
1127205.0 5.8
1128364.0 6.7
1128500.0 7.5
1129032.0 6.4
1129212.0 7.0
1129272.0 6.2
1130164.0 6.0
1130772.0 6.6
1130952.0 6.0
1132902.0 6.7
1133712.0 5.5
1135092.0 5.6
1135272.0 6.3
1136156.0 6.0
1136220.0 6.5
1136292.0 5.8
1136652.0 6.0
1136876.0 7.4
1136932.0 5.9
1136972.0 7.1
1136995.0 6.5
1137039.0 6.4
1137105.0 7.4
1137172.0 7.1
1137222.0 6.3
1137382.0 6.6
1137474.0 5.6
1137496.0 7.1
1137550.0 5.6
1137597.0 7.5
1137672.0 6.1
1137732.0 6.4
1137912.0 6.0
1137972.0 6.9
1138032.0 6.4
1138092.0 6.5
1138249.0 5.6
1138260.0 5.5
1138307.0 6.4
1138354.0 7.0
1138422.0 5.6
1138628.0 6.5
1138641.0 7.2
1138696.0 6.7
1138732.0 6.6
1138814.0 7.4
1138842.0 7.2
1138870.0 6.1
1138890.0 5.8
1138934.0 6.4
1139030.0 6.2
1139112.0 7.1
1139172.0 6.3
1139232.0 6.0
1139292.0 7.2
1139352.0 5.7
1139472.0 5.9
1139532.0 6.9
1139694.0 7.3
1139730.0 6.1
1139970.0 6.6
1140000.0 6.2
1140040.0 6.2
1140218.0 6.4
1140282.0 7.3
1140342.0 5.8
1140432.0 5.6
1140470.0 6.4
1140507.0 6.0
1140672.0 6.1
1141448.0 6.5
1141724.0 6.6
1141982.0 5.5
1142790.0 7.0
1142822.0 6.3
1143137.0 5.7
1143492.0 5.9
1143732.0 6.0
1143852.0 6.6
1144306.0 5.5
1144334.0 5.9
1144374.0 5.9
1144471.0 6.1
1144564.0 5.6
1144605.0 7.0
1144778.0 6.8
1155072.0 6.8
1156392.0 6.6
1156452.0 6.3
1156960.0 6.4
1157759.0 6.3
1157826.0 6.9
1157952.0 5.9
1158802.0 5.8
1158930.0 7.2
1159043.0 7.0
1159272.0 6.0
1159632.0 5.9
1159890.0 5.8
1161867.0 6.2
1162332.0 6.5
1162452.0 6.9
1162766.0 6.9
1162942.0 5.8
1163120.0 5.8
1163952.0 5.7
1164638.0 6.8
1164672.0 7.0
1165212.0 6.5
1165272.0 5.9
1165332.0 6.8
1165392.0 6.1
1165452.0 5.5
1165512.0 7.3
1165685.0 6.6
1165794.0 7.5
1166079.0 5.8
1166333.0 7.2
1166652.0 6.7
1166772.0 5.6
1166832.0 6.0
1167356.0 7.2
1167541.0 5.6
1167736.0 7.0
1168092.0 6.0
1168152.0 5.7
1168272.0 7.1
1168332.0 5.6
1168662.0 7.1
1168696.0 6.1
1168745.0 6.0
1168805.0 5.6
1169412.0 6.6
1169772.0 7.3
1169832.0 7.0
1169915.0 7.4
1170150.0 7.1
1170285.0 7.3
1170391.0 6.7
1170482.0 6.2
1170567.0 6.9
1171212.0 7.0
1171616.0 5.6
1171662.0 6.3
1171901.0 6.0
1171949.0 5.5
1172292.0 6.8
1172412.0 5.9
1172592.0 6.1
1172811.0 6.2
1173510.0 7.1
1174152.0 7.4
1174414.0 6.0
1174418.0 6.2
1174597.0 7.0
1174669.0 6.4
1174800.0 5.9
1174819.0 6.9
1174913.0 6.0
1175112.0 6.6
1175232.0 7.4
1175292.0 6.8
1175352.0 6.3
1175412.0 6.1
1175532.0 7.0
1175845.0 5.7
1175852.0 7.2
1176030.0 6.2
1176502.0 6.5
1176552.0 6.8
1176612.0 7.3
1176672.0 5.9
1176732.0 6.2
1176852.0 6.3
1176972.0 7.1
1177032.0 5.8
1177094.0 6.5
1177205.0 7.3
1177313.0 7.2
1177366.0 6.0
1177617.0 6.5
1177638.0 6.9
1177699.0 7.0
1177823.0 6.1
1177886.0 6.1
1177960.0 5.8
1177971.0 6.7
1177992.0 6.1
1178052.0 6.5
1178232.0 7.0
1178292.0 6.3
1178352.0 6.4
1178600.0 6.3
1178679.0 7.0
1178730.0 7.2
1178868.0 6.9
1178903.0 5.8
1178918.0 5.9
1178962.0 5.7
1179050.0 5.8
1179100.0 5.9
1179294.0 6.5
1179415.0 7.0
1179612.0 7.4
1179852.0 6.7
1179912.0 5.5
1180021.0 6.7
1180062.0 7.3
1180103.0 6.5
1180147.0 6.9
1180163.0 6.3
1180210.0 7.1
1180245.0 7.4
1180397.0 6.1
1180417.0 6.1
1180442.0 5.5
1180482.0 6.2
1180521.0 6.3
1180697.0 6.7
1180820.0 6.1
1180860.0 5.6
1180872.0 6.5
1180932.0 6.6
1181292.0 6.3
1181491.0 6.9
1181512.0 6.1
1181685.0 5.6
1182176.0 6.6
1182205.0 7.1
1182432.0 7.1
1182552.0 6.8
1182672.0 7.1
1182732.0 7.0
1183039.0 7.2
1183199.0 7.1
1183212.0 7.3
1183349.0 6.7
1183707.0 7.4
1183752.0 7.3
1183872.0 7.3
1184292.0 6.5
1184410.0 5.8
1184482.0 6.6
1184767.0 5.5
1185192.0 7.0
1185372.0 6.0
1185552.0 5.9
1185672.0 6.5
1185848.0 7.3
1185930.0 5.6
1186136.0 6.3
1186311.0 6.9
1186405.0 5.6
1196712.0 5.5
1196772.0 5.9
1196832.0 6.8
1197810.0 6.1
1198038.0 5.9
1198116.0 6.2
1198152.0 7.4
1198272.0 6.9
1198512.0 7.3
1198572.0 6.9
1198632.0 7.3
1198906.0 6.2
1199056.0 7.3
1199273.0 7.4
1199445.0 7.3
1199832.0 6.6
1199892.0 7.4
1200367.0 5.9
1200712.0 6.7
1200743.0 5.6
1200900.0 7.0
1200925.0 6.2
1201032.0 6.3
1201152.0 6.1
1201212.0 6.4
1201512.0 6.8
1201717.0 5.9
1201733.0 5.5
1201825.0 6.8
1201911.0 5.9
1202532.0 6.8
1203912.0 5.5
1204651.0 6.1
1205472.0 7.1
1205652.0 6.5
1205772.0 6.1
1206249.0 7.4
1206479.0 6.7
1206625.0 6.5
1206762.0 5.5
1207535.0 6.6
1208177.0 5.6
1208532.0 7.4
1208712.0 7.0
1209097.0 6.0
1209164.0 5.5
1209417.0 6.0
1209672.0 5.7
1209912.0 6.5
1209972.0 6.8
1210092.0 5.9
1210222.0 6.2
1210322.0 7.4
1210432.0 5.9
1211025.0 6.4
1211112.0 7.4
1211232.0 7.0
1211704.0 6.8
1211772.0 6.4
1211784.0 6.5
1212028.0 7.2
1212094.0 6.2
1212145.0 6.5
1212214.0 5.6
1212267.0 6.4
1212332.0 7.0
1212405.0 6.1
1212552.0 7.2
1212672.0 7.1
1212792.0 7.3
1212852.0 6.0
1212972.0 5.9
1213107.0 6.7
1213172.0 6.6
1213195.0 7.1
1213253.0 5.5
1213317.0 6.4
1213382.0 6.9
1213447.0 5.6
1213498.0 5.8
1213574.0 5.8
1213594.0 6.8
1213689.0 5.8
1213830.0 5.8
1213836.0 7.4
1213958.0 5.9
1214172.0 6.9
1214232.0 7.0
1214292.0 5.6
1214412.0 5.8
1214685.0 5.6
1214713.0 6.9
1215059.0 7.4
1215099.0 6.0
1215184.0 6.1
1215207.0 7.3
1215291.0 5.9
1215552.0 6.4
1215672.0 5.5
1215732.0 6.6
1215792.0 6.5
1216097.0 7.2
1216122.0 5.7
1216543.0 7.0
1216742.0 6.1
1216790.0 5.9
1216809.0 6.5
1217112.0 6.1
1217172.0 7.4
1217352.0 6.6
1217850.0 6.1
1217931.0 6.8
1218024.0 7.1
1218099.0 7.0
1218132.0 6.3
1218152.0 6.2
1218189.0 5.5
1218261.0 5.8
1218295.0 6.0
1218312.0 7.1
1218372.0 6.0
1218432.0 6.6
1218492.0 7.3
1218552.0 5.5
1218612.0 7.1
1218672.0 7.3
1218732.0 6.6
1218867.0 7.3
1218929.0 6.3
1218992.0 6.2
1219030.0 7.1
1219084.0 7.3
1219107.0 6.8
1219189.0 6.9
1219253.0 7.0
1219282.0 6.5
1219329.0 6.8
1219405.0 7.4
1219477.0 6.5
1219744.0 6.9
1220112.0 7.4
1220312.0 7.5
1220665.0 6.1
1220682.0 5.7
1220810.0 7.0
1220955.0 6.8
1221005.0 6.8
1221040.0 6.5
1221127.0 5.7
1221192.0 6.9
1221432.0 6.3
1221672.0 5.7
1222009.0 6.7
1222080.0 7.4
1222197.0 6.7
1222692.0 6.4
1222872.0 7.4
1222932.0 7.1
1223300.0 7.5
1223441.0 7.4
1223850.0 5.8
1223949.0 7.3
1223998.0 6.3
1224252.0 6.0
1224372.0 7.3
1224552.0 6.0
1224657.0 5.9
1225014.0 6.1
1225692.0 6.9
1225812.0 6.1
1225872.0 7.0
1225932.0 5.6
1226207.0 5.7
1226326.0 5.9
1226397.0 5.7
1228752.0 6.3
1229164.0 7.3
1229597.0 5.8
1239912.0 5.7
1239972.0 7.3
1240212.0 6.9
1241174.0 7.2
1241252.0 6.4
1241337.0 5.7
1241352.0 6.7
1241772.0 6.3
1241832.0 6.8
1242005.0 7.2
1243212.0 6.4
1244043.0 7.1
1244836.0 7.2
1245029.0 6.8
1245099.0 6.2
1245732.0 6.0
1246152.0 6.5
1246310.0 6.7
1249672.0 5.6
1250232.0 6.6
1250412.0 7.3
1251394.0 6.6
1251552.0 6.8
1251732.0 6.5
1251852.0 7.0
1252119.0 5.9
1252450.0 5.6
1252472.0 7.4
1252490.0 7.2
1252521.0 7.0
1252561.0 6.7
1252932.0 5.9
1253232.0 6.0
1253352.0 7.4
1253448.0 7.3
1253478.0 6.8
1253488.0 6.6
1253527.0 5.9
1253657.0 5.6
1253702.0 6.2
1253761.0 6.6
1254023.0 6.4
1254032.0 6.9
1254162.0 6.9
1254212.0 7.1
1254256.0 6.1
1254274.0 7.1
1254281.0 5.8
1254372.0 5.6
1254432.0 6.9
1254552.0 6.7
1254732.0 7.1
1254873.0 7.5
1254923.0 6.9
1255024.0 6.1
1255122.0 7.2
1255197.0 5.5
1255249.0 6.2
1255326.0 7.5
1255812.0 7.2
1256172.0 5.9
1256232.0 5.6
1256292.0 6.7
1256403.0 6.1
1256660.0 6.6
1256703.0 6.5
1256720.0 6.4
1256747.0 7.0
1256763.0 6.9
1256910.0 6.6
1257104.0 7.1
1257192.0 7.0
1257492.0 5.7
1257552.0 5.7
1257672.0 6.4
1257823.0 6.8
1257903.0 7.2
1257953.0 7.1
1257987.0 6.0
1258311.0 5.9
1258373.0 6.0
1258437.0 5.8
1258692.0 7.0
1258812.0 5.8
1258992.0 6.2
1259112.0 5.6
1259234.0 7.3
1259283.0 6.7
1259308.0 6.0
1259333.0 7.1
1259390.0 5.8
1259404.0 6.1
1259442.0 7.0
1259461.0 6.0
1259542.0 7.2
1259584.0 7.2
1259642.0 6.4
1259804.0 7.1
1259877.0 7.0
1259893.0 7.3
1260002.0 6.8
1260252.0 6.2
1260372.0 6.2
1260432.0 5.7
1260552.0 6.7
1260808.0 7.0
1261333.0 6.0
1261505.0 7.1
1261992.0 5.5
1262240.0 6.8
1262344.0 6.0
1262380.0 7.3
1262457.0 7.2
1262590.0 6.5
1262890.0 5.6
1262925.0 6.3
1262952.0 6.7
1263012.0 7.3
1263192.0 5.6
1263432.0 6.5
1263777.0 6.9
1263807.0 7.5
1264353.0 7.0
1264752.0 7.2
1265454.0 6.5
1265670.0 5.5
1266457.0 6.7
1268952.0 5.9
1271284.0 6.1
1282483.0 7.3
1284301.0 7.1
1285961.0 7.1
1285992.0 6.8
1286412.0 5.9
1287342.0 6.8
1287672.0 6.4
1287732.0 5.8
1288232.0 6.0
1290552.0 6.1
1292172.0 6.1
1292707.0 7.5
1292812.0 6.3
1292937.0 7.5
1293068.0 6.6
1293252.0 5.9
1293312.0 6.0
1293372.0 7.2
1293432.0 5.6
1293492.0 5.7
1293612.0 6.7
1293672.0 5.6
1293874.0 6.5
1294061.0 6.9
1294282.0 6.3
1294308.0 5.6
1294393.0 6.7
1294577.0 5.8
1294692.0 5.8
1294812.0 6.1
1294872.0 6.5
1295052.0 6.8
1295112.0 6.0
1295305.0 6.7
1295315.0 5.6
1295396.0 5.7
1295570.0 6.8
1295725.0 6.9
1295820.0 6.4
1295869.0 6.1
1296252.0 7.4
1296312.0 5.6
1296837.0 6.2
1296902.0 5.6
1297041.0 6.1
1297140.0 6.9
1297512.0 7.3
1297572.0 6.0
1297692.0 6.4
1298105.0 5.8
1298217.0 6.2
1298587.0 6.6
1299012.0 7.4
1299072.0 7.3
1299132.0 5.5
1299657.0 5.7
1300392.0 7.4
1300572.0 5.9
1300632.0 6.0
1300692.0 6.4
1302615.0 5.9
1302640.0 6.0
1302940.0 6.1
1303008.0 6.2
1303088.0 6.5
1303572.0 6.5
1305132.0 6.4
1306512.0 7.3
1307058.0 6.3
1308287.0 7.1
1309789.0 5.9
1311896.0 7.3
1312682.0 6.0
1314287.0 5.6
1325052.0 6.9
1325172.0 6.2
1326312.0 6.6
1326857.0 6.2
1329552.0 5.6
1330459.0 7.3
1333217.0 7.1
1333360.0 6.8
1333387.0 7.3
1333442.0 5.6
1333504.0 6.4
1333572.0 7.2
1333692.0 7.1
1334111.0 6.5
1334139.0 7.2
1334337.0 7.0
1334522.0 6.8
1334559.0 6.3
1334902.0 5.8
1334922.0 6.7
1334942.0 7.1
1334952.0 7.0
1335432.0 6.4
1335512.0 5.6
1335587.0 5.9
1335714.0 6.2
1335899.0 6.7
1335912.0 6.4
1336122.0 5.9
1336241.0 5.8
1336572.0 7.4
1336632.0 6.9
1337742.0 6.2
1338072.0 5.5
1338702.0 6.9
1338747.0 5.8
1338816.0 6.6
1338858.0 7.0
1339452.0 7.4
1339830.0 6.1
1340496.0 6.4
1341577.0 5.8
1341758.0 6.7
1342005.0 6.5
1342272.0 6.5
1343464.0 6.2
1343772.0 5.6
1344481.0 6.9
1344508.0 7.1
1347285.0 6.5
1347403.0 7.4
1351272.0 6.5
1351422.0 6.5
1351790.0 6.6
1352042.0 6.8
1352087.0 5.9
1352172.0 7.0
1352532.0 7.2
1352652.0 5.6
1352712.0 6.8
1352967.0 7.0
1353002.0 6.5
1353399.0 6.3
1353672.0 7.0
1353852.0 6.5
1354320.0 6.6
1354569.0 6.3
1354615.0 6.2
1354628.0 6.8
1354652.0 6.7
1354760.0 5.7
1356088.0 6.8
1356224.0 7.2
1356792.0 6.3
1357240.0 6.8
1357689.0 6.6
1357901.0 7.4
1369376.0 6.7
1370934.0 7.1
1371252.0 5.7
1371957.0 7.4
1371980.0 6.2
1373144.0 6.3
1373770.0 6.9
1373832.0 7.1
1373892.0 5.6
1373952.0 6.8
1374390.0 6.8
1374415.0 6.7
1374535.0 5.6
1375512.0 6.8
1375572.0 5.6
1375632.0 6.0
1375752.0 6.7
1375913.0 6.8
1376374.0 6.3
1376487.0 6.8
1376527.0 6.5
1376952.0 7.2
1377510.0 6.0
1377687.0 7.1
1377914.0 7.0
1378272.0 6.7
1378452.0 6.8
1378512.0 6.8
1378706.0 6.3
1378815.0 6.9
1379652.0 6.4
1380301.0 5.6
1380694.0 7.2
1380949.0 5.6
1381212.0 5.8
1381452.0 6.7
1381633.0 6.9
1381759.0 6.6
1382253.0 7.2
1382532.0 5.8
1382712.0 6.2
1383804.0 6.7
1383912.0 5.9
1388292.0 5.9
1388592.0 6.6
1389010.0 6.3
1389216.0 6.0
1389732.0 7.4
1389852.0 7.3
1390032.0 6.9
1390092.0 6.0
1390933.0 6.4
1391109.0 6.7
1391172.0 6.1
1391352.0 6.8
1391412.0 7.2
1391472.0 6.6
1391592.0 7.2
1391817.0 5.9
1391982.0 6.8
1392009.0 6.4
1392076.0 6.8
1392173.0 6.4
1392179.0 7.1
1392250.0 6.3
1392552.0 6.5
1392612.0 6.5
1392732.0 6.5
1392972.0 7.4
1393110.0 6.1
1393153.0 5.5
1393383.0 6.1
1393387.0 6.8
1393949.0 6.9
1394472.0 7.5
1395259.0 6.1
1395343.0 6.8
1395552.0 5.7
1395672.0 7.5
1396643.0 6.8
1397112.0 6.5
1397533.0 6.9
1399317.0 5.7
1410312.0 6.5
1410482.0 6.2
1411189.0 6.8
1411692.0 7.1
1411886.0 6.5
1412644.0 6.0
1412712.0 6.3
1412772.0 6.8
1412952.0 6.3
1413012.0 7.3
1413072.0 6.6
1413433.0 6.9
1413788.0 6.3
1414011.0 7.1
1414272.0 6.7
1414452.0 6.1
1414572.0 6.1
1414632.0 5.8
1415378.0 6.1
1416235.0 7.1
1416624.0 5.9
1416719.0 5.9
1417852.0 6.4
1418413.0 6.0
1418472.0 7.5
1418592.0 6.2
1419603.0 6.7
1420332.0 6.6
1420513.0 6.2
1420689.0 6.0
1421310.0 7.1
1421412.0 5.8
1421532.0 7.3
1421652.0 6.3
1422160.0 6.2
1422371.0 5.8
1422580.0 5.5
1423032.0 7.1
1423413.0 6.1
1423510.0 5.8
1423748.0 6.7
1423820.0 5.8
1424070.0 6.4
1424103.0 7.0
1424472.0 7.5
1424592.0 5.7
1424652.0 7.5
1424787.0 7.4
1424810.0 7.1
1424870.0 5.7
1425501.0 6.1
1425792.0 5.7
1427472.0 7.5
1427661.0 5.7
1428106.0 5.5

| | |
|---|---|
| 1428912.0 | 6.0 |
| 1429032.0 | 7.2 |
| 1429467.0 | 5.9 |
| 1429562.0 | 6.6 |
| 1429573.0 | 7.2 |
| 1429742.0 | 7.2 |
| 1429767.0 | 6.6 |
| 1430172.0 | 6.0 |
| 1430232.0 | 6.8 |
| 1430352.0 | 7.4 |
| 1431267.0 | 5.8 |
| 1431288.0 | 6.2 |
| 1431395.0 | 6.3 |
| 1431912.0 | 6.4 |
| 1432665.0 | 6.8 |
| 1432795.0 | 6.1 |
| 1432932.0 | 6.2 |
| 1433052.0 | 6.4 |
| 1433569.0 | 6.4 |
| 1433599.0 | 5.5 |
| 1433628.0 | 6.0 |
| 1433661.0 | 5.5 |
| 1433676.0 | 7.3 |
| 1433774.0 | 6.4 |
| 1433995.0 | 6.8 |
| 1434432.0 | 6.3 |
| 1434492.0 | 7.4 |
| 1434899.0 | 6.7 |
| 1435078.0 | 7.2 |
| 1435342.0 | 6.7 |
| 1435482.0 | 7.0 |
| 1435496.0 | 7.4 |
| 1435752.0 | 6.2 |
| 1435872.0 | 6.4 |
| 1435932.0 | 5.7 |
| 1436172.0 | 5.5 |
| 1436232.0 | 6.0 |
| 1436309.0 | 6.3 |
| 1436374.0 | 7.3 |
| 1436385.0 | 7.1 |
| 1436732.0 | 6.9 |
| 1436843.0 | 6.0 |
| 1436922.0 | 6.0 |
| 1437192.0 | 7.3 |
| 1437252.0 | 7.4 |
| 1437312.0 | 7.3 |
| 1437432.0 | 6.8 |
| 1437492.0 | 6.9 |
| 1437552.0 | 7.2 |
| 1437612.0 | 6.6 |
| 1437779.0 | 6.8 |
| 1437801.0 | 6.6 |
| 1437870.0 | 6.8 |
| 1437914.0 | 6.9 |
| 1438050.0 | 5.9 |
| 1438106.0 | 7.1 |
| 1438348.0 | 7.3 |
| 1438444.0 | 5.5 |
| 1438531.0 | 6.7 |
| 1438620.0 | 5.9 |
| 1438632.0 | 5.7 |
| 1438692.0 | 6.6 |
| 1439112.0 | 7.4 |
| 1439204.0 | 6.3 |
| 1439239.0 | 5.8 |
| 1439490.0 | 5.8 |
| 1439808.0 | 6.7 |
| 1442374.0 | 6.6 |
| 1442597.0 | 6.0 |
| 1453152.0 | 5.7 |
| 1453626.0 | 6.1 |
| 1454034.0 | 6.3 |
| 1454152.0 | 7.5 |
| 1454274.0 | 6.0 |
| 1454308.0 | 5.7 |
| 1454411.0 | 5.9 |
| 1454532.0 | 7.0 |
| 1454772.0 | 5.6 |
| 1454832.0 | 6.7 |
| 1455094.0 | 7.0 |
| 1455356.0 | 7.1 |
| 1455744.0 | 6.9 |
| 1455753.0 | 7.3 |
| 1456492.0 | 7.0 |
| 1456528.0 | 5.5 |
| 1456708.0 | 5.9 |
| 1456894.0 | 6.9 |
| 1456922.0 | 5.7 |
| 1457279.0 | 6.2 |
| 1457352.0 | 6.7 |
| 1457712.0 | 7.1 |
| 1457772.0 | 6.0 |
| 1457954.0 | 6.7 |
| 1457981.0 | 6.6 |
| 1458049.0 | 7.2 |
| 1458057.0 | 6.0 |
| 1458143.0 | 7.4 |
| 1458286.0 | 7.1 |
| 1458415.0 | 7.4 |
| 1458588.0 | 5.7 |
| 1458852.0 | 5.9 |
| 1458912.0 | 6.2 |
| 1459032.0 | 6.5 |
| 1459212.0 | 7.3 |
| 1459272.0 | 6.4 |
| 1459457.0 | 6.0 |
| 1459690.0 | 6.6 |
| 1459902.0 | 6.3 |
| 1459958.0 | 7.1 |
| 1460352.0 | 7.2 |
| 1460412.0 | 7.3 |
| 1460472.0 | 6.1 |
| 1460652.0 | 7.5 |
| 1460712.0 | 5.8 |
| 1460790.0 | 6.2 |
| 1460863.0 | 6.3 |
| 1460921.0 | 5.8 |
| 1460960.0 | 6.7 |
| 1460994.0 | 7.3 |
| 1461085.0 | 7.3 |
| 1461311.0 | 7.4 |
| 1461452.0 | 6.9 |
| 1461482.0 | 7.4 |
| 1461620.0 | 6.4 |
| 1461648.0 | 7.3 |
| 1461852.0 | 7.4 |
| 1462032.0 | 6.5 |
| 1462152.0 | 6.8 |
| 1462319.0 | 6.1 |
| 1462355.0 | 5.7 |
| 1462423.0 | 5.9 |
| 1462596.0 | 6.5 |
| 1462745.0 | 5.8 |
| 1462882.0 | 5.6 |
| 1462935.0 | 5.9 |
| 1463093.0 | 6.5 |
| 1463172.0 | 5.8 |
| 1463232.0 | 6.8 |
| 1463292.0 | 7.5 |
| 1463412.0 | 6.4 |
| 1463742.0 | 7.2 |
| 1463786.0 | 7.3 |
| 1463814.0 | 5.9 |
| 1463876.0 | 6.2 |
| 1463903.0 | 7.3 |
| 1464048.0 | 6.7 |
| 1464170.0 | 7.4 |
| 1464320.0 | 6.9 |
| 1464367.0 | 7.4 |
| 1464417.0 | 7.2 |
| 1464462.0 | 7.4 |
| 1464538.0 | 7.1 |
| 1464852.0 | 6.6 |
| 1464912.0 | 6.9 |
| 1464972.0 | 7.4 |
| 1465234.0 | 7.0 |
| 1465378.0 | 5.7 |
| 1465531.0 | 6.5 |
| 1465690.0 | 7.4 |
| 1465798.0 | 6.4 |
| 1465854.0 | 5.7 |
| 1465883.0 | 7.2 |
| 1465992.0 | 6.4 |
| 1466232.0 | 7.3 |
| 1466292.0 | 7.3 |
| 1466352.0 | 7.4 |
| 1466612.0 | 7.3 |
| 1466643.0 | 6.1 |
| 1466713.0 | 6.7 |
| 1466896.0 | 6.7 |
| 1467240.0 | 6.4 |
| 1467432.0 | 6.9 |
| 1467492.0 | 6.8 |
| 1467672.0 | 6.6 |
| 1467732.0 | 6.5 |
| 1467792.0 | 5.9 |
| 1467852.0 | 6.8 |
| 1467912.0 | 7.3 |
| 1468018.0 | 6.3 |
| 1468092.0 | 6.8 |
| 1468217.0 | 5.9 |
| 1468305.0 | 6.5 |
| 1468332.0 | 6.0 |
| 1468434.0 | 7.5 |
| 1468492.0 | 7.1 |
| 1468596.0 | 6.8 |
| 1468673.0 | 6.3 |
| 1468701.0 | 6.7 |
| 1468792.0 | 6.0 |
| 1468833.0 | 5.7 |
| 1468992.0 | 7.2 |
| 1469052.0 | 6.8 |
| 1469112.0 | 6.2 |
| 1469172.0 | 6.5 |
| 1469232.0 | 5.6 |
| 1469554.0 | 6.6 |
| 1469734.0 | 7.1 |
| 1469824.0 | 5.9 |
| 1469835.0 | 6.7 |
| 1469894.0 | 5.8 |
| 1470142.0 | 5.7 |
| 1470312.0 | 5.8 |
| 1470432.0 | 5.6 |
| 1470873.0 | 5.8 |
| 1470984.0 | 7.0 |
| 1471099.0 | 7.1 |
| 1471204.0 | 5.7 |
| 1471254.0 | 7.0 |
| 1471331.0 | 7.4 |
| 1471425.0 | 6.1 |
| 1471560.0 | 7.2 |
| 1471570.0 | 5.8 |
| 1471658.0 | 6.6 |
| 1471722.0 | 6.7 |
| 1472172.0 | 5.9 |
| 1472232.0 | 6.1 |
| 1472367.0 | 6.9 |
| 1472387.0 | 5.9 |
| 1472446.0 | 6.7 |
| 1472596.0 | 5.8 |
| 1472733.0 | 5.7 |
| 1472778.0 | 6.4 |
| 1472784.0 | 5.9 |
| 1472822.0 | 6.4 |
| 1472904.0 | 7.4 |
| 1473135.0 | 6.1 |
| 1473155.0 | 6.7 |
| 1473192.0 | 7.1 |
| 1473252.0 | 6.7 |
| 1473312.0 | 6.3 |
| 1473552.0 | 6.4 |
| 1473612.0 | 7.1 |
| 1473672.0 | 7.2 |
| 1473759.0 | 6.1 |
| 1473775.0 | 5.9 |
| 1473804.0 | 6.6 |
| 1473850.0 | 6.4 |
| 1473982.0 | 6.8 |
| 1474097.0 | 7.1 |
| 1474135.0 | 7.0 |
| 1474215.0 | 7.2 |
| 1474344.0 | 5.7 |
| 1474506.0 | 5.7 |
| 1474565.0 | 5.8 |
| 1474614.0 | 6.9 |
| 1474632.0 | 6.9 |
| 1474752.0 | 7.4 |
| 1475507.0 | 7.1 |
| 1475524.0 | 6.8 |
| 1475712.0 | 6.8 |
| 1475952.0 | 7.4 |
| 1475984.0 | 6.9 |
| 1476072.0 | 6.3 |
| 1476132.0 | 6.2 |
| 1476192.0 | 5.7 |
| 1476312.0 | 6.0 |
| 1476432.0 | 7.3 |
| 1476727.0 | 5.6 |
| 1477239.0 | 6.3 |
| 1477252.0 | 7.5 |
| 1477264.0 | 5.8 |
| 1477512.0 | 6.9 |
| 1477572.0 | 5.8 |
| 1477692.0 | 6.7 |
| 1477752.0 | 6.0 |
| 1477872.0 | 6.4 |
| 1477932.0 | 7.4 |
| 1478206.0 | 7.2 |
| 1478422.0 | 6.1 |
| 1478712.0 | 6.3 |
| 1478735.0 | 6.3 |
| 1479252.0 | 7.2 |
| 1479432.0 | 6.6 |
| 1479716.0 | 6.3 |
| 1480002.0 | 6.4 |
| 1480632.0 | 5.5 |
| 1481327.0 | 5.9 |
| 1481358.0 | 7.4 |
| 1481761.0 | 6.8 |
| 1481952.0 | 6.9 |
| 1482252.0 | 6.8 |
| 1482634.0 | 6.6 |
| 1482769.0 | 5.6 |
| 1482872.0 | 7.3 |
| 1482974.0 | 6.2 |
| 1483039.0 | 7.2 |
| 1483165.0 | 5.8 |
| 1483272.0 | 6.1 |
| 1483752.0 | 6.5 |
| 1483889.0 | 7.2 |
| 1483917.0 | 6.0 |
| 1484073.0 | 6.3 |
| 1484219.0 | 5.9 |
| 1484272.0 | 7.3 |
| 1484285.0 | 5.8 |
| 1484342.0 | 7.5 |
| 1484438.0 | 5.9 |
| 1484528.0 | 7.0 |
| 1484612.0 | 7.4 |
| 1484637.0 | 5.7 |
| 1484712.0 | 6.7 |
| 1484772.0 | 6.7 |
| 1484832.0 | 6.6 |
| 1484952.0 | 5.8 |
| 1485192.0 | 7.4 |
| 1485467.0 | 6.0 |
| 1485507.0 | 5.9 |
| 1485708.0 | 7.0 |
| 1485783.0 | 6.8 |
| 1485972.0 | 7.5 |
| 1486002.0 | 7.0 |
| 1496712.0 | 7.4 |
| 1496796.0 | 6.7 |
| 1496808.0 | 7.4 |
| 1496825.0 | 5.6 |
| 1497215.0 | 6.4 |
| 1497512.0 | 6.9 |
| 1497792.0 | 6.6 |
| 1497852.0 | 6.7 |
| 1498092.0 | 7.4 |
| 1498339.0 | 5.6 |
| 1498477.0 | 6.0 |
| 1498909.0 | 6.0 |
| 1499412.0 | 6.6 |
| 1499472.0 | 7.0 |
| 1499652.0 | 6.5 |
| 1499719.0 | 6.8 |
| 1499757.0 | 6.0 |
| 1499855.0 | 7.0 |
| 1500150.0 | 6.9 |
| 1500172.0 | 5.9 |
| 1500392.0 | 6.0 |
| 1500421.0 | 5.8 |
| 1500492.0 | 5.7 |
| 1500537.0 | 7.0 |
| 1500612.0 | 7.2 |
| 1500672.0 | 5.9 |
| 1500972.0 | 7.4 |
| 1501032.0 | 7.1 |
| 1501182.0 | 5.7 |
| 1501472.0 | 5.7 |
| 1501565.0 | 6.1 |
| 1501658.0 | 5.7 |
| 1501724.0 | 6.0 |
| 1501874.0 | 6.2 |
| 1502232.0 | 6.7 |
| 1502292.0 | 6.4 |
| 1502816.0 | 6.9 |
| 1502909.0 | 6.6 |
| 1503142.0 | 6.9 |
| 1503172.0 | 7.0 |
| 1503256.0 | 7.3 |
| 1503432.0 | 7.1 |
| 1503612.0 | 7.3 |
| 1503672.0 | 7.1 |
| 1503732.0 | 6.1 |
| 1503792.0 | 7.4 |
| 1504164.0 | 5.9 |
| 1504197.0 | 7.3 |
| 1504284.0 | 6.8 |
| 1504351.0 | 6.6 |
| 1504403.0 | 6.3 |
| 1504459.0 | 5.6 |
| 1504592.0 | 6.3 |
| 1504625.0 | 7.2 |
| 1504678.0 | 6.9 |
| 1504710.0 | 5.8 |
| 1504872.0 | 7.4 |
| 1504932.0 | 6.5 |
| 1505172.0 | 6.5 |
| 1505292.0 | 6.4 |
| 1505352.0 | 5.9 |
| 1505594.0 | 5.6 |
| 1505665.0 | 5.6 |
| 1506040.0 | 6.6 |
| 1506191.0 | 7.5 |
| 1506672.0 | 5.6 |
| 1506922.0 | 7.2 |
| 1506949.0 | 5.7 |
| 1506959.0 | 6.2 |
| 1506996.0 | 7.1 |
| 1507062.0 | 5.6 |
| 1507231.0 | 6.9 |
| 1507255.0 | 7.2 |
| 1507306.0 | 7.3 |
| 1507471.0 | 7.2 |
| 1507545.0 | 5.7 |
| 1507595.0 | 6.2 |
| 1507717.0 | 6.4 |
| 1507740.0 | 6.5 |
| 1507812.0 | 7.0 |
| 1507872.0 | 6.6 |
| 1507932.0 | 7.1 |
| 1507992.0 | 6.8 |
| 1508112.0 | 6.2 |
| 1508172.0 | 7.2 |
| 1508232.0 | 6.0 |
| 1508312.0 | 6.5 |
| 1508345.0 | 6.2 |
| 1508407.0 | 5.8 |
| 1508451.0 | 6.2 |
| 1508497.0 | 5.6 |
| 1508509.0 | 6.7 |
| 1508547.0 | 7.4 |
| 1508572.0 | 5.7 |
| 1508615.0 | 5.6 |
| 1508622.0 | 7.2 |
| 1508704.0 | 6.9 |
| 1508752.0 | 6.6 |
| 1508886.0 | 6.9 |
| 1508912.0 | 6.0 |
| 1508937.0 | 6.2 |
| 1509042.0 | 6.9 |
| 1509095.0 | 7.3 |
| 1509130.0 | 7.2 |
| 1509252.0 | 7.0 |
| 1509312.0 | 7.4 |
| 1509372.0 | 6.8 |
| 1509432.0 | 6.5 |
| 1509492.0 | 6.4 |
| 1509552.0 | 7.4 |
| 1509612.0 | 7.4 |
| 1509672.0 | 5.7 |
| 1509740.0 | 6.3 |
| 1509916.0 | 7.4 |
| 1509950.0 | 5.6 |
| 1509989.0 | 7.0 |
| 1510074.0 | 6.1 |
| 1510119.0 | 6.8 |
| 1510172.0 | 7.0 |
| 1510544.0 | 6.0 |
| 1510692.0 | 7.4 |
| 1510752.0 | 6.3 |
| 1510812.0 | 5.8 |
| 1510872.0 | 7.2 |
| 1510992.0 | 6.4 |
| 1511310.0 | 6.5 |
| 1511325.0 | 6.1 |
| 1512020.0 | 6.0 |
| 1512372.0 | 7.1 |
| 1512492.0 | 7.4 |
| 1512552.0 | 6.9 |
| 1512907.0 | 5.9 |
| 1513191.0 | 7.3 |
| 1513440.0 | 6.7 |
| 1513572.0 | 6.9 |
| 1513752.0 | 6.6 |
| 1513812.0 | 6.3 |
| 1513872.0 | 6.9 |
| 1513932.0 | 5.8 |
| 1514060.0 | 7.2 |
| 1514172.0 | 6.2 |
| 1514200.0 | 5.5 |
| 1514304.0 | 6.5 |
| 1514361.0 | 7.5 |
| 1514487.0 | 6.3 |
| 1514507.0 | 6.1 |
| 1514715.0 | 6.7 |
| 1514762.0 | 5.8 |
| 1514824.0 | 6.5 |
| 1514946.0 | 6.1 |
| 1515192.0 | 7.2 |
| 1515252.0 | 7.0 |
| 1515842.0 | 5.7 |
| 1516008.0 | 6.4 |
| 1516139.0 | 6.5 |
| 1516239.0 | 6.5 |
| 1516572.0 | 5.5 |
| 1517408.0 | 7.1 |
| 1517432.0 | 6.5 |
| 1518012.0 | 5.9 |
| 1518192.0 | 5.6 |
| 1518493.0 | 6.1 |
| 1518590.0 | 6.6 |
| 1518873.0 | 7.2 |
| 1518992.0 | 6.4 |
| 1519692.0 | 5.8 |
| 1519886.0 | 5.5 |
| 1520183.0 | 7.0 |
| 1521192.0 | 6.2 |
| 1523405.0 | 6.8 |
| 1523712.0 | 5.6 |
| 1524642.0 | 6.6 |
| 1525092.0 | 7.0 |
| 1525624.0 | 6.4 |
| 1525631.0 | 5.5 |
| 1525755.0 | 7.1 |
| 1525904.0 | 5.6 |
| 1526095.0 | 6.2 |
| 1526177.0 | 7.3 |
| 1526239.0 | 6.9 |
| 1526389.0 | 5.6 |
| 1526472.0 | 7.3 |
| 1526532.0 | 5.6 |
| 1526652.0 | 6.3 |
| 1526712.0 | 5.7 |
| 1526772.0 | 6.3 |
| 1526892.0 | 6.1 |
| 1526952.0 | 6.4 |
| 1527061.0 | 5.8 |
| 1527093.0 | 6.1 |
| 1527281.0 | 6.1 |
| 1527377.0 | 6.5 |
| 1527399.0 | 6.8 |
| 1527481.0 | 6.8 |
| 1527501.0 | 6.5 |
| 1527563.0 | 5.8 |
| 1527594.0 | 7.1 |
| 1527658.0 | 6.2 |
| 1527704.0 | 6.7 |
| 1527790.0 | 6.3 |
| 1538877.0 | 5.9 |
| 1538900.0 | 6.9 |
| 1539120.0 | 5.9 |
| 1539145.0 | 7.4 |
| 1539279.0 | 6.0 |
| 1540587.0 | 6.3 |
| 1540932.0 | 7.5 |
| 1541352.0 | 7.0 |
| 1541752.0 | 5.7 |
| 1542058.0 | 7.2 |
| 1542261.0 | 5.6 |
| 1542372.0 | 7.4 |
| 1542492.0 | 6.5 |
| 1542959.0 | 5.8 |
| 1543342.0 | 6.3 |
| 1543752.0 | 7.0 |
| 1543812.0 | 5.5 |
| 1543872.0 | 5.9 |
| 1543992.0 | 6.4 |
| 1544112.0 | 7.0 |
| 1544172.0 | 6.7 |
| 1544545.0 | 6.1 |
| 1544637.0 | 7.2 |
| 1544802.0 | 7.5 |
| 1544823.0 | 5.8 |
| 1544946.0 | 5.5 |
| 1545012.0 | 7.1 |
| 1545160.0 | 5.7 |
| 1545372.0 | 7.0 |
| 1545492.0 | 6.2 |
| 1545552.0 | 6.5 |
| 1545778.0 | 5.9 |
| 1545897.0 | 7.1 |
| 1546030.0 | 7.1 |
| 1546197.0 | 6.3 |
| 1546568.0 | 7.0 |
| 1546692.0 | 6.6 |
| 1546932.0 | 7.3 |
| 1547112.0 | 6.6 |
| 1547272.0 | 5.9 |
| 1547433.0 | 7.3 |
| 1547509.0 | 7.4 |
| 1547519.0 | 6.4 |
| 1547635.0 | 6.3 |
| 1547767.0 | 7.0 |
| 1547797.0 | 6.2 |
| 1547835.0 | 5.8 |
| 1547862.0 | 6.6 |
| 1547899.0 | 7.4 |
| 1548000.0 | 6.6 |
| 1548072.0 | 7.0 |
| 1548132.0 | 6.7 |
| 1548192.0 | 6.2 |
| 1548252.0 | 6.7 |
| 1548432.0 | 7.1 |
| 1548678.0 | 6.4 |
| 1548871.0 | 7.5 |
| 1548901.0 | 5.5 |
| 1548966.0 | 6.3 |
| 1548989.0 | 6.2 |
| 1549066.0 | 6.1 |
| 1549125.0 | 5.9 |
| 1549192.0 | 6.7 |
| 1549399.0 | 6.6 |
| 1549472.0 | 7.5 |
| 1549499.0 | 5.7 |
| 1549572.0 | 5.8 |
| 1549632.0 | 5.8 |
| 1549812.0 | 7.2 |
| 1549872.0 | 6.5 |
| 1549932.0 | 7.0 |
| 1549992.0 | 5.8 |
| 1550239.0 | 5.5 |
| 1550314.0 | 5.5 |
| 1550407.0 | 6.1 |
| 1550544.0 | 7.0 |
| 1550774.0 | 7.5 |
| 1550792.0 | 6.3 |
| 1550946.0 | 6.6 |
| 1550952.0 | 6.3 |
| 1551132.0 | 5.7 |
| 1551192.0 | 5.6 |
| 1551252.0 | 6.3 |
| 1551372.0 | 7.0 |
| 1551564.0 | 5.8 |
| 1551717.0 | 6.2 |
| 1551950.0 | 6.3 |
| 1551974.0 | 5.8 |
| 1552076.0 | 6.8 |
| 1552217.0 | 6.6 |
| 1552247.0 | 7.4 |
| 1552632.0 | 6.7 |
| 1553122.0 | 6.5 |
| 1553257.0 | 5.9 |
| 1553337.0 | 5.6 |
| 1553401.0 | 6.7 |
| 1553427.0 | 6.8 |
| 1553599.0 | 5.8 |
| 1553629.0 | 5.8 |
| 1553952.0 | 6.8 |
| 1554132.0 | 6.5 |
| 1554192.0 | 5.9 |
| 1554420.0 | 7.2 |
| 1554627.0 | 6.3 |
| 1554884.0 | 7.2 |
| 1554991.0 | 6.9 |
| 1555095.0 | 7.2 |
| 1555392.0 | 7.3 |
| 1555857.0 | 5.9 |
| 1556038.0 | 6.1 |
| 1556065.0 | 7.0 |
| 1556114.0 | 7.0 |
| 1556222.0 | 6.8 |
| 1556366.0 | 5.5 |
| 1556707.0 | 5.8 |
| 1556952.0 | 7.0 |
| 1557335.0 | 7.4 |
| 1557461.0 | 7.3 |
| 1557552.0 | 6.9 |
| 1557912.0 | 6.1 |
| 1558043.0 | 5.6 |
| 1558212.0 | 6.3 |
| 1559044.0 | 6.7 |
| 1559064.0 | 5.5 |
| 1559357.0 | 5.6 |
| 1559832.0 | 7.0 |
| 1560860.0 | 6.9 |
| 1561634.0 | 7.1 |
| 1562592.0 | 6.9 |
| 1564212.0 | 6.2 |
| 1564604.0 | 5.9 |
| 1565412.0 | 5.9 |
| 1565772.0 | 6.1 |
| 1566972.0 | 6.0 |
| 1567092.0 | 7.4 |
| 1567152.0 | 5.5 |
| 1567272.0 | 6.4 |
| 1567424.0 | 6.4 |
| 1567472.0 | 5.7 |
| 1567504.0 | 5.6 |
| 1567592.0 | 7.5 |
| 1567720.0 | 5.7 |
| 1567972.0 | 6.6 |
| 1568047.0 | 7.1 |
| 1568087.0 | 6.3 |
| 1568141.0 | 6.4 |
| 1568163.0 | 7.0 |
| 1568185.0 | 6.1 |
| 1568217.0 | 7.1 |
| 1568292.0 | 5.6 |
| 1568352.0 | 7.3 |
| 1568472.0 | 6.1 |
| 1568782.0 | 5.9 |
| 1568817.0 | 6.1 |
| 1568840.0 | 5.9 |
| 1568863.0 | 7.4 |
| 1568936.0 | 6.7 |
| 1568974.0 | 7.0 |
| 1569054.0 | 5.7 |
| 1569058.0 | 7.1 |
| 1569093.0 | 5.5 |
| 1569150.0 | 6.6 |
| 1569181.0 | 5.7 |
| 1569263.0 | 6.4 |
| 1569274.0 | 5.7 |
| 1569296.0 | 7.2 |
| 1569304.0 | 6.1 |
| 1569333.0 | 5.6 |
| 1569437.0 | 5.8 |
| 1569577.0 | 5.7 |
| 1569672.0 | 6.6 |
| 1569852.0 | 7.2 |
| 1569912.0 | 7.1 |
| 1570032.0 | 5.5 |
| 1570152.0 | 6.8 |
| 1570257.0 | 5.5 |
| 1570344.0 | 6.0 |
| 1570452.0 | 7.3 |
| 1570556.0 | 5.7 |
| 1570590.0 | 7.3 |
| 1570704.0 | 6.5 |
| 1570712.0 | 5.8 |
| 1570734.0 | 6.6 |
| 1570855.0 | 5.7 |
| 1570974.0 | 5.9 |
| 1570999.0 | 6.4 |
| 1581192.0 | 7.1 |
| 1581252.0 | 7.1 |
| 1583970.0 | 5.6 |
| 1584372.0 | 7.3 |
| 1584931.0 | 5.6 |
| 1584984.0 | 5.7 |
| 1585156.0 | 6.1 |
| 1585381.0 | 7.0 |
| 1585632.0 | 6.6 |
| 1585752.0 | 6.8 |
| 1585812.0 | 5.7 |
| 1586245.0 | 6.3 |
| 1586308.0 | 6.8 |
| 1586607.0 | 6.6 |
| 1586623.0 | 6.7 |
| 1586641.0 | 5.9 |
| 1587072.0 | 6.0 |
| 1588169.0 | 5.5 |
| 1588244.0 | 6.5 |
| 1588325.0 | 5.9 |
| 1588334.0 | 6.1 |
| 1588452.0 | 6.8 |
| 1588572.0 | 5.7 |
| 1588812.0 | 7.3 |
| 1589099.0 | 6.6 |
| 1589143.0 | 7.3 |
| 1589152.0 | 6.8 |
| 1589230.0 | 5.8 |
| 1589315.0 | 6.4 |
| 1589390.0 | 7.2 |
| 1589558.0 | 7.0 |
| 1589832.0 | 6.5 |
| 1590072.0 | 6.3 |
| 1590132.0 | 6.9 |
| 1590192.0 | 6.5 |
| 1590312.0 | 6.5 |
| 1590383.0 | 6.0 |
| 1590442.0 | 7.1 |
| 1590549.0 | 6.3 |
| 1590638.0 | 6.3 |
| 1590784.0 | 5.6 |
| 1590801.0 | 5.7 |
| 1590856.0 | 7.1 |
| 1590931.0 | 5.9 |
| 1591062.0 | 5.7 |
| 1591127.0 | 6.1 |
| 1591812.0 | 7.5 |
| 1592079.0 | 6.8 |
| 1592132.0 | 5.9 |
| 1592202.0 | 6.2 |
| 1592332.0 | 6.5 |
| 1592574.0 | 5.9 |
| 1592712.0 | 6.8 |
| 1592832.0 | 6.1 |
| 1592952.0 | 7.0 |
| 1593012.0 | 5.7 |
| 1593321.0 | 7.0 |
| 1593333.0 | 5.5 |
| 1593771.0 | 7.1 |
| 1593815.0 | 5.7 |
| 1593838.0 | 6.1 |
| 1593942.0 | 6.0 |
| 1593998.0 | 5.9 |
| 1594090.0 | 6.4 |
| 1594332.0 | 7.4 |
| 1594779.0 | 6.7 |
| 1594857.0 | 6.4 |
| 1594935.0 | 6.6 |
| 1595179.0 | 6.1 |
| 1595237.0 | 6.0 |
| 1595275.0 | 7.2 |
| 1595382.0 | 6.7 |
| 1595453.0 | 7.1 |
| 1595652.0 | 5.6 |
| 1595892.0 | 5.9 |
| 1596012.0 | 5.6 |
| 1596180.0 | 6.4 |
| 1596269.0 | 6.8 |
| 1596547.0 | 6.4 |
| 1597092.0 | 5.6 |
| 1597272.0 | 6.1 |
| 1597392.0 | 5.7 |
| 1597662.0 | 5.9 |
| 1597679.0 | 7.4 |
| 1598100.0 | 6.9 |
| 1598302.0 | 6.7 |
| 1598352.0 | 6.2 |
| 1598451.0 | 6.9 |
| 1598592.0 | 6.0 |
| 1598652.0 | 6.7 |
| 1599036.0 | 6.4 |
| 1599452.0 | 6.2 |
| 1599640.0 | 6.9 |
| 1600272.0 | 6.4 |
| 1601002.0 | 6.6 |
| 1601472.0 | 7.4 |
| 1602295.0 | 6.1 |
| 1602560.0 | 6.7 |
| 1602972.0 | 6.8 |
| 1603649.0 | 6.9 |
| 1604073.0 | 6.2 |
| 1604185.0 | 7.2 |
| 1604221.0 | 6.2 |
| 1604352.0 | 7.4 |
| 1604592.0 | 5.8 |
| 1604712.0 | 5.7 |
| 1605113.0 | 6.1 |
| 1605571.0 | 7.2 |
| 1605972.0 | 6.9 |
| 1606032.0 | 7.1 |
| 1606361.0 | 5.8 |
| 1606415.0 | 5.9 |
| 1606672.0 | 6.7 |
| 1606712.0 | 5.8 |
| 1606834.0 | 7.4 |
| 1606856.0 | 6.6 |
| 1606907.0 | 7.2 |
| 1606989.0 | 5.5 |
| 1607059.0 | 7.2 |
| 1607079.0 | 6.0 |
| 1607112.0 | 7.5 |
| 1607172.0 | 7.2 |
| 1607232.0 | 6.9 |
| 1607292.0 | 6.2 |
| 1607352.0 | 6.9 |
| 1607412.0 | 7.1 |
| 1607472.0 | 5.8 |
| 1607657.0 | 7.3 |
| 1607715.0 | 6.7 |
| 1607730.0 | 6.1 |
| 1607811.0 | 7.0 |
| 1607837.0 | 6.7 |
| 1607867.0 | 6.9 |
| 1607918.0 | 7.2 |
| 1607950.0 | 6.0 |
| 1607968.0 | 5.6 |
| 1607987.0 | 5.6 |
| 1607990.0 | 5.6 |
| 1608224.0 | 5.5 |
| 1608477.0 | 5.6 |
| 1608552.0 | 6.5 |
| 1608672.0 | 6.9 |
| 1608732.0 | 6.4 |
| 1608792.0 | 6.3 |
| 1609220.0 | 7.4 |
| 1609272.0 | 6.7 |
| 1609340.0 | 7.1 |
| 1609460.0 | 6.9 |
| 1609482.0 | 6.6 |
| 1609817.0 | 6.5 |
| 1610352.0 | 6.8 |
| 1610970.0 | 7.2 |
| 1611024.0 | 7.5 |
| 1611136.0 | 6.2 |
| 1611250.0 | 6.0 |
| 1611282.0 | 6.2 |
| 1611432.0 | 5.7 |
| 1611552.0 | 6.3 |
| 1611612.0 | 7.1 |
| 1611672.0 | 6.5 |
| 1611732.0 | 6.9 |
| 1611792.0 | 5.6 |
| 1611852.0 | 6.3 |
| 1611987.0 | 6.9 |
| 1612109.0 | 6.1 |
| 1612135.0 | 5.5 |
| 1612181.0 | 6.2 |
| 1612314.0 | 7.0 |
| 1612355.0 | 6.0 |
| 1612390.0 | 6.1 |
| 1612521.0 | 6.4 |
| 1612635.0 | 7.4 |
| 1612779.0 | 5.6 |
| 1623132.0 | 6.6 |
| 1623252.0 | 7.4 |
| 1623602.0 | 6.7 |
| 1623882.0 | 7.0 |
| 1623955.0 | 6.1 |
| 1624007.0 | 6.3 |
| 1624632.0 | 6.0 |
| 1624812.0 | 6.9 |
| 1625062.0 | 6.6 |
| 1625115.0 | 7.4 |
| 1626192.0 | 7.1 |
| 1626378.0 | 6.4 |
| 1627141.0 | 6.7 |
| 1627252.0 | 6.2 |
| 1627392.0 | 5.5 |
| 1627888.0 | 6.8 |
| 1627960.0 | 7.0 |
| 1628077.0 | 6.6 |
| 1628134.0 | 7.5 |
| 1628178.0 | 5.9 |
| 1628217.0 | 6.9 |
| 1628243.0 | 7.0 |
| 1628304.0 | 7.1 |
| 1628398.0 | 6.1 |
| 1628626.0 | 6.7 |
| 1628659.0 | 7.5 |
| 1629012.0 | 6.7 |
| 1629430.0 | 5.7 |
| 1629542.0 | 7.3 |
| 1630045.0 | 5.5 |
| 1630212.0 | 7.2 |
| 1630743.0 | 5.8 |
| 1631104.0 | 7.5 |
| 1631352.0 | 6.3 |
| 1631528.0 | 6.9 |
| 1631652.0 | 6.7 |
| 1631772.0 | 6.0 |
| 1632072.0 | 6.3 |
| 1632992.0 | 6.7 |
| 1633272.0 | 6.5 |
| 1633512.0 | 6.8 |
| 1633608.0 | 7.2 |
| 1634278.0 | 7.3 |
| 1634362.0 | 6.5 |
| 1634652.0 | 7.0 |
| 1634712.0 | 6.6 |
| 1635209.0 | 5.8 |
| 1635285.0 | 5.7 |
| 1635668.0 | 5.7 |
| 1635886.0 | 6.5 |
| 1635912.0 | 6.9 |
| 1636572.0 | 6.9 |
| 1637172.0 | 6.8 |
| 1637226.0 | 7.0 |
| 1637954.0 | 7.2 |
| 1638057.0 | 6.4 |
| 1639152.0 | 7.4 |
| 1639486.0 | 7.4 |
| 1639508.0 | 7.3 |
| 1639598.0 | 6.8 |
| 1639999.0 | 6.7 |
| 1640155.0 | 5.5 |
| 1640292.0 | 5.9 |
| 1640592.0 | 7.3 |
| 1640972.0 | 5.5 |
| 1642032.0 | 7.4 |
| 1642252.0 | 5.8 |
| 1643014.0 | 6.7 |
| 1643733.0 | 5.7 |
| 1644612.0 | 6.9 |
| 1644792.0 | 6.6 |
| 1644852.0 | 7.5 |
| 1645196.0 | 6.9 |
| 1645279.0 | 7.0 |
| 1645652.0 | 6.6 |
| 1645992.0 | 6.2 |
| 1646052.0 | 6.3 |
| 1646112.0 | 5.5 |
| 1646292.0 | 6.6 |
| 1646352.0 | 6.6 |
| 1646472.0 | 7.1 |
| 1646580.0 | 6.5 |
| 1646607.0 | 6.5 |
| 1646632.0 | 5.8 |
| 1646763.0 | 7.4 |
| 1647432.0 | 5.8 |
| 1647612.0 | 6.0 |
| 1648133.0 | 7.4 |
| 1648455.0 | 6.8 |
| 1648681.0 | 5.7 |
| 1648769.0 | 7.5 |
| 1649052.0 | 5.7 |
| 1649519.0 | 6.0 |
| 1650063.0 | 7.5 |
| 1650672.0 | 5.6 |
| 1651056.0 | 5.5 |
| 1651269.0 | 5.7 |
| 1652052.0 | 7.4 |
| 1652112.0 | 7.2 |
| 1652435.0 | 6.8 |
| 1652545.0 | 7.1 |
| 1652570.0 | 6.1 |
| 1652971.0 | 5.9 |
| 1653069.0 | 7.2 |
| 1653104.0 | 6.6 |
| 1653192.0 | 6.4 |
| 1653252.0 | 6.7 |
| 1653372.0 | 5.8 |
| 1653432.0 | 6.5 |
| 1653552.0 | 7.0 |
| 1653672.0 | 7.1 |
| 1653757.0 | 6.6 |
| 1653805.0 | 6.7 |
| 1653857.0 | 6.8 |
| 1654032.0 | 7.5 |
| 1654051.0 | 6.0 |
| 1654221.0 | 6.0 |
| 1654309.0 | 5.9 |
| 1654400.0 | 6.4 |
| 1654531.0 | 7.3 |
| 1654589.0 | 5.8 |
| 1654627.0 | 6.7 |
| 1654632.0 | 6.3 |
| 1654752.0 | 7.4 |
| 1654872.0 | 7.0 |
| 1654992.0 | 7.3 |
| 1655172.0 | 7.4 |
| 1655282.0 | 6.1 |
| 1655533.0 | 6.8 |
| 1655600.0 | 5.9 |
| 1655667.0 | 5.6 |

1655813.0 6.3
1655865.0 7.1
1666152.0 6.2
1666212.0 7.5
1666452.0 6.9
1666572.0 5.8
1666849.0 7.3
1666966.0 7.4
1667832.0 6.0
1668012.0 6.0
1668408.0 5.7
1668591.0 6.2
1668638.0 6.8
1668812.0 6.7
1669332.0 6.1
1669747.0 6.1
1669868.0 7.1
1670472.0 7.0
1670592.0 5.7
1670652.0 7.3
1671064.0 5.9
1671121.0 5.6
1671462.0 6.3
1671546.0 5.6
1671912.0 7.4
1672332.0 6.7
1672392.0 7.3
1672627.0 5.9
1672674.0 5.9
1672915.0 5.6
1673107.0 6.9
1673192.0 6.0
1674317.0 5.6
1674780.0 6.0
1675092.0 6.9
1675152.0 6.5
1675399.0 7.1
1675763.0 5.9
1676950.0 5.7
1677477.0 7.0
1678292.0 6.0
1678636.0 6.4
1679232.0 5.5
1680044.0 6.4
1681414.0 6.5
1681514.0 6.3
1682172.0 5.6
1682472.0 6.8
1682687.0 7.3
1682794.0 5.9
1682857.0 6.2
1683032.0 7.3
1683127.0 5.7
1683287.0 6.7
1683330.0 5.9
1683612.0 6.8
1683792.0 5.7
1684233.0 5.5
1684300.0 7.0
1684326.0 7.1
1684800.0 7.3
1684932.0 7.1
1685052.0 5.7
1685232.0 6.9
1685502.0 5.9
1685650.0 6.7
1685789.0 7.0
1685870.0 5.9
1686094.0 6.9
1686152.0 6.7
1686202.0 7.3
1686260.0 6.0
1686732.0 6.8
1686792.0 7.0
1687431.0 6.5
1687752.0 6.1
1687992.0 7.1
1688052.0 6.4
1688112.0 6.2
1688232.0 5.6
1688450.0 6.1
1688484.0 7.4
1689005.0 6.4
1689164.0 5.6
1689192.0 6.7
1689672.0 6.1
1689737.0 6.3
1689817.0 7.2
1689887.0 5.9
1690078.0 6.7
1690120.0 6.8
1690165.0 5.6
1690692.0 6.7
1691445.0 7.3
1691528.0 7.0
1692649.0 7.3
1692739.0 5.7
1692820.0 5.8
1693331.0 5.9
1693377.0 7.4
1693632.0 7.4
1693812.0 5.7
1693872.0 7.2
1693992.0 7.1
1694080.0 7.2
1694133.0 6.3
1694265.0 6.5
1694350.0 7.0
1694415.0 5.8
1694794.0 6.8
1695132.0 6.5
1695694.0 6.1
1696320.0 5.5
1696392.0 7.3
1696452.0 6.1
1696632.0 5.8
1696812.0 7.1
1696942.0 5.9
1697003.0 7.4
1697220.0 5.6
1697279.0 5.8
1697562.0 6.0
1707912.0 5.8
1708032.0 5.9
1708479.0 6.7
1709197.0 7.4
1709242.0 6.8
1709292.0 7.1
1709592.0 6.7
1709772.0 6.7
1709917.0 6.6
1709965.0 6.6
1710342.0 7.2
1710489.0 6.7
1710792.0 7.2
1711092.0 6.3
1711566.0 7.0
1712292.0 7.4
1712412.0 6.6
1712532.0 6.2
1712652.0 6.3
1712853.0 6.6
1713171.0 6.3
1713223.0 5.6
1713311.0 6.1
1713732.0 6.0
1713792.0 7.0
1713912.0 6.3
1714032.0 5.7
1714152.0 6.7
1714347.0 6.4
1714467.0 6.0
1714577.0 5.8
1714989.0 5.8
1715097.0 6.5
1715172.0 6.9
1715472.0 6.0
1715739.0 7.1
1715825.0 7.0
1715980.0 5.7
1716540.0 5.7
1716612.0 6.3
1716672.0 7.3
1716732.0 7.4
1716972.0 6.3
1717328.0 6.7
1717355.0 7.5
1717992.0 6.8
1718052.0 7.2
1718472.0 7.4
1718781.0 6.3
1718929.0 5.8
1719047.0 5.8
1719386.0 5.7
1719552.0 6.6
1719852.0 5.9
1720154.0 6.4
1720195.0 5.9
1720281.0 6.4
1720484.0 7.0
1720750.0 6.2
1720872.0 6.4
1721232.0 6.6
1721352.0 6.1
1721739.0 5.5
1721806.0 7.3
1722022.0 6.4
1722142.0 5.6
1722170.0 6.6
1722279.0 6.0
1722300.0 6.1
1722672.0 7.0
1722907.0 5.7
1723122.0 6.3
1723440.0 6.5
1723472.0 7.1
1723497.0 7.0
1723577.0 6.8
1723584.0 6.0
1723608.0 6.4
1723691.0 6.5
1723735.0 6.3
1723752.0 5.7
1723812.0 7.0
1723932.0 7.5
1724052.0 7.3
1724172.0 7.4
1724232.0 7.4
1724431.0 5.6
1724464.0 6.4
1724509.0 6.1
1724539.0 6.4
1724596.0 7.0
1724672.0 5.8
1724690.0 5.6
1724782.0 6.5
1724854.0 6.3
1724958.0 6.6
1725022.0 6.8
1725080.0 6.5
1725125.0 5.6
1725192.0 6.1
1725312.0 5.9
1725432.0 6.0
1725612.0 6.8
1725948.0 6.4
1725988.0 6.9
1726110.0 5.5
1726139.0 5.6
1726216.0 7.4
1726325.0 7.4
1726386.0 6.5
1726430.0 7.3
1726872.0 7.1
1727112.0 5.9
1727414.0 6.1
1727485.0 6.8
1727520.0 6.2
1727647.0 7.0
1727758.0 6.0
1727788.0 6.4
1727878.0 7.2
1727982.0 6.0
1728072.0 6.8
1728132.0 6.7
1728252.0 6.6
1728312.0 5.7
1728372.0 6.6
1728432.0 6.6
1728492.0 6.8
1728552.0 6.5
1728686.0 6.9
1729000.0 6.7
1729112.0 6.0
1729132.0 6.7
1729182.0 5.9
1729197.0 6.7
1729301.0 7.1
1729692.0 5.9
1729752.0 6.2
1729872.0 7.3
1729932.0 6.4
1729992.0 5.9
1730108.0 5.7
1730160.0 6.0
1730190.0 5.7
1730242.0 6.2
1730468.0 7.4
1730500.0 6.3
1730544.0 7.3
1730566.0 7.3
1730760.0 7.0
1730844.0 6.4
1730952.0 6.1
1731012.0 7.4
1731132.0 6.9
1731252.0 6.5
1731312.0 5.8
1731432.0 5.6
1731510.0 5.8
1731638.0 7.2
1731837.0 7.5
1732124.0 7.3
1732156.0 6.5
1732392.0 7.1
1732452.0 7.0
1732572.0 6.3
1732752.0 7.2
1732872.0 7.1
1732996.0 6.6
1733226.0 7.4
1733544.0 7.0
1733600.0 5.7
1733637.0 6.2
1733724.0 6.6
1733767.0 5.7
1733892.0 7.4
1733952.0 6.8
1734012.0 7.3
1734072.0 6.5
1734252.0 6.6
1734312.0 6.3
1734418.0 7.0
1734577.0 7.5
1734682.0 7.0
1734781.0 7.3
1734818.0 6.9
1734979.0 6.7
1735332.0 6.7
1735392.0 6.9
1735452.0 5.8
1735512.0 6.6
1735905.0 6.5
1736510.0 6.9
1736612.0 6.0
1736772.0 6.1
1737279.0 6.5
1737613.0 6.6
1737648.0 7.2
1737882.0 6.3
1737954.0 5.8
1738051.0 6.0
1738152.0 5.9
1738272.0 6.3
1738392.0 5.6
1738512.0 7.1
1738572.0 6.8
1738632.0 6.9
1738801.0 6.7
1738932.0 7.3
1739050.0 6.2
1739100.0 6.7
1739214.0 6.9
1739298.0 5.9
1739386.0 5.6
1739449.0 5.9
1749644.0 7.4
1749972.0 6.5
1750289.0 7.5
1750859.0 6.8
1751018.0 6.6
1751112.0 6.7
1751172.0 6.8
1752060.0 6.4
1752518.0 7.2
1753322.0 5.6
1753794.0 5.9
1753890.0 6.2
1755283.0 6.0
1755326.0 6.9
1755416.0 6.2
1756431.0 6.2
1756638.0 6.2
1756823.0 6.1
1756872.0 6.2
1758144.0 6.6
1758179.0 7.2
1759193.0 6.1
1759245.0 6.9
1760232.0 6.3
1762632.0 5.9
1762812.0 7.0
1762872.0 6.6
1762932.0 6.1
1762992.0 6.0
1763175.0 6.3
1763209.0 6.9
1763245.0 6.7
1763503.0 6.6
1763677.0 5.9
1764034.0 7.0
1764132.0 6.6
1764192.0 5.7
1764312.0 5.7
1764372.0 5.8
1764432.0 6.7
1764492.0 5.7
1764552.0 6.4
1764907.0 6.3
1765028.0 6.1
1765123.0 6.2
1765289.0 5.5
1765346.0 6.3
1765441.0 6.0
1765692.0 6.7
1765752.0 6.9
1765812.0 6.0
1765992.0 6.3
1766091.0 5.6
1766455.0 7.4
1766659.0 6.9
1767012.0 5.5
1767072.0 7.4
1767312.0 6.2
1767372.0 7.0
1767432.0 7.4
1767602.0 7.5
1767701.0 7.1
1767841.0 7.4
1767854.0 6.7
1768282.0 6.6
1768315.0 6.1
1768872.0 7.5
1769273.0 5.8
1769395.0 6.1
1769488.0 7.1
1769532.0 6.2
1769587.0 6.6
1769607.0 5.9
1769952.0 7.2
1770012.0 7.4
1770072.0 6.1
1770132.0 6.9
1770477.0 5.9
1770610.0 7.2
1770972.0 6.0
1771142.0 6.6
1771272.0 5.7
1771332.0 6.3
1771392.0 6.7
1771512.0 7.3
1771632.0 6.3
1772041.0 5.7
1772162.0 6.9
1772415.0 6.0
1772952.0 6.7
1773012.0 7.1
1773314.0 7.2
1773387.0 5.8
1773720.0 6.1
1773732.0 7.1
1774047.0 6.6
1774119.0 5.6
1774152.0 6.8
1774452.0 5.6
1774816.0 6.1
1775074.0 5.8
1775554.0 5.6
1775652.0 7.1
1776559.0 6.2
1777212.0 6.3
1777607.0 6.9
1778042.0 5.6
1778379.0 5.8
1778411.0 7.0
1778832.0 7.1
1778892.0 5.5
1779395.0 6.6
1779578.0 7.2
1779632.0 5.7
1779672.0 6.4
1780272.0 7.3
1780766.0 6.7
1780781.0 7.2
1780811.0 6.6
1780897.0 7.1
1781032.0 5.8
1781204.0 7.0
1781532.0 6.3
1781592.0 5.7
1781712.0 7.1
1781955.0 6.9
1782130.0 6.6
1782327.0 7.1
1782525.0 5.6
1782610.0 6.8
1782731.0 5.8
1793112.0 6.7
1793172.0 7.3
1793451.0 6.1
1793742.0 6.1
1793824.0 7.4
1793917.0 7.1
1795049.0 5.7
1795160.0 6.8
1797252.0 5.9
1798992.0 6.7
1799422.0 7.1
1800372.0 6.5
1800432.0 5.9
1800640.0 6.7
1800730.0 6.4
1800789.0 6.7
1800816.0 7.1
1801009.0 6.1
1801152.0 5.6
1801267.0 7.5
1801433.0 7.3
1801692.0 6.3
1801932.0 5.9
1802160.0 6.4
1802331.0 6.0
1802442.0 5.7
1802538.0 7.1
1802578.0 7.4
1802689.0 6.3
1802797.0 7.4
1802862.0 5.8
1802910.0 7.3
1803012.0 7.2
1803192.0 6.1
1803252.0 6.4
1803312.0 7.0
1803372.0 5.7
1803432.0 6.7
1803508.0 6.2
1803512.0 7.2
1803584.0 5.7
1803620.0 7.3
1803634.0 6.9
1803714.0 6.2
1803800.0 5.8
1803870.0 6.7
1803892.0 6.3
1803907.0 6.7
1804009.0 6.7
1804041.0 7.5
1804060.0 7.2
1804114.0 7.4
1804132.0 5.5
1804234.0 6.6
1804254.0 6.0
1804512.0 6.5
1804572.0 7.5
1804632.0 7.4
1804752.0 6.9
1804812.0 7.3
1805025.0 6.7
1805095.0 6.6
1805207.0 6.8
1805324.0 6.9
1805400.0 6.9
1805648.0 6.5
1805721.0 5.7
1805754.0 6.2
1805832.0 5.7
1805892.0 5.7
1806192.0 6.7
1806410.0 6.2
1806495.0 6.6
1806720.0 6.3
1806779.0 7.2
1806868.0 7.4
1806917.0 7.1
1807240.0 6.0
1807272.0 5.6
1807392.0 6.0
1807452.0 7.2
1807572.0 7.3
1807632.0 5.7
1807692.0 6.8
1807867.0 7.3
1808122.0 6.8
1808522.0 6.2
1808618.0 7.3
1808647.0 5.9
1808772.0 6.0
1809072.0 7.0
1809132.0 6.7
1809192.0 6.9
1809294.0 5.6
1809360.0 6.1
1809957.0 6.7
1809985.0 5.5
1810024.0 6.4
1810392.0 7.4
1810452.0 7.4
1810572.0 6.8
1810937.0 6.5
1810994.0 5.7
1811073.0 7.0
1811217.0 6.0
1811290.0 5.7
1811364.0 6.9
1811832.0 6.0
1812072.0 6.4
1812287.0 5.9
1812565.0 6.8
1813002.0 6.7
1813092.0 5.9
1813332.0 5.9
1813512.0 7.3
1813662.0 7.0
1814652.0 6.5
1815874.0 6.7
1815912.0 6.9
1816152.0 7.1
1816332.0 6.1
1817964.0 7.0
1818199.0 6.7
1818387.0 5.5
1818560.0 5.8
1818683.0 6.2
1819152.0 5.7
1819212.0 5.6
1819272.0 6.4
1819424.0 6.8
1819499.0 7.0
1819734.0 7.1
1819952.0 7.5
1820178.0 5.8
1820232.0 7.4
1820352.0 7.1
1820472.0 6.2
1820838.0 6.5
1821386.0 6.1
1821441.0 6.8
1821672.0 7.4
1821792.0 6.6
1821912.0 6.5
1822092.0 6.4
1822266.0 6.7
1822399.0 7.3
1822458.0 7.1
1822681.0 5.9
1822731.0 6.0
1822802.0 6.8
1822961.0 6.4
1833432.0 6.2
1833800.0 7.1
1834000.0 6.6
1834055.0 6.5
1834752.0 6.9
1834932.0 6.4
1835617.0 7.3
1835627.0 7.1
1839132.0 6.3
1839312.0 6.2
1839498.0 6.4
1840212.0 6.3
1840572.0 6.8
1840632.0 6.7
1841031.0 6.2
1841196.0 7.4
1841639.0 5.6
1841689.0 5.7
1841952.0 6.5
1842012.0 6.9
1842597.0 5.9
1842645.0 5.6
1843632.0 7.3
1843752.0 6.5
1844016.0 6.6
1844289.0 6.6
1844772.0 7.0
1844832.0 7.5
1844892.0 6.3
1845262.0 7.0
1845621.0 6.7
1845898.0 5.8
1846272.0 7.1
1846392.0 6.9
1846512.0 6.6
1846780.0 7.5
1847079.0 7.1
1847201.0 6.6
1847400.0 5.9
1847580.0 6.0
1848363.0 6.2
1848429.0 6.8
1848671.0 6.2
1848769.0 7.2
1849092.0 6.2
1849801.0 5.9
1849965.0 6.3
1850021.0 5.9
1850092.0 6.0
1850232.0 7.0
1850467.0 7.4
1851024.0 7.2
1851176.0 6.9
1851202.0 7.0
1851345.0 5.5
1851498.0 6.7
1851863.0 6.2
1851910.0 5.5
1853472.0 7.1
1853592.0 6.8
1853712.0 5.6
1854051.0 5.7
1854264.0 6.3
1854912.0 6.5
1855092.0 7.0
1855212.0 6.1
1855517.0 5.5
1855529.0 7.2
1856850.0 6.0
1858985.0 6.3
1859172.0 6.7
1859868.0 6.1
1859961.0 5.7
1860016.0 6.9
1860141.0 6.3
1860167.0 5.9
1860304.0 5.6
1860409.0 6.8
1860532.0 6.0
1860552.0 6.7
1860672.0 6.7
1860792.0 6.6
1861032.0 6.5
1861149.0 5.6
1861192.0 5.5
1861328.0 6.3
1861473.0 5.8
1861667.0 7.1
1861762.0 6.0
1861800.0 6.0
1861846.0 7.1
1862412.0 5.8
1862551.0 7.1
1862675.0 6.4
1862939.0 6.6
1863232.0 7.2
1863299.0 5.7
1863792.0 5.6
1863992.0 6.4
1864168.0 5.9
1864179.0 6.8
1864251.0 5.8
1864329.0 6.1
1864452.0 6.5
1864580.0 7.5
1864643.0 5.7
1864712.0 5.9
1864813.0 7.5
1864847.0 6.7
1864992.0 7.1
1865112.0 7.0
1865292.0 6.8
1865352.0 6.0
1865444.0 5.8
1865541.0 7.0
1865598.0 6.4
1865738.0 6.4
1865760.0 7.3
1866113.0 7.1
1877060.0 6.8
1878132.0 6.4
1878192.0 7.4
1879572.0 5.5
1880772.0 6.7
1880892.0 6.5
1881072.0 6.7
1881132.0 7.5
1881427.0 6.6
1882332.0 7.1
1882572.0 7.5
1882632.0 6.4
1883302.0 7.3
1883464.0 7.4
1883536.0 6.4
1883892.0 5.6
1884322.0 6.0
1884343.0 6.8
1885152.0 6.6
1885332.0 6.9
1885512.0 7.1
1885812.0 7.2
1885945.0 6.2
1886340.0 6.4
1886472.0 5.9
1886592.0 6.5
1886652.0 7.3
1887125.0 5.8
1887483.0 6.0
1888152.0 6.4
1888272.0 5.8
1889652.0 6.6
1889922.0 6.8
1890545.0 7.2
1890852.0 6.6
1894511.0 7.4
1899036.0 5.6
1899081.0 6.6
1899171.0 7.1
1900142.0 6.6
1900189.0 7.5
1900490.0 5.8
1900590.0 5.6
1903077.0 5.6
1903682.0 6.3
1903932.0 6.0
1903992.0 6.2
1904172.0 6.1
1904655.0 6.9
1905008.0 6.1
1905192.0 6.0
1905612.0 6.8
1906584.0 6.7
1906692.0 7.0
1906812.0 6.0
1906932.0 5.9
1907052.0 6.2
1907112.0 7.3
1907264.0 6.1
1907396.0 5.9
1907536.0 6.1
1907680.0 6.6
1907864.0 6.9
1918572.0 6.0
1918833.0 7.3
1919115.0 7.3
1919280.0 5.6
1919324.0 7.0
1919712.0 5.9
1919772.0 6.8
1920072.0 6.7
1920909.0 6.2
1921332.0 6.6
1922652.0 5.6
1922952.0 6.2
1923972.0 6.4
1924092.0 6.3
1924152.0 6.8
1924502.0 6.2
1924530.0 6.4
1924615.0 6.4
1924712.0 7.3
1924846.0 7.3
1924922.0 6.8
1925000.0 5.6
1925057.0 5.6
1925155.0 7.3
1925224.0 6.8
1925275.0 6.1
1925352.0 6.4
1925412.0 6.1
1925592.0 6.3
1925652.0 6.9
1925899.0 6.6
1925984.0 7.4
1926079.0 7.1
1926601.0 6.0
1926728.0 6.8
1926972.0 6.6
1927032.0 6.4
1927152.0 6.6
1927212.0 5.6
1927373.0 6.0
1927555.0 7.0
1927652.0 5.9
1927753.0 7.3
1928152.0 5.8
1928232.0 5.8
1928532.0 6.6
1928652.0 6.8
1928776.0 6.5
1928784.0 5.6
1928840.0 6.5
1928854.0 6.8
1929633.0 7.0
1929658.0 7.3
1929672.0 6.8
1929792.0 7.3
1929852.0 6.0
1930385.0 5.6
1930848.0 7.0
1931051.0 7.0
1931992.0 6.8
1933274.0 7.3
1933306.0 6.9
1933409.0 7.2
1933622.0 7.4
1933679.0 6.9
1933964.0 7.0
1934292.0 6.6
1935672.0 7.2
1935912.0 6.1
1936412.0 5.6
1936572.0 6.4
1936582.0 6.6
1936704.0 5.6
1937292.0 6.7
1937532.0 7.2
1937552.0 6.9
1937591.0 6.9
1937715.0 7.1
1937942.0 6.2
1938046.0 5.9
1938252.0 6.8
1938372.0 5.6
1938432.0 7.3
1938492.0 7.0
1938552.0 6.3
1938612.0 6.2
1938672.0 5.9
1938792.0 5.8
1938977.0 7.1
1939365.0 7.3
1939872.0 6.2
1940932.0 7.1
1942192.0 5.5
1942369.0 6.4
1942632.0 6.3
1942992.0 7.0
1943182.0 6.6
1943204.0 5.7
1943383.0 5.6
1943714.0 7.0
1944192.0 5.7
1944372.0 6.2
1944733.0 6.8
1951101.0 5.9
1951146.0 7.2
1962162.0 6.9
1962384.0 6.4
1962520.0 6.7
1963272.0 5.7
1965852.0 6.6
1966092.0 6.2
1966152.0 5.7
1966352.0 7.0
1966365.0 5.9
1966481.0 6.6
1966879.0 5.6
1967232.0 6.9
1967412.0 6.9
1967472.0 5.8
1968000.0 7.0
1968042.0 7.3
1968672.0 7.2
1968732.0 6.9
1968792.0 7.4
1968912.0 5.6
1968972.0 6.1
1969149.0 7.1
1969262.0 6.6
1969359.0 7.1
1969380.0 6.0
1969732.0 6.2
1969782.0 7.2
1969912.0 5.6
1970052.0 7.5
1970659.0 6.4
1970776.0 6.1
1970804.0 6.2
1970832.0 7.3
1971364.0 5.7
1971672.0 7.0
1971792.0 7.1
1972043.0 6.2
1972331.0 6.9
1972450.0 6.2
1972728.0 7.3
1972776.0 6.9
1972932.0 7.1
1973112.0 6.3
1973172.0 5.7
1973232.0 6.5
1973647.0 6.6
1973680.0 6.1
1973795.0 6.9
1973841.0 5.5
1973984.0 6.7
1974123.0 6.1
1974312.0 7.0
1974732.0 6.7
1975078.0 6.3
1975172.0 7.4
1975245.0 5.9
1975686.0 7.0
1975932.0 7.4
1976112.0 6.5
1976416.0 7.1
1976656.0 5.5
1976916.0 7.5
1977063.0 5.8
1977612.0 7.0
1977759.0 5.7
1977940.0 7.5
1978059.0 6.2
1978442.0 6.5
1978522.0 6.1
1978579.0 7.0
1978632.0 6.0
1978812.0 7.1
1978992.0 5.8
1979112.0 5.8
1979327.0 5.7
1979454.0 6.0
1979516.0 7.3
1979718.0 7.2
1979828.0 7.3
1979876.0 5.7
1980062.0 6.1
1980072.0 6.3
1980132.0 6.3
1980192.0 6.4
1980552.0 6.5
1981812.0 5.7
1981932.0 6.8
1982626.0 5.7
1982720.0 7.1
1983012.0 5.8
1984452.0 5.6
1985479.0 7.3
1986392.0 6.7
1986702.0 6.5
1987055.0 6.3
1993032.0 7.1

1993212.0 6.8
1993452.0 7.4
1993512.0 5.6
1993754.0 6.0
1993961.0 5.6
1993973.0 6.8
1994106.0 7.3
1994205.0 6.2
2005824.0 5.7
2007912.0 7.0
2008752.0 7.5
2009052.0 5.6
2009352.0 6.6
2009604.0 6.4
2009622.0 6.0
2009717.0 6.6
2009865.0 7.3
2009995.0 6.3
2010372.0 6.1
2010552.0 6.1
2010882.0 6.0
2011170.0 6.3
2011812.0 7.0
2011932.0 5.8
2012112.0 6.8
2013023.0 6.8
2013735.0 7.0
2014107.0 6.9
2014136.0 7.0
2014396.0 6.8
2014425.0 6.0
2014584.0 7.0
2014632.0 6.8
2014692.0 6.2
2014872.0 5.8
2015112.0 6.1
2015365.0 6.3
2015628.0 5.8
2015708.0 7.0
2015826.0 5.8
2015982.0 6.9
2016132.0 6.6
2016192.0 7.1
2016252.0 6.6
2016312.0 6.8
2016745.0 7.2
2016762.0 5.9
2016805.0 5.7
2017017.0 6.2
2017044.0 7.0
2017124.0 6.8
2017149.0 5.7
2017482.0 6.7
2017512.0 5.9
2017632.0 6.7
2018072.0 5.6
2018311.0 5.9
2019192.0 6.9
2019312.0 7.1
2019514.0 6.8
2020304.0 7.3
2020372.0 5.5
2021014.0 6.9
2021832.0 7.1
2021952.0 6.7
2022012.0 6.1
2022072.0 5.5
2022470.0 6.6
2023272.0 6.0
2025012.0 5.6
2025541.0 6.3
2026392.0 7.1
2026762.0 7.0
2028782.0 6.8
2030052.0 5.9
2030532.0 7.0
2030652.0 6.6
2032624.0 6.2
2033247.0 6.3
2047167.0 6.1
2047565.0 7.2
2047703.0 5.6
2048917.0 6.0
2049778.0 6.0
2051316.0 6.9
2051428.0 6.6
2051662.0 6.7
2051776.0 7.1
2051956.0 6.5
2052192.0 5.5
2052752.0 6.8
2053512.0 6.5
2053992.0 6.6
2054795.0 6.7
2054852.0 6.6
2054952.0 6.0
2055012.0 5.7
2055072.0 7.4
2055132.0 7.1
2055252.0 6.6
2055432.0 6.3
2055551.0 5.8
2055782.0 7.5
2055947.0 7.4
2056095.0 5.6
2056213.0 6.3
2056240.0 6.9
2056260.0 6.3
2056281.0 6.9
2056329.0 5.8
2056392.0 7.2
2056452.0 5.8
2056632.0 7.0
2056932.0 6.3
2056957.0 6.6
2057037.0 7.4
2057130.0 6.7
2057247.0 6.0
2057341.0 6.0
2057349.0 7.1
2057393.0 7.4
2057583.0 5.8
2057676.0 7.5
2057722.0 6.4
2057745.0 6.7
2058072.0 6.5
2058192.0 6.1
2058433.0 7.5
2059022.0 5.9
2059045.0 7.5
2059203.0 5.8
2059230.0 7.3
2059272.0 6.3
2059332.0 6.4
2059452.0 7.4
2059512.0 7.1
2059692.0 6.7
2059964.0 7.2
2060577.0 7.3
2060772.0 6.1
2061556.0 7.0
2061912.0 7.1
2062007.0 6.0
2062152.0 7.2
2062272.0 7.5
2062993.0 6.6
2063588.0 6.0
2063592.0 6.0
2063652.0 6.2
2063772.0 5.6
2064151.0 6.1
2064230.0 5.7
2064494.0 6.2
2065092.0 6.4
2065452.0 5.7
2065512.0 5.9
2065802.0 7.0
2065873.0 6.3
2066592.0 6.1
2066712.0 6.0
2066772.0 7.1
2067809.0 5.9
2068681.0 6.0
2069472.0 7.1
2069712.0 7.0
2070202.0 5.8
2071032.0 6.7
2071092.0 6.6
2071494.0 7.2
2071726.0 6.3
2075052.0 6.2
2075112.0 6.3
2075172.0 7.4
2075697.0 6.7
2076102.0 6.3
2076672.0 6.2
2077153.0 6.3
2077304.0 5.7
2077479.0 6.8
2077706.0 6.7
2077830.0 6.8
2077916.0 6.6
2078052.0 6.5
2078352.0 7.4
2078412.0 6.9
2078472.0 5.6
2078970.0 6.8
2079197.0 6.5
2089752.0 7.0
2090292.0 5.9
2091312.0 5.8
2091432.0 5.8
2092034.0 6.5
2093087.0 6.7
2093435.0 5.7
2093446.0 6.5
2093609.0 5.6
2093832.0 5.9
2094312.0 7.0
2094380.0 6.2
2094629.0 6.2
2094642.0 7.4
2095019.0 7.0
2095156.0 5.8
2095200.0 5.7
2095512.0 7.3
2095632.0 6.0
2095692.0 6.8
2095752.0 7.1
2095972.0 7.2
2096337.0 5.7
2096354.0 6.1
2096522.0 7.5
2096772.0 5.5
2096892.0 6.3
2097535.0 6.6
2097550.0 6.1
2097642.0 5.7
2097967.0 7.2
2098332.0 7.2
2098512.0 6.6
2099033.0 7.2
2099554.0 5.8
2099564.0 5.6
2099592.0 6.9
2099652.0 6.5
2099892.0 6.4
2099952.0 6.2
2100483.0 7.4
2101392.0 6.5
2101452.0 6.8
2101585.0 6.4
2101820.0 6.8
2102233.0 6.7
2102592.0 6.8
2102952.0 6.1
2104754.0 5.7
2104772.0 6.6
2106792.0 5.6
2107032.0 5.8
2107272.0 7.4
2107350.0 5.8
2107448.0 7.2
2107515.0 7.3
2108068.0 6.0
2108089.0 5.9
2108412.0 7.0
2108532.0 7.5
2108592.0 7.0
2108933.0 5.6
2109114.0 6.4
2109162.0 7.3
2109792.0 5.8
2109852.0 5.6
2109972.0 6.2
2110032.0 7.0
2110294.0 5.8
2110422.0 6.0
2110500.0 5.9
2110885.0 6.4
2111232.0 6.7
2113967.0 6.5
2113992.0 7.0
2114472.0 6.6
2114544.0 6.3
2115306.0 6.3
2115552.0 7.1
2115672.0 6.1
2115912.0 7.1
2116200.0 6.5
2116494.0 5.7
2116517.0 5.5
2116537.0 6.1
2116687.0 7.1
2116809.0 7.2
2117705.0 7.4
2117817.0 7.3
2118055.0 6.4
2118127.0 5.8
2118173.0 6.7
2118264.0 6.4
2119597.0 7.0
2119752.0 6.0
2119992.0 5.8
2120455.0 6.4
2120802.0 5.7
2121552.0 7.5
2122299.0 7.1
2122490.0 6.2
2122500.0 5.7
2132772.0 7.4
2134120.0 5.5
2134332.0 7.0
2135304.0 5.9
2135652.0 6.6
2135772.0 6.7
2135832.0 6.4
2135952.0 6.7
2136136.0 5.8
2136862.0 7.2
2139542.0 6.9
2140272.0 6.9
2141201.0 7.1
2141532.0 6.6
2142030.0 6.5
2142070.0 5.6
2142107.0 6.2
2146705.0 5.7
2148547.0 7.5
2148852.0 6.8
2149992.0 6.5
2150954.0 7.1
2151234.0 6.5
2151374.0 6.4
2152761.0 5.7
2152992.0 6.3
2153172.0 5.7
2153292.0 6.9
2153452.0 6.5
2153557.0 6.7
2153909.0 6.9
2154290.0 6.7
2154312.0 6.0
2154432.0 6.3
2154492.0 5.6
2154672.0 5.8
2154732.0 5.5
2154877.0 5.9
2155349.0 6.0
2155660.0 6.6
2156627.0 5.7
2156652.0 6.0
2157192.0 6.6
2157492.0 6.7
2157552.0 6.7
2157672.0 6.0
2158380.0 5.8
2158422.0 6.7
2158450.0 6.8
2159209.0 7.1
2159527.0 5.8
2159880.0 6.5
2159955.0 7.3
2160132.0 5.7
2161572.0 6.2
2161992.0 5.6
2162776.0 6.1
2162898.0 7.0
2162919.0 6.9
2162939.0 6.7
2163012.0 7.0
2163072.0 7.1
2163252.0 5.9
2163312.0 5.9
2163497.0 6.4
2163566.0 6.6
2163640.0 7.3
2163706.0 6.8
2163783.0 6.4
2163932.0 7.3
2164082.0 7.1
2164168.0 6.0
2164309.0 6.0
2174532.0 6.9
2175182.0 5.6
2175360.0 6.6
2175492.0 7.2
2175686.0 6.7
2175902.0 6.1
2175912.0 5.6
2177320.0 6.0
2177934.0 5.8
2178555.0 6.4
2178634.0 6.0
2178649.0 6.6
2179032.0 6.2
2179575.0 5.8
2180160.0 6.5
2180352.0 5.9
2180592.0 6.4
2181732.0 5.8
2182458.0 6.4
2184792.0 5.8
2185556.0 7.2
2191317.0 6.4
2193914.0 7.2
2193999.0 7.4
2194026.0 5.5
2194277.0 6.4
2194364.0 6.7
2194510.0 7.3
2194812.0 6.1
2195289.0 6.5
2195349.0 6.3
2196252.0 6.8
2196635.0 6.4
2196738.0 5.9
2196882.0 7.0
2196892.0 7.1
2197127.0 5.7
2197172.0 5.6
2197179.0 6.1
2197238.0 6.4
2197363.0 7.4
2197632.0 7.3
2198147.0 6.4
2198420.0 7.5
2198803.0 6.3
2198825.0 5.6
2198878.0 6.5
2198952.0 6.7
2199372.0 7.0
2199713.0 6.2
2200262.0 6.8
2200364.0 6.4
2202397.0 7.0
2202428.0 7.4
2202830.0 5.9
2203332.0 5.5
2203572.0 6.5
2203692.0 7.0
2203752.0 6.0
2203856.0 7.0
2203921.0 6.8
2203972.0 6.9
2204037.0 6.0
2204158.0 6.5
2204285.0 7.4
2204396.0 7.1
2204409.0 6.5
2204464.0 6.0
2204475.0 5.9
2204712.0 5.9
2204772.0 6.1
2204832.0 6.2
2204952.0 6.4
2205132.0 6.0
2205192.0 7.1
2205320.0 6.0
2205354.0 6.6
2205520.0 6.0
2205590.0 7.0
2205694.0 6.3
2205812.0 7.2
2205882.0 7.1
2206002.0 7.1
2206084.0 5.8
2206272.0 7.2
2206452.0 6.2
2206512.0 7.0
2206632.0 7.2
2206802.0 7.3
2206852.0 6.0
2206929.0 6.1
2206987.0 5.7
2207075.0 7.5
2207361.0 6.6
2218152.0 7.1
2218314.0 7.5
2218357.0 5.6
2218760.0 6.5
2219078.0 7.1
2219893.0 7.0
2220023.0 6.9
2220672.0 7.1
2221202.0 6.6
2222052.0 6.2
2223314.0 5.7
2223357.0 7.2
2223399.0 6.2
2223852.0 7.2
2225773.0 6.7
2227689.0 6.2
2227812.0 7.3
2228052.0 7.1
2229492.0 5.7
2230465.0 6.8
2231610.0 6.6
2232132.0 6.4
2232252.0 6.6
2233243.0 5.6
2233572.0 5.7
2233752.0 6.3
2234327.0 6.6
2234571.0 6.9
2234662.0 5.7
2236812.0 6.3
2237832.0 6.5
2238192.0 6.0
2239098.0 5.8
2239194.0 7.4
2239242.0 7.4
2239392.0 6.2
2240352.0 7.4
2243712.0 5.7
2245392.0 6.1
2245629.0 7.2
2245665.0 6.2
2246712.0 6.8
2247121.0 6.0
2247783.0 5.8
2248461.0 5.7
2264112.0 6.8
2271840.0 7.5
2272079.0 6.7
2272342.0 6.2
2272752.0 7.0
2272812.0 5.5
2274072.0 6.4
2274192.0 6.8
2274664.0 6.3
2274760.0 5.8
2275992.0 6.1
2276193.0 6.0
2276772.0 7.2
2276892.0 7.2
2276952.0 7.5
2277012.0 6.5
2277530.0 6.0
2278452.0 6.1
2278751.0 5.6
2279237.0 5.7
2283737.0 6.7
2284546.0 7.1
2284693.0 5.9
2284718.0 6.9
2285352.0 7.1
2286553.0 6.2
2286648.0 5.9
2286724.0 6.0
2286767.0 5.8
2287032.0 5.8
2287391.0 6.6
2287497.0 6.9
2287552.0 6.1
2287800.0 7.4
2288232.0 6.1
2288824.0 6.3
2289393.0 5.8
2289732.0 7.4
2289792.0 6.4
2290032.0 6.2
2290224.0 6.9
2290316.0 6.7
2290372.0 5.5
2290388.0 7.0
2290413.0 5.7
2290525.0 5.5
2290690.0 6.0
2290874.0 5.6
2290909.0 7.3
2291112.0 6.3
2291232.0 5.6
2291696.0 6.8
2291724.0 6.5
2291752.0 5.7
2291779.0 7.4
2291982.0 6.7
2292017.0 6.1
2292136.0 6.3
2292238.0 6.4
2292251.0 6.8
2292404.0 6.6
2302632.0 6.6
2305812.0 6.9
2307372.0 6.8
2307580.0 6.8
2309344.0 5.5
2310072.0 7.3
2310469.0 5.9
2310552.0 5.7
2311268.0 7.0
2311272.0 5.6
2311332.0 5.5
2311392.0 5.7
2311512.0 6.1
2311632.0 6.0
2311849.0 6.3
2312237.0 5.9
2312340.0 6.8
2312483.0 7.2
2322752.0 6.7
2323704.0 6.0
2323715.0 5.6
2323720.0 6.0
2324127.0 7.1
2324412.0 7.5
2325050.0 7.1
2325080.0 5.8
2325467.0 6.2
2325672.0 6.2
2326533.0 6.4
2327412.0 6.3
2327744.0 5.8
2327753.0 6.3
2328143.0 7.3
2328200.0 7.0
2328341.0 7.2
2328374.0 7.1
2328426.0 7.3
2328672.0 6.1
2329339.0 5.9
2330112.0 7.0
2330292.0 6.8
2330646.0 6.4
2330865.0 7.2
2331492.0 7.0
2332056.0 6.8
2332140.0 6.9
2332222.0 5.9
2332376.0 5.7
2332576.0 5.8
2332693.0 5.7
2332707.0 6.6
2332740.0 7.0
2332824.0 5.9
2332872.0 6.9
2332992.0 6.7
2333052.0 5.6
2333172.0 6.0
2333505.0 6.8
2333516.0 5.9
2333577.0 6.8
2333692.0 7.0
2333883.0 6.9
2333949.0 5.5
2334102.0 6.1
2344692.0 6.1
2346525.0 7.2
2346958.0 6.0
2347997.0 5.8
2348002.0 6.9
2349780.0 7.4
2353512.0 6.0
2353702.0 5.6
2354421.0 6.6
2355115.0 6.9
2355224.0 6.8
2355555.0 5.7
2356212.0 6.8
2356788.0 7.0
2357948.0 6.5
2359152.0 7.2
2359212.0 6.1
2359362.0 5.5
2359422.0 7.2
2360977.0 7.4
2361534.0 7.3
2363232.0 7.3
2364487.0 5.8
2364612.0 6.5
2366785.0 5.9
2367365.0 6.8
2368099.0 5.7
2368216.0 7.3
2368242.0 5.8
2368869.0 6.3
2369352.0 5.6
2369819.0 5.8
2370247.0 5.9
2371477.0 6.5
2371699.0 6.9
2371872.0 6.6
2372052.0 6.5
2372172.0 6.8
2372232.0 6.2
2372361.0 6.1
2372434.0 6.6
2372497.0 5.7
2372540.0 7.3
2372688.0 5.7
2372904.0 6.7
2373182.0 6.6
2373492.0 7.5
2374574.0 6.5
2375982.0 5.5
2386212.0 6.3
2387121.0 7.3
2387185.0 6.5
2387712.0 5.6
2387832.0 6.7
2387952.0 6.4
2388527.0 7.3
2388833.0 7.1
2389032.0 6.8
2389152.0 6.2
2389452.0 6.3
2389815.0 5.5
2390832.0 6.3
2390952.0 6.3
2391912.0 7.0
2392032.0 6.4
2392332.0 6.3
2392392.0 5.8
2392647.0 6.5
2392802.0 5.5
2392830.0 5.6
2392912.0 5.5
2392960.0 5.6
2393352.0 6.6
2393412.0 6.3
2393712.0 5.6
2393896.0 5.6
2394041.0 6.9
2394138.0 5.9
2394413.0 6.8
2394518.0 6.8
2394587.0 6.1
2394702.0 6.6
2394762.0 7.2
2394852.0 7.3
2394972.0 7.0
2395032.0 5.8
2395366.0 5.8
2395681.0 5.8
2395786.0 6.0
2396007.0 7.4
2396126.0 7.3
2396224.0 7.2
2396232.0 7.0
2396292.0 6.1
2396352.0 6.2
2396592.0 6.6
2396652.0 6.2
2396712.0 7.1
2396868.0 6.3
2396966.0 6.5
2397167.0 6.5
2397317.0 7.0
2397387.0 5.6
2397553.0 7.4
2397615.0 6.4
2397672.0 6.4
2397912.0 5.5
2398092.0 7.4
2398626.0 7.1
2398653.0 6.6
2399112.0 7.3
2399412.0 6.8
2399472.0 6.3
2400475.0 6.0
2400792.0 5.8
2401125.0 5.5
2402052.0 6.7
2402232.0 7.0
2402753.0 6.9
2403155.0 5.9
2403270.0 7.2
2403552.0 6.3
2406089.0 7.1
2406970.0 6.6
2407056.0 5.9
2407588.0 6.7
2407812.0 6.0
2408112.0 6.5
2409192.0 7.2
2409252.0 5.8
2409492.0 7.4
2409847.0 6.4
2409907.0 5.8
2409977.0 6.2
2410692.0 6.5
2410752.0 7.2
2410932.0 6.7
2410992.0 6.1
2411052.0 6.5
2411307.0 6.3
2411347.0 6.9
2411541.0 5.9
2411562.0 6.7
2411975.0 6.6
2412072.0 7.3
2412719.0 6.1
2412980.0 6.2
2413080.0 7.0
2413512.0 7.2
2414616.0 6.7
2414647.0 7.1
2415012.0 6.6
2415132.0 7.4
2415252.0 6.3
2415312.0 5.5
2415512.0 7.5
2415892.0 6.6
2415987.0 5.9
2416037.0 5.8
2417437.0 7.1
2419003.0 7.0
2419042.0 6.4
2430193.0 6.5
2430792.0 5.7
2430972.0 6.9
2431092.0 7.4
2431212.0 6.3
2431395.0 6.2
2432352.0 6.6
2433574.0 6.3
2433579.0 6.1
2433912.0 6.9
2434317.0 6.0
2434500.0 5.6
2434808.0 6.6
2434882.0 7.0
2435472.0 5.9
2435592.0 7.0
2435681.0 7.0
2435918.0 6.6
2436038.0 5.8
2436174.0 6.2
2436382.0 6.5
2437032.0 7.1
2437288.0 6.6
2437464.0 6.7
2437602.0 7.0
2437860.0 6.2
2437982.0 6.5
2438172.0 7.3
2438232.0 6.8
2438292.0 7.0
2438753.0 5.8
2438799.0 7.4
2438989.0 7.2
2440628.0 6.6
2440847.0 7.1
2441052.0 6.8
2442099.0 6.3
2442168.0 5.7
2442252.0 7.0
2443128.0 5.6
2444770.0 6.3
2445492.0 7.1
2446632.0 6.6
2446752.0 6.3
2448732.0 7.2
2448881.0 7.3
2449132.0 7.2
2449632.0 7.0
2449752.0 7.0
2449992.0 6.9
2451252.0 6.0
2451432.0 6.7
2451524.0 6.6
2454012.0 6.7
2454252.0 6.8
2456399.0 5.8
2458452.0 6.6
2471057.0 6.0
2472672.0 6.4
2473530.0 6.0
2473565.0 6.3
2473768.0 6.8
2475612.0 7.3
2475912.0 6.6
2476208.0 5.7
2476357.0 6.1
2476589.0 5.9
2476801.0 6.7
2477352.0 6.8
2477491.0 5.5
2477570.0 6.8
2477885.0 6.7
2478050.0 5.8
2478312.0 6.6
2478612.0 5.9
2478732.0 5.9
2479189.0 6.7
2480572.0 5.7
2480725.0 6.6
2480770.0 6.7
2480905.0 7.5
2480950.0 7.3
2481033.0 7.0
2481192.0 6.4
2481312.0 6.0
2482872.0 7.1
2482992.0 7.2
2483564.0 6.6
2483829.0 6.2
2484054.0 6.2
2484831.0 6.1
2484952.0 7.0
2485372.0 7.5
2485436.0 6.6
2485484.0 6.9
2485752.0 6.1
2485932.0 7.2
2485992.0 5.8
2486461.0 5.9
2486577.0 7.2
2486639.0 7.1
2486740.0 6.7
2486876.0 6.6
2486952.0 5.6
2487252.0 6.4
2487608.0 6.7
2487682.0 7.2
2488123.0 6.5
2488194.0 6.3
2488452.0 6.6
2488512.0 6.7
2488692.0 5.8
2488872.0 7.0
2489216.0 5.8
2489275.0 6.9
2489491.0 7.1
2492688.0 6.8
2492892.0 5.9
2493452.0 7.0
2493769.0 6.2
2494452.0 5.8
2494632.0 6.2
2496226.0 7.4
2496288.0 6.0
2503622.0 7.3
2514612.0 6.0
2514905.0 7.2
2515050.0 5.6
2515305.0 6.5
2515338.0 6.6
2515992.0 5.6
2516052.0 6.5
2518932.0 7.4
2519112.0 6.2
2519379.0 5.8
2519850.0 7.5
2520132.0 6.7
2520192.0 7.1
2520552.0 6.9
2520665.0 7.0
2520721.0 5.7
2521118.0 6.6
2521126.0 5.9
2521512.0 5.5
2521932.0 6.6
2522221.0 6.0
2522503.0 6.0
2522663.0 5.7
2522952.0 6.5
2523012.0 5.8
2523685.0 5.9
2524115.0 6.1
2524216.0 5.7
2524340.0 7.5
2524392.0 7.1
2524632.0 6.3
2524692.0 6.3
2524812.0 7.0
2525030.0 7.3
2525175.0 6.8
2525223.0 5.6
2525450.0 6.2
2525559.0 6.1
2526132.0 7.3
2526393.0 6.3
2526466.0 6.7
2526553.0 5.7
2526573.0 5.6
2526700.0 5.7
2526731.0 7.0
2527067.0 6.4
2527135.0 6.3
2527392.0 5.9
2527512.0 7.3
2528116.0 5.8
2528134.0 5.5
2528225.0 7.0
2528459.0 7.5
2528609.0 6.4
2528952.0 6.7
2529012.0 6.0
2529436.0 5.9
2529480.0 5.9
2530016.0 6.2
2530057.0 6.8
2530152.0 5.5
2530512.0 7.3
2530572.0 6.0
2530632.0 5.9
2531712.0 6.9
2531832.0 6.4
2532012.0 5.7
2532072.0 6.1
2532381.0 7.1
2532476.0 7.3
2532502.0 6.7
2532749.0 7.4
2532813.0 6.1
2533003.0 5.6
2533092.0 7.5
2533332.0 7.0
2533392.0 5.6
2533452.0 6.9
2533512.0 6.0
2533900.0 6.2
2534472.0 6.5
2535523.0 5.5
2535912.0 6.2
2537532.0 6.3
2538642.0 6.2
2539272.0 7.0
2542857.0 7.2

2543352.0 6.8
2544069.0 6.5
2545200.0 6.3
2545362.0 6.1
2545561.0 6.0
2561319.0 6.6
2562673.0 5.8
2563632.0 5.6
2564565.0 7.2
2565132.0 6.8
2565674.0 7.4
2566893.0 7.1
2567184.0 6.5
2567449.0 6.0
2567553.0 5.7
2567652.0 7.1
2568072.0 5.6
2568230.0 6.5
2568310.0 6.0
2568340.0 5.6
2568495.0 6.8
2569092.0 5.7
2569332.0 6.3
2569854.0 6.5
2570200.0 6.7
2570592.0 6.2
2571681.0 7.1
2571972.0 6.5
2572872.0 5.8
2572944.0 6.9
2573070.0 7.2
2573970.0 5.8
2574069.0 6.8
2574912.0 6.1
2579352.0 7.3
2581399.0 6.8
2581992.0 7.1
2582412.0 6.8
2582840.0 6.3
2584092.0 5.8
2584400.0 7.3
2584509.0 6.7
2584580.0 5.7
2584720.0 6.6
2584868.0 6.1
2584872.0 5.6
2584932.0 6.5
2585052.0 5.9
2585112.0 7.1
2585232.0 6.0
2585292.0 5.9
2585352.0 6.5
2585588.0 5.5
2585620.0 6.2
2585641.0 6.5
2585867.0 6.8
2585947.0 5.6
2585997.0 7.5
2586024.0 6.9
2587408.0 5.6
2587520.0 6.9
2588112.0 7.5
2588232.0 6.5
2599332.0 6.5
2599392.0 6.7
2601192.0 7.2
2601535.0 5.9
2603584.0 6.8
2604288.0 6.4
2604474.0 7.0
2604610.0 6.0
2606197.0 6.3
2606712.0 5.6
2607381.0 6.6
2607572.0 5.6
2608152.0 7.2
2608212.0 7.2
2608272.0 5.5
2608780.0 5.8
2608982.0 7.3
2610004.0 7.1
2610047.0 6.5
2610367.0 5.7
2610490.0 7.0
2611546.0 5.8
2612074.0 7.3
2612209.0 6.1
2612472.0 7.5
2612969.0 5.9
2613479.0 6.9
2613792.0 5.8
2614032.0 7.0
2614504.0 6.0
2614732.0 5.5
2615472.0 7.4
2616972.0 5.7
2617107.0 7.0
2617137.0 6.0
2618472.0 6.4
2620525.0 6.1
2621854.0 7.0
2622552.0 5.9
2622672.0 6.1
2622732.0 5.9
2622792.0 7.2
2622868.0 6.4
2623377.0 6.6
2623538.0 5.9
2623872.0 7.3
2623992.0 7.3
2624112.0 5.6
2625492.0 7.2
2626872.0 6.6
2626992.0 5.9
2627299.0 6.6
2627737.0 7.2
2629632.0 6.9
2630556.0 6.3
2630824.0 6.3
2631192.0 5.6
2631252.0 7.0
2643104.0 5.7
2643414.0 6.2
2643844.0 6.2
2644032.0 6.1
2644092.0 6.2
2644212.0 7.0
2644332.0 6.7
2644722.0 7.4
2644882.0 6.8
2645160.0 5.8
2645412.0 7.2
2645592.0 5.9
2645652.0 7.0
2646261.0 6.8
2646302.0 5.9
2646320.0 6.7
2646436.0 7.3
2646586.0 6.2
2647152.0 5.7
2647362.0 6.5
2647480.0 6.7
2647563.0 6.1
2647989.0 5.6
2648072.0 6.4
2648098.0 6.3
2648472.0 6.0
2648917.0 7.4
2649004.0 6.9
2649732.0 6.1
2650940.0 6.5
2651532.0 7.2
2651938.0 6.5
2652203.0 5.7
2652436.0 6.7
2653095.0 6.9
2654292.0 6.3
2654352.0 7.0
2654412.0 7.1
2654880.0 5.6
2655992.0 5.5
2656857.0 6.5
2660052.0 6.3
2660442.0 6.9
2660574.0 6.4
2660706.0 5.9
2660849.0 6.6
2661252.0 6.4
2661907.0 7.0
2663491.0 5.9
2664192.0 7.2
2664767.0 6.6
2664859.0 6.3
2664914.0 7.1
2665222.0 5.7
2665362.0 5.8
2665692.0 6.6
2666164.0 6.9
2666207.0 6.3
2666430.0 5.7
2666540.0 7.3
2666560.0 7.3
2667132.0 6.0
2669492.0 5.7
2670748.0 7.2
2670793.0 6.7
2670832.0 6.1
2671512.0 6.6
2671979.0 6.8
2672023.0 6.0
2672402.0 7.2
2672544.0 6.2
2672772.0 5.7
2672832.0 5.5
2673012.0 6.1
2673072.0 7.1
2673273.0 6.9
2673777.0 7.2
2684177.0 6.9
2684712.0 6.1
2684883.0 6.5
2685076.0 7.4
2685165.0 6.8
2685229.0 6.3
2685248.0 7.2
2685732.0 7.3
2685852.0 6.6
2685972.0 5.6
2686464.0 6.6
2686690.0 7.0
2686876.0 6.0
2686900.0 6.1
2687111.0 6.1
2687352.0 5.8
2687472.0 6.1
2688672.0 6.9
2688792.0 7.5
2689901.0 6.5
2690112.0 7.4
2691091.0 6.4
2691552.0 7.4
2692420.0 6.7
2692474.0 6.3
2692556.0 7.0
2693052.0 6.9
2693693.0 5.7
2694182.0 6.5
2694312.0 5.7
2694792.0 6.1
2695111.0 6.3
2698872.0 6.2
2700492.0 6.9
2700834.0 7.2
2700985.0 6.7
2701162.0 6.5
2701209.0 7.0
2701512.0 7.1
2701572.0 6.6
2701632.0 7.1
2701692.0 7.2
2701872.0 6.3
2702052.0 7.1
2702150.0 7.3
2702227.0 7.0
2702530.0 7.2
2702577.0 6.7
2702940.0 6.2
2704000.0 6.9
2704126.0 7.0
2704512.0 6.3
2705009.0 5.6
2705202.0 5.9
2705316.0 7.2
2706012.0 5.9
2706509.0 6.4
2707980.0 7.2
2708550.0 7.4
2708639.0 7.4
2712644.0 6.6
2712892.0 6.7
2715785.0 6.7
2716092.0 6.4
2716512.0 6.6
2727492.0 6.6
2730854.0 7.1
2731237.0 6.8
2732384.0 6.9
2732408.0 5.9
2732486.0 5.9
2734752.0 6.4
2735112.0 5.6
2736950.0 6.3
2738729.0 7.3
2739924.0 6.3
2739962.0 6.6
2740392.0 5.5
2740572.0 6.1
2740752.0 6.8
2740954.0 5.7
2741304.0 6.7
2741686.0 5.9
2742192.0 6.1
2742312.0 7.4
2742379.0 6.9
2743157.0 5.5
2743268.0 7.1
2743946.0 6.8
2744625.0 6.4
2745012.0 7.0
2745521.0 5.9
2745751.0 6.2
2745840.0 6.8
2747196.0 7.3
2747892.0 7.0
2748072.0 7.1
2749272.0 6.4
2749512.0 6.7
2750372.0 6.1
2750394.0 5.7
2751674.0 6.8
2751746.0 6.6
2751912.0 6.1
2752212.0 7.0
2752332.0 7.2
2753057.0 6.8
2753278.0 6.5
2753350.0 6.6
2754912.0 5.9
2757619.0 6.5
2757792.0 6.7
2758152.0 5.8
2758602.0 7.4
2758771.0 6.5
2759033.0 7.3
2759592.0 6.2
2759844.0 7.4
2760075.0 5.5
2770578.0 6.2
2770992.0 7.1
2778312.0 6.6
2779122.0 7.4
2779332.0 5.8
2779452.0 7.5
2779572.0 5.6
2779692.0 6.7
2779910.0 6.4
2780159.0 7.4
2780210.0 7.5
2780511.0 5.9
2780712.0 7.1
2780832.0 5.8
2780892.0 6.8
2781386.0 7.5
2781565.0 6.6
2782332.0 6.9
2782714.0 6.8
2782825.0 6.4
2782880.0 5.9
2782979.0 6.9
2783652.0 6.2
2783712.0 6.4
2785452.0 7.2
2785626.0 6.5
2786066.0 6.3
2786343.0 5.9
2791385.0 5.8
2791968.0 6.4
2792352.0 7.0
2793151.0 5.5
2793414.0 7.2
2793596.0 5.5
2793972.0 5.8
2794152.0 7.3
2794508.0 6.9
2796464.0 6.3
2800449.0 5.9
2800548.0 6.7
2801352.0 7.2
2812384.0 5.8
2816210.0 6.4
2816712.0 6.5
2817539.0 6.8
2818958.0 6.5
2819772.0 5.9
2820187.0 7.2
2820745.0 5.8
2820779.0 6.4
2820856.0 7.5
2821032.0 6.6
2821092.0 6.2
2821272.0 5.8
2821452.0 5.7
2821838.0 6.6
2822022.0 5.7
2822363.0 5.6
2822398.0 5.9
2822772.0 6.4
2822892.0 5.8
2823342.0 7.2
2823405.0 6.4
2823617.0 6.4
2823653.0 6.3
2823802.0 5.5
2825010.0 6.2
2825146.0 6.3
2825177.0 7.4
2825412.0 5.7
2825832.0 7.1
2826486.0 6.8
2827152.0 6.7
2828592.0 6.0
2829732.0 6.2
2829972.0 6.4
2830032.0 5.8
2831106.0 7.1
2831592.0 6.5
2833218.0 5.9
2835981.0 6.3
2839812.0 6.4
2839932.0 5.6
2840232.0 6.7
2841252.0 6.4
2841789.0 7.3
2842752.0 5.8
2843292.0 7.2
2857032.0 7.2
2858772.0 6.7
2860720.0 6.0
2862113.0 7.3
2862604.0 7.2
2862702.0 7.5
2864970.0 5.6
2866092.0 7.4
2866630.0 7.2
2866906.0 7.0
2868090.0 6.7
2868338.0 6.8
2869167.0 5.8
2869619.0 6.1
2869712.0 6.5
2870112.0 5.9
2870412.0 5.7
2870809.0 6.0
2871400.0 6.7
2872172.0 5.8
2877017.0 6.9
2877672.0 7.3
2881366.0 5.9
2881512.0 5.7
2881752.0 7.3
2882160.0 6.3
2882168.0 5.5
2882360.0 6.8
2882525.0 7.1
2882846.0 6.9
2882916.0 6.6
2883192.0 6.3
2883312.0 7.0
2883432.0 5.9
2883680.0 7.3
2884268.0 6.0
2884452.0 7.4
2885219.0 6.7
2885892.0 6.6
2885952.0 5.6
2886252.0 6.5
2886312.0 7.4
2897302.0 6.8
2897772.0 5.8
2898676.0 6.7
2899396.0 5.9
2901384.0 6.9
2901912.0 6.9
2902379.0 7.0
2905406.0 6.9
2906782.0 5.8
2907912.0 6.3
2908000.0 7.2
2908671.0 7.3
2908755.0 7.4
2908872.0 5.7
2908932.0 5.7
2909424.0 5.8
2909755.0 7.2
2909927.0 7.3
2910169.0 6.6
2910312.0 5.6
2911164.0 7.3
2912542.0 7.0
2913192.0 6.5
2914566.0 7.4
2915589.0 5.7
2915909.0 6.6
2915923.0 7.2
2916851.0 6.6
2917337.0 6.4
2917872.0 5.6
2919372.0 5.9
2919432.0 7.1
2919557.0 5.7
2920452.0 5.5
2920512.0 6.9
2920949.0 6.3
2921558.0 7.1
2921824.0 7.4
2922445.0 5.9
2922501.0 7.3
2922539.0 5.5
2922645.0 6.2
2922792.0 5.9
2922959.0 6.4
2923271.0 6.4
2923752.0 7.3
2924333.0 5.6
2924712.0 7.2
2925132.0 6.1
2925567.0 6.7
2926152.0 7.0
2926452.0 6.0
2926949.0 6.0
2928769.0 6.3
2929650.0 6.7
2929954.0 7.0
2930352.0 6.1
2943492.0 5.5
2945112.0 6.4
2945172.0 6.3
2949010.0 5.8
2949612.0 6.5
2949834.0 6.3
2951220.0 6.0
2953872.0 6.6
2955312.0 6.8
2955372.0 7.2
2955872.0 6.3
2955882.0 6.8
2957044.0 7.4
2957718.0 5.6
2958012.0 5.7
2958192.0 6.2
2958746.0 7.0
2961072.0 6.3
2961442.0 6.1
2962332.0 6.0
2962452.0 7.3
2962826.0 6.8
2963284.0 5.9
2963515.0 6.7
2964137.0 6.0
2964306.0 6.6
2964558.0 6.7
2964697.0 7.1
2964799.0 5.7
2968890.0 6.6
2969135.0 5.7
2970912.0 6.3
2971212.0 5.9
2972005.0 6.0
2985192.0 6.6
2985552.0 6.6
2989632.0 6.9
2989692.0 6.5
2992223.0 5.8
2993450.0 5.6
2993690.0 7.0
2993751.0 6.5
2996629.0 6.1
2997615.0 6.4
2997667.0 7.4
2998135.0 7.4
2998152.0 5.9
2998392.0 7.1
2998512.0 5.7
2998632.0 5.6
2999249.0 7.2
2999429.0 6.0
2999772.0 5.8
3001212.0 6.8
3001272.0 6.3
3002532.0 7.4
3002592.0 5.6
3004032.0 6.3
3004588.0 5.6
3005268.0 5.7
3005333.0 6.2
3005592.0 6.4
3006076.0 5.6
3007032.0 7.0
3008292.0 6.4
3010923.0 6.5
3010982.0 7.2
3013840.0 6.0
3024874.0 6.8
3025063.0 6.5
3029100.0 5.8
3033995.0 6.8
3034272.0 6.5
3035077.0 6.7
3035497.0 6.6
3035592.0 6.2
3035892.0 6.9
3036290.0 6.8
3036488.0 5.7
3036817.0 5.5
3036982.0 6.6
3037642.0 6.5
3037702.0 6.6
3037742.0 6.3
3037862.0 7.3
3038398.0 6.4
3038532.0 6.6
3038652.0 6.1
3038712.0 5.8
3039185.0 6.6
3039527.0 5.6
3040152.0 6.4
3040748.0 5.9
3041081.0 5.5
3041197.0 6.4
3041712.0 5.9
3042664.0 7.4
3042792.0 6.5
3043754.0 6.1
3045313.0 7.2
3046152.0 6.4
3050172.0 7.3
3054131.0 7.0
3054219.0 5.9
3054956.0 6.0
3055300.0 6.8
3055307.0 7.4
3055624.0 5.8
3056052.0 5.8
3067572.0 5.5
3068050.0 7.5
3074044.0 7.1
3075460.0 7.4
3075972.0 6.5
3076032.0 6.5
3076392.0 6.8
3076850.0 7.2
3077472.0 6.8
3078527.0 5.6
3078792.0 5.6
3079032.0 6.1
3079092.0 7.3
3079272.0 6.4
3079438.0 6.0
3079454.0 7.0
3079905.0 7.2
3080472.0 6.4
3081076.0 6.2
3081214.0 7.5
3081792.0 5.8
3082460.0 6.5
3082747.0 7.2
3083592.0 6.8
3085111.0 7.1
3085990.0 6.9
3087094.0 6.8
3088402.0 6.4
3091264.0 5.6
3093097.0 6.6
3093312.0 6.1
3093432.0 7.2
3093803.0 6.1
3093845.0 6.4
3094568.0 5.7
3094872.0 7.3
3095307.0 5.9
3095922.0 6.0
3098730.0 7.2
3114852.0 7.2
3115032.0 5.8
3115973.0 5.7
3117395.0 6.7
3118255.0 6.3
3118525.0 7.5
3118604.0 5.8
3118679.0 6.2
3118901.0 6.2
3118949.0 7.3
3119232.0 5.6
3119472.0 6.1
3120388.0 6.8
3120414.0 7.0
3120552.0 6.1
3120732.0 6.8
3121032.0 6.8
3121226.0 6.1
3122292.0 7.0
3122472.0 6.2
3123037.0 7.0
3123492.0 6.9
3123732.0 7.0
3124216.0 6.5
3124736.0 6.7
3125052.0 6.9
3129312.0 5.7
3130383.0 6.4
3130600.0 7.0
3130632.0 5.6
3131582.0 6.2
3132072.0 7.0
3132432.0 7.0
3132552.0 6.0
3132632.0 7.0
3132849.0 6.6
3134205.0 6.7
3134488.0 7.3
3134909.0 5.9
3134952.0 6.9
3136392.0 6.9
3137892.0 6.6
3138689.0 6.7
3155532.0 6.0
3155738.0 6.3
3156612.0 6.2
3157112.0 7.4
3157421.0 7.0
3157764.0 6.4
3158292.0 6.2
3158352.0 6.4
3158561.0 5.5
3159236.0 7.3
3159792.0 7.5
3159912.0 6.7
3160188.0 6.1
3160252.0 6.3
3160582.0 5.6
3161112.0 6.4
3161426.0 6.4
3161826.0 7.5
3161870.0 5.7
3162311.0 6.2
3162492.0 6.4
3162905.0 6.1
3164649.0 6.1
3164842.0 7.3
3165192.0 7.5
3165252.0 6.7
3165672.0 6.4
3166599.0 6.7
3168730.0 7.2
3169230.0 6.3
3169692.0 7.0
3170167.0 5.7
3171432.0 5.8
3171676.0 7.5
3171821.0 5.6
3171967.0 6.8
3172036.0 6.1
3172263.0 5.8
3172452.0 6.9
3173038.0 6.6
3173215.0 6.6
3173892.0 6.2
3174980.0 6.6
3175168.0 5.5
3175752.0 7.0
3176437.0 6.3
3176680.0 7.2
3177561.0 5.9
3177624.0 6.8
3177746.0 6.4
3177934.0 5.5
3178272.0 6.5
3178864.0 5.6
3180338.0 6.6
3180421.0 6.8
3194706.0 7.0
3195229.0 5.9
3195424.0 6.6
3196025.0 5.5
3196407.0 7.3
3196932.0 6.7
3197352.0 6.4
3197487.0 6.5
3198066.0 6.5
3198492.0 6.9
3198672.0 5.7
3198792.0 7.3
3198962.0 6.2
3199272.0 7.3
3199932.0 7.3
3200112.0 5.8
3200306.0 6.2
3200410.0 6.8
3201552.0 6.7
3202396.0 6.8
3202812.0 7.4
3203218.0 6.9
3203430.0 6.3
3203958.0 7.3
3205192.0 7.2
3205512.0 6.0
3205572.0 6.5
3205632.0 6.5
3205752.0 7.5
3206422.0 6.5
3206492.0 5.7
3206578.0 6.7
3207132.0 6.3
3207539.0 6.4
3209348.0 6.7
3209385.0 7.3
3209452.0 5.5
3209590.0 7.1
3210192.0 6.2
3210674.0 5.6
3210696.0 5.9
3211123.0 5.9
3212162.0 6.5
3212276.0 7.4
3212712.0 6.6
3212832.0 6.0
3213192.0 6.0
3213539.0 6.6
3213656.0 6.3
3213693.0 6.2
3214771.0 6.9
3214826.0 6.4
3215216.0 5.9
3215257.0 6.7
3215712.0 7.3
3216072.0 5.8
3217032.0 6.8
3217706.0 6.2
3218124.0 6.3
3218314.0 7.0
3220462.0 7.3
3221200.0 6.4
3221352.0 7.2
3222235.0 6.0
3223759.0 7.1
3223810.0 5.7
3225275.0 7.0
3225912.0 6.5
3238042.0 5.8
3238378.0 6.2
3238409.0 7.5
3239052.0 7.4
3240699.0 5.5
3240949.0 6.3
3241692.0 5.9
3242082.0 6.9
3242505.0 6.8
3242705.0 6.2
3243432.0 7.1
3243822.0 5.8
3244692.0 6.4
3246012.0 7.3
3246312.0 6.0
3246800.0 7.0
3247130.0 7.2
3248024.0 6.1
3248354.0 6.6
3248629.0 6.9
3249012.0 6.3
3249072.0 5.7
3249974.0 6.4
3250272.0 5.7
3250392.0 7.1
3250452.0 6.7
3250572.0 7.1
3250632.0 6.7
3250712.0 7.1
3250844.0 6.5
3251017.0 7.3
3251034.0 7.0
3251488.0 6.2
3251505.0 6.0
3251652.0 7.1
3252248.0 7.1
3252418.0 6.9
3252649.0 6.5
3252915.0 7.4
3253392.0 7.1
3253650.0 6.2
3253824.0 7.4
3254090.0 7.5
3254170.0 5.5
3254772.0 5.5
3255207.0 7.1
3255864.0 7.4
3256392.0 7.3
3257532.0 5.8
3260232.0 6.4
3261792.0 6.3
3267732.0 5.7
3268413.0 6.7
3269848.0 7.0
3280993.0 6.0
3284712.0 6.9
3286572.0 6.1
3286862.0 7.1
3287712.0 7.0
3287772.0 7.3
3287892.0 7.3
3288661.0 6.3
3288726.0 6.3
3289092.0 6.7
3289332.0 5.8
3289392.0 7.1
3289452.0 5.5
3289512.0 7.2
3289984.0 6.9
3290019.0 6.2
3290042.0 6.4
3290049.0 6.3
3290142.0 6.9
3290652.0 6.6
3290772.0 6.6
3290832.0 6.6
3290892.0 7.3
3290952.0 6.2
3291016.0 7.4
3291141.0 6.1
3291154.0 6.3
3291553.0 6.6
3291668.0 5.8
3291800.0 7.0
3292032.0 5.5
3292332.0 5.9
3292758.0 6.2
3292929.0 6.3
3293034.0 7.4
3293412.0 6.2
3293532.0 6.3
3293903.0 6.1
3294228.0 6.4
3294430.0 5.6
3295152.0 6.4
3295212.0 5.8
3295661.0 5.7
3295832.0 6.7
3296167.0 6.6
3296292.0 6.4
3297096.0 5.7
3297502.0 6.7
3297792.0 7.0
3299532.0 7.1
3300467.0 6.0
3301468.0 6.8
3304446.0 5.5
3305232.0 6.8
3305824.0 5.9
3310410.0 6.3
3310692.0 5.9
3310872.0 6.8
3311112.0 6.3
3322741.0 5.9
3322814.0 7.3
3322858.0 5.6
3325152.0 6.1
3327550.0 7.0
3327812.0 7.4
3327972.0 6.0
3328152.0 6.4
3328332.0 6.3
3328643.0 6.3
3328862.0 7.0
3328902.0 7.3
3329652.0 6.5
3330054.0 6.6
3330100.0 7.2
3330226.0 7.3
3330326.0 6.2
3330364.0 6.0
3330438.0 7.2
3330509.0 6.3
3331092.0 5.8
3331506.0 6.4
3331661.0 5.5
3331839.0 5.9
3332204.0 6.6
3332292.0 7.4
3332652.0 5.9
3332950.0 5.6
3332979.0 7.3
3333262.0 7.2
3333449.0 7.4
3333672.0 7.1
3333972.0 5.7
3334092.0 6.7
3334422.0 6.9
3334875.0 7.2
3335112.0 5.5
3336017.0 5.9
3336174.0 6.8
3336852.0 5.9
3337626.0 7.4
3338232.0 5.7
3338957.0 7.3
3339552.0 6.4
3341232.0 6.8
3341692.0 5.9
3341790.0 6.9
3342792.0 7.1
3344774.0 7.1
3348647.0 5.6
3348975.0 7.3
3349049.0 7.3
3349572.0 7.2
3349932.0 6.7
3350123.0 6.4
3350207.0 6.4
3350412.0 7.3
3351915.0 6.0
3353010.0 7.1
3353253.0 5.7
3354505.0 5.5
3354764.0 6.9
3366332.0 6.9
3368532.0 7.3
3368712.0 5.6
3368876.0 6.3
3371084.0 5.8
3371172.0 7.0
3372107.0 6.2
3372386.0 7.0

3372792.0 6.1
3373969.0 7.0
3374617.0 5.9
3374698.0 5.5
3377441.0 5.8
3378432.0 5.8
3378880.0 7.4
3379320.0 6.8
3380353.0 7.1
3383265.0 7.4
3383617.0 7.1
3384492.0 7.0
3384692.0 6.4
3384885.0 6.2
3384958.0 5.9
3387812.0 6.4
3387939.0 7.2
3388310.0 6.0
3389137.0 6.6
3390012.0 6.0
3390502.0 7.2
3390992.0 6.7
3391087.0 6.4
3391114.0 6.2
3391272.0 6.0
3392017.0 6.7
3392633.0 7.2
3392772.0 5.8
3393589.0 6.4
3393872.0 6.0
3394699.0 5.8
3395652.0 6.4
3395832.0 6.0
3395952.0 6.0
3396012.0 6.6
3396143.0 6.2
3396256.0 6.7
3396294.0 7.4
3396598.0 6.5
3396617.0 7.0
3396986.0 6.3
3408549.0 5.9
3409992.0 5.8
3410292.0 6.2
3410727.0 5.6
3412046.0 6.1
3412932.0 6.0
3413352.0 6.0
3415723.0 5.7
3415872.0 6.7
3417192.0 7.1
3417813.0 6.0
3418297.0 5.5
3418752.0 6.0
3419289.0 6.1
3419619.0 6.2
3420905.0 7.3
3424392.0 5.8
3424812.0 6.0
3425832.0 7.3
3426423.0 7.1
3427572.0 7.0
3427865.0 7.0
3428952.0 6.9
3429305.0 6.2
3429617.0 6.5
3429702.0 7.0
3429707.0 6.9
3429978.0 7.0
3430212.0 5.9
3430512.0 5.9
3430572.0 6.1
3431309.0 6.6
3431464.0 6.0
3431537.0 5.8
3431591.0 6.5
3431952.0 6.3
3432532.0 5.6
3432726.0 6.5
3433212.0 5.7
3433904.0 6.7
3434325.0 6.5
3434532.0 6.5
3435086.0 5.5
3437079.0 7.1
3437652.0 6.5
3437712.0 5.6
3438792.0 6.8
3438912.0 6.5
3439960.0 6.4
3450492.0 6.0
3450884.0 6.7
3451495.0 7.2
3452384.0 6.6
3452475.0 7.4
3452818.0 7.2
3453492.0 7.1
3457572.0 6.7
3457932.0 7.2
3458682.0 6.0
3458716.0 6.3
3460392.0 5.8
3461127.0 5.7
3464440.0 5.9
3465817.0 7.5
3466125.0 6.2
3466152.0 5.9
3466332.0 6.8
3466632.0 7.4
3466764.0 6.6
3466894.0 6.6
3467113.0 6.1
3467178.0 6.4
3467322.0 6.5
3467404.0 6.7
3467592.0 6.6
3467712.0 7.0
3467832.0 5.8
3468850.0 6.8
3469032.0 7.0
3469152.0 7.3
3469392.0 7.3
3469512.0 6.6
3469644.0 6.1
3470386.0 6.7
3470532.0 5.9
3470772.0 6.4
3471045.0 5.7
3471105.0 6.4
3471114.0 6.2
3471355.0 7.5
3471422.0 6.2
3471655.0 5.7
3471777.0 6.3
3471904.0 5.7
3472721.0 7.3
3472899.0 6.5
3473171.0 7.4
3473293.0 7.0
3473652.0 7.3
3473832.0 5.6
3473908.0 6.0
3474243.0 7.1
3474562.0 6.1
3474972.0 7.0
3475032.0 6.3
3475690.0 6.9
3475892.0 7.3
3476015.0 6.3
3476122.0 6.5
3476151.0 6.7
3476352.0 6.1
3476850.0 6.2
3476873.0 6.5
3477056.0 6.1
3477158.0 7.3
3477792.0 7.0
3477912.0 6.7
3478032.0 6.1
3478763.0 7.3
3478987.0 6.1
3479089.0 5.7
3479112.0 7.5
3479352.0 6.6
3479992.0 5.8
3480552.0 6.6
3480792.0 7.2
3481102.0 5.7
3481289.0 6.9
3481327.0 5.7
3481802.0 7.3
3481972.0 5.7
3481992.0 7.1
3482112.0 7.1
3482292.0 7.3
3482352.0 6.5
3482804.0 6.9
3482930.0 6.0
3483343.0 6.3
3483352.0 5.6
3494064.0 6.8
3498488.0 7.3
3500242.0 6.2
3502272.0 7.2
3504252.0 7.1
3504293.0 6.9
3505512.0 7.0
3505745.0 7.4
3506712.0 6.7
3507635.0 5.8
3507846.0 5.9
3508092.0 5.9
3508272.0 5.9
3508332.0 6.2
3508784.0 7.2
3509068.0 6.7
3509330.0 7.2
3509925.0 6.1
3510470.0 7.1
3510581.0 6.2
3510704.0 6.3
3511587.0 6.0
3512292.0 6.3
3512472.0 7.5
3512532.0 7.1
3512592.0 6.8
3512974.0 5.8
3513103.0 7.3
3513543.0 5.8
3514316.0 7.4
3514494.0 6.5
3514735.0 6.9
3515854.0 6.0
3516166.0 6.4
3516257.0 7.1
3516405.0 6.8
3516552.0 7.2
3519792.0 7.0
3520631.0 6.7
3523117.0 6.0
3523380.0 7.4
3523493.0 7.4
3523752.0 6.0
3523812.0 6.3
3524052.0 5.9
3524232.0 5.6
3524731.0 7.5
3524802.0 6.3
3524902.0 7.2
3537012.0 6.4
3537072.0 5.9
3537498.0 5.6
3539482.0 6.1
3543710.0 6.9
3544032.0 5.6
3544392.0 7.3
3544507.0 6.4
3545352.0 6.0
3545412.0 5.9
3545652.0 6.3
3545832.0 6.8
3546412.0 5.8
3546584.0 6.2
3546657.0 6.9
3546912.0 7.1
3547404.0 7.2
3547597.0 7.3
3547944.0 5.6
3548352.0 6.1
3548532.0 7.3
3548939.0 5.9
3548955.0 5.9
3549014.0 7.3
3549044.0 7.1
3549175.0 6.1
3549190.0 7.1
3549236.0 7.1
3549732.0 7.2
3549912.0 6.2
3550376.0 7.0
3550812.0 7.0
3551412.0 6.8
3552457.0 5.5
3552478.0 6.7
3552494.0 5.7
3552612.0 6.7
3552852.0 6.0
3553514.0 6.8
3553992.0 5.9
3554607.0 5.8
3555279.0 5.8
3555409.0 5.6
3556255.0 7.5
3556514.0 7.0
3557352.0 6.9
3558119.0 5.6
3558792.0 6.8
3560232.0 6.3
3560725.0 6.5
3560967.0 6.8
3561139.0 6.2
3561312.0 5.8
3561672.0 6.4
3562632.0 6.7
3562692.0 6.4
3563112.0 7.0
3563361.0 5.5
3563775.0 5.8
3563787.0 7.1
3564072.0 7.0
3564192.0 5.6
3564252.0 7.4
3564492.0 6.0
3564767.0 7.1
3564991.0 6.1
3565407.0 6.5
3565444.0 6.9
3565572.0 5.8
3565692.0 6.2
3565812.0 5.8
3565872.0 5.6
3566084.0 6.2
3566154.0 5.6
3566251.0 7.4
3566396.0 6.9
3566519.0 7.1
3566534.0 6.0
3566670.0 6.9
3580212.0 7.4
3582912.0 6.9
3583909.0 7.1
3584292.0 5.8
3585100.0 7.3
3585184.0 6.6
3585599.0 7.4
3585732.0 6.3
3585792.0 6.4
3585852.0 6.8
3586375.0 6.3
3586436.0 5.6
3586584.0 7.3
3586620.0 6.8
3586710.0 5.7
3587718.0 7.2
3587928.0 5.6
3587946.0 6.8
3588046.0 6.6
3588113.0 6.8
3588246.0 6.0
3588309.0 7.3
3588432.0 6.5
3588672.0 5.7
3588792.0 6.8
3588972.0 5.8
3589032.0 6.7
3589246.0 7.3
3589355.0 7.5
3589411.0 6.7
3589482.0 6.0
3589530.0 6.6
3589714.0 6.8
3589933.0 5.8
3589992.0 5.6
3590052.0 7.0
3590472.0 5.7
3590902.0 6.5
3591053.0 6.1
3591379.0 6.0
3591492.0 6.6
3592112.0 6.2
3592153.0 6.5
3592170.0 7.3
3592315.0 5.9
3593232.0 5.8
3593352.0 7.1
3594148.0 6.2
3594279.0 6.4
3594312.0 6.0
3594732.0 6.8
3595494.0 6.5
3595517.0 7.4
3595662.0 7.0
3596503.0 6.6
3596911.0 6.9
3596933.0 5.7
3597113.0 5.5
3597143.0 6.8
3597192.0 6.9
3597312.0 6.8
3598040.0 7.3
3598059.0 5.8
3598425.0 6.0
3598632.0 5.9
3601445.0 7.2
3602500.0 6.6
3602790.0 5.5
3602918.0 7.4
3603799.0 5.9
3603897.0 7.3
3604077.0 7.4
3604162.0 5.8
3604304.0 6.6
3604383.0 6.5
3604632.0 5.7
3604812.0 6.3
3604983.0 6.7
3605027.0 6.3
3605222.0 7.1
3605267.0 7.2
3605425.0 6.8
3605529.0 6.8
3605598.0 7.3
3605662.0 6.0
3605892.0 6.9
3606427.0 6.4
3606462.0 7.4
3606916.0 6.7
3607059.0 6.1
3607272.0 7.3
3607332.0 6.7
3607632.0 7.5
3608185.0 6.3
3608712.0 7.1
3608892.0 7.5
3609192.0 6.5
3609686.0 6.2
3609810.0 6.6
3621320.0 7.2
3622384.0 6.2
3622534.0 7.3
3622839.0 6.6
3622850.0 7.3
3623172.0 6.4
3623532.0 7.4
3623825.0 6.6
3623973.0 6.2
3624282.0 5.7
3624612.0 5.6
3625032.0 6.8
3625662.0 6.7
3626112.0 7.0
3626232.0 7.2
3626764.0 5.9
3627228.0 6.7
3627552.0 6.2
3628283.0 5.5
3628306.0 6.5
3628329.0 5.7
3628369.0 6.8
3628406.0 6.8
3628484.0 5.7
3628872.0 7.3
3629052.0 6.7
3629232.0 6.6
3629737.0 5.7
3629807.0 5.6
3630076.0 5.6
3630118.0 6.8
3630218.0 6.9
3630262.0 6.4
3630312.0 5.7
3630612.0 7.2
3631256.0 5.7
3631425.0 7.3
3631697.0 7.0
3631872.0 7.0
3632052.0 6.5
3632232.0 5.7
3632436.0 7.4
3632524.0 6.7
3633038.0 6.5
3633252.0 7.0
3633372.0 6.9
3633790.0 6.1
3634042.0 6.9
3634090.0 6.5
3634106.0 5.6
3634342.0 6.1
3634569.0 6.9
3634932.0 6.1
3635675.0 7.1
3636034.0 6.1
3638167.0 6.1
3639012.0 6.8
3639997.0 6.6
3642252.0 6.6
3642598.0 7.0
3642625.0 6.5
3642664.0 6.1
3642679.0 5.6
3642704.0 5.9
3642777.0 6.0
3643086.0 5.8
3643272.0 6.7
3643392.0 7.2
3643632.0 6.1
3643752.0 6.1
3643867.0 5.8
3644094.0 5.9
3644131.0 6.5
3644145.0 6.3
3644172.0 7.4
3644395.0 7.0
3644407.0 6.0
3644502.0 5.6
3644533.0 7.4
3644555.0 7.1
3644599.0 5.9
3644952.0 6.3
3645192.0 6.3
3645307.0 6.9
3645724.0 5.8
3645865.0 5.7
3645921.0 7.5
3646332.0 7.4
3646512.0 6.1
3646632.0 7.1
3647100.0 7.3
3647379.0 5.8
3647413.0 7.2
3647485.0 5.6
3648442.0 5.7
3648611.0 6.4
3648638.0 5.6
3648715.0 6.1
3648889.0 7.3
3650202.0 6.0
3650422.0 7.1
3661975.0 7.5
3662472.0 7.4
3662613.0 6.3
3662728.0 6.0
3662956.0 6.3
3663366.0 5.9
3664215.0 5.7
3664327.0 7.0
3664362.0 5.6
3664629.0 6.7
3665112.0 6.3
3665172.0 7.2
3665537.0 5.7
3666104.0 7.5
3666235.0 6.1
3667353.0 7.5
3667932.0 6.3
3668112.0 6.9
3668511.0 6.4
3669108.0 6.6
3669432.0 5.6
3669492.0 6.2
3669612.0 5.6
3669745.0 6.8
3669900.0 7.3
3670512.0 6.8
3670932.0 7.3
3670992.0 6.0
3671537.0 6.6
3671609.0 6.3
3671886.0 6.9
3671964.0 7.1
3672739.0 7.0
3673057.0 6.6
3673201.0 6.2
3673417.0 6.0
3675012.0 6.4
3675132.0 5.7
3675372.0 6.5
3675432.0 6.6
3675977.0 5.9
3676452.0 5.7
3676512.0 6.7
3677258.0 5.6
3677398.0 6.2
3677463.0 7.4
3678132.0 5.5
3680484.0 6.9
3680772.0 7.3
3680952.0 5.7
3681384.0 6.6
3681397.0 6.8
3681473.0 6.3
3681508.0 6.7
3681600.0 7.3
3681617.0 7.5
3681726.0 6.5
3682332.0 7.3
3682452.0 5.6
3682572.0 5.6
3682632.0 6.5
3682845.0 6.6
3682904.0 7.0
3683019.0 6.7
3683036.0 7.2
3683054.0 6.7
3683094.0 7.3
3683128.0 5.9
3683144.0 6.0
3683183.0 5.8
3683227.0 6.0
3683247.0 6.5
3683265.0 5.5
3683338.0 5.9
3683489.0 6.6
3683592.0 6.2
3683712.0 6.7
3683832.0 5.7
3683892.0 6.0
3683952.0 7.2
3684012.0 7.1
3684216.0 7.1
3684327.0 6.5
3684356.0 6.3
3684472.0 7.0
3684707.0 6.6
3684892.0 5.7
3684924.0 6.9
3684958.0 6.5
3685092.0 7.2
3685152.0 5.8
3685212.0 6.1
3685608.0 6.8
3685959.0 6.2
3686407.0 6.2
3687036.0 7.4
3687128.0 6.5
3687172.0 6.3
3687470.0 7.1
3687618.0 6.1
3688092.0 7.2
3688651.0 7.0
3689532.0 6.1
3690248.0 5.7
3691780.0 7.1
3692652.0 6.3
3692780.0 6.9
3692860.0 7.4
3693590.0 5.7
3705998.0 7.2
3706044.0 6.5
3706203.0 6.5
3706247.0 7.5
3706632.0 7.1
3706812.0 5.8
3707705.0 7.1
3708072.0 6.5
3708647.0 7.1
3709410.0 7.1
3709418.0 7.3
3709512.0 7.3
3710739.0 6.3
3710909.0 6.1
3710952.0 6.2
3712288.0 6.8
3712392.0 6.1
3712572.0 5.7
3713105.0 6.0
3713760.0 7.4
3714930.0 6.0
3715512.0 5.6
3715572.0 6.0
3716447.0 6.9
3716712.0 7.3
3717847.0 5.6
3718804.0 6.7
3719053.0 5.9
3719419.0 5.5
3719832.0 6.1
3720242.0 6.0
3720314.0 5.7
3720435.0 5.8
3720590.0 6.3
3721452.0 7.2
3721653.0 6.7
3721838.0 7.2
3721953.0 6.8
3722048.0 6.5
3722163.0 7.4
3722267.0 5.7
3722384.0 6.9
3722472.0 6.7
3722532.0 7.3
3722712.0 5.6
3722772.0 6.8
3722892.0 6.3
3722952.0 7.1
3723176.0 5.9
3723282.0 6.1
3723307.0 7.1
3723526.0 5.6
3723688.0 5.7
3723703.0 7.1
3723972.0 5.9
3724092.0 6.3
3724570.0 7.1
3724786.0 5.7
3724889.0 6.7
3725068.0 6.0
3725089.0 7.0
3725472.0 7.3
3725532.0 6.0
3725592.0 7.2
3726117.0 5.7
3726280.0 6.0
3726451.0 6.4
3726587.0 7.4
3726852.0 7.5
3727032.0 6.3
3727092.0 7.4
3727651.0 7.4
3727826.0 5.6
3727842.0 7.0
3728944.0 6.0
3729214.0 6.4
3729972.0 6.6
3730648.0 6.8
3730908.0 7.5
3731592.0 6.1
3732732.0 6.9
3733635.0 5.7
3735302.0 5.6
3748083.0 7.0
3748572.0 5.9
3749552.0 7.4
3750132.0 5.6
3750502.0 7.3
3750577.0 5.9
3751512.0 7.1
3752127.0 6.6
3754143.0 6.3
3754512.0 6.3
3754572.0 7.0
3756473.0 6.4
3757092.0 5.9
3757332.0 5.6
3757795.0 6.0
3757906.0 7.2
3758008.0 5.8
3758083.0 5.9
3758114.0 6.6
3758140.0 6.3
3758224.0 7.5
3758532.0 6.0
3758952.0 7.2
3759077.0 6.3
3759146.0 7.0
3759339.0 7.0
3759426.0 7.3
3759714.0 7.3
3759972.0 6.9
3760152.0 5.8
3760212.0 7.4
3760478.0 6.7
3760597.0 5.6
3760640.0 7.4
3760665.0 6.6
3760768.0 5.6
3761128.0 6.9
3761160.0 6.8
3761179.0 5.5
3761310.0 6.3
3761352.0 5.9
3761472.0 5.8
3761532.0 6.0
3761652.0 6.6
3761712.0 6.7
3761772.0 5.8
3761832.0 5.9
3761945.0 7.4
3762112.0 6.5
3762153.0 7.1
3762270.0 6.5
3762352.0 6.6
3762448.0 7.3
3762510.0 6.4
3762589.0 6.1
3762722.0 6.7
3762739.0 6.2
3762747.0 7.0
3762792.0 6.0
3762912.0 5.9
3762972.0 6.4
3763032.0 6.6
3763272.0 5.8
3763398.0 7.5
3763422.0 6.3
3763748.0 5.8
3763915.0 6.3
3764352.0 6.7
3764652.0 6.5
3765159.0 5.7
3765289.0 6.9
3765577.0 6.3
3765792.0 7.4
3765972.0 5.7
3766092.0 7.0
3766842.0 6.9
3767172.0 6.1
3767412.0 5.9
3767665.0 7.2
3767759.0 5.7
3768309.0 6.4
3768330.0 7.4
3768445.0 5.9
3768552.0 7.2
3769890.0 7.0
3771308.0 5.7
3771432.0 6.9
3771672.0 6.7
3777672.0 6.8
3778812.0 6.2
3778872.0 6.7
3779052.0 6.8
3779363.0 7.2
3779700.0 6.1
3795239.0 5.9
3797532.0 6.9
3797592.0 6.4
3797832.0 7.2
3797892.0 6.2
3798557.0 6.6
3798912.0 6.6
3798972.0 7.1
3799092.0 5.8
3799425.0 5.9
3799548.0 7.2
3799657.0 6.2
3799701.0 7.3
3799769.0 5.9
3799848.0 6.8
3799919.0 6.2
3799942.0 6.2
3800056.0 6.3
3800072.0 5.6
3800232.0 5.5
3800532.0 7.1
3800652.0 7.1
3800779.0 5.9
3800793.0 6.7
3800813.0 6.3
3801069.0 5.8
3801142.0 5.5
3801245.0 6.7
3801448.0 6.0
3801498.0 6.1
3801562.0 5.8
3801732.0 6.4
3801972.0 7.3
3802032.0 6.5
3802092.0 6.3
3802243.0 6.1
3802380.0 6.2
3802474.0 5.7
3802530.0 7.4
3802788.0 6.9
3802949.0 6.7
3803104.0 6.5
3803112.0 6.2
3803292.0 6.4
3803412.0 7.0
3804155.0 5.8
3804271.0 6.0
3804344.0 6.1
3804552.0 7.4
3804612.0 7.3
3805032.0 6.1
3805159.0 7.5
3805370.0 5.9
3805585.0 6.7
3805672.0 5.8
3805789.0 5.8
3805992.0 7.3
3806352.0 6.0
3807075.0 7.2
3807404.0 6.9
3807552.0 6.2
3808103.0 7.0
3808992.0 6.9
3811167.0 6.4
3811428.0 6.9
3811659.0 6.1
3811752.0 7.3
3812628.0 6.8
3812702.0 6.2
3815877.0 6.5
3816432.0 5.8
3816552.0 7.2
3816669.0 6.1
3817287.0 5.5
3817752.0 6.5
3818375.0 6.4
3819132.0 7.4
3819724.0 5.7
3819774.0 6.6
3820089.0 7.2
3820289.0 6.6
3821089.0 7.3
3821469.0 7.3
3821555.0 6.1
3821678.0 5.9
3834149.0 7.5
3836098.0 5.8
3837732.0 6.4
3837972.0 6.5
3838032.0 6.8
3838657.0 6.4
3838692.0 5.6
3838755.0 7.4
3838763.0 6.6
3839172.0 7.1
3839292.0 5.7
3839352.0 7.3
3839412.0 6.1
3839472.0 7.2
3839532.0 7.0
3839592.0 6.3
3839757.0 6.6
3839846.0 7.0
3839952.0 7.0
3840044.0 7.2
3840147.0 6.1
3840174.0 5.7
3840325.0 5.6
3840452.0 5.7
3840515.0 5.6
3840732.0 5.9
3840852.0 6.0
3840912.0 6.7
3840972.0 7.0
3841032.0 6.7
3841142.0 7.0
3841457.0 7.1
3841586.0 5.9
3841764.0 7.2
3841992.0 7.2
3842172.0 7.4
3842232.0 6.6
3842292.0 6.0
3842412.0 5.9
3842754.0 6.6
3843328.0 7.1
3843392.0 5.8
3843552.0 7.2
3843612.0 5.9
3843732.0 5.7
3843792.0 6.7
3844012.0 5.6
3844332.0 5.6
3844537.0 6.6
3845232.0 5.7
3845352.0 6.8
3845477.0 6.8
3845644.0 6.2
3845809.0 6.2
3845913.0 6.6
3846082.0 7.4
3846172.0 5.9
3846285.0 7.4
3846492.0 7.0
3847812.0 6.7
3849801.0 5.9
3850586.0 7.3
3851865.0 5.9
3852132.0 6.3
3852252.0 7.4
3854872.0 6.6
3854930.0 6.7
3856010.0 6.0
3856872.0 7.4
3857002.0 7.3
3857055.0 6.4
3857832.0 6.2
3857892.0 5.6
3858584.0 6.0
3858606.0 5.5
3858656.0 7.1
3858696.0 6.6
3858719.0 5.6
3858746.0 7.1
3858798.0 7.3
3859006.0 7.2
3859042.0 7.5
3859217.0 6.2
3859272.0 6.6
3859332.0 6.3
3859752.0 5.7
3859874.0 7.5
3860186.0 6.9
3860712.0 5.8
3860832.0 7.2
3860952.0 5.5
3862918.0 5.5
3863712.0 7.2
3863892.0 6.0
3864012.0 6.6
3864194.0 6.4
3864207.0 6.2
3864766.0 6.2
3864970.0 5.5
3875922.0 5.8
3877910.0 6.3
3878172.0 5.9
3878292.0 5.9
3879092.0 6.2
3879139.0 5.6
3879160.0 6.7
3879432.0 5.6
3879792.0 6.1
3880115.0 7.2
3880335.0 5.8
3880538.0 7.2
3880700.0 6.4
3880932.0 6.5
3881052.0 7.5
3881112.0 6.5
3881172.0 7.4
3881232.0 6.9
3881352.0 7.2
3881780.0 5.7
3881810.0 5.6
3881939.0 7.0
3882161.0 6.4

| | |
|---|---|
| 3882298.0 | 6.0 |
| 3882492.0 | 6.2 |
| 3882552.0 | 7.2 |
| 3882732.0 | 6.4 |
| 3882792.0 | 6.2 |
| 3883044.0 | 6.2 |
| 3883134.0 | 6.0 |
| 3883199.0 | 6.8 |
| 3883528.0 | 5.8 |
| 3884490.0 | 5.7 |
| 3885042.0 | 6.2 |
| 3886274.0 | 6.4 |
| 3886397.0 | 7.3 |
| 3886872.0 | 7.4 |
| 3886992.0 | 5.8 |
| 3887230.0 | 6.7 |
| 3887449.0 | 6.8 |
| 3887669.0 | 7.5 |
| 3888312.0 | 7.0 |
| 3888432.0 | 5.8 |
| 3889203.0 | 5.9 |
| 3893697.0 | 7.1 |
| 3895512.0 | 6.9 |
| 3895632.0 | 6.8 |
| 3896439.0 | 6.4 |
| 3896772.0 | 6.5 |
| 3896952.0 | 7.1 |
| 3897012.0 | 6.4 |
| 3897488.0 | 7.0 |
| 3897899.0 | 5.6 |
| 3898933.0 | 7.4 |
| 3899014.0 | 5.5 |
| 3899131.0 | 7.4 |
| 3899277.0 | 7.5 |
| 3899350.0 | 5.6 |
| 3899532.0 | 6.7 |
| 3899712.0 | 6.2 |
| 3900162.0 | 7.0 |
| 3900311.0 | 7.3 |
| 3900370.0 | 6.2 |
| 3900548.0 | 6.0 |
| 3900591.0 | 6.0 |
| 3900700.0 | 6.4 |
| 3900746.0 | 6.2 |
| 3900776.0 | 6.8 |
| 3901092.0 | 7.0 |
| 3901152.0 | 5.5 |
| 3901605.0 | 7.1 |
| 3901823.0 | 6.9 |
| 3902952.0 | 5.6 |
| 3903152.0 | 7.5 |
| 3903519.0 | 7.3 |
| 3903794.0 | 6.8 |
| 3903816.0 | 6.7 |
| 3903859.0 | 7.2 |
| 3905309.0 | 6.9 |
| 3905532.0 | 6.9 |
| 3905592.0 | 6.6 |
| 3905652.0 | 6.7 |
| 3906043.0 | 6.7 |
| 3906140.0 | 6.0 |
| 3906234.0 | 7.2 |
| 3906319.0 | 7.1 |
| 3906609.0 | 5.8 |
| 3906665.0 | 6.1 |
| 3906702.0 | 5.8 |
| 3917112.0 | 7.1 |
| 3917172.0 | 6.8 |
| 3917232.0 | 7.4 |
| 3917292.0 | 6.0 |
| 3917352.0 | 6.1 |
| 3917484.0 | 5.9 |
| 3917534.0 | 6.2 |
| 3917625.0 | 7.3 |
| 3917666.0 | 7.1 |
| 3918219.0 | 6.1 |
| 3918296.0 | 7.2 |
| 3918895.0 | 6.9 |
| 3918920.0 | 6.5 |
| 3919652.0 | 6.9 |
| 3919682.0 | 7.3 |
| 3920112.0 | 7.4 |
| 3920735.0 | 6.9 |
| 3920886.0 | 6.2 |
| 3920934.0 | 6.6 |
| 3921372.0 | 6.5 |
| 3921492.0 | 6.5 |
| 3921552.0 | 7.4 |
| 3921944.0 | 5.6 |
| 3921957.0 | 6.4 |
| 3922333.0 | 7.5 |
| 3922342.0 | 5.8 |
| 3922692.0 | 6.0 |
| 3923747.0 | 7.5 |
| 3923822.0 | 6.8 |
| 3924072.0 | 6.9 |
| 3924432.0 | 7.2 |
| 3924492.0 | 7.2 |
| 3924552.0 | 5.6 |
| 3926555.0 | 6.2 |
| 3926623.0 | 7.5 |
| 3933012.0 | 7.0 |
| 3933907.0 | 5.8 |
| 3934512.0 | 5.7 |
| 3935592.0 | 6.5 |
| 3935772.0 | 6.9 |
| 3936584.0 | 6.1 |
| 3936688.0 | 5.9 |
| 3936975.0 | 7.4 |
| 3937212.0 | 7.1 |
| 3937392.0 | 5.8 |
| 3937512.0 | 6.4 |
| 3937809.0 | 6.1 |
| 3938532.0 | 5.6 |
| 3938712.0 | 6.9 |
| 3938952.0 | 5.8 |
| 3939276.0 | 5.6 |
| 3939364.0 | 7.4 |
| 3939455.0 | 6.3 |
| 3939665.0 | 5.9 |
| 3939871.0 | 6.6 |
| 3940392.0 | 7.5 |
| 3940615.0 | 7.1 |
| 3940643.0 | 7.5 |
| 3940731.0 | 6.9 |
| 3941155.0 | 6.5 |
| 3941301.0 | 6.0 |
| 3941412.0 | 7.4 |
| 3941592.0 | 7.0 |
| 3942296.0 | 6.5 |
| 3942431.0 | 5.6 |
| 3942551.0 | 6.2 |
| 3942852.0 | 6.9 |
| 3943866.0 | 7.5 |
| 3944652.0 | 6.5 |
| 3944712.0 | 6.0 |
| 3945002.0 | 6.7 |
| 3946364.0 | 5.9 |
| 3946533.0 | 7.0 |
| 3947412.0 | 7.1 |
| 3947592.0 | 6.1 |
| 3947744.0 | 7.0 |
| 3947904.0 | 5.8 |
| 3948075.0 | 7.4 |
| 3948469.0 | 5.5 |
| 3948672.0 | 5.6 |
| 3948852.0 | 7.0 |
| 3949130.0 | 7.1 |
| 3960192.0 | 7.3 |
| 3960667.0 | 6.9 |
| 3960897.0 | 5.7 |
| 3961028.0 | 6.0 |
| 3962617.0 | 6.0 |
| 3963942.0 | 5.6 |
| 3966192.0 | 6.2 |
| 3967070.0 | 7.3 |
| 3967122.0 | 6.2 |
| 3970272.0 | 7.0 |
| 3970739.0 | 7.2 |
| 3972687.0 | 6.4 |
| 3974712.0 | 5.7 |
| 3974832.0 | 5.7 |
| 3975097.0 | 7.1 |
| 3975254.0 | 6.0 |
| 3975287.0 | 6.4 |
| 3975407.0 | 6.1 |
| 3975804.0 | 7.1 |
| 3975912.0 | 7.0 |
| 3975972.0 | 5.8 |
| 3976272.0 | 5.8 |
| 3976332.0 | 7.1 |
| 3976472.0 | 6.6 |
| 3976497.0 | 7.3 |
| 3976666.0 | 6.6 |
| 3976797.0 | 6.0 |
| 3976911.0 | 7.3 |
| 3977002.0 | 7.0 |
| 3977412.0 | 5.9 |
| 3977532.0 | 6.2 |
| 3978012.0 | 6.2 |
| 3978099.0 | 7.5 |
| 3978504.0 | 5.8 |
| 3978560.0 | 6.9 |
| 3978852.0 | 7.4 |
| 3978912.0 | 6.4 |
| 3979032.0 | 5.7 |
| 3979391.0 | 7.0 |
| 3979447.0 | 7.2 |
| 3979708.0 | 6.2 |
| 3979807.0 | 7.0 |
| 3980292.0 | 6.9 |
| 3981574.0 | 5.7 |
| 3983719.0 | 7.3 |
| 3985165.0 | 6.0 |
| 3985547.0 | 6.8 |
| 3987261.0 | 6.1 |
| 3987732.0 | 7.5 |
| 3990552.0 | 5.8 |
| 3991170.0 | 7.4 |
| 3991191.0 | 7.1 |
| 3991560.0 | 6.3 |
| 3991638.0 | 6.4 |
| 4003572.0 | 5.8 |
| 4013592.0 | 6.8 |
| 4013912.0 | 7.2 |
| 4014717.0 | 6.2 |
| 4015152.0 | 6.3 |
| 4015704.0 | 5.9 |
| 4015763.0 | 6.2 |
| 4015949.0 | 6.9 |
| 4016027.0 | 6.7 |
| 4016232.0 | 6.8 |
| 4016292.0 | 6.4 |
| 4017182.0 | 6.0 |
| 4017261.0 | 5.8 |
| 4017612.0 | 6.2 |
| 4017732.0 | 6.8 |
| 4017912.0 | 6.9 |
| 4018032.0 | 6.0 |
| 4018092.0 | 5.8 |
| 4018291.0 | 6.7 |
| 4018378.0 | 7.4 |
| 4018427.0 | 6.0 |
| 4018513.0 | 6.1 |
| 4018672.0 | 7.1 |
| 4018905.0 | 6.5 |
| 4020852.0 | 5.6 |
| 4024601.0 | 6.5 |
| 4025052.0 | 6.4 |
| 4027618.0 | 7.4 |
| 4027732.0 | 5.8 |
| 4028657.0 | 7.0 |
| 4029151.0 | 6.2 |
| 4029312.0 | 5.7 |
| 4030387.0 | 6.5 |
| 4030475.0 | 6.0 |
| 4030490.0 | 7.1 |
| 4030565.0 | 6.4 |
| 4030622.0 | 6.1 |
| 4030872.0 | 6.6 |
| 4031237.0 | 5.7 |
| 4031705.0 | 6.8 |
| 4032132.0 | 6.9 |
| 4032192.0 | 6.0 |
| 4032492.0 | 6.3 |
| 4032552.0 | 7.5 |
| 4032834.0 | 6.5 |
| 4032947.0 | 5.8 |
| 4032968.0 | 6.9 |
| 4032989.0 | 6.8 |
| 4033195.0 | 7.0 |
| 4033260.0 | 7.4 |
| 4033512.0 | 6.7 |
| 4033692.0 | 6.5 |
| 4033752.0 | 5.9 |
| 4033812.0 | 5.9 |
| 4033992.0 | 6.1 |
| 4034189.0 | 7.3 |
| 4034259.0 | 7.0 |
| 4034269.0 | 5.8 |
| 4034389.0 | 6.7 |
| 4034624.0 | 6.8 |
| 4034633.0 | 6.6 |
| 4034737.0 | 5.6 |
| 4034829.0 | 6.7 |
| 4034868.0 | 6.4 |
| 4045669.0 | 6.3 |
| 4045790.0 | 6.1 |
| 4045806.0 | 6.3 |
| 4046319.0 | 7.1 |
| 4047972.0 | 6.9 |
| 4048032.0 | 7.3 |
| 4048756.0 | 5.6 |
| 4050792.0 | 7.0 |
| 4052980.0 | 6.1 |
| 4053136.0 | 6.4 |
| 4053347.0 | 6.8 |
| 4053353.0 | 5.6 |
| 4053504.0 | 5.6 |
| 4053672.0 | 6.7 |
| 4053792.0 | 6.8 |
| 4054032.0 | 7.5 |
| 4054092.0 | 7.2 |
| 4054398.0 | 5.7 |
| 4054500.0 | 7.3 |
| 4054524.0 | 7.5 |
| 4054572.0 | 6.7 |
| 4054625.0 | 7.2 |
| 4054725.0 | 5.9 |
| 4054739.0 | 6.8 |
| 4054826.0 | 6.0 |
| 4055046.0 | 6.3 |
| 4055081.0 | 5.9 |
| 4055097.0 | 6.8 |
| 4055172.0 | 5.6 |
| 4055767.0 | 6.8 |
| 4056013.0 | 6.5 |
| 4056507.0 | 6.6 |
| 4056792.0 | 6.1 |
| 4056912.0 | 6.3 |
| 4057032.0 | 5.9 |
| 4057121.0 | 6.7 |
| 4057158.0 | 7.3 |
| 4057694.0 | 7.2 |
| 4057716.0 | 7.2 |
| 4057864.0 | 6.1 |
| 4058112.0 | 6.2 |
| 4058232.0 | 7.5 |
| 4058412.0 | 6.9 |
| 4058472.0 | 6.2 |
| 4058565.0 | 7.2 |
| 4058760.0 | 6.5 |
| 4058897.0 | 6.6 |
| 4058964.0 | 6.9 |
| 4059612.0 | 5.6 |
| 4059732.0 | 6.0 |
| 4059792.0 | 6.7 |
| 4060400.0 | 6.0 |
| 4060480.0 | 5.5 |
| 4060605.0 | 5.7 |
| 4060622.0 | 6.8 |
| 4060872.0 | 5.8 |
| 4061112.0 | 6.5 |
| 4061292.0 | 7.4 |
| 4061513.0 | 6.8 |
| 4061739.0 | 5.8 |
| 4062492.0 | 6.2 |
| 4062969.0 | 6.7 |
| 4063007.0 | 5.6 |
| 4063537.0 | 5.9 |
| 4064786.0 | 7.3 |
| 4066632.0 | 6.2 |
| 4066932.0 | 6.6 |
| 4067525.0 | 5.8 |
| 4067543.0 | 6.7 |
| 4067598.0 | 6.3 |
| 4068719.0 | 5.7 |
| 4069692.0 | 5.6 |
| 4070198.0 | 6.9 |
| 4070297.0 | 5.5 |
| 4070590.0 | 6.9 |
| 4070826.0 | 7.0 |
| 4070873.0 | 5.8 |
| 4070943.0 | 6.4 |
| 4071505.0 | 5.9 |
| 4071658.0 | 6.1 |
| 4071759.0 | 7.5 |
| 4071848.0 | 7.3 |
| 4072013.0 | 6.7 |
| 4072192.0 | 6.2 |
| 4072240.0 | 5.6 |
| 4072269.0 | 6.3 |
| 4072344.0 | 6.0 |
| 4072374.0 | 6.7 |
| 4072452.0 | 5.7 |
| 4072512.0 | 5.5 |
| 4072632.0 | 6.7 |
| 4072692.0 | 7.1 |
| 4072752.0 | 7.0 |
| 4072872.0 | 6.5 |
| 4073013.0 | 5.6 |
| 4073052.0 | 7.4 |
| 4073079.0 | 7.4 |
| 4073083.0 | 6.3 |
| 4073145.0 | 7.1 |
| 4073152.0 | 7.1 |
| 4073273.0 | 5.6 |
| 4073338.0 | 6.8 |
| 4073392.0 | 7.2 |
| 4073506.0 | 7.4 |
| 4073575.0 | 5.6 |
| 4073752.0 | 6.2 |
| 4073808.0 | 7.4 |
| 4073832.0 | 6.4 |
| 4073892.0 | 6.6 |
| 4073952.0 | 6.4 |
| 4074012.0 | 6.7 |
| 4074312.0 | 7.2 |
| 4074382.0 | 5.6 |
| 4074409.0 | 6.2 |
| 4074449.0 | 5.6 |
| 4074496.0 | 5.6 |
| 4074555.0 | 6.1 |
| 4074584.0 | 6.9 |
| 4074677.0 | 6.8 |
| 4074798.0 | 5.8 |
| 4074855.0 | 5.9 |
| 4074944.0 | 5.9 |
| 4075392.0 | 7.2 |
| 4075452.0 | 6.4 |
| 4075572.0 | 6.6 |
| 4076173.0 | 7.3 |
| 4076409.0 | 7.1 |
| 4076589.0 | 6.6 |
| 4086852.0 | 6.4 |
| 4086912.0 | 7.1 |
| 4087272.0 | 5.7 |
| 4089792.0 | 5.8 |
| 4090309.0 | 6.8 |
| 4090340.0 | 6.8 |
| 4090864.0 | 6.1 |
| 4091085.0 | 7.3 |
| 4091106.0 | 7.4 |
| 4091232.0 | 6.1 |
| 4092045.0 | 6.8 |
| 4092404.0 | 6.2 |
| 4092469.0 | 6.2 |
| 4092672.0 | 5.6 |
| 4093182.0 | 6.8 |
| 4093216.0 | 7.3 |
| 4093295.0 | 7.4 |
| 4093992.0 | 6.2 |
| 4094352.0 | 5.7 |
| 4094472.0 | 6.5 |
| 4094557.0 | 7.3 |
| 4094571.0 | 6.5 |
| 4094699.0 | 5.8 |
| 4094713.0 | 6.7 |
| 4094744.0 | 7.1 |
| 4095040.0 | 7.3 |
| 4095240.0 | 7.4 |
| 4095309.0 | 6.6 |
| 4095995.0 | 6.5 |
| 4096225.0 | 6.1 |
| 4096269.0 | 5.8 |
| 4096537.0 | 7.1 |
| 4097172.0 | 7.2 |
| 4097484.0 | 7.2 |
| 4097582.0 | 7.0 |
| 4097604.0 | 7.0 |
| 4097765.0 | 6.2 |
| 4097823.0 | 6.0 |
| 4097895.0 | 7.5 |
| 4098025.0 | 6.9 |
| 4098190.0 | 6.0 |
| 4098372.0 | 6.7 |
| 4098432.0 | 7.4 |
| 4098492.0 | 5.6 |
| 4098552.0 | 5.5 |
| 4098792.0 | 6.8 |
| 4098886.0 | 6.5 |
| 4099246.0 | 5.7 |
| 4099489.0 | 5.9 |
| 4099696.0 | 5.7 |
| 4099752.0 | 6.2 |
| 4099812.0 | 5.9 |
| 4099932.0 | 6.8 |
| 4100052.0 | 7.0 |
| 4100172.0 | 6.8 |
| 4100347.0 | 6.0 |
| 4100413.0 | 5.6 |
| 4100450.0 | 7.3 |
| 4100497.0 | 6.9 |
| 4100729.0 | 6.1 |
| 4100766.0 | 6.6 |
| 4100834.0 | 6.7 |
| 4100849.0 | 6.5 |
| 4101014.0 | 6.7 |
| 4101372.0 | 7.4 |
| 4101672.0 | 5.6 |
| 4101733.0 | 6.2 |
| 4101848.0 | 6.2 |
| 4102066.0 | 6.2 |
| 4102176.0 | 5.6 |
| 4102380.0 | 7.0 |
| 4102433.0 | 5.9 |
| 4102496.0 | 6.0 |
| 4102992.0 | 6.0 |
| 4103581.0 | 6.0 |
| 4103609.0 | 5.9 |
| 4103733.0 | 6.3 |
| 4103967.0 | 6.3 |
| 4104037.0 | 7.4 |
| 4104072.0 | 5.5 |
| 4104372.0 | 7.2 |
| 4104492.0 | 6.7 |
| 4104699.0 | 6.8 |
| 4106400.0 | 6.7 |
| 4106512.0 | 7.3 |
| 4107132.0 | 6.5 |
| 4108872.0 | 7.3 |
| 4109122.0 | 6.7 |
| 4109150.0 | 7.1 |
| 4109337.0 | 5.5 |
| 4109892.0 | 6.4 |
| 4110489.0 | 6.5 |
| 4111198.0 | 7.3 |
| 4111632.0 | 6.6 |
| 4111908.0 | 6.6 |
| 4111950.0 | 6.7 |
| 4111980.0 | 5.9 |
| 4112064.0 | 7.2 |
| 4112092.0 | 6.1 |
| 4112183.0 | 6.8 |
| 4112227.0 | 6.1 |
| 4112374.0 | 6.2 |
| 4112430.0 | 6.9 |
| 4112619.0 | 6.9 |
| 4112642.0 | 7.1 |
| 4112679.0 | 5.9 |
| 4112772.0 | 6.9 |
| 4112892.0 | 7.3 |
| 4112952.0 | 7.0 |
| 4113346.0 | 6.2 |
| 4113379.0 | 6.6 |
| 4113502.0 | 7.3 |
| 4114082.0 | 6.9 |
| 4114138.0 | 5.9 |
| 4114152.0 | 5.6 |
| 4114272.0 | 7.5 |
| 4114392.0 | 6.5 |
| 4114452.0 | 6.7 |
| 4114512.0 | 6.9 |
| 4114935.0 | 6.6 |
| 4115106.0 | 6.1 |
| 4115425.0 | 6.2 |
| 4115499.0 | 6.6 |
| 4115529.0 | 6.2 |
| 4115952.0 | 6.8 |
| 4116258.0 | 5.5 |
| 4116386.0 | 6.2 |
| 4116655.0 | 6.6 |
| 4116932.0 | 6.2 |
| 4116940.0 | 7.2 |
| 4117611.0 | 7.3 |
| 4117692.0 | 5.8 |
| 4117715.0 | 6.4 |
| 4117969.0 | 6.6 |
| 4118269.0 | 5.6 |
| 4128549.0 | 7.5 |
| 4128552.0 | 7.2 |
| 4128972.0 | 7.3 |
| 4129516.0 | 7.0 |
| 4130847.0 | 7.4 |
| 4131193.0 | 7.4 |
| 4131367.0 | 6.9 |
| 4132187.0 | 6.5 |
| 4132693.0 | 6.8 |
| 4132932.0 | 6.9 |
| 4133594.0 | 6.5 |
| 4133633.0 | 6.0 |
| 4133779.0 | 6.5 |
| 4133962.0 | 7.2 |
| 4134027.0 | 6.8 |
| 4134060.0 | 5.7 |
| 4134372.0 | 6.7 |
| 4134552.0 | 6.4 |
| 4135470.0 | 7.2 |
| 4135589.0 | 7.4 |
| 4136052.0 | 5.5 |
| 4136172.0 | 7.4 |
| 4136397.0 | 6.3 |
| 4136713.0 | 6.7 |
| 4137612.0 | 7.0 |
| 4137672.0 | 6.8 |
| 4137755.0 | 5.8 |
| 4137868.0 | 5.9 |
| 4137885.0 | 6.3 |
| 4138255.0 | 6.3 |
| 4138275.0 | 7.4 |
| 4138430.0 | 6.8 |
| 4138506.0 | 5.9 |
| 4138692.0 | 7.3 |
| 4138872.0 | 7.2 |
| 4138932.0 | 6.0 |
| 4138992.0 | 7.4 |
| 4139112.0 | 5.8 |
| 4139251.0 | 6.9 |
| 4139366.0 | 5.6 |
| 4139397.0 | 5.8 |
| 4139462.0 | 6.7 |
| 4139674.0 | 6.6 |
| 4139783.0 | 5.7 |
| 4140192.0 | 6.8 |
| 4140252.0 | 6.5 |
| 4140492.0 | 7.1 |
| 4140742.0 | 6.5 |
| 4140774.0 | 6.8 |
| 4140825.0 | 6.0 |
| 4140832.0 | 6.5 |
| 4141015.0 | 5.6 |
| 4141085.0 | 5.8 |
| 4141199.0 | 6.8 |
| 4141326.0 | 6.4 |
| 4141476.0 | 5.7 |
| 4141752.0 | 5.8 |
| 4141872.0 | 6.6 |
| 4141992.0 | 6.3 |
| 4142191.0 | 7.1 |
| 4142287.0 | 7.3 |
| 4142362.0 | 6.2 |
| 4142654.0 | 6.3 |
| 4142782.0 | 7.4 |
| 4142799.0 | 5.8 |
| 4142884.0 | 5.5 |
| 4142952.0 | 6.1 |
| 4143132.0 | 7.3 |
| 4143432.0 | 5.9 |
| 4143604.0 | 6.0 |
| 4143653.0 | 5.7 |
| 4143689.0 | 6.1 |
| 4143907.0 | 7.0 |
| 4144019.0 | 5.9 |
| 4144247.0 | 6.7 |
| 4144752.0 | 6.6 |
| 4147272.0 | 7.2 |
| 4147512.0 | 6.0 |
| 4148499.0 | 7.4 |
| 4148657.0 | 5.9 |
| 4149667.0 | 7.3 |
| 4150056.0 | 5.8 |
| 4150272.0 | 6.9 |
| 4150392.0 | 6.1 |
| 4150696.0 | 6.1 |
| 4150986.0 | 7.0 |
| 4151264.0 | 7.0 |
| 4151297.0 | 7.3 |
| 4151577.0 | 6.7 |
| 4151592.0 | 6.1 |
| 4151892.0 | 6.1 |
| 4152012.0 | 6.5 |
| 4152072.0 | 6.7 |
| 4152159.0 | 6.9 |
| 4152177.0 | 6.1 |
| 4152476.0 | 7.1 |
| 4152552.0 | 6.7 |
| 4152677.0 | 6.3 |
| 4152806.0 | 6.6 |
| 4152920.0 | 6.1 |
| 4152960.0 | 6.8 |
| 4153032.0 | 6.7 |
| 4153272.0 | 6.6 |
| 4153332.0 | 6.6 |
| 4153392.0 | 6.8 |
| 4153452.0 | 5.6 |
| 4153657.0 | 7.4 |
| 4154104.0 | 6.6 |
| 4154187.0 | 6.1 |
| 4154532.0 | 7.3 |
| 4154592.0 | 7.0 |
| 4155377.0 | 5.9 |
| 4155434.0 | 7.1 |
| 4155519.0 | 7.4 |
| 4155912.0 | 6.6 |
| 4156332.0 | 7.1 |
| 4156741.0 | 6.3 |
| 4157652.0 | 6.7 |
| 4158197.0 | 6.7 |
| 4158705.0 | 7.1 |
| 4158792.0 | 6.5 |
| 4158912.0 | 6.0 |
| 4159337.0 | 6.5 |
| 4159366.0 | 5.9 |
| 4159739.0 | 7.3 |
| 4160232.0 | 5.5 |
| 4160292.0 | 6.7 |
| 4160352.0 | 6.4 |
| 4160412.0 | 5.8 |
| 4160472.0 | 7.4 |
| 4160532.0 | 7.0 |
| 4160592.0 | 6.5 |
| 4172052.0 | 7.5 |
| 4172172.0 | 6.3 |
| 4172232.0 | 6.6 |
| 4172859.0 | 6.9 |
| 4172933.0 | 5.8 |
| 4173372.0 | 6.1 |
| 4173492.0 | 6.6 |
| 4174111.0 | 5.5 |
| 4174312.0 | 5.5 |
| 4174752.0 | 7.2 |
| 4175730.0 | 6.9 |
| 4175973.0 | 5.6 |
| 4176252.0 | 5.5 |
| 4176492.0 | 6.6 |
| 4176859.0 | 6.7 |
| 4177418.0 | 5.6 |
| 4177632.0 | 5.7 |
| 4177752.0 | 6.8 |
| 4178743.0 | 7.0 |
| 4179252.0 | 5.9 |
| 4180033.0 | 7.2 |
| 4180138.0 | 6.6 |
| 4180392.0 | 6.8 |
| 4180872.0 | 7.0 |
| 4181050.0 | 7.3 |
| 4181553.0 | 5.6 |
| 4181635.0 | 5.8 |
| 4182456.0 | 6.0 |
| 4182543.0 | 7.3 |
| 4182828.0 | 7.4 |
| 4183139.0 | 6.1 |
| 4183392.0 | 6.7 |
| 4183512.0 | 5.8 |
| 4184031.0 | 7.1 |
| 4184048.0 | 7.2 |
| 4184078.0 | 7.4 |
| 4184399.0 | 6.6 |
| 4184516.0 | 7.2 |
| 4184689.0 | 6.1 |
| 4184832.0 | 7.2 |
| 4184952.0 | 5.9 |
| 4185497.0 | 6.2 |
| 4186452.0 | 5.9 |
| 4186703.0 | 6.0 |
| 4186824.0 | 6.3 |
| 4187029.0 | 7.5 |
| 4187224.0 | 6.4 |
| 4187279.0 | 6.4 |
| 4187422.0 | 6.0 |
| 4187447.0 | 7.3 |
| 4187529.0 | 6.1 |
| 4187772.0 | 7.1 |
| 4187832.0 | 5.6 |
| 4187952.0 | 7.3 |
| 4188467.0 | 7.3 |
| 4188868.0 | 5.7 |
| 4189588.0 | 6.4 |
| 4189875.0 | 6.2 |
| 4189967.0 | 5.8 |
| 4189984.0 | 6.0 |
| 4190017.0 | 6.5 |
| 4190040.0 | 6.9 |
| 4190062.0 | 6.8 |
| 4190130.0 | 7.1 |
| 4190532.0 | 5.7 |
| 4191099.0 | 6.5 |
| 4191135.0 | 7.4 |
| 4191446.0 | 6.6 |
| 4191514.0 | 6.2 |
| 4191753.0 | 5.8 |
| 4191892.0 | 5.8 |
| 4191912.0 | 5.5 |
| 4192032.0 | 6.2 |
| 4192272.0 | 7.2 |
| 4192332.0 | 6.8 |
| 4192392.0 | 5.6 |
| 4192493.0 | 7.2 |
| 4192527.0 | 5.6 |
| 4192619.0 | 6.0 |
| 4192805.0 | 6.2 |
| 4192925.0 | 6.5 |
| 4192980.0 | 5.7 |
| 4193092.0 | 6.2 |
| 4193205.0 | 6.5 |
| 4193239.0 | 6.0 |
| 4193652.0 | 7.3 |
| 4193914.0 | 5.7 |
| 4194040.0 | 6.0 |
| 4194485.0 | 6.5 |
| 4194500.0 | 7.3 |
| 4194539.0 | 7.1 |
| 4194567.0 | 7.1 |
| 4194622.0 | 6.3 |
| 4194662.0 | 6.3 |
| 4194715.0 | 7.2 |
| 4194742.0 | 5.8 |
| 4194912.0 | 7.0 |
| 4194972.0 | 6.0 |
| 4195032.0 | 7.4 |
| 4195212.0 | 6.1 |
| 4195373.0 | 7.1 |
| 4195406.0 | 6.5 |
| 4195431.0 | 6.1 |
| 4195450.0 | 6.2 |
| 4195650.0 | 5.9 |
| 4195858.0 | 6.4 |
| 4195883.0 | 6.5 |
| 4195968.0 | 6.2 |
| 4196016.0 | 6.8 |
| 4196097.0 | 6.9 |
| 4196352.0 | 6.1 |
| 4196652.0 | 7.3 |
| 4196712.0 | 6.3 |
| 4197200.0 | 6.5 |
| 4197557.0 | 6.3 |
| 4197596.0 | 6.9 |
| 4198032.0 | 7.0 |
| 4198535.0 | 6.1 |
| 4199112.0 | 7.4 |
| 4199172.0 | 5.7 |
| 4199232.0 | 6.6 |
| 4199532.0 | 7.0 |
| 4199807.0 | 6.6 |
| 4199829.0 | 6.4 |
| 4199899.0 | 7.3 |
| 4200535.0 | 6.0 |
| 4200792.0 | 5.6 |
| 4201207.0 | 5.7 |
| 4201598.0 | 7.1 |
| 4215192.0 | 5.9 |
| 4215788.0 | 5.8 |
| 4218012.0 | 7.2 |
| 4219392.0 | 7.0 |
| 4219572.0 | 6.2 |
| 4220712.0 | 5.6 |
| 4222452.0 | 7.3 |
| 4222572.0 | 7.3 |
| 4222632.0 | 6.7 |
| 4222744.0 | 6.6 |
| 4223772.0 | 7.4 |
| 4223952.0 | 7.0 |
| 4224337.0 | 6.9 |
| 4224439.0 | 7.1 |
| 4224973.0 | 5.6 |
| 4225512.0 | 6.7 |
| 4226048.0 | 5.6 |
| 4226207.0 | 7.1 |
| 4226305.0 | 7.0 |
| 4226333.0 | 6.2 |
| 4226357.0 | 6.5 |
| 4227040.0 | 6.8 |
| 4228092.0 | 5.5 |
| 4228212.0 | 5.7 |
| 4228332.0 | 5.6 |
| 4228864.0 | 6.1 |
| 4228913.0 | 6.6 |
| 4229192.0 | 6.1 |
| 4229232.0 | 7.5 |
| 4229315.0 | 6.0 |
| 4229352.0 | 6.6 |
| 4229592.0 | 7.2 |
| 4229937.0 | 5.8 |
| 4230134.0 | 6.3 |
| 4230192.0 | 6.2 |
| 4230464.0 | 6.4 |
| 4230525.0 | 5.6 |
| 4230580.0 | 6.7 |
| 4230617.0 | 6.8 |
| 4230679.0 | 6.1 |
| 4230780.0 | 7.3 |
| 4230972.0 | 6.0 |
| 4231032.0 | 5.6 |
| 4231092.0 | 6.5 |
| 4232064.0 | 7.1 |
| 4232232.0 | 6.8 |
| 4232889.0 | 6.7 |
| 4232920.0 | 7.2 |
| 4233191.0 | 7.2 |
| 4233237.0 | 5.7 |
| 4233259.0 | 6.5 |
| 4233325.0 | 5.8 |
| 4233452.0 | 6.1 |
| 4233487.0 | 6.3 |
| 4233732.0 | 7.2 |
| 4233912.0 | 6.1 |
| 4234032.0 | 5.7 |
| 4234542.0 | 6.5 |
| 4234622.0 | 6.9 |
| 4234682.0 | 7.3 |
| 4234980.0 | 7.1 |
| 4235019.0 | 7.2 |
| 4235352.0 | 6.9 |
| 4235412.0 | 6.3 |
| 4235532.0 | 6.7 |
| 4235944.0 | 5.8 |
| 4235961.0 | 6.8 |
| 4236480.0 | 5.7 |
| 4236520.0 | 7.2 |
| 4236552.0 | 7.4 |
| 4236792.0 | 7.4 |
| 4237484.0 | 6.7 |
| 4237704.0 | 7.2 |
| 4237846.0 | 6.5 |
| 4237992.0 | 7.3 |
| 4239120.0 | 5.7 |
| 4240169.0 | 6.1 |
| 4240216.0 | 6.7 |
| 4242312.0 | 5.9 |
| 4258392.0 | 6.4 |
| 4259452.0 | 6.0 |
| 4260169.0 | 7.1 |
| 4262832.0 | 5.9 |
| 4262952.0 | 7.5 |
| 4263255.0 | 7.4 |
| 4266094.0 | 7.4 |
| 4266270.0 | 6.7 |
| 4266280.0 | 7.3 |
| 4266296.0 | 7.3 |
| 4266582.0 | 6.8 |
| 4267092.0 | 7.4 |
| 4267152.0 | 6.4 |
| 4267212.0 | 5.8 |
| 4267272.0 | 6.0 |
| 4267369.0 | 6.3 |
| 4267415.0 | 5.8 |
| 4267439.0 | 7.3 |
| 4267602.0 | 5.8 |
| 4267677.0 | 6.1 |
| 4267719.0 | 6.5 |
| 4267735.0 | 6.7 |
| 4267764.0 | 6.6 |
| 4268073.0 | 7.1 |
| 4268191.0 | 6.2 |
| 4268208.0 | 5.6 |
| 4268232.0 | 7.3 |
| 4268292.0 | 6.0 |
| 4268532.0 | 7.4 |
| 4268712.0 | 5.7 |
| 4268782.0 | 5.9 |
| 4269208.0 | 7.3 |
| 4269341.0 | 6.1 |
| 4269394.0 | 6.1 |
| 4269418.0 | 5.9 |
| 4269481.0 | 6.1 |
| 4269547.0 | 5.6 |
| 4269573.0 | 6.1 |
| 4269589.0 | 5.5 |
| 4269619.0 | 6.1 |
| 4269732.0 | 6.4 |
| 4269912.0 | 6.5 |
| 4269972.0 | 6.9 |
| 4270032.0 | 5.9 |
| 4270092.0 | 6.3 |
| 4270152.0 | 6.3 |
| 4270229.0 | 6.2 |
| 4270258.0 | 6.2 |
| 4270415.0 | 5.7 |
| 4270529.0 | 7.3 |
| 4270908.0 | 7.4 |
| 4271110.0 | 7.4 |
| 4271292.0 | 6.2 |
| 4271472.0 | 6.2 |
| 4271908.0 | 6.4 |
| 4272015.0 | 7.3 |
| 4272098.0 | 5.9 |
| 4272168.0 | 5.8 |
| 4272235.0 | 6.4 |
| 4272488.0 | 6.6 |
| 4272672.0 | 7.0 |
| 4272792.0 | 5.9 |
| 4272852.0 | 7.4 |
| 4272912.0 | 6.5 |
| 4273032.0 | 5.7 |
| 4273189.0 | 7.4 |
| 4273271.0 | 6.2 |
| 4273684.0 | 6.6 |
| 4273879.0 | 6.0 |
| 4274352.0 | 7.3 |
| 4274412.0 | 6.3 |
| 4274915.0 | 6.9 |
| 4275045.0 | 7.1 |
| 4275158.0 | 6.5 |
| 4276402.0 | 6.6 |
| 4276525.0 | 7.2 |
| 4276868.0 | 5.7 |
| 4277232.0 | 7.1 |
| 4278019.0 | 6.1 |
| 4278138.0 | 7.1 |
| 4278164.0 | 6.9 |
| 4278312.0 | 6.1 |
| 4278372.0 | 6.0 |
| 4278552.0 | 6.5 |
| 4278612.0 | 6.6 |
| 4279992.0 | 6.0 |
| 4286290.0 | 6.5 |
| 4287839.0 | 6.4 |
| 4287962.0 | 5.5 |

Table A-3:  0-reduced time series of 7950 moonquakes from Table A-1 that occurred during the Moon's traversals of the interplanetary magnetic field, i.e., the events listed in Table A-1 but not in Table A-2.

Supplement B – earthquakes Tables

| T(s) | Mag |
|---|---|
| 0.0 | 5.5 |
| 57741.0 | 6.1 |
| 334022.0 | 6.0 |
| 614223.0 | 5.8 |
| 717749.0 | 6.0 |
| 910557.0 | 7.1 |
| 1097040.0 | 6.1 |
| 1105692.0 | 6.0 |
| 1170437.0 | 5.7 |
| 1317771.0 | 5.7 |
| 1383206.0 | 7.5 |
| 1597835.0 | 5.8 |
| 1923610.0 | 5.9 |
| 1984768.0 | 5.8 |
| 2141289.0 | 6.5 |
| 2414137.0 | 6.8 |
| 2426257.0 | 5.7 |
| 2507932.0 | 5.7 |
| 2533855.0 | 6.5 |
| 2617659.0 | 6.4 |
| 2739512.0 | 6.8 |
| 2817613.0 | 5.8 |
| 2980531.0 | 6.7 |
| 3061455.0 | 5.7 |
| 3167007.0 | 5.8 |
| 3225189.0 | 5.9 |
| 3276843.0 | 6.5 |
| 3404098.0 | 7.0 |
| 3629408.0 | 6.1 |
| 3748799.0 | 5.7 |
| 3903930.0 | 6.0 |
| 3938088.0 | 7.6 |
| 4049352.0 | 6.7 |
| 4190656.0 | 6.2 |
| 4211701.0 | 5.9 |
| 4355804.0 | 5.9 |
| 4522340.0 | 5.8 |
| 4800528.0 | 7.1 |
| 5007241.0 | 7.2 |
| 5189145.0 | 6.8 |
| 5198515.0 | 6.0 |
| 5281749.0 | 5.7 |
| 5914428.0 | 6.5 |
| 6023688.0 | 6.0 |
| 6085738.0 | 5.8 |
| 6169890.0 | 6.0 |
| 6422344.0 | 5.8 |
| 6518877.0 | 6.1 |
| 6531292.0 | 5.8 |
| 6598919.0 | 5.7 |
| 6603521.0 | 6.3 |
| 6886538.0 | 5.8 |
| 7092054.0 | 5.8 |
| 7146274.0 | 6.3 |
| 7193169.0 | 5.8 |
| 7241173.0 | 5.8 |
| 7394954.0 | 6.7 |
| 7493107.0 | 5.9 |
| 7519089.0 | 6.0 |
| 7563824.0 | 5.7 |
| 7748155.0 | 5.6 |
| 7960557.0 | 5.7 |
| 8062920.0 | 6.5 |
| 8064717.0 | 6.2 |
| 8124545.0 | 5.9 |
| 8161734.0 | 5.7 |
| 8197195.0 | 5.8 |
| 8274561.0 | 6.1 |
| 8274568.0 | 6.6 |
| 8302096.0 | 5.7 |
| 8674100.0 | 5.9 |
| 8842628.0 | 5.9 |
| 8932252.0 | 6.5 |
| 9164064.0 | 7.1 |
| 9212644.0 | 5.7 |
| 9228356.0 | 6.3 |
| 9310457.0 | 6.0 |
| 9440113.0 | 5.7 |
| 9656944.0 | 7.2 |
| 9794575.0 | 6.0 |
| 9885879.0 | 6.2 |
| 9955396.0 | 5.7 |
| 10234881.0 | 6.4 |
| 10343249.0 | 5.8 |
| 10480989.0 | 6.4 |
| 10597019.0 | 6.3 |
| 10803978.0 | 6.2 |
| 10941457.0 | 5.8 |
| 11006000.0 | 5.7 |
| 11081187.0 | 6.0 |
| 11097118.0 | 5.9 |
| 11199149.0 | 6.1 |
| 11262597.0 | 6.0 |
| 11516117.0 | 5.8 |
| 11655658.0 | 5.9 |
| 11783560.0 | 5.8 |
| 12141218.0 | 6.1 |
| 12455622.0 | 7.8 |
| 12574304.0 | 5.8 |
| 12948021.0 | 6.0 |
| 13338639.0 | 6.3 |
| 13530574.0 | 5.7 |
| 13883946.0 | 6.0 |
| 13919425.0 | 6.0 |
| 14046854.0 | 6.4 |
| 14104366.0 | 5.6 |
| 14495298.0 | 5.7 |
| 14634325.0 | 5.8 |
| 14717930.0 | 5.6 |
| 15010981.0 | 5.9 |
| 15071330.0 | 6.1 |
| 15108834.0 | 6.2 |
| 15204468.0 | 6.9 |
| 15377596.0 | 5.9 |
| 15458327.0 | 5.7 |
| 15458563.0 | 6.7 |
| 15462091.0 | 6.0 |
| 15486568.0 | 6.0 |
| 15532608.0 | 6.6 |
| 15787110.0 | 5.7 |
| 15804918.0 | 5.6 |
| 15816772.0 | 6.6 |
| 15964001.0 | 5.8 |
| 16088351.0 | 6.9 |
| 16104349.0 | 5.9 |
| 16135803.0 | 5.8 |
| 16168859.0 | 5.9 |
| 16169430.0 | 6.1 |
| 16170382.0 | 5.6 |
| 16178862.0 | 6.0 |
| 16203261.0 | 6.3 |
| 16208344.0 | 5.6 |
| 16262123.0 | 6.1 |
| 16270140.0 | 7.0 |
| 16271281.0 | 5.7 |
| 16271390.0 | 5.7 |
| 16383751.0 | 7.8 |
| 16409873.0 | 5.7 |
| 16410894.0 | 5.9 |
| 16517405.0 | 5.9 |
| 16576175.0 | 6.2 |
| 16673861.0 | 6.1 |
| 16673944.0 | 6.0 |
| 16687419.0 | 5.8 |
| 16826855.0 | 6.1 |
| 16827847.0 | 5.9 |
| 16907308.0 | 5.7 |
| 17069308.0 | 5.7 |
| 17100663.0 | 6.0 |
| 17177026.0 | 5.6 |
| 17294116.0 | 6.0 |
| 17404640.0 | 7.0 |
| 17426252.0 | 6.6 |
| 17537977.0 | 5.8 |
| 17695518.0 | 5.7 |
| 17766285.0 | 5.7 |
| 17905219.0 | 5.7 |
| 18225476.0 | 5.8 |
| 18555670.0 | 5.8 |
| 18782700.0 | 5.8 |
| 18891947.0 | 5.8 |
| 19090859.0 | 6.7 |
| 19122638.0 | 6.9 |
| 19300678.0 | 6.0 |
| 19853303.0 | 5.9 |
| 19854758.0 | 6.4 |
| 19946564.0 | 6.9 |
| 19961453.0 | 7.2 |
| 20204861.0 | 6.3 |
| 20221673.0 | 5.9 |
| 20260516.0 | 5.7 |
| 20354396.0 | 6.4 |
| 20366821.0 | 5.7 |
| 20638504.0 | 5.7 |
| 20676567.0 | 6.3 |
| 20713161.0 | 6.0 |
| 20829332.0 | 6.3 |
| 20859550.0 | 6.3 |
| 20907334.0 | 5.8 |
| 20978222.0 | 6.1 |
| 21061756.0 | 6.0 |
| 21064898.0 | 6.1 |
| 21116029.0 | 5.8 |
| 21312213.0 | 5.7 |
| 21444798.0 | 6.2 |
| 21731412.0 | 5.7 |
| 21803617.0 | 5.7 |
| 21862276.0 | 6.3 |
| 21927288.0 | 5.9 |
| 22059029.0 | 6.1 |
| 22061762.0 | 6.3 |
| 22248564.0 | 5.8 |
| 22472465.0 | 6.4 |
| 22686641.0 | 5.8 |
| 22818868.0 | 5.9 |
| 23156974.0 | 5.8 |
| 23299552.0 | 5.8 |
| 23690747.0 | 5.8 |
| 23690915.0 | 5.9 |
| 23735105.0 | 5.9 |
| 23735700.0 | 6.3 |
| 23943652.0 | 6.0 |
| 23944507.0 | 6.3 |
| 23944946.0 | 5.9 |
| 23947699.0 | 5.8 |
| 24438154.0 | 5.8 |
| 24560231.0 | 6.0 |
| 24768255.0 | 5.8 |
| 24837197.0 | 5.9 |
| 24902086.0 | 5.8 |
| 24969744.0 | 5.8 |
| 25000244.0 | 6.1 |
| 25008159.0 | 6.4 |
| 25044794.0 | 5.9 |
| 25052894.0 | 5.7 |
| 25115346.0 | 5.9 |
| 25359743.0 | 7.7 |
| 25497436.0 | 5.8 |
| 25570004.0 | 5.7 |
| 25852745.0 | 6.2 |
| 25860508.0 | 6.3 |
| 26497829.0 | 7.1 |
| 26505203.0 | 6.1 |
| 26705950.0 | 5.8 |
| 27027243.0 | 5.7 |
| 27076418.0 | 6.0 |
| 27124577.0 | 7.4 |
| 27160655.0 | 5.8 |
| 27216320.0 | 6.0 |
| 27241318.0 | 5.8 |
| 27283757.0 | 6.1 |
| 27513818.0 | 6.0 |
| 27535227.0 | 6.2 |
| 27567529.0 | 6.8 |
| 27579159.0 | 5.7 |
| 27677317.0 | 5.9 |
| 27977632.0 | 7.1 |
| 28145730.0 | 6.7 |
| 28213709.0 | 5.8 |
| 28280512.0 | 7.0 |
| 28282679.0 | 6.0 |
| 28436798.0 | 5.7 |
| 28489326.0 | 5.7 |
| 28621376.0 | 5.8 |
| 28648679.0 | 6.1 |
| 28903814.0 | 6.1 |
| 29034734.0 | 6.1 |
| 29043087.0 | 5.8 |
| 29195748.0 | 5.8 |
| 29350945.0 | 5.9 |
| 29370534.0 | 6.0 |
| 29451678.0 | 5.7 |
| 29607851.0 | 6.0 |
| 29612152.0 | 5.8 |
| 29715690.0 | 5.7 |
| 29921110.0 | 6.1 |
| 30122308.0 | 6.3 |
| 30203823.0 | 6.3 |
| 30247703.0 | 5.9 |
| 30283868.0 | 6.4 |
| 30285155.0 | 6.9 |
| 30399832.0 | 5.6 |
| 31122281.0 | 5.7 |
| 31233454.0 | 5.7 |
| 31301754.0 | 5.7 |
| 31488313.0 | 5.7 |
| 31557021.0 | 5.8 |
| 31775118.0 | 5.7 |
| 32051252.0 | 6.3 |
| 32217532.0 | 6.8 |
| 32221122.0 | 5.9 |
| 32341235.0 | 5.9 |
| 32369395.0 | 6.6 |
| 32454800.0 | 5.6 |
| 32477390.0 | 5.6 |
| 32559077.0 | 6.2 |
| 32786960.0 | 5.8 |
| 32944437.0 | 5.8 |
| 32991824.0 | 6.0 |
| 33042122.0 | 6.1 |
| 33088905.0 | 5.8 |
| 33101832.0 | 5.9 |
| 33133009.0 | 6.0 |
| 33165051.0 | 5.9 |
| 33217603.0 | 5.8 |
| 33342253.0 | 6.5 |
| 33559643.0 | 6.0 |
| 33809078.0 | 6.3 |
| 34086919.0 | 5.7 |
| 34107199.0 | 5.8 |
| 34113580.0 | 6.0 |
| 34235693.0 | 5.9 |
| 34313283.0 | 5.8 |
| 34433215.0 | 6.1 |
| 34499985.0 | 5.9 |
| 34567612.0 | 7.8 |
| 34569362.0 | 6.4 |
| 34570599.0 | 6.1 |
| 34576521.0 | 6.2 |
| 34578295.0 | 5.8 |
| 34597972.0 | 5.7 |
| 34616296.0 | 6.4 |
| 34788865.0 | 5.8 |
| 34841646.0 | 5.7 |
| 35208098.0 | 6.4 |
| 35294623.0 | 6.9 |
| 35306616.0 | 5.9 |
| 35427832.0 | 5.8 |
| 35468851.0 | 5.8 |
| 35545662.0 | 6.9 |
| 35616504.0 | 6.6 |
| 36164660.0 | 6.2 |
| 36289649.0 | 5.9 |
| 36361681.0 | 5.7 |
| 36433018.0 | 6.3 |
| 36593176.0 | 5.9 |
| 36594447.0 | 6.5 |
| 36706738.0 | 6.0 |
| 36741221.0 | 6.6 |
| 36751361.0 | 7.8 |
| 36766346.0 | 5.9 |
| 36766801.0 | 6.5 |
| 36843240.0 | 6.9 |
| 36845238.0 | 5.9 |
| 36852136.0 | 5.8 |
| 36870574.0 | 5.8 |
| 36919710.0 | 6.0 |
| 36999426.0 | 5.8 |
| 37213516.0 | 6.0 |
| 37504504.0 | 7.9 |
| 37506693.0 | 6.4 |
| 37513588.0 | 6.7 |
| 37572619.0 | 5.9 |
| 37575817.0 | 5.9 |
| 37587058.0 | 6.2 |
| 37600445.0 | 6.4 |
| 37740322.0 | 6.4 |
| 37769829.0 | 6.0 |
| 37812070.0 | 6.7 |
| 37871272.0 | 5.9 |
| 38075770.0 | 5.9 |
| 38084569.0 | 5.8 |
| 38208380.0 | 7.6 |
| 38461366.0 | 5.9 |
| 38583253.0 | 6.2 |
| 38848621.0 | 5.8 |
| 38884750.0 | 5.9 |
| 38895463.0 | 5.9 |
| 38895477.0 | 6.3 |
| 39012985.0 | 6.9 |
| 39015846.0 | 6.0 |
| 39092867.0 | 5.7 |
| 39216869.0 | 5.7 |
| 39398162.0 | 5.9 |
| 39451866.0 | 6.0 |
| 39561461.0 | 7.3 |
| 39594669.0 | 6.3 |
| 39704532.0 | 5.8 |
| 39858199.0 | 5.7 |
| 39906935.0 | 6.0 |
| 39950978.0 | 5.7 |
| 40062032.0 | 5.8 |
| 40132292.0 | 5.9 |
| 40176716.0 | 5.9 |
| 40267086.0 | 5.7 |
| 40399690.0 | 6.5 |
| 40592058.0 | 7.9 |
| 41234625.0 | 5.9 |
| 41351958.0 | 5.9 |
| 41684297.0 | 5.7 |
| 42037669.0 | 6.2 |
| 42268057.0 | 6.5 |
| 42952449.0 | 6.4 |
| 43218778.0 | 6.5 |
| 43428971.0 | 5.9 |
| 43489958.0 | 6.9 |
| 43512639.0 | 5.9 |
| 43800776.0 | 5.7 |
| 44128938.0 | 6.0 |
| 44205132.0 | 5.7 |
| 44249171.0 | 5.7 |
| 44286707.0 | 6.3 |
| 44435533.0 | 5.7 |
| 44590700.0 | 5.9 |
| 44992510.0 | 6.0 |
| 45036454.0 | 5.7 |
| 45470840.0 | 6.0 |
| 46144453.0 | 6.2 |
| 46293198.0 | 6.6 |
| 46333160.0 | 5.7 |
| 46773852.0 | 6.5 |
| 46905186.0 | 6.1 |
| 47168000.0 | 5.9 |
| 47317668.0 | 5.7 |
| 47353365.0 | 5.8 |
| 47454370.0 | 5.7 |
| 47478295.0 | 5.8 |
| 47635788.0 | 5.8 |
| 47777017.0 | 6.3 |
| 47928324.0 | 5.7 |
| 48068142.0 | 6.0 |
| 48070429.0 | 5.9 |
| 48447600.0 | 6.0 |
| 48602544.0 | 6.9 |
| 48837027.0 | 5.7 |
| 48926040.0 | 5.9 |
| 48943633.0 | 6.8 |
| 49001787.0 | 5.7 |
| 49174548.0 | 6.2 |
| 49180931.0 | 6.3 |
| 49270456.0 | 5.9 |
| 49319467.0 | 6.0 |
| 49493610.0 | 5.8 |
| 49562402.0 | 5.9 |
| 49790120.0 | 5.7 |
| 49791084.0 | 5.9 |
| 49795481.0 | 6.2 |
| 49827489.0 | 6.0 |
| 49870566.0 | 6.8 |
| 49935831.0 | 5.8 |
| 49991250.0 | 6.5 |
| 49997617.0 | 5.7 |
| 50049907.0 | 5.7 |
| 50118319.0 | 6.2 |
| 50323697.0 | 5.7 |
| 50387192.0 | 6.2 |
| 50775008.0 | 5.9 |
| 50835761.0 | 5.7 |
| 51149651.0 | 5.7 |
| 51176395.0 | 5.8 |
| 51601157.0 | 6.6 |
| 51602508.0 | 5.8 |
| 51676425.0 | 5.7 |
| 51974921.0 | 6.8 |
| 51979932.0 | 5.7 |
| 52050088.0 | 6.0 |
| 52605968.0 | 5.7 |
| 52755545.0 | 5.7 |
| 52803153.0 | 6.3 |
| 52932574.0 | 5.8 |
| 52957979.0 | 6.9 |
| 53019011.0 | 5.9 |
| 53059570.0 | 5.8 |
| 53174482.0 | 5.7 |
| 53270996.0 | 6.1 |
| 53396377.0 | 5.8 |
| 53667298.0 | 6.8 |
| 53853186.0 | 5.7 |
| 53945184.0 | 6.0 |
| 54206687.0 | 6.0 |
| 54252424.0 | 6.0 |
| 54319132.0 | 5.9 |
| 54394418.0 | 6.0 |
| 54522165.0 | 5.8 |
| 54716993.0 | 5.9 |
| 54855251.0 | 5.8 |
| 54860871.0 | 6.5 |
| 54931605.0 | 5.7 |
| 55031493.0 | 5.7 |
| 55183544.0 | 5.8 |
| 55289042.0 | 6.6 |
| 55421521.0 | 5.9 |
| 55449603.0 | 6.4 |
| 55653375.0 | 5.9 |
| 55787070.0 | 5.7 |
| 55822143.0 | 6.3 |
| 55867089.0 | 7.7 |
| 55876153.0 | 6.4 |
| 55999219.0 | 6.0 |
| 56042119.0 | 5.8 |
| 56122505.0 | 6.6 |
| 56597811.0 | 5.9 |
| 56689952.0 | 5.8 |
| 56710639.0 | 5.9 |
| 56981083.0 | 5.7 |
| 57425675.0 | 5.8 |
| 57511144.0 | 5.7 |
| 57619837.0 | 5.8 |
| 57640565.0 | 5.7 |
| 57731024.0 | 6.5 |
| 57767507.0 | 6.3 |
| 57961941.0 | 6.2 |
| 58075087.0 | 5.8 |
| 58126327.0 | 6.4 |
| 58175718.0 | 5.7 |
| 58309542.0 | 5.7 |
| 58420519.0 | 5.7 |
| 58557796.0 | 6.6 |
| 58640686.0 | 6.4 |
| 59340104.0 | 6.3 |
| 59731850.0 | 6.3 |
| 60045093.0 | 6.1 |
| 60337844.0 | 6.1 |
| 60378795.0 | 8.1 |
| 60545479.0 | 5.8 |
| 60612255.0 | 5.8 |
| 60683943.0 | 5.7 |
| 60943523.0 | 5.8 |
| 61191002.0 | 5.7 |
| 61275053.0 | 5.8 |
| 61377512.0 | 7.1 |
| 61404444.0 | 6.1 |
| 61458070.0 | 6.1 |
| 61470824.0 | 6.4 |
| 61478250.0 | 5.9 |
| 61484598.0 | 5.7 |
| 61642449.0 | 5.7 |
| 61695721.0 | 5.8 |
| 61703816.0 | 6.1 |
| 61732263.0 | 5.7 |
| 61780206.0 | 5.7 |
| 61903998.0 | 5.9 |
| 61932234.0 | 6.4 |
| 61989966.0 | 5.8 |
| 62595779.0 | 5.7 |
| 62809407.0 | 6.2 |
| 62844320.0 | 5.7 |
| 63004114.0 | 6.1 |
| 63028373.0 | 6.2 |
| 63034707.0 | 6.5 |
| 63149774.0 | 6.3 |
| 63194248.0 | 6.7 |
| 63860693.0 | 6.1 |
| 63936355.0 | 5.9 |
| 64280187.0 | 5.9 |
| 64374701.0 | 6.7 |
| 64740000.0 | 5.8 |
| 64750496.0 | 5.9 |
| 64943163.0 | 6.8 |
| 64949080.0 | 5.9 |
| 64983285.0 | 6.1 |
| 65035671.0 | 6.6 |
| 65318654.0 | 6.8 |
| 65337171.0 | 5.7 |
| 65577934.0 | 5.7 |
| 65622633.0 | 6.5 |
| 65659145.0 | 5.7 |
| 65745965.0 | 6.0 |
| 65847927.0 | 5.8 |
| 65893209.0 | 6.1 |
| 66043331.0 | 7.3 |
| 66058081.0 | 5.9 |
| 66072737.0 | 6.6 |
| 66238107.0 | 5.8 |
| 66348392.0 | 5.9 |
| 66358050.0 | 5.8 |
| 66490694.0 | 6.4 |
| 66553856.0 | 5.8 |
| 66616181.0 | 6.3 |
| 66636777.0 | 6.6 |
| 66640321.0 | 5.9 |
| 66664044.0 | 7.0 |
| 66669198.0 | 6.0 |
| 66736501.0 | 6.0 |
| 67163636.0 | 5.8 |
| 67299306.0 | 6.0 |
| 67469580.0 | 5.7 |
| 67556204.0 | 6.5 |
| 67628200.0 | 6.1 |
| 67629234.0 | 6.0 |
| 67832580.0 | 6.1 |
| 67878719.0 | 5.7 |
| 68225208.0 | 6.4 |
| 68231632.0 | 5.9 |
| 68259302.0 | 6.3 |
| 68309166.0 | 5.7 |
| 68365099.0 | 6.1 |
| 68522034.0 | 5.7 |
| 68600831.0 | 6.0 |
| 68647525.0 | 6.0 |
| 68720857.0 | 6.5 |
| 68889113.0 | 6.5 |
| 69357852.0 | 5.8 |
| 70014255.0 | 5.7 |
| 70115087.0 | 5.7 |
| 70124391.0 | 5.7 |
| 70437673.0 | 5.9 |
| 70988988.0 | 5.8 |
| 71085326.0 | 7.5 |
| 71227818.0 | 6.0 |
| 71375310.0 | 5.8 |
| 71454159.0 | 7.1 |
| 71647667.0 | 5.9 |
| 71665272.0 | 5.9 |
| 71762452.0 | 5.8 |
| 71809967.0 | 5.8 |
| 71911300.0 | 6.3 |
| 72029435.0 | 6.3 |
| 72221929.0 | 6.0 |
| 72232537.0 | 7.9 |
| 72323715.0 | 6.3 |
| 72375592.0 | 5.8 |
| 72378868.0 | 6.2 |
| 72409508.0 | 5.8 |
| 72431018.0 | 5.8 |
| 72539512.0 | 6.3 |
| 72688019.0 | 6.5 |
| 72915054.0 | 6.2 |
| 73015785.0 | 6.0 |
| 73104107.0 | 5.8 |
| 73285236.0 | 6.1 |
| 73464877.0 | 6.4 |
| 73709269.0 | 5.9 |
| 73717783.0 | 5.8 |
| 73835683.0 | 5.8 |
| 73923488.0 | 6.0 |
| 74016937.0 | 5.8 |
| 74020018.0 | 5.7 |
| 74357014.0 | 7.2 |
| 74360446.0 | 5.8 |
| 74556055.0 | 5.9 |
| 75113317.0 | 7.5 |
| 75166252.0 | 5.9 |
| 75178802.0 | 5.8 |
| 75184729.0 | 6.1 |
| 75190913.0 | 6.3 |
| 75244986.0 | 5.7 |
| 75285197.0 | 6.0 |
| 75318578.0 | 6.1 |
| 75489846.0 | 5.9 |
| 75706635.0 | 5.7 |
| 75726012.0 | 6.0 |
| 75878221.0 | 6.7 |
| 75930246.0 | 5.9 |
| 76063427.0 | 6.8 |
| 76126203.0 | 5.8 |
| 76145237.0 | 5.7 |
| 76250970.0 | 5.9 |
| 76798443.0 | 5.7 |
| 77423246.0 | 6.3 |
| 77454147.0 | 6.1 |
| 77500926.0 | 5.7 |
| 77521278.0 | 5.7 |
| 77541522.0 | 6.4 |
| 77590495.0 | 6.7 |
| 77882106.0 | 5.8 |
| 77891371.0 | 6.9 |
| 77896687.0 | 5.7 |
| 78181287.0 | 6.1 |
| 78209068.0 | 6.8 |
| 78244034.0 | 5.9 |
| 78433055.0 | 6.0 |
| 78451913.0 | 5.8 |
| 78612753.0 | 6.3 |
| 78737786.0 | 5.7 |
| 78765890.0 | 5.7 |
| 78888208.0 | 6.2 |
| 79353278.0 | 5.9 |
| 79704791.0 | 6.0 |
| 79872490.0 | 5.8 |
| 79948990.0 | 5.7 |
| 80573220.0 | 5.8 |
| 80645327.0 | 5.8 |
| 80727572.0 | 5.9 |
| 80736708.0 | 5.9 |
| 80775403.0 | 6.0 |
| 81005810.0 | 6.9 |
| 81033778.0 | 6.1 |
| 81385309.0 | 6.0 |
| 81395139.0 | 6.2 |
| 81486050.0 | 5.7 |
| 81507983.0 | 5.9 |
| 81931923.0 | 5.9 |
| 82023968.0 | 5.8 |
| 82115164.0 | 5.7 |
| 82126026.0 | 5.7 |
| 82140565.0 | 6.1 |
| 82481400.0 | 5.7 |
| 83473022.0 | 5.9 |
| 83489577.0 | 5.7 |
| 83519523.0 | 5.7 |
| 84377543.0 | 5.9 |
| 84652046.0 | 6.0 |
| 84838415.0 | 5.8 |
| 85129914.0 | 5.7 |
| 85148246.0 | 6.1 |
| 85398457.0 | 5.9 |
| 85863644.0 | 5.9 |
| 86373842.0 | 6.1 |
| 86495264.0 | 5.8 |
| 86531863.0 | 6.0 |
| 87007842.0 | 6.4 |
| 87152473.0 | 6.0 |
| 87192790.0 | 6.0 |
| 87279190.0 | 5.8 |
| 87343607.0 | 5.9 |
| 87539747.0 | 5.8 |
| 87542422.0 | 5.8 |
| 87557667.0 | 6.0 |
| 87739213.0 | 5.9 |
| 87786681.0 | 5.8 |
| 87874794.0 | 5.9 |
| 88330277.0 | 6.0 |
| 88350691.0 | 6.4 |
| 88854888.0 | 5.8 |
| 89002231.0 | 6.9 |
| 89182864.0 | 5.7 |
| 89324966.0 | 5.9 |
| 89457360.0 | 5.9 |
| 89618569.0 | 6.4 |
| 89641136.0 | 6.0 |
| 89750026.0 | 6.1 |
| 89902849.0 | 6.6 |
| 89976325.0 | 5.8 |
| 89976407.0 | 6.2 |
| 90052736.0 | 6.5 |
| 90080780.0 | 6.1 |
| 90170612.0 | 8.2 |
| 90184457.0 | 6.3 |
| 90185573.0 | 6.8 |
| 90223222.0 | 6.9 |
| 90419738.0 | 7.3 |
| 90423382.0 | 6.5 |
| 90462941.0 | 6.2 |
| 90477103.0 | 5.8 |
| 90496411.0 | 5.8 |
| 90527950.0 | 6.3 |
| 90634080.0 | 7.1 |
| 90748438.0 | 5.9 |
| 90767840.0 | 6.0 |
| 90972724.0 | 6.2 |
| 90994385.0 | 5.7 |
| 91028347.0 | 6.4 |
| 91038530.0 | 5.8 |
| 91047350.0 | 7.1 |
| 91550472.0 | 5.9 |
| 91703513.0 | 6.6 |
| 91781591.0 | 7.8 |
| 91818958.0 | 6.2 |
| 91820591.0 | 5.8 |
| 91923643.0 | 6.1 |
| 91933603.0 | 5.8 |
| 92054130.0 | 6.5 |
| 92085775.0 | 6.9 |
| 92140533.0 | 6.3 |
| 92472850.0 | 5.8 |
| 92502735.0 | 5.7 |
| 92664943.0 | 6.5 |
| 92788295.0 | 6.0 |
| 92804506.0 | 5.8 |
| 93033873.0 | 5.8 |
| 93033935.0 | 5.8 |
| 93214565.0 | 5.8 |
| 93454887.0 | 5.8 |
| 93576581.0 | 5.7 |
| 93650635.0 | 6.1 |
| 93661598.0 | 7.5 |
| 93837377.0 | 6.6 |
| 93971015.0 | 6.0 |
| 94284218.0 | 5.7 |
| 94403638.0 | 5.7 |
| 94403745.0 | 5.9 |
| 94603747.0 | 6.1 |
| 94729729.0 | 6.0 |
| 94737134.0 | 7.0 |
| 94745996.0 | 6.5 |
| 94911223.0 | 5.7 |
| 94922259.0 | 5.7 |
| 94961656.0 | 6.7 |
| 95026169.0 | 5.7 |
| 95053508.0 | 5.9 |
| 95184456.0 | 6.5 |
| 95618671.0 | 6.0 |
| 95721623.0 | 6.8 |
| 95801932.0 | 5.8 |
| 95843279.0 | 5.7 |
| 96025205.0 | 5.7 |
| 96040726.0 | 6.8 |
| 96077373.0 | 5.9 |
| 96298067.0 | 6.1 |
| 96328695.0 | 6.3 |
| 96388202.0 | 5.8 |
| 96398253.0 | 6.1 |
| 96444955.0 | 5.7 |
| 96633255.0 | 5.8 |
| 96643425.0 | 6.2 |
| 96689109.0 | 6.0 |
| 96850763.0 | 6.0 |
| 96892198.0 | 5.9 |
| 97260814.0 | 6.8 |
| 97282726.0 | 5.7 |
| 97371432.0 | 6.1 |
| 97477674.0 | 6.3 |
| 97763145.0 | 6.1 |
| 97844776.0 | 6.4 |
| 97855975.0 | 6.3 |
| 97871449.0 | 6.1 |
| 98105379.0 | 6.7 |
| 98435459.0 | 5.9 |
| 98635593.0 | 5.8 |
| 98649885.0 | 6.0 |
| 98696486.0 | 6.2 |
| 98722878.0 | 5.7 |
| 98852842.0 | 5.7 |
| 98865914.0 | 5.7 |
| 99055539.0 | 5.8 |
| 99131602.0 | 7.0 |
| 99203475.0 | 6.4 |
| 99516122.0 | 7.5 |
| 99524822.0 | 6.6 |
| 99671453.0 | 5.9 |
| 100027825.0 | 7.1 |
| 100153070.0 | 6.3 |
| 100485991.0 | 6.1 |
| 100503014.0 | 5.8 |
| 100716094.0 | 6.3 |
| 100788102.0 | 5.8 |
| 100857949.0 | 7.3 |
| 100906667.0 | 5.9 |
| 101021335.0 | 6.0 |
| 101126176.0 | 5.7 |
| 101139157.0 | 6.4 |
| 101187913.0 | 6.1 |
| 101366111.0 | 5.8 |
| 101498847.0 | 5.8 |
| 101587381.0 | 7.0 |
| 101691786.0 | 5.7 |
| 101756371.0 | 6.0 |
| 102246664.0 | 5.9 |
| 102248972.0 | 6.8 |
| 102261487.0 | 5.8 |
| 102314752.0 | 5.7 |
| 102328273.0 | 6.6 |
| 102483805.0 | 6.3 |
| 102672273.0 | 5.8 |
| 103108228.0 | 6.6 |
| 103115234.0 | 5.7 |
| 103270230.0 | 6.2 |
| 103350145.0 | 6.0 |
| 103362276.0 | 5.9 |
| 103480605.0 | 6.7 |
| 103567229.0 | 5.7 |
| 103604472.0 | 5.7 |
| 103647796.0 | 6.0 |
| 103666437.0 | 6.4 |
| 103716338.0 | 6.7 |
| 103924155.0 | 5.7 |
| 104001076.0 | 5.7 |
| 104007332.0 | 6.2 |
| 104023002.0 | 5.9 |
| 104065238.0 | 6.2 |
| 104116821.0 | 5.8 |
| 104394928.0 | 5.9 |
| 104570286.0 | 6.5 |
| 104632289.0 | 6.1 |

Table B-1: 0-reduced time series of 845 $M_w$5.6+ earthquakes that occurred from 01 Oct 2015–02 Feb 2019, declustered by excluding 21 events that had occurred within minutes of time and location of another, where stronger event was kept. Event magnitudes are robust (outliers discarded) means from the respective USGS, EMSC, and GFZ final moment magnitudes.

| T(s) | Mag |
|---|---|
| 1317771.0 | 5.7 |
| 1383206.0 | 7.5 |
| 1597835.0 | 5.8 |
| 3748799.0 | 5.7 |
| 3903930.0 | 6.0 |
| 3938088.0 | 7.6 |
| 4049352.0 | 6.7 |
| 4190656.0 | 6.2 |
| 4211701.0 | 5.9 |
| 6422344.0 | 5.8 |
| 6518877.0 | 6.1 |
| 6531292.0 | 5.8 |
| 6598919.0 | 5.7 |
| 6603521.0 | 6.3 |
| 8932252.0 | 6.5 |
| 9164064.0 | 7.1 |
| 9212644.0 | 5.7 |
| 9228356.0 | 6.3 |
| 9310457.0 | 6.0 |
| 9440113.0 | 5.7 |
| 11516117.0 | 5.8 |
| 11655658.0 | 5.9 |
| 11783560.0 | 5.8 |
| 14046854.0 | 6.4 |
| 14104366.0 | 5.6 |
| 14495298.0 | 5.7 |
| 16576175.0 | 6.2 |
| 16673861.0 | 6.1 |
| 16673944.0 | 6.0 |
| 16687419.0 | 5.8 |
| 16826855.0 | 6.1 |
| 16827847.0 | 5.9 |
| 16907308.0 | 5.7 |
| 17069308.0 | 5.7 |
| 17100663.0 | 6.0 |
| 19090859.0 | 6.7 |
| 19122638.0 | 6.9 |
| 19300678.0 | 6.0 |
| 21731412.0 | 5.7 |
| 21803617.0 | 5.7 |
| 21862276.0 | 6.3 |
| 21927288.0 | 5.9 |
| 22059029.0 | 6.1 |
| 22061762.0 | 6.3 |
| 22248564.0 | 5.8 |
| 24438154.0 | 5.8 |
| 24560231.0 | 6.0 |
| 26705950.0 | 5.8 |
| 27027243.0 | 5.7 |
| 27076418.0 | 6.0 |
| 27124577.0 | 7.4 |
| 27160655.0 | 5.8 |
| 27216320.0 | 6.0 |
| 27241318.0 | 5.8 |
| 27283757.0 | 6.1 |
| 29350945.0 | 5.9 |
| 29370534.0 | 6.0 |
| 29451678.0 | 5.7 |
| 29607851.0 | 6.0 |
| 29612152.0 | 5.8 |
| 29715690.0 | 5.7 |
| 32051252.0 | 6.3 |
| 32217532.0 | 6.8 |
| 32221122.0 | 5.9 |
| 32341235.0 | 5.9 |
| 32369395.0 | 6.6 |
| 34433215.0 | 6.1 |
| 34499985.0 | 5.9 |
| 34567612.0 | 7.8 |
| 34569362.0 | 6.4 |
| 34570599.0 | 6.1 |
| 34576521.0 | 6.2 |
| 34578295.0 | 5.8 |
| 34597972.0 | 5.7 |
| 34616296.0 | 6.4 |
| 34788865.0 | 5.8 |
| 34841646.0 | 5.7 |
| 36999426.0 | 5.8 |
| 37213516.0 | 6.0 |
| 37504504.0 | 7.9 |
| 37506693.0 | 6.4 |
| 37513588.0 | 6.7 |
| 39561461.0 | 7.3 |
| 39594669.0 | 6.3 |
| 39704532.0 | 5.8 |
| 39858199.0 | 5.7 |
| 39906935.0 | 6.0 |
| 39950978.0 | 5.7 |
| 42268057.0 | 6.5 |
| 44590700.0 | 5.9 |
| 44992510.0 | 6.0 |
| 45036454.0 | 5.7 |
| 47168000.0 | 5.9 |
| 47317668.0 | 5.7 |
| 47353365.0 | 5.8 |
| 47454370.0 | 5.7 |
| 47478295.0 | 5.8 |
| 47635788.0 | 5.8 |
| 49790120.0 | 5.7 |
| 49791084.0 | 5.9 |
| 49795481.0 | 6.2 |
| 49827489.0 | 6.0 |
| 49870566.0 | 6.8 |
| 49935831.0 | 5.8 |
| 49991250.0 | 6.5 |
| 49997617.0 | 5.7 |
| 50049907.0 | 5.7 |
| 50118319.0 | 6.2 |
| 52605968.0 | 5.7 |
| 52755545.0 | 5.7 |
| 52803153.0 | 6.3 |
| 54855251.0 | 5.8 |
| 54860871.0 | 6.5 |
| 54931605.0 | 5.7 |
| 55031493.0 | 5.7 |
| 55183544.0 | 5.8 |
| 55289042.0 | 6.6 |
| 55421521.0 | 5.9 |
| 57425675.0 | 5.8 |
| 57511144.0 | 5.7 |
| 57619837.0 | 5.8 |
| 57640565.0 | 5.7 |
| 57731024.0 | 6.5 |
| 57767507.0 | 6.3 |
| 60045093.0 | 6.1 |
| 60337844.0 | 6.1 |
| 60378795.0 | 8.1 |
| 62595779.0 | 5.7 |
| 62809407.0 | 6.2 |
| 62844320.0 | 5.7 |
| 63004114.0 | 6.1 |
| 63028373.0 | 6.2 |
| 63034707.0 | 6.5 |
| 65035671.0 | 6.6 |
| 65318654.0 | 6.8 |
| 65337171.0 | 5.7 |
| 65577934.0 | 5.7 |
| 65622633.0 | 6.5 |
| 67556204.0 | 6.5 |
| 67628200.0 | 6.1 |
| 67629234.0 | 6.0 |
| 67832580.0 | 6.1 |
| 67878719.0 | 5.7 |
| 70437673.0 | 5.9 |
| 72688019.0 | 6.5 |
| 72915054.0 | 6.2 |
| 73015785.0 | 6.0 |
| 73104107.0 | 5.8 |
| 75244986.0 | 5.7 |
| 75285197.0 | 6.0 |
| 75318578.0 | 6.1 |
| 75489846.0 | 5.9 |
| 75706635.0 | 5.7 |
| 75726012.0 | 6.0 |
| 77882106.0 | 5.8 |
| 77891371.0 | 6.9 |
| 77896687.0 | 5.7 |
| 78181287.0 | 6.1 |
| 78209068.0 | 6.8 |
| 78244034.0 | 5.9 |
| 80573220.0 | 5.8 |
| 80645327.0 | 5.8 |
| 80727572.0 | 5.9 |
| 80736708.0 | 5.9 |
| 80775403.0 | 6.0 |
| 85863644.0 | 5.9 |
| 88330277.0 | 6.0 |
| 88350691.0 | 6.4 |
| 90527950.0 | 6.3 |
| 90634080.0 | 7.1 |
| 90748438.0 | 5.9 |
| 90767840.0 | 6.0 |
| 90972724.0 | 6.2 |
| 90994385.0 | 5.7 |
| 91028347.0 | 6.4 |
| 91038530.0 | 5.8 |
| 91047350.0 | 7.1 |
| 93214565.0 | 5.8 |
| 93454887.0 | 5.8 |
| 93576581.0 | 5.7 |
| 93650635.0 | 6.1 |
| 93661598.0 | 7.5 |
| 95618671.0 | 6.0 |
| 95721623.0 | 6.8 |
| 95801932.0 | 5.8 |
| 95843279.0 | 5.7 |
| 96025205.0 | 5.7 |
| 96040726.0 | 6.8 |
| 96077373.0 | 5.9 |
| 98435459.0 | 5.9 |
| 98635593.0 | 5.8 |
| 98649885.0 | 6.0 |
| 98696486.0 | 6.2 |
| 98722878.0 | 5.7 |
| 100716094.0 | 6.3 |
| 100788102.0 | 5.8 |
| 100857949.0 | 7.3 |
| 100906667.0 | 5.9 |
| 101021335.0 | 6.0 |
| 101126176.0 | 5.7 |
| 101139157.0 | 6.4 |
| 101187913.0 | 6.1 |
| 103350145.0 | 6.0 |
| 103362276.0 | 5.9 |
| 103480605.0 | 6.7 |
| 103567229.0 | 5.7 |
| 103604472.0 | 5.7 |
| 103647796.0 | 6.0 |
| 103666437.0 | 6.4 |
| 103716338.0 | 6.7 |

Table B-2: 0-reduced time series of 210 earthquakes from Table B-1 that occurred during the Moon's traversals of the Earth's magnetotail, i.e., within three days from the full Moon.

| T(s) | Mag |
|---|---|
| 0.0 | 5.5 |
| 57741.0 | 6.1 |
| 334022.0 | 6.0 |
| 614223.0 | 5.8 |
| 717749.0 | 6.0 |
| 910557.0 | 7.1 |
| 1097040.0 | 6.1 |
| 1105692.0 | 6.0 |
| 1170437.0 | 5.7 |
| 1923610.0 | 5.9 |
| 1984768.0 | 5.8 |
| 2141289.0 | 6.5 |
| 2414137.0 | 6.8 |
| 2426257.0 | 5.7 |
| 2507932.0 | 5.7 |
| 2533855.0 | 6.5 |
| 2617659.0 | 6.4 |
| 2739512.0 | 6.8 |
| 2817613.0 | 5.8 |
| 2980531.0 | 6.7 |
| 3061455.0 | 5.7 |
| 3167007.0 | 5.8 |
| 3225189.0 | 5.9 |
| 3276843.0 | 6.5 |
| 3404098.0 | 7.0 |
| 3629408.0 | 6.1 |
| 4355804.0 | 5.9 |
| 4522340.0 | 5.8 |
| 4800528.0 | 7.1 |
| 5007241.0 | 7.2 |
| 5189145.0 | 6.8 |
| 5198515.0 | 6.0 |
| 5281749.0 | 5.7 |
| 5914428.0 | 6.5 |
| 6023688.0 | 6.0 |
| 6085738.0 | 5.8 |
| 6169890.0 | 6.0 |
| 6886538.0 | 5.8 |
| 7092054.0 | 5.8 |
| 7146274.0 | 6.3 |
| 7193169.0 | 5.8 |
| 7241173.0 | 5.8 |
| 7394954.0 | 6.7 |
| 7493107.0 | 5.9 |
| 7519089.0 | 6.0 |
| 7563824.0 | 5.7 |
| 7748155.0 | 5.6 |
| 7960557.0 | 5.7 |
| 8062920.0 | 6.5 |
| 8064717.0 | 6.2 |
| 8124545.0 | 5.9 |
| 8161734.0 | 5.7 |
| 8197195.0 | 5.8 |
| 8274561.0 | 6.1 |
| 8274568.0 | 6.6 |
| 8302096.0 | 5.7 |
| 8674100.0 | 5.9 |
| 8842628.0 | 5.9 |
| 9656944.0 | 7.2 |
| 9794575.0 | 6.0 |
| 9885879.0 | 6.2 |
| 9955396.0 | 5.7 |
| 10234881.0 | 6.4 |
| 10343249.0 | 5.8 |
| 10480989.0 | 6.4 |
| 10597019.0 | 6.3 |
| 10803978.0 | 6.2 |
| 10941457.0 | 5.8 |
| 11006000.0 | 5.7 |
| 11081187.0 | 6.0 |
| 11097118.0 | 5.9 |
| 11199149.0 | 6.1 |
| 11262597.0 | 6.0 |
| 12141218.0 | 6.1 |
| 12455622.0 | 7.8 |
| 12574304.0 | 5.8 |
| 12948021.0 | 6.0 |
| 13338639.0 | 6.3 |
| 13530574.0 | 5.7 |
| 13883946.0 | 6.0 |
| 13919425.0 | 6.0 |
| 14634325.0 | 5.8 |
| 14717930.0 | 5.6 |
| 15010981.0 | 5.9 |
| 15071330.0 | 6.1 |
| 15108834.0 | 6.2 |
| 15204468.0 | 6.9 |
| 15377596.0 | 5.9 |
| 15458327.0 | 5.7 |
| 15458563.0 | 6.7 |
| 15462091.0 | 6.0 |
| 15486568.0 | 6.0 |
| 15532608.0 | 6.6 |
| 15787110.0 | 5.7 |
| 15804918.0 | 5.6 |
| 15816772.0 | 6.6 |
| 15964001.0 | 5.8 |
| 16088351.0 | 6.9 |
| 16104349.0 | 5.9 |
| 16135803.0 | 5.8 |
| 16168859.0 | 5.9 |
| 16169430.0 | 6.1 |
| 16170382.0 | 5.6 |
| 16178862.0 | 6.0 |
| 16203261.0 | 6.3 |
| 16208344.0 | 5.6 |
| 16262123.0 | 6.1 |
| 16270140.0 | 7.0 |
| 16271281.0 | 5.7 |
| 16271390.0 | 5.7 |
| 16383751.0 | 7.8 |
| 16409873.0 | 5.7 |
| 16410894.0 | 5.9 |
| 16517405.0 | 5.9 |
| 17177026.0 | 5.6 |
| 17294116.0 | 6.0 |
| 17404640.0 | 7.0 |
| 17426252.0 | 6.6 |
| 17537977.0 | 5.8 |
| 17695518.0 | 5.7 |
| 17766285.0 | 5.7 |
| 17905219.0 | 5.7 |
| 18225476.0 | 5.8 |
| 18555670.0 | 5.8 |
| 18782700.0 | 5.8 |
| 18891947.0 | 5.8 |
| 19853303.0 | 5.9 |
| 19854758.0 | 6.4 |
| 19946564.0 | 6.9 |
| 19961453.0 | 7.2 |
| 20204861.0 | 6.3 |
| 20221673.0 | 5.9 |
| 20260516.0 | 5.7 |
| 20354396.0 | 6.4 |
| 20366821.0 | 5.7 |
| 20638504.0 | 5.7 |
| 20676567.0 | 6.3 |
| 20713161.0 | 6.0 |
| 20829332.0 | 6.3 |
| 20859550.0 | 6.3 |
| 20907334.0 | 5.8 |
| 20978222.0 | 6.1 |
| 21061756.0 | 6.0 |
| 21064898.0 | 6.1 |
| 21116029.0 | 5.8 |
| 21312213.0 | 5.7 |
| 21444798.0 | 6.2 |
| 22472465.0 | 6.4 |
| 22686641.0 | 5.8 |
| 22818868.0 | 5.9 |
| 23156974.0 | 5.8 |
| 23299552.0 | 5.8 |
| 23690747.0 | 5.8 |
| 23690915.0 | 5.9 |
| 23735105.0 | 5.9 |
| 23735700.0 | 6.3 |
| 23943652.0 | 6.0 |
| 23944507.0 | 6.3 |
| 23944946.0 | 5.9 |
| 23947699.0 | 5.8 |
| 24768255.0 | 5.8 |
| 24837197.0 | 5.9 |
| 24902086.0 | 5.8 |
| 24969744.0 | 5.8 |
| 25000244.0 | 6.1 |
| 25008159.0 | 6.4 |
| 25044794.0 | 5.9 |
| 25052894.0 | 5.7 |
| 25115346.0 | 5.9 |
| 25359743.0 | 7.7 |
| 25497436.0 | 5.8 |
| 25570004.0 | 5.7 |
| 25852745.0 | 6.2 |
| 25860508.0 | 6.3 |
| 26497829.0 | 7.1 |
| 26505203.0 | 6.1 |
| 27513818.0 | 6.0 |
| 27535227.0 | 6.2 |
| 27567529.0 | 6.8 |
| 27579159.0 | 5.7 |
| 27677317.0 | 5.9 |
| 27977632.0 | 7.1 |
| 28145730.0 | 6.7 |
| 28213709.0 | 5.8 |
| 28280512.0 | 7.0 |
| 28282679.0 | 6.0 |
| 28436798.0 | 5.7 |
| 28489326.0 | 5.7 |
| 28621376.0 | 5.8 |
| 28648679.0 | 6.1 |
| 28903814.0 | 6.1 |
| 29034734.0 | 6.1 |
| 29043087.0 | 5.8 |
| 29195748.0 | 5.8 |
| 29921110.0 | 6.1 |
| 30122308.0 | 6.3 |
| 30203823.0 | 6.3 |
| 30247703.0 | 5.9 |
| 30283868.0 | 6.4 |
| 30285155.0 | 6.9 |
| 30399832.0 | 5.6 |
| 31122281.0 | 5.7 |
| 31233454.0 | 5.7 |
| 31301754.0 | 5.7 |
| 31488313.0 | 5.7 |
| 31557021.0 | 5.8 |
| 31775118.0 | 5.7 |
| 32454800.0 | 5.6 |
| 32477390.0 | 5.6 |
| 32559077.0 | 6.2 |
| 32786960.0 | 5.8 |
| 32944437.0 | 5.8 |
| 32991824.0 | 6.0 |
| 33042122.0 | 6.1 |
| 33088905.0 | 5.8 |
| 33101832.0 | 5.9 |
| 33133009.0 | 6.0 |
| 33165051.0 | 5.9 |
| 33217603.0 | 5.8 |
| 33342253.0 | 6.5 |
| 33559643.0 | 6.0 |
| 33809078.0 | 6.3 |
| 34086919.0 | 5.7 |
| 34107199.0 | 5.8 |
| 34113580.0 | 6.0 |
| 34235693.0 | 5.9 |
| 34313283.0 | 5.8 |
| 35208098.0 | 6.4 |
| 35294623.0 | 6.9 |
| 35306616.0 | 5.9 |
| 35427832.0 | 5.8 |
| 35468851.0 | 5.8 |
| 35545662.0 | 6.9 |
| 35616504.0 | 6.6 |
| 36164660.0 | 6.2 |
| 36289649.0 | 5.9 |
| 36361681.0 | 5.7 |
| 36433018.0 | 6.3 |
| 36593176.0 | 5.9 |
| 36594447.0 | 6.5 |
| 36706738.0 | 6.0 |
| 36741221.0 | 6.6 |
| 36751361.0 | 7.8 |
| 36766346.0 | 5.9 |
| 36766801.0 | 6.5 |
| 36843240.0 | 6.9 |
| 36845238.0 | 5.9 |
| 36852136.0 | 5.8 |
| 36870574.0 | 5.8 |
| 36919710.0 | 6.0 |
| 37572619.0 | 5.9 |
| 37575817.0 | 5.9 |
| 37587058.0 | 6.2 |
| 37600445.0 | 6.4 |
| 37740322.0 | 6.4 |
| 37769829.0 | 6.0 |
| 37812070.0 | 6.7 |
| 37871272.0 | 5.9 |
| 38075770.0 | 5.9 |
| 38084569.0 | 5.8 |
| 38208380.0 | 7.6 |
| 38461366.0 | 5.9 |
| 38583253.0 | 6.2 |
| 38848621.0 | 5.8 |
| 38884750.0 | 5.9 |
| 38895463.0 | 5.9 |
| 38895477.0 | 6.3 |
| 39012985.0 | 6.9 |
| 39015846.0 | 6.0 |
| 39092867.0 | 5.7 |
| 39216869.0 | 5.7 |
| 39398162.0 | 5.9 |
| 39451866.0 | 6.0 |
| 40062032.0 | 5.8 |
| 40132292.0 | 5.9 |
| 40176716.0 | 5.9 |
| 40267086.0 | 5.7 |
| 40399690.0 | 6.5 |
| 40592058.0 | 7.9 |
| 41234625.0 | 5.9 |
| 41351958.0 | 5.9 |
| 41684297.0 | 5.7 |
| 42037669.0 | 6.2 |
| 42952449.0 | 6.4 |
| 43218778.0 | 6.5 |
| 43428971.0 | 5.9 |
| 43489958.0 | 6.9 |
| 43512639.0 | 5.9 |
| 43800776.0 | 5.7 |
| 44128938.0 | 6.0 |
| 44205132.0 | 5.7 |
| 44249171.0 | 5.7 |
| 44286707.0 | 6.3 |
| 44435533.0 | 5.7 |
| 45470840.0 | 6.0 |
| 46144453.0 | 6.2 |
| 46293198.0 | 6.6 |
| 46333160.0 | 5.7 |
| 46773852.0 | 6.5 |
| 46905186.0 | 6.1 |
| 47777017.0 | 6.3 |
| 47928324.0 | 5.7 |
| 48068142.0 | 6.0 |
| 48070429.0 | 5.9 |
| 48447600.0 | 6.0 |
| 48602544.0 | 6.9 |
| 48837027.0 | 5.7 |
| 48926040.0 | 5.9 |
| 48943633.0 | 6.8 |
| 49001787.0 | 5.7 |
| 49174548.0 | 6.2 |
| 49180931.0 | 6.3 |
| 49270456.0 | 5.9 |
| 49319467.0 | 6.0 |
| 49493610.0 | 5.8 |
| 49562402.0 | 5.9 |
| 50323697.0 | 5.7 |
| 50387192.0 | 6.2 |
| 50775008.0 | 5.9 |
| 50835761.0 | 5.7 |
| 51149651.0 | 5.7 |
| 51176395.0 | 5.8 |
| 51601157.0 | 6.6 |
| 51602508.0 | 5.8 |
| 51676425.0 | 5.7 |
| 51974921.0 | 6.8 |
| 51979932.0 | 5.7 |
| 52050088.0 | 6.0 |
| 52932574.0 | 5.8 |
| 52957979.0 | 6.9 |
| 53019011.0 | 5.9 |
| 53059570.0 | 5.8 |
| 53174482.0 | 5.7 |
| 53270996.0 | 6.1 |
| 53396377.0 | 5.8 |
| 53667298.0 | 6.8 |
| 53853186.0 | 5.7 |
| 53945184.0 | 6.0 |
| 54206687.0 | 6.0 |
| 54252424.0 | 6.0 |
| 54319132.0 | 5.9 |
| 54394418.0 | 6.0 |
| 54522165.0 | 5.8 |
| 54716993.0 | 5.9 |
| 55449603.0 | 6.4 |
| 55653375.0 | 5.9 |
| 55787070.0 | 5.7 |
| 55822143.0 | 6.3 |
| 55867089.0 | 7.7 |
| 55876153.0 | 6.4 |
| 55999219.0 | 6.0 |
| 56042119.0 | 5.8 |
| 56122505.0 | 6.6 |
| 56597811.0 | 5.9 |
| 56689952.0 | 5.8 |
| 56710639.0 | 5.9 |
| 56981083.0 | 5.7 |
| 57961941.0 | 6.2 |
| 58075087.0 | 5.8 |
| 58126327.0 | 6.4 |
| 58175718.0 | 5.7 |
| 58309542.0 | 5.7 |
| 58420519.0 | 5.7 |
| 58557796.0 | 6.6 |
| 58640686.0 | 6.4 |
| 59340104.0 | 6.3 |
| 59731850.0 | 6.3 |
| 60545479.0 | 5.8 |
| 60612255.0 | 5.8 |
| 60683943.0 | 5.7 |
| 60943523.0 | 5.8 |
| 61191002.0 | 5.7 |
| 61275053.0 | 5.8 |
| 61377512.0 | 7.1 |
| 61404444.0 | 6.1 |
| 61458070.0 | 6.1 |
| 61470824.0 | 6.4 |
| 61478250.0 | 5.9 |
| 61484598.0 | 5.7 |
| 61642449.0 | 5.7 |
| 61695721.0 | 5.8 |
| 61703816.0 | 6.1 |
| 61732263.0 | 5.7 |
| 61780206.0 | 5.7 |
| 61903998.0 | 5.9 |
| 61932234.0 | 6.4 |
| 61989966.0 | 5.8 |
| 63149774.0 | 6.3 |
| 63194248.0 | 6.7 |
| 63860693.0 | 6.1 |
| 63936355.0 | 5.9 |
| 64280187.0 | 5.9 |
| 64374701.0 | 6.7 |
| 64740000.0 | 5.8 |
| 64750496.0 | 5.9 |
| 64943163.0 | 6.8 |
| 64949080.0 | 5.9 |
| 64983285.0 | 6.1 |
| 65659145.0 | 5.7 |
| 65745965.0 | 6.0 |
| 65847927.0 | 5.8 |
| 65893209.0 | 6.1 |
| 66043331.0 | 7.3 |
| 66058081.0 | 5.9 |
| 66072737.0 | 6.6 |
| 66238107.0 | 5.8 |
| 66348392.0 | 5.9 |
| 66358050.0 | 5.8 |
| 66490694.0 | 6.4 |
| 66553856.0 | 5.8 |
| 66616181.0 | 6.3 |
| 66636777.0 | 6.6 |
| 66640321.0 | 5.9 |
| 66664044.0 | 7.0 |
| 66669198.0 | 6.0 |
| 66736501.0 | 6.0 |
| 67163636.0 | 5.8 |
| 67299306.0 | 6.0 |
| 67469580.0 | 5.7 |
| 68225208.0 | 6.4 |
| 68231632.0 | 5.9 |
| 68259302.0 | 6.3 |
| 68309166.0 | 5.7 |
| 68365099.0 | 6.1 |
| 68522034.0 | 5.7 |
| 68600831.0 | 6.0 |
| 68647525.0 | 6.0 |
| 68720857.0 | 6.5 |
| 68889113.0 | 6.5 |
| 69357852.0 | 5.8 |
| 70014255.0 | 5.7 |
| 70115087.0 | 5.7 |
| 70124391.0 | 5.7 |
| 70988988.0 | 5.8 |
| 71085326.0 | 7.5 |
| 71227818.0 | 6.0 |
| 71375310.0 | 5.8 |
| 71454159.0 | 7.1 |
| 71647667.0 | 5.9 |
| 71665272.0 | 5.9 |
| 71762452.0 | 5.8 |
| 71809967.0 | 5.8 |
| 71911300.0 | 6.3 |
| 72029435.0 | 6.3 |
| 72221929.0 | 6.0 |
| 72232537.0 | 7.9 |
| 72323715.0 | 6.3 |
| 72375592.0 | 5.8 |
| 72378868.0 | 6.2 |
| 72409508.0 | 5.8 |
| 72431018.0 | 5.8 |
| 72539512.0 | 6.3 |
| 73285236.0 | 6.1 |
| 73464877.0 | 6.4 |
| 73709269.0 | 5.9 |
| 73717783.0 | 5.8 |
| 73835683.0 | 5.8 |
| 73923488.0 | 6.0 |
| 74016937.0 | 5.8 |
| 74020018.0 | 5.7 |
| 74357014.0 | 7.2 |
| 74360446.0 | 5.8 |
| 74556055.0 | 5.9 |
| 75113317.0 | 7.5 |
| 75166252.0 | 5.9 |
| 75178802.0 | 5.8 |
| 75184729.0 | 6.1 |
| 75190913.0 | 6.3 |
| 75878221.0 | 6.7 |
| 75930246.0 | 5.9 |
| 76063427.0 | 6.8 |
| 76126203.0 | 5.8 |
| 76145237.0 | 5.7 |
| 76250970.0 | 5.9 |
| 76798443.0 | 5.7 |
| 77423246.0 | 6.3 |
| 77454147.0 | 6.1 |
| 77500926.0 | 5.7 |
| 77521278.0 | 5.7 |
| 77541522.0 | 6.4 |
| 77590495.0 | 6.7 |
| 78433055.0 | 6.0 |
| 78451913.0 | 5.8 |
| 78612753.0 | 6.3 |
| 78737786.0 | 5.7 |
| 78765890.0 | 5.7 |
| 78888208.0 | 6.2 |
| 79353278.0 | 5.9 |
| 79704791.0 | 6.0 |
| 79872490.0 | 5.8 |
| 79948990.0 | 5.7 |
| 81005810.0 | 6.9 |
| 81033778.0 | 6.1 |
| 81385309.0 | 6.0 |
| 81395139.0 | 6.2 |
| 81486050.0 | 5.7 |
| 81507983.0 | 5.9 |
| 81931923.0 | 5.9 |
| 82023968.0 | 5.8 |
| 82115164.0 | 5.7 |
| 82126026.0 | 5.7 |
| 82140565.0 | 6.1 |
| 82481400.0 | 5.7 |
| 83473022.0 | 5.9 |
| 83489577.0 | 5.7 |
| 83519523.0 | 5.7 |
| 84377543.0 | 5.9 |
| 84652046.0 | 6.0 |
| 84838415.0 | 5.8 |
| 85129914.0 | 5.7 |
| 85148246.0 | 6.1 |
| 85398457.0 | 5.9 |
| 86373842.0 | 6.1 |
| 86495264.0 | 5.8 |
| 86531863.0 | 6.0 |
| 87007842.0 | 6.4 |
| 87152473.0 | 6.0 |
| 87192790.0 | 6.0 |
| 87279190.0 | 5.8 |
| 87343607.0 | 5.9 |
| 87539747.0 | 5.8 |
| 87542422.0 | 5.8 |
| 87557667.0 | 6.0 |
| 87739213.0 | 5.9 |
| 87786681.0 | 5.8 |
| 87874794.0 | 5.9 |
| 88854888.0 | 5.8 |
| 89002231.0 | 6.9 |
| 89182864.0 | 5.7 |
| 89324966.0 | 5.9 |
| 89457360.0 | 5.9 |
| 89618569.0 | 6.4 |
| 89641136.0 | 6.0 |
| 89750026.0 | 6.1 |
| 89902849.0 | 6.6 |
| 89976325.0 | 5.8 |
| 89976407.0 | 6.2 |
| 90052736.0 | 6.5 |
| 90080780.0 | 6.1 |
| 90170612.0 | 8.2 |
| 90184457.0 | 6.3 |
| 90185573.0 | 6.8 |
| 90223222.0 | 6.9 |
| 90419738.0 | 7.3 |
| 90423382.0 | 6.5 |
| 90462941.0 | 6.2 |
| 90477103.0 | 5.8 |
| 90496411.0 | 5.8 |
| 91550472.0 | 5.9 |
| 91703513.0 | 6.6 |
| 91781591.0 | 7.8 |
| 91818958.0 | 6.2 |
| 91820591.0 | 5.8 |
| 91923643.0 | 6.1 |
| 91933603.0 | 5.8 |
| 92054130.0 | 6.5 |
| 92085775.0 | 6.9 |
| 92140533.0 | 6.3 |
| 92472850.0 | 5.8 |
| 92502735.0 | 5.7 |
| 92664943.0 | 6.5 |
| 92788295.0 | 6.0 |
| 92804506.0 | 5.8 |
| 93033873.0 | 5.8 |
| 93033935.0 | 5.8 |
| 93837377.0 | 6.6 |
| 93971015.0 | 6.0 |
| 94284218.0 | 5.7 |
| 94403638.0 | 5.7 |
| 94403745.0 | 5.9 |
| 94603747.0 | 6.1 |
| 94729729.0 | 6.0 |
| 94737134.0 | 7.0 |
| 94745996.0 | 6.5 |
| 94911223.0 | 5.7 |
| 94922259.0 | 5.7 |
| 94961656.0 | 6.7 |
| 95026169.0 | 5.7 |
| 95053508.0 | 5.9 |
| 95184456.0 | 6.5 |
| 96298067.0 | 6.1 |
| 96328695.0 | 6.3 |
| 96388202.0 | 5.8 |
| 96398253.0 | 6.1 |
| 96444955.0 | 5.7 |
| 96633255.0 | 5.8 |
| 96643425.0 | 6.2 |
| 96689109.0 | 6.0 |
| 96850763.0 | 6.0 |
| 96892198.0 | 5.9 |
| 97260814.0 | 6.8 |
| 97282726.0 | 5.7 |
| 97371432.0 | 6.1 |
| 97477674.0 | 6.3 |
| 97763145.0 | 6.1 |
| 97844776.0 | 6.4 |
| 97855975.0 | 6.3 |
| 97871449.0 | 6.1 |
| 98105379.0 | 6.7 |
| 98852842.0 | 5.7 |
| 98865914.0 | 5.7 |
| 99055539.0 | 5.8 |
| 99131602.0 | 7.0 |
| 99203475.0 | 6.4 |
| 99516122.0 | 7.5 |
| 99524822.0 | 6.6 |
| 99671453.0 | 5.9 |
| 100027825.0 | 7.1 |
| 100153070.0 | 6.3 |
| 100485991.0 | 6.1 |
| 100503014.0 | 5.8 |
| 101366111.0 | 5.8 |
| 101498847.0 | 5.8 |
| 101587381.0 | 7.0 |
| 101691786.0 | 5.7 |
| 101756371.0 | 6.0 |
| 102246664.0 | 5.9 |
| 102248972.0 | 6.8 |
| 102261487.0 | 5.8 |
| 102314752.0 | 5.7 |
| 102328273.0 | 6.6 |
| 102483805.0 | 6.3 |
| 102672273.0 | 5.8 |
| 103108228.0 | 6.6 |
| 103115234.0 | 5.7 |
| 103270230.0 | 6.2 |
| 103924155.0 | 5.7 |
| 104001076.0 | 5.7 |
| 104007332.0 | 6.2 |
| 104023002.0 | 5.9 |
| 104065238.0 | 6.2 |
| 104116821.0 | 5.8 |
| 104394928.0 | 5.9 |
| 104570286.0 | 6.5 |
| 104632289.0 | 6.1 |

Table B-3: 0-reduced time series of 635 earthquakes from Table B-1 that occurred during the Moon's traversals of the interplanetary magnetic field, i.e., the events listed in Table B-1 but not in Table B-2.

Supplement C – marsquakes Table

| T(d) | Mrnd |
|---|---|
| 0.0000000000 | 2.1 |
| 60.6288888889 | 1.8 |
| 72.0198379630 | 1.9 |
| 84.1513425926 | 2.8 |
| 88.3154166667 | 2.6 |
| 89.5043981482 | 2.4 |
| 107.8825231482 | 2.4 |
| 109.0596527778 | 1.0 |
| 111.0485416667 | 2.0 |
| 123.2143634259 | 1.8 |
| 124.4442708334 | 2.4 |
| 124.5648032408 | 2.1 |
| 125.5033101852 | 1.0 |
| 128.5417476852 | 1.9 |
| 129.8509722222 | 2.7 |
| 131.6465046297 | 1.4 |
| 139.8496990741 | 1.2 |
| 140.8529629630 | 2.1 |
| 140.9139351852 | 1.3 |
| 142.8422685185 | 1.7 |
| 142.8949768519 | 1.3 |
| 143.2216550926 | 1.2 |
| 143.8960995371 | 1.1 |
| 146.9442708334 | 1.3 |
| 146.9876157408 | 2.8 |
| 147.9689236111 | 2.0 |
| 148.9988773148 | 2.6 |
| 149.0429166667 | 1.5 |
| 150.0238194445 | 1.7 |
| 151.0405439815 | 2.1 |
| 151.2247453704 | 2.9 |
| 152.0665509259 | 1.8 |
| 152.0900000000 | 1.0 |
| 152.1868865741 | 1.2 |
| 152.2740625000 | 2.9 |
| 153.0958564815 | 1.9 |
| 153.1109027778 | 2.7 |
| 154.1335532408 | 1.3 |
| 155.0959606482 | 2.8 |
| 155.2623263889 | 2.1 |
| 155.6916435185 | 1.3 |
| 156.1902662037 | 2.3 |
| 157.2197685185 | 1.8 |
| 158.1844328704 | 2.1 |
| 158.2466782408 | 2.4 |
| 158.3770138889 | 2.6 |
| 159.2748726852 | 2.4 |
| 159.2891782408 | 2.5 |
| 159.3849305556 | 1.2 |
| 159.8044560185 | 2.0 |
| 160.3028935185 | 1.6 |
| 160.3363657408 | 2.8 |
| 160.4097916667 | 2.8 |
| 161.3305439815 | 2.5 |
| 161.4079282408 | 1.1 |
| 161.4692592593 | 1.9 |
| 161.7488425926 | 1.7 |
| 162.3639351852 | 1.0 |
| 163.3924537037 | 1.7 |
| 163.4656481482 | 2.0 |
| 164.4175810185 | 2.1 |
| 165.4453125000 | 2.8 |
| 166.4776851852 | 2.8 |
| 167.5015972222 | 2.5 |
| 168.5307986111 | 2.3 |
| 169.5587615741 | 1.3 |
| 169.5768055556 | 2.5 |
| 169.6533333334 | 1.9 |
| 169.9077662037 | 2.8 |
| 170.0213657408 | 2.1 |
| 170.0649305556 | 2.4 |
| 170.5887847222 | 2.9 |
| 171.0302199074 | 1.2 |
| 171.6126967593 | 2.5 |
| 172.6375810185 | 2.8 |
| 173.6649421297 | 1.6 |
| 173.7162615741 | 2.7 |
| 173.7426620371 | 2.6 |
| 174.1678935185 | 2.4 |
| 174.6922453704 | 1.6 |
| 174.7446875000 | 1.6 |
| 174.7784027778 | 1.3 |
| 175.7178703704 | 2.8 |
| 175.7721180556 | 1.7 |
| 176.7456134259 | 2.3 |
| 176.9594444445 | 2.1 |
| 177.7645717593 | 1.9 |
| 177.7759953704 | 2.7 |
| 177.8485532408 | 1.6 |
| 177.8888194445 | 2.6 |
| 178.8039930556 | 1.1 |
| 179.0858680556 | 1.9 |
| 179.8313657408 | 2.7 |
| 179.8812731482 | 2.1 |
| 179.9041203704 | 2.4 |
| 180.8583449074 | 2.8 |
| 180.8985763889 | 2.1 |
| 181.8847106482 | 2.0 |
| 182.4158680556 | 1.6 |
| 182.9131134260 | 2.8 |
| 183.9134143519 | 2.2 |
| 183.9427083334 | 2.8 |
| 184.0168865741 | 1.4 |
| 184.9424189815 | 1.0 |
| 184.9717708334 | 2.7 |
| 184.9822337963 | 2.6 |
| 185.0259490741 | 2.4 |
| 185.9967476852 | 1.7 |
| 186.0048958334 | 1.9 |
| 186.0399537037 | 1.8 |
| 186.0767708334 | 1.9 |
| 186.1145254630 | 2.6 |
| 186.9849305556 | 1.2 |
| 187.0251388889 | 2.2 |
| 187.0556481482 | 2.9 |
| 187.1523842593 | 1.2 |
| 188.1182175926 | 1.5 |
| 190.0606134260 | 1.2 |
| 190.0907407408 | 2.6 |
| 190.1074305556 | 1.8 |
| 191.1335069445 | 1.6 |
| 192.1597800926 | 2.3 |
| 192.2399421297 | 1.2 |
| 193.1426388889 | 1.8 |
| 193.1817476852 | 2.1 |
| 193.1890740741 | 2.7 |
| 193.2765740741 | 1.2 |
| 193.2857291667 | 2.5 |
| 193.3454166667 | 1.6 |
| 194.2006134260 | 1.8 |
| 194.2145138889 | 1.6 |
| 194.2651967593 | 1.5 |
| 194.2891782408 | 2.1 |
| 194.3821875000 | 1.1 |
| 195.1981134259 | 1.1 |
| 195.2474421297 | 2.4 |
| 195.3271759260 | 2.5 |
| 196.2762731482 | 1.2 |
| 196.3437037037 | 1.6 |
| 197.3001041667 | 2.2 |
| 197.3121527778 | 1.7 |
| 198.2754050926 | 1.9 |
| 198.3285532408 | 2.1 |
| 198.3449884260 | 2.4 |
| 198.5665625000 | 1.4 |
| 199.3414004630 | 2.6 |
| 199.3528587963 | 1.7 |
| 199.3672800926 | 2.9 |
| 200.3784375000 | 2.5 |
| 200.4081944445 | 1.2 |
| 200.5320254630 | 1.9 |
| 200.6539814815 | 1.1 |
| 201.3836111111 | 1.2 |
| 201.4076157408 | 1.2 |
| 201.4206481482 | 2.7 |
| 201.4904050926 | 2.8 |
| 202.4340856482 | 1.3 |
| 203.4601157408 | 1.7 |
| 203.4658449074 | 2.6 |
| 203.4901736111 | 2.6 |
| 203.5050810185 | 2.3 |
| 203.5533680556 | 1.4 |
| 203.6184375000 | 1.4 |
| 204.4938310185 | 1.6 |
| 205.5114930556 | 2.5 |
| 205.5156712963 | 2.7 |
| 205.6208564815 | 1.9 |
| 206.5443750000 | 2.1 |
| 206.6541319445 | 1.0 |
| 206.6660416667 | 1.7 |
| 206.9160532408 | 1.4 |
| 207.5698148148 | 1.2 |
| 207.6409722222 | 1.1 |
| 207.6616550926 | 1.8 |
| 207.8041203704 | 2.6 |
| 208.5810185185 | 1.7 |
| 208.6490162037 | 1.1 |
| 209.6229629630 | 2.3 |
| 209.6451388889 | 1.2 |
| 209.7620138889 | 1.8 |
| 210.6150347223 | 2.7 |
| 210.6248726852 | 2.2 |
| 210.7156250000 | 2.1 |
| 211.6373495371 | 1.5 |
| 211.6766898148 | 1.7 |
| 211.7453125000 | 2.5 |
| 212.7073032408 | 2.0 |
| 212.7345486111 | 1.8 |
| 212.7638888889 | 2.8 |
| 213.7530902778 | 1.7 |
| 213.8615393519 | 1.6 |
| 213.8728472223 | 2.6 |
| 214.7608449074 | 1.5 |
| 214.7686111111 | 1.4 |
| 214.9179513889 | 1.5 |
| 215.1611921297 | 1.3 |
| 215.7908680556 | 1.4 |
| 215.8485069445 | 2.8 |
| 215.9504282408 | 2.9 |
| 216.7651967593 | 1.5 |
| 216.7955787037 | 2.3 |
| 216.9005208334 | 1.1 |
| 217.8438425926 | 1.7 |
| 218.8680208334 | 1.3 |
| 219.8977314815 | 2.2 |
| 219.9031018519 | 2.1 |
| 219.9662268519 | 1.1 |
| 220.0510763889 | 2.6 |
| 220.9254282408 | 2.0 |
| 220.9751273148 | 1.1 |
| 221.0478009260 | 2.9 |
| 221.9311921297 | 2.6 |
| 221.9536689815 | 1.2 |
| 222.1111689815 | 2.1 |
| 222.2184375000 | 2.5 |
| 222.9789467593 | 1.4 |
| 223.0086805556 | 1.1 |
| 223.0453356482 | 2.9 |
| 223.4075925926 | 1.5 |
| 224.0131481482 | 2.0 |
| 224.0143171297 | 2.5 |
| 224.0568171297 | 1.2 |
| 224.1030092593 | 2.3 |
| 224.1870370371 | 1.9 |
| 224.9858912037 | 1.5 |
| 225.0104976852 | 1.3 |
| 225.0344328704 | 1.4 |
| 225.0360069445 | 2.3 |
| 225.0895254630 | 2.3 |
| 225.1524305556 | 2.5 |
| 225.1861111111 | 1.3 |
| 226.0488888889 | 1.1 |
| 226.0643171297 | 2.0 |
| 226.0700462963 | 2.8 |
| 226.1538657408 | 2.5 |
| 226.1901273148 | 1.5 |
| 226.2158217593 | 2.3 |
| 227.0848148148 | 1.6 |
| 249.6911342593 | 2.7 |
| 249.7512731482 | 2.8 |
| 249.7813425926 | 2.6 |
| 250.7196296297 | 2.9 |
| 250.8382175926 | 1.6 |
| 250.8899074074 | 1.5 |
| 251.0096296297 | 1.8 |
| 251.7455671297 | 1.1 |
| 251.8811805556 | 2.7 |
| 251.9075462963 | 1.2 |
| 251.9871296297 | 2.3 |
| 252.7744212963 | 2.2 |
| 252.7969560185 | 1.7 |
| 252.8188078704 | 2.8 |
| 252.9059953704 | 1.1 |
| 253.7995717593 | 2.8 |
| 254.8280671297 | 1.0 |
| 254.9170833334 | 1.6 |
| 255.0198263889 | 2.2 |
| 255.3072222222 | 1.9 |
| 255.8568634260 | 2.8 |
| 256.0097569445 | 2.6 |
| 256.8828587963 | 2.5 |
| 257.2936689815 | 2.9 |
| 257.3786342593 | 2.7 |
| 257.9106712963 | 2.2 |
| 257.9775231482 | 2.0 |
| 258.0351273148 | 1.8 |
| 258.9382175926 | 1.2 |
| 259.0307638889 | 1.6 |
| 259.1789583334 | 2.1 |
| 259.3015509259 | 2.8 |
| 259.9685416667 | 1.1 |
| 260.0224421297 | 2.6 |
| 260.9925578704 | 1.4 |
| 261.0571064815 | 2.3 |
| 261.3758912037 | 1.3 |
| 262.0198032408 | 1.6 |
| 262.1031481482 | 1.1 |
| 263.0476041667 | 2.7 |
| 263.0973726852 | 2.9 |
| 263.1469212963 | 1.3 |
| 263.1635532408 | 2.5 |
| 263.2059606482 | 1.4 |
| 263.4682638889 | 1.4 |
| 264.0719791667 | 1.8 |
| 264.0770949074 | 2.2 |
| 264.4996180556 | 1.6 |
| 265.1029629630 | 1.7 |
| 265.1172453704 | 1.5 |
| 265.2832986111 | 2.5 |
| 266.1250810185 | 2.5 |
| 266.2483680556 | 1.3 |
| 266.4854282408 | 1.8 |
| 266.5327662037 | 2.2 |
| 267.1576273148 | 2.1 |
| 267.3765277778 | 2.6 |
| 268.1911921297 | 2.4 |
| 269.1674421297 | 2.4 |
| 269.2143171297 | 1.7 |
| 269.4067939815 | 1.3 |
| 270.2079166667 | 2.5 |
| 270.2268981482 | 2.0 |
| 270.2412847222 | 2.7 |
| 270.3867824074 | 1.8 |
| 270.4008101852 | 2.3 |
| 271.2280902778 | 2.0 |
| 271.2703009259 | 2.9 |
| 272.2950578704 | 1.4 |
| 272.2970486111 | 1.6 |
| 272.2994675926 | 2.9 |
| 272.4292013889 | 1.6 |
| 273.3237152778 | 1.9 |
| 273.3705671297 | 1.6 |
| 273.4468055556 | 1.8 |
| 274.2956365741 | 2.8 |
| 274.3515162037 | 1.5 |
| 275.3522916667 | 2.1 |
| 275.3763194445 | 1.2 |
| 275.4043634260 | 1.7 |
| 275.7671180556 | 1.3 |
| 276.3842129630 | 2.0 |
| 276.4033333334 | 2.5 |
| 276.4033912037 | 2.6 |
| 276.4707175926 | 2.9 |
| 277.4289120371 | 1.0 |
| 278.4556597222 | 2.9 |
| 279.4850925926 | 2.9 |
| 279.5592129630 | 2.2 |
| 279.8934027778 | 2.8 |
| 280.4888541667 | 1.1 |
| 280.5151041667 | 1.0 |
| 280.9531597222 | 2.5 |
| 281.5393171297 | 1.1 |
| 281.5953587963 | 2.0 |
| 282.5672916667 | 2.9 |
| 282.5711805556 | 2.0 |
| 282.8396527778 | 1.6 |
| 283.5926273148 | 2.1 |
| 284.6202314815 | 1.4 |
| 284.6351273148 | 2.6 |
| 284.7545949074 | 1.8 |
| 284.7678703704 | 1.1 |
| 284.7911111111 | 2.5 |
| 284.9740625000 | 1.9 |
| 285.0169328704 | 2.8 |
| 285.0901273148 | 1.5 |
| 285.6392824074 | 1.0 |
| 285.6519328704 | 2.4 |
| 285.8464351852 | 1.5 |
| 285.9453472222 | 2.1 |
| 286.0418865741 | 2.7 |
| 286.6774305556 | 1.9 |
| 286.8105208334 | 1.5 |
| 286.8347106482 | 2.8 |
| 286.9171643519 | 1.3 |
| 287.7004398148 | 2.5 |
| 288.7273842593 | 2.1 |
| 288.7300810185 | 2.1 |
| 288.8205555556 | 1.0 |
| 288.9200925926 | 1.1 |
| 288.9335300926 | 2.1 |
| 289.7543865741 | 2.6 |
| 290.7831828704 | 2.1 |
| 290.9449537037 | 2.0 |
| 291.1843518519 | 2.2 |
| 291.8118171297 | 2.4 |
| 292.1119560185 | 2.5 |
| 292.8257638889 | 1.4 |
| 292.8360763889 | 1.1 |
| 293.0387268519 | 2.9 |
| 293.8299537037 | 2.6 |
| 293.8396759259 | 2.4 |
| 293.8660532408 | 2.2 |
| 294.8886921297 | 2.8 |
| 294.9896180556 | 1.3 |
| 295.9149652778 | 1.7 |
| 295.9559837963 | 1.8 |
| 295.9774074074 | 1.2 |
| 296.0070370371 | 1.0 |
| 296.9452430556 | 2.2 |
| 296.9565509259 | 1.8 |
| 297.0614004630 | 1.2 |
| 297.1319328704 | 2.5 |
| 297.9694444445 | 2.9 |
| 298.9986342593 | 1.0 |
| 299.0668634259 | 1.0 |
| 299.2338888889 | 1.5 |
| 299.2603703704 | 1.1 |
| 300.0075810185 | 1.1 |
| 300.0262500000 | 2.0 |
| 300.0582407408 | 2.3 |
| 301.0220370371 | 2.8 |
| 301.0348379630 | 2.0 |
| 301.0510879630 | 1.5 |
| 301.0594675926 | 1.9 |
| 301.0946412037 | 1.7 |
| 301.1404282408 | 2.1 |
| 301.1720486111 | 1.7 |
| 301.2170254630 | 1.9 |
| 302.0800462963 | 1.8 |
| 302.0853935185 | 2.5 |
| 302.1650462963 | 1.0 |
| 302.2172106482 | 1.0 |
| 303.0710995371 | 2.0 |
| 303.1096180556 | 2.2 |
| 303.2685069445 | 1.5 |
| 304.1213078704 | 2.6 |
| 304.1304050926 | 2.4 |
| 304.1726967593 | 2.9 |
| 304.2965162037 | 1.7 |
| 305.1238541667 | 1.9 |
| 305.1346064815 | 1.5 |
| 305.2062268519 | 2.8 |
| 305.2377662037 | 2.6 |
| 305.2498842593 | 2.4 |
| 306.1496990741 | 1.2 |
| 306.1595949074 | 2.0 |
| 306.1689467593 | 1.1 |
| 307.2490162037 | 1.1 |
| 307.2706944445 | 1.6 |
| 307.3361342593 | 1.7 |
| 307.3659143519 | 2.3 |
| 307.4049537037 | 2.9 |
| 308.2320717593 | 1.0 |
| 308.2344907408 | 1.4 |
| 308.2739236111 | 2.3 |
| 308.3552662037 | 1.9 |
| 308.3855671297 | 1.1 |
| 308.3946180556 | 1.3 |
| 309.3688310185 | 2.7 |
| 310.2709837963 | 1.4 |
| 310.3359143519 | 1.1 |
| 310.3694328704 | 2.0 |
| 310.4087962963 | 1.5 |
| 310.4901273148 | 2.9 |
| 310.7032060185 | 1.3 |
| 311.4672453704 | 2.1 |
| 312.3920023148 | 2.3 |
| 312.4164930556 | 2.8 |
| 313.4861805556 | 2.3 |
| 314.3701620371 | 1.0 |
| 314.4889930556 | 1.9 |
| 314.5123032408 | 1.5 |
| 314.5622106482 | 2.2 |
| 314.8505208334 | 2.0 |
| 315.3790972222 | 2.3 |
| 315.4439583334 | 1.4 |
| 315.4635995371 | 2.6 |
| 315.4907754630 | 1.4 |
| 315.5396990741 | 2.6 |
| 315.5853240741 | 1.4 |
| 315.6665162037 | 2.3 |
| 315.8088888889 | 1.6 |
| 316.4140972222 | 2.7 |
| 317.4501851852 | 1.3 |
| 317.5254745371 | 2.5 |
| 318.5457291667 | 2.6 |
| 319.4946990741 | 1.3 |
| 319.5601620371 | 2.9 |
| 319.5906712963 | 1.9 |
| 319.7652083334 | 2.4 |
| 320.6066666667 | 2.6 |
| 320.6600115741 | 2.0 |
| 320.7082060185 | 1.0 |
| 320.7848726852 | 1.2 |
| 321.5740972222 | 1.5 |
| 321.6329629630 | 1.8 |
| 322.5895833334 | 2.3 |
| 322.6718171297 | 1.1 |
| 322.7473148148 | 1.7 |
| 322.8920601852 | 1.3 |
| 323.5828703704 | 2.6 |
| 323.6244328704 | 2.9 |
| 323.6990509259 | 1.1 |
| 323.7195370371 | 2.4 |
| 323.7812268519 | 1.9 |
| 324.5793287037 | 1.3 |
| 324.6720254630 | 2.3 |
| 324.7257638889 | 1.5 |
| 324.7339467593 | 2.6 |
| 324.8050810185 | 2.3 |
| 325.7083564815 | 1.8 |
| 325.7581134260 | 2.0 |
| 325.7587500000 | 1.5 |
| 325.7857291667 | 2.7 |
| 325.8183217593 | 2.5 |
| 326.8693634260 | 2.3 |
| 326.9581944445 | 2.7 |
| 327.7417476852 | 1.8 |
| 327.7856597222 | 2.9 |
| 327.8228240741 | 1.3 |
| 327.9794560185 | 1.2 |
| 328.0013425926 | 2.2 |
| 328.7695949074 | 2.5 |
| 328.8254861111 | 2.5 |
| 328.8472800926 | 1.2 |
| 328.8593865741 | 2.6 |
| 328.9010416667 | 2.0 |
| 328.9861458334 | 1.1 |
| 329.7741087963 | 1.9 |
| 329.8241319445 | 2.7 |
| 329.8624189815 | 1.6 |
| 329.9076851852 | 1.3 |
| 329.9483217593 | 2.5 |
| 330.8604976852 | 2.3 |
| 330.9004513889 | 1.8 |
| 331.0486111111 | 2.5 |
| 331.8302893519 | 2.8 |
| 331.8384490741 | 2.2 |
| 331.8611111111 | 2.7 |
| 331.8901157408 | 2.1 |
| 331.9153240741 | 1.8 |
| 331.9314236111 | 1.7 |
| 332.0873148148 | 1.1 |
| 332.8915972222 | 2.3 |
| 332.9382638889 | 1.9 |
| 333.9283449074 | 2.3 |
| 333.9764004630 | 2.9 |
| 334.0112962963 | 1.3 |
| 334.0311226852 | 2.2 |
| 334.1052314815 | 2.3 |
| 334.9915972222 | 1.2 |
| 335.0159027778 | 2.2 |
| 335.0407870371 | 2.3 |
| 335.0682638889 | 2.4 |
| 336.0010995371 | 2.4 |
| 336.0493055556 | 2.3 |
| 336.0630555556 | 2.4 |
| 336.1820833334 | 2.3 |
| 337.0358680556 | 1.4 |
| 337.0885185185 | 2.5 |
| 337.0891435185 | 2.0 |
| 337.1165740741 | 2.1 |
| 337.1694328704 | 1.4 |
| 337.2405902778 | 2.6 |
| 338.1203356482 | 1.0 |
| 338.1251504630 | 2.0 |
| 338.1760532408 | 1.8 |
| 338.4424537037 | 2.2 |
| 339.0933333334 | 2.2 |
| 339.1468518519 | 2.9 |
| 339.2465509259 | 2.0 |
| 340.0521527778 | 2.3 |
| 340.1712384260 | 1.8 |
| 340.2056365741 | 1.4 |
| 340.2517824074 | 2.3 |
| 341.0802430556 | 1.9 |
| 341.1503587963 | 1.3 |
| 341.1962268519 | 2.4 |
| 341.3006597222 | 1.7 |
| 342.1914351852 | 2.0 |
| 342.2211689815 | 1.5 |
| 342.2706018519 | 1.9 |
| 342.3264351852 | 1.1 |
| 342.5210069445 | 1.1 |
| 343.2039120371 | 2.3 |
| 343.3932638889 | 2.1 |
| 343.5176967593 | 2.4 |
| 343.5387384260 | 1.3 |
| 344.1687847222 | 2.7 |
| 344.2057523148 | 1.3 |
| 344.2143865741 | 1.9 |
| 344.2400578704 | 1.1 |
| 344.2448032408 | 1.6 |
| 344.3151620371 | 2.6 |
| 344.3282754630 | 2.6 |
| 345.2076157408 | 1.2 |
| 345.2749074074 | 2.4 |
| 345.3017592593 | 2.7 |
| 345.3579745371 | 2.6 |
| 346.3027546297 | 1.0 |
| 346.3292592593 | 2.5 |
| 346.3600694445 | 1.2 |
| 346.3943287037 | 1.5 |
| 346.4073148148 | 1.0 |
| 347.2461921297 | 1.9 |
| 347.2987847222 | 1.7 |
| 347.3224189815 | 1.1 |
| 347.3400694445 | 2.8 |
| 347.3549189815 | 1.0 |
| 347.3571990741 | 2.0 |
| 347.3684606482 | 2.0 |
| 347.4478819445 | 2.3 |
| 347.6506134260 | 1.6 |
| 348.2788773148 | 2.9 |
| 348.3319907408 | 1.4 |
| 348.3336458334 | 2.6 |
| 348.3748726852 | 1.2 |
| 348.3823842593 | 1.1 |
| 348.3859143519 | 2.9 |
| 348.4187384259 | 2.7 |
| 349.3568981482 | 1.6 |
| 349.3722453704 | 1.1 |
| 349.3774884259 | 2.2 |
| 349.3968518519 | 2.8 |
| 349.4099074074 | 1.8 |
| 349.5042245371 | 2.2 |
| 349.7054976852 | 1.7 |
| 350.3605439815 | 2.9 |
| 350.4175578704 | 2.0 |
| 350.4319212963 | 2.9 |
| 351.4104745371 | 1.3 |
| 351.4332638889 | 1.3 |
| 351.4615046297 | 2.3 |
| 351.4985416667 | 1.5 |
| 351.5186689815 | 1.3 |
| 351.7556018519 | 1.5 |
| 352.4730671297 | 2.2 |
| 352.4822800926 | 2.4 |
| 352.4869907408 | 1.3 |
| 352.5795601852 | 1.8 |
| 353.4002662037 | 1.7 |
| 353.4242245371 | 2.8 |
| 353.4994328704 | 1.7 |
| 353.5103356482 | 1.5 |
| 353.5125115741 | 2.3 |
| 353.5183912037 | 1.9 |
| 353.5864467593 | 1.0 |
| 354.5300810185 | 1.3 |
| 354.5375810185 | 1.6 |
| 354.5385300926 | 2.0 |
| 354.5967476852 | 1.8 |
| 355.4756481482 | 2.7 |
| 355.5247106482 | 1.0 |
| 355.5461689815 | 2.1 |
| 355.5681597223 | 1.5 |
| 355.5721643519 | 1.2 |
| 356.5507407408 | 2.0 |
| 356.5545023148 | 2.2 |
| 356.5734259259 | 1.6 |
| 356.5928935185 | 1.6 |
| 356.6103125000 | 1.9 |
| 356.8019907408 | 2.2 |
| 357.5795717593 | 2.3 |
| 357.5964120371 | 2.2 |
| 357.6191898148 | 2.0 |
| 357.6212847222 | 2.8 |
| 358.6403125000 | 1.2 |
| 358.6438888889 | 1.5 |
| 358.6612037037 | 1.0 |
| 358.6844675926 | 2.1 |
| 358.7115740741 | 1.9 |
| 358.9892129630 | 2.1 |
| 359.6686342593 | 2.3 |
| 359.6747800926 | 2.6 |
| 359.7261226852 | 2.6 |
| 359.8254282408 | 1.2 |
| 360.6322685185 | 1.1 |
| 360.6575578704 | 1.6 |
| 360.6947569445 | 1.8 |
| 360.6953819445 | 1.0 |
| 360.6980092593 | 2.6 |
| 360.6984375000 | 2.1 |
| 360.7596527778 | 1.1 |
| 360.8221412037 | 1.4 |
| 361.6488425926 | 1.6 |
| 361.6881365741 | 1.8 |
| 361.7122222222 | 1.2 |
| 361.7322569445 | 2.0 |
| 361.7333101852 | 1.4 |
| 361.7482291667 | 2.0 |
| 361.8791087963 | 1.6 |
| 362.7273379630 | 1.6 |
| 362.7491203704 | 1.1 |
| 362.7597453704 | 1.6 |
| 362.7782291667 | 1.2 |
| 363.6823495371 | 1.8 |
| 363.7850578704 | 2.0 |
| 363.7867592593 | 2.0 |
| 363.7903125000 | 1.3 |
| 364.7650810185 | 2.7 |
| 364.7964351852 | 2.0 |
| 364.8054282408 | 2.5 |
| 364.8077893519 | 1.8 |
| 364.8155439815 | 2.8 |
| 364.8506828704 | 2.6 |
| 365.0886342593 | 1.5 |
| 365.8170486111 | 1.8 |
| 365.8376157408 | 2.9 |
| 365.8510648148 | 2.0 |
| 366.1158796297 | 2.9 |
| 366.1367708334 | 1.8 |
| 366.1802199074 | 1.2 |
| 366.8517361111 | 2.7 |
| 366.8536805556 | 2.8 |
| 366.8604398148 | 1.8 |
| 366.8631712963 | 1.8 |
| 367.1360300926 | 1.0 |
| 367.1579861111 | 2.8 |
| 367.1811111111 | 2.9 |
| 367.2034259260 | 1.4 |
| 367.8630902778 | 1.3 |
| 367.8890393519 | 1.9 |
| 367.8946412037 | 1.2 |
| 368.1892708334 | 1.9 |
| 368.1914699074 | 1.3 |
| 368.2138078704 | 1.8 |
| 368.2193750000 | 2.4 |
| 368.8216782408 | 1.6 |
| 368.8906712963 | 1.5 |
| 368.9104629630 | 1.5 |
| 368.9135648148 | 2.8 |
| 368.9147337963 | 1.7 |
| 368.9390277778 | 2.7 |
| 368.9922453704 | 2.3 |
| 369.1914120371 | 2.1 |
| 369.2125578704 | 2.6 |
| 369.2356597222 | 2.4 |
| 369.2579745371 | 2.2 |
| 369.8676851852 | 1.8 |
| 369.9202662037 | 1.9 |
| 369.9247800926 | 2.6 |
| 369.9391087963 | 1.1 |
| 369.9598842593 | 1.3 |
| 370.2202893519 | 2.0 |
| 370.2429513889 | 2.1 |
| 370.2670601852 | 1.1 |
| 370.2905555556 | 2.2 |
| 370.9275000000 | 2.4 |
| 370.9532986111 | 1.8 |
| 370.9561921297 | 1.7 |
| 370.9683796297 | 1.0 |
| 371.1668634259 | 1.6 |
| 371.2038194445 | 2.6 |
| 371.2684027778 | 2.8 |
| 371.3111805556 | 2.3 |
| 371.9678935185 | 1.4 |
| 371.9814236111 | 1.0 |
| 371.9975000000 | 2.5 |
| 371.9994097222 | 2.3 |
| 372.0043750000 | 2.6 |
| 372.9997453704 | 2.6 |
| 373.0076388889 | 1.4 |
| 373.0155092593 | 2.4 |
| 373.0244328704 | 1.7 |
| 373.0494907408 | 2.7 |
| 373.1855208334 | 2.0 |
| 373.2314467593 | 2.9 |
| 374.0376041667 | 1.1 |
| 374.0534143519 | 1.7 |
| 374.0988194445 | 1.8 |
| 374.9847106482 | 1.7 |
| 375.0214351852 | 1.2 |
| 375.0609837963 | 2.3 |
| 375.0659953704 | 1.5 |
| 375.0773842593 | 1.0 |
| 375.0900578704 | 1.3 |
| 375.3000115741 | 1.6 |
| 376.0852314815 | 1.6 |
| 376.0950462963 | 1.7 |
| 376.4247916667 | 1.1 |
| 376.4426273148 | 1.0 |
| 377.0816319445 | 2.3 |
| 377.1138194445 | 1.1 |
| 377.1245023148 | 1.3 |
| 377.1343750000 | 2.0 |
| 377.1529513889 | 2.7 |
| 377.3585648148 | 1.6 |
| 378.1472916667 | 2.5 |
| 378.1617013889 | 2.2 |
| 378.1665393519 | 1.0 |
| 378.2589699074 | 1.5 |
| 378.4392013889 | 2.2 |
| 379.1903240741 | 2.2 |
| 379.2028125000 | 2.8 |
| 379.2575462963 | 1.9 |
| 380.1920717593 | 1.1 |
| 380.1986458334 | 1.6 |
| 380.2113078704 | 2.6 |
| 380.2173263889 | 1.1 |
| 380.2659953704 | 1.4 |
| 380.4854629630 | 1.9 |
| 381.2108217593 | 2.0 |
| 381.2300000000 | 2.8 |
| 381.2339699074 | 1.9 |
| 381.2386574074 | 2.1 |
| 381.2459259259 | 1.9 |
| 382.2341782408 | 1.5 |
| 382.2462962963 | 1.5 |
| 382.2516319445 | 2.8 |
| 382.2570254630 | 2.6 |
| 382.2654629630 | 1.2 |
| 382.3309259259 | 1.7 |
| 383.2701388889 | 2.4 |
| 383.2840046297 | 2.6 |
| 383.2915277778 | 2.9 |
| 383.3085532408 | 1.3 |
| 383.3941087963 | 2.9 |
| 383.5113541667 | 1.2 |
| 383.5823032408 | 2.3 |
| 384.3014467593 | 2.2 |
| 384.3067592593 | 1.3 |
| 384.3194444445 | 1.3 |
| 384.3310300926 | 1.8 |
| 384.3452430556 | 2.6 |
| 384.3909606482 | 2.6 |
| 384.5307870371 | 2.3 |
| 385.3419212963 | 1.8 |
| 385.3560069445 | 1.3 |
| 385.3595254630 | 1.7 |
| 385.3872222222 | 2.8 |
| 385.4151273148 | 1.4 |
| 385.4504282408 | 1.6 |
| 385.4829398148 | 2.0 |
| 386.2805208334 | 2.4 |
| 386.3767476852 | 1.1 |
| 386.3889236111 | 1.7 |
| 386.6318750000 | 2.9 |
| 387.3801273148 | 1.6 |
| 387.4031828704 | 1.8 |
| 387.4381944445 | 2.7 |
| 387.4457754630 | 2.8 |
| 387.5621643519 | 1.5 |
| 387.6125925926 | 2.2 |
| 388.3769907408 | 1.4 |
| 388.3921643519 | 1.8 |
| 388.4242708334 | 1.7 |
| 388.4479282408 | 1.5 |
| 388.4679166667 | 2.5 |
| 388.4694560185 | 2.3 |
| 388.4781944445 | 2.0 |
| 389.4468634260 | 1.8 |
| 389.4995717593 | 1.0 |
| 390.4270601852 | 2.1 |
| 390.4758680556 | 2.0 |
| 391.4988425926 | 1.8 |
| 391.5053703704 | 1.7 |
| 391.5118518519 | 1.9 |
| 391.6910879630 | 2.0 |
| 391.7218981482 | 1.6 |
| 391.7840856482 | 1.6 |
| 392.4257060185 | 1.3 |
| 392.4450925926 | 2.0 |
| 392.5291898148 | 2.0 |
| 392.5299074074 | 1.6 |
| 392.5322453704 | 2.7 |
| 392.5510069445 | 2.2 |
| 392.5617129630 | 2.7 |
| 392.6041087963 | 2.4 |
| 393.5414351852 | 2.5 |
| 393.5440740741 | 2.8 |
| 393.5544560185 | 2.3 |
| 393.5832523148 | 1.2 |
| 393.6124189815 | 1.5 |
| 393.7378472222 | 1.0 |
| 393.7798148148 | 1.6 |
| 393.8584606482 | 1.5 |
| 394.5493518519 | 2.6 |
| 394.5508101852 | 1.9 |
| 394.5526504630 | 1.9 |
| 394.5666550926 | 2.5 |
| 394.5791087963 | 1.6 |
| 394.6153819445 | 2.8 |
| 394.6254629630 | 2.0 |
| 394.6459837963 | 2.7 |
| 394.6764120371 | 2.1 |
| 395.5823958334 | 2.5 |
| 395.6068287037 | 1.0 |
| 395.6115625000 | 1.3 |
| 395.6400231482 | 2.7 |
| 395.6503587963 | 2.3 |
| 395.6673263889 | 1.0 |
| 395.7717592593 | 1.1 |
| 396.6036458334 | 2.6 |
| 396.6048032408 | 1.9 |
| 396.6328819445 | 1.1 |
| 396.6784837963 | 2.1 |
| 396.7877893519 | 2.3 |
| 396.8288888889 | 2.4 |
| 397.5566087963 | 2.8 |
| 397.5731944445 | 1.9 |
| 397.6310995371 | 1.9 |
| 397.6322800926 | 1.0 |
| 397.6623726852 | 2.3 |
| 397.6631481482 | 2.5 |
| 397.7083101852 | 1.6 |
| 398.6540625000 | 2.8 |
| 398.6644675926 | 1.1 |
| 398.6889930556 | 2.9 |
| 398.7442245371 | 2.7 |
| 399.6348032408 | 2.1 |
| 399.6826041667 | 2.0 |
| 399.6865509260 | 2.0 |
| 399.7160648148 | 2.6 |
| 399.7165509260 | 2.9 |
| 399.7523495371 | 1.6 |
| 399.7616435185 | 1.0 |
| 399.7639814815 | 2.9 |
| 400.0217245371 | 2.8 |
| 400.6804513889 | 2.4 |
| 400.7120138889 | 1.5 |
| 400.7176157408 | 1.2 |
| 400.7474305556 | 2.2 |
| 400.7695949074 | 2.8 |
| 400.7939930556 | 1.7 |
| 400.8373958334 | 1.1 |
| 401.6511805556 | 1.8 |
| 401.6695370371 | 1.4 |
| 401.7461226852 | 2.0 |
| 401.7639120371 | 2.6 |
| 401.7678587963 | 1.8 |
| 401.7699074074 | 1.1 |
| 401.8028240741 | 2.8 |
| 402.6865625000 | 1.8 |
| 402.7772453704 | 2.6 |
| 402.7837268519 | 2.2 |
| 402.7960995371 | 1.3 |
| 402.8179861111 | 2.4 |
| 403.1112847222 | 2.4 |
| 403.8043981482 | 1.2 |
| 403.8223032408 | 2.5 |
| 403.8258217593 | 1.4 |
| 404.8382175926 | 2.9 |
| 404.8431365741 | 2.9 |
| 404.8487500000 | 1.9 |
| 404.8647916667 | 2.4 |
| 404.8807407408 | 1.1 |
| 405.8663888889 | 2.1 |
| 405.8757407408 | 2.1 |
| 405.9030902778 | 1.8 |
| 405.9482060185 | 2.6 |
| 405.9676967593 | 2.5 |
| 406.0047569445 | 2.0 |
| 406.8756134260 | 1.0 |
| 406.9043981482 | 1.3 |
| 406.9060763889 | 2.8 |
| 406.9154976852 | 2.1 |
| 406.9162731482 | 2.5 |
| 406.9456018519 | 2.0 |
| 407.1795023148 | 2.5 |
| 407.8738310185 | 1.3 |
| 407.9032754630 | 2.2 |
| 407.9319560185 | 1.5 |
| 407.9457291667 | 1.0 |
| 407.9720138889 | 2.1 |
| 408.9279282408 | 1.8 |
| 408.9574537037 | 1.8 |
| 408.9833333334 | 2.0 |
| 409.0003472222 | 2.7 |
| 409.9660648148 | 1.1 |
| 410.0323032408 | 2.8 |
| 410.2417245371 | 2.5 |
| 410.2959143519 | 1.1 |
| 410.9844328704 | 2.2 |
| 411.0508333334 | 2.9 |
| 411.0526967593 | 1.6 |
| 412.0487962963 | 2.2 |
| 412.0967939815 | 2.2 |
| 413.0431250000 | 1.6 |
| 413.0542013889 | 2.3 |
| 413.1078935185 | 2.6 |
| 413.2711921297 | 2.9 |
| 414.0801504630 | 2.7 |
| 414.0909259260 | 1.0 |
| 414.1134490741 | 1.6 |
| 414.1453472222 | 1.4 |
| 414.1670949074 | 1.1 |
| 415.0963773148 | 2.2 |
| 415.1207291667 | 2.5 |
| 415.1599768519 | 1.4 |
| 415.2035532408 | 1.1 |
| 416.1716435185 | 2.4 |
| 416.1887152778 | 2.1 |
| 417.2065162037 | 2.4 |
| 417.2698726852 | 2.7 |
| 418.2001851852 | 1.9 |
| 418.2226273148 | 1.3 |
| 418.2312384259 | 2.9 |
| 418.2357060185 | 2.0 |
| 418.2569675926 | 2.7 |
| 418.3570717593 | 1.7 |
| 419.2159606482 | 1.5 |
| 419.2460648148 | 1.3 |
| 419.2599189815 | 2.0 |
| 419.2905671297 | 1.8 |
| 420.2752083334 | 1.6 |
| 421.2891087963 | 2.5 |
| 421.2916782408 | 2.4 |
| 421.3314699074 | 1.4 |
| 422.3120370371 | 2.3 |
| 422.3206712963 | 2.0 |
| 423.3389699074 | 1.0 |
| 423.3517476852 | 2.2 |
| 423.3640277778 | 1.8 |
| 423.3953703704 | 2.8 |
| 423.4967361111 | 2.9 |
| 423.5171527778 | 2.8 |
| 424.3629861111 | 2.9 |
| 424.4023495371 | 1.7 |
| 424.4181134259 | 2.4 |
| 424.4428356482 | 2.8 |
| 424.6191087963 | 2.1 |
| 425.2770023148 | 1.2 |
| 425.3875000000 | 1.4 |
| 425.4493865741 | 1.0 |
| 426.3332754630 | 1.1 |
| 426.3927546297 | 2.9 |
| 426.4107407408 | 1.8 |
| 426.4174884259 | 2.6 |
| 426.4232638889 | 2.7 |
| 426.4415625000 | 1.2 |
| 427.4252314815 | 1.5 |
| 427.4467013889 | 2.5 |
| 427.4519791667 | 2.2 |
| 427.5084259260 | 1.0 |
| 427.5321296297 | 2.2 |
| 428.4722337963 | 2.8 |
| 429.5372453704 | 1.5 |
| 430.5553009260 | 2.2 |
| 431.6188425926 | 2.8 |
| 431.6264814815 | 2.7 |
| 431.6424421297 | 1.7 |
| 432.5094791667 | 2.7 |
| 432.5440972223 | 1.3 |
| 432.5672222222 | 2.1 |
| 432.6268865741 | 2.4 |
| 432.6592361111 | 1.2 |
| 433.5996064815 | 1.2 |
| 433.6101273148 | 2.4 |
| 433.6610185185 | 1.3 |
| 433.6935995371 | 2.9 |
| 434.6166319445 | 2.2 |
| 434.6274305556 | 1.2 |
| 435.6018171297 | 2.8 |
| 435.6330787037 | 1.4 |
| 435.6845601852 | 1.2 |
| 436.6754629630 | 2.1 |
| 436.6769328704 | 2.3 |
| 436.6963541667 | 1.7 |
| 436.7435995371 | 1.0 |
| 436.9488888889 | 1.9 |
| 437.7166087963 | 2.6 |
| 437.7624189815 | 2.2 |
| 438.7134722222 | 1.3 |
| 438.7355787037 | 2.2 |
| 438.7949652778 | 2.1 |
| 439.7554745371 | 2.8 |
| 439.7771180556 | 2.5 |
| 439.8031828704 | 1.3 |
| 439.8923379630 | 1.9 |
| 439.9727199074 | 1.5 |
| 440.7736226852 | 2.9 |
| 440.7806597222 | 1.1 |
| 440.8268981482 | 2.4 |
| 441.8572685185 | 2.4 |
| 441.9578240741 | 2.4 |
| 442.7838310185 | 1.8 |
| 442.8256597223 | 1.2 |
| 443.8836689815 | 2.1 |
| 444.9148379630 | 2.3 |
| 444.9329166667 | 2.5 |
| 444.9563194445 | 2.7 |
| 445.7850462963 | 1.8 |
| 445.8998495371 | 2.3 |
| 445.9105208334 | 1.4 |
| 445.9362731482 | 2.6 |
| 445.9665393519 | 2.7 |
| 446.8723611111 | 1.7 |
| 446.9618750000 | 2.1 |
| 447.8666319445 | 1.0 |
| 447.9921990741 | 1.3 |
| 447.9934606482 | 1.8 |
| 448.0213773148 | 2.9 |
| 448.9954745371 | 1.5 |
| 449.0483912037 | 2.2 |
| 450.0296064815 | 1.2 |
| 450.0310069445 | 1.7 |
| 450.0892592593 | 1.2 |
| 450.1212384259 | 1.5 |
| 453.0370949074 | 1.7 |
| 453.1531597222 | 2.0 |
| 455.2187731482 | 1.8 |
| 456.1520717593 | 2.9 |
| 456.2243865741 | 2.9 |
| 457.2699189815 | 2.4 |
| 458.2785648148 | 1.9 |
| 458.2834722222 | 1.0 |
| 459.2883680556 | 1.5 |
| 459.3085879630 | 1.5 |
| 459.4291782408 | 2.1 |
| 460.2746412037 | 1.8 |
| 460.3188078704 | 1.0 |
| 461.3439467593 | 2.6 |
| 461.3464930556 | 1.8 |
| 463.2973611111 | 1.2 |
| 464.4516898148 | 2.8 |
| 465.3691666667 | 1.7 |
| 466.3985763889 | 2.8 |
| 466.5746412037 | 2.5 |
| 466.6987152778 | 2.9 |
| 470.6263773148 | 2.4 |
| 473.7106712963 | 1.5 |
| 473.7204629630 | 1.7 |
| 473.7705439815 | 2.2 |
| 473.8473958334 | 1.3 |
| 473.9383449074 | 2.0 |
| 476.9326851852 | 2.8 |
| 479.8884837963 | 1.3 |
| 484.2557870371 | 2.2 |
| 500.3285532408 | 2.2 |
| 501.3070023148 | 1.0 |
| 509.2997916667 | 2.8 |
| 529.9758564815 | 1.9 |
| 536.2896412037 | 1.1 |
| 537.3024305556 | 1.8 |
| 623.7226273148 | 1.2 |
| 631.4461574074 | 1.6 |
| 643.2324652778 | 1.7 |
| 683.2887847222 | 2.6 |
| 684.1301736111 | 2.5 |
| 685.2605208334 | 2.0 |
| 701.5182291667 | 2.7 |
| 701.6160879630 | 2.4 |
| 706.1793634260 | 1.9 |
| 721.3465162037 | 1.8 |
| 722.3220486111 | 1.1 |
| 728.8012847222 | 2.4 |
| 731.7663425926 | 2.8 |
| 737.7575925926 | 1.4 |
| 740.7993865741 | 1.9 |
| 755.1982407408 | 1.4 |
| 755.2012152778 | 2.9 |
| 758.2623495371 | 2.5 |
| 758.2807175926 | 1.2 |
| 760.5973032408 | 1.2 |
| 766.5989236111 | 1.7 |
| 767.5708101852 | 2.7 |
| 767.8112962963 | 1.1 |
| 769.5890509259 | 2.0 |
| 770.6089699074 | 2.1 |
| 773.8520370371 | 1.0 |
| 774.7530902778 | 1.5 |
| 776.6967476852 | 2.6 |
| 776.9191087963 | 2.3 |
| 777.0094444445 | 1.6 |
| 784.2192245371 | 2.5 |
| 786.1136458334 | 2.8 |
| 790.2035416667 | 2.5 |
| 790.2406944445 | 1.0 |
| 791.0834143519 | 2.0 |
| 791.4944791667 | 1.4 |
| 791.5181018519 | 2.3 |
| 792.4350115741 | 1.6 |
| 793.3515046297 | 1.6 |
| 795.1560532408 | 2.5 |
| 795.3734722223 | 2.1 |
| 795.3780902778 | 2.5 |
| 798.3759027778 | 2.6 |
| 798.4440162037 | 2.2 |
| 799.4074652778 | 2.0 |
| 799.4791087963 | 2.7 |
| 799.7599884260 | 2.4 |
| 800.4330787037 | 1.4 |
| 801.5500694445 | 2.5 |
| 801.5791782408 | 2.4 |
| 802.4894675926 | 2.1 |
| 802.5084490741 | 1.3 |
| 803.5160995371 | 2.0 |
| 803.6366666667 | 2.2 |
| 803.9927199074 | 2.9 |
| 805.5687962963 | 2.9 |
| 805.7162731482 | 1.7 |
| 806.5899074074 | 1.5 |
| 806.7333680556 | 1.9 |
| 806.7542592593 | 2.6 |
| 807.6221412037 | 2.9 |
| 807.7099074074 | 1.2 |
| 807.7358912037 | 2.9 |
| 808.6433449074 | 2.7 |
| 808.7155555556 | 1.5 |
| 809.6646875000 | 1.6 |
| 809.7408912037 | 2.8 |
| 809.7972916667 | 1.8 |
| 810.6871643519 | 2.3 |
| 810.6965393519 | 2.3 |
| 810.7038541667 | 1.5 |
| 810.8360416667 | 2.7 |
| 811.7263425926 | 1.6 |
| 811.7803009260 | 2.7 |
| 811.7933912037 | 1.0 |
| 812.7565856482 | 1.1 |
| 813.7818402778 | 2.3 |
| 814.8056481482 | 1.6 |
| 815.2136805556 | 2.3 |
| 815.8308449074 | 2.9 |
| 816.6960879630 | 1.5 |
| 817.2614699074 | 1.7 |
| 817.2766087963 | 1.6 |
| 819.0640972222 | 2.7 |
| 819.9340393519 | 2.7 |
| 819.9439004630 | 2.9 |
| 825.5522106482 | 2.8 |
| 826.1441666667 | 2.3 |
| 826.2237500000 | 1.2 |
| 826.4929861111 | 1.2 |
| 828.1475578704 | 2.3 |
| 829.1920601852 | 1.3 |
| 829.3245833334 | 2.5 |
| 829.5950462963 | 1.4 |
| 830.2267939815 | 1.5 |
| 831.2506481482 | 2.2 |
| 832.2769444445 | 2.8 |
| 833.2902314815 | 2.4 |
| 833.3004976852 | 1.7 |
| 833.3272916667 | 2.6 |
| 834.5044444445 | 1.1 |
| 836.3450694445 | 2.5 |
| 837.5233796297 | 1.4 |
| 838.4402777778 | 2.1 |
| 839.7943287037 | 1.1 |
| 839.8108796297 | 1.5 |
| 840.5382175926 | 1.2 |
| 840.6222453704 | 1.5 |
| 841.5190625000 | 2.0 |
| 841.6071990741 | 1.1 |
| 841.6535995371 | 1.5 |
| 842.5968518519 | 2.8 |
| 842.7115393519 | 2.3 |
| 843.6318055556 | 2.0 |
| 843.6714583334 | 2.0 |
| 844.6064004630 | 2.0 |
| 844.6292476852 | 1.1 |
| 844.6578935185 | 1.8 |
| 844.6662152778 | 1.6 |
| 844.7794097223 | 2.0 |
| 845.6100925926 | 1.2 |
| 845.6260069445 | 2.7 |
| 845.6744675926 | 2.5 |
| 845.8064236111 | 1.7 |
| 845.8593402778 | 2.8 |
| 846.6750000000 | 2.4 |
| 846.8254513889 | 2.9 |
| 847.6775462963 | 1.3 |
| 847.7689930556 | 1.0 |
| 847.9006250000 | 1.2 |
| 848.7155208334 | 1.9 |
| 848.8860416667 | 2.3 |
| 849.8606828704 | 2.0 |
| 850.7625231482 | 1.1 |
| 851.0527893519 | 1.0 |
| 851.7945370371 | 1.5 |
| 851.9133564815 | 1.8 |
| 851.9497106482 | 1.0 |
| 852.8200231482 | 2.1 |
| 852.9780439815 | 1.6 |
| 853.0013773148 | 1.5 |
| 853.8470833334 | 1.6 |
| 853.9781481482 | 2.2 |
| 854.8708564815 | 2.7 |
| 855.0183564815 | 2.9 |
| 856.0503935185 | 1.8 |
| 856.1408217593 | 1.6 |
| 857.1117939815 | 2.4 |
| 857.9576851852 | 1.6 |
| 858.0751851852 | 2.4 |
| 858.1384143519 | 2.6 |
| 858.9843865741 | 1.6 |
| 859.0156481482 | 1.5 |
| 859.1148958334 | 2.2 |
| 860.0057870371 | 1.1 |
| 860.0117129630 | 2.6 |
| 860.1021296297 | 1.0 |
| 860.1212384259 | 1.0 |
| 860.1467476852 | 2.0 |
| 862.0690625000 | 2.5 |
| 862.1091087963 | 2.8 |
| 863.1909027778 | 1.4 |
| 863.2035069445 | 1.8 |
| 863.2187615741 | 1.5 |
| 864.1250578704 | 2.2 |
| 864.2159027778 | 1.0 |
| 864.2469444445 | 2.6 |
| 865.1512152778 | 2.1 |
| 865.2532870371 | 1.4 |
| 865.3374652778 | 1.5 |
| 866.1817592593 | 2.9 |
| 866.2805324074 | 1.4 |
| 866.3205324074 | 2.8 |
| 866.3708796297 | 2.1 |
| 867.1896180556 | 2.8 |
| 867.2105208334 | 2.6 |
| 867.3651736111 | 1.7 |
| 868.2430555556 | 2.1 |
| 868.4037731482 | 1.9 |
| 869.2728472222 | 1.2 |
| 870.2642592593 | 1.8 |
| 870.2902546297 | 2.1 |
| 870.2913425926 | 2.0 |
| 870.4746064815 | 2.1 |
| 871.3224074074 | 1.1 |
| 871.3423032408 | 2.4 |
| 873.3795023148 | 1.1 |
| 873.4460648148 | 1.2 |
| 873.5461226852 | 2.9 |
| 874.4042824074 | 2.1 |
| 875.4319560185 | 1.8 |
| 875.6102893519 | 1.4 |
| 876.4593287037 | 1.7 |
| 876.5021875000 | 1.7 |
| 876.5204398148 | 2.3 |
| 876.5440740741 | 1.3 |
| 876.5929861111 | 1.9 |
| 877.4887500000 | 2.8 |
| 877.4930902778 | 2.8 |
| 877.5343634259 | 2.8 |
| 878.5088657408 | 2.1 |
| 878.5092476852 | 1.7 |
| 878.6633101852 | 2.2 |
| 879.6302546297 | 2.9 |
| 879.6565740741 | 2.9 |
| 879.7030439815 | 2.3 |
| 880.5681828704 | 2.9 |
| 880.6057175926 | 1.2 |
| 881.5960300926 | 2.0 |
| 881.6914467593 | 2.5 |
| 882.6265509260 | 2.2 |
| 882.6725462963 | 1.6 |
| 882.7067245371 | 1.1 |
| 882.7278935185 | 2.7 |
| 882.7498148148 | 1.4 |
| 882.7700115741 | 2.9 |
| 883.6479861111 | 2.5 |
| 883.7161458334 | 1.1 |
| 883.7605671297 | 1.6 |
| 883.7845254630 | 2.4 |
| 884.6721064815 | 2.7 |
| 884.6804629630 | 2.3 |
| 884.7987731482 | 2.9 |
| 885.7018402778 | 1.7 |
| 885.8161805556 | 1.4 |
| 885.8873379630 | 2.4 |
| 886.1800000000 | 1.6 |
| 886.7011574074 | 1.2 |
| 886.7350578704 | 2.6 |
| 886.7551736111 | 1.0 |
| 886.7774884259 | 1.4 |
| 886.8352893519 | 1.1 |
| 887.7817824074 | 2.4 |
| 887.8224074074 | 1.5 |
| 887.8345486111 | 2.4 |
| 887.9323263889 | 1.2 |
| 888.7830092593 | 1.0 |
| 888.8462268519 | 1.0 |
| 888.9593750000 | 1.3 |
| 889.8165972222 | 2.6 |
| 889.9760763889 | 1.1 |
| 890.0041898148 | 2.8 |
| 890.8793634259 | 2.8 |
| 891.8472222222 | 1.4 |
| 891.8756365741 | 2.8 |
| 891.8879282408 | 2.3 |
| 891.9071875000 | 2.5 |
| 891.9439814815 | 2.8 |
| 891.9965625000 | 2.7 |
| 892.0399305556 | 1.5 |
| 892.0518634260 | 1.8 |
| 892.8635069445 | 2.0 |
| 892.9127314815 | 1.1 |
| 892.9553240741 | 1.4 |
| 893.8764699074 | 1.5 |
| 893.8810069445 | 1.2 |
| 893.8977546297 | 2.9 |
| 893.9252546297 | 1.0 |
| 893.9581944445 | 2.1 |
| 893.9721875000 | 1.8 |
| 894.0704398148 | 2.8 |
| 894.9492476852 | 1.2 |
| 894.9590625000 | 1.3 |
| 895.0636226852 | 1.3 |
| 895.9848379630 | 1.7 |
| 895.9857291667 | 1.7 |
| 896.0002777778 | 2.4 |
| 896.1048032408 | 1.9 |
| 897.0063541667 | 1.6 |
| 897.0869907408 | 1.5 |
| 897.1177546297 | 2.5 |
| 897.1824768519 | 1.5 |
| 898.0374421297 | 2.2 |
| 898.1510069445 | 1.5 |
| 898.1718981482 | 1.9 |
| 898.1987152778 | 2.3 |
| 898.2302199074 | 1.7 |
| 899.0387847222 | 1.7 |
| 899.0448611111 | 1.6 |
| 899.0632523148 | 1.4 |
| 899.0678819445 | 2.2 |
| 899.1332175926 | 2.9 |
| 899.1828935185 | 2.7 |
| 899.1987615741 | 1.7 |
| 899.2065277778 | 2.5 |
| 900.0847685185 | 1.9 |
| 900.1208101852 | 1.1 |
| 900.1384722223 | 1.8 |
| 900.1523148148 | 1.6 |
| 900.2306597222 | 2.6 |
| 901.0911921297 | 2.4 |
| 901.1103125000 | 1.5 |
| 901.1151504630 | 2.7 |
| 901.2182407408 | 1.5 |
| 901.2602546297 | 2.2 |
| 901.2776504630 | 2.3 |
| 901.2995254630 | 2.5 |
| 902.1472569445 | 2.8 |
| 902.1814467593 | 2.8 |
| 902.2597337963 | 1.3 |
| 902.5756481482 | 1.1 |
| 903.3045717593 | 1.3 |
| 903.3299537037 | 2.3 |
| 903.6525578704 | 1.8 |
| 903.6717129630 | 2.0 |
| 904.2036226852 | 1.3 |
| 904.2107638889 | 1.6 |
| 904.2925347222 | 2.5 |
| 904.3570833334 | 2.0 |
| 904.6131597222 | 1.7 |
| 905.2498958334 | 2.7 |
| 905.3025462963 | 1.3 |
| 905.3862384259 | 1.5 |
| 905.4102546297 | 1.0 |
| 906.2917708334 | 1.4 |
| 906.4107407408 | 1.5 |
| 906.4498726852 | 2.4 |
| 907.3215162037 | 2.4 |
| 907.3891087963 | 1.3 |
| 907.4379282408 | 1.2 |
| 907.4639236111 | 2.0 |
| 908.3086574074 | 2.3 |
| 908.3118518519 | 1.5 |
| 908.3474768519 | 2.2 |
| 908.4076967593 | 2.0 |
| 908.4471527778 | 2.4 |
| 908.4713773148 | 2.4 |
| 909.3268402778 | 2.2 |
| 909.3427314815 | 1.9 |
| 909.4428009260 | 1.3 |
| 909.7114699074 | 2.5 |
| 910.4638541667 | 2.9 |
| 910.4795949074 | 2.8 |
| 911.3871527778 | 2.5 |
| 911.4666666667 | 2.1 |
| 911.5725000000 | 1.8 |
| 911.7217476852 | 1.8 |
| 912.4251273148 | 2.2 |
| 912.5915972223 | 1.7 |
| 913.4051736111 | 2.8 |
| 913.4147685185 | 2.3 |
| 913.4473032408 | 2.8 |
| 913.4545254630 | 1.8 |
| 913.4741203704 | 2.2 |
| 913.5396643519 | 1.4 |
| 913.6065393519 | 2.7 |
| 914.4754282408 | 1.1 |
| 914.5694212963 | 2.4 |
| 914.5712384260 | 2.6 |
| 915.5032523148 | 1.0 |
| 915.5165046297 | 2.4 |
| 915.6184953704 | 1.6 |
| 915.6298958334 | 1.1 |
| 915.8059606482 | 2.5 |
| 916.6010532408 | 1.3 |
| 916.7882523148 | 1.6 |
| 917.5742013889 | 1.2 |
| 917.6188425926 | 1.5 |
| 918.5879745371 | 2.1 |
| 918.6576157408 | 2.9 |
| 918.7166203704 | 1.4 |
| 918.7680555556 | 2.3 |
| 918.7906018519 | 2.1 |
| 919.6118981482 | 2.6 |
| 920.1052662037 | 2.5 |
| 920.5977777778 | 1.4 |
| 920.6419907408 | 1.8 |
| 920.6750578704 | 1.8 |
| 920.7490856482 | 1.1 |
| 921.1039930556 | 2.4 |
| 921.1363078704 | 2.3 |
| 921.6364236111 | 2.2 |
| 921.6567129630 | 1.5 |
| 921.6638657408 | 1.2 |
| 921.7077893519 | 2.2 |
| 921.7786342593 | 1.4 |
| 921.8152546297 | 1.7 |
| 922.6969097222 | 1.6 |
| 922.6979166667 | 2.8 |
| 922.7259027778 | 2.5 |
| 922.8273379630 | 1.1 |
| 923.7252430556 | 2.0 |
| 923.7602314815 | 2.5 |
| 923.8994328704 | 1.9 |
| 924.7487615741 | 2.4 |
| 924.7540277778 | 2.6 |
| 924.7988425926 | 1.2 |
| 924.8895254630 | 2.6 |
| 925.7537268519 | 1.4 |
| 925.9245254630 | 2.9 |
| 926.7737152778 | 1.2 |
| 926.8037615741 | 2.1 |
| 926.8361111111 | 2.9 |
| 926.9674652778 | 2.2 |
| 926.9871875000 | 1.9 |
| 927.2947916667 | 2.1 |
| 927.7944560185 | 2.5 |
| 927.8300231482 | 2.5 |
| 927.8540856482 | 2.3 |
| 928.2651388889 | 1.4 |
| 928.8618171297 | 2.1 |
| 928.8659606482 | 1.1 |
| 928.8669097222 | 2.2 |
| 928.9272800926 | 2.6 |
| 929.3232291667 | 2.0 |
| 929.8841087963 | 1.1 |
| 929.8851388889 | 2.7 |
| 929.9813425926 | 1.3 |
| 930.0471527778 | 2.4 |
| 930.1397222222 | 1.1 |
| 930.9144328704 | 2.4 |
| 930.9263888889 | 2.7 |
| 931.0025810185 | 2.6 |
| 931.0780208334 | 1.2 |
| 931.3194675926 | 1.4 |
| 931.9270370371 | 1.5 |
| 931.9589236111 | 1.7 |
| 932.9716782408 | 1.1 |
| 933.0304166667 | 2.7 |
| 933.0763541667 | 2.3 |
| 933.1018750000 | 1.9 |
| 933.4308912037 | 1.2 |
| 933.9749189815 | 2.8 |
| 933.9932175926 | 2.4 |
| 933.9943055556 | 2.7 |
| 934.1745949074 | 1.4 |
| 934.1977777778 | 2.6 |
| 935.0037268519 | 1.5 |
| 935.0238657408 | 1.4 |
| 935.0469097222 | 2.5 |
| 935.1158101852 | 2.8 |
| 935.1399768519 | 1.4 |
| 935.1793981482 | 2.8 |
| 935.1983912037 | 2.8 |
| 936.0191666667 | 1.8 |
| 936.0218055556 | 1.6 |
| 936.0536342593 | 1.4 |
| 936.1376041667 | 2.6 |
| 936.2365277778 | 2.6 |
| 936.4127777778 | 1.9 |
| 936.5142013889 | 2.5 |
| 937.0810069445 | 2.6 |
| 937.1629629630 | 1.0 |
| 937.1788310185 | 1.8 |
| 937.2141319445 | 1.1 |
| 938.0671990741 | 2.1 |
| 938.0694675926 | 2.7 |
| 938.1317361111 | 2.0 |
| 938.2134953704 | 2.2 |
| 939.1095949074 | 1.4 |
| 939.1355902778 | 2.8 |
| 939.1641898148 | 1.6 |
| 939.1758101852 | 1.3 |
| 939.2654166667 | 2.2 |
| 940.1655902778 | 2.6 |
| 940.1896990741 | 2.5 |
| 940.2928587963 | 2.9 |
| 941.1928819445 | 2.0 |
| 941.2578703704 | 2.1 |
| 941.3366203704 | 2.2 |
| 942.2165972223 | 2.2 |
| 942.2226273148 | 2.8 |
| 942.2440740741 | 1.7 |
| 942.3937152778 | 1.2 |
| 943.1932638889 | 2.7 |
| 943.2268750000 | 1.6 |
| 944.2265277778 | 2.1 |
| 944.3138888889 | 2.3 |
| 945.2896527778 | 1.8 |
| 945.3008680556 | 2.8 |
| 945.4888888889 | 2.9 |
| 946.3285763889 | 1.6 |
| 946.3340856482 | 2.9 |
| 946.5212500000 | 2.6 |
| 946.6896527778 | 2.2 |
| 947.3483680556 | 1.0 |
| 947.3528819445 | 2.6 |
| 947.4410069445 | 1.6 |
| 947.4835300926 | 1.1 |
| 947.8265509260 | 1.2 |
| 948.4306597222 | 1.9 |
| 948.5792476852 | 1.9 |
| 948.8546296297 | 1.4 |
| 949.4085185185 | 2.1 |
| 949.8640277778 | 2.3 |
| 950.4384722222 | 1.2 |
| 950.6017824074 | 2.0 |
| 951.4558680556 | 1.5 |
| 951.4647800926 | 1.6 |
| 951.5171527778 | 1.0 |
| 951.5825000000 | 1.3 |
| 952.5233680556 | 1.3 |
| 952.5873032408 | 2.4 |
| 952.6370370371 | 1.3 |
| 952.6695601852 | 2.1 |
| 953.4711226852 | 2.8 |
| 953.4910185185 | 2.0 |
| 953.5151157408 | 2.6 |
| 953.5315972222 | 1.9 |
| 953.6450578704 | 2.5 |
| 953.9534953704 | 2.1 |
| 954.5348842593 | 1.0 |
| 954.5414583334 | 2.2 |
| 954.6437152778 | 1.3 |
| 954.6960763889 | 1.7 |
| 954.9017939815 | 2.4 |
| 955.4567708334 | 1.1 |
| 955.5696064815 | 1.3 |
| 955.5897800926 | 1.6 |
| 955.6755555556 | 1.2 |
| 956.6011342593 | 1.4 |
| 956.6612615741 | 2.3 |
| 957.6280902778 | 1.6 |
| 957.6497569445 | 2.4 |
| 957.6865277778 | 2.2 |
| 958.6386458334 | 2.3 |
| 958.6399074074 | 1.3 |
| 958.6536342593 | 2.8 |
| 958.6710648148 | 2.0 |
| 958.7057060185 | 1.2 |
| 958.7825694445 | 2.2 |
| 959.6786226852 | 1.4 |
| 959.7829282408 | 1.0 |
| 960.7072916667 | 2.0 |
| 960.7206134260 | 1.0 |
| 960.8410069445 | 2.8 |
| 960.9232291667 | 1.0 |
| 961.7211689815 | 2.0 |
| 961.7711689815 | 2.4 |
| 961.8290625000 | 2.3 |
| 961.8675578704 | 1.5 |
| 961.8997569445 | 1.9 |
| 962.1747337963 | 2.4 |
| 962.7882175926 | 2.0 |
| 962.9353240741 | 1.1 |
| 963.7885185185 | 2.2 |
| 963.8156250000 | 2.8 |
| 963.8219328704 | 1.5 |
| 963.9243518519 | 2.0 |
| 964.7873148148 | 1.1 |
| 964.8193981482 | 2.7 |
| 964.9527314815 | 1.6 |
| 965.8440856482 | 1.3 |
| 965.8698495371 | 2.6 |
| 965.9791319445 | 2.6 |
| 966.0187384259 | 2.5 |
| 966.1842013889 | 2.4 |
| 966.8273726852 | 1.8 |
| 966.8706481482 | 1.7 |
| 966.9165162037 | 1.1 |
| 966.9349074074 | 1.4 |
| 966.9643750000 | 2.9 |
| 967.0640509260 | 2.5 |
| 967.8954166667 | 2.4 |
| 967.9019907408 | 2.2 |
| 967.9333796297 | 2.0 |
| 967.9528125000 | 2.4 |
| 968.0253356482 | 1.0 |
| 968.9246643519 | 2.4 |
| 968.9572337963 | 2.1 |
| 969.0778935185 | 2.1 |
| 969.0844675926 | 2.0 |
| 969.9294675926 | 2.4 |
| 969.9474884260 | 2.6 |
| 969.9482986111 | 2.7 |
| 970.0649421297 | 2.7 |
| 970.2246064815 | 1.3 |
| 970.9750578704 | 2.6 |
| 971.0319328704 | 1.0 |
| 971.0866435185 | 2.7 |
| 971.1007175926 | 2.3 |
| 971.1457638889 | 2.7 |
| 971.1666435185 | 2.9 |
| 972.0039814815 | 1.0 |
| 972.0777777778 | 2.9 |
| 972.1246527778 | 1.5 |
| 972.1901851852 | 2.2 |
| 972.9920601852 | 2.3 |
| 973.0319560185 | 1.4 |
| 973.0531365741 | 2.8 |
| 973.1096759260 | 1.9 |
| 974.0208101852 | 2.0 |
| 974.0497337963 | 2.6 |
| 974.0607986111 | 1.9 |
| 974.0897337963 | 2.8 |
| 974.1777893519 | 1.9 |
| 974.2151620371 | 1.1 |
| 974.2258333334 | 2.3 |
| 974.4635648148 | 2.3 |
| 974.5675115741 | 2.8 |
| 975.0751736111 | 2.8 |
| 975.0827430556 | 1.3 |
| 976.0773842593 | 2.9 |
| 976.1111226852 | 1.3 |
| 977.2688888889 | 1.6 |
| 977.3011111111 | 1.1 |
| 977.3653240741 | 2.8 |
| 978.1291435185 | 1.3 |
| 978.1686921297 | 1.5 |
| 978.2533217593 | 2.6 |
| 978.3579745371 | 1.9 |
| 979.1813541667 | 1.7 |
| 979.1936342593 | 2.9 |
| 979.2420601852 | 2.2 |
| 979.3159606482 | 1.9 |
| 979.3790277778 | 1.7 |
| 979.4960069445 | 2.5 |
| 979.5528819445 | 2.8 |
| 980.1982754630 | 1.3 |
| 980.3103240741 | 1.5 |
| 980.3243402778 | 1.9 |
| 980.3489351852 | 2.1 |
| 980.4404513889 | 1.0 |
| 981.2524305556 | 2.6 |
| 981.2870138889 | 1.4 |
| 982.2792592593 | 1.0 |
| 983.3748958334 | 1.7 |
| 983.4029976852 | 2.6 |
| 983.4392708334 | 2.2 |
| 983.5144791667 | 2.1 |
| 983.7693865741 | 1.9 |
| 984.3317013889 | 2.1 |
| 984.4041666667 | 2.5 |
| 985.3262615741 | 1.6 |
| 985.3355671297 | 1.3 |
| 985.3618518519 | 1.3 |
| 985.4735763889 | 2.9 |
| 985.4888657408 | 1.3 |
| 985.5166550926 | 1.6 |
| 986.3878356482 | 1.4 |
| 987.3752314815 | 2.8 |
| 987.4465277778 | 2.0 |
| 987.4652662037 | 2.6 |
| 987.5276157408 | 2.1 |
| 987.5895717593 | 1.3 |
| 987.7257523148 | 2.6 |
| 988.5016782408 | 1.6 |
| 988.5319907408 | 1.7 |
| 989.4313888889 | 1.8 |
| 989.4705092593 | 1.1 |
| 989.4792476852 | 2.9 |
| 989.5170486111 | 2.6 |
| 989.6362615741 | 1.1 |
| 989.6795717593 | 2.3 |
| 990.5028703704 | 2.3 |
| 990.5081481482 | 2.2 |
| 990.6526041667 | 2.5 |
| 991.5243518519 | 2.0 |
| 991.5408564815 | 1.3 |
| 991.5760532408 | 1.6 |
| 991.7236689815 | 1.9 |

Table C: 0-reduced time series of 1755 mixed-quality marsquakes of unspecified magnitudes, 13 Jan 2019–30 Sep 2021 (v.9 release of 1 April 2022; see statements on data). Consult the text for explanations on dates and random seismic magnitudes.